# System Identification and Model-based Robust Nonlinear Disturbance Rejection Control

Dissertation

zur

Erlangung des Grades

Doktor-Ingenieur

der

Fakultät für Maschinenbau

der Ruhr-Universität Bochum

von

Atta Oveisi

Aus Tabriz, Iran

Bochum 2021

**Dissertation eingereicht am:** 19.10.2020

**Tag der mündlichen Prüfung:** 27.04.2021

**Erstgutachter:** Prof. Dr.-Ing. Tamara Nestorović

**Zweitgutachter:** Prof. Dr.-Ing. D. Söffker

# Zusammenfassung

Der robuste Stör- oder Führungsregler war aufgrund der unbestreitbaren Wichtigkeit für die Automatisierung Gegenstand intensiver Forschung. Die moderne Regelungstheorie tendiert dazu, modelbasierte Ansätze gegenüber modelfreien zu verwenden, insbesondere wenn es um hoch-moderne Anwendungen geht. Dies liegt daran, dass diese Anwendungen ein gewisses Maß an Robustheit und Zuverlässigkeit aufweisen sollten, bevor sie als Produkt zertifiziert werden. In dieser Dissertation habe ich mich mit der Verbesserung veschiedener Aspekte der Störungsregler befasst. Das Backbone der Dissertation ist auf der systematische Modelierung dynamischer Systeme sowie der Entwicklung fortgeschritter Regelungsmethoden basiert. Dementsprechend beginnt die Dissertation mit der Untersuchung von Nichtlinearitäten in dynamischen Systemen. Hierfür wird die Polynom-Nichtlinearität als allgemeiner Ansatz zur Erfassung von (sogar nicht-) Polynom-Nichtlinearitäten sowohl in der Zeit- als auch im Frequenzbereichssystemmodelierung berücksichtigt. Durch die Analyse der mathematischen Form von Nichtlinearitäten (in geschlossenen Lösungen) wird diesen Ansatz gerechtfertigt. Insbesondere werden strukturelle Schwingungen als Benchmark-Problem in Simulationen und Experimenten verwendet. Obwohl die analytischen Ansätze (geschlossenen Lösungen) zwar mit der Physik des Problems übereinstimmen, sind sie jedoch rechenintensiv. Außerdem können sie nur für einfache Geometrien und nur unter bestimmten Annahmen von Randbedingungen hergestellt werden. Folglich hat der Autor die Systemidentifikationsansätze angewendet, d. H. Subspace- und Maximum-Likelihood-Techniken im Frequenzbereich. Dementsprechend wird die Erweiterung klassischer Subspacealgorithmen für lineare Systeme im Frequenzbereich unter Verwendung des neuartigen lokalen Polynomansatzes angegangen. Das Nebenprodukt des letzteren ermöglicht es, das Rauschen und die Nichtlinearität zu quantifizieren. Die systematische Systemidentifizierung im Frequenzbereich wird eingehend analysiert, indem das Design der Eingangsanregung, die Systemkopplung (z. B. der Kopplungseffekt der Aktuatoren), die Erkennung und Quantifizierung der Nichtlinearität, die Spektralanalyse, die nichtparametrische und parametrische Modelierung untersucht werden. Als nächstes wird (durch das genaue Model) das Problem der Störungsregelung zerlegt, nämlich werden Modelierungsunsicherheit und nicht modelierte Dynamik hoher Ordnung, Fragilität des Regler- und Beobachter-Systeme, die Nichtlinearitäten von (wie Sättigung) seperat analysiert. Für jeden Aspekt werden die verfügbaren Ansätze in der Literatur untersucht und neue Ansätze als Verbesserung für die Probleme vorgeschlagen und in Echt-Zeit Test validiert.

# Abstract

The robust disturbance rejection control in regulation and tracking problems has been the subject of intensive research due to its undeniable importance in automation. The modern control theory tends to use model-based approaches over model-free ones especially when it comes to real life high-tech applications. This is because these applications should satisfy a degree of robustness and reliability before being certified as a product which can only be achieved by basing the control algorithm on accurate models of the underlying real-life process. In this dissertation, I have looked into perfecting several aspects of the disturbance rejection control. The backbone of the dissertation is based on revisiting modeling of dynamical systems, as well as development of advanced control techniques to tackle various aspects of imperfections in disturbance rejection control, and finally perfecting the results in simulation and real-time test. Accordingly, the journey starts from looking into nonlinearities in dynamical systems and more specifically the natural way to model them in system theory. Accordingly, the polynomial nonlinearity as the general trend to capture (even) non-polynomial nonlinearities is tackled in both time- and frequency-domain system modeling. This is justified not only by looking into the physical behavior of nonlinearities in closed-form mathematical solutions but also is proven to be in agreement with test results from the real-life plant. Here, I have specifically looked into structural vibrations as this was later served as the benchmark to perform simulations and experiments. The analytical approaches although may agree the best with the physics of the problem, are indeed also computationally expensive. Needless to mention other than simple geometries and only under assumptions such as simply-supported boundary conditions, no closed-form solution can be produced in an arbitrary general case. Consequently, the author has moved towards the system identification approaches i.e., subspace and maximum likelihood techniques in frequency-domain. Accordingly, the extension of classical subspace algorithms for linear systems in frequency-domain is tackled employing the novel local polynomial approach. The byproduct of the latter makes it possible to qualify and quantify the available noise and nonlinearity is system output. The perfection of frequency-domain system identification is deeply analyzed by looking into input excitation design, system coupling e.g., coupling effect of the actuators in system output, nonlinearity detection and quantification, spectral analysis, non-parametric and parametric modeling. Next, by having an accurate model, the problem of disturbance rejection is broken down into its essence, namely, disturbance as an external stimuli on system, modeling uncertainty, and unmodeled dynamics of high-order nature, fragility of the controller and observer systems, actuator nonlinearities such as saturation, the windup problem, and the real-time implementation of all of these issues on a benchmark setup. For each aspect, the available approaches in the literature are studied and novel approaches are proposed as remedy for the persistent issues. In the course of the dissertation, I came upon intriguing practical applications where a real-time test was not an option for instance because the plant under study is not available due to costs. Consequently, having this deficiency in mind, a platform where the control engineer may implement and test arbitrary control system in a software-in-the-loop system is developed. All of the aforementioned aspects for disturbance rejection control can be accordingly simulated on this platform.

# Contents

# List of Figures

## List of Tables

# Abbreviation

| | |
|---|---|
| MRI | Magnetic resonance imaging |
| LQR/G | Linear quadratic regulator/Gaussian |
| FEM | Finite element method |
| ACLD | Active constrained layer damping |
| PFRC | Piezoelectric fiber-reinforced composite |
| SNR | Signal-to-noise ratio |
| CF | Crest factor |
| SISO | Single-input single-output |
| MIMO | Multi-input multi-output |
| FRF | Frequency response function |
| FRM | Frequency response matrix |
| SMC | Sliding mode control |
| LKF | Lyapunov-Krasovski functional |
| DoF | Degree of freedom |
| NN | Neural network |
| DSP | Digital signal processing |
| FEBPNN | Filtered-error back propagating NN |
| AVC | Active vibration control |
| SMO | Sliding mode control |
| EKF | Extended Kalman filter |
| GA | Genetic algorithm |
| MPC | Model predictive control |
| IO | Input-Output |
| DUEA | Disturbance/uncertainty estimation and attenuation |
| DOBC | Disturbance observer-based control |
| DAC | Disturbance accommodation control |
| UDE | Uncertainty and disturbance estimator |
| LFT | Linear fractional transformation |
| BLA | Best linear approximation |
| LPM | Local polynomial method |
| STD | Standard deviation |
| PEM | Predictive error method |
| DRC | Disturbance rejection control |
| UIO | Unknown input observer |
| HiL | Hardware-in-the-loop |
| SiL | Software-in-the-loop |
| MFC | Micro fiber composite |
| ESO | Extended state observer |
| SFO | Strong functional observer |
| FSO | Functional state observer |
| LMI | Linear matrix inequality |
| LME | Linear matrix equality |
| BMI | Bilinear matrix inequality |
| BRL | Bounded real lemma |
| QP | Quadratic programming |
| RWNN | Recurrent wavelet NN |
| PID | Proportional-integral-derivative controller |
| PDE | Partial differential equation |
| ODE | Ordinary differential equation |
| CBC | Control-based continuation |
| PNLSS | Polynomial nonlinear state-space |
| RPM | Reverse path method |
| LDV | Laser Doppler vibrometer |
| FDI | Fault detection and isolation |

# Introduction

## 1.1 Motivation

The cancellation/suppression of the unwanted vibrations as a benchmark is analyzed in this dissertation which represents a regulation problem in control theory. Most of the proposed methods in disturbance rejection control are validated therefore in performance on an active vibration suppression setup introduced later (see Figure 6.1). Hence, it is motivated to study disturbance-induced vibration closely and find analogies to the corresponding issues in control theory. Each chapter is concentrated on a single aspect of model-based disturbance rejection both in time- and frequency-domain while verifying the proposed solutions in the vibration control framework.

Vibrations with high amplitude due to excitations at or close to resonance frequencies of mechanical structures are one of the main reasons for mechanical malfunctions and failure. One of the first steps in the modeling of these dynamics is to describe the system's behavior in the form of some mathematical equations that are not necessarily of a linear nature [1]. Various sources of nonlinearities such as geometric nonlinearities, material nonlinearities, and disturbing nonlinear excitations are the primary sources of uncertainties that deteriorate light-weighted mechanical structures' performance. Henceforth, in order to reduce the uncertainties and also to achieve robust structural configurations, particular attention has to be paid to the modeling of these nonlinearities [2]–[5] as well as to their suppression in system output. This two-step paradigm is the backbone of the dissertation. It is important to note that the high-gain approaches where the nonlinearities are not explicitly modeled such as sliding mode controller/observer (see Chapter 9 for instance) and high-gain techniques e.g. [6] are also investigated and compared in the course of the dissertation.

In the course of this dissertation, a specific test bench is employed to look into the practical aspects of the system identification and control, e.g., actuator saturation nonlinearity especially based on state-of-the-art time-domain system identification and robust control algorithm, respectively. This structure is the funnel part of a magnetic resonance imaging (MRI) device, which is schematically shown in Figure 1. 1 and Figure 5.3. It should be noted that the second benchmark problem which is also investigated in the dissertation namely the cantilever beam shown in Figure 6.1 is not shown here for the sake of brevity.

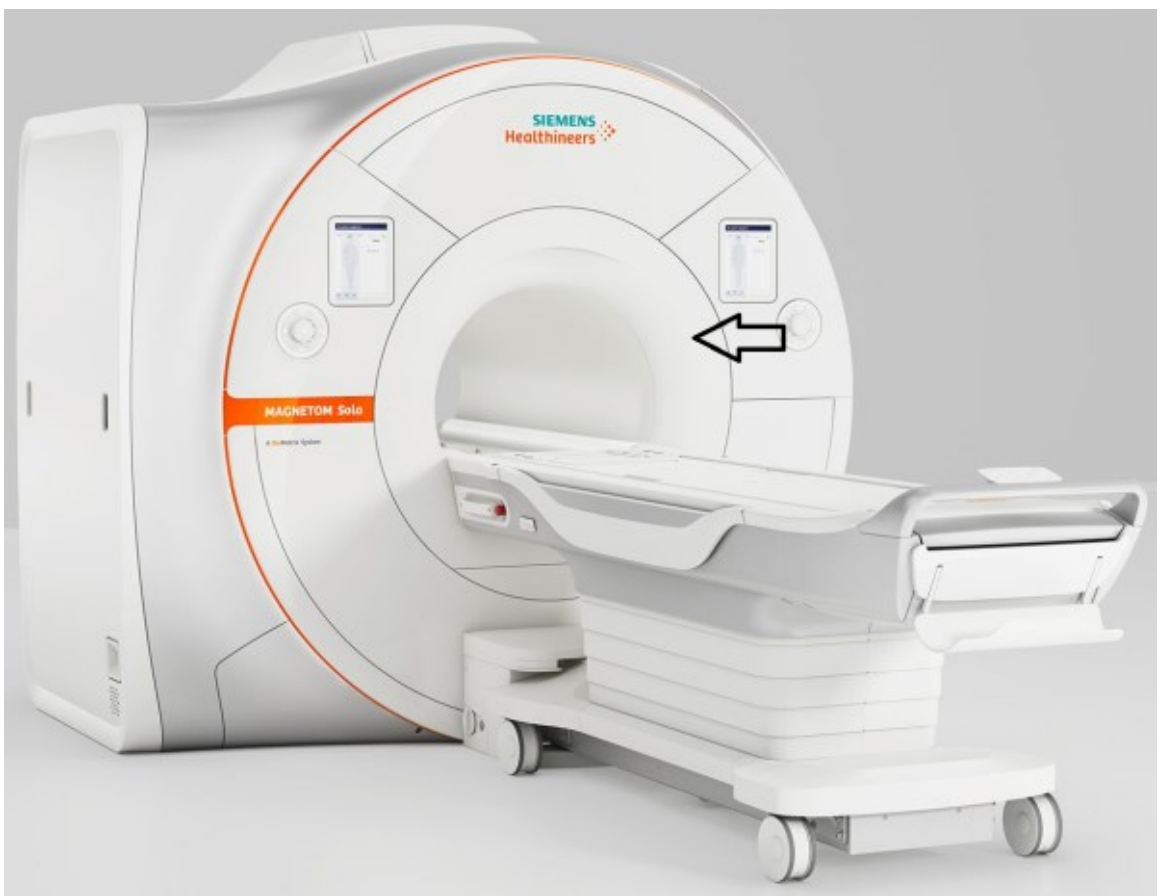

Figure 1. 1 Magnetic resonance imaging (MRI) device (a product of the Siemens company [7])

MRI is an influential method that is broadly used in medical inquiries and hospital clinics. The current research in the development of MRI technology is partially concentrated on improving the high static magnetic field strength while keeping the gradient coil's current switching at considerable levels. This re-sults in a higher scanning quality of the internal organs of the patient [8]. A key drawback in the development of these medical instruments is identified to be the acoustic noise that is propagated within the operational scanning. One of the sources of acoustic noise that is reported by Mechefske et al. is the gradient coil [9]. The gradient coil is used to propagate measured variations in the 3D magnetic field, which permits localized image slices. High acoustical noise levels can engender heightened anxiety, noise-induced hearing loss, and possibly permanent hearing impairment [10]. The gradient coil's vibration can also affect image quality and resolution [11].

Active vibration control as a practical solution can be employed to block the pathways that carry and transmit the noise in the whole MRI structure [12], [13]. The MRI shell is subjected to the excitation of the uninterrupted pulse Lorentz force of the coils. Qiu and Tani tackled the active vibration suppression due to the Lorentz force between the pulse current applied to the coil and the main magnetic field of the circular cylindrical shell that is equipped with MRI devices [14]. Nestorović et al. implemented an optimal LQ tracking controller with additional dynamics for vibration suppression of the funnel-shaped shell MRI structure [15]. They designed an additional model reference adaptive controller in order to compare the performance of the LQ controller experimentally [16]. In all of the studies above, the vibration and acoustic noise cancellation are carried out using passive methods, model-free controllers, or deterministic model-based control approaches.

In the context of control theory, one would regard these noise-induced vibrations as external disturbance which should be suppressed at the system output level. As it is motivated before, the first step towards suppressing the disturbance in model-based control theory is to generate an accurate *model* which represents the system accurately.

## 1.2 Parametric and non-parametric system modeling and identification

The unwanted vibrations of a system can only be suppressed by employing active elements (as actuators) which defines the concept of adaptive mechanical systems. Such elements are realized by the multi-domain material-based transducers which simply enable the passive structure to react to the environmental inputs [17]. The active vibration control is recognized to be effective especially in case that the additional masses should be avoided, the system is highly nonlinear, and in the case of time-varying disturbance [18]–[20].

Piezo-transducers, due to their capability of coupling strain and electric field and due to the easiness of binding with host structure, are the common candidates of active elements and therefore, a large number of researches are devoted to dynamic modeling of these structures [21]–[23]. Moita et al. used the finite element method (FEM) to study the geometrical non-linearity of thin laminated structures with embedded piezoelectric actuator/sensor elements [24]. Gao and Shen used the active damping method to control the nonlinear vibration of smart structures. They employed the incremental FE equations using the total Lagrange technique by utilizing virtual velocity incremental variation [25]. Belouettar et al. used the harmonic balance method to investigate the nonlinear vibration behavior of a smart beam under large deformation. Their formulation is based on the implementation of a simple PD controller and the use of the single-mode Galerkin method [26]. Ray and Shivakumar tackled the investigation of active constrained layer damping (ACLD) of geometrically nonlinear vibrations of laminated thin composite plates using piezoelectric fiber-reinforced composite (PFRC) materials [16]. Warminski et al. implemented different control algorithms on a nonlinear beam based on the model that they obtained by the use of multiple scale method [27]. Oveisi and Gudarzi studied the vibration control of a nonlinear plate using a piezoelectric actuator based on an adaptive robust control algorithm. They evaluated the vibration suppression performance of the closed-loop system based on various control parameters, modeling uncertainty, and initial conditions [28]. In addition, they implemented an adaptive Ziegler Nichols PID (AZNPID) on a nonlinear beam and compared the vibration attenuation performance with the previous adaptive robust method [29]. Damanpack et al. studied the dynamic response of sandwich beams impacted by blast pulses with an integrated piezoelectric sensor/actuator. They used von Karman theory and the first-order shear deformation theory to extract the nonlinear dynamics of the system [30]. The models employed in the majority of the referred research are semi-/analytical which are therefore limited to simple geometries. In contrast, system identification is a more general platform for modeling systems based on acquired data from real-time tests.

The freedom in input excitation design for system identification purposes, which may be subsequently employed in controller design and disturbance rejection, is profoundly exploited in Goodwin and Payne's pioneering work as *optimal experiments* [31]. Accordingly, other than the operational bandwidth of interest, the maximum permissible excitation amplitude on actuators, and the sampling time limitations before the task-overrun error are more than a few variables which can be optimized with regards to the input excitation signal. Several signals in terms of measurement duration, accuracy, and sensitivity to noise/nonlinearity are compared in [32] and it is concluded that the multisine signal has a minimum time-factor, i.e., minimum time per frequency for reaching a specified signal-to-noise ratio (SNR). Since the work of Schoukens et al., several attempts have been made to improve the energy content of the signal, i.e., the property of having a low crest factor (CF) for the input/output data. Consequently, CF minimization is shown to secure a desirable quality in signal processing, namely, a high SNR [33]. Then, having a specific excitation vector (for MIMO systems) in hand, the problem of non-parametric modeling e.g., frequency response function comes next.

Frequency response functions (FRFs) are proven to confer in-depth insight into the complex dynamical systems' behavior. In lightly-damped dynamical systems, modal tests using the spectral analysis technique are well-studied for parametric estimation of the single-input single-output dynamic systems [34], [35]. However, the estimation procedure of the FR matrix (FRM) for the multivariable systems is technically much more involved since it is obtained via cross-correlation techniques, which yield the input excitations to be uncorrelated. Apart from the low-frequency resolution method implemented by [36], which is generally not intended for lightly-damped smart structures, one of the major issues of random excitations is that no differentiation between noise and nonlinearity can be deduced from the results.

Contrary to $H_1$/$H_2$ functions in the spectral analysis, FRM and its covariance matrices can be calculated following [37]. Several consecutive periods and several independent realizations of the multi-reference tests should be performed, which in turn provides the means for acquiring the sample means and sample (co-)variances of the input/output spectra. The latter is realized while attenuating the stochastic noise and transients. Naturally, a model obtained via white- or black-box system identification serves as the basis for model-based *robust* control development in the aforementioned active vibration suppression problem.

## 1.3 Classical robust control in active vibration suppression

The current developments in robust control design for active structural dynamic control demand accurate uncertainty quantification [38]. For instance, in the state space modeling of nonlinear structures, $\dot{x} = \mathcal{A}x + \mathcal{B}u + \mathcal{G}$, where $\mathcal{A}$, $\mathcal{B}$, $x$, and $u$ are respectively the state matrix, input matrix, state vector, and control input vector with appropriate dimensions, one possible form of including the modeling uncertainties is the lumped norm-bounded term $\mathcal{G}$. Such uncertainty quantification may include the unmodeled dynamics of nonlinear nature. Obtaining an accurate estimation of the $H_\infty$-norm of $\mathcal{G}$ provides less conservative results in final closed-loop arrangement based on worst case analysis [19]–[21]. In practice, some approaches are proposed for calculating the $H_\infty$-norm of $\mathcal{G}$, however, with the method proposed in Chapter 5 and similar views for other geometries a good approximation is achievable [3]. Next, we look into more sophisticated issues in disturbance rejection control, namely actuator nonlinearities and predictive approaches in robust control.

## 1.4 Model predictive control in active vibration suppression

A model predictive control (MPC) algorithm is an optimization-based method in which the control inputs are obtained by solving a set of finite-horizon optimal control problems successively over the sampling instants. The optimal control formulation of the MPC explicitly takes into account model-based predictions of the controlled system trajectories. The optimal control problems' obtained solutions are concatenated in a receding horizon framework to construct the MPC feedback law. With the advances of embedded computer processors along with the enhanced optimization solvers, the MPC methods are appearing more and more in the emerging industrial applications [39]. In comparison with the other industrial control technologies, such as proportional-integral-derivative (PID) control, the distinguishing features of MPC are [40]: *i*) MPC has an inherent capability of taking into account the hard constraints on the system, like actuator saturation. *ii*) MPC algorithms offer an optimal systematic framework for handling multi-input-multi-output (MIMO) systems. The practical advantages of MPC over the model-free approach were investigated widely in the literature [41]. Model-free control is another technique to control systems by using a simplified model of the system (ultralocal representation) and consequent algebraic estimate techniques to design a simple, but operational, controller [42]. In the scope of this dissertation, it is assumed that a model of the plant and/or disturbance is/are available, and correspondingly, solutions have been proposed to extract these models (see Chapters 2, 4, and 11).

MPC can handle multi-objective performance requirements in MIMO systems, whereas the multiple PID compensators are hard to tune. Additionally, compared to the classic optimal, robust, and robust multi-objective controllers, MPC is the only candidate that efficiently incorporates actuator limitations such as saturation, the rate-of-change constraints, and even plant output amplitude constraints. For instance, as pointed out by Wai et al., in structural control of wind turbines, maximizing the power production while keeping the fatigue forces minimal are contradictive objectives on the actuation force which cannot be handled by PID, LQG, and classical $H_\infty$ controllers [43].

In the application of AVC of the cantilever beam (the benchmark of this dissertation), constraints on the system outputs such as the maximum oscillation level will imply the reduction of the overall structural stress level, which can be realized by MPC [44]. In these manipulators, the lightweight elastic links are replacing the bulky ones that are important in robotics and aerospace applications. For instance, the International Space Station has significantly contributed to this area e.g., Canadarm2 and European Robotic Arm [45]. Although both systems have been manufactured from carbon fiber composite materials, their beam-like geometries cause consequent oscillations in their arm tip. Owing to the fast dynamics of such lightly damped mechanical systems, the implementation of an MPC requires large optimization horizons. However, a broad horizon increases the MPC optimization vector's dimension, precluding the method's application under high sampling speeds. To address this issue, the control variables along the optimization horizon are parameterized by a finite set of orthonormal basis functions [40], [46]. On the other hand, asymmetric dynamical loads in practical applications resulting from periodic, repetitive disturbances are unavoidable. Consequently, the information on the repetitive pattern of the disturbance signal is incorporated into the synthesis process of MPC [47].

Following the aforementioned constraints on the flexible links, the so-called *End-effector regulation* problem, as well as constraint-time Rest-to-rest motion, is of interest in the scope of the benchmark AVC problem of Figure 4.1 [48]. It should be emphasized that in the context of this dissertation, we are dealing with local vibration suppression in contrast to the selective mode shape control scheme e.g. [49]. However, considering the cantilever geometry of the structure (see Figure 6.1 (a) Experimental rig of the control loop), the transient response mostly depends on the dominant set of transverse modes. It should be noted that the flexible manipulator benchmark is considered as a substitute model represented by a smart cantilever beam. Therefore, suppressing the vibration at the free end of the beam results in global attenuation. This justifies the use of end-effector regulation in the context of this dissertation. The constraints on the end-effector regulation problem are directly related to the global damping ratio and the closed-loop system's exponential decay rate. Despite the limiting requirement for fast sampling, which is known as the main drawback of MPC in AVC applications, it is shown the MPC is fast enough to overcome this problem even in high frequencies of disturbance signal. Finally, the application of MPC in nonlinear systems is mostly neglected in literature [50], [51].

The high computational overhead of available solvers in MPC motivates to find alternative approaches in addressing the hard constraints in robust disturbance rejection control i.e., *anti-windup* techniques.

## 1.5 Actuator nonlinearities and windup problem

The literature of the linear quadratic regulator (LQR), linear quadratic Gaussian (LQG), and robust control ($H_\infty$- and $\mu$-synthesis) has been reached to a mature level. However, the application of such benchmarks is rarely adopted in the industry [43]. Actuator saturation or bandwidth limitations in the nominal frequency range of the plant, which are ignored in these methods, are two main reasons for this matter. As pointed out in [52], violations of these constraints may cause various technical difficulties e.g., windup problem. To the modest knowledge of the author, other than MPC, only anti-windup compensation in augmentation with linear control is available as a systematical solution for actuator saturation problems [53], [54].

In the context of smart structures, it may often occur that the difference between the designed controller's output signal and the actual implemented control signal on the system is nonzero. Other than the control system's fragility due to the finite-word length in the digital system and round-off errors in binary arithmetics [55], which are mostly permanent, two foremost temporary sources of such imperfections are actuation constraints and control law substitutions. Control law substitution refers to switching control laws where no smooth transition is available for instance in the case of multi-modal controllers.

The calamitous influence of substitution appears when a multi-mode switching controller is responsible for several neighborhoods around several distinct linear operating points, and repetitive changes between the modes are needed [56]. The offline behavior of switching (where each control mode is unaware of the others), while the controller is in an online loop, introduces overshoot and eventually windup. Although a smooth transition between the controller modes may compensate for the mismatch control law, detailed mathematical treatment of this problem is not discussed here in the case of substitution. This is justified by considering the application of smart structures in active vibration control, where the system model is mostly linearized around a single operating point. However, the actuator nonlinearities such as saturation and hysteresis may play critical roles in driving the system states to non-ideal trajectories. As a constraint on the amplitude of the control signal, saturation is defined by a particular limit that is eligible to be applied to the active elements. This limit includes the output voltage and electrical current range of a piezo-amplifier, the maximum achievable displacement of a shaker's baffle, and the depolarization voltage of the piezo-actuator patches.

A good deal of control systems in vibrating smart structures use an oversimplified linear model neglecting these nonlinearities. In addition, it is well-established that controllers with sluggish or unstable dynamics eventually suffer from the windup. A bumpless transition in such a case is realizable by compensating the controller input, states, and outputs by means of various anti-windup strategies. The state-of-the-art for anti-windup compensators is a two-step paradigm where the nominal controller is synthesized prior to the development of the compensator which accounts for actuator nonlinearities. Here, compensation refers to the manipulation of the nominal controller's inputs, states, and outputs, correspondingly. The needed manipulation signals are generated by the (static) dynamical functions which are referred to as anti-windup compensators. In this regards, a standard approach is intentionally employed for the nominal control design so that the main goal, namely unification and comparison of the second step (refers to the step after the nominal controller design while neglecting the actuator nonlinearities) in the decoupled *anti-windup compensator technology* is not overshadowed by the nominal controller. A typical windup scenario in active vibration control appears when the observer-based controller experiences actuator saturation in which the integrated linear state observer is unaware of the actuator nonlinearity chopping off the control signal. A windup in the linear controller's output may happen in this case leading to diverging system output which is catastrophic.

In comparison to MPC, in which the hard constraints are included in the online optimization of the control synthesis, in the anti-windup framework, the system should be subjected to actuation nonlinearities in real-time so that an additional compensator becomes active [57]. Moreover, updating the transducers to include a greater dynamic range decreases the signal-to-noise ratio (SNR), not mentioning the physical size limitations as well as the cost of new transducers [58]. Consequently, the MPC is known to be the control strategy that can include various explicit process constraints while an anti-windup compensator can serve as an ad-hoc solution for practical problems [59]–[61].

In the aforementioned literature, it is assumed that the implementation of the controller can perfectly be realized. However, in real-life realizations, this is often not the case. Accordingly, the practical implementation of a control system is investigated considering the implementation imperfections known as the *fragility* of controller/state-observer systems.

## 1.6 The fragility of the controller and sliding mode controller

The fragility of an observer-based control system as a practical issue should be investigated by evaluating the sensitivity of the controller and observer structure regarding the control feedback and observer gain, respectively [62]. The uncertainty in controller gain as a key design limitation is inspected by [63]. They extended the classical robust control design procedure, which is unable to handle the control gain variations, to address the robust resilient static output feedback regulator for systems with controller gain uncertainty. They considered structured uncertainty only in the state matrix of the plant. [64] proposed a non-fragile observer design technique by use of multiple Lyapunov function methods and delta-operator theory for systems without plant uncertainties and disturbance. Du et al. tackled the application of non-fragile $H_\infty$ robust controller on an uncertain four degree of freedom (DoF) building model without input uncertainties by appropriate use of bounded real lemma (BRL) [65]. Yazici et al. proposed a robust delay-dependent controller by using the Lyapunov–Krasovskii functional (LKF) and BRL in order to reach a minimization algorithm that gives a sub-optimal solution for a deterministic linear system [66]. Ramakrishan and Ray presented a delay-dependent non-fragile $H_\infty$ controller for a nonlinear system with time-varying delay using LKF [67].

Although the controller/observer's sensitivity to the gain variations is an essential issue for non-fragile control systems, model uncertainties remain the key parameter in robust control analysis, especially because the obtained control system is developed based on the modeling uncertainties. This is referred to as uncertainty propagation [68], [69]. However, in most of the previous studies [70], [71], the main emphasis was placed on the control gain variations in either additive/multiplicative lumped uncertainty or the matched uncertainty. For instance, Yang and Wang presented their non-fragile $H_\infty$ observer-based controller for LTI systems without any source of uncertainties in system model and the results then lead to symmetric positive-definite solutions of algebraic Riccati inequalities [72], [73]; Liu et al. guaranteed the $H_\infty$ performance of the non-fragility of the observer-based sliding mode control (SMC) method for delayed systems with uncertainty only in the state matrix using Lyapunov method [74]; Pourgholi and Majd limited their study to the fragility of the PI-filtered error adaptive observer in Lipschitz systems that are linear with respect to the unknown parameters which is in contrast with the unmodeled nonlinearities in system dynamics such as geometrical nonlinearities in structural mechanics [75]. Especially in structural vibration, where the system contains infinite number of eigenvalues, the reduced order model should be able to handle the unmodeled dynamics since it is impossible to perfectly include all system mode shapes in the control design procedure.

While most of the cited researches are dealing with $H_\infty$-performance of the control system, the $H_2$-normed performance of non-fragile state/output feedback system, where the optimality of control system w.r.t. energy consumption is aimed, is mostly neglected. The advantage of a multi-objective $H_2/H_\infty$ the nonfragile scheme as shown in Chapter 8 lies in the optimality of the closed-loop system, especially for frequency range where the plant model could be less accurate or invalid. A principally practical issue of such methods based on solely $H_\infty$-performance is the vague specification of the worst-case bound of the transfer function from disturbance to the output performance index. It should be pointed out that Famularo et al. considered the LQ/$H_2$ with multiplicative structured uncertainty only for state matrix and control gain in non-fragile state-feedback controller [76]. Changwei and Jie proposed the non-fragile $H_2/H_\infty$ for the regulation problem in state feedback systems without DRC [77], and similarly, Xin et al. reached a non-fragile frame-work for LTI systems with coprime factor uncertainties [78]. The latter two references are developed very similarly to the robust resilient approach proposed in [79].

The sliding mode controller (SMC) is regarded as a robust technique due to the rejection of external disturbance and insensitivity with respect to bounded perturbation [80]. In addition, SMC is proven to have a fast reaction with low control order [81]. An important issue of the SM technique is the sensitivity of the closed-loop system to the perturbation in the reaching initial period, during which the dynamic trajectory of the system moves toward the sliding surface [82]. An alternative approach known as integral SMC (ISMC) is contributed with aiming at solving this issue throughout the elimination of the reaching phase [83]. However, even ISMC method cannot generally solve the matching problem and therefore it is used in combination with other robust techniques. Recently, some researchers donated their attention toward the effects of these mismatch uncertainties in the robust performance of ISMC. Cao and Xu limited their control development to the case of mismatch uncertainty only in the state matrix [84]. Choi proposed a new ISMC in an LMI form for the systems with mismatch uncertainty in both state and input matrices. The controller is proven to be able to guaranty the asymptotic stability as well as satisfying the $\alpha$-stability constraint [85]. Plestana et al. presented a new methodology for adaptive SMC without the requirement of the prior knowledge of the numerical values for the bounds of the uncertainties [86]. Their algorithms are limited to single-input single-output (SISO) systems with a limited order adaptive law. Ha et al., combined the fuzzy logic with SMC in order to obtain a robust disturbance rejection controller on mismatch systems. A fuzzy system is introduced to address the chattering problem of the SMC [87]. Oveisi and Gudarzi used an ideal controller based on SMC and introduced a fuzzy system to mimic this controller on the real plant. The robust stability is guaranteed by designing the controller based on compensation of the difference between the fuzzy controller and the ideal controller by use of Lyapunov theory [28].

Most of the mentioned researches have limited their case study to some numerical examples. However, Wai et al. adopted an AFISMC to control the position of an electrical servo drive [88]. Gholami and Markazi introduced a new AFSM observer for a class of nonlinear MIMO systems and implemented their observer on a modular and reconfigurable robot (MRR) system [89].

The scope of the model-based techniques in almost all of the aforementioned literature is limited to models which have nominal time-invariant nature. However, in systems where the disturbance has unknown bandwidth an adaptive model which can be valid extensively over the excited dynamics of the real system is necessary. Artificial intelligence-based (AI-based) techniques such as *neural-network* models in combination with classical system identification-based reduced-order models are dedicated to addressing such scenarios.

## 1.7 Artificial neural network in control theory

Neural networks (NN) are structures of massively interconnected cells which as a whole can model complex dynamical systems. These structures function by imitating the composition and capacity of the human brain. Each of the simplistic processing unit cells receives weighted input signals, passes the weighted summation of those signals through a nonlinear operator, and emits an output to be transmitted to the next level processing elements along the outgoing pathways. NN's application as a controller (neuro-controllers) in active vibration control has been studied in the last two decades as the practical implementation of disturbance rejection control. However, only a handful of real-time implementations of nonlinear neuro-active vibration control systems is reported in the literature [90]. For instance, Li et al. proposed a new filtered-error back-propagation NN (FEBPNN) algorithm for vibration control of flexible piezolaminated structures [91]. They implemented their FEBPNN algorithm on a digital signal processor (DSP), and it was shown that the proposed active vibration/noise control method (AVC/ANC) is effective for the system in the presence of modeling nonlinearities.

The conventional controllers are often synthesized to deliver only a specific control performance. In contrast, the neuro-controllers adapt in an online fashion to satisfy the time-varying performance objectives in a supervised/unsupervised form depending on the available computational power. This distinguishing factor enables the NN to detect and learn extremely involved and nonlinear mappings [92]. Moreover, various sources of nonlinearities in elastic light-weight mechanical structures make it impossible to use simplistic feedforward neural network, excluding the tapped delay [93]. In other words, especially in any ANC scheme and AVC of non-collocated input/output (IO) configuration, where the measurement delay is the natural feature of the plant, an adaptive mechanism should be employed in synthesizing the control input signal associated with nonlinear output measurements. Two practical realizations of these nonlinearities in smart structures are the geometrical nonlinearities due to high vibration amplitudes where the linear models are not valid anymore [5] and the non-collocated sensor/actuators configurations where an inherent delay is an inevitable feature of the IO relation [94].

The nonlinear model-free robust control schemes based on the upper-bounds of the norm of the disturbance signals and matched-/mismatched-uncertainties such as high-gain, variable-structure, and fuzzy methods can be used in combination with NN to address nonlinearities in system dynamics [6]. For instance, Jnifene and Andrews proposed a combination of a fuzzy logic controller and neural networks to regulate the end-effector vibration in a flexible smart beam positioned on a two DoF platform [95]. He et al. employed a neural network for modeling the dynamics of a flexible robot manipulator subjected to input deadzone. They have used radial basis function NN to capture the deadzone and designed a high-gain observer-based NN controller [96]. Moreover, it is reported that one of the major deficiencies of standard control systems based on linear time-invariant nominal models is the evolution of plant dynamics w.r.t. time and the actuator windup problem both of which add to the nonlinearity of the system [97]. Accordingly, Li et al. introduced a genetic algorithm (GA) based back-propagation neural network suboptimal controller to address the vibration attenuation of a nine DoF modular robot [91].

## 1.8 Disturbance observer-based control

Research in the area of state observation and disturbance estimation in linear multivariable systems has reached a relatively mature level, and various existence and sufficient conditions have been proposed in the last few decades [98], [99]. In the case of measurable disturbances, feedforward, and feedback/feedforward strategies have been used to prevent performance degradation through optimization of some user-defined performance indices. This concept is also referred to as state reconstruction, as in [100], where unknown exogenous input signals are the erroneous system data source. For such systems, a simple and well-established approach is the sliding mode observer (SMO), which allows for robust reconstruction of the states under unknown inputs [101]–[103]. Considering the disturbance as a bounded unknown mismatched exogenous input instead of opting for the classical stochastic approach based on a statistical distribution of the plant's signal seems to be more practical in control applications and system theory. The classical extended Kalman filter (EKF) is not referred to here because the closed-loop stability analysis is not easy to follow. The observer error propagation in the control system may lead to instability unless the stability is proven by means of some Lyapunov function. However, in the literature, the stability analysis is mostly done for the deterministic part of the plant. The mismatched unknown inputs are those signals that penetrate the system states through channels that are independent of the control input signals. Additionally, due to the lack of information about the external disturbances, intuitively high gain controllers are designed in the early development of DRC. It is well-known however that high gain controllers may lead to the actuator windup problem and peaking phenomenon [104].

To a large extent, although categorized under a separate control synthesis approach, disturbances (and modeling uncertainties) may be handled by classical (robust) techniques. However, often conflicting performance and stability constraints on the closed-loop system require more advanced alternative approaches in estimating and then rejecting the disturbance. Recently, linear and nonlinear disturbance and uncertainty estimation and attenuation (LDUEA/NDUEA) methods have attracted considerable interest. Disturbance-observer-based control (DOBC) method as one candidate of DUEA, in general, delivers high robustness with respect to modeling uncertainties and matched/mismatched disturbances without performance loss [105]. Additionally, more rigorous stability analysis and synthesis can be obtained through the use of DOBC [106] compared to the methods of extended state observer (ESO), disturbance accommodation control (DAC) [107], and uncertainty and disturbance estimator (UDE) [108]. The unknown input excitations acting on the system also known as the lumped disturbance represent a generalized form of external stimuli that also takes into account the unmodeled dynamics of high-order nature and parameter variations [109], [110].

## Contributions of this dissertation

This thesis reports the findings achieved within this doctoral work. Large pieces of this dissertation are reported in the below articles:

Nestorović, T., Hassw, K., & Oveisi, A. (2021). Software-in-the-loop optimization of actuator and sensor placement for a smart piezoelectric funnel-shaped inlet of a magnetic resonance imaging tomography. *Mechanical Systems and Signal Processing*, *147*, 107097. https://doi.org/10.1016/j.ymssp.2020.107097

Oveisi, A., Aldeen, M., & Nestorović, T. (2017). Disturbance rejection control based on state-reconstruction and persistence disturbance estimation. *Journal of the Franklin Institute*, *354*(18), 8015–8037. https://doi.org/10.1016/j.jfranklin.2017.08.049

Oveisi, A., Hosseini-Pishrobat, M., Nestorović, T., & Keighobadi, J. (2018). Observer-based repetitive model predictive control in active vibration suppression. *Structural Control and Health Monitoring*, 25(5), 1–23. https://doi.org/10.1002/stc.2149

Oveisi, A., Jeronimo, M. B., & Nestorović, T. (2018). Nonlinear observer-based recurrent wavelet neuro-controller in disturbance rejection control of flexible structures. *Engineering Applications of Artificial Intelligence*, 69, 50–64. https://doi.org/https://doi.org/10.1016/j.engappai.2017.12.009

## Technical organization of the dissertation

The dissertation is organized as follows:

**Chapter 2** proposes a new analytical formulation for predicting the nonlinear dynamic behavior of composite piezolaminated plates for simply supported boundary conditions. For this purpose, the equation of motion of the coupled electromechanical system is derived considering the geometrical nonlinearities (strain-displacement relationship of von Karman type) and satisfying the appropriate electrostatic piezo-sensor/-actuator boundary conditions. Hamilton's principle is used to extract the strong form of the equation of motion in the form of an NPDE. Galerkin method and orthogonality of shape functions for simply supported boundary conditions in the spatial domain are employed and then the obtained nonlinear system of ODEs of motion is solved numerically. The vibration suppression problem is investigated by the implementation of the model-free control law. The presented method as a semi-analytical approach is much faster than FE package ABAQUS and, therefore, it can be used as a benchmark in designing smart plates under large vibration excitation amplitudes (where the higher-order nonlinear strain terms are not negligible). High order strain terms play an important role in the dynamic response of the closed-loop system, and the presented model describes the system dynamics more accurately compared to the several simplifications used in linear models. Two important observations compared to classical linear assumptions are: 1) The sensitivity of the transient behavior of the structure to the magnitude of the excitation in large vibration amplitudes, i.e., there exists a nonlinear correspondence between the excitation and vibration amplitudes and frequencies. 2) Identifying the realistic bounds of proportional feedback gain that may result in structural instability.

The content, figures, and tables in this chapter are mainly based on the publications in peer-reviewed journal paper:

Oveisi, A., & Nestorović, T. (2017). Transient response of an active nonlinear sandwich piezolaminated plate. Communications in Nonlinear Science and Numerical Simulation, 45, 158–175. https://doi.org/10.1016/j.cnsns.2016.09.012

**Chapter 3** presents the systematic modeling of a piezolaminated sandwich plate considering the geometrical nonlinearity of the von Karman type. Then, for the simply supported case, a semi-analytical solution for the nonlinear PDE is proposed by using the harmonic balance method along with the single-mode Galerkin approach. The algebraic equation obtained is employed to find the amplitude equation with an analogy to the modal analysis, where the nonlinear modal parameters are equated analytically. The sensitivity of the response of the closed-loop system based on a proportional derivative control law is investigated for different parameters for both the linear and nonlinear cases.

The content, figures, and tables in this chapter are mainly based on the publications in peer-reviewed journal paper:

Oveisi, A., Nestorović, T., & Nguyen, N. L. (2017). Semi-analytical modeling and vibration control of a geometrically nonlinear plate. International Journal of Structural Stability and Dynamics, 17(4), 1–12. https://doi.org/10.1142/S0219455417710031

**Chapter 4** investigates the impact of various input excitation scenarios on two different MIMO linear non-parametric modeling schemes in the frequency domain. It is intended to provide an insight into the optimal experiment design that not only provides the best linear approximation (BLA) of the FRFs but also delivers the means for assessing the variance of the estimations. Finding the mathematical representations of the variances in terms of the estimation coherence and noise/nonlinearity contributions are of critical importance for the frequency-domain system identification where the objective function needs to be weighted in the parametrization step. The input excitation signal design is tackled in two cases, i.e., multiple single-reference experiments based on the zero-mean Gaussian and the colored noise signal, the random-phase multisine, the Schroeder multisine, and minimized crest factor multisine; and multi-reference experiments based on the Hadamard matrix, and the so-called orthogonal multisine approach, which additionally examines the coupling between the input channels. The time-domain data from both cases are taken into the classical H1 spectral analysis as well as the robust local polynomial method (LPM) to extract the BLAs. The results are applied for data-driven modeling of a flexible beam as a model of a flexible robotic arm.

Black-box system identification is tackled next subjected to the modeling uncertainties that are propagated from the non-parametric model of the system in the time/frequency domain. Unlike classical H1/H2 spectral analysis, in the recent robust LPM, the modeling variances are separated to noise contribution and nonlinear contribution while suppressing the transient noise. On the other hand, without an appropriate weighting on the objective function in the system identification methods, the acquired model is subjected to bias. Consequently, in this chapter, the weighted regression problem in subspace frequency-domain system identification is revisited in order to have an unbiased estimate of the frequency response matrix of a flexible manipulator as a multi-input multi-output lightly-damped system. Although the unbiased parametric model representing the best linear approximation of the system in this combination is a reliable framework for the control design, it is limited for a specific SNR ratio and standard deviation (STD) of the involved input excitations. As a result, in this chapter, an additional step is carried out to investigate the sensitivity of the identified model w.r.t. SNR/STD in order to provide an uncertainty interval for robust control design.

The content, figures, and tables in this chapter are mainly based on the publications in peer-reviewed conference papers:

Oveisi, A., Montazeri, A., Anderson, A., & Nestorović, T. (2018). Optimal input excitation design for non-parametric uncertainty quantification of multi-input multi-output systems. In 18th IFAC Symposium on System Identification (SYSID 2018). Stockholm, Sweden. https://doi.org/accepted contribution

Oveisi, A., Montazeri, A., & Nestorović, T. (2018). Frequency-domain subspace identification of multivariable dynamical systems for robust control design. In 18th IFAC Symposium on System Identification (SYSID 2018). Stockholm, Sweden: IFAC. https://doi.org/accepted contribution

**Chapter 5** proposes a robust control technique based on $\mu$-synthesis in order to investigate the vibration control of a funnel-shaped structure that is used as the inlet of a magnetic resonance imaging (MRI) device. In order to address this issue and as a part of noise cancellation study in MRI devices, distributed piezo-transducers are bounded on the top surface of the funnel as functional sensor/actuator modules. Then, a reduced-order linear time-invariant model of the piezolaminated structure in state-space representation is estimated by means of the predictive error minimization (PEM) algorithm as a subspace identification method based on the trust-region-reflective technique. The reduced-order model is expanded by the introduction of appropriate frequency-dependent weighting functions that address the unmodeled dynamics and the augmented multiplicative modeling uncertainties of the system. Then, the standard D-K iteration algorithm as an output-feedback control method is used based on the nominal model with the subordinate uncertainty elements from the previous step. Finally, the proposed control system was implemented experimentally on the real structure to evaluate the robust vibration attenuation performance of the closed-loop system.

The content, figures, and tables in this chapter are mainly based on the publications in peer-reviewed journal papers:

Nestorović, T., Hassw, K., & Oveisi, A. (2021). Software-in-the-loop optimization of actuator and sensor placement for a smart piezoelectric funnel-shaped inlet of a magnetic resonance imaging tomograph. Mechanical Systems and Signal Processing, 147, 107097. https://doi.org/10.1016/j.ymssp.2020.107097

Oveisi, A., & Nestorović, T. (2016a). Mu-synthesis based active robust vibration control of an MRI inlet. Facta Universitatis, Series: Mechanical Engineering, 14(1).

**Chapter 6** represents the development of an observer-based feedback/feedforward model predictive control algorithm. For this purpose, the feedback control design process is formulated in the framework of disturbance rejection control, and a repetitive MPC is adapted to guarantee the robust asymptotic stability of the closed-loop system. To this end, a recursive least square algorithm is engaged for online estimation of disturbance signal, and the estimated disturbance is feedforwarded through the control channels. Next, the internal model principle is adjusted to account for the disturbance dynamics in the active vibration control method. A broad frequency range is considered for unknown energy bounded disturbances as an independent input to the nominal plant by means of the shaker-test in the experimental modal analysis. For the sake of relieving the computational burden of online optimization within the broad prediction horizons, a set of orthonormal Laguerre functions is utilized. Additionally, a nominal frequency range is defined to identify a reduced-order plant for control development purposes. A series of comprehensive experimental analyses are carried out to address the robustness of the AVC system with respect to unmodeled dynamics, parametric uncertainties of modeling, and external noise both in time- and frequency-domain. The dynamical properties of AVC systems subjected to external mechanical disturbances pose additional challenges such as the spillover effect of the actuation authorities and saturation of the active elements both of which are discussed carefully.

The content, figures, and tables in this chapter are mainly based on the publications in peer-reviewed journal paper:

Oveisi, A., Hosseini-Pishrobat, M., Nestorović, T., & Keighobadi, J. (2018). Observer-based repetitive model predictive control in active vibration suppression. Structural Control and Health Monitoring, 25(5), 1–23. https://doi.org/10.1002/stc.2149

**Chapter 7** studies the windup problem in active vibration control systematically. Instead of evaluating the performance of several anti-windup compensators implemented on independent abstract simulation problems, a unified benchmark setup in AVC (active-damping experiment) is used. The investigated anti-windup schemes (analysis and synthesis) are adapted to the disturbance rejection control. Large attention is given to capture the similarities and differences of the methods in dealing with the windup event in a practical context. Therefore, instead of categorizing the methods into static and non-static methods or model recovery and direct linear anti-windup schemes, a logical route is followed to highlight the significance of each method. The mathematical interpretations of the methods are provided for the vibration engineer while delivering forthright implementation algorithms for AVC. The tackled methods are unified on a state-space model obtained from the frequency-domain subspace system identification approach. Practical issues that may arise for each technique are mentioned, and detained guidelines are provided for tuning each algorithm. Finally, in order to compare the compensated system's performance, comprehensive time-domain studies are carried out by separating the transient response of the compensated systems into three modes: linear mode, where the actuation nonlinearity is inactive; the nonlinear mode, where the windup event is in progress, and finally, the output mismatch rejection mode, where the windup incident is over, but performance degradation is still present.

The content, figures, and tables in this chapter are mainly based on the publications in peer-reviewed journal paper:

Oveisi, A., & Nestorović, T. (2019). Vibration control subjected to windup problem: An applied view on analysis and synthesis with convex formulation. Control Engineering Practice, 82. https://doi.org/10.1016/j.conengprac.2018.09.020

**Chapter 8** introduces a robust non-fragile observer-based controller for a linear time-invariant system with structured uncertainty. The $H_\infty$ robust stability of the closed-loop system is guaranteed by the use of the Lyapunov theorem in the presence of undesirable disturbance. For the sake of addressing the fragility problem, independent sets of time-dependent gain-uncertainties are assumed to be existing for the controller and the observer elements. In order to satisfy the arbitrary $H_2$-normed constraints for the control system and to enable automatic determination of the optimal $H_2/H_\infty$ bound of the performance functions in disturbance rejection control, additional necessary and sufficient conditions are presented in a linear matrix equality/inequality (LME/LMI) framework. The $H_2/H_\infty$ observer-based controller is then transformed into an optimization problem of a coupled set of LMIs/LME that can be solved iteratively by the use of numerical software such as Scilab. Finally, concerning the evaluation of the performance of the controller, the control system is implemented in real-time on a mechanical system aiming at vibration suppression. The nominal mathematical reduced-order model of the beam with piezo-actuators is used to design the proposed controller and then the control system is implemented experimentally on the full-order real-time system. The results show that the closed-loop system has a robust performance in rejecting the disturbance in the presence of the structured uncertainty and the presence of the unmodeled dynamics.

The content, figures, and tables in this chapter are mainly based on the publications in peer-reviewed journal paper:

Oveisi, A., & Nestorović, T. (2018). Robust nonfragile observer-based H2/H∞ controller. Journal of Vibration and Control, 24(4), 1–17. https://doi.org/https://doi.org/10.1177/1077546316651548

**Chapter 9** proposes a new observer-based adaptive fuzzy integral sliding mode controller based on the Lyapunov stability theorem. The plant under study is subjected to a square-integrable disturbance and is assumed to have mismatch uncertainties both in the state- and input matrices. In addition, a norm-bounded time-varying term is introduced to address the possible existence of un-modeled/nonlinear dynamics. Based on the classical sliding mode controller, the equivalent control effort is obtained to satisfy the sufficiency requirement of the sliding mode controller and then the control law is modified to guarantee the reachability of the system trajectory to the sliding manifold. The sliding surface is compensated based on the observed states in the form of linear matrix inequality. In order to relax the norm-bounded constraints on the control law and solve the chattering problem of the sliding mode controller, a fuzzy logic inference mechanism is combined with the controller. An adaptive law is then introduced to tune the parameters of the fuzzy system online. Finally, by aiming at evaluating the validity of the controller and the robust performance of the closed-loop system, the proposed regulator is implemented on a real-time mechanical vibrating system.

The content, figures, and tables in this chapter are mainly based on the publications in peer-reviewed journal paper:

Oveisi, A., & Nestorović, T. (2016c). Robust observer-based adaptive fuzzy sliding mode controller. Mechanical Systems and Signal Processing, 76–77. https://doi.org/10.1016/j.ymssp.2016.01.015

**Chapter 10** proposes a control scheme in terms of an observer-based recurrent wavelet neural network (RWNN) for nonlinear MIMO systems with time-varying matched/mismatched uncertainties. The proposed control system employs an adaptive neural network bound estimator aiming at reducing the conservatism associated with the standard worst-case-based algorithms. In other words, unlike the robust controllers where the uncertainties are dealt with based on small gain theorem, here we identify a NN on the mismatch between the model and the real system's output. Moreover, in order to generalize the subjected class of systems, the echo-state feature of adaptive RWNN is used to contribute to the performance of nonminimum phase systems. Accordingly, in the context of flexible smart structures with non-collocated sensor/actuator configuration, delayed feedback is added in the network which brings about a better model match with the real data from the system. Then, the proposed emulator NN adaptively trains to follow an ideal state-feedback controller from the immediate history of the system's response. As a result, even for systems with an unknown Lipschitz constant of lumped uncertainty, the neuro-controller can be trained online to compensate with an additional revision of the control law following some Lyapunov-based adaptive stabilizing rules. Additionally, the current investigation is proposed as an alternative to the hot topic of nonlinear system identification-based control synthesis where the exact structure of the nonlinearity is required.

The content, figures, and tables in this chapter are mainly based on the publications in peer-reviewed journal paper:

Oveisi, A., Jeronimo, M. B., & Nestorović, T. (2018). Nonlinear observer-based recurrent wavelet neuro-controller in disturbance rejection control of flexible structures. Engineering Applications of Artificial Intelligence, 69, 50–64. https://doi.org/https://doi.org/10.1016/j.engappai.2017.12.009

**Chapter 11** revisits the DRC based on unknown input observation (UIO), and DOBC methods for a class of MIMO systems with mismatch disturbance conditions. In both of these methods, the estimated disturbance is considered to be in the feedback channel. The disturbance term could represent either unknown mismatched signals penetrating the states, or unknown dynamics not captured in the modeling process, or physical parameter variations not accounted for in the mathematical model of the plant. Unlike the high-gain approaches and variable structure methods, a systematic synthesis of the state/disturbance observer-based controller is carried out. For this purpose, first, using a series of singular value decompositions, the linearized plant is transformed into disturbance-free and disturbance-dependent subsystems. Then, functional state reconstruction based on the generalized detectability concept is proposed for the disturbance-free part. Then, a DRC based on quadratic stability theorem is employed to guarantee the performance of the closed-loop system. An important contribution offered in this chapter is the independence of the estimated disturbance from the control input which seems to be missing in the literature for disturbance decoupling problems. In the second method, DOBC is reconsidered to achieve a high level of robustness against modeling uncertainties and matched/mismatched disturbances, while at the same time retaining performance. Accordingly, unlike the first method, DRC, full information state observation is developed independently of the disturbance estimation. An advantage of such a combination is that disturbance estimation does not involve output derivatives. Finally, the case of systems with matched disturbances is presented as a corollary of the main results.

The content, figures, and tables in this chapter are mainly based on the publications in peer-reviewed journal paper:

Oveisi, A., Aldeen, M., & Nestorović, T. (2017). Disturbance rejection control based on state-reconstruction and persistence disturbance estimation. Journal of the Franklin Institute, 354(18), 8015–8037. https://doi.org/10.1016/j.jfranklin.2017.08.049

**Chapter 12** introduces a new framework for running the FE packages inside an online Loop together with MATLAB. Contrary to the Hardware-in-the-Loop techniques (HiL) used widely in all of the previous chapters, in the proposed Software-in-the-Loop framework (SiL), the FE package represents a simulation platform replicating the real system which can be out of access due to several strategic reasons, e.g., costs and accessibility. Practically, SiL for sophisticated structural design and multi-physical simulations provides a platform for preliminary tests before prototyping and mass production. This feature may reduce the new product's costs significantly and may add several flexibilities in implementing different instruments with the goal of shortlisting the most cost-effective ones before moving to real-time experiments for the civil and mechanical systems. The proposed SiL interconnection is not limited to ABAQUS as long as the host FE package is capable of executing user-defined commands in FORTRAN language. The focal point of this chapter is on using the compiled FORTRAN subroutine as a messenger between ABAQUS/CAE kernel and MATLAB Engine. In order to show the generality of the proposed scheme, the limitations of the available SiL schemes in the literature are addressed in this chapter. Additionally, all technical details for establishing the connection between FEM and MATLAB are provided for the interested reader.

The content, figures, and tables in this chapter are mainly based on the publications in peer-reviewed journal paper:

Oveisi, A., Sukhairi, T. A., & Nestorović, T. (2017). Finite element-based software-in-the-loop for offline post-processing and real-time simulation. Structural Engineering and Mechanics, An International Journal.

# 2 Analytical approach for nonlinear modeling in time-domain

Following the structure of the dissertation described in the previous chapter, in this chapter, the dynamic modeling and control of a piezolaminated plate with geometrical nonlinearities are investigated using a semi-analytical approach. In the grand scheme of the dissertation, this chapter serves as the first step towards systematic modeling of nonlinear vibrating structures (here: geometrical nonlinearity). In this regard, the chapter aims at the proper formulation of the nonlinear strain terms in the equation of motion which is shown to be a nonlinear partial differential equation (NPDE). Naturally, these nonlinear terms are propagated in the solution (regardless of time-/frequency-domain approach) into the resulting coupled ordinary differential equations (ODEs). This relation is intentionally expressed in an explicit manner to point out how such an active nonlinear system would come into an eventual state-space model that may be used in output feedback control design. Although the core aspect of the controller development is based on the (nonlinear) model-based techniques, in this chapter the main focus is rather lighted on the modeling and the controller is limited to a *linear* model-free technique. The intention was to separate the focus of modeling from the model-based control method. As a side advantage, an opportunity revealed itself in that the performance of such a closed-loop nonlinear system is evaluated.

## 2.1 Nonlinear modeling and control: a short review

The unwanted vibrations with high amplitudes due to excitations at or close to resonance frequencies of mechanical structures are one of the main reasons for mechanical malfunctions and failure. One of the first steps in the modeling procedure is to describe the dynamical behavior of the system in the form of mathematical equations which are not necessarily of a linear nature [1]. Various sources of nonlinearities such as geometric nonlinearities, material nonlinearities, and nonlinear disturbing excitations are the primary sources of uncertainties that deteriorate the performance of mechanical structures. Hence, in order to reduce the uncertainties and in order to achieve robust structural configurations, particular attention has to be paid to the modeling of these nonlinearities [2]–[4].

To put it in a nutshell, the following steps are carried out in this chapter. First, Hamilton's principle is used to get the strong form of the nonlinear dynamical equation with the reflection of the second-order strain terms by utilizing the von Karman strain-displacement correlation in form of an NPDE. In the grand scheme of the dissertation, this serves as the mathematical form of the nonlinearities which may be used later in the controller synthesis process. Then, by introducing the proper and applicable feedback terms, the dynamic behavior of the open-loop system is potentially enhanced in rejecting the possible mechanical disturbances. For this purpose and in order to identify the behavior of the closed-loop structure for various proportional and derivative gains, the obtained NPDE is first converted to a system of NODEs by employing the Galerkin method and using the orthogonality of shape functions for simply supported boundary conditions. It should be pointed out that as long as the shape functions for a particular boundary condition are available, the current approach can be modified and then can be conducted to cover more general problem-specific boundary conditions. In the next step, the vibration attenuation and sensitivity of the dynamic response are assessed by the variation of the controller gains and the magnitude of external disturbances. It is assumed that the plate is under large-amplitude vibration due to the existence of an arbitrary mechanical disturbance. Moreover, to distinguish between attenuation performances of the closed-loop system in the case of static and dynamic loadings, the simulation examples are selected to cover both static and moving loads with a complex profile.

## 2.2 Mathematical model

The sandwich piezolaminated plate, as shown in Figure 2.1, consists of an elastic core, piezo-sensor layer, and piezo-actuator layer which are perfectly bounded on top and bottom of the core layer, respectively. Based on the Kirchhoff-Love plate theory, the displacement functions are described as

$$
\begin{aligned}
u(x,y,z,t) &= u_0(x,y,z,t) - z\frac{\partial w_0(x,y,t)}{\partial x},\\
v(x,y,z,t) &= v_0(x,y,z,t) - z\frac{\partial w_0(x,y,t)}{\partial y},\\
w(x,y,z,t) &= w_0(x,y,t).
\end{aligned}
\tag{2.1}
$$

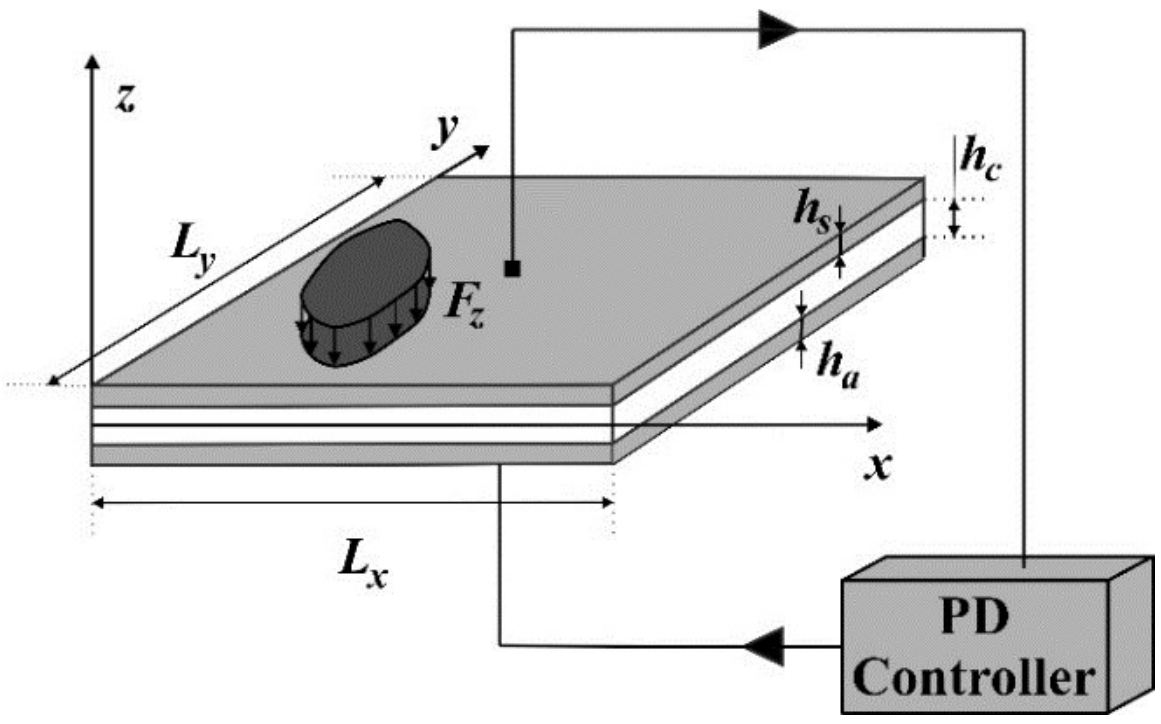

Figure 2.1 Configuration of the sandwich smart plate [111]

Considering the symmetricity of the general strain tensor and then applying von Karman's theory for small strain and moderate rotation and neglecting all higher-order terms except for $(\partial w/\partial x)^2$, $(\partial w/\partial y)^2$, and $(\partial w/\partial x)(\partial w/\partial y)$, the strain components can be obtained as

$$
\begin{aligned}
\varepsilon_{xx} &= \varepsilon_{xx}^0 + z\varepsilon_{xx}^1,\\
\varepsilon_{yy} &= \varepsilon_{yy}^0 + z\varepsilon_{yy}^1,\\
\varepsilon_{xy} &= \varepsilon_{xy}^0 + z\varepsilon_{xy}^1,
\end{aligned}
\tag{2.2}
$$

where,

$$
\begin{aligned}
\varepsilon_{xx}^0 &= \frac{\partial u_0}{\partial x} + \frac{1}{2}\left(\frac{\partial w_0}{\partial x}\right)^2,\\
\varepsilon_{yy}^0 &= \frac{\partial u_0}{\partial y} + \frac{1}{2}\left(\frac{\partial w_0}{\partial y}\right)^2,\\
\varepsilon_{xx}^1 &= -\frac{\partial^2 w_0}{\partial x^2},\\
\varepsilon_{yy}^1 &= -\frac{\partial^2 w_0}{\partial y^2},\\
\varepsilon_{xy}^0 &= \frac{1}{2}\left(\frac{\partial u_0}{\partial y} + \frac{\partial v_0}{\partial x} + \left(\frac{\partial w_0}{\partial x}\right)\left(\frac{\partial w_0}{\partial y}\right)\right),\\
\varepsilon_{xy}^1 &= -\frac{\partial^2 w_0}{\partial x \partial y}.
\end{aligned}
\tag{2.3}
$$

By assuming the polarization of both of the piezo-layers in $z$-direction, the constitutive equation of the piezoelectricity for orthotropic material and the case of plane strain can be written as [112]

$$\begin{Bmatrix} \sigma_{xx} \\ \sigma_{yy} \\ \sigma_{xy} \\ D_3 \end{Bmatrix} = \begin{bmatrix} c_{11}^{*i} & c_{12}^{*i} & 0 & -e_{31}^{*i} \\ c_{12}^{*i} & c_{22}^{*i} & 0 & -e_{32}^{*i} \\ 0 & 0 & c_{66}^{i} & 0 \\ e_{31}^{*i} & e_{32}^{*i} & 0 & \epsilon_{33}^{*i} \end{bmatrix} \begin{Bmatrix} \varepsilon_{xx} \\ \varepsilon_{yy} \\ \varepsilon_{xy} \\ E_3 \end{Bmatrix}, \tag{2.4}$$

in which, $(\sigma_{xx}, \sigma_{yy}, \sigma_{xy})^T$, $D_3$, and $E_3$ are correspondingly the stress vector, electric displacement, and electric field in $z$-direction with $i = s, a$ representing the sensor and actuator material properties with following definitions [113]

$$c_{11}^{*i} = c_{11}^{i} - \frac{{c_{13}^{i}}^2}{c_{33}^{i}}, \qquad c_{12}^{*i} = c_{12}^{i} - \frac{c_{13}^{i} c_{23}^{i}}{c_{33}^{i}}, \qquad c_{22}^{*i} = c_{22}^{i} - \frac{{c_{23}^{i}}^2}{c_{33}},$$
$$e_{31}^{*i} = e_{31}^{i} - \frac{c_{13}^{i} e_{33}^{i}}{c_{33}^{i}}, \qquad e_{32}^{*i} = e_{32}^{i} - \frac{c_{22}^{i} e_{33}^{i}}{c_{33}^{i}}, \qquad \epsilon_{33}^{*i} = \epsilon_{33}^{i} + \frac{{e_{33}^{i}}^2}{c_{33}^{i}}, \tag{2.5}$$

where $c_{\iota\kappa}^{i}$, $e_{3\kappa}^{i}$ $(\iota = 1,2,6, \kappa = 1,2)$ and $\epsilon_{33}$ are the components of elasticity-, piezoelectricity-tensor, and dielectric constant, respectively. In this notation, it is assumed that the surfaces of the piezo-layers are parallel to the host plate and consequently the shear stress and strain terms are negligible. Therefore, $\sigma_{zz}, \varepsilon_{xz}, \varepsilon_{yz}$ are assumed to be zero. The index 6 in (2.5) represents the $xy$-direction and following [113], (2.4) can be obtained. The electrostatic equation $(\partial D_3/\partial z = 0)$, with zero electric displacement boundary conditions at both sides of piezo-layers results in $D(z) \equiv 0$ [26]. Considering $\varphi$ as electrical potential and using (2.4) for electrical field together with $E_3(z) = -\partial\varphi/\partial z$, one can obtain the electrical field of piezo-sensor layer as

$$E_3^s = -\frac{\Delta\varphi_s}{h_s} - \frac{e_{31}^{*s}}{\epsilon_{33}^{*s}}(z - z_s)\varepsilon_{xx}^1 - \frac{e_{32}^{*s}}{\epsilon_{33}^{*s}}(z - z_s)\varepsilon_{yy}^1, \tag{2.6}$$

where,

$$\Delta\varphi_s = \varphi_s = \frac{e_{31}^{*s}}{\epsilon_{33}^{*s}} h_s(\varepsilon_{xx}^0 + z_s\varepsilon_{xx}^1) + \frac{e_{32}^{*s}}{\epsilon_{33}^{*s}} h_s(\varepsilon_{yy}^0 + z_s\varepsilon_{yy}^1), \tag{2.7}$$

and $z_s = (h_c + h_s)/2$, $z_a = -(h_c + h_a)/2$, with $h_s$ being the thickness of the piezo-sensor. For vibration control of the system, model-free proportional-derivative (PD) feedback control law in terms of the measurement from the sensor layer is considered within the dynamic modeling as $\varphi_a = G_p\varphi_s + G_d\dot{\varphi}_s$, where, $G_p$ and $G_d$ are the constant proportional and velocity feedback gains, respectively.

**Remark 2.1.** It should be noted that the system modeling is carried out considering the nonlinearity, however, the controller is selected to be a model-free approach. This choice is two-fold: (1) by selecting a PD controller, the key aspect of the closed-loop system behavior is easier to interpret and the introduction of a more complicated controller, may complicate the upcoming numerical analysis. (2) This choice also provides an insight into the capabilities of the linear controller in suppressing the vibrations of a nonlinear system as is discussed in Chapter 2.4. The latter should be interpreted with care as the system under analysis here is stable in nature (at its equilibrium state).

## 2.3 Dynamic equations

The focal point of this formulation is on utilizing Hamilton's principle in order to obtain the dynamic coupled equation of motion. The kinetic energy can be written as

$$K = \frac{1}{2}\int_V \left[\left(\frac{\partial u}{\partial t}\right)^2 + \left(\frac{\partial v}{\partial t}\right)^2 + \left(\frac{\partial w}{\partial t}\right)^2\right] dV, \tag{2.8}$$

where $V$ is the volume of the media. By taking the variation of the kinetic energy with respect to the velocity vector and using (2.1), one can obtain

$$\delta K = \int_{\Omega_0} \left[ I_0(\dot{u}_0\delta\dot{u}_0 + \dot{v}_0\delta\dot{v}_0 + \dot{w}_0\delta\dot{w}_0) - I_1\left(\dot{u}_0\frac{\partial\delta\dot{w}_0}{\partial x} + \delta\dot{u}_0\frac{\partial\dot{w}_0}{\partial x} + \dot{v}_0\frac{\partial\delta\dot{w}_0}{\partial y} + \delta\dot{v}_0\frac{\partial\dot{w}_0}{\partial y}\right) + I_2\left(\frac{\partial\dot{w}_0}{\partial x}\frac{\partial\delta\dot{w}_0}{\partial x} + \frac{\partial\dot{w}_0}{\partial y}\frac{\partial\delta\dot{w}_0}{\partial y}\right)\right] dxdy, \tag{2.9}$$

where $\Omega_0$ is the area of the cross-section parallel with $oxy$ (see Figure 2.1) and $I_j = \int_{-\frac{h}{2}}^{\frac{h}{2}} \rho z^j dz,\ j = 0,1,2$ (Appendix A.). $\rho$ represents the material density. $I_1$ represents the coupling effect between the axial and transverse vibration which is introduced by the piezoelectric layers. Strain energy can be written as

$$U = \frac{1}{2}\int_V (\sigma_{xx}\varepsilon_{xx} + \sigma_{yy}\varepsilon_{yy} + 2\sigma_{xy}\varepsilon_{xy})dV, \tag{2.10}$$

by taking the variation of the strain energy (2.10) with respect to strain tensor and using Eqs. (2.2)-(2.4), (2.11) is formulated as

$$\delta U = \int_{\Omega_0} \left[N_{xx}\delta\varepsilon_{xx}^0 + M_{xx}\delta\varepsilon_{xx}^1 + N_{yy}\delta\varepsilon_{yy}^0 + M_{yy}\delta\varepsilon_{yy}^1 + 2N_{xy}\delta\varepsilon_{xy}^0 + 2M_{xy}\delta\varepsilon_{xy}^1\right]dxdy, \tag{2.11}$$

where, $N_{kl} = \int_{-\frac{h}{2}}^{\frac{h}{2}} \sigma_{kl}dz$, and $M_{kl} = \int_{-\frac{h}{2}}^{\frac{h}{2}} z\sigma_{kl}dz$ (Appendix A.). The variation of the external work due to the bending force $F_z(x,y,t)$ (see Figure 2.1) can be calculated as

$$\delta W = \int_{\Omega_0} F_z\delta w_0 dxdy. \tag{2.12}$$

By applying Hamilton's principle to the piezo-plate, the time-dependent integral $\int_{t_1}^{t_2}(\delta K - \delta U + \delta W)\,dt = 0$ should be satisfied, in which, the effect of internal dielectric energy due to the boundary conditions for sensor and actuator layers is considered to be zero [114]. By taking the variation from the strain tensor ((2.2)) and substituting Eqs. (2.9), (2.11), and (2.12) in the time-domain integral equation and then using integration by part technique, the following integral equation is obtained

$$\begin{aligned}\int_{t_1}^{t_2} \Bigg\{ \int_{\Omega_0} \Bigg[ & -I_0(\ddot{u}_0\delta u_0 + \ddot{v}_0\delta v_0 + \ddot{w}_0\delta w_0) - I_1\left(\frac{\partial\ddot{u}_0}{\partial x}\delta w_0 - \frac{\partial\ddot{w}_0}{\partial x}\delta u_0 + \frac{\partial\ddot{v}_0}{\partial x}\delta w_0 - \frac{\partial\ddot{w}_0}{\partial x}\delta v_0\right) \\ & + I_2\left(\frac{\partial^2\ddot{w}_0}{\partial x^2}\delta w_0 + \frac{\partial^2\ddot{w}_0}{\partial y^2}\delta w_0\right) + \frac{\partial N_{xx}}{\partial x}\delta u_0 + \frac{\partial N_{yy}}{\partial y}\delta v_0 \\ & + \left(\frac{\partial N_{xx}}{\partial x}\frac{\partial w_0}{\partial x} + N_{xx}\frac{\partial^2 w_0}{\partial x^2} + \frac{\partial N_{yy}}{\partial y}\frac{\partial w_0}{\partial y} + N_{yy}\frac{\partial^2 w_0}{\partial y^2}\right)\delta w_0 \\ & + \left(\frac{\partial^2 M_{xx}}{\partial x^2} + \frac{\partial^2 M_{yy}}{\partial x^2}\right)\delta w_0 + \frac{\partial N_{xy}}{\partial y}\delta u_0 + \frac{\partial N_{xy}}{\partial x}\delta v_0 \\ & + \left(\frac{\partial N_{xy}}{\partial x}\frac{\partial w_0}{\partial y} + 2N_{xy}\frac{\partial^2 w_0}{\partial x\partial y} + \frac{\partial N_{xy}}{\partial y}\frac{\partial w_0}{\partial x}\right)\delta w_0 + 2\frac{\partial^2 M_{xy}}{\partial x\partial y}\delta w_0 \\ & + F_z\delta w_0 \Bigg] dxdy \Bigg\} dt.\end{aligned} \tag{2.13}$$

Since the nonlinear transverse vibration of the sandwich smart plate is studied here, the axial inertial terms are ignored and then by separating the coefficients of $\delta u_0$, $\delta v_0$, and $\delta w_0$ in integral (2.13) and setting them to zero individually, the following coupled nonlinear PDE in terms of lateral displacement can be written

$$I_1\frac{\partial \ddot{w}_0}{\partial x}+\frac{\partial N_{xx}}{\partial x}+\frac{\partial N_{xy}}{\partial y}=0,$$

$$I_1\frac{\partial \ddot{w}_0}{\partial y}+\frac{\partial N_{yy}}{\partial y}+\frac{\partial N_{xy}}{\partial x}=0,$$

$$-I_0\ddot{w}_0+I_2\left(\frac{\partial^2\ddot{w}_0}{\partial x^2}+\frac{\partial^2\ddot{w}_0}{\partial y^2}\right)+\frac{\partial N_{xx}}{\partial x}\frac{\partial w_0}{\partial x}+N_{xx}\frac{\partial^2 w_0}{\partial x^2}+\frac{\partial N_{yy}}{\partial y}\frac{\partial w_0}{\partial y}+N_{yy}\frac{\partial^2 w_0}{\partial y^2}+\frac{\partial^2 M_{xx}}{\partial x^2}$$
$$+\frac{\partial^2 M_{yy}}{\partial y^2}+\frac{\partial N_{xy}}{\partial x}\frac{\partial w_0}{\partial y}+2N_{xy}\frac{\partial^2 w_0}{\partial x\partial y}+\frac{\partial N_{xy}}{\partial y}\frac{\partial w_0}{\partial x}+2\frac{\partial^2 M_{xy}}{\partial x\partial y}+F_z=0, \tag{2.14}$$

by using (2.14) and the obtained results for the in-plane force resultants and moment resultants (see Appendix), the final form of the dynamic equation of bending motion of the piezolaminated plate is obtained as

$$\begin{aligned}
&-I_0\ddot{w}_0+I_2\left(\frac{\partial^2\ddot{w}_0}{\partial x^2}+\frac{\partial^2\ddot{w}_0}{\partial y^2}\right)-I_1\left(\frac{\partial \ddot{w}_0}{\partial x}\frac{\partial w_0}{\partial x}+\frac{\partial \ddot{w}_0}{\partial y}\frac{\partial w_0}{\partial y}\right)\\
&+\left[J_{xx1}\left(\frac{\partial w_0}{\partial x}\right)^2+J_{xx2}\frac{\partial^2 w_0}{\partial x^2}+J_{xx3}\left(\frac{\partial w_0}{\partial y}\right)^2+J_{xx4}\frac{\partial^2 w_0}{\partial y^2}+J_{xx5}\left(\frac{\partial w_0}{\partial x}\right)\left(\frac{\partial \dot{w}_0}{\partial x}\right)\right.\\
&\left.+J_{xx6}\frac{\partial^2 \dot{w}_0}{\partial x^2}+J_{xx7}\left(\frac{\partial w_0}{\partial y}\right)\left(\frac{\partial \dot{w}_0}{\partial y}\right)+J_{xx8}\frac{\partial^2 \dot{w}_0}{\partial y^2}\right]\left(\frac{\partial^2 w_0}{\partial x^2}\right)\\
&+\left[J_{yy1}\left(\frac{\partial w_0}{\partial x}\right)^2+J_{yy2}\frac{\partial^2 w_0}{\partial x^2}+J_{yy3}\left(\frac{\partial w_0}{\partial y}\right)^2+J_{yy4}\frac{\partial^2 w_0}{\partial y^2}+J_{yy5}\left(\frac{\partial w_0}{\partial x}\right)\left(\frac{\partial \dot{w}_0}{\partial x}\right)\right.\\
&\left.+J_{yy6}\frac{\partial^2 \dot{w}_0}{\partial x^2}+J_{yy7}\left(\frac{\partial w_0}{\partial y}\right)\left(\frac{\partial \dot{w}_0}{\partial y}\right)+J_{yy8}\frac{\partial^2 \dot{w}_0}{\partial y^2}\right]\left(\frac{\partial^2 w_0}{\partial y^2}\right)\\
&+2\left[J_{xy1}\left(\frac{\partial w_0}{\partial x}\right)\left(\frac{\partial w_0}{\partial y}\right)+J_{xy2}\frac{\partial^2 w_0}{\partial x\partial y}\right]\left(\frac{\partial^2 w_0}{\partial x\partial y}\right)\\
&+2K_{xx1}\left[\left(\frac{\partial^2 w_0}{\partial x^2}\right)^2+\left(\frac{\partial w_0}{\partial x}\right)\left(\frac{\partial^3 w_0}{\partial x^3}\right)\right]+K_{xx2}\frac{\partial^4 w_0}{\partial x^4}\\
&+2K_{xx3}\left[\left(\frac{\partial^2 w_0}{\partial x\partial y}\right)^2+\left(\frac{\partial w_0}{\partial y}\right)\left(\frac{\partial^3 w_0}{\partial x^2\partial y}\right)\right]+K_{xx4}\frac{\partial^4 w_0}{\partial x^2\partial y^2}\\
&+K_{xx5}\left[\left(\frac{\partial^3 w_0}{\partial x^3}\right)\left(\frac{\partial \dot{w}_0}{\partial x}\right)+2\left(\frac{\partial^2 w_0}{\partial x^2}\right)\left(\frac{\partial^2 \dot{w}_0}{\partial x^2}\right)+\left(\frac{\partial^3 \dot{w}_0}{\partial x^3}\right)\left(\frac{\partial w_0}{\partial x}\right)\right]+K_{xx6}\frac{\partial^4 \dot{w}_0}{\partial x^4}\\
&+K_{xx7}\left[\left(\frac{\partial^3 w_0}{\partial x^2\partial y}\right)\left(\frac{\partial \dot{w}_0}{\partial y}\right)+2\left(\frac{\partial^2 w_0}{\partial x\partial y}\right)\left(\frac{\partial^2 \dot{w}_0}{\partial x\partial y}\right)+\left(\frac{\partial^3 \dot{w}_0}{\partial x^2\partial y}\right)\left(\frac{\partial w_0}{\partial y}\right)\right]+K_{xx8}\frac{\partial^4 \dot{w}_0}{\partial x^2\partial y^2}\\
&+2K_{yy1}\left[\left(\frac{\partial^2 w_0}{\partial x\partial y}\right)^2+\left(\frac{\partial w_0}{\partial x}\right)\left(\frac{\partial^3 w_0}{\partial x\partial y^2}\right)\right]+K_{yy2}\frac{\partial^4 w_0}{\partial x^2\partial y^2}\\
&+2K_{yy3}\left[\left(\frac{\partial^2 w_0}{\partial y^2}\right)^2+\left(\frac{\partial w_0}{\partial y}\right)\left(\frac{\partial^3 w_0}{\partial y^3}\right)\right]+K_{yy4}\frac{\partial^4 w_0}{\partial y^4}\\
&+K_{yy5}\left[\left(\frac{\partial^3 w_0}{\partial x\partial y^2}\right)\left(\frac{\partial \dot{w}_0}{\partial x}\right)+2\left(\frac{\partial^2 w_0}{\partial x\partial y}\right)\left(\frac{\partial^2 \dot{w}_0}{\partial x\partial y}\right)+\left(\frac{\partial^3 \dot{w}_0}{\partial x\partial y^2}\right)\left(\frac{\partial w_0}{\partial x}\right)\right]+K_{yy6}\frac{\partial^4 \dot{w}_0}{\partial x^2\partial y^2}\\
&+K_{yy7}\left[\left(\frac{\partial^3 w_0}{\partial y^3}\right)\left(\frac{\partial \dot{w}_0}{\partial y}\right)+2\left(\frac{\partial^2 w_0}{\partial y^2}\right)\left(\frac{\partial^2 \dot{w}_0}{\partial y^2}\right)+\left(\frac{\partial^3 \dot{w}_0}{\partial y^3}\right)\left(\frac{\partial w_0}{\partial y}\right)\right]+K_{yy8}\frac{\partial^4 \dot{w}_0}{\partial y^4}\\
&+2K_{xy1}\left[\left(\frac{\partial^3 w_0}{\partial x^2\partial y}\right)\left(\frac{\partial w_0}{\partial y}\right)+\left(\frac{\partial^2 w_0}{\partial x^2}\right)\left(\frac{\partial^2 w_0}{\partial y^2}\right)+\left(\frac{\partial^2 w_0}{\partial x\partial y}\right)^2+\left(\frac{\partial w_0}{\partial x}\right)\left(\frac{\partial^3 w_0}{\partial x\partial y^2}\right)\right]\\
&+2K_{xy2}\frac{\partial^4 w_0}{\partial x^2\partial y^2}+F_z=0.
\end{aligned} \tag{2.15}$$

For studying the transient response of the system, it is assumed that the structure is excited by a transverse distributed force $F_z=F_z(x,y,t)$. In order to solve (2.15), Galerkin approximation in the spatial domain is adopted which converts the nonlinear PDE of motion (2.15) into a system of nonlinear ODEs. For this purpose, the transverse deflection function by considering the simply-supported boundary conditions is defined as

$$w_0(x,y,t) = \sum_{i=1}^{\infty}\sum_{j=1}^{\infty} f_{ij}(t)\sin\left(\frac{i\pi x}{L_x}\right)\sin\left(\frac{j\pi y}{L_y}\right), \tag{2.16}$$

where $i$ and $j$ represent the mode number in $x$- and $y$-direction, respectively. $f_{ij}(t)$ represents the time-dependent unknown transient coefficients of the solution (2.16). Substituting (2.16) in (2.15) and then using the orthogonality of mode-shapes results in the following system of nonlinear ODEs

$$\begin{aligned}
&c_1 f_{ij}(t) + f_{ij}(t)\sum_{\substack{n\\ n\neq i}}\sum_{\substack{m\\ m\neq j}} c_2 f_{nm}(t) + \sum_{\substack{n\\ n\neq i}}\sum_{\substack{m\\ m\neq j}}\sum_{\substack{k\\ k\neq i,n}}\sum_{\substack{l\\ l\neq j,m}} c_3 f_{nm}(t) f_{kl}(t) + c_4 f_{ij}^2(t)\\
&\quad+\sum_{\substack{n\\ n\neq i}}\sum_{\substack{m\\ m\neq j}} c_5 f_{nm}^2(t) + c_6 f_{ij}^3(t) + c_7 f_{ij}^2(t) f_{ij}'(t) + f_{ij}\sum_{\substack{n\\ n\neq i}}\sum_{\substack{m\\ m\neq j}} c_8 f_{nm}^2(t)\\
&\quad+ c_9 f_{ij}'(t) + c_{10} f_{ij}'(t) f_{ij}(t) + f_{ij}'(t)\sum_{\substack{n\\ n\neq i}}\sum_{\substack{m\\ m\neq j}} c_{11} f_{nm}(t)\\
&\quad+ f_{ij}(t)\sum_{\substack{n\\ n\neq i}}\sum_{\substack{m\\ m\neq j}} c_{12} f_{nm}'(t) + f_{ij}(t)\sum_{\substack{n\\ n\neq i}}\sum_{\substack{m\\ m\neq j}} c_{13} f_{nm}'(t) f_{nm}(t)\\
&\quad+\sum_{\substack{n\\ n\neq i}}\sum_{\substack{m\\ m\neq j}}\sum_{\substack{k\\ k\neq i,n}}\sum_{\substack{l\\ l\neq j,m}} c_{14} f_{nm}'(t) f_{kl}(t) + \sum_{\substack{n\\ n\neq i}}\sum_{\substack{m\\ m\neq j}} c_{15} f_{nm}'(t) f_{nm}(t) + c_{16} f_{ij}''(t)\\
&\quad+ c_{17} f_{ij}''(t) f_{ij}(t) + f_{ij}''(t)\sum_{\substack{n\\ n\neq i}}\sum_{\substack{m\\ m\neq j}} c_{18} f_{nm}(t) + \sum_{\substack{n\\ n\neq i}}\sum_{\substack{m\\ m\neq j}} c_{19} f_{nm}''(t) f_{nm}(t)\\
&\quad+\sum_{\substack{n\\ n\neq i}}\sum_{\substack{m\\ m\neq j}}\sum_{\substack{k\\ k\neq i,n}}\sum_{\substack{l\\ l\neq j,m}} c_{20} f_{nm}''(t) f_{kl}(t)\\
&\quad+\int_0^{L_x}\int_0^{L_y} F_z(x,y,t)\sin\left(\frac{i\pi x}{L_x}\right)\sin\left(\frac{j\pi y}{L_y}\right)\mathrm{d}x\mathrm{d}y = 0,
\end{aligned} \tag{2.17}$$

where $c_i, i = 1 \dots 20$ are given in Appendix. In order to obtain the transient response of the system for arbitrary external disturbance $F_z$ and control gain $(G_p, G_d)$ the built-in Mathematica function, "NDSolve" is used to solve the nonlinear system of ODEs. In the sequel, the numerical results are presented in accordance with the numerical solution of NODEs.

## 2.4 Numerical results

In this section, a parametric study is conducted to investigate the impact of various modeling parameters on the transient response of the system and the stability of the closed-loop configuration. A Mathemathica code is developed to compute the dynamic response of the coupled system. Numerical results are obtained based on the piezo-elasto-piezo plate that is depicted in Figure 2.1. The host layer is an aluminum rectangular plate with $h_c = 0.003$ m, $L_x = 0.41$ m, $L_y = 0.386$ m, and $\rho_c = 2700$ kg/m$^3$. The sensor and the actuator layers are respectively assumed to be perfectly bounded on the upper and lower surfaces of the host layer ($h_a = h_c = 0.0005$ m) and fabricated from PZT4/$Ba_2NaNb_5O_{15}$ with $\rho_a = 5300$ kg/m$^3$ and $\rho_s = 7500$ kg/m$^3$. All other material properties are given in Table 2.1.

Two kinds of loading are considered in the simulation results; the first one is a static point force on the top surface of the plate (load 1 in Figure 2.2) and the second one is the moving concentrated force along the second-order trajectory (load 2 in Figure 2.2).

The sensitivity of the open-loop transient response under a point force excitation with a constant amplitude of 10 kN at $x_0 = L_x/8, y_0 = L_y/8$ is shown in Figure 2.3. This figure shows the displacement of the center of the smart plate. In order to alleviate the computational burden of the numerical solution of (2.17) the minimum number of required shape functions is determined. Accordingly, it can be seen that for static point force at least **49** modes are required to guarantee the convergence of the response to a single transient response.

**Remark 2.2.** It should be highlighted that such a large number of modes and additionally considering the coupling effects due to high-order of strain introduces a mathematically accurate model which is not controller-development-friendly despite its accuracy.

It should be also mentioned that the response of the system by assuming a single-mode plant cannot describe the time-dependent behavior of the system which is tackled by Belouettar et al. for simply supported beam [26].

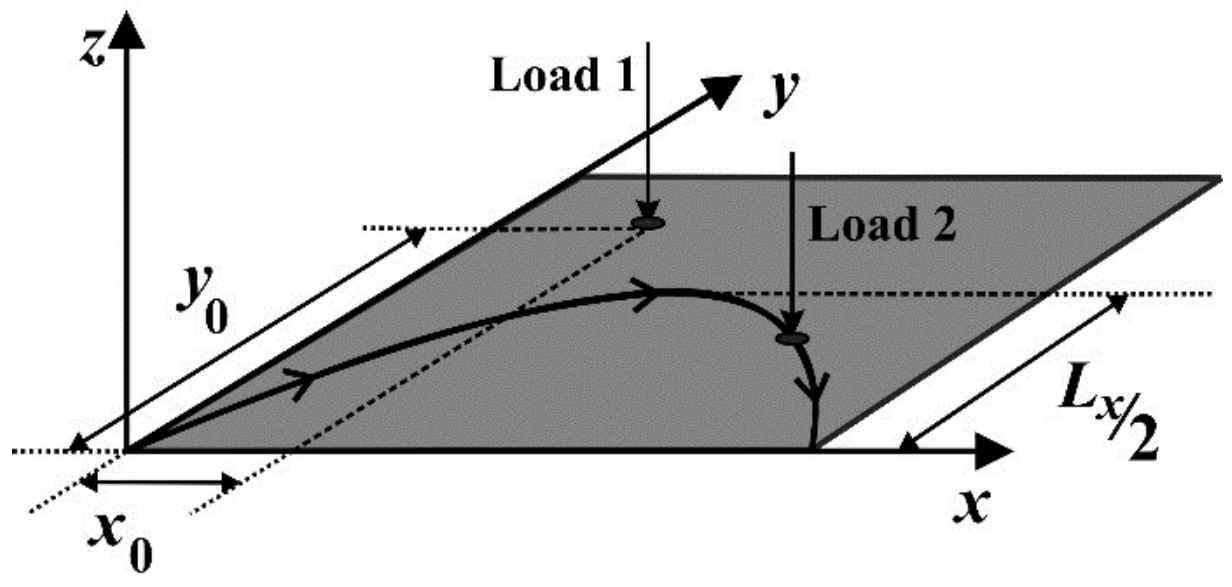

Figure 2.2 The two external applied disturbance on the top surface of the plate [111]

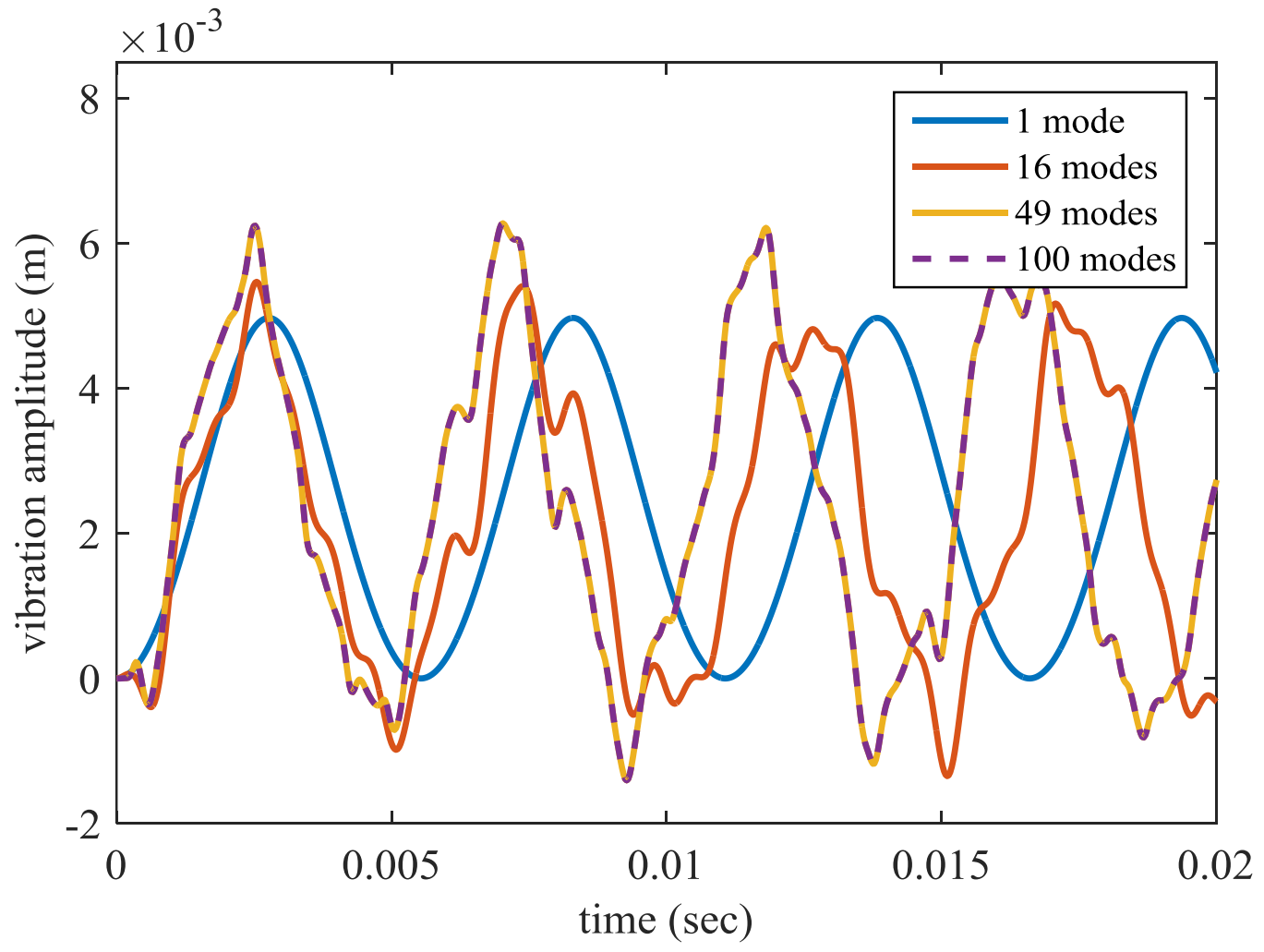

Figure 2.3 Sensitivity of the solution to the number of shape functions that are included in the Galerkin method for a simply supported case under point loading [111]

In order to study the effectiveness of the feedback control system, the transient response of the system is compared for various values of the proportional and derivative gain ($G_p$ and $G_d$). For this purpose, the linear form of (2.17) is obtained by setting all high order terms equal to zero. All of the remaining linear terms are considered as the dynamics of the open-loop system. In order to investigate the effect of excitation amplitude as well as control parameter ($G_d$) in a representation independent from the nature of the force, a static point force at $x_0 = L_x/5, y_0 = L_y/5$ is applied on the system for two different values of the excitation magnitude as depicted in Figure 2.4. Figure 2.4 shows that in contrast to the nonlinear case, this system has a linear relationship between the excitation and vibration amplitude. In addition, the vibration suppression performance is independent of the magnitude of the excitation which is less realistic and this emphasizes the importance of this study.

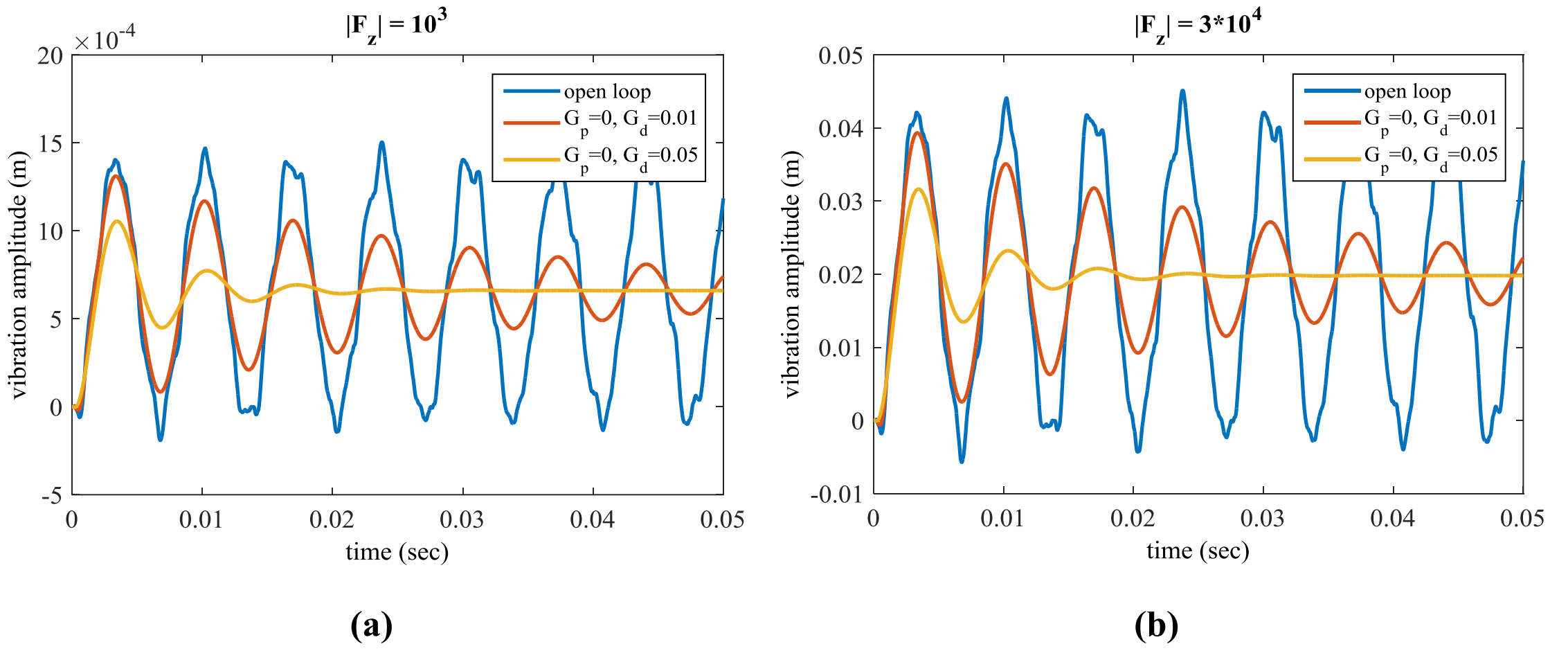

Figure 2.4 Vibration amplitude of the linear controlled system under excitation force with two different magnitudes [111]

Table 2.1 Material properties of each layer of sandwich plate [111]

| | Actuator layer | Sensor layer | Host layer |
|---|---|---|---|
| $c_{ij}(\mathrm{N/m^2})$ | | | |
| $c_{11}$ | $23.9 \times 10^{10}$ | $13.9 \times 10^{10}$ | $11.2 \times 10^{10}$ |
| $c_{12}$ | $10.4 \times 10^{10}$ | $7.8 \times 10^{10}$ | $6.04 \times 10^{10}$ |
| $c_{12}$ | $5.2 \times 10^{10}$ | $7.43 \times 10^{10}$ | $6.04 \times 10^{10}$ |
| $c_{22}$ | $24.7 \times 10^{10}$ | $13.9 \times 10^{10}$ | $11.2 \times 10^{10}$ |
| $c_{23}$ | $5.2 \times 10^{10}$ | $7.43 \times 10^{10}$ | $6.04 \times 10^{10}$ |
| $c_{33}$ | $13.5 \times 10^{10}$ | $11.5 \times 10^{10}$ | $11.2 \times 10^{10}$ |
| $c_{66}$ | $7.6 \times 10^{10}$ | $3.06 \times 10^{10}$ | $2.59 \times 10^{10}$ |
| $e_{ij}(\mathrm{C/m^2})$ | | | |
| $e_{31}$ | $-0.4$ | $-5.2$ | |
| $e_{32}$ | $-0.3$ | $-5.2$ | |
| $e_{33}$ | $4.3$ | $15.1$ | |
| $\epsilon_{ij}(\mathrm{F/m})$ | | | |
| $\epsilon_{11}$ | $196 \times 10^{-11}$ | $650 \times 10^{-11}$ | |
| $\epsilon_{22}$ | $201 \times 10^{-11}$ | $650 \times 10^{-11}$ | |
| $\epsilon_{33}$ | $28 \times 10^{-11}$ | $560 \times 10^{-11}$ | |

The nonlinear vibration control of the sandwich plate ($|F_z| = 10^4$) is presented in Figure 2.5. This figure consists of two sub-plots in which the effect of the proportional and derivative gains are studied individually due to the nonlinear nature of the problem. In the first sub-plot (Figure 2.5a), although increasing $G_p$ attenuates the vibration amplitude, it also increases the frequency of the vibration. This is due to the fact that the proportional gain of the feedback channel can change the global stiffness of the structure (see (2.17) and $c_1$ in Appendix.

In contrast to the linear case and owing to higher-order coupled strain term, the closed-loop system is highly sensitive with respect to $G_p$ and increasing proportional gain $G_p > 40$ in our simulations, leads to performance degradation and then instability. However, increasing $G_d$ similar to the linear behavior improves the settling time and stability of the system (see Figure 2.5b). Moreover, it can be seen in Figure 2.5b that the derivative term of the controller can reject the higher-order terms.

Figure 2.6 displays the time series of snap-shots of displacement propagation in the piezolaminated sandwich plate under a step-like loading at $x_0 = L_x/5$, $y_0 = L_y/5$. Regions having lower vibrating levels are depicted in darker colors. These figures are obtained by setting the magnitude of the external input, the proportional, and the derivative gain equal to $10^5$, $25$, and $0.001$, respectively. Figure 2.6 shows that, since the external force continues to act on the coupled structure, although the feedback control system attenuates the dynamical part of the vibration, the static displacement remains uncontrolled.

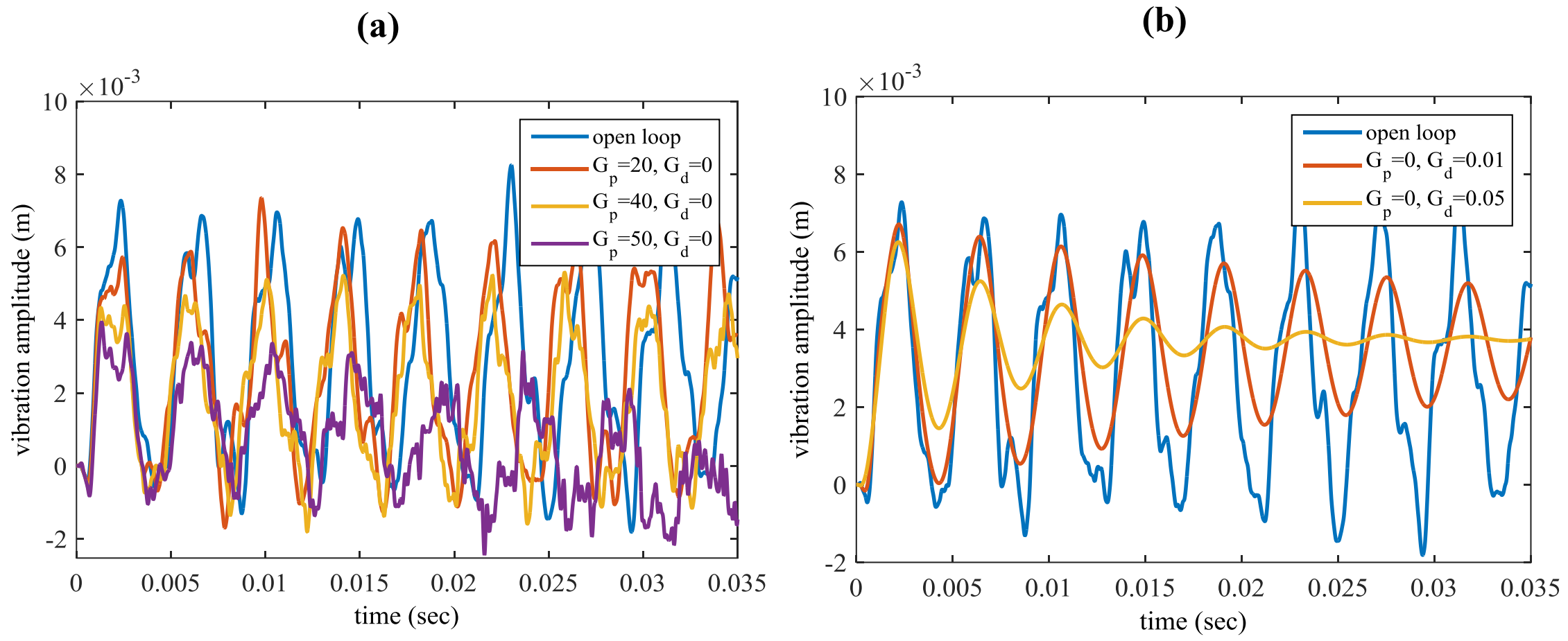

Figure 2.5 Effectiveness of the vibration control system in suppressing the vibration [111]

In order to investigate the performance of the control system after removing the force from the structure, the moving concentrated load $F_z(x,y,t) = F_0\delta\left(x - \frac{L_x}{t_0}t\right)\delta\left(y + \frac{2L_y}{t_0^2}t^2 - \frac{2L_y}{t_0}t\right)[H(t) - H(t-t_0)]$ is applied as the external time-varying disturbance in the time interval $[0 \quad t_0]$ sec on second-order trajectory (see Figure 2.2). In this function, $\delta(.)$ and $H(.)$ stand for the Dirac's delta and Heaviside step functions, respectively. Because of the nonlinearity of the problem, a similar analysis to Figure 2.3 is conducted to determine the minimum required the number of mode shapes that satisfy the convergence and the first **100** mode-shapes of the coupled piezolaminated structure are taken as the plant model (see

**Remark 2.2**). Figure 2.7 illustrates the dynamic response of the system under moving load on a second-order trajectory that is depicted in Figure 2.2. The loading phase lasts $0.05$ sec during which the concentrated force reaches $x = L_x/2$ and $y = L_y/2$ at $t_0 = 0.025$ sec. The control system is unable to damp the static deformation of the structure during the loading phase, however, it can be seen that the oscillations around the static deflection are suppressed by activating the PD control in the feedback channel. In addition, by comparing two controlled outputs, it is obvious that the undershoot of the controlled system with $G_p = 40$ is more than in the case with $G_p = 10$ which is the price for decreasing the steady state error.

Finally, in order to validate the proposed analytical model in this chapter, the six fundamental natural frequencies of the system are compared with results calculated by means of the commercial finite element package ABAQUS 6.13.

Table 2.2 Comparison of the six fundamental natural frequencies of the coupled structure (Hz) [111]

| Method | Mode number | | | | | |
|---|---|---|---|---|---|---|
| | (1,1) | (2,1) | (1,2) | (2,2) | (3,1) | (1,3) |
| Present | 147.54 | 343.39 | 392.35 | 589.92 | 668.72 | 799.45 |
| FEM | 146.90 | 342.44 | 390.83 | 586.53 | 667.40 | 796.15 |
| Error (%) | 0.43 | 0.28 | 0.39 | 0.57 | 0.20 | 0.41 |

It is worth mentioning that in the ABAQUS model, 113160 twenty-node quadratic piezoelectric brick (C3D20RE) elements were employed as meshing configuration for the sensor and actuator sheets, and 226320 twenty-node quadratic brick (C3D20R) elements were used to model the host structure. The results, as shown in Table 2.2, demonstrate very good agreements with those obtained from ABAQUS. For the sake of brevity, the mesh convergence analysis is suppressed here. Moreover, no additional numerical damping component is introduced for FE simulations.

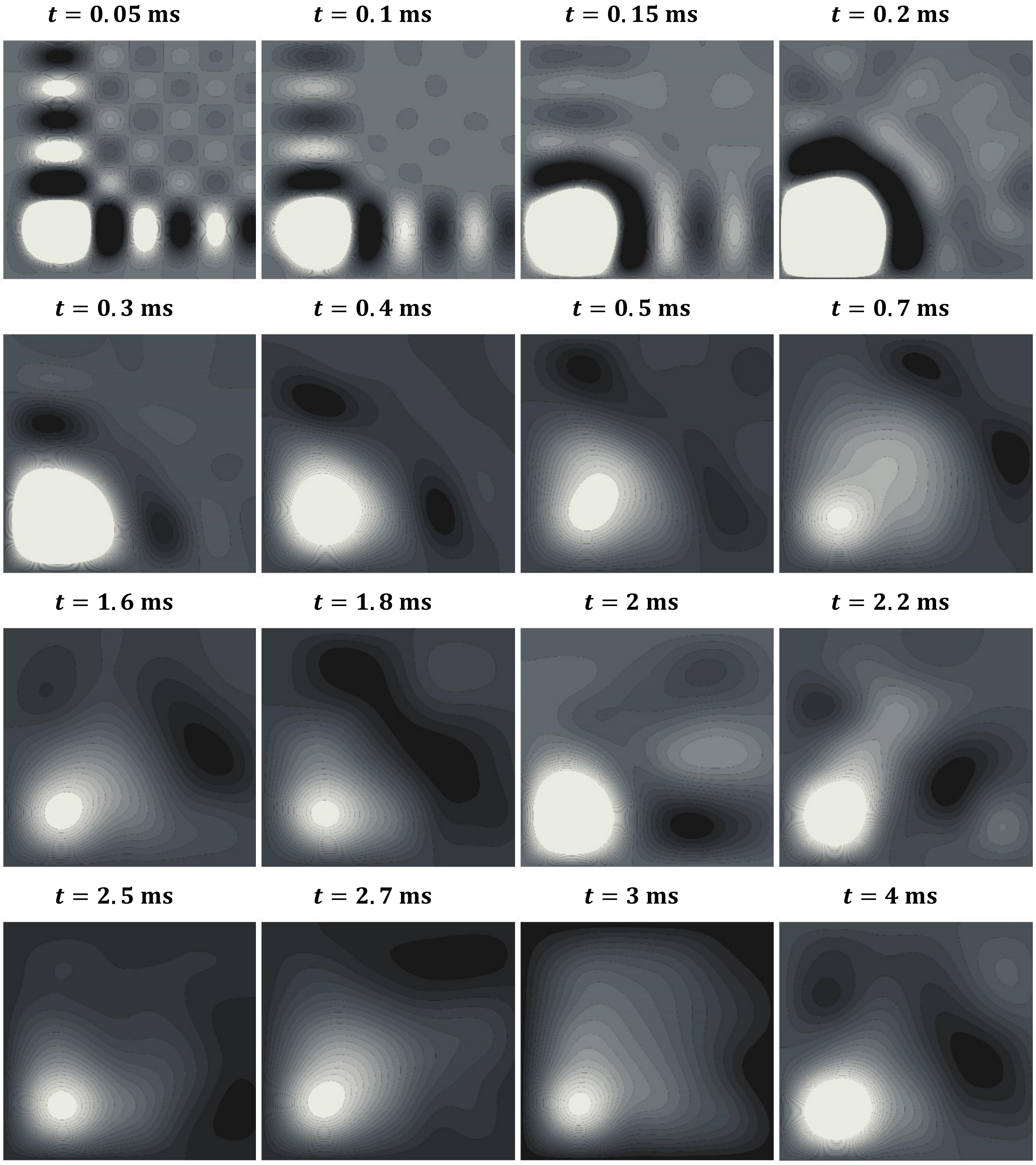

Figure 2.6 Time snap-shots of the vibration propagation in the piezoelectric laminated plate due to the mechanical point load [111]

The magnitude of FRF of the sandwich piezolaminated plate based on the proposed model is compared with the result that is obtained from ABAQUS by steady-state dynamic, Direct method within the frequency range of $[100 \quad 420]$ Hz as shown in Figure 2.8. The FRF of the system is obtained by exciting the plate with a harmonic constant point force at $x_0 = 3L_x/4$ and $y_0 = L_y/4$ and collecting the response of the system at the nodal point of $x_1 = L_x/4$ and $y_1 = 3L_y/4$. The analytical solution is carried out by taking fast Fourier transformation (FFT) from the transient response of the analytical model under the same excitation.

Finally, in order to validate the analytical model in the time-domain, dynamic explicit analysis is performed in FEM package (ABAQUS) under a point force excitation that acts at the center of the plate with a transient form of $t/t_0\,(U(t) - U(t - t_0)) + (-t/t_0 + 2)\big(U(t - t_0) - U(t - t_1)\big)$ with $t_0 = 1e^{-4}$ sec and $t_1 = 2e^{-4}$ sec and $U(.)$ being the step function. The results for the deformation of the piezolaminated plate are shown in two different time steps in Figure 2.9.

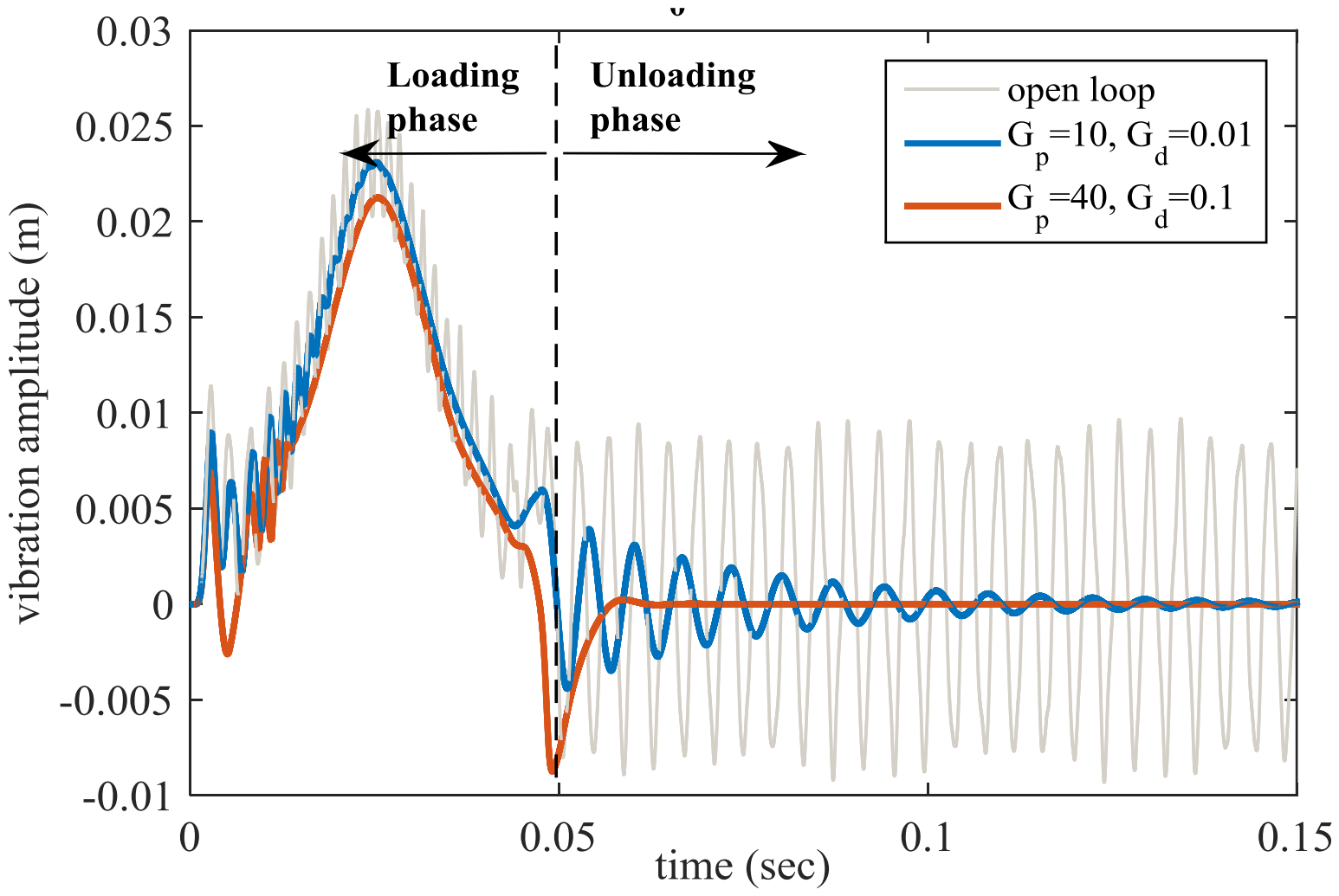

Figure 2.7 Transient response of the system under moving load with amplitude of $F_0$ =50 kN [111]

Based on the new analytical formulation, a key insight into the form of the equation of motion in NODE form is obtained.

Note that the geometrically nonlinear systems with arbitrary structural geometry cannot be treated in an analytical framework that is presented in (2.17). However, for the alternative *nonlinear system identification*, this can be seen as a characterization step. The important aspects of this chapter are listed as:

1) Hamilton's principle is used to extract the strong form of the equation of motion in the form of an NPDE.

2) Galerkin method and orthogonality of shape functions for simply supported boundary conditions in the spatial domain are employed and then the obtained nonlinear system of ODEs of motion is solved numerically.

3) Vibration suppression problem is investigated by the implementation of the model-free control law. Model-free indicates that the development of the controller structure and its parameterization is done without a need for a system (plant) model.

4) The presented method as a semi-analytical approach is much faster than FEM and, therefore, it can be used as a benchmark in designing smart plates under large deformation. Although in the grand scheme of the dissertation, there would be no place for comparison of these two numerical approaches (except for validation), the closed-form formulation enables a model-based controller development in the nonlinear domain.

5) High order strain terms play an important role in the dynamic response of the closed-loop system and the presented model shows many accurate results for system response compared to the linear simplification. Two important observations compared to classical linear assumptions are a) The sensitivity of the transient behavior of the structure to the magnitude of the excitation in large vibration amplitudes. b) Identifying the realistic bounds of proportional feedback gain that may result in structural instability.

The numerical solution of this chapter provides a framework for structural analysis in the time domain as well as provides an insight into the form of geometrical nonlinearity that may be employed later in nonlinear black-box system identification. Since the ultimate goal of deriving a model in the scope of the dissertation is for the model-based controller synthesis, a more general approach for geometrically nonlinear structures with arbitrary geometry needs to be foreseen. Hence, in the following Chapter 4, the interest of the author is shifted towards the black-box models (eventually the nonlinear identification can be interpreted as a gray-box because of the characterization step of this chapter). In order to have the possibility of employing both time- and frequency-domain system identification techniques, in the upcoming chapter the frequency-domain response of the system is also investigated.

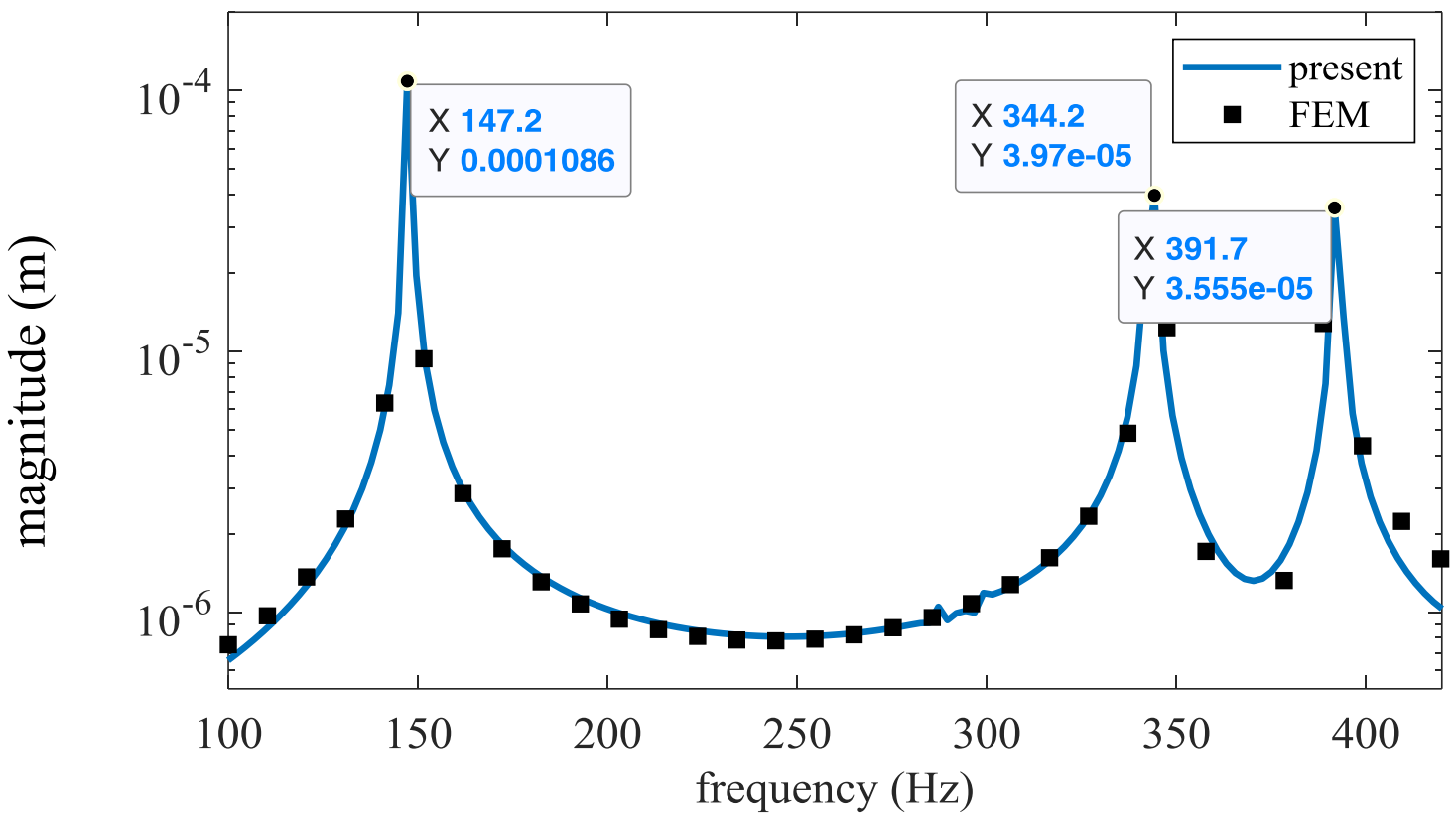

Figure 2.8 FRF of the sandwich plate under a point force at $x_0 = 3L_x/4$, $y_0 = L_y/4$ [111]

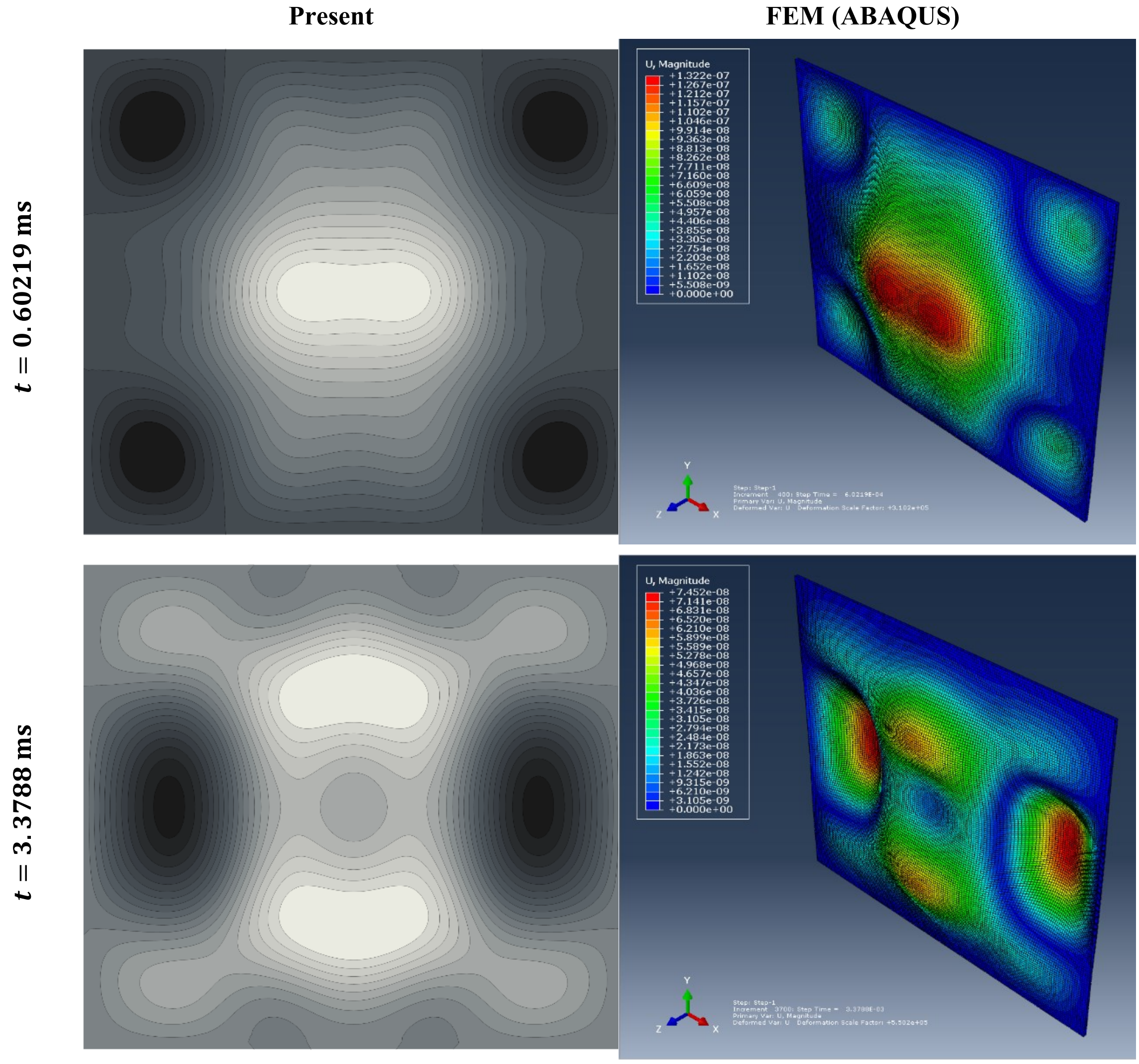

Figure 2.9 Validation of the deformation of the coupled plate with FEM [111]

# 3 Analytical approach for nonlinear modeling in frequency-domain

In this chapter, following the previous chapter, the main idea is continued in frequency-domain. The main motivation here is to approach the black-box nonlinear system identification. Such a connection can be established by combining the findings of this chapter with the so-called control-based continuation (CBC) techniques [115]. In this approach, a set of discrete points of the frequency response function is extracted in the real-time experiments using CBC similar to [116], [117]. Then, having the form of nonlinear terms as of (2.17), a parameter identification can be performed in order to estimate the coefficients of the unknown equation of motion. This will be clarified once we narrow down the frequency range to the area around the natural frequency.

The systematic formulation of the nonlinear dynamic equation of motion, as well as an appropriate semi-analytical solution, is presented in the frequency domain.

## 3.1 Reformulation of equation of motion for single mode analysis

To study the harmonic response of the system, it is assumed that the structure is excited by a transverse distributed force $F_z = f_z(x, y)\cos(\omega t)$ in Figure 2.1. In order to solve the nonlinear dynamic equation of motion, the harmonic balance method in frequency-domain, along with the single-mode Galerkin approximation in the spatial-domain, is adopted, which converts the nonlinear PDE of motion to a nonlinear algebraic equation. Accordingly, the transverse deflection for the simply support boundary conditions is defined as $w_0(x, y, t) = (Ae^{i\omega t} + \bar{A}e^{-i\omega t})\sin\left(\frac{\pi x}{L_x}\right)\sin\left(\frac{\pi y}{L_y}\right)$, in which $A$ represents the complex amplitude of vibration to be calculated, $\bar{A}$ is the complex conjugate of $A$, and $\omega$ is the nonlinear vibration frequency. Without loss of generality, the external disturbance is assumed to be distributed on the upper layer uniformly ($F_z = Be^{i\omega t} + \bar{B}e^{-i\omega t}$). Using the orthogonality of mode shapes and setting the secular terms equal to zero, one can obtain the following algebraic equation:

$$-\omega^2 MA + K(\omega)A + K_{NL}(\omega)\bar{A}A^2 - Q = 0, \tag{3.1}$$

where

$$M = -\frac{1}{4}I_0L_xL_y - \frac{I_2(L_x^2 + L_y^2)\pi^2}{4L_xL_y},$$

$$K(\omega) = K^R + iK^I(\omega) = |K|e^{i\alpha},$$

$$K_{NL} = K_{NL}^R + iK_{NL}^I(\omega) = |K_{NL}|e^{i\beta},$$

$$K^R = \frac{\left(K_{yy4}L_x^4 + \left(K_{xx4} + 2K_{xy2} + K_{yy2}\right)L_x^2L_y^2 + K_{xx2}L_y^4\right)\pi^4}{4L_x^3L_y^3},$$

$$K^I(\omega) = \frac{(K_{yy8}L_x^4 + (K_{xx8} + K_{yy6})L_x^2L_y^2 + K_{xx6}L_y^4)\pi^4\omega}{4L_x^3L_y^3},$$

$$K_{NL}^R = -\frac{3\left(3J_{yy3}L_x^4 + \left(3J_{xx3} - 2J_{xy1} + 3J_{yy1}\right)L_x^2L_y^2 + 3J_{xx1}L_y^4\right)\pi^4}{64L_x^3L_y^3},$$

$$K_{NL}^I(\omega) = -\frac{3\left(J_{yy7}L_x^4 + \left(J_{xx7} + J_{yy5}\right)L_x^2L_y^2 + J_{xx5}L_y^4\right)\pi^4\omega}{64L_x^3L_y^3},$$

$$Q = -\frac{4BL_xL_y}{\pi^2},$$

in which $K_{mni}$ and $J_{mni}$ $mn = xx, yy, xy$ and $i = 1 \dots 8$ represent the constants that can be calculated after the integration. By using $A = re^{i\theta}$ and decomposing (3.1) in its real and imaginary parts, one can derive the following relation for the magnitude of vibration with respect to nonlinear frequency change and external input

$$r^6|K_{NL}|^2 + r^4[2(K^RK_{NL}^R + K^IK_{NL}^I) - 2\omega^2MK_{NL}^R] + r^2(\omega^4M^2 + |K|^2 - 2\omega^2MK^R) - Q^2 = 0. \tag{3.2}$$

## 3.2 Numerical simulation in frequency-domain

Numerical results are obtained based on the piezo-elasto-piezo plate shown in Figure 2.1. The geometry of the system is the same as the piezo-laminated plate given in Chapter 2.4. The material properties are given in Table 2.1 based on [20].

The effect of the proportional gain on the fundamental natural frequencies of the "closed-loop" system is presented in Figure 3.1. It should be noted that as the control system is active, the closed-loop system can be seen as an open-loop one considering the external mechanical excitation (disturbance) as the system input. In other words, the inclusion of $G_p$ should here rather be considered as an additional boundary condition. In this figure, the roots of the characteristic equation are calculated for various values of the $G_p$.

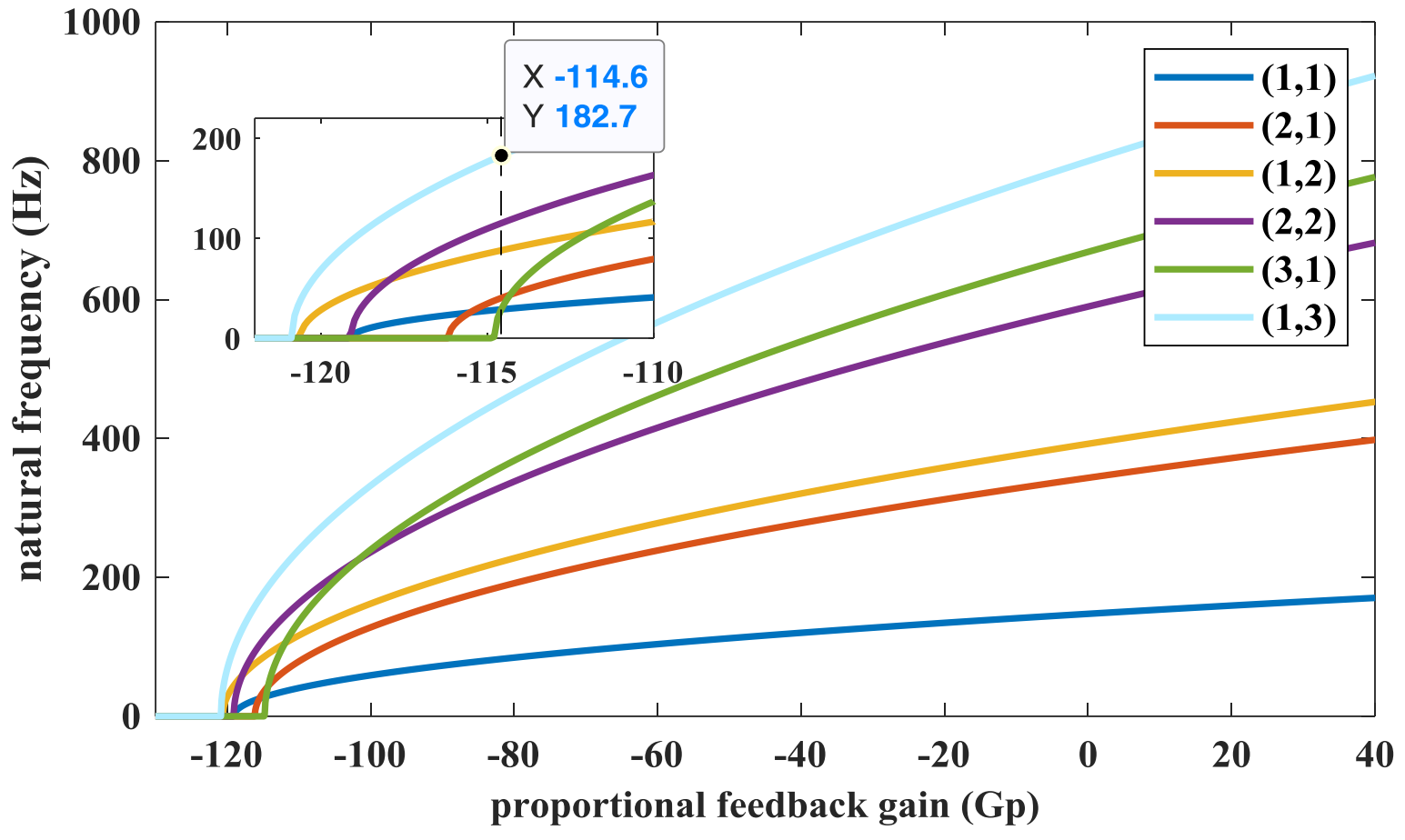

Figure 3.1 Effect of $G_p$ on natural frequencies [5]

Obviously, by increasing the proportional gain ($G_p$), the natural frequencies of the system tend to increase. This is due to the fact that the proportional term of the feedback channel changes the global stiffness of the system drastically. In addition, there exists a value of $G_p^*$ for each of the mode shapes that set the real part of the global linear stiffness of the system equal to zero. These values for the six first natural frequencies of the system are given in Table 3.1.

Table 3.1 Values of $G_p^*$ for the six fundamental mode shapes of the system [5]

| | Mode number | | | | | |
|---|---|---|---|---|---|---|
| | (1,1) | (2,1) | (1,2) | (2,2) | (3,1) | (1,3) |
| $G_p^*$ | -119.112 | -116.178 | -120.634 | -119.112 | -114.8 | -120.897 |

It can be observed from the zoomed area in Figure 3.1 that for the values of the proportional gain near to $G_p^*$ for each mode shape, the order of the natural frequencies changes. For instance, considering $G_p =$ $-114.6$, mode order (1,3) will be the fundamental natural frequency of the system instead of (1,1). In other words, the proportional gain can even change the fundamental vibration mode. In order to observe the effect of $G_p^*$ on the single-mode transient response of the system, the decoupled equation of motion is solved by assuming $w_0(x,y,t) = A(t)\sin(\pi x/L_x)\sin(\pi y/L_y)$, in which $A(t)$ represents an unknown time dependent scalar variable that should be calculated numerically. The built-in Mathematica function, "NDSolve" is employed for this purpose and the transient response of the system is calculated numerically as depicted in Figure 3.2. It can be seen that with the proportional gain equal to $G_P^*$ of mode number (1,1), the system vibrates in a harmonic manner. However, by increasing the magnitude of the proportional gain until the condition of zero linear global stiffness matrix is reached, the structure vibrates in a nonlinear trend, during which the system tends to stay at the rest mode (static equilibrium) longer than the rise mode. This can be explained by the introduction of anti-resonance modes which occur by increasing the magnitude of $G_p$.

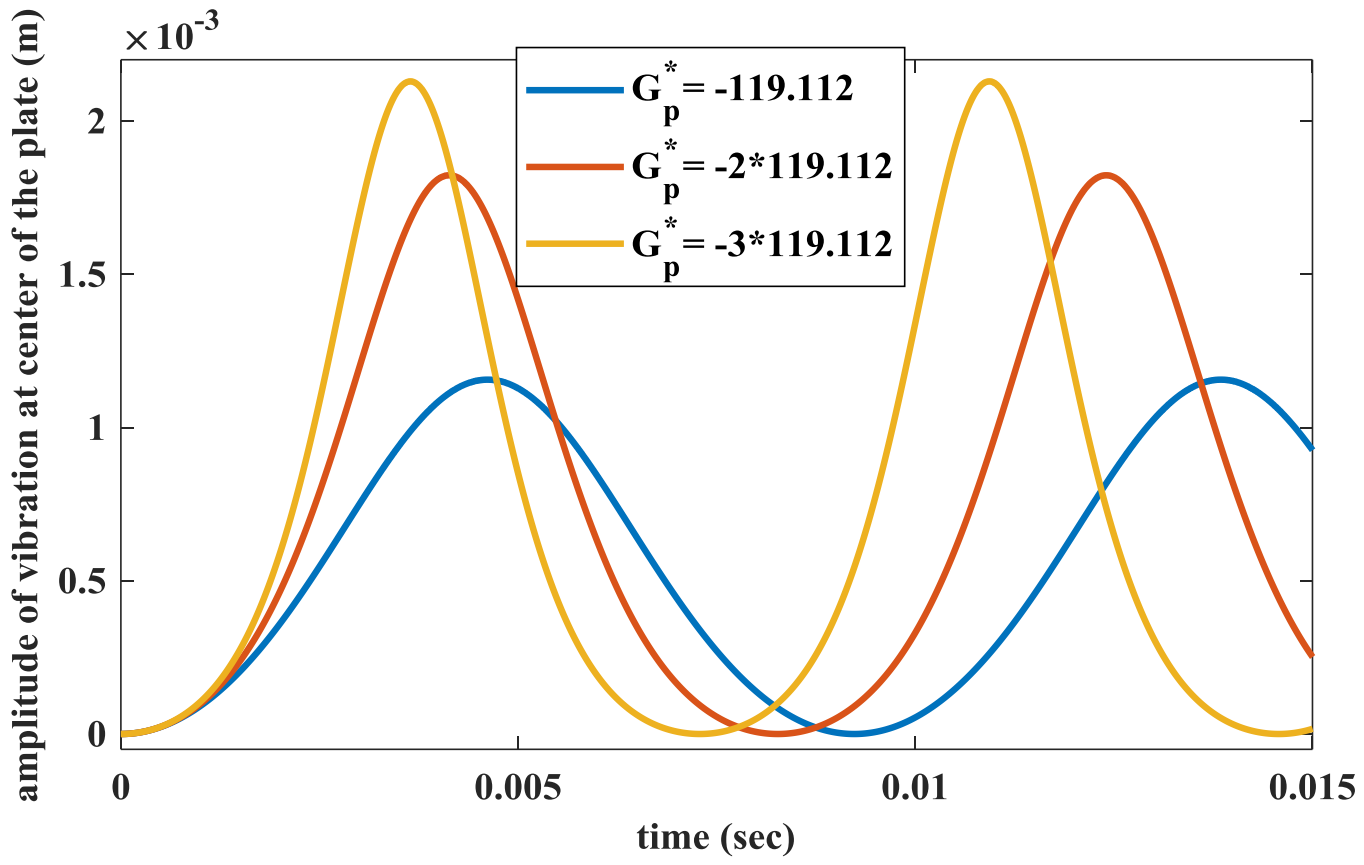

Figure 3.2 Effect of $G_p^*$ on transient response of the system with the first dominant natural frequency [5]

Further investigations are focused on the nonlinear frequency response of the simply supported plate based on the amplitude relation in (3.2) with $w_0$ and $h$ being the transverse displacement at the center of the host plate and overall thickness of the smart structure, respectively. The results are mostly emphasized in the normalized frequency with respect to the linear natural frequency $\omega_L^2 = K^R/M$.

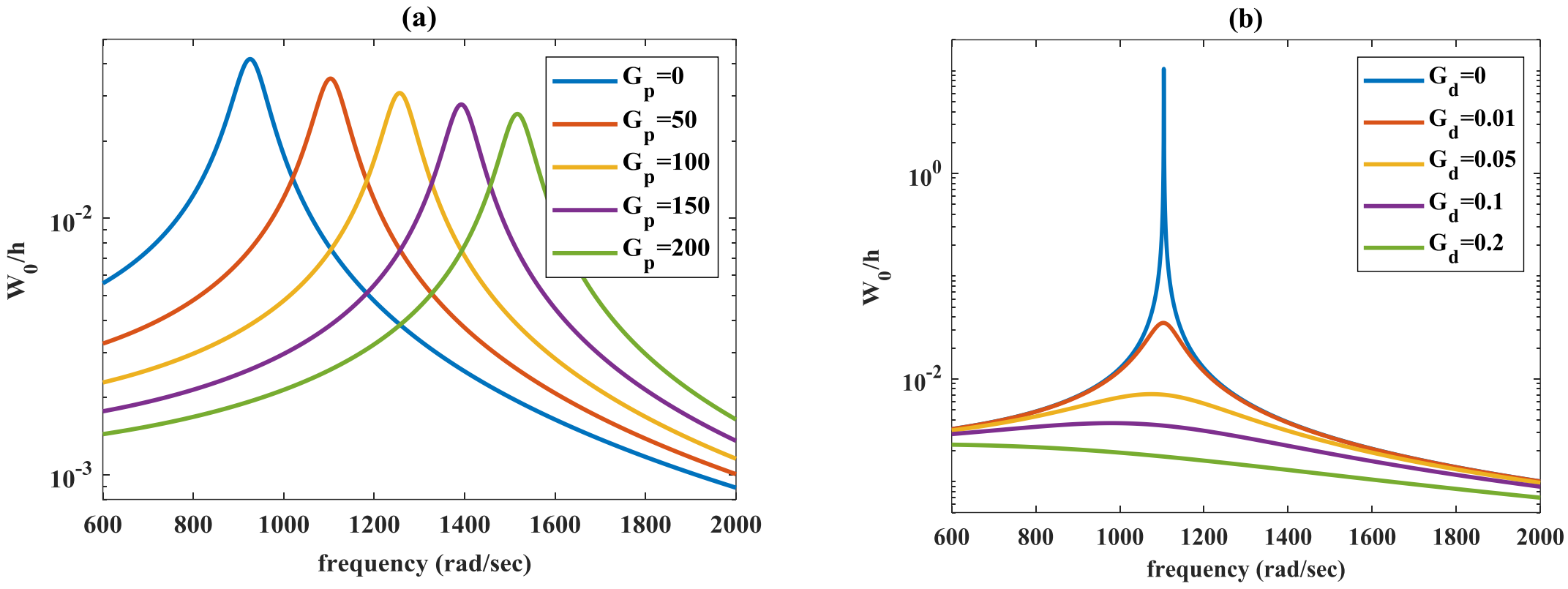

Figure 3.3 Linear frequency response sensitivity to (a) $G_p$ and (b) $G_d$ [5]

Figure 3.3 shows the effects of proportional and derivative gain on the linear frequency response of the system at the center of the plate around the first natural frequency. In Figure 3.3(a), $G_p$ is increased from 0 to 200, while $G_d$ is kept equal to 0.01 and in Figure 3.3(b), $G_d$ is increased from 0 to 0.2, but with $Gp =$

50. As it can be seen from Figure 3.3(a), due to the effect of proportional gain on $K^R$ ((3.1)) and the amplitude equation (3.2) the peak of the non-normalized frequency response of the controlled linear system shifts to higher frequencies by increasing $G_p$. In contrast, as can be seen from Figure 3.3(b), the characteristic equation of the model is independent of $G_d$ in linear case, but by increasing the derivative gain, the amplitude of the vibration damps out. The effect of the proportional gain ($G_p$) on the bandwidth of the system is demonstrated in Figure 3.4. The response of the system is normalized with respect to the first controlled natural frequency. It is obvious from Figure 3.3 and Figure 3.4 that the bandwidth of the closed-loop system can be controlled by $G_p$ in the proposed controller.

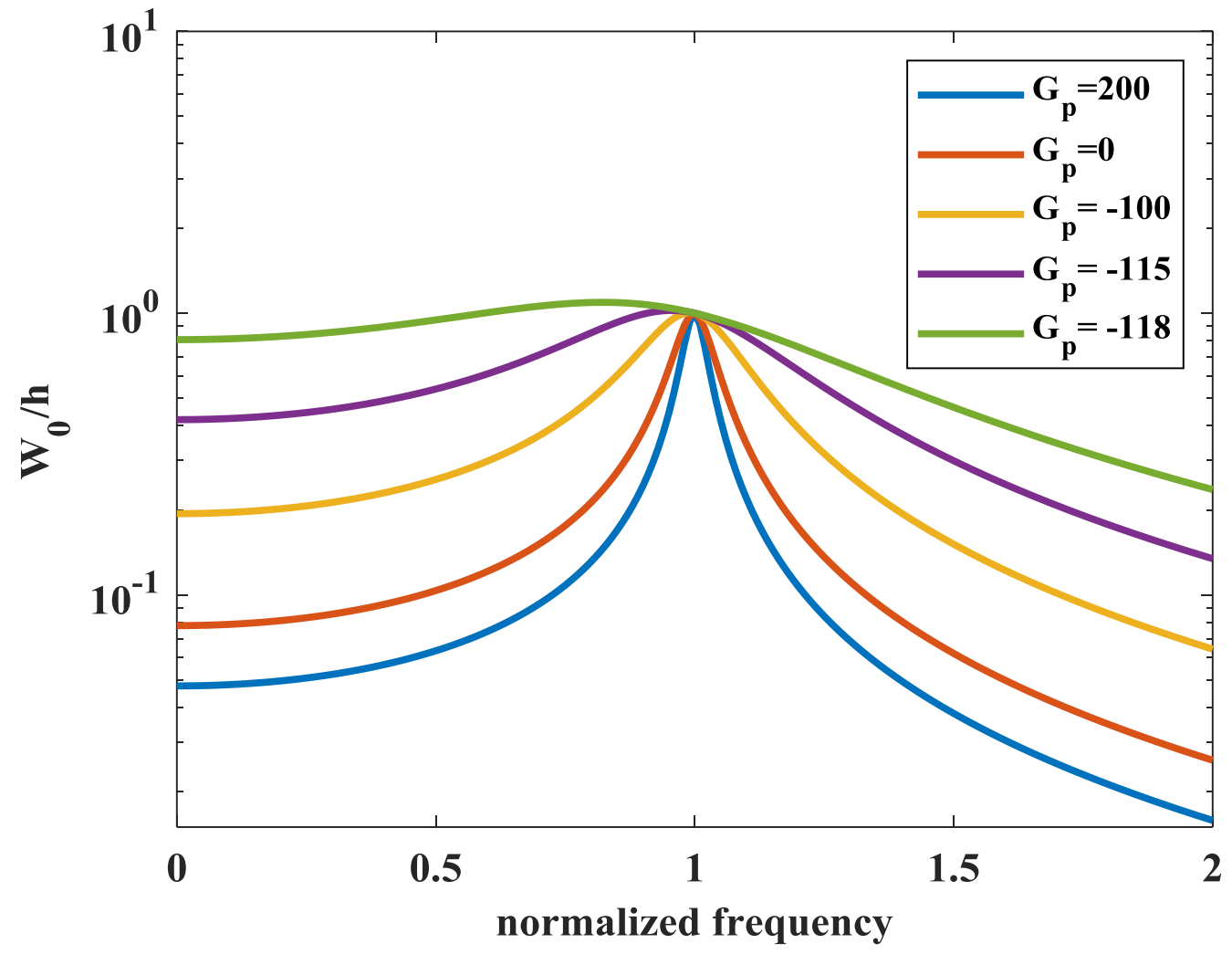

Figure 3.4 Effect of proportional gain on the bandwidth of dynamic response [5]

The sensitivity concerning the excitation magnitude ($|f_z|$) is investigated in Figure 3.5 for both linear and nonlinear controlled systems. For this purpose, the normalized response of the system under three different load magnitudes $|f_z| = 0, 100$, and 400 are compared by assuming the constant control gains of $G_p = 110$ and $G_d = 0.005$. In this figure, superscripts $l$ and $n$ in the plot legend represent the linear and nonlinear response, respectively. The first observation is that the linear frequency response is more sensitive with respect to $G_d$ than the nonlinear response.

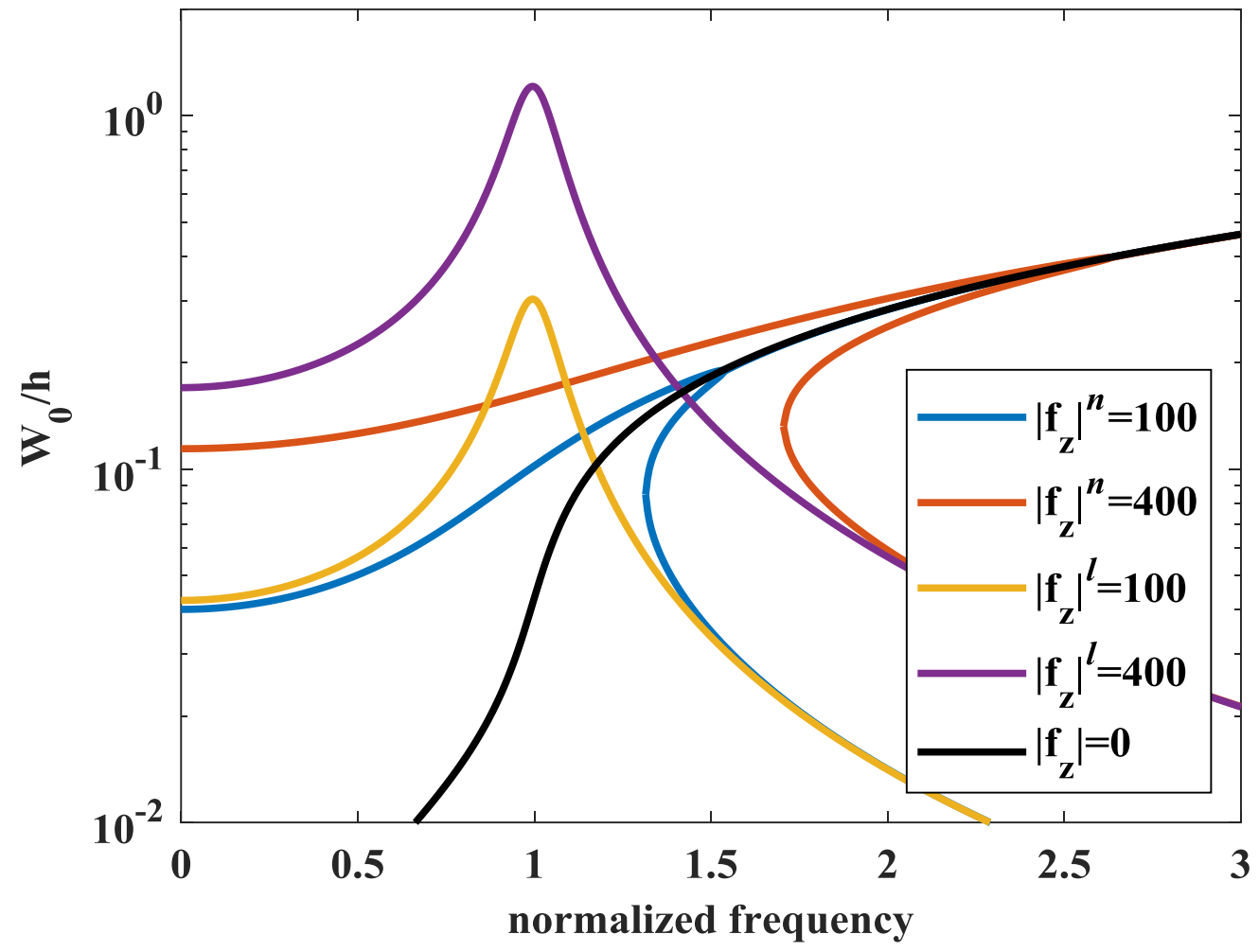

Figure 3.5 Frequency response for two excitation amplitude $G_p = 110$, $G_d = 0.005$ [5]

Also, it can be seen that in the low-frequency range, as the excitation magnitude increases, the linear and nonlinear responses tend to diverge, which reveals the importance of the nonlinear model. The nonlinear controlled response is highly sensitive to the control gain. This phenomenon is analyzed by keeping a constant (small) value for $G_d$, while varying the proportional gain. The backbone arc for each case, corresponding to $|f_z| = 0$, is also presented with the dashed line in this comparison. As it can be seen, in the nonlinear case, there is a critical gain of $G_p$, for which the response of the system is changed from hardening to softening. In addition, Figure 3.6 demonstrates that the increase in proportional gain alone despite the linear case cannot suppress the vibration amplitude.

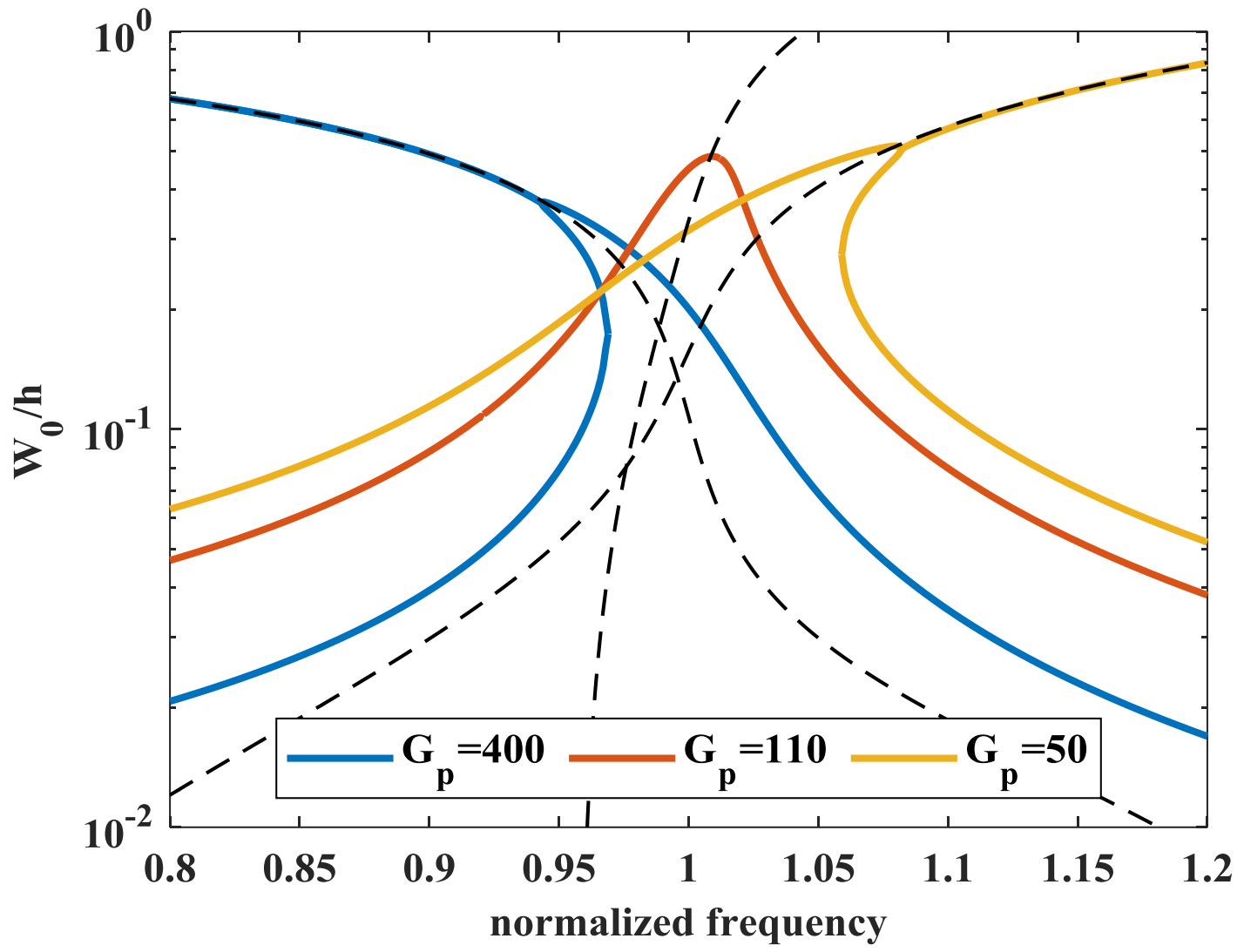

Figure 3.6 Free and forced nonlinear frequency sensitivity with respect to $G_p$ [5]

The algebraic equation of (3.2) is employed to find the amplitude equation with an analogy to the modal analysis, where the nonlinear modal parameters are equated analytically. The sensitivity of the response of the closed-loop system based on a proportional derivative control law is investigated for different parameters for both the linear and nonlinear cases.

Since the proposed framework of the dissertation is given in the geometrically independent form, a systematical approach should be employed in modeling the vibrating system. This problem is tacked through the black-box subspace method and predictive error method (PEM) in the upcoming chapter. The two employed methods are shown in frequency-domain. However, in the controller development, the time-domain representation of these two methods is shown. The connection between the time- and frequency approaches are given for example in [118] and the derivations are not replicated here.

The second issue that should be addressed here is the fact that in the classical form, both methods are developed in the linear domain. The main motivation for selecting this system identification approach is twofold: 1) the model-based robust control techniques that are proposed in the upcoming chapters deal with nonlinearity in locally Lipschitz sense. 2) the extended version of the subspace technique for nonlinear system identification is given based on the linear one used in this dissertation. In fact, in the optimization method of the polynomial state space method (a gradient-based approach), the process is initialized by solution of the linear subspace technique. Consequently, an immediate extension to nonlinear ansatz is available.

# 4 Real-time experimental modeling of MIMO systems

In this chapter, first, the impact of various input excitation scenarios on two different MIMO linear non-parametric modeling schemes is investigated in the frequency domain. It is intended to provide insight into the optimal experiment design that not only provides the BLA of the FRFs but also delivers the means for assessing the variance of the estimations. Finding the mathematical representations of the variances in terms of the estimation coherence and noise/nonlinearity contributions are of critical importance for the frequency-domain system identification where the objective function needs to be weighted in the parametrization step. The input excitation signal design is tackled in two cases, i.e., multiple *single-reference experiments* based on the zero-mean Gaussian and the colored noise signal, the random-phase multisine, the Schroeder multisine, and minimized crest factor multisine; and *multi-reference experiments* based on the Hadamard matrix, and the so-called orthogonal multisine approach, which additionally examines the coupling between the input channels. The time-domain data from both cases are taken into the classical $H_1$ spectral analysis as well as the robust LPM to extract the BLAs.

## 4.1 Alternatives in experiment design

The different choices in selecting an input excitation for system identification purposes, which may be subsequently employed in (model-based) controller design, are profoundly exploited in the pioneering work by Goodwin and Payne as *optimal experiments* [31]. Accordingly, several aspects of input excitation design can be exploited such as the operational bandwidth of interest, the maximum permissible excitation amplitude on actuators, and the sampling time (before running into the task-overrun error). Several signals in terms of measurement duration, accuracy, and sensitivity to noise/nonlinearity are compared in [32] and it is concluded that the multisine signal has a minimum time-factor, i.e., minimum time per frequency for reaching a specified SNR. Since the work of Schoukens et al., several attempts have been made to improve the energy content of the signal, i.e., the property of having a low crest factor (CF) for the input/output data. Mathematically, the CF of a time-dependent vector $x(t)$ is defined as $CF_{x(t)} = \|x(t)\|_\infty / \|x(t)\|_2$ which is the peak-value over the signal root mean square. Consequently, CF minimization is shown to secure a desirable quality in signal processing namely, a high SNR [33].

FRFs are proven to confer in-depth insight into the behavior of complex dynamical systems. In the case of lightly damped dynamical systems, modal tests using the spectral analysis technique is well-studied for parametric estimation of the single-input single-output dynamic systems [34], [35]. However, the estimation procedure of the FR matrix (FRM) for the multivariable systems is technically much more involved since it is obtained via cross-correlation techniques which yield the input excitations to be uncorrelated. Apart from the low-frequency resolution method implemented by [36], which is generally not intended for lightly damped smart structures, one of the major issues of random excitations is that no differentiation between noise and nonlinearity can be deduced from the results.

Contrary to $H_1$/$H_2$ functions in the spectral analysis, FRM and its covariance matrices can be calculated following [37]. Several consecutive periods and several independent realizations of the multi-reference tests should be performed, which in turn provides the means for acquiring the sample means and sample (co-)variances of the input/output spectra. The latter is realized while attenuating the stochastic noise and transients.

In the rest of the chapter, the optimal experiment design problem is revisited for lightly-damped mechanical structures. To show the practicality of the obtained results, a benchmark problem in AVC of smart structures is considered as a case study. Accordingly, first, the input excitation design is considered in both single- and multi-reference scenarios to shed some light on input-channel coupling. Additionally, the effect of various excitation signals is investigated in terms of the accuracy of the obtained FRM, the time factor, and the CF. Consequently, some guidelines are outlined for the user in selecting the appropriate experimental setting.

Moreover, the classical spectral analysis is compared experimentally against the robust LPM in nonparametric modeling. To this end, several advantages of extracting the statistical properties of the obtained FRM, e.g. estimation variance in regards to the noise/nonlinearity, in the latter method are highlighted, which are crucial for the parametric system identification, the state estimation problem, and the robust control design. The proposed procedure amounts to a substantial reduction in the experimental time which may be very expensive in the scope of the lightweight aerospace systems [119].

The plant is shown in Figure 4.1, while geometric dimensions, the material properties of the four piezo-actuator patches and the aluminum host, the technical details of the measurement setup, and sensor and actuator placement optimization are all referred to [103]. The detailed explanation of the controller performance-evaluation setup based on the dSPACE toolchain is postponed to the upcoming chapters where the control design problem is analyzed. In the scope of this chapter, the plant (cantilever beam with piezo-actuators) properties and plant dynamics are investigated in detail.

The system has five inputs, including four collocated piezo-patches (two on each side of the beam) and a shaker, as well as two outputs viz. a laser Doppler vibrometer (LDV) collocated with a 1D accelerometer at the free end of the beam.

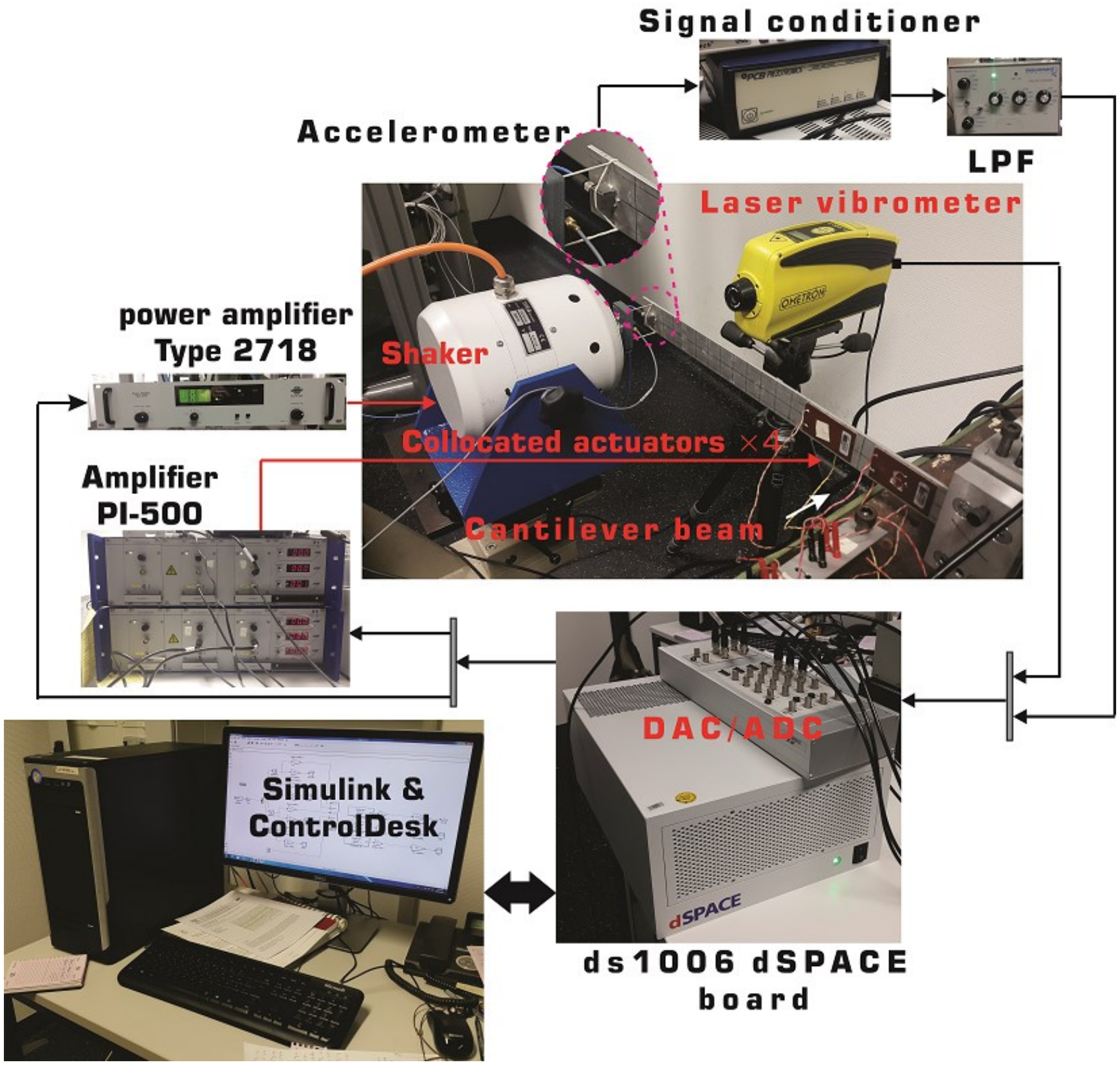

Figure 4.1 The experimental rig for the system identification [120], [121]

In the rest of Chapter 4, $j$ indicates the unit imaginary number, $\omega_k$ is the frequency of the line $k$ in the continuous spectrum of the signal, and the hat operator $(\hat{\cdot})$ and zero subscripts like in $G_0$ symbolize the estimated and unknown correct values of the associated variables, respectively. Capital letters are used for the frequency-domain data obtained using FFT while lower case variables are reserved for the sampled time-domain signals.

## 4.2 Experiment design

The measured sensor signal every so often would not be suitable for a direct connection to ADC due to high impedance and low power. Therefore an additional preprocessing step i.e., conditioning is required before sending to ADC. The twofold goal of the signal conditioning consists of reducing the effect of the acquisition channel on the sensor data as well as scaling the acquired data to cover the range of the ADC for clear reasons.

The first point is to guarantee the matching of the measured impedance to that of the signal. Amplification, as well as rejection, are done here and while the rejection task attenuation is carried out by a passive divider, the amplification task is performed by an active circuit. Trivially both of the additional preprocessing tasks should be done in a frequency-independent form because the data are manipulated otherwise. Also, the hardware is guaranteed to be safe in case of a user error by implementing an over-voltage protection system.

The sampling time is assumed to be equidistant in the experiments and control design and therefore no special treatment is performed in an anti-aliasing system. It is expected that measurement data are affected by erroneous stochastic noise, the extent of which depends on the measurement methods and the instruments involved. Additionally, morphing dynamic systems often have long-lasting transient behavior under periodic loading, which contributes as an additional source of imperfection in the results of the post-processing phase. Unlike the stochastic noise and transients in the response, the effect of nonlinearities in the system output persists even after long measurements for suppressing the non-steady-state history and after averaging over multiple periods.

### 4.2.1 Single-reference experiment

In the single-reference (SR) modal analysis scheme, the input excitation design is initiated with a random Gaussian zero-mean signal (RGS). In order to be able to capture the higher-order nonlinearities, the sampling time is set to 122.07 μs (≈10 times higher than the maximum operational frequency) following the literature in the nonlinear system identification [122]. $2^{16}$ lines are considered in the Fourier analysis which guarantees a sufficient set of samples in the 3dB range of the resonance frequencies in the framework of LPM [123].

As one would expect, the coupling between the input channels is neglected in the single-reference experiments. The FRFs of the system based on the RGS excitations are then compared with the colored noise excitation and the random-phase multisine. To this end, individual experiments are performed for ten periods of each excitation signal to quantify the contribution of the transient noise. The distorted periods in the input/output data are then discarded, followed by spectral analysis of the remaining time history. In Figure 4.2(a), the frequency content of the two random signals is compared to the periodic random-phase multisine excitation. Unlike the RGS, the multisine excitation has a slight advantage in terms of band-limited analysis since it has an insignificant contribution over other frequencies. Though the colored noise signal also satisfies the band-limited constraint of the desired excitation, it is categorized as a non-periodic signal and is thus subjected to leakage errors. The contribution of the transients under multisine periodic excitation for the two measurement outputs is evaluated in Figure 4.2(b). It can be seen that the transient response falls below the noise floor within two consecutive periods. Consequently, the response of the first period is discarded in the classical spectral analysis based on the $H_1$ function.

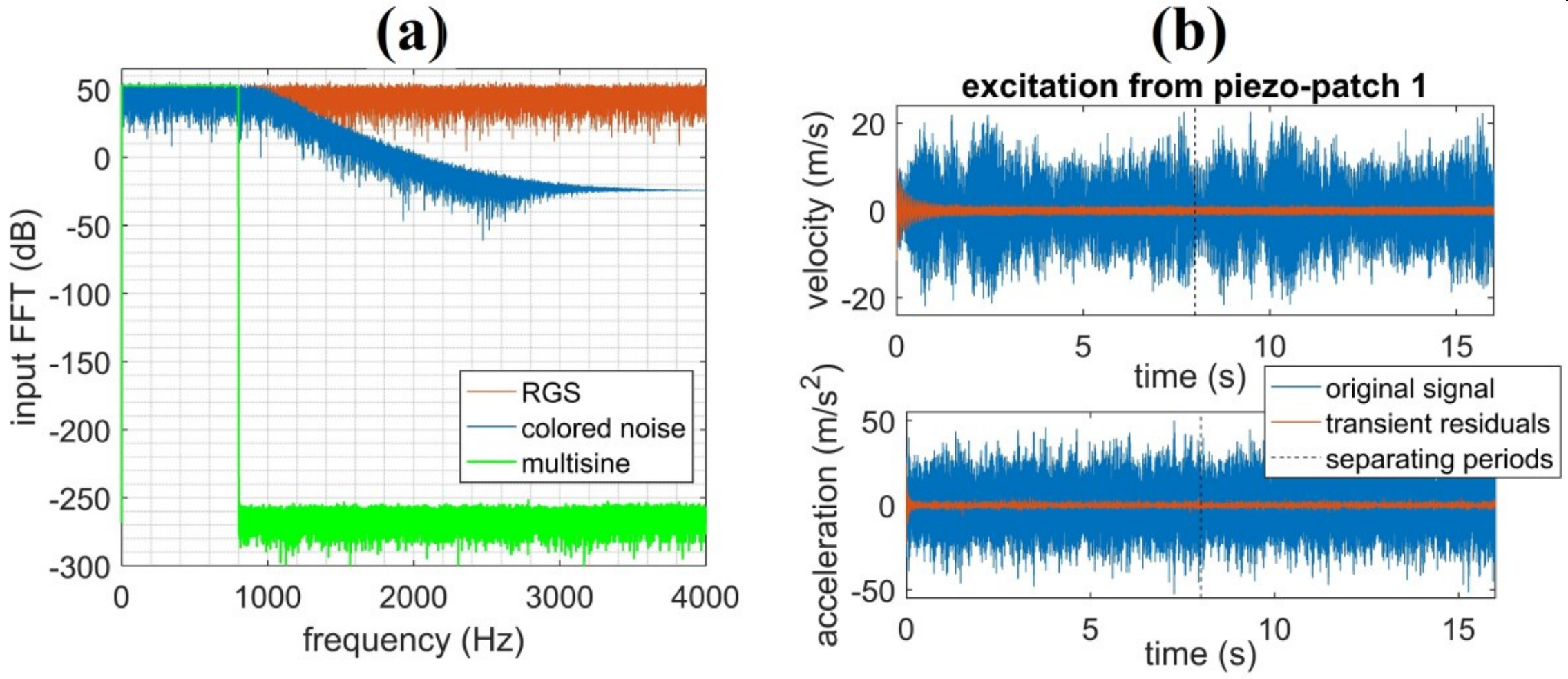

Figure 4.2 (a) Frequency content of the input excitation signals. (b) Contribution of transient noise over two periods [120], [121]

For a periodic multisine signal with a high SNR, it can be shown that under reasonable experimental conditions, e.g., averaging over the periods, we have $\hat{G}(j\omega_k) \approx G_0(j\omega_k)$ where $G_0$ is the unknown unpolluted transfer function. However, for random excitations, even neglecting the effect of persistent structural nonlinearities, a biased estimation of the FRFs, given by $\hat{G}(j\omega_k) = G_0(j\omega_k)(1 + N_Y(k)/Y_0(k) - N_U(k)/U_0(k))$, is still expected since the measurement output $y$ and excitation input $u$ are polluted by stochastic output and input noise ($n_y$, and $n_u$), respectively. The Capital notation e.g., $N_U$ represents the Fourier transform of the corresponding time variable e.g. $n_u$, The systematic error in random excitation cannot be resolved unless the SNR for the input signal generator ($|u| \gg |n_u|$) is high.

Under the assumption that the SNR at the input is higher than that of the output, $H_1$ (as opposed to $H_2$) estimation is used to calculate the FRFs as $\hat{G}(j\omega_k) = \hat{S}_{YU}(k)/\hat{S}_{UU}$ with $\hat{S}_{YU}$ and $\hat{S}_{UU}$ representing the sampled estimation of the cross-spectrum, and auto-spectrum of the output and input. The FRFs of the system under the three loading scenarios are plotted in Figure 4.3(a). In Figure 4.3(a) and Figure 4.3(b) (as well as all subsequent best linear approximation (BLA) figures throughout the remainder of the chapter), each subplot indicates the FRF from the piezo-actuators and shaker to the measurement outputs, (accelerometer and LDV), represented in columns 1-4 or 1-5, and rows 1-2, respectively. As a measure of the FRF quality, the coherence of the captured output w.r.t. the excitation input as a measure of the FRF quality is shown in Figure 4.3(b) and is defined as $\gamma^2 = |\hat{S}_{YU}|^2/\hat{S}_{UU}\hat{S}_{YY}$ with $\hat{S}_{YY}$ representing the sampled estimation of the auto-spectrum of the output. In Figure 4.3(a) and Figure 4.3(b), the first period of the periodic excitation is discarded to suppress transient distortions (see [124]), and a Hann-window with 80 averages is used to reduce the leakage errors for the two non-periodic inputs. The transient distortions refer to the transient part of the system response to a periodic excitation which dies out gradually on each period depending on the system's inherent damping. It is well-known that these transients distort the FRFs.

The observations are as follows: 1) The RGS signal has the worst coherence, and as a result, the obtained FRFs are unreliable. As pointed out in [124], although the clipped random noise with uniform distribution has a better performance than the RGS, it is recommended to pre-filter (blue lines in Figure 4.3) the excitation signal with/without sign operation (random binary excitation).

2) Despite the similarity between the FRFs of the MIMO system in Figure 4.3(a) for both filtered noise and multisine (multisine: SR) signals, the coherence comparison in Figure 4.3(b) reveals the superiority of the estimation quality of the multisine signal.

3) The coherence of all signals in the vicinity of the anti-resonance frequencies drops significantly below 1. In the case of multisine, this can be remedied by replacing the uniform distribution of the excitation lines with a spectrum configuration with populated lines around anti-resonances. This guarantees the injection of enough energy at those frequency ranges, resulting in a higher SNR.

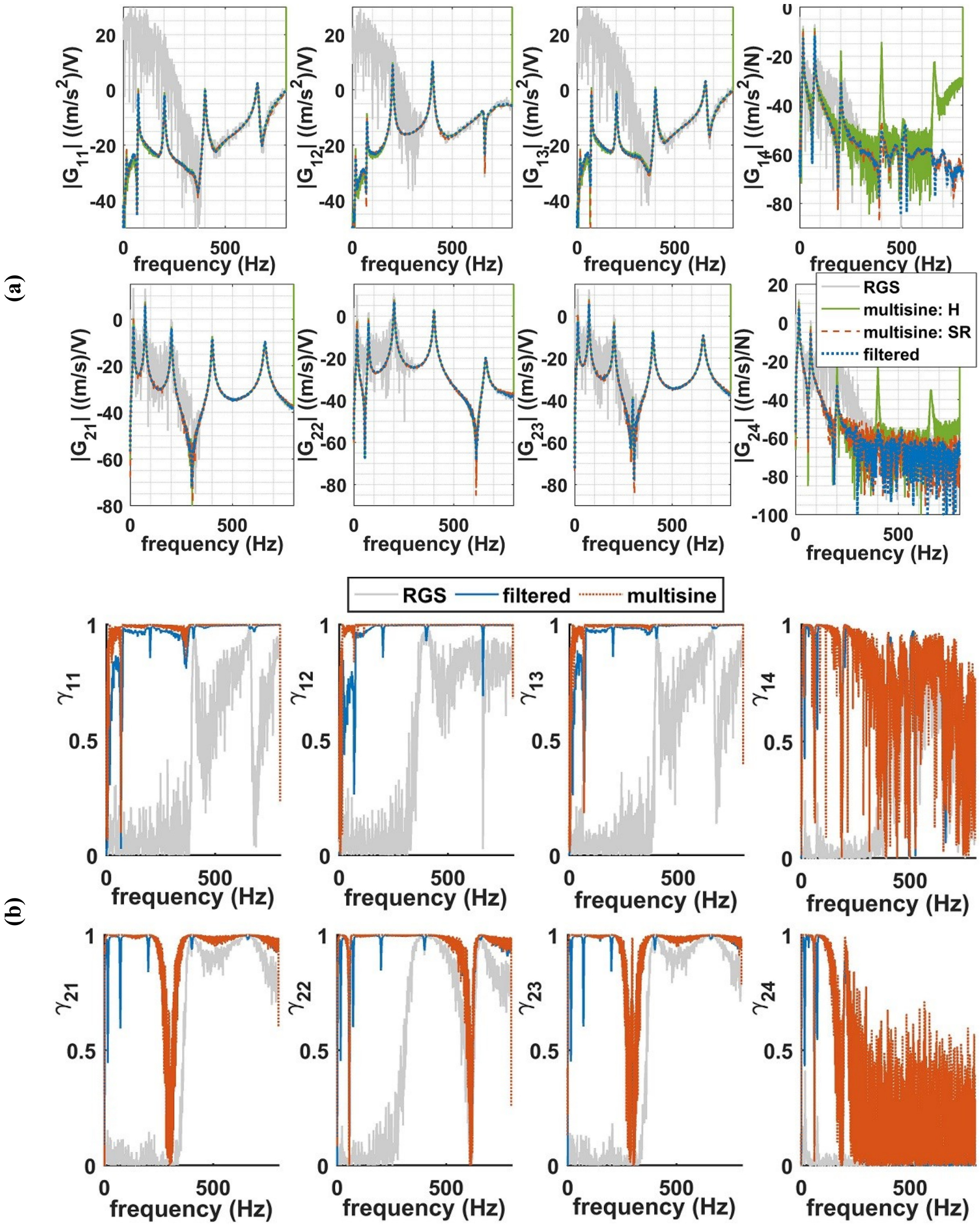

Figure 4.3 (a) Magnitude of FRFs in dB range based on three single-reference loading and Hadamard multi-reference scenarios. (b) Coherence diagram of the measured FRFs [120], [121]

However, there is a trade-off in the form of reduced resampling time of the experiment, i.e., a reduction in resampling time may result in hardware memory shortage which is of particular concern in practical situations involving lightly-damped structures with long transient behavior and consequently lengthy measurements.

4) In order to investigate the input coupling, or in other words, the essence of performing multi-reference modal analysis instead of multiple single-reference ones, the BLA based on the Hadamard multisine method is compared with single-reference results. Figure 4.3(a) shows that the multi-reference excitation (the green line, multisine: H) and the calculated FRFs, excluding the shaker channel, have similar outcomes. Although no apparent coupling between the piezo-patches is observed, the distorted FRF associated with the shaker (and both outputs) is evidence of input coupling between the shaker and the piezo-patches. This behavior is due to the flexible assembly of the shaker and the beam which are connected to each other via rubber band as shown in Figure 4.1. It should be noted that the aforementioned coupling is no longer observed when the shaker and beam are connected via a screw [13].

The technical details of the Hadamard matrix are briefly discussed in the following subsection. It is important to note here that the crest factor of the excitation signals is not changed in the multi-reference scenario unless, as is the case for the shaker input coupling, similar non-parametric modeling results are expected.

### 4.2.2 Multi-reference experiments

In the multi-reference analyses (for $n_u$ number of input channels), two scenarios are investigated based on the number of input signals:

1) The case where three piezo-actuators (realizing the control inputs) along with the shaker (realizing the mismatch disturbance channel), are included in the Hadamard multisine approach, i.e., four channels to satisfy the radix-2 condition, i.e., $n_u = 2^m$ [124]. In this approach, the single-reference signal is multiplied by the Hadamard matrix $T = 1/\sqrt{n_u} H_{2^m}$ where $H_{2^m} = H_2 \otimes H_{2^{m-1}}$, $H_2 = [1,1; 1, -1]$, (in MATLAB matrix notation). The Kronecker matrix product is represented with $\otimes$. Experimentally, Hadamard multisine is realized by employing a set of inverters (see [33], [125]).

2) The case of an orthogonal multisine based on the multi-generator approach for an odd number of inputs (four piezo-actuators and the shaker) is carried out according to the procedure outlined in [125]. Unlike the Hadamard multisine where the number of input channels must satisfy the radix-2 condition, the so-called orthogonal multisine can be used for an arbitrary number of input channels with the orthogonal elements of the matrix given by $T_{p,q} = n_u^{-1/2} \exp j(p-1)(q-1)/n_u$ for $p, q = 1, \dots, n_u$.

To this end, the experiments are performed with the sampling frequency of 8192 Hz, and the results are generated in a compatible form of robust LPM. Each realization encompasses $n_u$ number of individual experiments (equal to the number of inputs) which are performed for ten consecutive periods. Relying on the minimum number of realizations in the robust LPM [124], four and five realizations of the multi-reference random-phase multisine signals are applied in 16 and 25 individual experiments for the Hadamard and orthogonal multisine approaches, respectively. Unlike the Hadamard multisine approach where the multi-reference excitation can be produced by a single generator and a set of inverters, orthogonal multisine experiments require $n_u$ independent generators. The standard deviation of the excitation signals in both cases is retained at 0.75 to keep the actual implemented signals on the piezo-actuators beneath 250 V in amplitude. Figure 4.4 presents the results of the robust LPM based on the two multi-reference schemes.

The following observations can be made based on the results shown in Figure 4.4: 1) As expected, unlike the accelerometer, the contactless LDV is less prone to noise. This can be deduced from the matching quality between the two multi-reference schemes as well as from the significant difference between the acceleration measurements.

2) For the same RMS value of the excitation signal, BLAs obtained from the Hadamard single-generator matrix method have higher total variance in comparison to the orthogonal multisine scheme. This is justifiable by examining the noise floor in the two cases, i.e., comparing NV:D and NV:H for the accelerometer, which indicates high achievable SNR in the orthogonal multisine approach. Consequently, the orthogonal approach is preferred for parameterized modeling.

3) The FRFs associated with the shaker input is severely distorted at frequencies higher than 100 Hz. Unlike the classical $H_1$ function in Figure 4.3(a) and its coherence in Figure 4.3(b) which indicate unreliable modeling quality at these frequencies, the source of distortions in the robust MIMO LPM method of [123] are associated with noise/nonlinear contributions. Physically, the nonlinear distortion is due to the nature of the connection between the shaker and the beam, i.e., the rubber band (see Figure 4.1). The reason for using a rubber band instead of a direct connection (adhesive wax, screws, etc.) is to match the impedance of the (control) input signals realized by the piezo-patches with the (disturbance) signal generated by the electromagnetic shaker [124].

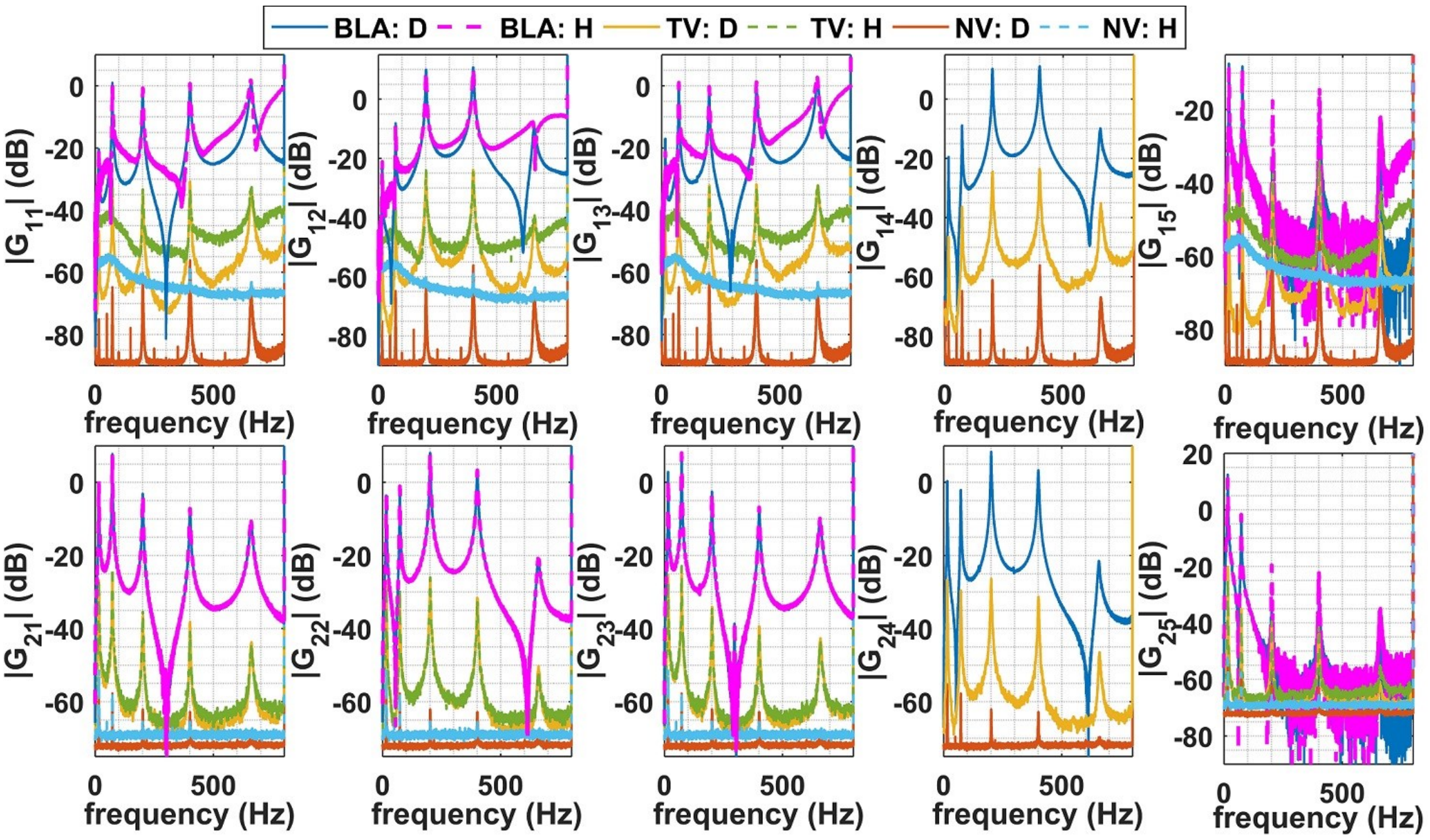

Figure 4.4 BLA of the FRF using Hadamard matrix approach (H) and orthogonal multisine (D) as well as noise variance (NV) and total variance (TV) [120], [121]

4) In the results of the orthogonal multisine method, the contribution of nonlinear distortions can be neglected since the total variance is 40 dB below the BLA. However, depending on the application, this may become non-negligible. Before proceeding to the assessment of input excitation optimization in MIMO smart structures, two remarks should be made regarding the importance of the MIMO robust LPM: a) The total variance not only reflects the quality of the estimated FRFs but also provides a tool for quantifying the modeling uncertainty which is crucial in robust control design. Additionally, it provides the means for state observer design techniques, e.g., Kalman filter that may be used in output feedback control by quantifying the process/measurement noise's covariance matrices. b) The parameterization of the calculated BLAs based on the black/grey-box subspace method reduces to a weighted regression problem where the obtained total variance from the MIMO LPM serves as the frequency-dependent weighing without which the estimated linear model would be biased [126], [127].

## 4.3 Input excitation optimization

### 4.3.1 Crest factor minimization

Since the spectrum of the designed excitation signal is known *a priory*, the minimization of the CF is defined in terms of the random phases associated with the active line in the multisine excitation. Active lines, as one of the input design parameters, are the frequencies at which significant energy is being injected to the system. In time-domain realization of multi-sine excitation, the random phases are the phase part of each sine function in the summation that is involved over all sine functions. Since the CF-minimized multisine is unique, the contribution of the inherent system nonlinearities is not quantifiable in the framework of the robust LPM [124]. On the other hand, as an advantage of CF minimization, the number of required averages for a specific accuracy regarding the SNR at low-frequency ranges, where the experiment durations are lengthy, is proportional to the square of the CF.

The analysis here is only concerned with optimizing the phases (excluding the zero line) of each line for a given auto-power spectrum. Since the objective function ($CF_{x(t)}$) is nondifferentiable, an analytical optimization solution is unavailable. As a result, the Quasi-Newton (QN) algorithm, in which the Hessian matrix is estimated (updated) from the gradient vector, is implemented. Accordingly, the Hessian matrix is calculated by the Davidon-Fletcher-Powell (DFP) formula to mimic the Newton algorithm in computing the search direction [128]. The algorithm is initialized by the Schroeder multisine, i.e., for line number $i$ and $n_l$ number of nonzero lines in the spectrum, the associated phase is initialized with $-i(i-1)/n_l$.

For the AVC benchmark problem shown in Figure 4.1, the frequency range of [0 800] Hz is discretized with all radix-2 number of active lines between 512 and 131072. This is done to show the explosion of the computational time when moving towards high-frequency resolution. For optimization purposes, a parallel computing Linux workstation with 28 cores @ 2.40 GHz (Intel Xenon E5-2680 v4) and 96 GB RAM is employed.

The objective function (CF) of the optimization is defined in terms of a wide range of parameters, namely, achievable CF for the optimized signal, variations of the CF for the Schroeder multisine, the number of required function evaluation in QN optimization scheme, and the CPU time, as shown in Figure 4.5.

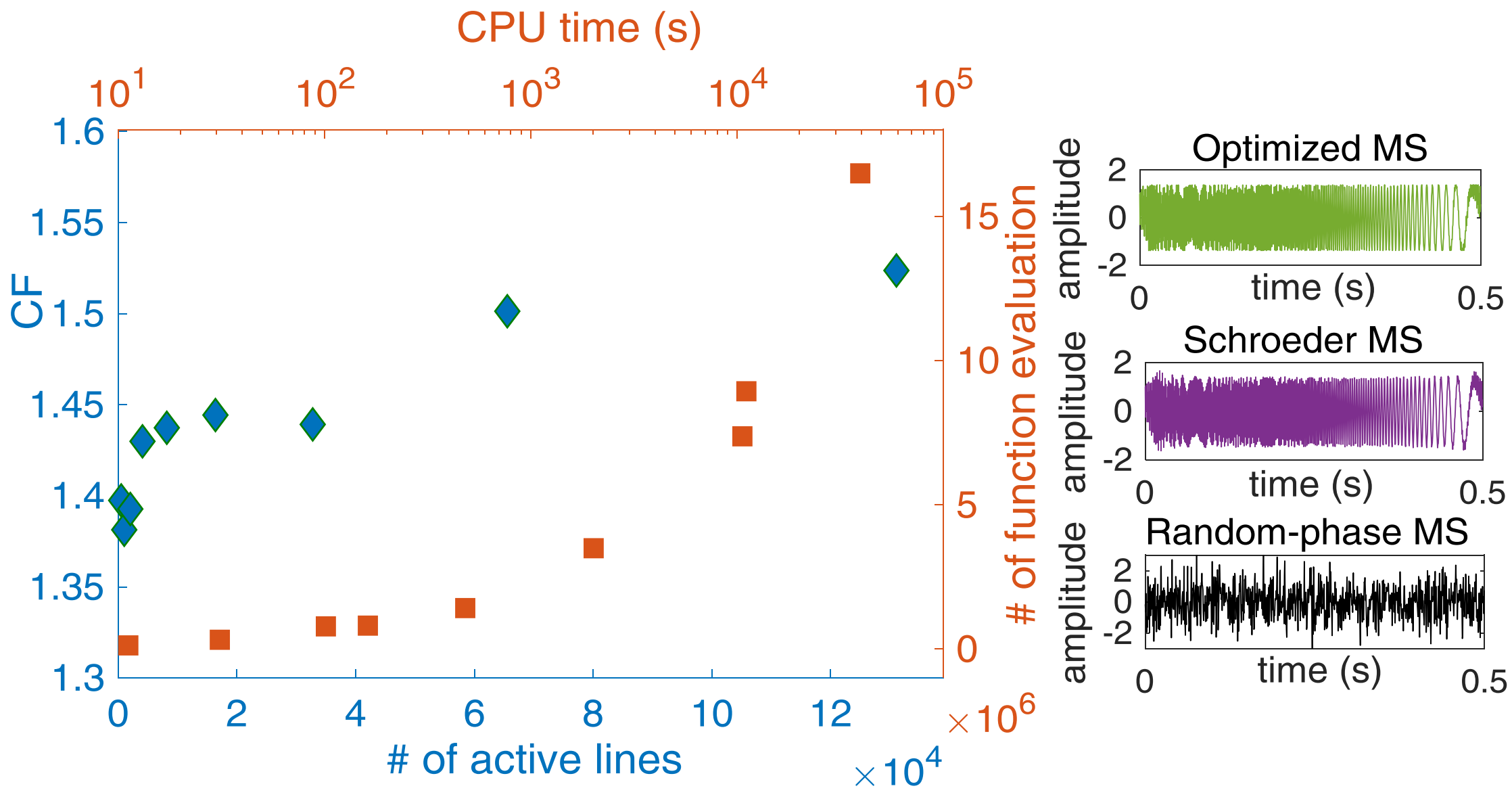

Figure 4.5 Effect of the frequency resolution of the multisine excitation on the CF and subsequent computational overhead [120], [121]

Figure 4.5 consists of two data set, i.e. the dependency of crest factor on the number of active lines (frequency resolution) shown with blue axis and blue rhomboids and the dependency on the number of function evaluation w.r.t. CPU time needed to be shown with orange axis and squares. It can be seen that for higher frequency resolution, the crest factor increases. As the demanded resolution in the frequency domain increases, the number of required function evaluations (and as a result, the CPU-time) also increases due to the number of involved optimization variables. The typical optimized CF for the multisine is estimated between 1.4 and 1.5, while the CF calculated for the Schroeder multisine, which is suppressed in Figure 4.5 for the sake of briefness, is typically between 1.65-1.7 [124]. Additionally, the CF associated with the random-phase multisine for $10^6$ realizations varied between 2.5 and 6.3, further emphasizing the importance of testing various random-phase multisine signals before deciding to use one in an experiment. Moreover, three subplots are added to Figure 4.5 to compare the time history of the random-phase multisine, the Schroeder multisine, and the CF-optimized multisine for 2048 lines in the frequency range interest. Although the random-phase multisine exemplifies a stochastic-like behavior, its amplitude spectrum is deterministic.

On the other hand, comparing the time histories of the Schroeder and minimized-CF multisine signals indicates the reduction in magnitude (despite higher injected energy per line) of the optimized signals. Finally, it should be noted that the optimization should be done over each line in the frequency domain since time domain optimization induces additional computational burden. This is because a line in the frequency domain has a trigonometric nonlinear realization in the time domain.

### 4.3.2 Multi-Reference Non-parametric Modeling with CF-minimized Multisine Excitation

In this section, first, the system in Figure 4.1 is excited through the input channels by the multi-reference (CF-minimized) signal. Then, the obtained time histories are analyzed with the robust LPM. In this regard, the modal analyses are performed for the optimized signal with 65536 number of lines for the Nyquist frequency band of 2048 Hz for eight consecutive periods of the Hadamard multisine and orthogonal multisine frameworks, the results of which are illustrated in Figure 4.6.

From Figure 4.6, one observes that although the frequencies of the resonant states are the same for both the CF-optimized and the random-phase multisine cases in the multi-reference experiments, the FRFs in the anti-resonances and transition frequencies between the resonances are significantly different. Noting that the noise floor is independent of the optimization (compare Figure 4.6 and Figure 4.4 for NV), it is essential to assess the introduced mismatch due to the CF minimization. To this end, a perturbation analysis of the clamped-free beam geometry in Figure 4.1 is carried out in the operational frequency range using ABAQUS finite element (FE) software. It is observed that imperfect boundary conditions and the attached sensor configurations lead to the excitation of the torsional and in-plane mode shapes. Due to the higher energy content that is injected through the active lines of the multisine signal, the first and second in-plane and torsional modes are significantly excited, consequently distorting the FRFs associated with transverse vibrations. Hence, the frequencies that are highlighted by 3, 5, 7, and 8 in Figure 4.6, indicate the first in-plane mode, the first and second torsional modes, and the second in-plane mode that is irrelevant to the transverse vibration. The transverse vibration modes associated with the frequencies 1, 2, 4, 6, and 9 in the BLAs of Figure 4.6 are also plotted on top of the figure, respectively. Unless the user tends to identify these modes (torsional/in-plane) and the actuators can control them, the optimization approach may cause an incorrect interpretation of the system. Additionally, this method provides no insight from the BLA regarding nonlinear variance. It should be noted that if the geometrical nonlinearities due to the large vibration amplitudes are relevant to the vibrations, CF minimization is not recommended since the total variance of the estimated BLAs is expected to increase significantly.

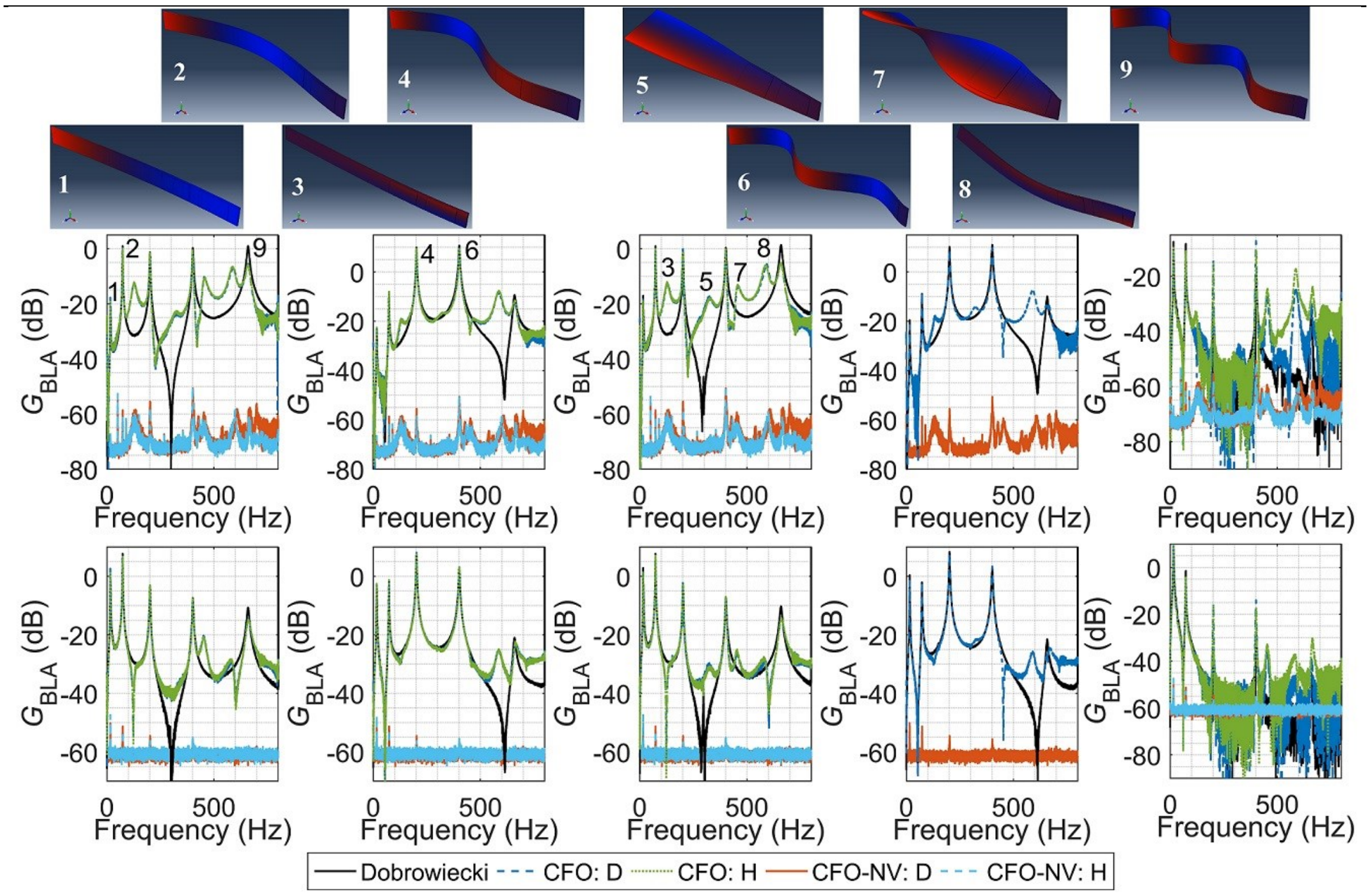

Figure 4.6 BLA based on optimized CF (CFO) using Hadamard multisine (H) and orthogonal multisine (D), the noise variance (NV) compared to random-phase multisine (black line) [120], [121]

This prediction is a direct result of invoking the higher-order strain terms in system dynamics [111], [129].

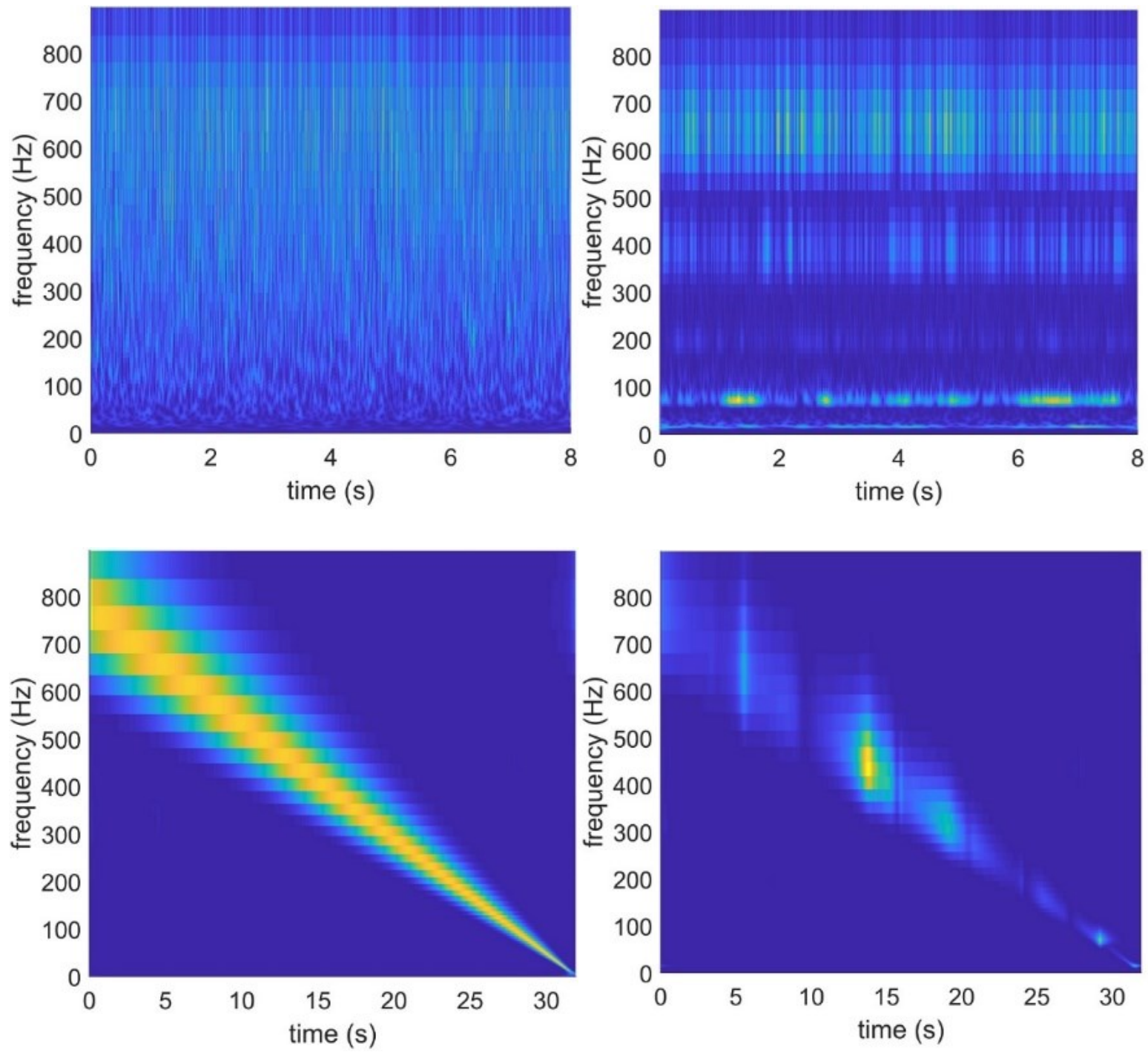

Figure 4.7 Time-frequency analysis of the input/output data [120], [121]

This can be justified by analyzing the time-frequency content of the two cases i.e., random-phase multisine and CF-minimized multisine. To this end, the time-frequency analysis based on the continuous Morlet wavelet transformation (the technical details of which is referred to [122]) is performed on the input/output data in the two cases, and the normalized results are shown in Figure 4.7. In this figure, the top row is reserved for random-phase multisine while the bottom row is dedicated to the minimized CF. The first and second columns represent input and output, correspondingly. Unlike the random-phase multisine, the spectrum of the input in the case of minimized CF follows a specific line of harmonics that resembles the sweep-sine excitation. Consequently, the time-frequency analysis over the measurement output of the minimized CF case reveals that the energy of the input signal is injected at a specific frequency range at each time increment. However, the output of the system under random-phase multisine input indicates the random distribution of energy at all frequencies at each time sample.

Notwithstanding the inability of the method to detect nonlinearities, CF minimization should be implemented with caution and is not recommended without appropriate insight into the system, especially in combination with black-box identification.

Several input excitation signals are tested experimentally and the results in combination with $H_1$ function and robust LPM in single-/multi-reference schemes are used to extract the FRM of the system. Moreover, quantified measures of the imperfections due to stochastic noise, transient distortion, and nonlinear structural behavior are calculated. In addition to the technical conclusions in each case, several guidelines are provided for the vibration engineer regarding the selection of the optimal experiment configurations depending on the accuracy of involved measurement devices and the potential insight that one may have regarding the nonlinearity/noise level. The results of the chapter enable not only the use of the estimated covariance matrices in both single-step, e.g., subspace method and iterative, e.g., predictive error method of parametric identification methods, but also facilitate lumped uncertainty quantification and state observer design.

## 4.4 Frequency-domain Subspace Identification of Dynamical Systems for Robust Control Design

Black-box system identification is subjected to the modeling uncertainties that are propagated from the non-parametric model of the system in the time/frequency domain. Unlike classical $H_1$/$H_2$ spectral analysis, in the recent robust Local Polynomial Method (LPM), the modeling variances are separated to noise contribution and nonlinear contribution while suppressing the transient noise. On the other hand, without an appropriate weighting on the objective function in the system identification methods, the acquired model is subjected to bias. Consequently, the weighted regression problem in subspace frequency-domain system identification is revisited in order to have an unbiased estimate of the frequency response matrix of a flexible manipulator as a multi-input multi-output lightly-damped system. Although the unbiased parametric model representing the BLA of the system in this combination is a reliable framework for the control design, it is limited for a specific SNR ratio and STD of the involved input excitations. As a result, in this chapter, an additional step is carried out to investigate the sensitivity of the identified model w.r.t. SNR/STD in order to provide an uncertainty interval for robust control design.

### 4.4.1 Background study

Motivated by the need for active vibration control of the hydraulically actuated 7 DOF manipulator modeled in [130], [131], the frequency-domain subspace system identification for robust control design is illustrated by this example. Compared to the typical robots driven by electric motors, hydraulic actuator robots are more flexible due to the higher loop gains, wider bandwidths, and lightly-damped nonlinear dynamics. This flexibility depends on the parameters such as the weight, the dimension, the payload, and the speed of the manipulator. This results in induced vibrations both during and after the motion of the robot preventing the precise functioning of the robot. This, in turn, degrades the repeatability of the manipulator, particularly in high-speed applications.

It is proven that the vibration of the single or multi-link robots can be suppressed using active vibration control techniques. Although the mathematical model of the flexible arm can be obtained using finite element methods, a more viable approach to cope with uncertainties and nonlinearities of the system for both deification and control relies on the data-driven models [34], [35]. Contrary to the classical identification methods, robust identification algorithms use *a priori* information on the system and its input-output data to produce a nominal model and its associated uncertainty. A comparative study of three primary robust identification approaches, i.e., Stochastic Embedding, Model Error Modeling, and Set Membership, for a lightly-damped system is studied in [132]. The main disadvantage of using the classical spectral analysis in these works for measuring the FRF is that the difference between the measurement noise and the system nonlinearity cannot be distinguished. The measurement noise contributes to the variance error of the estimated FRF while unmodeled dynamics result in the estimation bias. Variance error is uncorrelated with the input signal (in open loop data collection case), but the bias error strongly depends on the nominal model structure and the input signal design. To address this problem, a multi-sine input signal in the multi-reference orthogonal experiment setting is designed and tested for multivariable identification of the frequency response matrix of the system.

To avoid the shortcomings of the classic methods for estimating the FRM, a more advanced approach based on the so-called LPM is investigated [37]. This technique enables estimation of the FRM of the system using periodic excitations. In the LPM framework, the contribution of the noise leakage is approximated by a local least square problem (Pintelon, Vandersteen, et al. 2011). This results in suppression of the noise leakage (transient) errors in the calculation of the sample mean/covariance matrices. The technical details and MATLAB implementation of the robust and fast versions of the LPM method can be found in [124].

### 4.4.2 Robust local polynomial method

As a result of applying the multi-reference periodic signal designed in the previous section (see Figure 4.1), the structure exhibits both transient and steady-state responses. Since the transient response is undesirable for FRM identification, it is treated as noise. This is referred to as noise leakage and the robust LPM technique is utilized to suppress its effect in the calculation of the sample mean/covariance matrices. For this purpose, $n_u$ multi-reference modal analyses are performed under full random-phase multi-sine for several realizations. Here, each realization refers to an individual Hadamard multi-sine with several ($P$) consecutive periods; for which the noise variance is updated by comparing the different periods. The nonlinear variance is also obtained by comparing the response of different realizations [37]. This means that for the test rig shown in Figure 4.1, four realizations of the multi-reference Hadamard multi-sine signals are applied (in 16 individual experiments) and ten consecutive periods of the response are recorded. By setting the order of the local polynomial approximation to six and the degree of freedom of the (co-)variance estimates to eight, the robust LPM is used to estimate the FRM and the sample means/covariance matrices of the input/output DFT spectra. The results are summarized in Figure 4.8. In this figure, the subscript $i = 1,2, j = 1, \dots, 4$ in $G_{ij}$ indicate the output measurements and input excitation signals (three actuators and a shaker). In this figure, $|G_{ij}|$ represents the amplitude (phase diagram suppressed) of the corresponding FRFs. The following remarks are briefly deduced:

1) By comparing the noise floor for two measurement outputs, it can be seen that, unlike the accelerometer, the contactless LDV is less prone to noise.

2) It can be observed that the total variance (measurement noise & nonlinearity) is $\approx 40$ dB below the BLA of FRM. This indicates that under the ideal laboratory conditions, the variance of FRM is negligible.

3) The FRFs of the shaker are severely distorted at frequencies higher than 100 Hz. It can be seen that both the noise and nonlinearity levels are high in this case which can be physically justified by the nature of the connection between the shaker and the beam, i.e., rubber band (see Figure 4.1). The reason for using a rubber band instead of a direct connection (adhesive wax, screws, etc.) is to match the impedance of the (control) input signals realized by piezo-patches and the (disturbance) signal generated by the electromagnetic shaker [124].

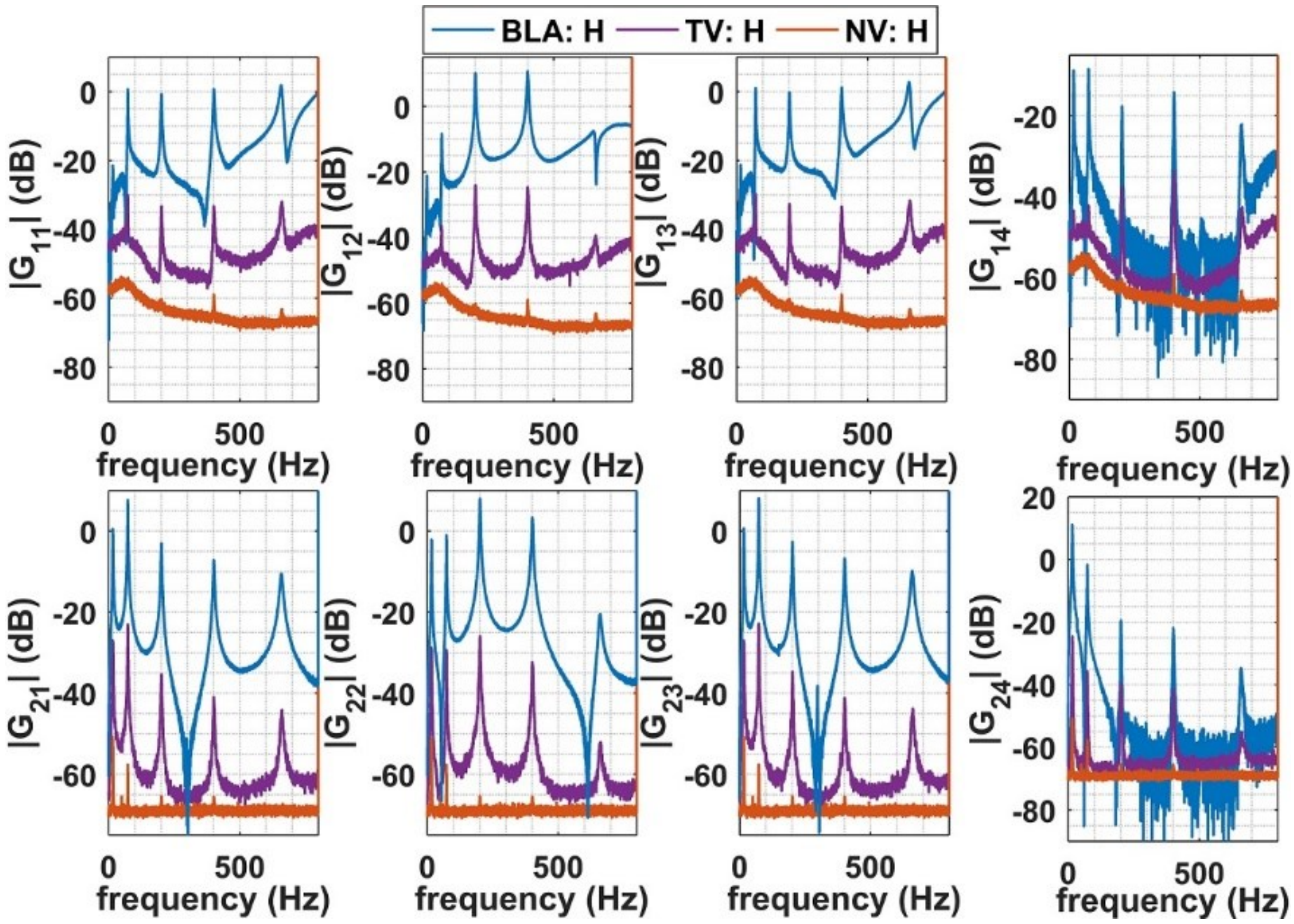

Figure 4.8 BLA of the FRF using Hadamard multi-sine (H), noise variance (NV), and total variance (TV) [120], [121]

### 4.4.3 Parametric state space modeling

#### 4.4.3.1 Black-box subspace system identification

The black-box subspace system identification is based on the estimation of the extended observability matrix $O$ from the discrete Fourier transformation (DFT) of the continuous state space representation $e^{j\omega}\mathbf{X} = A\mathbf{X} + B\mathbf{U}$, and $\mathbf{Y} = C\mathbf{X} + D\mathbf{U}$. If we measure the FRM of the system at $M + 1$ discrete normalized frequency $\omega_k = \pi k/M$ ; $k = 0, \dots, M$, then the extended sampled transfer matrix ($G_{M+k} = G^*_{M-k}$; $k = 1, \dots, M - 1$) is used to calculate 2M-point inverse discrete Fourier transformation (IDFT) as in [126]. For user-defined scalars $q, r \in \mathbb{N}^+$ that satisfy $q > n$ and $r \geq n$, the Hankel matrix is calculated as

$$\hat{H} = \begin{bmatrix} \hat{h}_1 & \hat{h}_2 & \cdots & \hat{h}_r \\ \hat{h}_2 & \hat{h}_3 & \cdots & \hat{h}_{r+1} \\ \vdots & \vdots & \ddots & \vdots \\ \hat{h}_q & \hat{h}_{q+1} & \cdots & \hat{h}_{q+r-1} \end{bmatrix}. \tag{4.1}$$

where $\hat{h}_i = {}^{1}\!/_{2M} \sum_{k=0}^{2M-1} G_k e^{\frac{j2\pi ik}{2M}}, i = 0, \dots, 2M - 1$. Here $n$ is the number of highest singular values of $\hat{\Sigma}$ that are gathered in $\hat{\Sigma}_s$ such that in the SVD of $\hat{H} = \hat{U}\hat{\Sigma}\hat{V}^T$ we have $\hat{H} = [\hat{U}_s \quad \hat{U}_o]\mathrm{diag}(\hat{\Sigma}_s, \hat{\Sigma}_o)[\hat{V}_s \quad \hat{V}_o]^T$. Moreover, diag(.) denotes the diagonalized vector of entities following the MATLAB notation. Then, the estimated state matrix $\hat{A} \in \mathbb{R}^{n \times n}$ and output matrix $\hat{C} \in \mathbb{R}^{p \times n}$ are correspondingly calculated as $\left(J_1\widehat{U}_s\right)^{\dagger}J_2\widehat{U}_s$ and $J_3\widehat{U}_s$ with $(.)^{\dagger}$ denoting the Moor-Penrose pseudoinverse and $J_1 = [\mathbf{I}_{(q-1)p} \quad \mathbf{0}_{(q-1)p\times p}]$, $J_2 = [\mathbf{0}_{(q-1)p\times p} \quad \mathbf{I}_{(q-1)p}]$, and $J_3 = [\mathbf{I}_p \quad \mathbf{0}_{p\times(q-1)p}]$.

The solution for the estimates of the input matrix $\widehat{B} \in \mathbb{R}^{n\times m}$ and feedthrough matrix $\widehat{D} \in \mathbb{R}^{p\times m}$ is determined by solving the batch least square problem in $\widehat{B}, \widehat{D} = \arg\min_{B,D} \sum_{k=0}^{M} \left\| G_k - D - \widehat{C}\left(e^{j\omega_k}\mathbf{I} - \widehat{A}\right)^{-1}B \right\|_F^2$ where $\|.\|_F$ represents the Frobenius norm [133].

If the noise model on the measurement output is known ($\mathbf{Y} = C\mathbf{X} + D\mathbf{U} + \mathbf{V}$), assuming that the elements of $\mathbf{U}$ are defined $U(\omega) = e_i$ (with $e_i$ being the unitary vector in orthonormal space), the DFT of the state-space equation can be derived as $e^{j\omega}\mathbf{X}^{\mathbf{C}} = A\mathbf{X}^{\mathbf{C}} + B$, and $\mathbf{G} = \mathcal{O}\mathbf{X}^{\mathbf{C}} + \Gamma\mathbf{W}_m + \boldsymbol{N}$, where $\mathbf{X}^{\mathbf{C}}$ is the state vector under $U(\omega) = e_i$ input and [126]

$$\mathbf{G} = \frac{1}{\sqrt{M}}\begin{bmatrix} G(1) & G(2) & \cdots & G(N) \\ \phi_1 G(1) & \phi_2 G(2) & \cdots & \phi_N G(N) \\ \vdots & \vdots & \ddots & \vdots \\ \phi_1^{q-1}G(1) & \phi_2^{q-1}G(2) & \cdots & \phi_N^{q-1}G(N) \end{bmatrix},$$

$$\mathcal{O} = [C^T \quad A^T C^T \quad \ldots \quad A^{T^{q-1}}C^T]^T,$$

$$\mathbf{W}_m = \frac{1}{\sqrt{M}}\begin{bmatrix} \mathbf{I}_m & \mathbf{I}_m & \cdots & \mathbf{I}_m \\ \phi_1\mathbf{I}_m & \phi_2\mathbf{I}_m & \cdots & \phi_N\mathbf{I}_m \\ \vdots & \vdots & \ddots & \vdots \\ \phi_1^{q-1}\mathbf{I}_m & \phi_2^{q-1}\mathbf{I}_m & \cdots & \phi_N^{q-1}\mathbf{I}_m \end{bmatrix}, \tag{4.2}$$

$$\Gamma = \begin{bmatrix} D & 0 & \cdots & 0 \\ CB & D & \cdots & 0 \\ \vdots & \vdots & \ddots & \vdots \\ CA^{q-2}B & CA^{q-3}B & \cdots & D \end{bmatrix},$$

$$\boldsymbol{N} = \mathbf{W}_p \operatorname{diag}(n_1, n_2, \ldots, n_M),$$

in which $\phi_k = e^{j\omega_k}$ and $n_i, i = 1, \ldots, M$ are noise samples on output. Then, for a strongly consistent estimate, while $M \to \infty$, we define $\mathcal{N} = [\operatorname{Re}\boldsymbol{N} \quad \operatorname{Im}\boldsymbol{N}]$ which satisfies $\operatorname{E}\{\mathcal{N}\mathcal{N}^T\} = \operatorname{Re}\left(\mathbf{W}_p \operatorname{diag}(R_1, R_2, \ldots, R_M)\mathbf{W}_p^*\right)$ in which $\mathbf{W}_p$ has the same form as $\mathbf{W}_m$ while replacing $\mathbf{I}_m$ with $\mathbf{I}_p$. Assuming that the covariance functions ($R_k$) are known (see the previous section), and applying the Cholesky factorization, the matrix $\boldsymbol{K} \in \mathbb{R}^{qp\times qp}$ is constructed that satisfies $\boldsymbol{K}\boldsymbol{K}^T = \alpha\operatorname{E}\{\mathcal{N}\mathcal{N}^T\}$ for some $\alpha \in \mathbb{R}$ [134]. Here $\operatorname{E}\{.\}$ and $\operatorname{Re}(.)$ represent the expectation operator and the real part of a complex-valued matrix, respectively. Then, having $\mathbf{G}$, $\mathbf{W}_m$, and $\boldsymbol{K}$ in hand, similar to the noise-free output, the system matrices can be estimated using QR factorization of the modified matrix shown in

$$\begin{bmatrix} \operatorname{Re}(\mathbf{W}_m) & \operatorname{Im}(\mathbf{W}_m) \\ \operatorname{Re}(\mathbf{G}) & \operatorname{Im}(\mathbf{G}) \end{bmatrix} = \begin{bmatrix} \mathbf{R}_{11} & \mathbf{0} \\ \mathbf{R}_{21} & \mathbf{R}_{22} \end{bmatrix} \begin{bmatrix} \mathbf{Q}_1^T \\ \mathbf{Q}_2^T \end{bmatrix}, \tag{4.3}$$

where $\operatorname{Im}(.)$ is the imaginary part of the entity. Then, applying the SVD on $\boldsymbol{K}^{-1}\mathbf{R}_{22}$ hands in $\widehat{U}_s$ and system matrices as

$$\widehat{A} = \left(J_1\boldsymbol{K}\widehat{U}_s\right)^{\dagger}J_2\boldsymbol{K}\widehat{U}_s,\ \widehat{C} = J_3\boldsymbol{K}\widehat{U}_s,$$

$$\widehat{B}, \widehat{D} = \arg\min_{B,D} \sum_{k=0}^{M} \left\| R_k\left(G_k - D - \widehat{C}\left(e^{j\omega_k}\mathbf{I} - \widehat{A}\right)^{-1}B\right) \right\|_F^2, \tag{4.4}$$

where $R_k$ is calculated by noise variance in the robust LPM of the previous section. For the implementation of (4.4), the non-parametric model based on the multi-reference experiments is employed.

### 4.4.3.2 Parametric model of AVC benchmark problem

Following the results obtained in (4.4), and realizing it on benchmark problem of Figure 4.1, the $n$ singular values in SVD operation is swept between 10 and 22. The process of selecting the correct $q$ in (4.1) is a nontrivial task and despite the rule of thumb ($q = 5n$), it is recommended to evaluate all candidates in the range of $n + 1 \leq q < 10n$. The best value of $q$ varies for a specific system order $n$. Additionally, the stability of the identified system is investigated in the stability diagram of Figure 4.9.

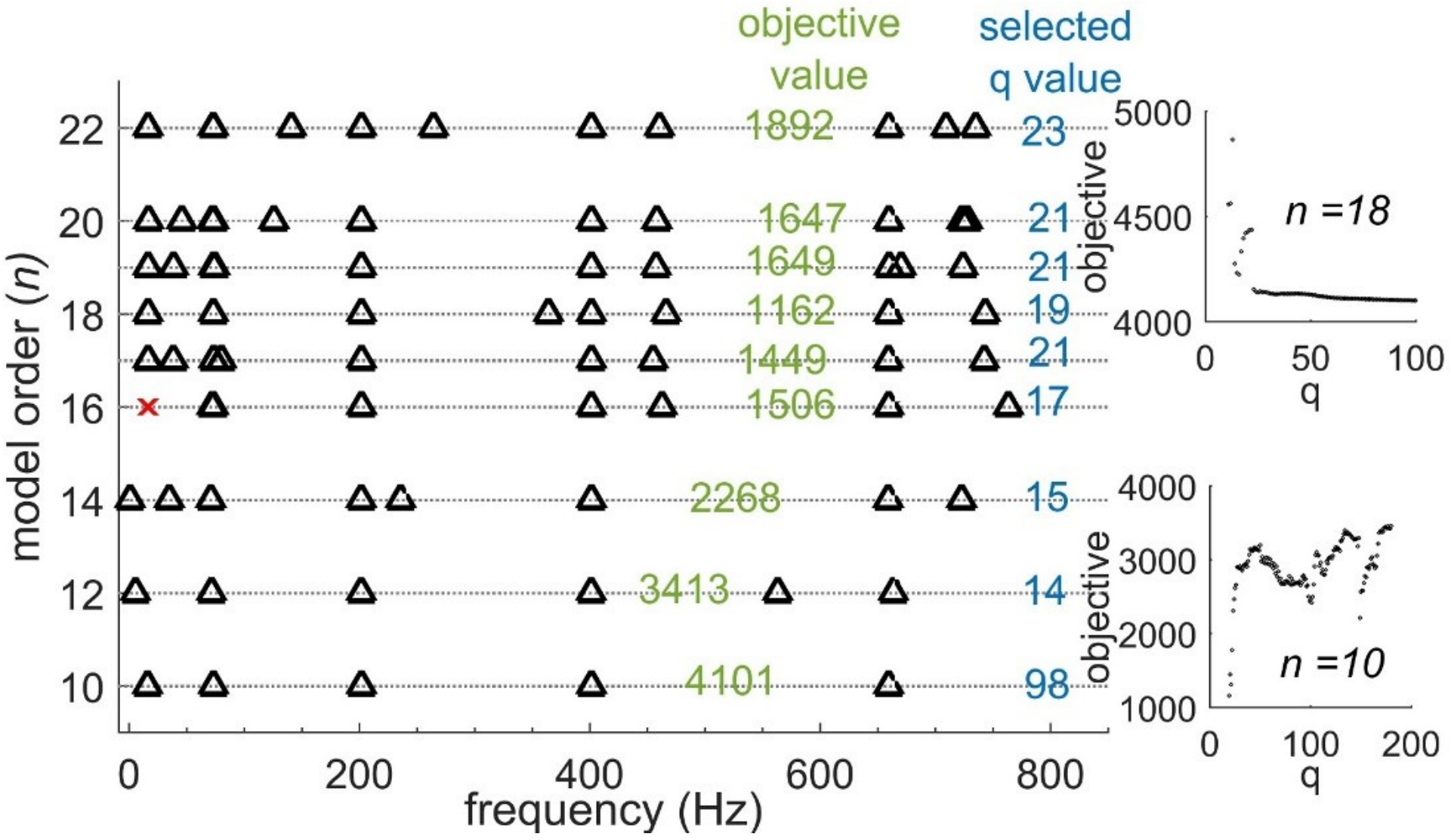

Figure 4.9 Stability diagram for black-box subspace system identification of benchmark problem in Figure 4.1: Δ represents stable poles, and x represents unstable poles [120], [121]

Figure 4.9 represents the location and stability of the identified poles for various model order $n$ accompanied with two important parameters: a) The best $q$ in (4.1) and (4.2). b) The associated objective function in (4.4) for the selected variables $q$ and $n$. For each system order $n$, $q$ is varied between $n + 1$ and $10n$ and the identification process is repeated. The result of two cases ($n = 10$ and $n = 18$) are shown in two subplots in Figure 4.9. This illustration is to emphasize the importance of the nontrivial process of selecting $q$. Accordingly, for each system order $n$, the minimum value of the objective function for swept values of $q$ is presented in blue font colour and the associated objective values in green font colour.

For the cantilever smart beam in Figure 4.1, the system order $n = 18$ has the minimum value of the objective function with completely stable poles. It should be noted that for the results in Figure 4.9, no stability constraints are added since no additional information about the structural damping of the system is assumed. On the other hand, if after some structural analysis, the modal damping of the system is calculated, for the price of biased estimation, the constraint identification instead of (4.4) can be performed in similar lines as [135]. The identified model based on the subspace method is compared against the non-parametric FRFs in Figure 4.10. In this figure, a third dotted line is also shown that represents the results of the predictive error method (PEM) which is initialized with the identified model from the subspace method [136].

In order to shed light on the effect of this model refinement, the pole of the state matrix of the identified models in the discrete domain is presented in Figure 4.11.

Both the subspace method, which is a single-step identification approach, and its refinement through PEM, which is based on an iterative optimization technique, have done a great job in fitting the non-parametric frequency-domain data. For output-feedback control design purposes, the rank of controllability and observability matrices are crucial and the identified models have the observability rank of six in both cases and controllability rank of ten and eight for subspace, and PEM, respectively.

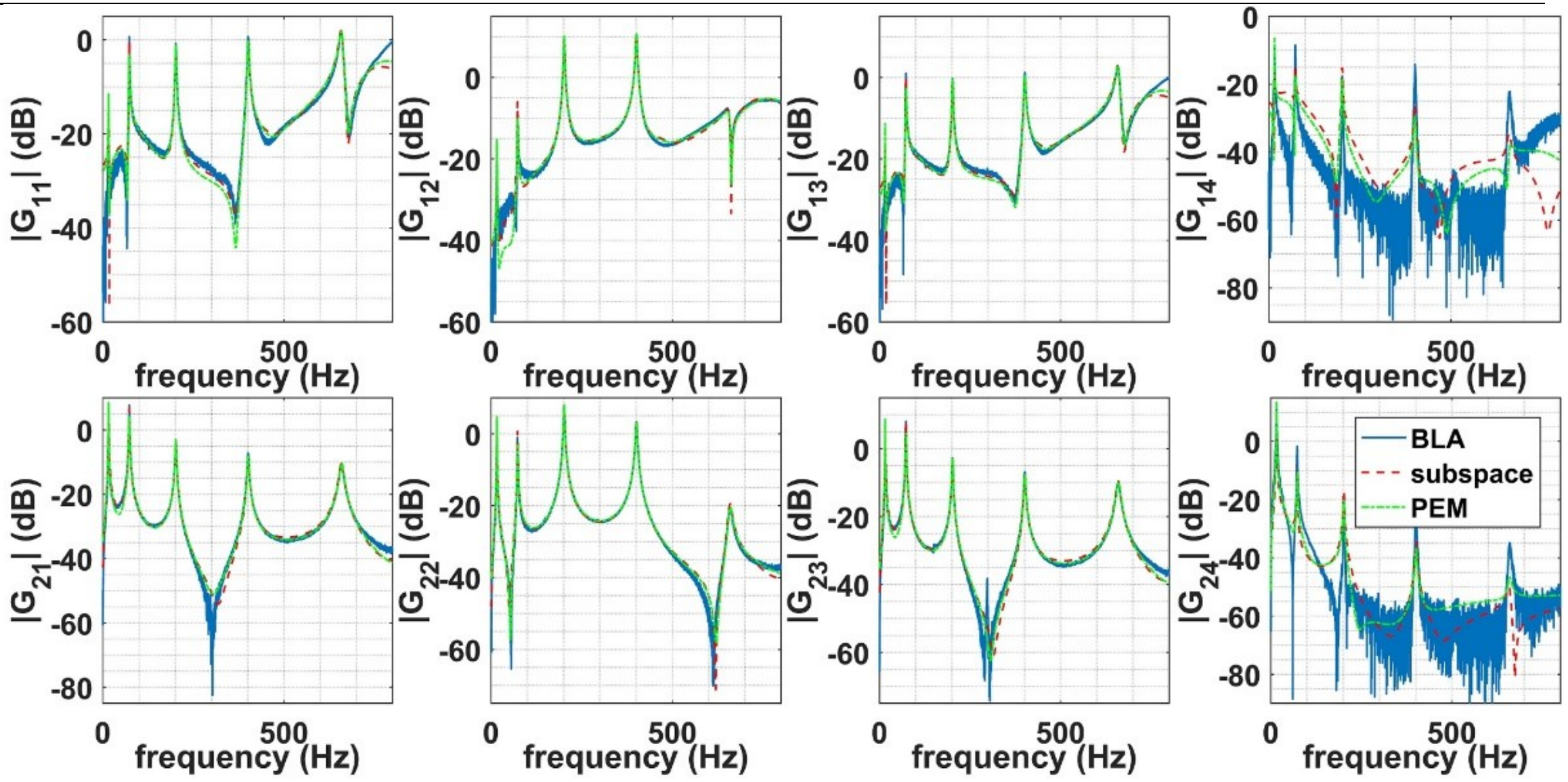

Figure 4.10 Parameterizing the FRFs of the system based on black-box subspace and PE methods [120], [121]

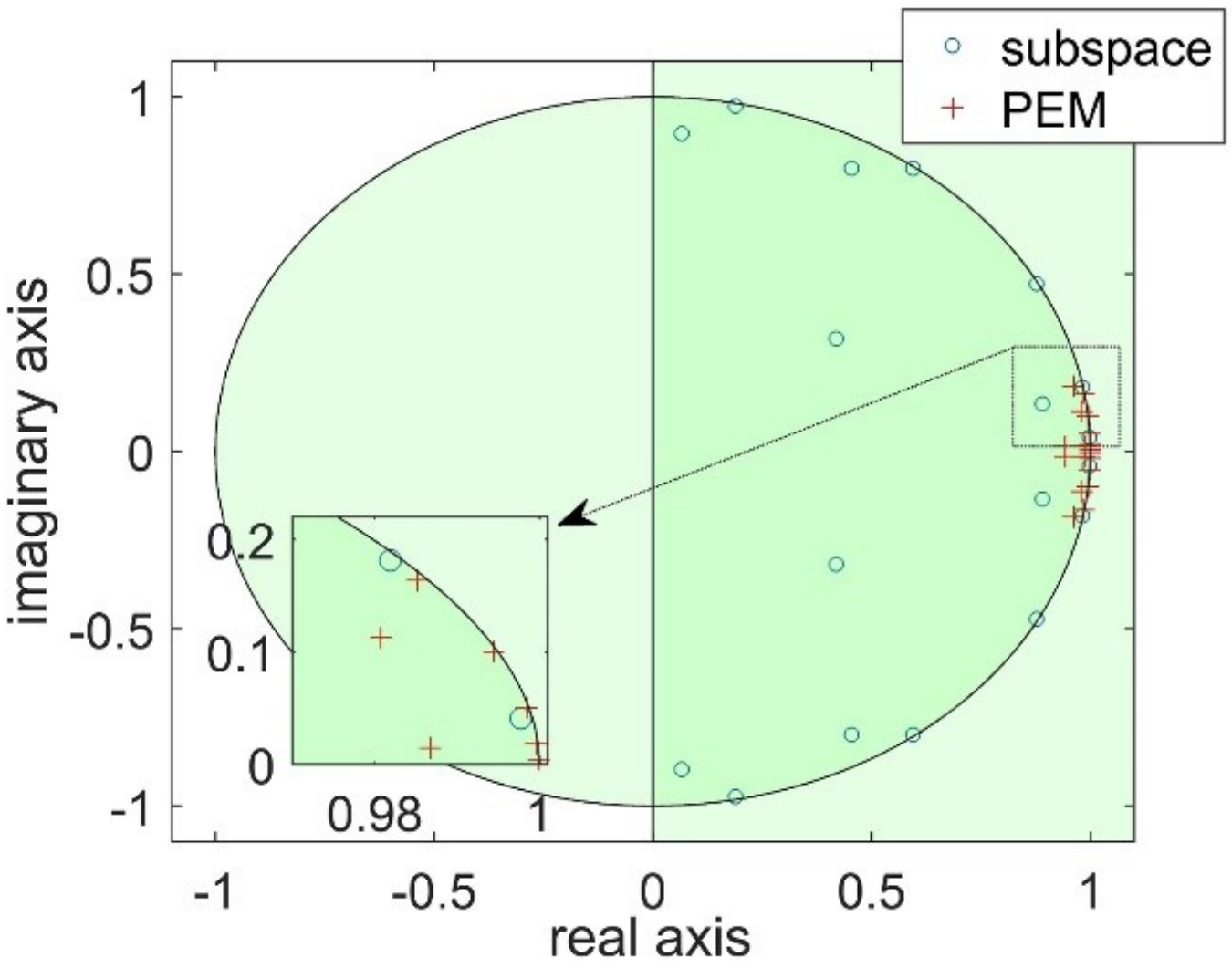

Figure 4.11 Locations of the poles of the identified models for benchmark cantilever beam in Figure 4.1 [120], [121]

### 4.4.4 Comments on uncertainty quantification

In the previous section, the parametric system identification is performed on a set of noisy data. Logically, one should describe the uncertainty bounds of the estimated model in terms of the so-called confidence intervals. Consequently, the variability of several identification parameters may come into mind including a) matched uncertainty over the matrices (and their elements) in the states space representation. b) The variance of model poles/zeros from the nominal values. c) The frequency-dependent uncertainty bounds (lumped) over the FRFs of the identified model. While both (a) and (c) can be used in the robust control design and under some reasonable conditions may be converted to one another, (b) has a visual advantage in terms of the stability margin and minimum-phase behavior of the system. In this section, both (b) and (c) are tackled based on the uncertainty ellipsoids given a probability level.

Since an approximation (sampled) of the covariance of the system in terms of the noise plus nonlinearity is extracted from Hadamard multi-sine experiments in combination with the robust LPM, a series of exhaustive Monte-Carlo simulations are performed in order to approximate these bounds.

A useful method in extracting the confidence intervals of a SISO transfer function model is reported in [137]. Two main features of this technique compared to the classical 95 % confidence bound which are also reported in [138] are

1) the less conservative estimation of the confidence interval for low signal-to-noise ratio (SNR).

2) More accurate bound estimates for high SNR where the uncertainty region associated with each pole/zero of the model may join the uncertainty bounds of others.

Developments in recent years regarding the computational power of the workstations resolve the problem associated with the exhaustive simulations for MIMO systems with large model dimensions.

Considering the results for the non-parametric modeling in Figure 4.8, where the total variances are ≈ 40 dB below the BLAs, and considering the matching quality in Figure 4.10, for the application of structural control in Figure 4.1, the identified model may be assumed as very accurate. However, in real applications of active vibration control, e.g., flutter control of airfoils in aerospace applications [139], the noise contributions can be significantly higher. Therefore, to investigate the sensitivity of the model, the estimated covariance of the FRM estimates can be used to perturb the non-parametric model with realistic random noise with the normal distribution. For this purpose, three levels of perturbation with magnitudes of 22 dB (case 1), 18 dB (case 2), and 8 dB (case 3) below the BLA are investigated.

Considering the system dimension, 500 Monte Carlo simulation is performed. This requires significantly large system memory and processing power. In each sample, the subspace system identification is carried out on a perturbed BLA with the same $q = 19, n = 18$.

In Figure 4.12, the variation of the poles and zeros of the system for two inputs (piezo-actuator 1 and shaker) and one measurement output (acceleration) is shown for the sake of conciseness. In this figure, the first, second, and third rows indicate cases 1, 2, and 3, respectively while the nominal values are in black color and the results from Monte Carlo simulations are shown in gray color.

Before analyzing the results, it should be pointed out that the poles/zeros with large stability margins are not shown in the range of the real axis (x-axis) since their variation are negligible. From Figure 4.12, it can be seen that for high SNR, i.e., case 1, the uncertainties regions on each pole and zero can be distinguished. However, for low SNR (case 3), the regions show interference and an explicit variance estimation of the system poles/zeros is not possible. Unlike the SISO case (not reported here due to the lack of space), the variation of the identified model is not necessarily close to the nominal values (see the uncertainty regions associated with shaker zeros). This issue is closely related to Tustin transformation that is involved in the identification and indicates the sensitivity of the algorithm w.r.t. distortions of poles/zeros close to the imaginary axis in discrete frequency-domain. As reported in [137], the 95 % confidence interval, although may be used as an approximation, represents an inaccurate estimation of the uncertainty bounds. Having the variance of BLA in Figure 4.8 in mind, it is naturally observed that the poles and zeros are significantly scattered for the shaker input.

In order to generate a frequency weighting function, for robust control design, which may encompass the modeling uncertainties for various SNR, the parametric FRM is investigated next. The variation of the perturbed model (red lines) is shown around the nominal model (black line) in Figure 4.13. The parametrization of the weighting function for $H_\infty$, min-max LQG, and $\mu$-synthesis can be done in a similar line as [13], [140]. It is clear from Figure 4.13, that due to the introduced perturbations, the identification algorithm is unable to capture the FRM of the system regarding the first natural frequency. This explains the bias of the results in Figure 4.12 where some of the poles/zeros are not distributed around the nominal values.

In the view of this dissertation, the current chapter delivers the best linear approximation of the vibration system. Since the geometry of the practical systems does not typically allow the use of beam/plate/panel theories and closed-form solutions, as of Chapter 2, the methodology explained in this chapter would deliver the means for the reduced-order model that eventually would be used in model-based control.

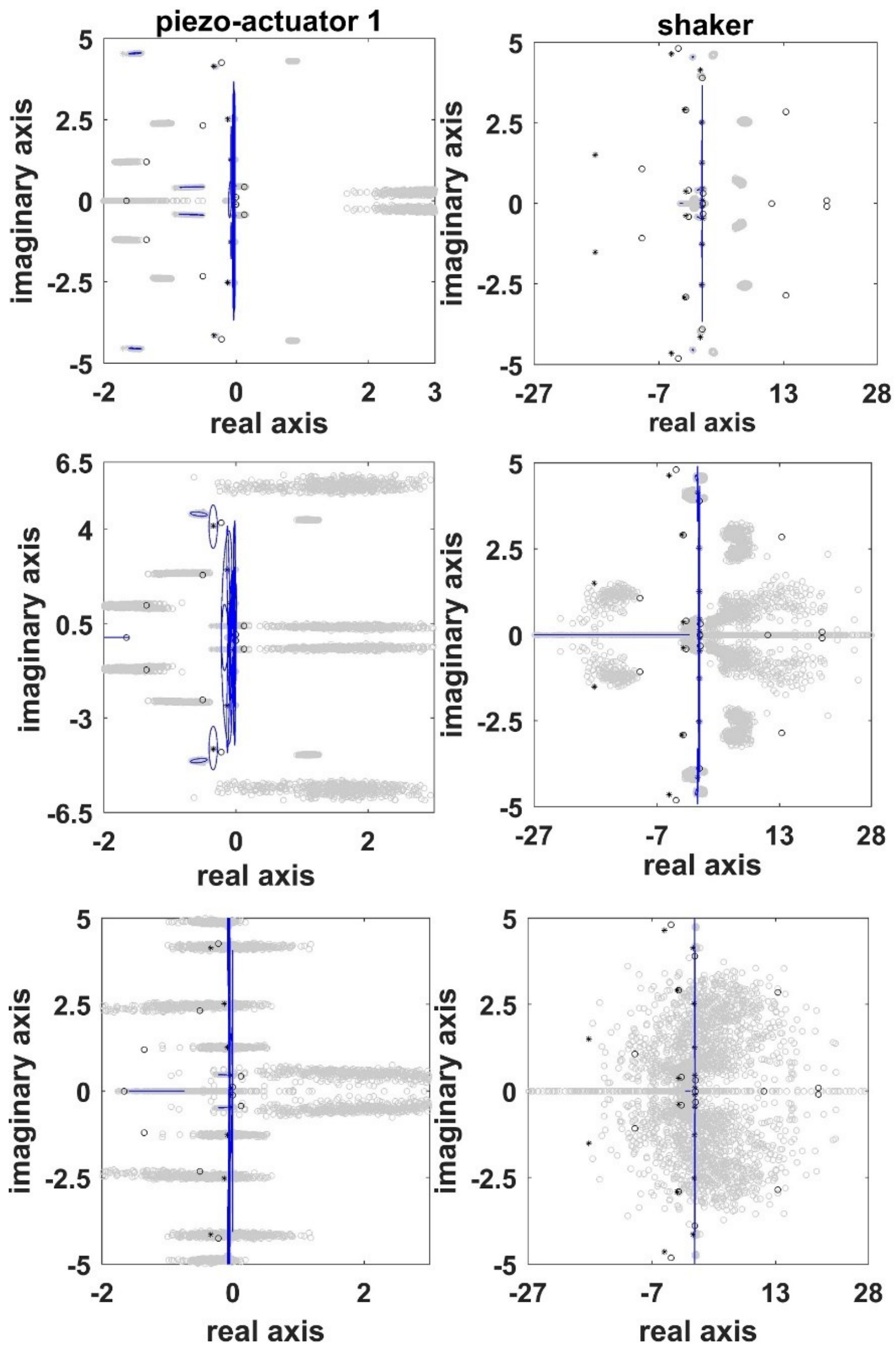

Figure 4.12 Variations of poles (*) and zeros (°) of the system and 95 % confidence ellipses for 500 simulations [120], [121]

In the upcoming chapter, the frequency-dependent weighting that was explained in Figure 4.13 is used to implement as a classical robust technique, i.e., μ-synthesis. In view of the experimental implementation, the sensor/actuator placement problem should be tackled to have respectively a proper observability/controllability condition [129]. The notions of controllability and observability are different for the linear system than those of (2.9). In the upcoming chapter, the assumption is done that the system is linear (low vibration amplitude) and the placement architecture would be valid for the nonlinear vibration domain. The validity of the first part of the assumption is shown in [141].

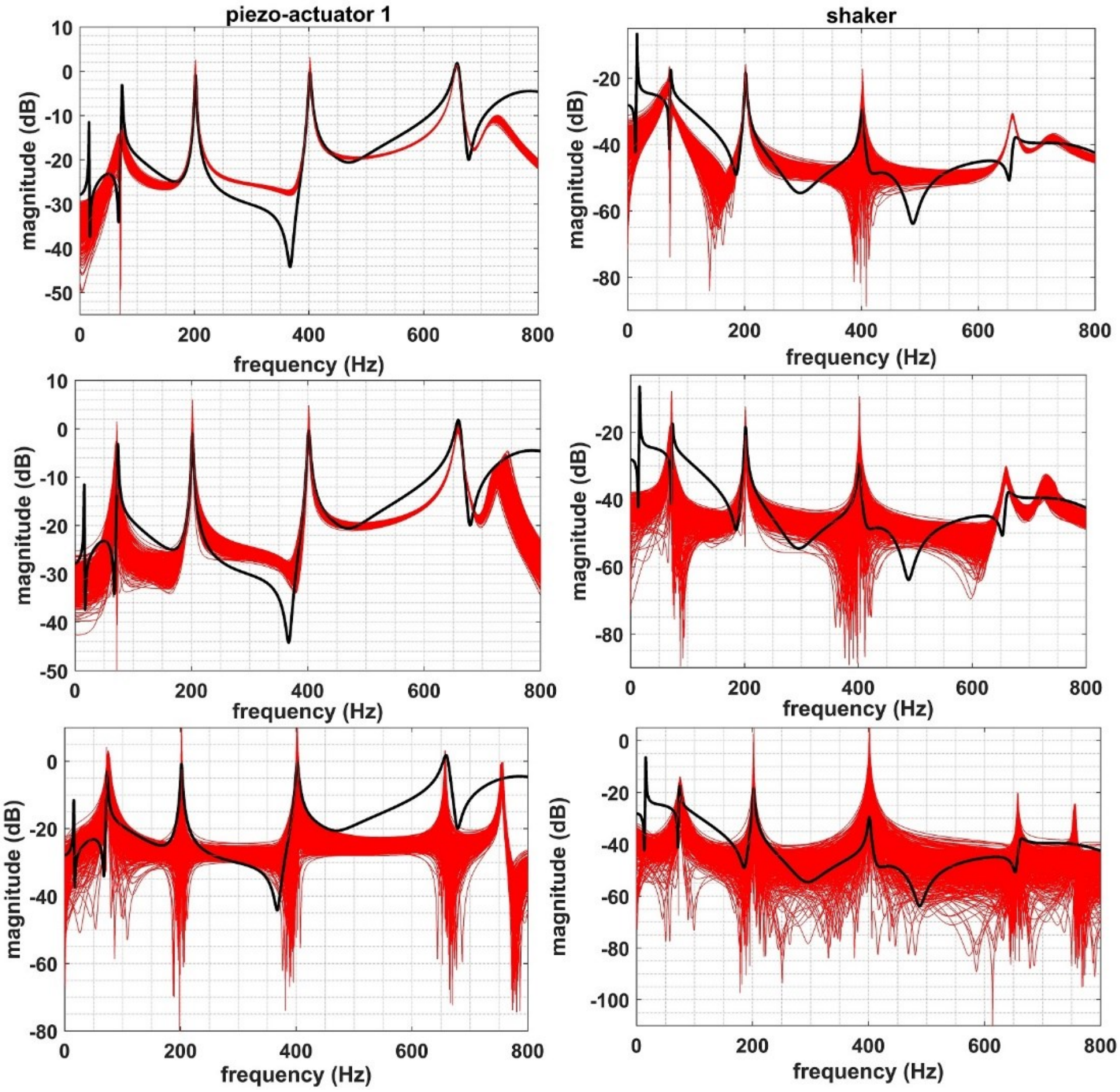

Figure 4.13 Variation of FRM for different SNR [120], [121]

# 5 Robust control of the piezo-actuated structure

This chapter serves as a transition from model development in Chapters 2, 3, and 4 towards model-based controller development in Chapters 6-11. Accordingly, it is intended to revisit a well-known classical robust control technique, namely $\mu$-synthesis approach, and look into the issues that may arise in developing such a robust controller for disturbance rejection. Apart from a reliable model which has been discussed thoroughly, issues such as actuator nonlinearity, uncertainty modeling, and modeling imperfection are highlighted. Accordingly, the current chapter not only serves as an AVC benchmark but also is a technical introduction that partially motivates the issues such as the fragility of the control system, actuator windup compensator, etc. Consequently, without loss of generality, the problem formulation is dedicated to the specific benchmark in Chapter 1.3, namely the vibration control of a funnel-shaped structure used as the inlet of a magnetic resonance imaging (MRI) device.

MRI devices are widely subjected to the vibration of the magnetic gradient coil which then propagates to acoustic noise and leads to series of clinical and mechanical problems. In order to address this issue and as a part of noise cancellation study in MRI devices, distributed piezo-transducers are bounded on the top surface of the funnel as functional sensor/actuator modules.

The influence of MRI devices in medical inquiries and the role of AVC in vibration suppression of them is discussed in 1.1.

Active vibration control as an effective solution can be employed to block the pathways that carry and transmit the noise in the whole MRI structure using piezo-actuators, also referred to as piezo-laminated structures, [12], [13]. The MRI shell is subjected to the excitation of the uninterrupted pulse Lorentz force of the coils. Qiu and Tani tackled the active vibration suppression due to the Lorentz force between the pulse current applied to the coil and the main magnetic field of the circular cylindrical shell that is equipped with MRI devices [14]. Nestorović et al. implemented an optimal LQ tracking controller with additional dynamics for vibration suppression of the funnel-shaped shell MRI structure [15]. They designed an additional model reference adaptive controller in order to compare the performance of the LQ controller experimentally[16]. In all of the aforementioned studies, the vibration and acoustic noise cancellation are carried out by the passive methods, model-free controllers, or deterministic model-based control approach.

A reduced-order linear time-invariant (LTI) model of the structure in state-space representation is estimated by means of the predictive error minimization (PEM) algorithm as a subspace identification method based on the trust-region-reflective technique. It should be noted that the PEM formulation used here is in the time domain and is fundamentally different from that of (4.4) (see [136]). The reduced-order model is expanded by the introduction of appropriate frequency-dependent weighting functions that address the unmodeled dynamics and the augmented multiplicative modeling uncertainties of the system. Then, the standard D-K iteration algorithm as an output-feedback control method is used based on the nominal model with the subordinate uncertainty elements from the previous step.

## 5.1 Funnel-shaped entrance of MRI device

### 5.1.1 The experimental setup

Since the ultimate goal of this chapter is vibration suppression (along with introducing the philosophy of placement indices), distributed piezo-transducers are bounded on the top surface of the funnel as is depicted in Figure 5.1. In order to address the actuator/sensor placement, the results are used based on the research conducted by DLR Braunschweig and the most recent research [129], [142].

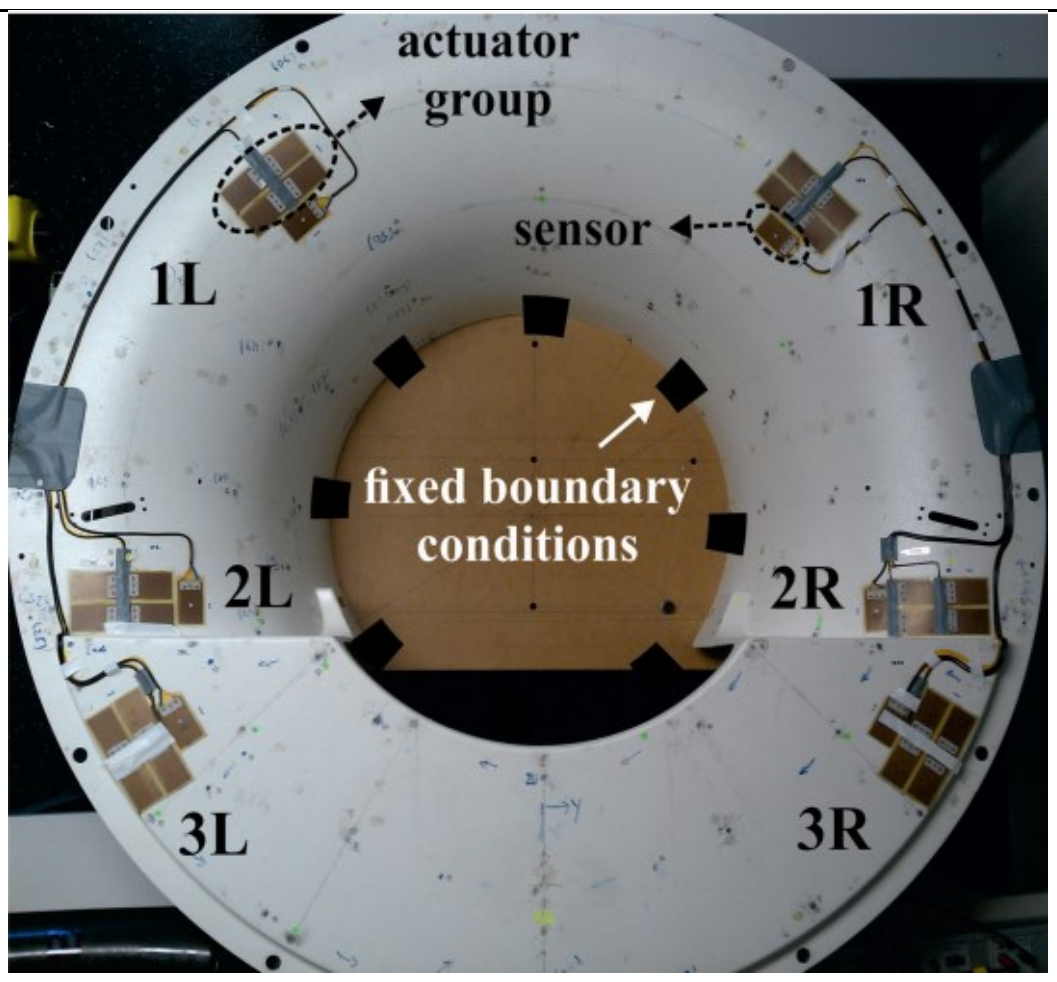

Figure 5.1 Distribution of the piezo-transducers on the surface of the funnel [13]

The structure consists of the funnel-shaped MRI inlet and six groups of sensor/actuator elements ($L_1$, $L_2$, $L_3$, $R_1$, $R_2$, and $R_3$). Each group of transducers consists of four individual paired actuator patches and one sensor element (PZT film Sonox P53). Each piezo-transducer is embedded in a polymer matrix for electrical insulation and respectively has an overall length and thickness of 50 *mm* and 0.2 *mm*. In order to reinforce the actuation power of the function modules, the second set of actuators are bounded on the top of the first set. The base of the funnel is fixed on a vibration isolation table with a set of screws. The measured output signals of the system are obtained through the six piezo-sensors that are placed on the top surface of the funnel as shown in Figure 5.2.

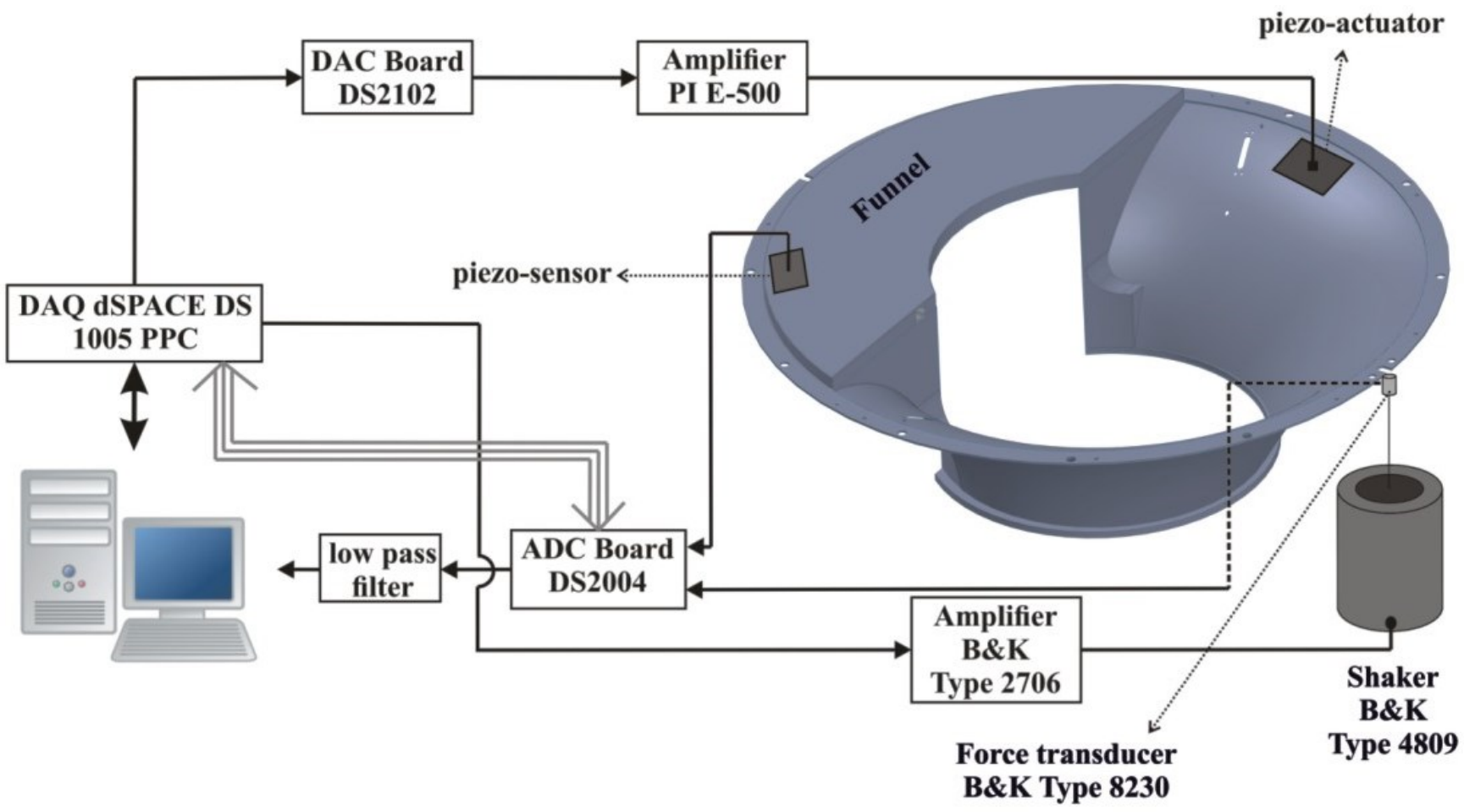

Figure 5.2 Sketch of the experimental setup [13]

dSPACE digital data acquisition (DAQ) with DS1005 PPC board is employed to compile the simulated disturbance and the designed control system in real-time analysis. The connection of the DAQ system with the actuators/sensors and the mainframe is provided by an analog to digital converter (ADC dSPACE DS2004) and a digital to analog converter (DAC dSPACE DS2102). Due to the different working voltage of dSPACE DAQ (±10 V) compared to the transducers ([-60, 200] V), an amplifier (PI E-500) is utilized to amplify the signal generated by DAQ with a constant gain of 100 and similarly, the second amplifier with maximum input current up to 5 A (Brüel&Kjær amplifier type 2706) is utilized to amplify the generated signal by DAQ system for the vibration exciter (Brüel&Kjær shaker type 4809). The control law of the active vibration system is realized on the SIMULINK platform and then compiled regulator law with standard fixed-step explicit ODE5 solver (Dormand-Prince method) is downloaded to the dSPACE DAQ in real-time by means of experimental rig shown in Figure 5.3. The simulated disturbance acts on the structure by the shaker which is connected to the funnel with a force transducer (Brüel&Kjær DeltaTron Type 8230-001) with the sensitivity of 22 mV/N. This transducer measures the applied mechanical force on the structure that is used as the definition of the uncertainty weighting function of input disturbance. Furthermore, it should be mentioned that due to the computational limitations, only four sets of actuators and sensors ($L_1$, $L_2$, $R_1$, and $R_2$) are activated for modeling and control design purposes.

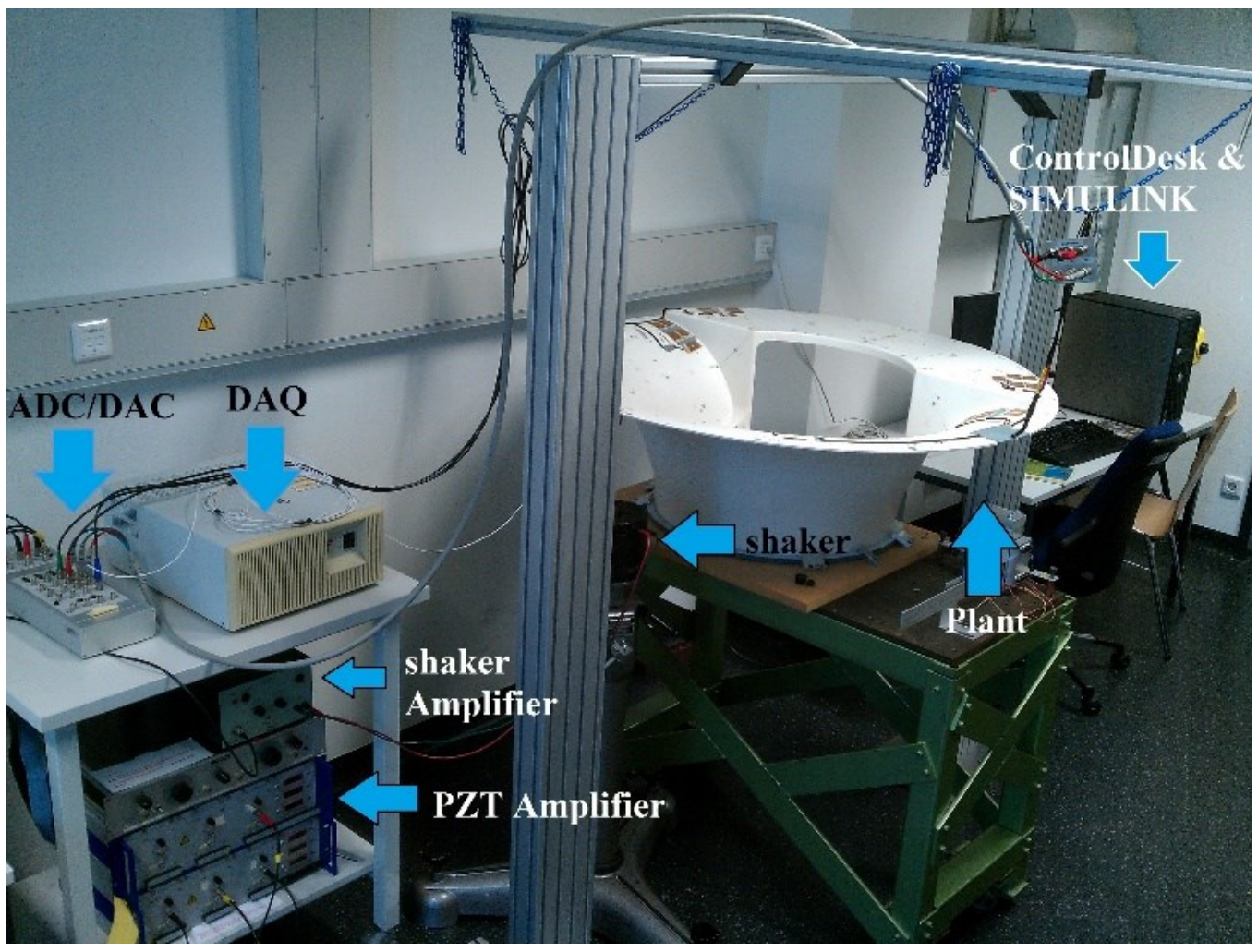

Figure 5.3 Experimental rig of the closed-loop system [13]

## 5.2 System model

Following the idea in Chapter 4.4, the predictive error minimization (PEM) identification algorithm in the *time-domain* for continuous LTI systems is utilized to estimate the system matrices from input-output data. The process begins with collecting data (voltage) from the input/output channels of the open-loop experimental setup. This includes the external input mechanical excitation channel that is realized by the shaker ($B_w$) and the input piezoelectrical excitation channel for applied control effort by means of the functional modules ($B$). (5.1) is obtained from system identification. Here, ($B_w$) represents the realization of an external disturbance channel. Accordingly, the reduced-order deterministic MIMO LTI system in the state-space framework which represents the dynamics of the funnel structure with distributed piezo-transducers in a constraint frequency range can be constructed through trust-region-reflective Newton technique as [13], [136]

$$\dot{x} = Ax + Bu + B_w d, y = Cx + Du + w_n, \tag{5.1}$$

where $d \in R^p$, $x \in R^n$is the disturbance input. $A \in \mathcal{R}^{2n\times 2n}$, $B \in \mathcal{R}^{2n\times n_u}$, and $C \in \mathcal{R}^{n_y\times 2n}$ are the state, control input, and output matrices, correspondingly. In addition, $B_w$, represents the input disturbance matrix. Furthermore, $\mathrm{w_n}$ stands for the measurement noise which is considered to be an uncorrelated zero-mean Gaussian stochastic process with constant power spectral density ($W_n$) such that $E\{w_n w_n^T\} = W_n\delta(t-\tau)$, in which $E$ is the expectation operator (see [143]).

### 5.2.1 $\mu$-Synthesis

The coupled piezo-laminated funnel shape structure as shown in Figure 5.4 has five inputs including one disturbance channel and four actuation signals. The measured output of the system consists of four channels associated with four piezo-sensors bounded on the structure. The modeling uncertainty for the input matrix and input disturbance matrix is assumed to be multiplicative unstructured uncertainty [144]. In Figure 5.4, *K(s)* represents the controller that is designed based on the output feedback $\mu$-synthesis (explained in [13] Algorithm 5.1). Here, $n$ stands for the white Gaussian output measurement noise. In addition, $W_{dist}$, $W_{Rdist}$, $W_n$, and $W_{Ri}$ ($i$ = 1, 2, 3, 4) are the frequency-dependent norm bounded weighting function for external input disturbance signal, the multiplicative uncertainty of disturbance, high pass filter representing the activation frequency of the noise signal, and weighting function of the actuation uncertainty, respectively. Also, as a standard multiplicative representation, it is assumed that $\|\Delta_{\text{dist}}\|_\infty \leq 1$ $\|\Delta_{\text{dist}}\|_\infty \leq 1$and $\|\Delta_{\text{act}i}\|_\infty \leq 1, i = 1, \dots, 4$ $\|\Delta_{\text{act}i}\|_\infty \leq 1, i = 1, \dots, 4$ [145]. For the sake of applying the structured singular value theory to the plant under study, the signal configuration in Figure 5.4 should be recast into linear fractional transformation (LFT) [146] representation that is shown in the standard form of Figure 5.5.

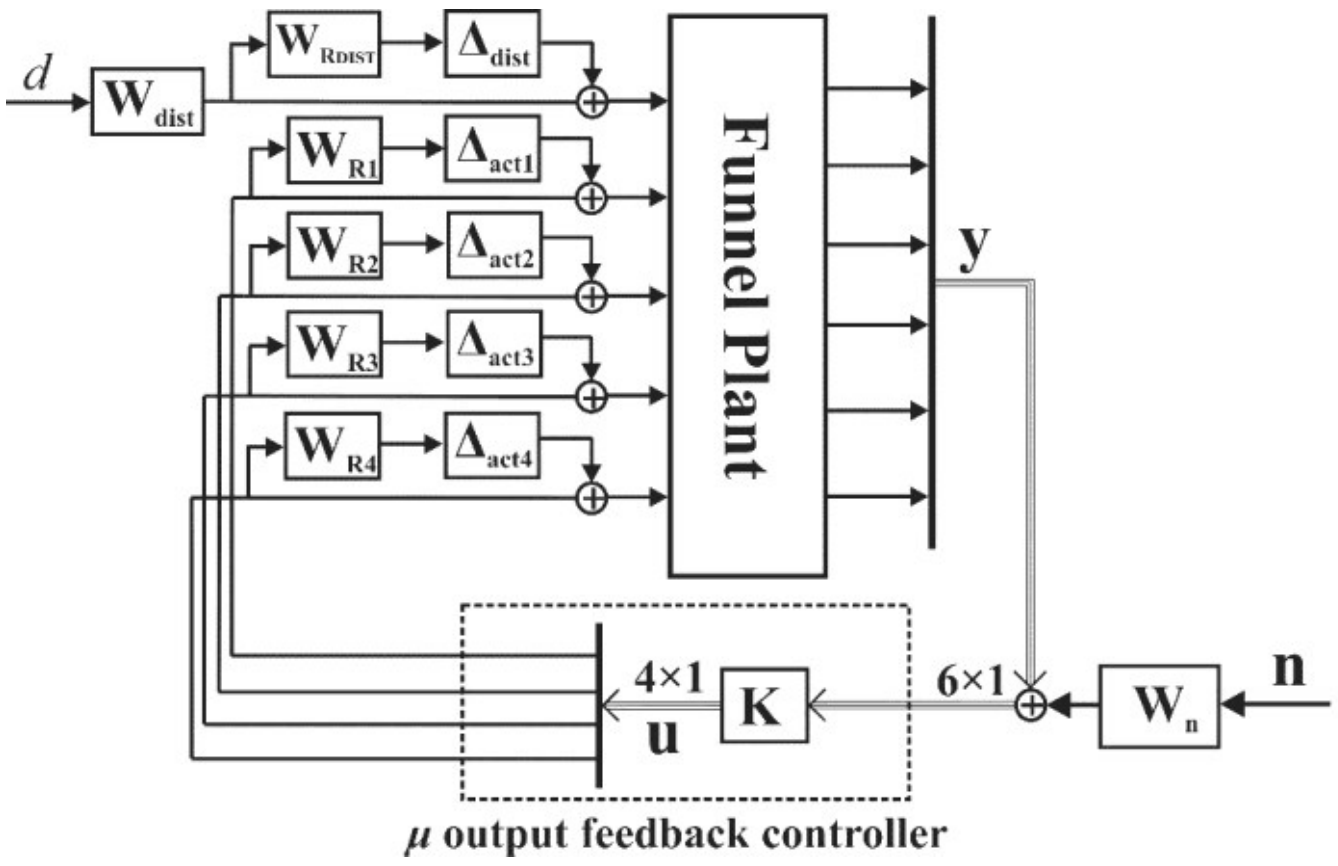

Figure 5.4 Closed-loop system with multiplicative uncertainty [13]

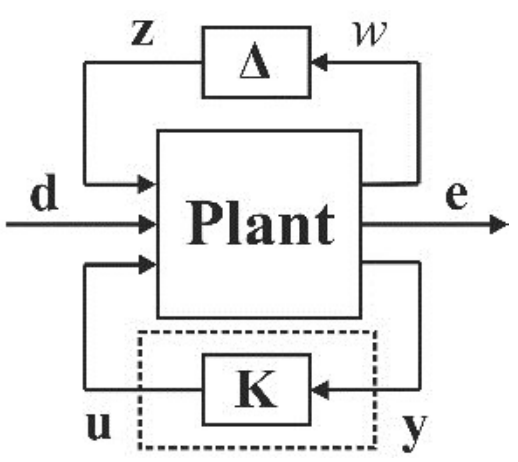

Figure 5.5 LFT representation of the control system [13]

In Figure 5.5, the plant is the open-loop interconnection which encloses certain elements such as the nominal structure model and all weighting functions. Δ-block parameterizes all of the defined uncertain elements in the previous representation of Figure 5.4. The inputs to the MIMO plant are allocated into three standard groups in which $w$ represents the perturbation and similarly, the outputs are grouped into three signals with e representing the error signal. The LFT representation covers the set of all control systems $F_U$ such that $F_U(P(s), \Delta)\colon \max_{\omega} \sigma[\Delta(j\omega) \leq 1]$ with $\omega$ being the frequency such that for perturbation Δ, a stabilizing controller $K(s)$ exists that satisfies the following equation

$$\|F_L[F_U(P(s), \Delta), K(s)]\|_\infty = \left\|F_U\left[F_L\left(P(s), K(s)\right), \Delta\right]\right\|_\infty < 1. \tag{5.2}$$

For a given controller K(s), the constrained $H_\infty$ index function of (5.2) with the predefined augmented uncertainty structure of LFT representation in Figure 5.5 can be checked by examining the structured singular values of the closed-loop system if and only if K(s) satisfies the following constraint

$$\max \quad \mu_\Delta[\mathrm{F}_L[\mathrm{F}_U(P(s),\Delta),K(s)]]<1. \tag{5.3}$$

The standard μ-synthesis minimizes the maximum structured singular value of the close-loop system over frequency. In order to solve the $\mu$-synthesis problem, $\mu_\Delta$ should be replaced with the upper bound.

**Definition 5.1.** Considering $M$ to be a constant matrix and the uncertainty δ, the upper bound of $\mu_\delta(M)$ is defined as $\inf \bar{\sigma}(DMD^{-1})$ for $D \in D_\delta$ with $D_\delta$ being the set of all matrices $D$ that satisfies $D\delta=\delta D$.

Using **Definition 5.1** for the optimization function of the $\mu$-synthesis and Δ, the optimization equation is reformatted as

$$\min_{K(s)} \quad \max_{\omega} \quad \min_{D_\omega \in D_\Delta} \bar{\sigma}[D_\omega F_L(P(s),K(s))(j\omega)D_\omega^{-1}], \tag{5.4}$$

by defining $D$ as a frequency-dependent function that satisfies $D_\omega \in D_\Delta$ for an arbitrary frequency $\omega$, and replacing $\max_{\omega} \bar{\sigma}[.]$ with $\|.\|_\infty$, the optimization problem in (5.4) can be rewritten as

$$\min_{K(s)} \quad \min_{D, D_\omega \in D_\Delta} \left\| DF_L(\mathrm{P}(s),\mathrm{K}(s))(j\omega)D^{-1} \right\|_\infty, \tag{5.5}$$

then, assuming $U$ as a block orthogonal complex matrix for complex $M$ (**Definition 5.1**) to have the same structure as $D \in D_\Delta$ and satisfying $U^*U = UU^* = I$ with $I$ being the identity matrix and superscript * denoting the complex conjugate of the original matrix, it is easy to obtain $\bar{\sigma}[(UD)M(UD)^{-1}] = \bar{\sigma}[DMD^{-1}]$.

This is due to the fact that the matrix multiplication with orthogonal matrix does not change $\bar{\sigma}$ and, as a result, one can transform $D$ into $UD$ without changing the maximum structured singular value of the system. By the introduction of an appropriate $U$, it is possible to restrict $D_\omega$ to be real-rational, stable, and minimum-phase transfer function symbolized as $\widehat{D}(s)$ [147]. The optimization index, (5.5), based on this transformation and due to the orthogonality of the introduced matrix is reformatted as

$$\min_{K(s)} \min_{\widehat{D}(s)} \left\| \widehat{D} F_L\big(P(s), K(s)\big) \widehat{D}^{-1} \right\|_\infty \tag{5.6}$$

The transformed optimization problem of (5.6), as shown in Figure 5.6, can be solved iteratively by means of *D-K* iteration approach [148].

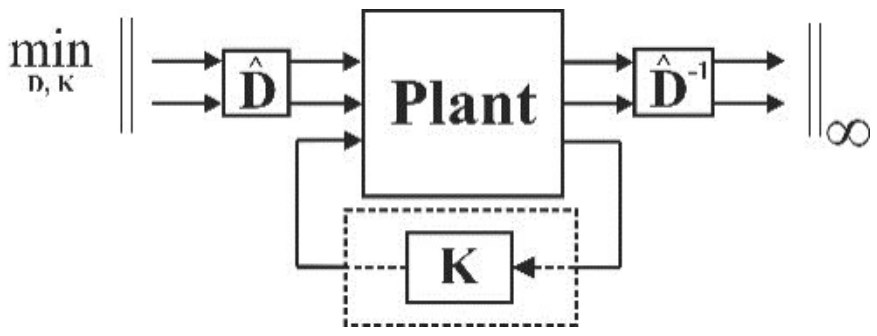

Figure 5.6 Replacing $\mu_\Delta$ with the upper bound [13]

Algorithm 5.1. *D-K* iteration algorithm:

1. Solve optimization problem $\min_{K(s)} \left\| \widehat{D} F_L\big(P(s), K(s)\big) \widehat{D}^{-1} \right\|_\infty$ with an initial guess of stable, minimum-phase, and real rational $\widehat{D}(s)$and define $P_D(s))$ as shown in Figure 5.7.

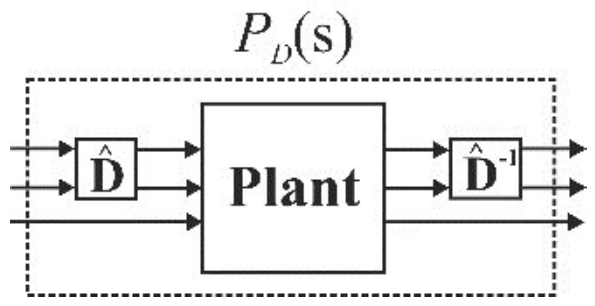

Figure 5.7 Real-ration $D$ scaling in $D$-$K$ iteration [13]

2. With known $P_D(s)$ from the previous step, solve the standard $H_\infty$ problem with the optimization function of $\min_{K(s)} \| F_L(P(s), K(s)) \|_\infty$.
3. With the fixed controller $K(s)$ from the previous step, move the optimization over $K(s)D(s)$ and look for an appropriate frequency-dependent function $K(s)$ in a certain range of frequency to satisfy the previous constraints on $\widehat{D}(s)$.

### 5.2.2 Experimental and Numerical Results

Considering the number of variables that are involved in the modeling of the structure and the control system and keeping in view the limitations on computation power, a systematic procedure in evaluating the closed-loop system is followed. For this purpose, first, a reduced-order model of the real structure with distributed piezoelectric sensor/actuator groups is obtained by means of system identification. The reduced-order model is assumed to cover the dynamics of the system within the frequency range of 0-18 Hz which includes two natural frequencies of the structure(9.72 Hz and 15.22 Hz) (also see [142]). In order to present the identification quality, the frequency response function (FRF) of the real structure is compared with the bode diagram of the identified model in the nominal frequency range of Figure 5.8(a, b, and c).

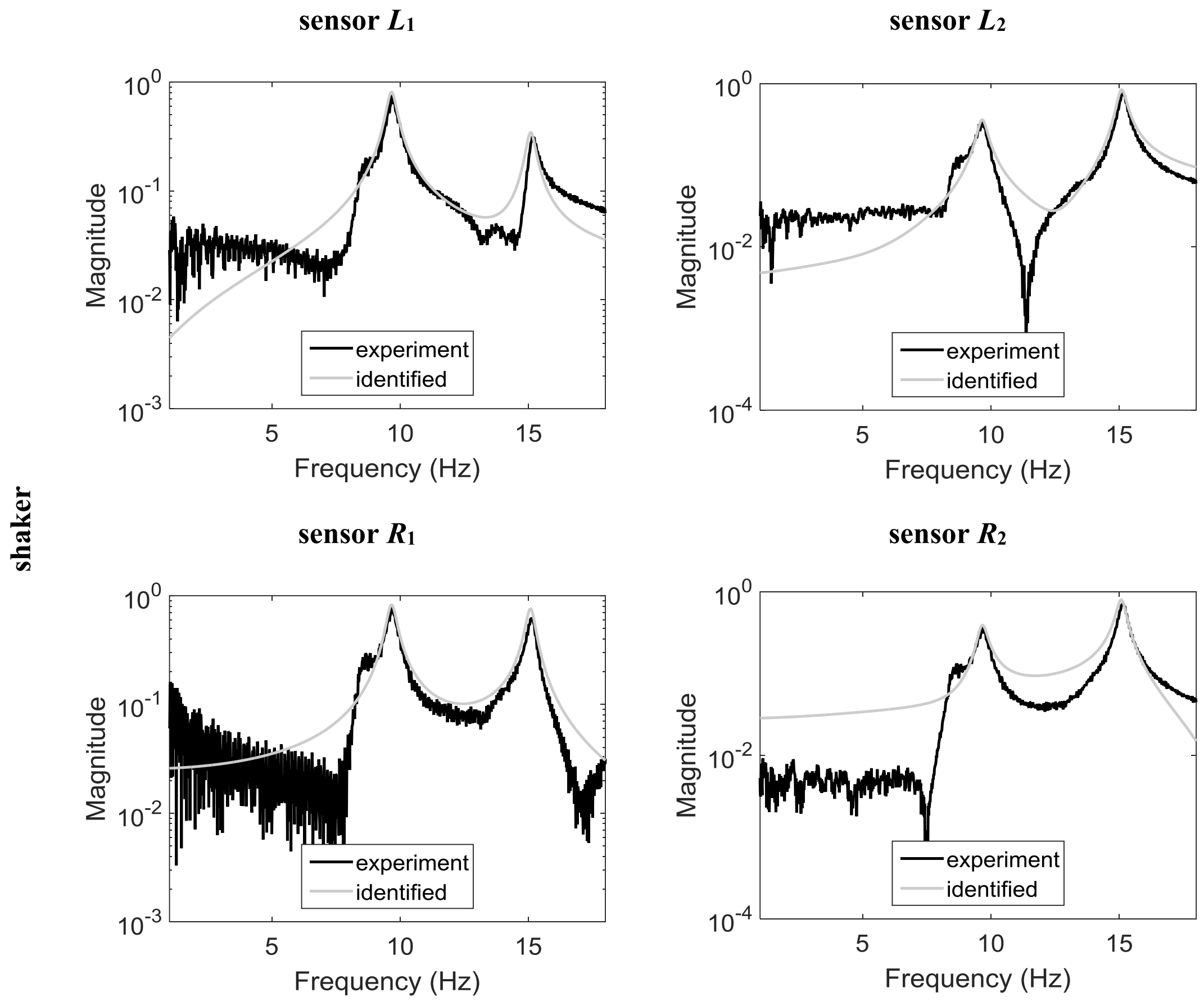

Figure 5.8a Comparison of FRF of the real structure with the identified model for disturbance channel [13]

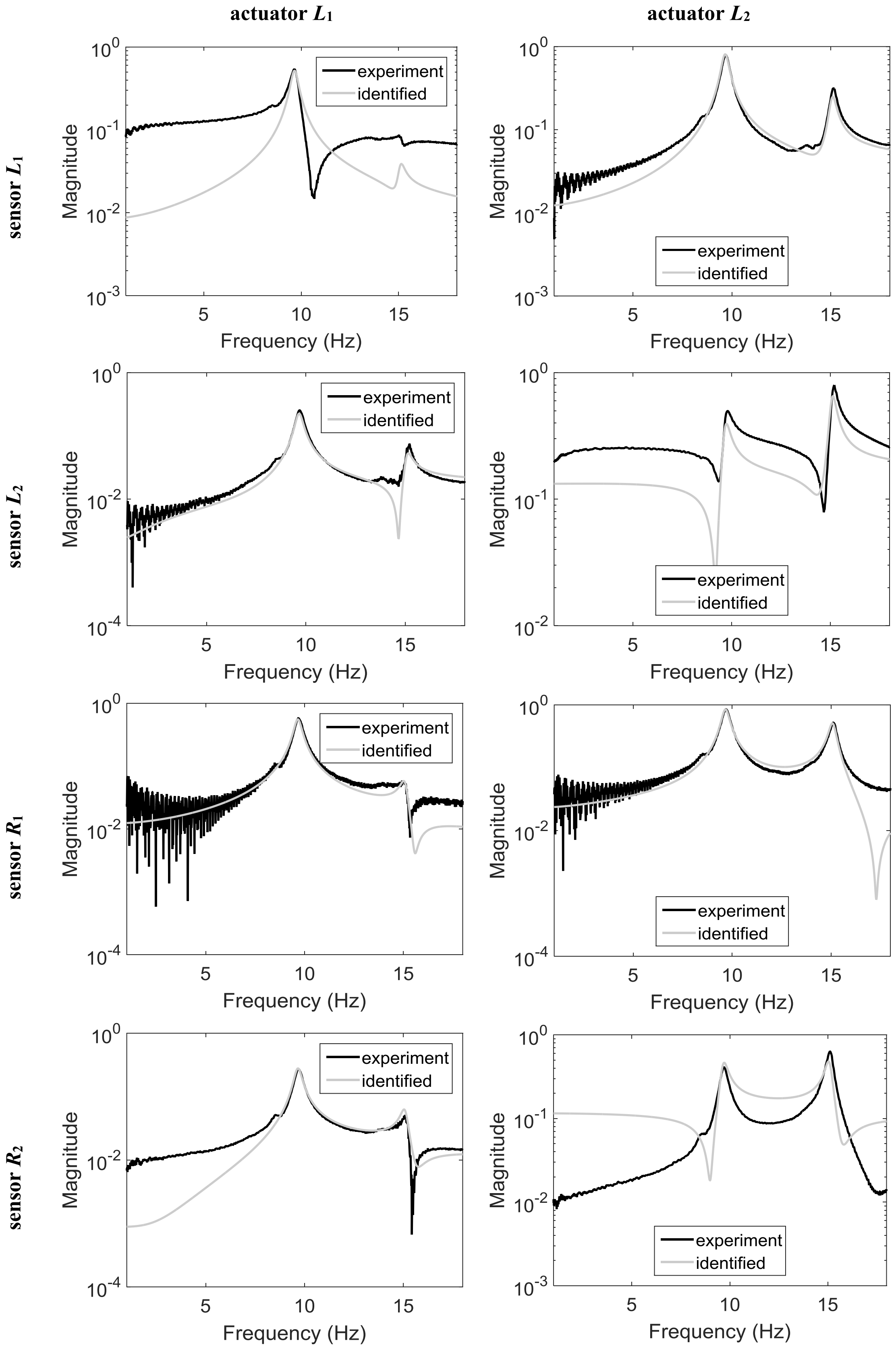

Figure 5.8b Comparison of FRF of the real structure with the identified model for actuator groups $L_1$ and $L_2$ [13]

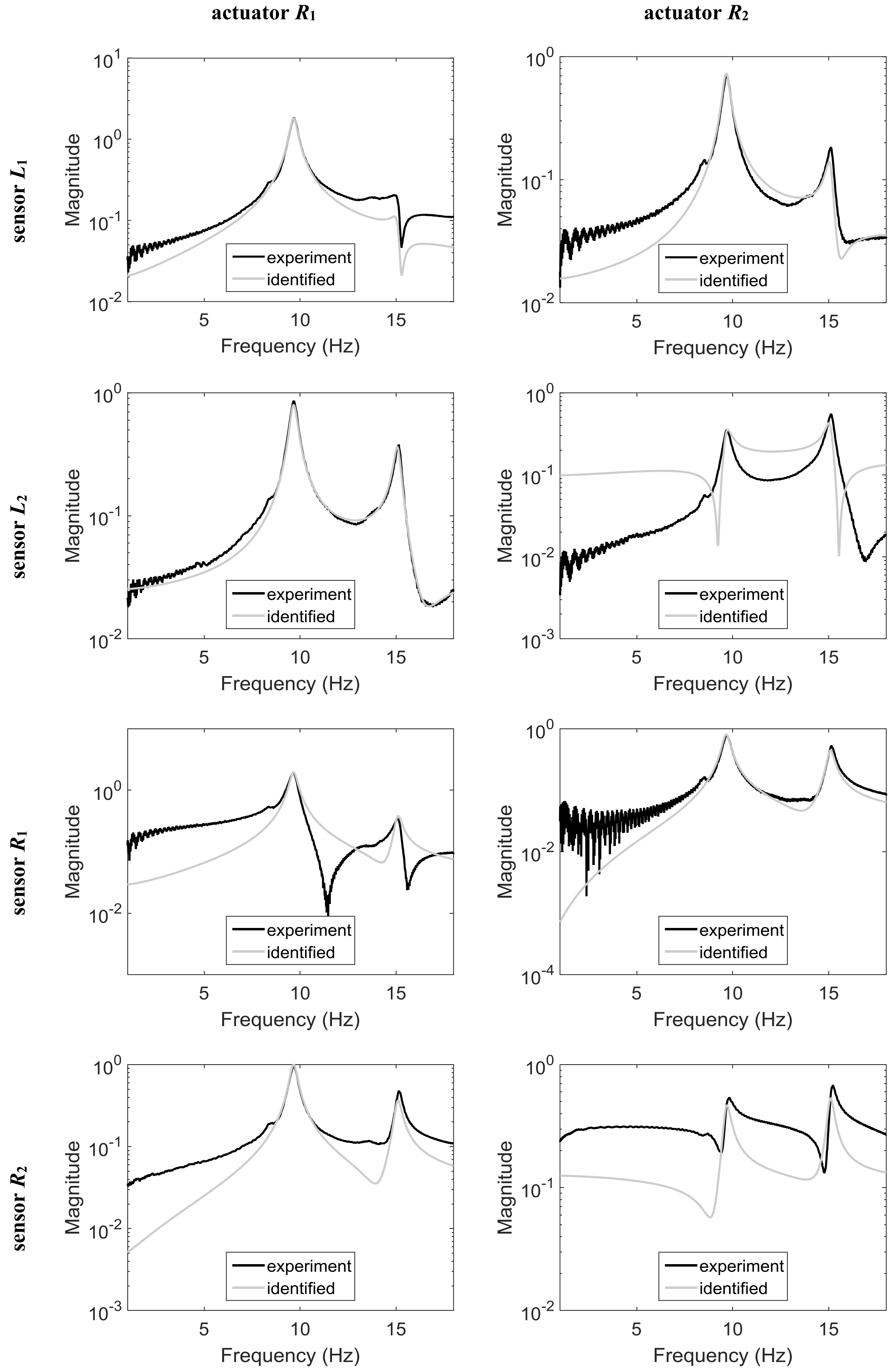

Figure 5.8c Comparison of FRF of the real structure with the identified model for actuator groups $R_1$ and $R_2$ [13]

It is worth mentioning that the reduced-order model is a sixth-order LTI object. The identified system is employed to design a robust controller on the feedback channel based on the output measurements that are collected by the four piezo-sensors ($L_1$, $L_2$, $R_1$, and $R_2$) and dSPACE DS2004 based on the *D-K* iteration approach. Based on the FRF of the system, appropriate weighting functions are selected on the input channels that represent the frequency-dependent uncertainty bounds and activation frequency of each channel. Due to the multiplicative nature of the augmented uncertainty, the weighting functions are selected as shown in Figure 5.9 to cover the higher-order mode-shapes of the structure which are neglected in the modeling procedure.

In order to investigate the performance of the output feedback control system, the structure is excited through the disturbance channel by a sweep sine signal with a frequency range of 0-18 Hz for both the open-loop and closed-loop systems. The applied mechanical disturbance signal is measured by the force transducer (B&K 8230) and represented in Figure 5.10. Accordingly, the output measurements are obtained by the piezo-sensors and are depicted in Figure 5.11. The transient response of the open-loop system is compared against the closed-loop system based on the designed controller. It can be observed that the μ-control system has successfully alleviated the vibration amplitude although the external disturbance continues to excite the structure.

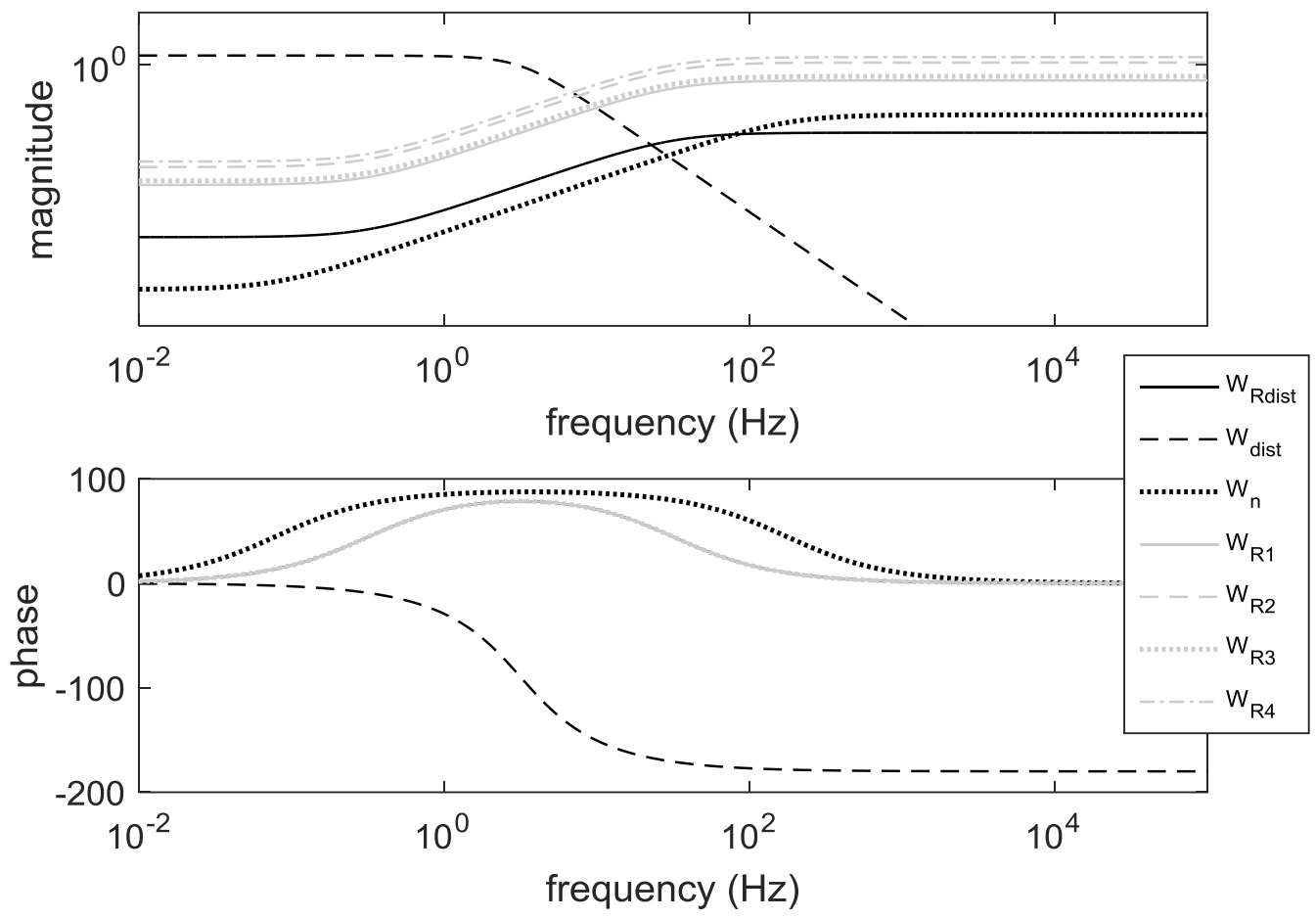

Figure 5.9 Weighting functions associated with uncertainty presentation in control design [13]

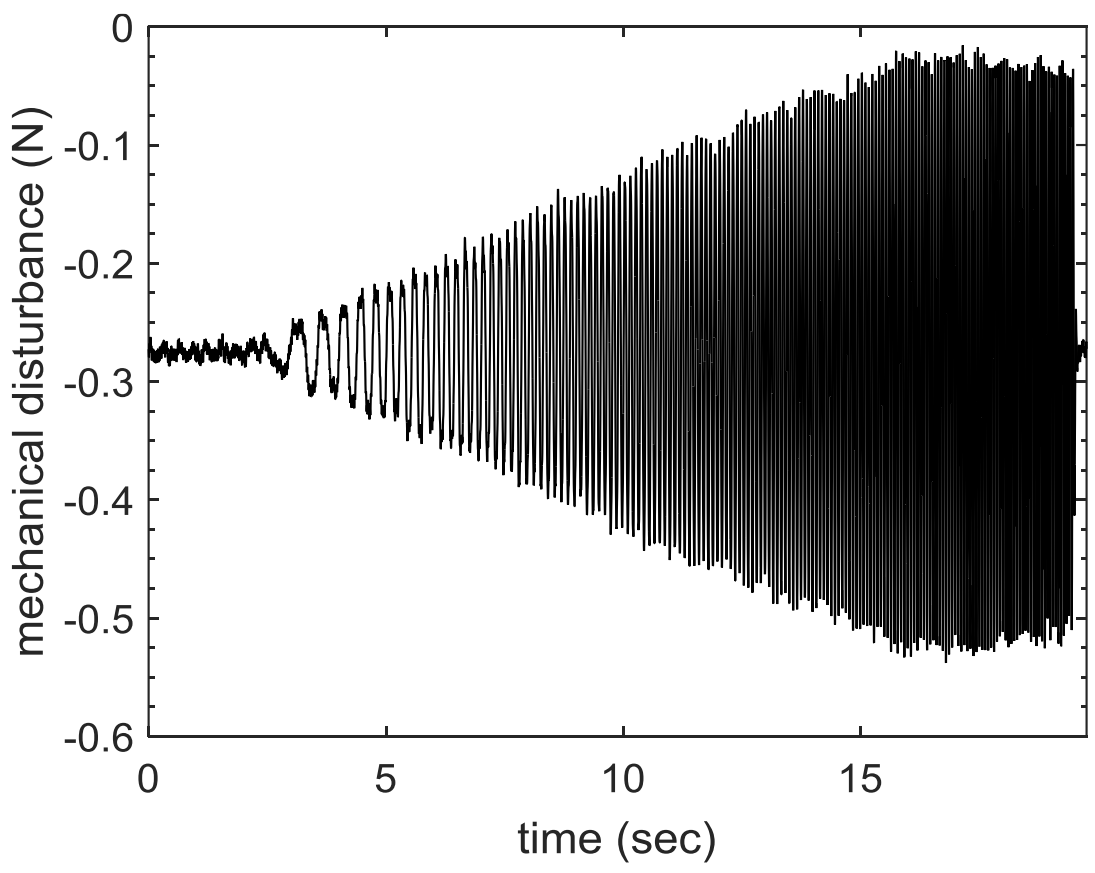

Figure 5.10 The amplitude of applied mechanical disturbance through shaker [13]

The applied control effort for the sweep sine signal is presented in Figure 5.12. As expected, the control voltage for the $L_2$ and $R_2$ actuator groups are less than for the $L_1$ and $R_1$ groups which are due to the small level of strain energy that reaches these actuators because of the distance between the disturbance source and actuator elements (see Figure 5.1 and Figure 5.2). In addition, by the introduction of the appropriate weighting functions (see Figure 5.9), that contain the frequency-dependent information of the actuator elements, the applied control efforts have a smooth behavior with a maximum amplitude of 55 V. In order to investigate the spillover effect of the controller, the Fast Fourier Transformation (FFT) analysis was performed on the control signal to clarify the frequency content of the applied. The result as depicted in Figure 5.13 shows that although the disturbance is acting within the maximum frequency of Hz, the applied control effort of actuator $R_1$ leads to excitation of higher-order mode-shape. However, significant reductions in the magnitude of the applied control effort can be observed for all of the actuation elements due to the introduction of the weighting functions. In addition, in the mechanical systems, the lower eigen-frequencies have a dominant impact, which is also confirmed here by the influence of the first, second, and third eigen-frequencies.

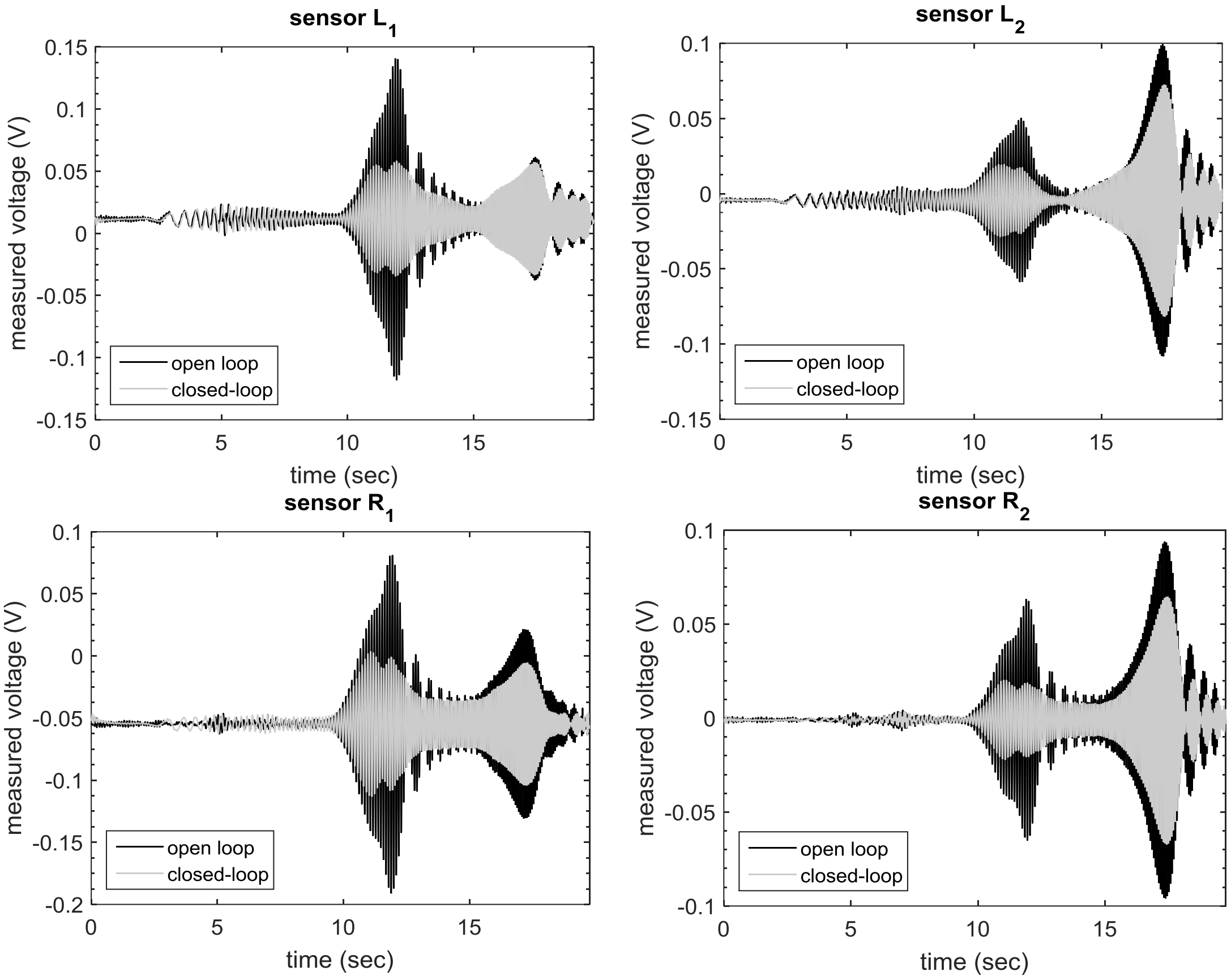

Figure 5.11 The vibration suppression performance of the control system in the nominal frequency range [13]

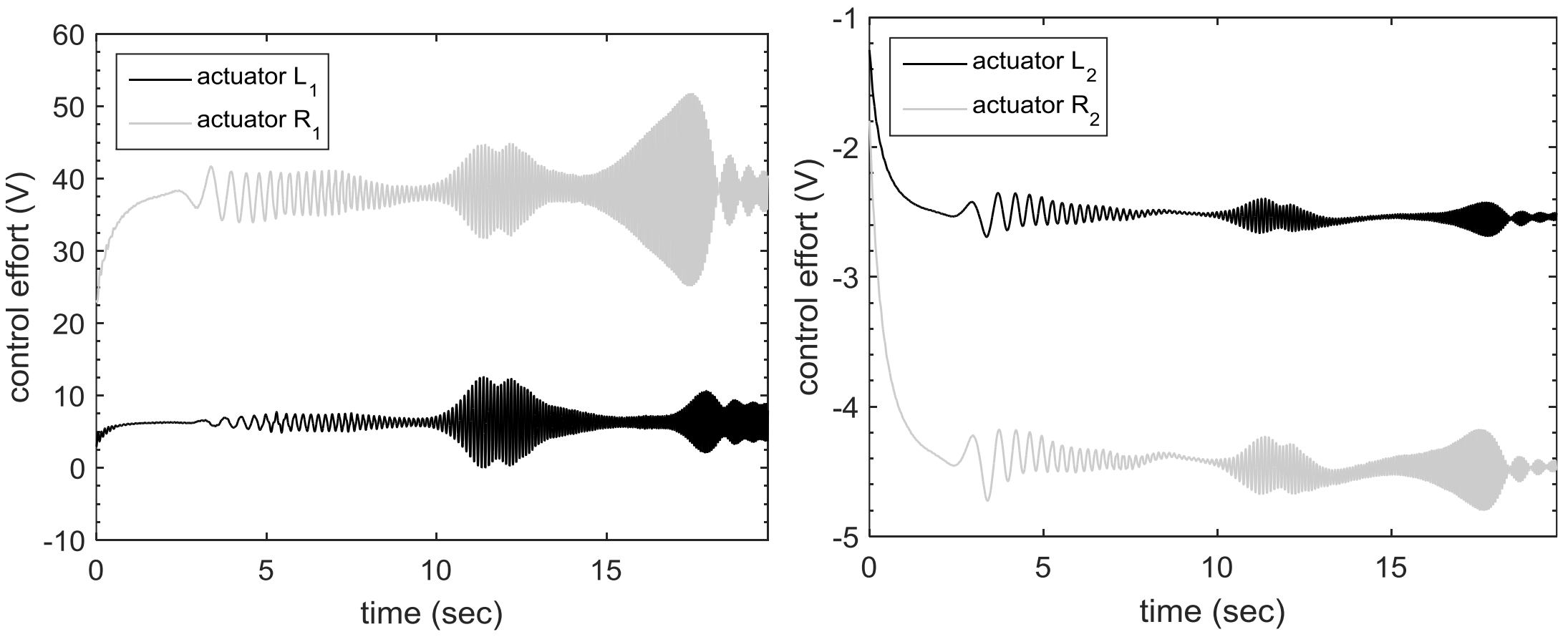

Figure 5.12 Applied control effort on the piezo-actuator groups [13]

Therefore, even if the high-frequency excitation would occur, it would not deteriorate the controller performance significantly due to the negligible influence of the higher modes with respect to the dominant mode-shapes.

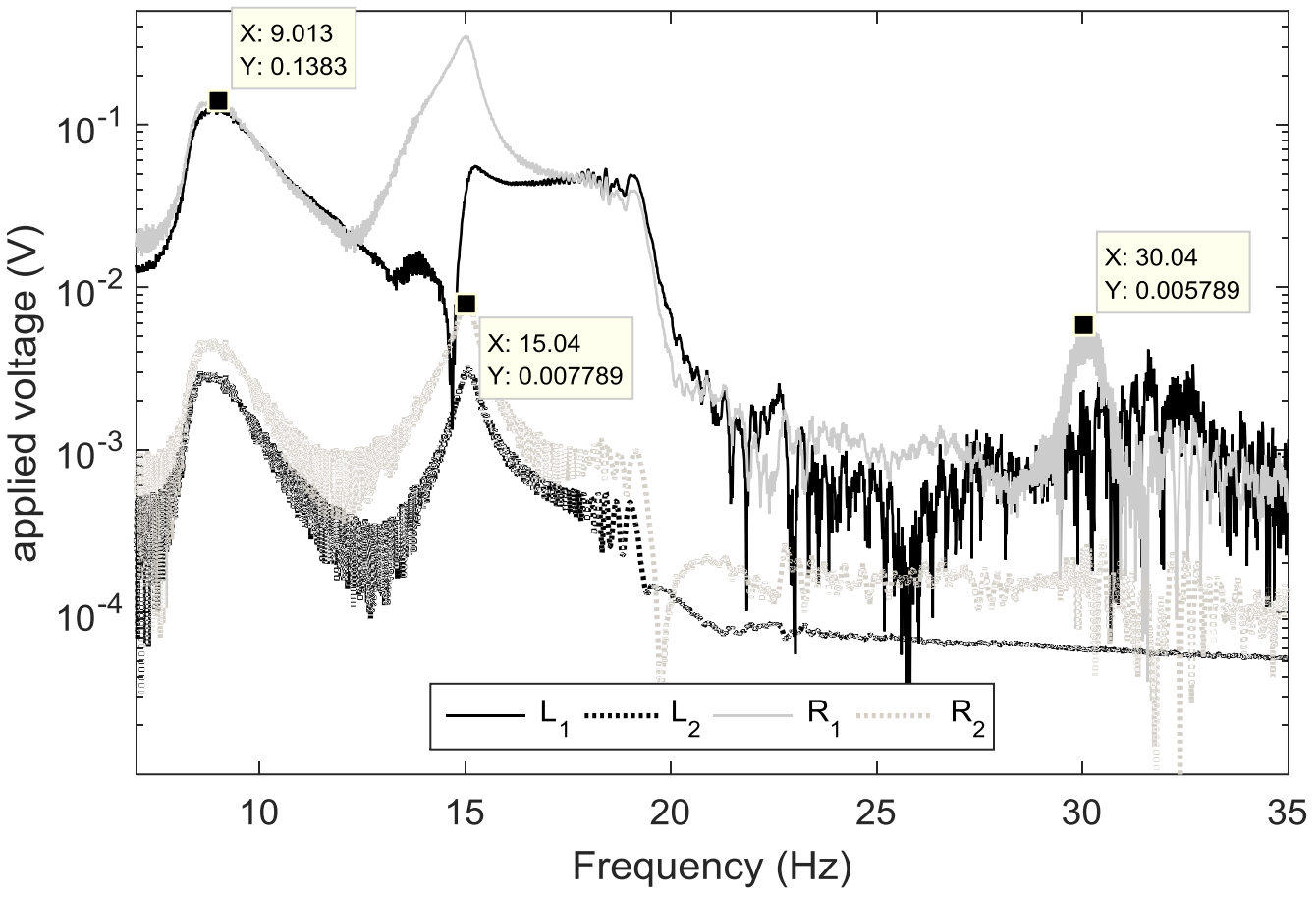

Figure 5.13 FFT analysis on the control effort [13]

Finally, by aiming at investigating the robust performance of the closed-loop system in stabilizing the unmodeled higher-order dynamics of the MRI funnel structure, the coupled system is excited through the disturbance channel with a sweep sine signal that charges the higher-order mode-shapes of the structure in the frequency range of 20-38 Hz. The applied mechanical force acts on the system within -sec and the experimental results for the output measurements are presented in Figure 5.14.

In this figure, the controlled and the uncontrolled cases are brought together in four subplots to evaluate the robustness of the control system. It can be observed that the attenuation performance is reduced compared to the nominal frequency range but the control system mostly suppressed the vibration. Furthermore, as it can be seen that the applied control effort on the actuator $R_1$ saturates for a short period between 6.8 and 7.7 sec which leads to performance loss that is mostly detected in sensor elements $L_1$ and $R_1 R_1$. This time period stands for the higher-order unmodeled dynamics with the natural frequency around 27 Hz. This issue motivates the consideration of actuator nonlinearity (here: saturation) in the design of robust control, i.e., (a) Model predictive control (MPC) in Chapter 6 and (b) Anti-windup compensation (AWC) in Chapter 7.

By comparing the control effort in the nominal frequency range (Figure 5.12) and the high-frequency area Figure 5.15, it can be seen that the control system loses its optimality since by using more actuation energy less performance is achieved. This emphasizes the importance of the order of the nominal system and the weighting functions in the control design procedure. Other than the issue with actuator saturation, the classical approach does **not** deliver a fixed-order controller and is computationally very expensive. In order to deal with the disturbance rejection problem in a more systematical approach, two new methods are proposed in Chapter 11. A very detailed analysis of the output feedback disturbance rejection control is given that solves the numerical issues reported in this chapter.

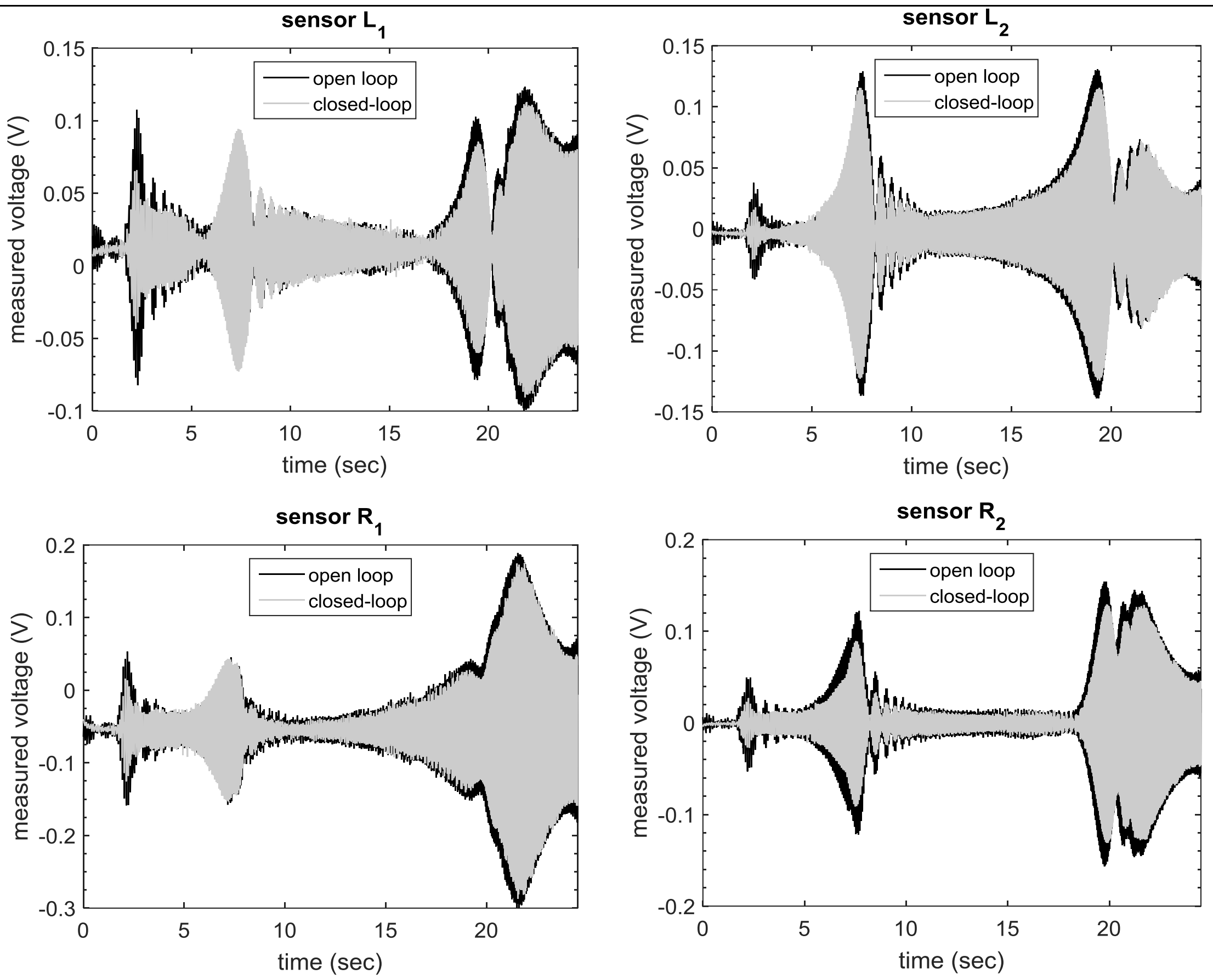

Figure 5.14 Robust vibration attenuation performance for unmodeled dynamics

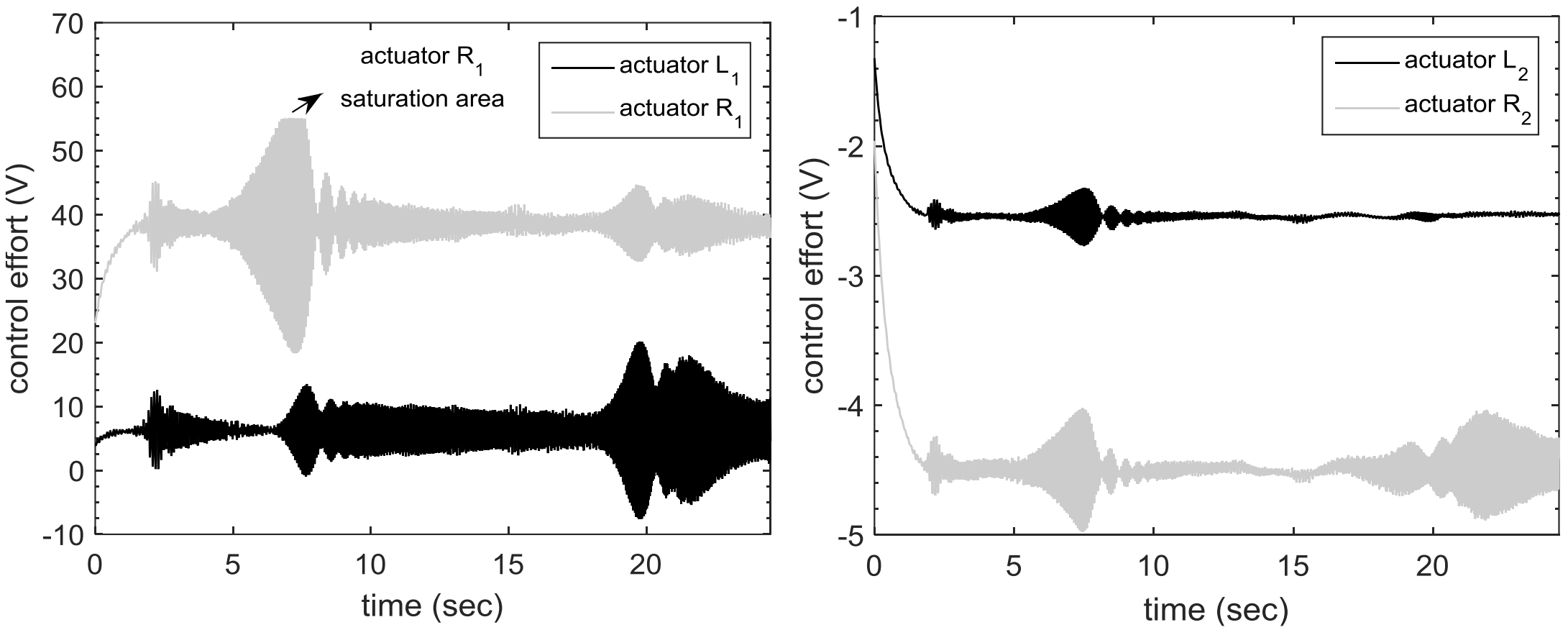

Figure 5.15 Applied voltage on the piezo-actuators in the high-frequency range [13]

# 6 Dealing with hard constraints using a model predictive approach

In this chapter, two fundamental contributions in dealing with hard constraints relevant for robust output feedback control are presented. MPC is capable of handling hard system constraints and nonlinearities such as actuator saturation. Additionally, the proposed internal model repetitive observer-based control system provides a useful framework for rejecting periodic disturbances common in AVC systems. The proposed MPC method in the rest of the chapter removes the need for robustification of the MPC algorithm. The Laguerre parameterization resulted in a significant reduction in the number of the optimization variables with respect to the conventional MPC algorithm. To be more specific, the MPC approach consists of a prediction step followed by an optimization step. In the optimization step, the predicted system states and outputs in the future horizon are used in the framework of an optimal controller to generate a vector of optimal control inputs while satisfying the system's hard constraints. Moreover, it is observed that for geometrically linear vibration amplitudes and for the disturbance frequency limited to the range in which the internal model is valid, the perturbations are guaranteed to be asymptotically rejected. Finally, the performance of the control system is proven to be robust against the high-order unmodeled dynamics of the system and against the spillover effect of the actuators.

In the upcoming chapter, as an alternative ad-hoc approach, special attention is given towards the so-called two-step paradigm of AWC and a novel unified framework for anti-windup compensator design is provided. The analysis and synthesis process of several AWC schemes are adapted for active vibration control problem and detailed experimental investigations are carried out for extraction of the true goal of anti-windup compensators. It is of great importance that by having the two families of methods (MPC and AWC), the advantages and disadvantages of the two methodologies are detailed because both deal with hard constraints.

## 6.1 Observer-based repetitive model predictive control

An observer-based feedback/feedforward MPC algorithm is developed for addressing the AVC of lightly damped structures. For this purpose, the feedback control design process is formulated in the framework of DRC and a repetitive MPC is adapted to guarantee the robust asymptotic stability of the closed-loop system. To this end, a recursive least squares (RLS) algorithm is engaged for online estimation of the disturbance signal, and the estimated disturbance is feed-forwarded through the control channels. The mismatch disturbance is considered as a broadband energy bounded unknown signal independent of the control input, and the internal model principle is adjusted to account for its governing dynamics. For the sake of relieving the computational burden of online optimization in MPC scheme, within the broad prediction horizons, a set of orthonormal Laguerre functions is utilized. The controller design is carried out on a reduced-order model of the experimental system in the nominal frequency range of operation.

It should be emphasized that in contrast to the selective mode shape control scheme e.g. [49], the so-called local vibration suppression is addressed in this chapter. However, considering the geometry of the structure i.e. smart cantilever beam (see Figure 6.1 (a) Experimental rig of the control loop), the transient response mostly depends on the dominant set of transverse modes. Therefore, suppressing the vibration at the free end of the beam results in global attenuation. This justifies the use of "end-effector regulation" in the context of this dissertation.

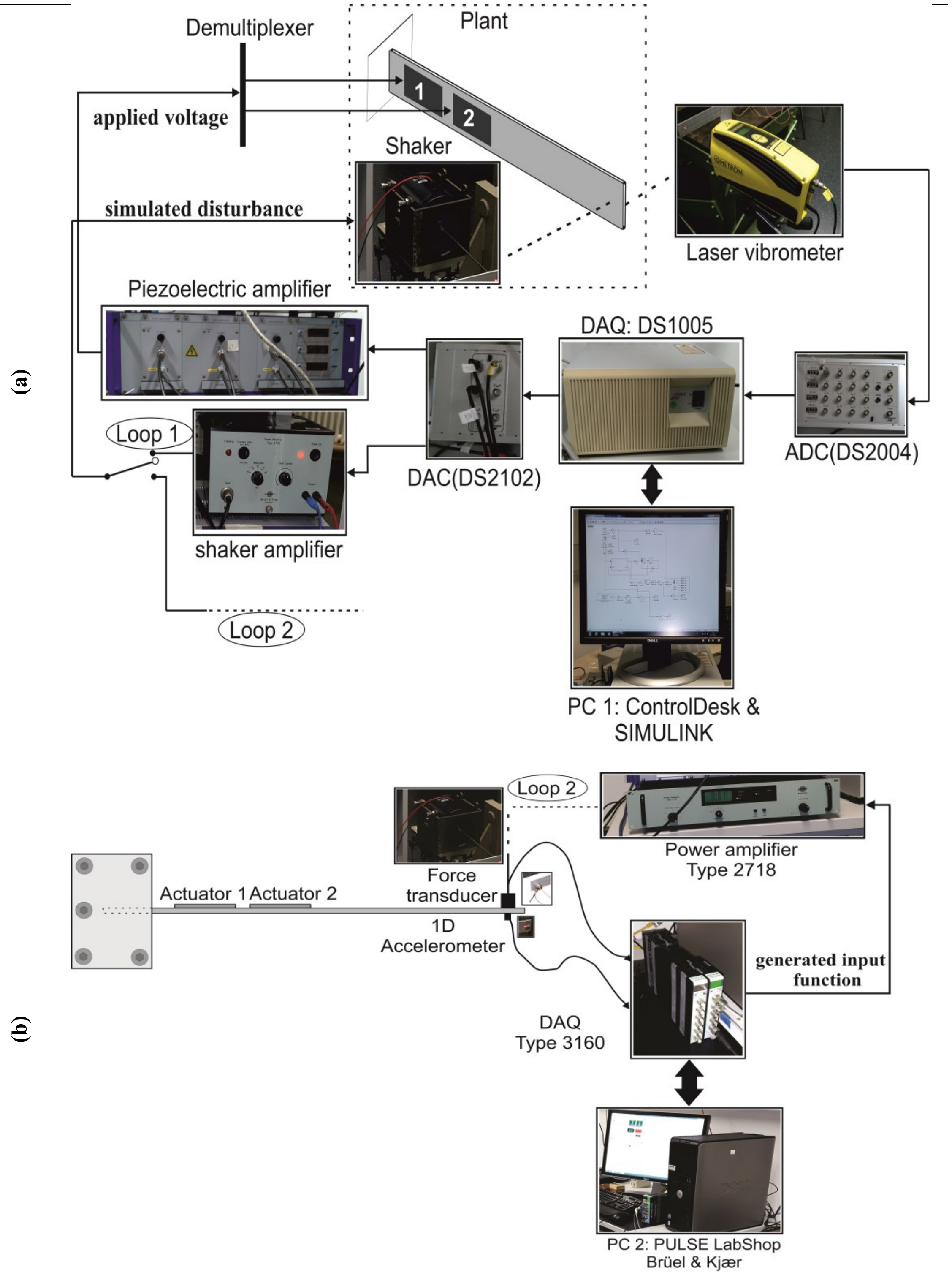

Figure 6.1 (a) Experimental rig of the control loop. (b) Experimental setup for modal analysis [149]

A system identification approach (as shown in Chapter 4) is employed to extract the dynamical model. As expected and shown in Figure 4.12 and Figure 4.13, the identification methods introduce various sorts of uncertainties such as structured uncertainty and unmodeled dynamics which should be addressed by the controller and observer systems [13].

This chapter contributes to the development of a repetitive MPC based on the internal model principle (IMP) and disturbance estimation together with practical consideration of implementing the MPC in AVC. Accordingly, a memory records the plant output, and a time-shifter shift the states of the estimator forwards in time to generate the appropriate control action in DRC. It should be mentioned that AVC effectively operates in an approximate bandwidth of 1 kHz (in this chapter: 800 Hz) and higher frequency ranges are mostly treated by passive methods [58].

### 6.1.1 The MPC strategy

In this section, first, the classical MPC method based on IMP is presented for a discrete LTI system in the state space form. Next, the repetitive control strategy is presented to account for periodic DRC. The control problem is then formulated in the form of optimal feedback controller synthesis for the augmented plant. The hard actuator constraints, as side conditions on the optimization problem, are enforced in each time interval in an online manner [150]. Various improvements are considered for enhancing the stability and performance of the standard control system as well as the numerical solution for the optimization problem.

### 6.1.2 Augmented system representation

An observer-based MPC as a model-based control strategy uses the current control input and the observed states of the plant (and possibly the estimation of disturbance) to simultaneously predict the future output of the plant and the control signal in a finite horizon. The underlying system is modeled by the following finite dimensional, discrete-time state-space representation through a lumped parameter approximation

Note that there is a slight notational mismatch between the system model of (4.4), (5.1), and those of AWC. This is due to the originality of the proposed schemes as the contribution of the author in the original form of the published articles that are accordingly cited in the dissertation. Consider the state space discrete model (6.1) as the system plant.

$$\begin{aligned} x(k+1) &= Ax(k) + Bu(k) + Hd(k), \\ y(k) &= Cx(k). \end{aligned} \tag{6.1}$$

The analogy between the continuous form (5.1) and (6.1) is realized through Tustin transformation of Chapter 4. The vector, $x(.) \in \mathbf{R}^n$ is the state, $u(.) \in \mathbf{R}^m$ is the control input, $y(.) \in \mathbf{R}^p$ is the measurement vector and, $d(.) \in \mathbf{R}^r$ denotes the unknown, mismatched disturbance signal. It is assumed that the pair $(A, B)$ is stabilizable and, $(C, A)$ is detectable. The dimensions of the state, input, and output vectors are selected based on the results of the system identification problem in 6.2. The admissible control inputs are characterized by a compact set, $\mathcal{U} \coloneqq \{u \in \mathbf{R}^m \mid |u_i| \leq u_{i0}, i = 1, \dots, m\}$ in which, $u_{i0}$ are strictly positive numbers specifying the saturation limits of the actuators.

The class of mechanical disturbances, $d(.)$ is assumed to be generated by an autonomous dynamic system of the form [151],

$$\begin{aligned} w(k+1) &= Sw(k), \\ d(k) &= Ew(k), \end{aligned} \tag{6.2}$$

where $w(.) \in \mathbf{R}^{n_w}$ is the state vector and, $S$ and $E$ are real matrices of appropriate dimensions. It is assumed that the eigenvalues of the matrix $S$ belong to the unit disk in the complex plane i.e. $|z| \leq 1$. Besides, the eigenvalues which lie on the boundary $|z| = 1$ are simple. This assumption ensures that the non-vanishing components of the disturbance are bounded and periodic [152]. Accordingly, the eigenvalues of $S$ capture the dominant frequency characteristics of the disturbance signal. The characteristic polynomial of $S$ is considered as

$$\Gamma(z) := \det(zI - S) = z^{n_w} + \sum_{i=1}^{n_w} \alpha_i z^{n_w - i}.$$

The repetitive control systems provide an operational means for the rejection of periodic disturbances based on IMP [153]. The term "repetitive" accentuates the periodic nature of the signals to be rejected. By IMP, asymptotic disturbance rejection in a control system amounts to incorporating a suitable dynamic generator (of the disturbances) in the closed-loop system [154]. Using an internal model, the controller replicates the repetitive characteristics of the disturbance signals in its structure. Consequently, the dynamic model of disturbance allows the controller to reject the disturbance signal with proper elimination of the counteracting energy. To this end, the following causal operator corresponding to the characteristic polynomial $\Gamma(q^{-1})$ is considered as

$$\Gamma(q^{-1}) = 1 + \sum_{i=1}^{n_w} \alpha_i q^{-i}, \tag{6.3}$$

where, $q^{-1}$ denotes the backward shift operator.

**Remark 6.1.** In accordance with the characteristic equation of $S$ in (6.2), the following equality is deduced $\Gamma(q^{-1})d(k) \equiv 0$.

The following filtered variables are introduced to embed the internal model of the disturbance signal into the control system

$$\begin{aligned} x_f(k) &:= \Gamma(q^{-1})x(k), \\ u_f(k) &:= \Gamma(q^{-1})u(k). \end{aligned} \tag{6.4}$$

In view of

**Remark 6.1**, applying the operator $\Gamma(q^{-1})$ to the system equations (6.1) yields (6.5);

$$\begin{aligned} x_f(k+1) &= Ax_f(k) + Bu_f(k), \\ y(k) + \textstyle\sum_{i=1}^{n_w} \alpha_i y(k-i) &= Cx_f(k). \end{aligned} \tag{6.5}$$

Setting $X(k) := \left[x_f^T(k), y^T(k-1), y^T(k-2), \ldots, y^T(k-n_w)\right]^T \in \mathbf{R}^{n+pn_w}$, we describe both filtered state and output equations by the following augmented system

$$\begin{aligned} X(k+1) &= A_{aug}X(k) + B_{aug}u_f(k), \\ y(k) &= C_{aug}X(k), \end{aligned} \tag{6.6}$$

where the system matrices are given by

$$A_{aug} = \begin{bmatrix} A & 0 & 0 & \cdots & 0 \\ C & -\alpha_1 I & -\alpha_2 I & \cdots & -\alpha_{n_w} I \\ 0 & I & 0 & \cdots & 0 \\ \vdots & \vdots & \ddots & & \vdots \\ 0 & 0 & \cdots & I & 0 \end{bmatrix}, B_{aug} = \begin{bmatrix} B \\ 0 \\ 0 \\ \vdots \\ 0 \end{bmatrix}, \quad C_{aug} = [C \quad -\alpha_1 I \quad -\alpha_2 I \quad \cdots \quad -\alpha_{n_w} I].$$

**Remark 6.2.** To guarantee the controllability and observability of the augmented system of (6.6), we assume that the matrix

$$\begin{bmatrix} zI - A & B \\ -C & 0 \end{bmatrix},$$

is full rank for all $z$ satisfying $\Gamma(z) = 0$. That is, the original system of (6.1) has no common transmission zeros with the eigenvalues of the disturbance model [155]. The eigenvalues of the matrix $A_{aug}$ comprise those of $A$ and a companion block matrix corresponding to the disturbance dynamics [46]. Thereby, in view of IMP, the repetitive DRC reduces to stabilization problem of the augmented system of (6.6) [151].

### 6.1.3 Repetitive MPC

To control the augmented system (6.6) an optimal feedback strategy is employed in the framework of predictive control. In this regard, at sampling instant $k$, the following finite horizon optimal control problem is solved to obtain the control action;

$$\min_{u_f(k+.|k)} \left\{ J(k) := \sum_{j=1}^{N} \beta^{-2j} ||X(k+j|k)||_Q^2 + \beta^{-2(j-1)} ||u_f(k+j-1|k)||_R^2 \right\},$$

subject to

$$\begin{gathered} X(k+j|k) = A_{aug} X(k+j-1|k) + B_{aug} u_f(k+j-1|k), \\ X(k|k) = \hat{X}(k), \\ u_f(k+j-1|k) \in \mathcal{U}_f, j = 1, \dots, N. \end{gathered} \tag{6.7}$$

It should be remarked that the notation convention $||X||_W^2 = norm(X^T W X)^2$ is used in (6.7). In the formulation of (6.7), $N$ is a positive integer specifying the optimization horizon, $\beta > 1$ is the data weighting parameter, $Q \succcurlyeq 0$ and $R \succ 0$ are performance weight matrices penalizing the pertinent variables. The set $\mathcal{U}_f$ denotes the image of $\mathcal{U}$ under the filtering map $\Gamma(q^{-1})$. The vectors, $X(k+.|k)$ and $u_f(k+.|k)$, respectively stand for the predicted state and control variables along the optimization horizon as shown in (6.6).

It should be pointed out that compared to the linear quadratic regulator on output (LQRY), which is mostly implemented in structural vibration control based on the steady-state solution of algebraic Riccati equation (ARE), the repetitive MPC is the solution of online optimization of the closed-loop trajectory. Trajectory refers to the evolution of system states and output over time in the prediction horizon. To sum up, the low computational burden of linear MPC and its robustness features make it a logical candidate compared to LQRY [156], [157]. The low computation burden of the proposed MPC is due to the use of orthonormal Laguerre functions. Also, the conservativeness of the LQR and LQG controllers by tuning the weighting matrices in solving the stage cost has been reported by Bossi et al. [49].

In (6.7), the predictions are obtained through system dynamics with the recent state vector information as the initial value. Since the state vector $X(.)$ is not measured, an observer is used to asymptotically reconstruct the state information. To this end, the Kalman filter is employed owing to its optimal performance in noisy environments. The filter dynamics is governed by the following equations [154]

$$\begin{gathered} \hat{X}(k+1) = A_{aug}\hat{X}(k) + B_{aug}u_f(k) + K_f(k)\left(y(k) - C_{aug}\hat{X}(k)\right), \\ \hat{X}(0) = 0, \\ K_f(k) = A_{aug}P_f(k)C_{aug}^T\left(C_{aug}P_f(k)C_{aug}^T + R_f\right)^{-1}, \\ P_f(k+1) = A_{aug}(I - K_f(k)C_{aug})P_f(k)A_{aug}^T + Q_f, \\ P_f(0) = P_{f0} \succ 0. \end{gathered} \tag{6.8}$$

The matrix $P_f(.) \in \mathbf{R}^{(n+pn_w)\times(n+pn_w)}$ is the estimation error covariance, $Q_f \in \mathbf{R}^{(n+pn_w)\times(n+pn_w)}$ is the process noise covariance, $R_f \in \mathbf{R}^{p\times p}$ is the measurement noise covariance and, $K_f(.) \in \mathbf{R}^{(n+pn_w)\times p}$ denotes the gain of the filter. To alleviate the computational burden of the control method for AVC applications, the gain matrix $K_f(.)$ can be computed offline and stored in the memory for real-time implementations [154]. Solving the open-loop optimal control problem (6.7) over the sampling instants yields a receding horizon output feedback given by

$$u_f(k) = u_f^*(k|k), \tag{6.9}$$

in which, $u_f^*(k|k)$ is the first element of

$$\{u_f^*(k+j-1|k)\}_{j=1}^{N} \coloneqq \arg \min_{u_f(k+.|k)\in\mathcal{U}_f} J(k). \tag{6.10}$$

The weighting parameter $\beta$ in the optimal control formulation (6.7) is used to assign an exponentially decreasing weight to the sequence of predicted system variables. The data weighting technique resolves the ill-conditioning problem of the Hessian matrix associated with $J(.)$ for large optimization horizons [158]. To render the optimization problem (6.7) to an optimal control problem with a time-invariant cost function, the following theorem without proof is given.

**Theorem 6.1** [158] *The optimization problem in* (6.11) *is equivalent to the one in* (6.7)*;*

$$\min_{u_f^\beta(k+.|k)} \left\{ J_\beta(k) \coloneqq \sum_{j=1}^{N} ||X_\beta(k+j|k)||_Q^2 + ||u_f^\beta(k+j-1|k)||_R^2 \right\},$$

*subject to*

$$\begin{gathered} X_\beta(k+j|k) = \frac{A_{aug}}{\beta} X_\beta(k+j-1|k) + \frac{B_{aug}}{\beta} u_f^\beta(k+j-1|k), \\ X_\beta(k|k) = \beta^{-1}\hat{X}(k), \\ \beta^{j-1} u_f^\beta(k+j-1|k) \in \mathcal{U}_f, j = 1, \dots, N, \end{gathered} \tag{6.11}$$

*where*

$$\begin{aligned} &X_\beta(k+j|k) = \beta^{-j} X(k+j|k), \\ &u_f^\beta(k+j-1|k) = \beta^{-(j-1)} u_f(k+j-1|k), j = 1, \dots, N. \end{aligned} \tag{6.12}$$

Owing to

**Theorem 6.1**, numerical optimization of the weighted repetitive MPC (6.7) is carried out based on the time-invariant cost function, $J_\beta(.)$ along with the associated scaled system dynamics. Scaled system dynamics in MPC terminology are the augmented system states and matrices weighted over the horizon. It should be noted that the presented method compared to [58] is developed based on the IMP and a set of orthonormal Laguerre functions that alleviate the computational burden of online optimization.

#### 6.1.3.1 Numerical optimization via input parameterization

It is often required to use a large optimization horizon in MPC systems to achieve a satisfactory transient response and stability margin. The exponential data weighting technique removes the numerical ill-conditioning issue of the associated optimization. However, the dimension of the problem still could be large enough to be solved in fast sampling rates. A possible approach to remedy this issue and to reduce the size of the MPC synthesis is to parameterize the predicted control variables by a finite set of basis functions [40]. By this method, solving the optimal control problem (6.7) for the minimizing sequence, $\{u_f^*(k+j-1|k)\}_{j=1}^{N}$ , boils down to finding the optimal parameterization coefficients in a lower dimension. Considering the linear and time-invariant characteristics of the underlying system (6.6), the exponential functions can approximate the stabilizing control inputs. In this regard, a class of orthonormal exponential basis functions, known as the Laguerre functions are used. The Laguerre functions are powerful approximation tools used in various fields such as system identification and signal processing [159]. A Laguerre basis of dimension $N_\ell$ and convergence rate of $a \in (0,1)$, also known as the scale factor of the Laguerre basis, is constructed in the frequency-domain as follows

$$\mathcal{L}_i(z) := \frac{\sqrt{1-a^2}}{z-a}\left(\frac{1-az}{z-a}\right)^{i-1}, i = 1, \dots, N_\ell. \tag{6.13}$$

For $a = 0$, we have a classical MPC i.e., without any Laguerre network. The choice of $a$ also affects how the Laguerre functions decay to zero. For higher values of convergence rate, say $a = 0.9$, the decay rate of the Laguerre functions is much slower in comparison for lower values, say $a = 0.5$ with a factor of more than three (see page 89 of [40]). Consequently, the selection of $a$ is non-trivial and has an impact on the selection of basis of dimension $N_\ell$. In time-domain, the Laguerre functions are calculated using the inverse z-transform as

$$\ell_i(k) := Z^{-1}\{\mathcal{L}_i(z)\}, i = 1, \dots, N_\ell. \tag{6.14}$$

The orthonormality of the Laguerre functions can be represented as

$$\sum_{k=0}^{\infty} \ell_i(k)\ell_j(k) = \delta_{ij}, \tag{6.15}$$

where, $\delta_{ij}$ is the Kronecker delta. To parameterize the predicted control sequence along the optimization horizon, a Laguerre basis of dimension $N_i$ and pole $a_i$, denoted as $\{\ell_m^i(.)\}_{m=0}^{N_i}$, is designated to the $i$-th input $u_{f,i}^{\beta}(.)$. $a_i$ is referred to as poles in the literature due to the form of (6.13). Accordingly, the following linear parameterization is considered

$$u_{f,i}^{\beta}(k+p|k) = L_i^T(p)\eta_i, \qquad p = 0, \dots, N-1, \tag{6.16}$$

in which, $L_i(.) := \begin{bmatrix}\ell_1^i(.) & \cdots & \ell_{N_i}^i(.)\end{bmatrix}^T \in \mathbf{R}^{N_i}$ and $\eta_i \in \mathbf{R}^{N_i}$ denotes the pertinent parameterization vector. Using the input parameterization (6.16), (6.11) is settled to the optimization problem in (6.17);

$$\min_{\eta}\{\eta^T R_\ell \eta + \textstyle\sum_{j=1}^{N} ||X_\beta(k+j|k)||_Q^2\},$$

subject to

$$\begin{gathered} X_\beta(k+j|k) = \frac{A_{aug}}{\beta} X_\beta(k+j-1|k) + \frac{B_{aug}}{\beta} L^T(j-1)\eta, \\ X_\beta(k|k) = \beta^{-1}\hat{X}(k), \\ \eta \in \Pi, j = 1, \dots, N, \end{gathered} \tag{6.17}$$

where,

$$L(.) := \operatorname{diag}\big(L_1(.), \dots, L_m(.)\big) \in \mathbf{R}^{(\sum_{i=1}^{m} N_i)m},$$

$$\eta := \operatorname{col}(\eta_1, \dots, \eta_m) \in \mathbf{R}^{\sum_{i=1}^{m} N_i},$$

$$R_\ell := \sum_{j=1}^{N} L(j-1)RL^T(j-1).$$

The compact set $\Pi \subset \mathbf{R}^{\sum_{i=1}^{m} N_i}$ characterizes the feasible region for the parameterization vector, determined by the control constraint set. The primary control input $u$ is confined to the box constraint set, $\mathcal{U}$ and thereby, the compact set $\Pi$ can be represented by finite numbers of linear inequalities. As a result, the optimization problem (6.17) becomes standard quadratic programming (QP) that can be solved efficiently in small sampling intervals. The MPC formulation (6.11) comprises an optimization problem of dimension $Nm$ to obtain the current control action. The parameterization procedure converts this optimization problem to that of (6.17) with the dimension of $\sum_{i=1}^{m} N_i$. By the proper selection of the locations of the poles $a_i$, the predicted control trajectories can be effectively captured by the Laguerre approximation (6.11) with a few terms. This, in turn, considerably reduces the computational burden of the repetitive MPC and enables fast sampling rates for AVC applications. The proof of such computational advantages is available in the fundamental work of [40], see pp. 92 for instance.

### 6.1.4 Stability analysis

A number of ways and means have been proposed in the literature to establish the Lyapunov stability of predictive optimal control schemes [160], [161]. A majority of these methods impose fictitious state constraints to ensure that the optimal value of the MPC cost function decreases monotonically over the sampling instants. Predictive control formulations with terminal equality/inequality constraints are the most prominent representatives of these methods. However, such techniques for addressing the closed-loop stability often increase the computational burden and complexity of the associated optimization problem, precluding their applicability for fast sampling rates. Primbs and Nevistic [162] showed that, for a given MPC formulation, there exists a finite horizon for which the stability properties of the associated infinite horizon optimal control problem are retrieved. From such a perspective, Wang discussed the stability of the Laguerre approximation-based MPC in the framework of dual-mode control [40]. Given the results from [162], for a sufficiently large prediction horizon, the unconstrained solution of the optimal control problem (6.11) converges to that of the associated infinite horizon LQ control. When the constraints are active, the optimal control sequence satisfies the constraints along the optimization horizon for $N_0$ number of samples with $N_0 < N$. Then, the control sequence returns to the infinite horizon optimal solution that guarantees the closed-loop stability.

### 6.1.5 Repetitive MPC with a prescribed degree of stability

The exponential data weighting combined with the Laguerre parameterization offers a practical computational framework for handling large optimization horizons and thereby, for recovering the infinite horizon control solutions. Accordingly, the stability results of LQR can be extended to the case of receding horizon control. In particular, to achieve a prescribed degree of stability, it is proposed to use the following weight matrices;

$$\begin{aligned} Q &= \frac{\gamma^2}{\beta^2}\bar{Q} + \left(1 - \frac{\gamma^2}{\beta^2}\right)P_\infty, \\ R &= \frac{\gamma^2}{\beta^2}, \end{aligned} \tag{6.18}$$

in which, $\gamma \in (0,1)$ is the required stability margin, $\bar{Q} \succcurlyeq 0$, $\bar{R} \succ 0$ and $P_\infty$ denotes the solution of the following Riccati equation

$$A_{aug}^T\left(P_\infty - \frac{1}{\beta^2}P_\infty B_{aug}\left(\bar{R}+\frac{1}{\gamma^2}B^T P_\infty B\right)^{-1} B_{aug}^T P_\infty\right)A_{aug} + \gamma^2(\bar{Q}-P_\infty) = 0.$$

Providing a sufficiently large optimization horizon $N$, the above modification ensures that [40]

$$||X_\beta(k+j|k)|| \leq \gamma^j ||X_\beta(k)||, \forall j \geq N_0. \tag{6.19}$$

From a linear algebraic point of view, the eigenvalues of the closed-loop system lie inside the circle $|z| \leq \gamma$ on the complex plane [46].

### 6.1.6 RLS disturbance estimation

In the framework of DRC, in order to improve the overall robustness of the AVC system, we use an RLS estimator to reduce the power of the disturbance signal that affects the repetitive MPC. Based on discrete-time state equation (6.1), the following regression model is considered for disturbance estimation;

$$\hat{x}(k+1) - A\hat{x}(k) - Bu(k) = Hd(k). \tag{6.20}$$

Here, $\hat{x}(.)$ is obtained by imposing the inverse operator $\Gamma^{-1}(q^{-1})$ on the estimated state vector of the Kalman filter, $\hat{X}(.)$. The RLS algorithm and a detailed discussion of its convergence/robustness properties can be found in [154]. To achieve a 2-norm attenuation of the disturbance, the output of the RLS is fed-forwarded by the gain $(B^TB)^{-1}B^TH$ through the control channel (see Figure 6.2).

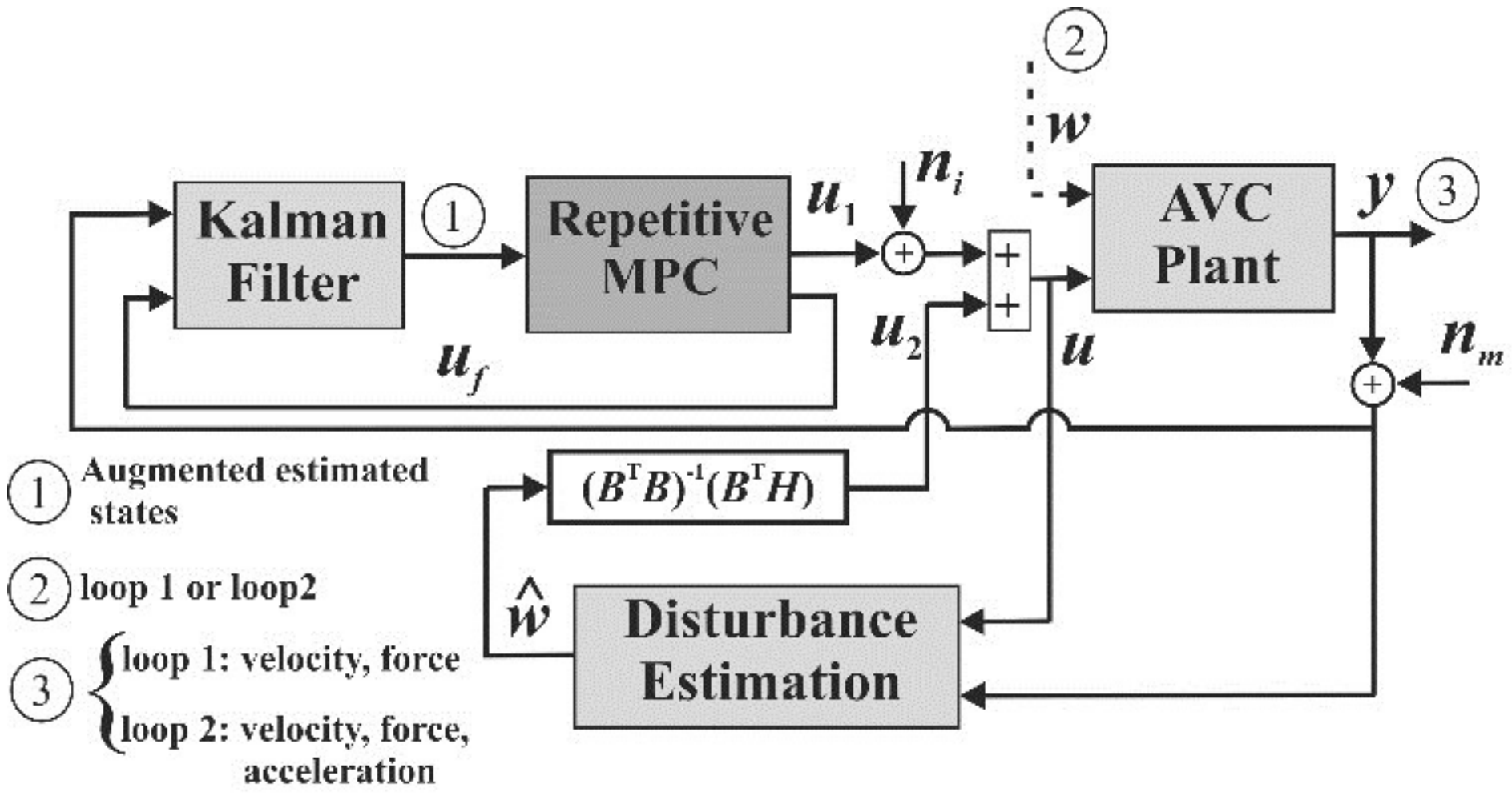

Figure 6.2 Schematic configuration of control system [149]

## 6.2 Experimental apparatus and the reduced-order plant

In this section, the experimental rigs that are used to obtain the reduced-order nominal plant for control design purposes, experimental modal analysis, and finally monitoring of the performance of the closed-loop system in real-time are introduced. The plant is a vibrating clamped-free smart aluminum beam with a geometrical dimension, 440×40×3 mm. The host structure is assumed to be isotropic with Young's modulus 70 Gpa and density 2.7 g/cm³. Two piezoelectric patches (DuraActTM P-876.A15) are attached on one side of the beam acting as the actuation elements. The optimality of the actuation power in controlling the mode shapes of the continuous structure is guaranteed by placing the piezo-patches according to robust-optimal actuator placement criterion that has been presented at [163], [164]. The numerical values of the system matrices $A$, $B$, $C$, and $H$ are omitted here for the sake of brevity and the interested reader may refer to [149].

The “optimal placement” terminology is based on the technique of balanced model reduction. The resulting system benefits equally from controllable and observable retained modes. Additionally, the algorithm in [163] relies on the assortment of the modes based on their simultaneous controllability/observability Gramians. The optimality is guaranteed over $H_2$ and $H_\infty$ norms of the transfer matrices of all placement candidates on the surface of the beam. As a result, the algorithm guarantees the globality of solution of sensor/actuator placement by evaluating all possible configuration. It should be pointed out that the non-collocated optimal structural configuration proposed in [163] introduces a time delay between the actuator and sensor elements which may be significant in large structures [165].

Two physical loops are involved in the implementation phase of modeling and control systems as Figure 6.1a and b. Figure 6.1a, shown as loop 1, presents the control execution circuit together with an external mechanical shaker. However, Figure 6.1b, which is indicated by loop 2, is used for obtaining the internal model of disturbance, (6.1), and later for evaluation of the closed-loop system in the frequency domain. It should be mentioned that in the last part of the experimental test both of the loop 1 and 2 are activated simultaneously to compare the FRF of the controlled system with the open-loop one. According to [55], as shown in Figure 6.1a, the first piezo-patch is at 20 mm distance from the clamped end and the second patch stands at 80 mm from the first one. The velocity and the external excitation are measured through the corresponding measurement channels of the acquisition system, connected to the laser vibrometer VH-1000-D, and the force sensor, respectively. The digital laser Doppler vibrometer supplies the signal to the feedback channel of the control system. The force transducer provides the input measurement data for modal analysis. Since the free end of the beam is in oscillation and plays a significant role in the overall response of the system, the scanning digital laser is placed at the free-end at 238 mm distance from the beam in its rest mode.

A mismatch disturbance is considered to be affecting the system behavior through an independent channel of the control input. For this purpose, a vibration exciter (Brüel&Kjær shaker Type 4809) is employed as the source of simulated disruption. The real-time analyses are carried out using dSPACE system with digital data acquisition (DAQ) DS1005 PPC Board with the working range of ±10 V. The front end of the DAQ system consists of an analog to digital converter (dSPACE ADC DS2004) and a digital to analog converter (dSPACE DAC DS2102). An amplifier (PI E-500) is utilized to amplify the generated signal by DAQ with a constant gain of 100 before sending it to the actuators. Similarly, another power amplifier (B&K Type 2706) is operated to amplify the signal generated by the dSPACE board for the vibration exciter (shaker in Figure 6.1a). The use of Brüel&Kjær 2706 amplifier separates loop 1 (control loop) from loop 2 (experimental modal analysis). Accordingly, loop 2 is activated by the third power amplifier (Brüel&Kjær amplifier Type 2718) with a maximum output current of 1.8 A. The experimental modal analyses in loop 2 are conducted on a LAN-XI DAQ Hardware (Brüel&Kjær Type 3160) with input/output of maximum 51.2 kHz DC channels. The acceleration is calculated using a 1D Piezoelectric CCLD accelerometer (Brüel&Kjær Type 4507) with the sensitivity of 100 mV/g which is collocated with the force sensor (see Figure 6.1b).

It should be mentioned that the piezo-elements are activated in the range of [-200 200] V. Both of the controller and estimator systems are created on SIMULINK platform, and then the model is compiled and uploaded to the DAQ system in real-time (see Figure 6.1a). Also, for experimental modal analyses PULSE, LabShop software is employed. The prediction of the controller’s trajectory in the MPC framework as a model-based technique widely depends on the minimal formulation of the plant dynamics. The interconnection of the proposed repetitive MPC with the real-time plant is schematically shown in Figure 6.2. In this figure, $w$ represents the disturbance realization on the control loop and the input data for the FFT which is used to calculate the FRF of the plant from disturbance channel to output (acceleration).

In the context of AVC mostly the reduced-order model of the real plant is a linearization of the system around a single operating point. It should be mentioned that the identification is carried out by neglecting the torsional and in-plain modes of the beam (see Chapter 4). The underlying linear model (ULM) is obtained through the two-step paradigm in the frequency domain system identification framework (see Chapter 4.4). For non-parametric, the multi-reference modal analysis is carried out in the frequency ranges of interest which are [0 250] Hz for the lumped state-space model, and [0 800] Hz for the autonomous disturbance model. Next, the obtained FRFs are parameterized using the frequency domain subspace system identification method proposed in [126]. In order to manifest the identification quality, the response of the real structure is compared with the identified model in the nominal frequency range of interest.

As it is depicted in Figure 6.3, an acceptable agreement exists between the system and the identified model. The state matrix is transformed into modal form, following the method proposed in [166], to assist the quantification of the weighting matrices later in designing procedure of the controller and Kalman gain. As a result, the interpretation of the states and output matrices can be explained in terms of physical variables such as modal velocity, and modal displacement.

On the other hand, it should be also noted that the employed subspace method of (4.4) in Chapter 4.4.3.1 is a black-box scheme and as a result, it may include superficial states without physical interpretation. In order to address this problem, the order of the lumped model is kept at six considering the fact that there exist three natural frequencies in the working range. This selection results in sacrificing accuracy in having a better match between the nonparametric model and the parametric one. This mismatch is due to 1) Neglecting the contribution of higher-order dynamics to the frequency range of operation. 2) Neglecting the small time delay in IO transfer matrix. 3) 80 dB difference between the resonance and anti-resonance amplitudes which makes it difficult to correctly capture both by the algorithm.

For the sake of briefness, further details are referred to the impressive contribution of Pintelon and Schoukens [124]. The obtained reduced-order system as the nominal plant for controller design and disturbance estimation is discretized using an abundantly small sampling time (500 $\mu s$). The sampling time is related to the maximum operational frequency of the plant and cannot be too small because it may be practically impossible to handle the computational efforts for each time increment. The nominal reduced-order system matrices are given in [149].

Next, in order to extract the disturbance dynamics (6.2) independent of the reduced-order model, an experimental modal analysis is performed by means of a shaker test in PULSE LabShop software. Effective prediction of the control action mostly depends on the accuracy of the nominal model as well as the profile of the future disturbances. Accordingly, the shaker is attached to one end of the beam through a rubber band. Similarly, the accelerometer is collocated with the force sensor as shown in Figure 6.1. While the piezo-actuators are inactive, the disturbance channel is triggered by a random signal in the frequency range of [0 800] Hz. A low-pass filter is considered to filter out the high frequency response of the transducer data before applying the FFT. Meanwhile, due to the nature of this excitation signal, leakage is a practical concern that leads to distortion of the measured FRF [120], [121].

To condense the leakage effect and simultaneously to keep the SNR high, averaging with Hann window is employed. Subsequently, the FFT parameters such as the frequency range (sampling time of B&K 3160: 488.3 μs) are considered as a broad range of 6400 lines (df = 125 mHz). The baseband analysis is carried out with 60 averages and 66.67 % overlap.

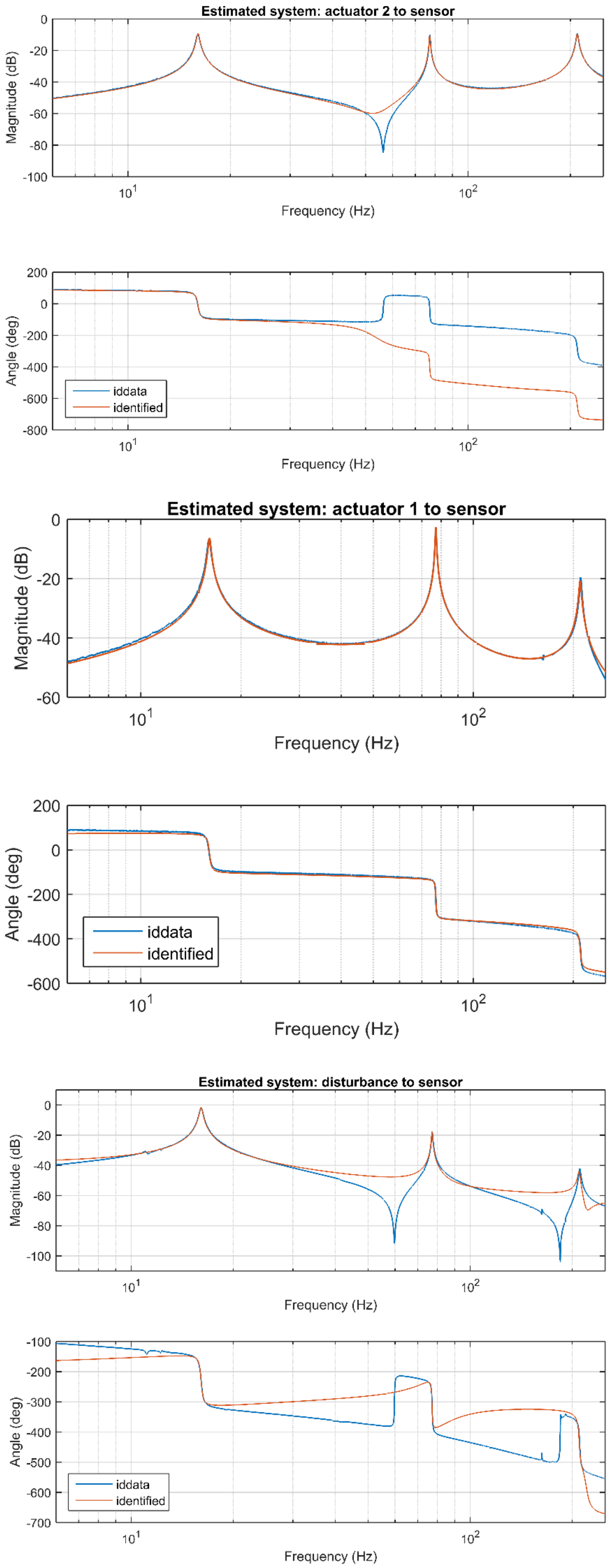

Figure 6.3 Frequency response of the real structure compared to the reduced-order system [149]

A 12th order the LTI object is adjusted to the obtained FRF to parameterize the autonomous exogenous system with a limited number of states in the augmented system of (6.8). The FRF of the identified model is examined over the FRF of the experimental system in Figure 6.4. The state matrix of the disturbance model is given in [149].

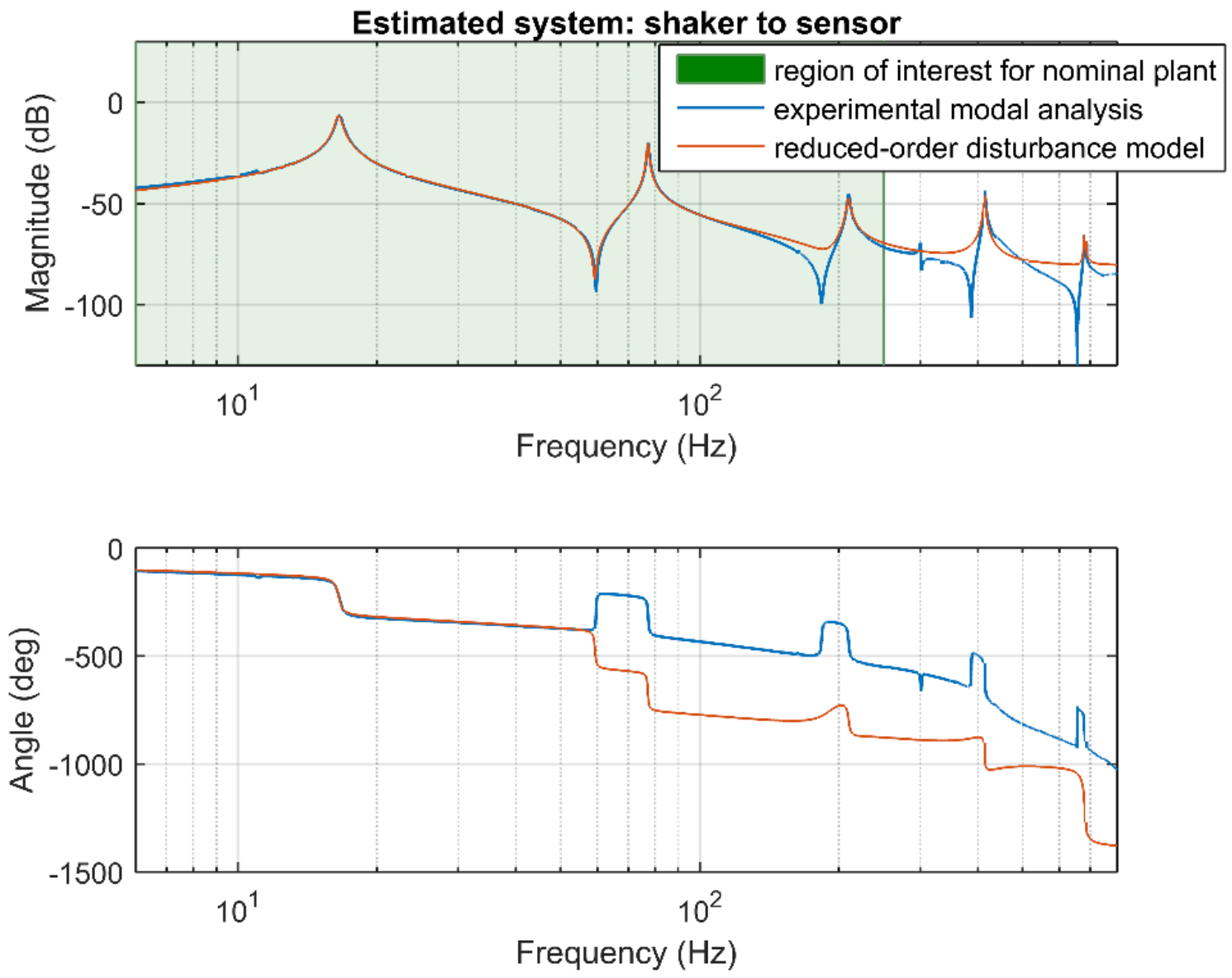

Figure 6.4 FRF of the real structure relative to the 12th disturbance model [149]

The disturbance model is in good agreement with the FRF of the actual structure. The quality of the measurement for identifying the internal model can be assessed by the coherence diagram in the nominal frequency range. Accordingly, as shown in Figure 6.5, the measurements have coherence closeness to one which indicates the linearity of the system response and the excitation. The low-frequency resolution of the measurement around the resonance and anti-resonance states can be improved by using zoom analysis and increasing the number of FFT points which are not necessary for the scope of this chapter. A more detailed investigation of the coherence diagram is given in the second part of this chapter.

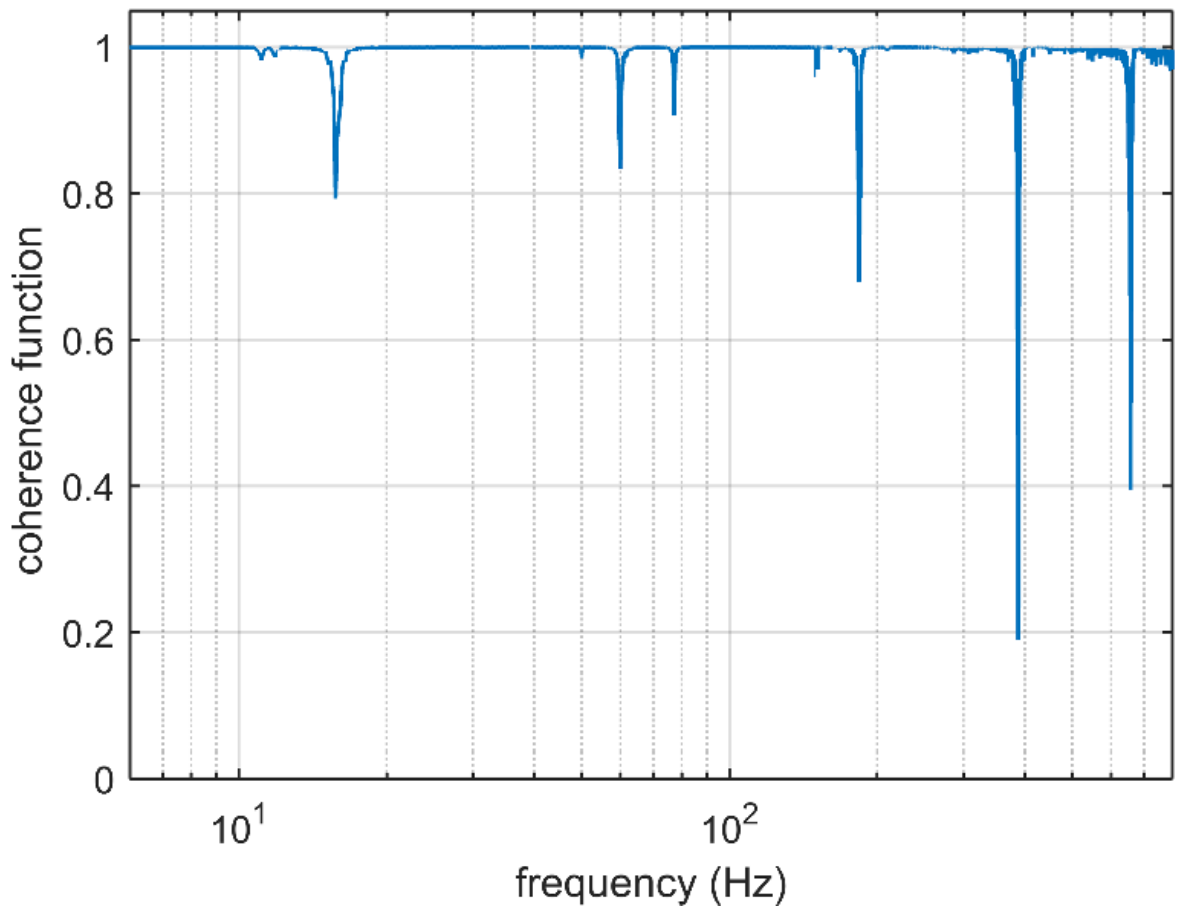

Figure 6.5 Frequency analysis of the coherence function for the disturbance model [149]

### 6.2.1 Experimental implementation

The robustness of the augmented AVC system in (6.6) concerning the bounded parametric uncertainties is achieved as follows: In AVC of the mechanical structures with small damping ratios, the 2-norm bound of the perturbations due to unknown variations and noises is much lower in magnitude compared to the nominal gains of the system in the respective frequencies. Therefore, assuming the vibrations remain in the linear region and the excitation frequency is limited to the range that the disturbance model is valid, due to the IMP, the perturbations will be rejected asymptotically by the asserted control arrangements. The control input noise ($n_i$; see Figure 6.2), as well as the measurement noise ($n_m$) are assumed bounded disturbances modeled as white noise signals with normal distributions. Thereby, the use of the Kalman filter as the estimator is justified. With computation demand being one of the primary sources of practical limitations, reducing excessive overhead for online implementations is achieved by calculating and storing the gain matrix $K_f(.)$, offline.

The higher mode shapes of the structure that lie out of the nominal frequency range are considered as the source of uncertainty and should not be activated unless the disturbance is focused on this area. The spill-over rejection is examined in detail in the last part of the *experimental results*. In the run-time phase of implementation, the following tasks are carried out sequentially:

At each sampling instant ($k$), the future trajectory of the plant state vector is predicted over the optimization horizon $N$ under the action of the predicted control trajectory. The predicted control trajectory along the optimization horizon is parameterized by a suitable Laguerre network. The predictive controller determines the Laguerre parameterization variables by solving an open-loop optimal control problem entailing the output regulation objectives. By the receding horizon policy, only the first component of the obtained optimal solution is applied, and then, the overall process is repeated in the next sampling instant, $k + 1$. The tuning factors of the MPC system are the optimization horizon $N$, the weighting matrices $Q$ and $R$, the data weighting parameter $\beta$, the prescribed degree of stability $\gamma$ and the scale factors $a_i$, and dimensions $N_i$ of the parametrization Laguerre networks. This section discusses the tuning process of the control system; The keywords “Online” and “Offline” at the beginning of each step indicate the state at which the procedure is carried out.

- Offline: The augmented model is constructed having the state-space matrices $A, B, H,$ and $C$, $K_f$ as the observation gain, prediction horizon, the weighting matrices in optimal control design, and matrices $S, E$ in (6.2). Note that matrices $S, E$ are also calculated using system identification where the only input is the external excitation (realized by shaker) and output is the measurements from the force transducer.
- Offline: Considering the MPC weight matrices, we use $Q = 10C_{aug}^T C_{aug}$ to minimize the output error and $R = 3 \times 10^{-3} I_2$ ($I_n$ is the $n \times n$ identity matrix) to achieve a fast transient response. To guarantee stability, the desired disk for the closed-loop eigenvalues is specified by $\gamma = 0.95$. The LQR problem associated with these parameters is considered as the reference LQ performance. The data weighting is selected by $\beta = 2.1$ to remedy the numerical ill-condition issue of the MPC optimization problem.
- Offline: The optimization horizon $N$ should be selected according to the AVC frequency range. Generally, faster dynamic systems require larger optimization horizons to achieve stability and optimality. To examine this criterion numerically, for each $N$, we compare the closed-loop eigenvalues of the unconstrained MPC (6.11) with those of the LQR. By this method, $N = 20$ produces a satisfactory LQ performance.
- Offline: The scale factors $a_1 = 0.76$ and $a_2 = 0.76$ of the parameterization Laguerre functions are selected to be close to the dominant closed-loop eigenvalues of the LQR [40]. Setting $N = 20$, we gradually increase the number of the Laguerre functions so that the closed-loop eigenvalues of the unconstrained parameterized MPC (6.17), converge to those of the LQR. The parameters $N_1 = 5$ and $N_2 = 5$ are selected as such to achieve an acceptable performance.
- Offline: The Kalman filter gain is calculated to estimate the augmented state vector and then it is stored in the memory. Regarding the stationary initial state of the system, the estimation covariance is initialized by a small value $P_{f0} = 10^{-4} I_8$. Referring to the datasheet of highly accurate LDV, a nominal value for the measurement noise covariance is considered. Owing to the internal model property of the operator $\Gamma(q^{-1})$, the disturbance signal $d(.)$ is eliminated from the augmented system in (6.6). Accordingly, from a practical Kalman filtering viewpoint, $Q_f$ is conceived as the covariance of two noise components: (i) a noise acting through the control channel due to the force transducers; (ii) a fictitious noise accounting for the unmodeled/remainder perturbations. Therefore, $Q_f = B_{aug} Q_u B_{aug}^{\mathrm{T}} + Q_{fic}$ where $Q_u$ is the covariance of force transducer noise (obtained from the datasheet and $Q_{fic}$ is the fictitious noise covariance which also makes $Q_f$ nonsingular. To guarantee both stability and optimality of the Kalman filter, an iterative tuning procedure is applied as follows. Starting from the nominal covariance matrices, the eigenvalues of the steady-state filter, that is $\mathrm{eig}\big(A_{aug} - K_f(k) C_{aug}\big)$ for large $k$, are compared to those of the corresponding LQR observer. Following this method, the covariance matrices are finely tuned as:$Q_f = 4.9 \times 10^{-9} B_{aug} B_{aug}^{T} + 10^{-9} I_8, R_f = 2 \times 10^{-4}$.

- Offline: The general interconnection of the state-observer and disturbance estimator are realized in SIMULINK based on Figure 6.1 and Figure 6.2.
- Offline: Then, the model from SIMULINK is built and a C++ code for dSPACE module is generated.
- Online: Using the estimated state vector $\hat{X}(.)$, the optimization problem in (6.17) is solved by a suitable QP solver under the control saturation constraint. Subsequently, the optimal Laguerre parameterization is calculated and the current control action is obtained.
- Online: The analog input data from the plant's output (Laser Doppler vibrometer) as well as the force transducer are read accordingly.
- Online: The states are reconstructed from the augmented observer and the disturbance is estimated.
- Online: Retrieving the control input generated by QP and writing the outcome in IO board to be converted to analog form, amplified, and applied on the active piezo-actuators are automatically performed next.
- Offline: All the required variables are saved.

In order to carry out the optimization involved in the predictive control formulation, a fast QP solver known as Hildreth's algorithm is employed [46]. A notable advantage of this algorithm is its ability to recover a sub-optimal solution in the case of conflicting constraints.

After constructing the control system based on Eqs. (6.17)-(6.18) in SIMULINK and compiling it to dSPACE RTI platform, the system is excited through the disturbance channel with a pulse period equal to 5 sec which is active for 2.5 sec (50 %). Each period of a periodic pulse includes a time window where the pulse is active (non-zero value) and a time window where it is inactive (zero value); in the literature, the ratio of the active to inactive phase presented in percentage represent the periodic pulse's characteristic. The uncontrolled and controlled systems are realized on the real-time DAQ of the dSPACE with a sampling frequency of 500 MHz. The captured velocity from the laser vibrometer for three cases of controlled (MPC), controlled (PID), and uncontrolled systems are compared in Figure 6.6.

The choice of PID is due to the fact that it is pertinent to implement PD and PID as a widely accepted technique in the industry. Additionally, as pointed out in [167], [168], the effective use of information on actuator constraints and behavioral prediction can avoid performance violations that are commonly happening in alternative techniques in real-time applications. The PID controllers are tuned on each of the input channels separately. It can be seen that the performance of the PID controller even around the fundamental natural frequency cannot match the Repetitive MPC controller. Moreover, the PID-based closed-loop system is unable to reject the disturbance of high-frequency nature as is shown in the later frequency analysis.

**Remark 6.3.** Comparing two independent control strategies is a critical matter because an analogy does not exist for tuning the two. In other words, in applied control problems, one method can be tuned better than the other. Nonetheless, in the context of this dissertation, a fair comparison is obtained by first tuning the PID controller to its best and then trying to match the same control level with repetitive MPC. However, better performances may be achieved by tuning MPC or including higher-order mode-shapes in the model of the reduced system for which the MPC will perform significantly better. Since the order of the PID as a model-free controller is limited, it is *intentionally* tried to limit the model order for a fair comparison. Additionally, the comparison of MPC with LQR has been tackled in the remarkable work of Bossi et al., [49]. Later, for the case, where the hard constraints are activated (here: actuator saturation), the enactment of the anticipated control system is compared with the LQG controller.

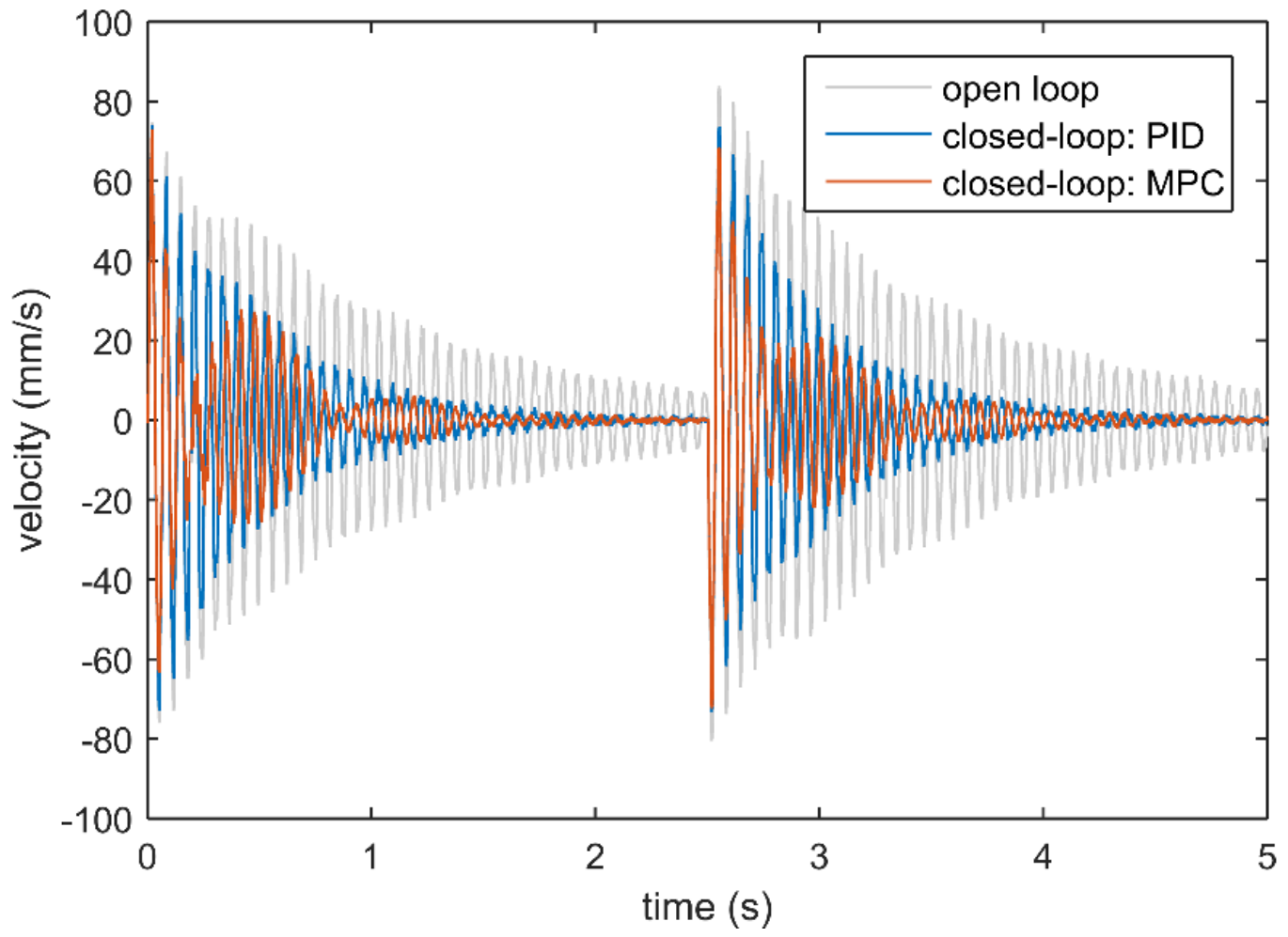

Figure 6.6 Disturbance rejection performance of the closed-loop system based on Repetitive MPC and PID [149]

The control efforts on both the piezo-actuators are depicted in Figure 6.7 for two closed-loop systems. The applied control signals on piezo-actuators are in the acceptable range of [−200 200] V with smooth variations. It is worth mentioning that, any disturbance signal with an abrupt change in amplitude such as pulse-type signals may excite the high-order dynamics of the system which are not included in the control development procedure.

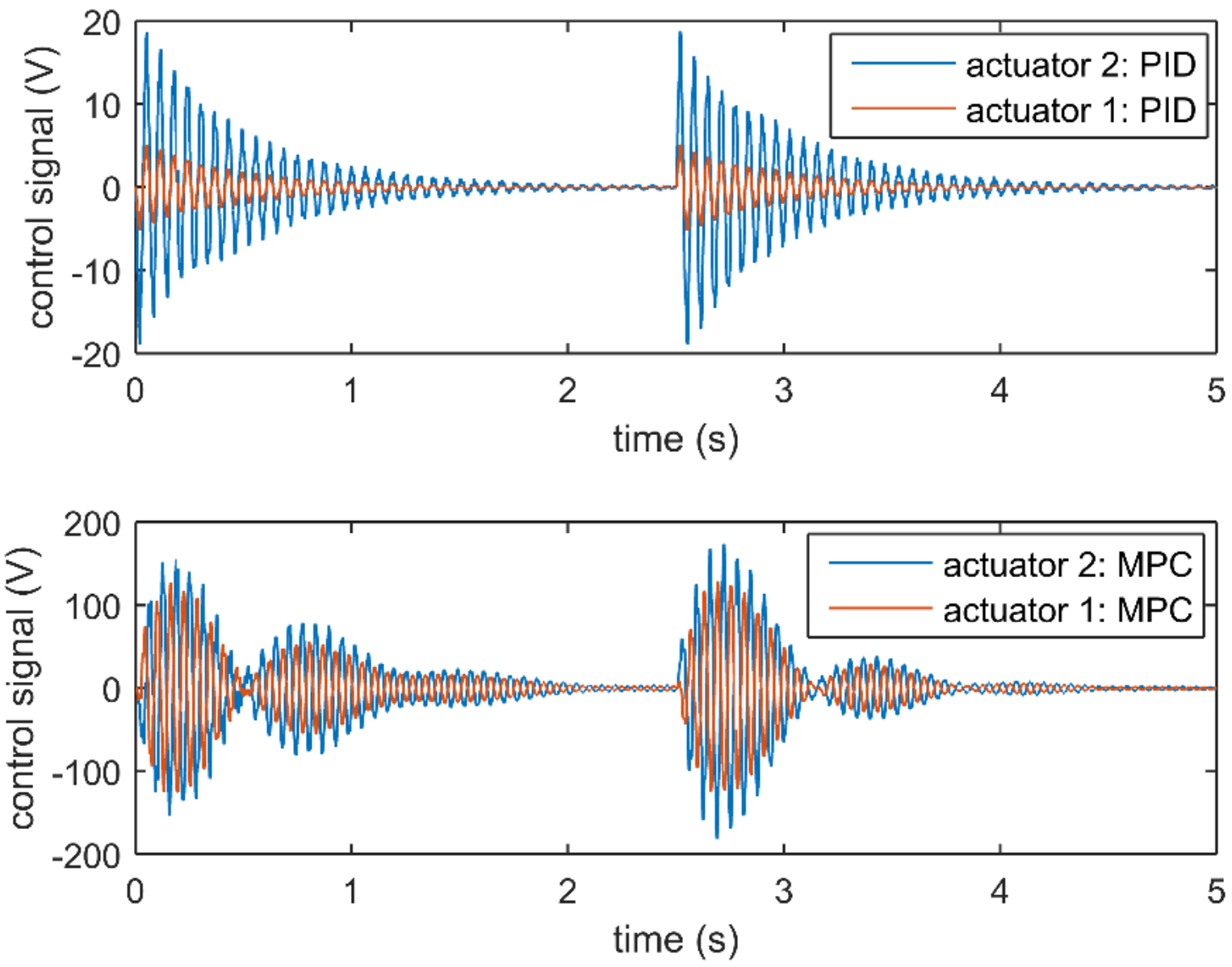

Figure 6.7 Applied control efforts on piezo-actuators [149]

The control effort of MPC is significantly higher than PID. This is because selecting a PID with higher gains simply leads to instability of the system. The output estimation error based on the Kalman filter is presented in Figure 6.8 which indicates the effectiveness of the state estimation system.

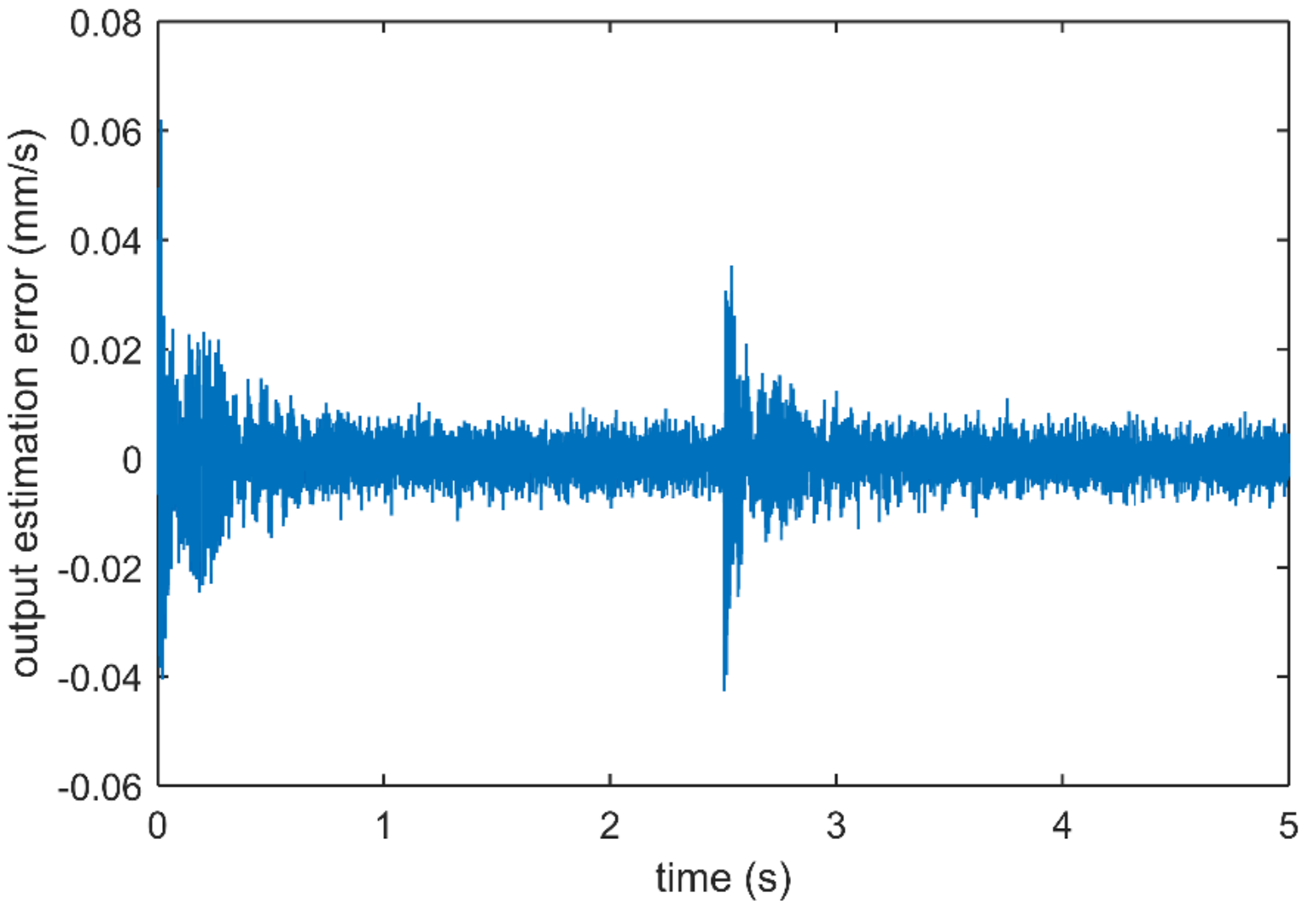

Figure 6.8 Observation error for the plant output [149]

In order to present the performance differences between the repetitive MPC and the repetitive MPC together with the feedforward disturbance rejection system, a chirp signal is realized in the disturbance channel. The frequency of the excitation is developed from 1 to 250 Hz in four cases of the open-loop and closed-loop systems: *i*) Repetitive MPC, *ii*) Repetitive MPC together with disturbance estimation/rejection system, and *iii*) PID compensator. The responses of the plant for the controlled and uncontrolled cases are shown in Figure 6.9 in the time-domain based on the measurement signal generated from the Doppler vibrometer, which is further fed to the dSPACE ADC board (DS2004).

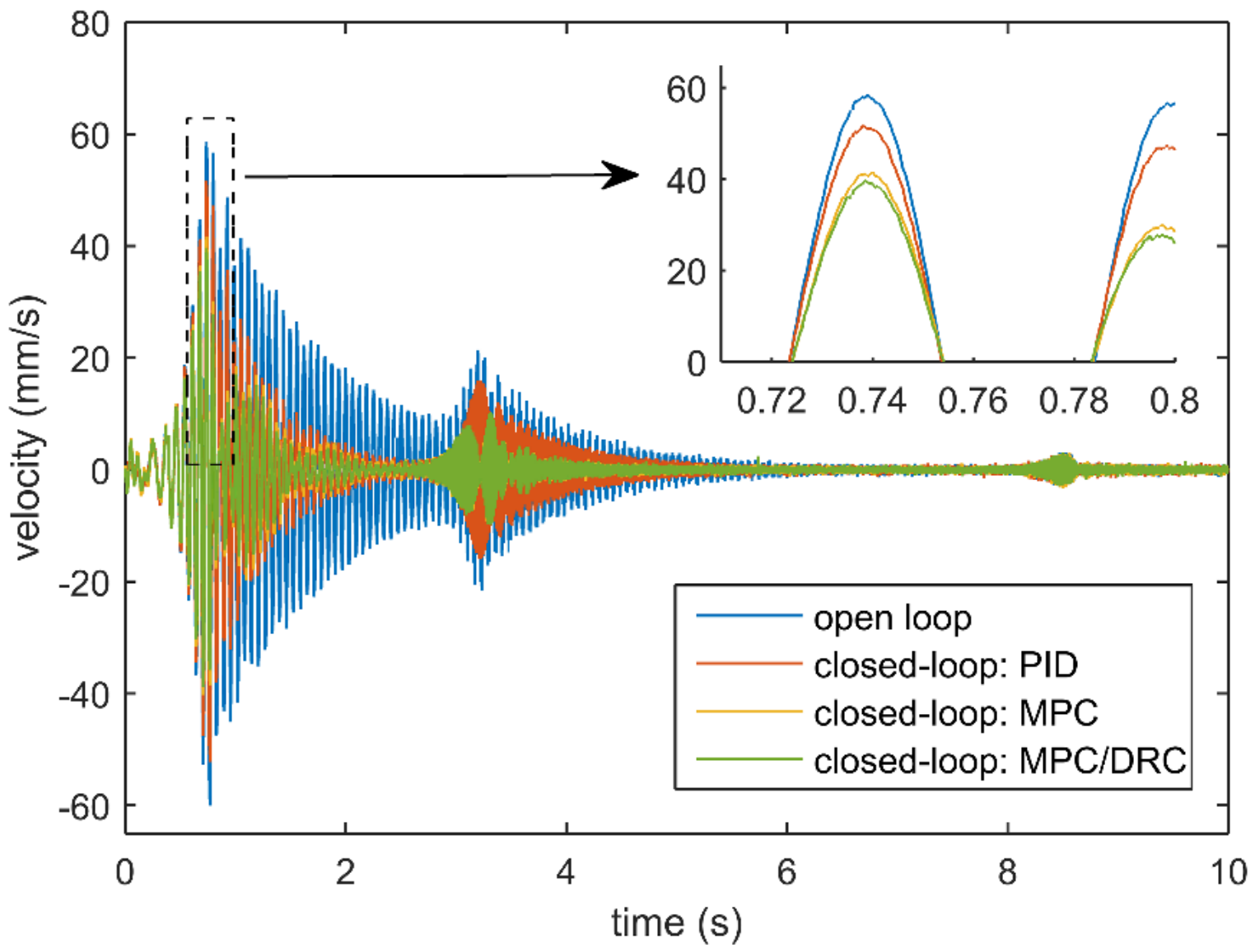

Figure 6.9 Experimental comparison of measured outputs in time-domain for chirp excitation [149]

Figure 6.9 displays a comprehensive evaluation of the vibration attenuation routine in the nominal frequency range. It can be observed that the MPC controller design, based on the identified model, suppressed the vibration magnitude within the considered frequency range. A distinct advantage of the repetitive MPC based on the modal realization of the identified model is the possibility of defining frequency-dependent (mode-dependent) weighting matrices that designate an emphasis on a restricted frequency range. The advantage of the repetitive MPC to the PID is revealed in the higher frequencies where substantial suppression can be achieved by MPC compared to PID. Additionally, the MPC based on the disturbance estimation has a better performance in attenuating the vibration amplitude. The corresponding control efforts that are generated for piezo modules by the dSPACE DAC board are shown in Figure 6.10.

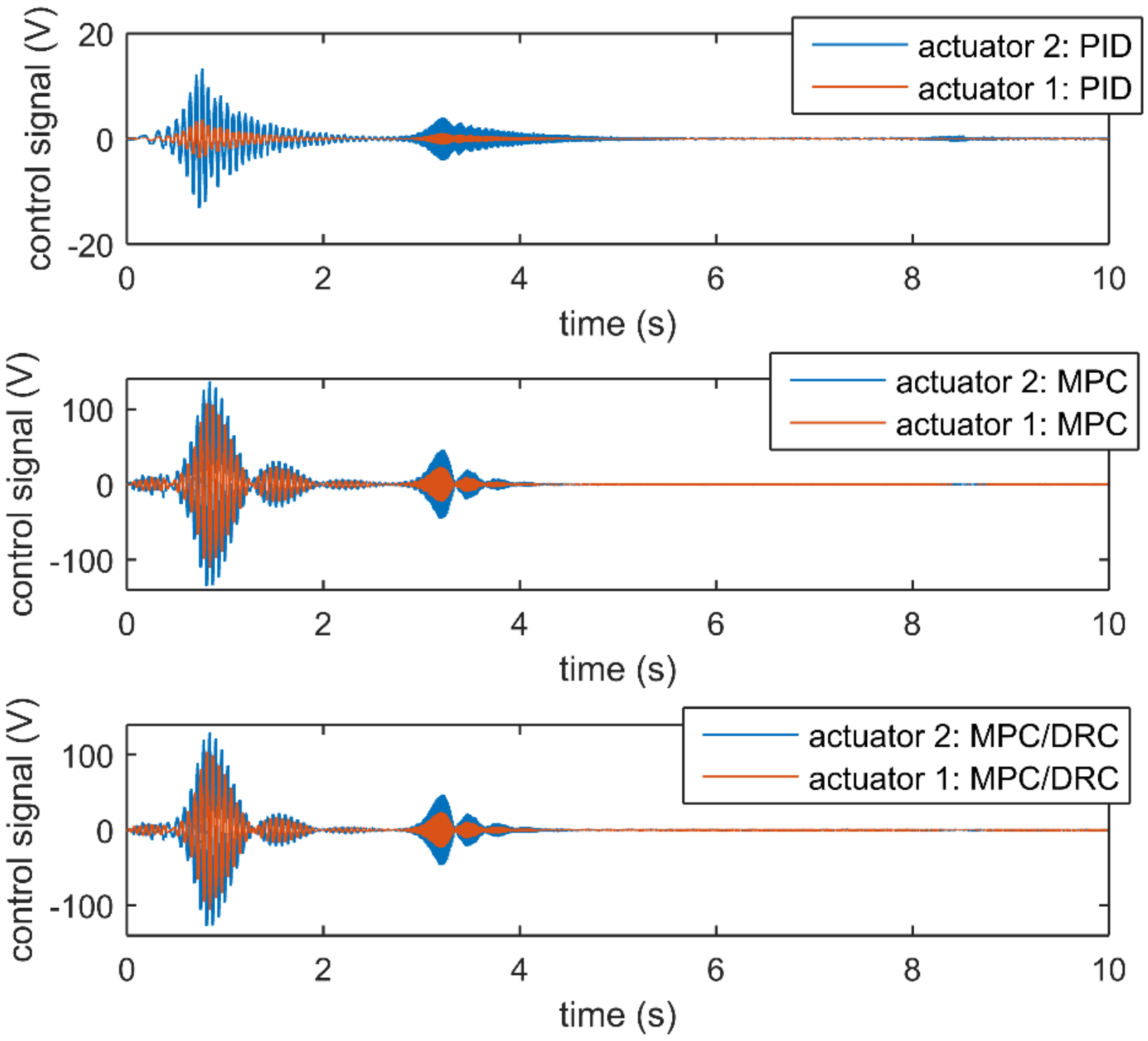

Figure 6.10 Control effort applied on the piezo patch actuators [149]

Figure 6.10 shows that the control efforts implemented to each of the piezo-actuators are restricted to a maximum 140 V with a smooth behavior. The experimental determinations confirm that the regulator system functions properly in attenuating the vibration amplitude in the presence of structured uncertainties in system elements. Also, the estimation error of the velocimeter is represented in Figure 6.11.

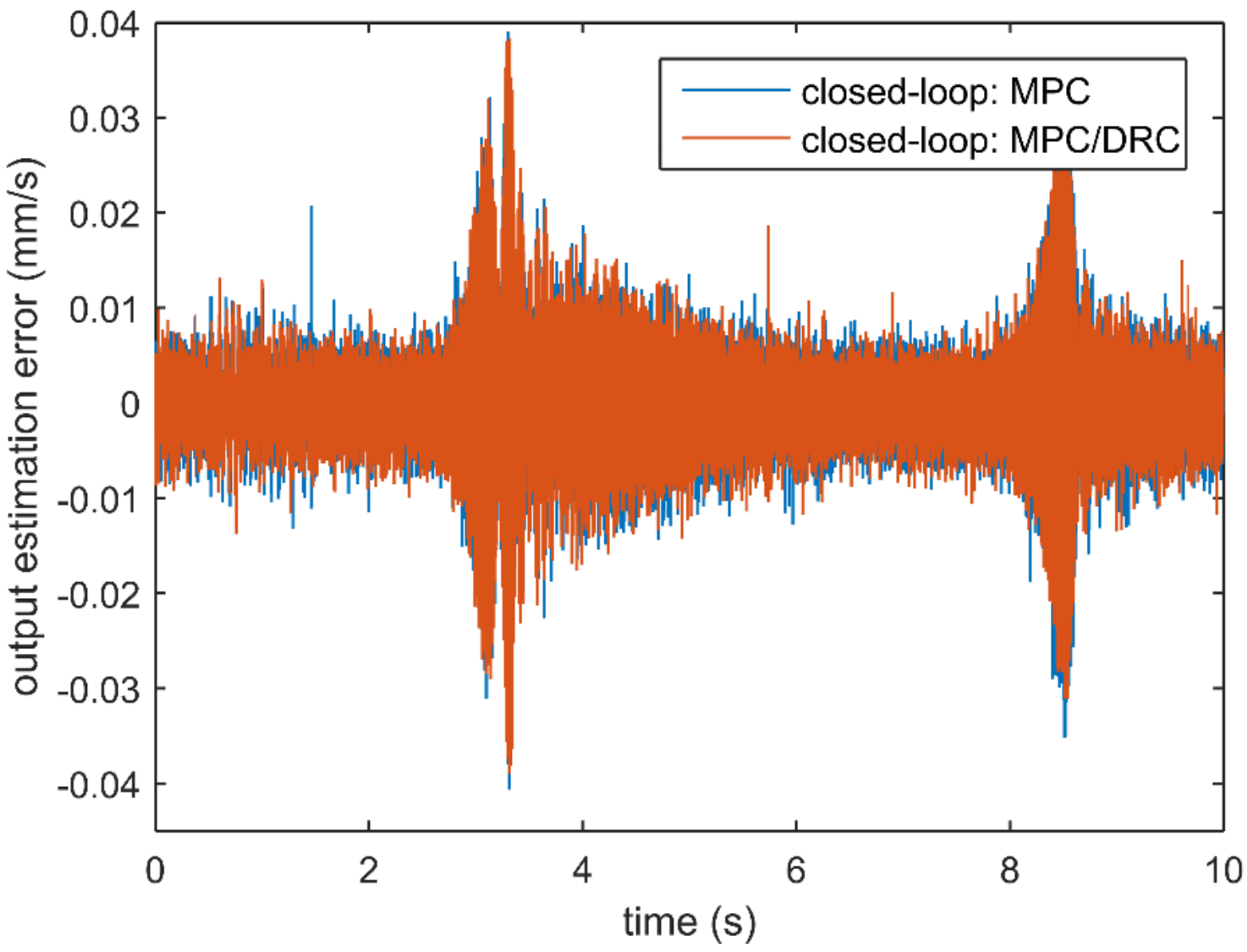

Figure 6.11 Estimation error of the output based on Kalman filter [149]

Next, the performance of the control system is evaluated in the frequency domain. For this purpose, loop 1 (the control loop) in Figure 6.1a is activated while the shaker is attached to loop 2 (the modal analysis loop). The generator is triggered by PULSE with a random signal in the range of [0 800] Hz. In order to assess the spillover effect, the cut-off frequency of the low-pass filter in the force transducer measurement channel is set to 800 Hz. As a result, based on FFT, the sampling time is calculated as 488.3 μs, and the number of lines is selected to be 6400 (sampling frequency of 125 mHz) to prevent long experimental duration. In order to address the leakage error and similarly to the case of disturbance dynamics, averaging with Hann-window is engaged. Accordingly, a baseband analysis is carried out with 20 averages and 66.67 % overlap for an overall duration of ≈ 60 s. The experimental configurations in the described form are examined by the same parameters four times: open-loop (loop 1 is deactivated), closed-loop based on Repetitive MPC, closed-loop based on Repetitive MPC disturbance estimation/rejection, and closed-loop based on PID controller. The FRF of the system, as well as the phase diagram, are shown in Figure 6.12.

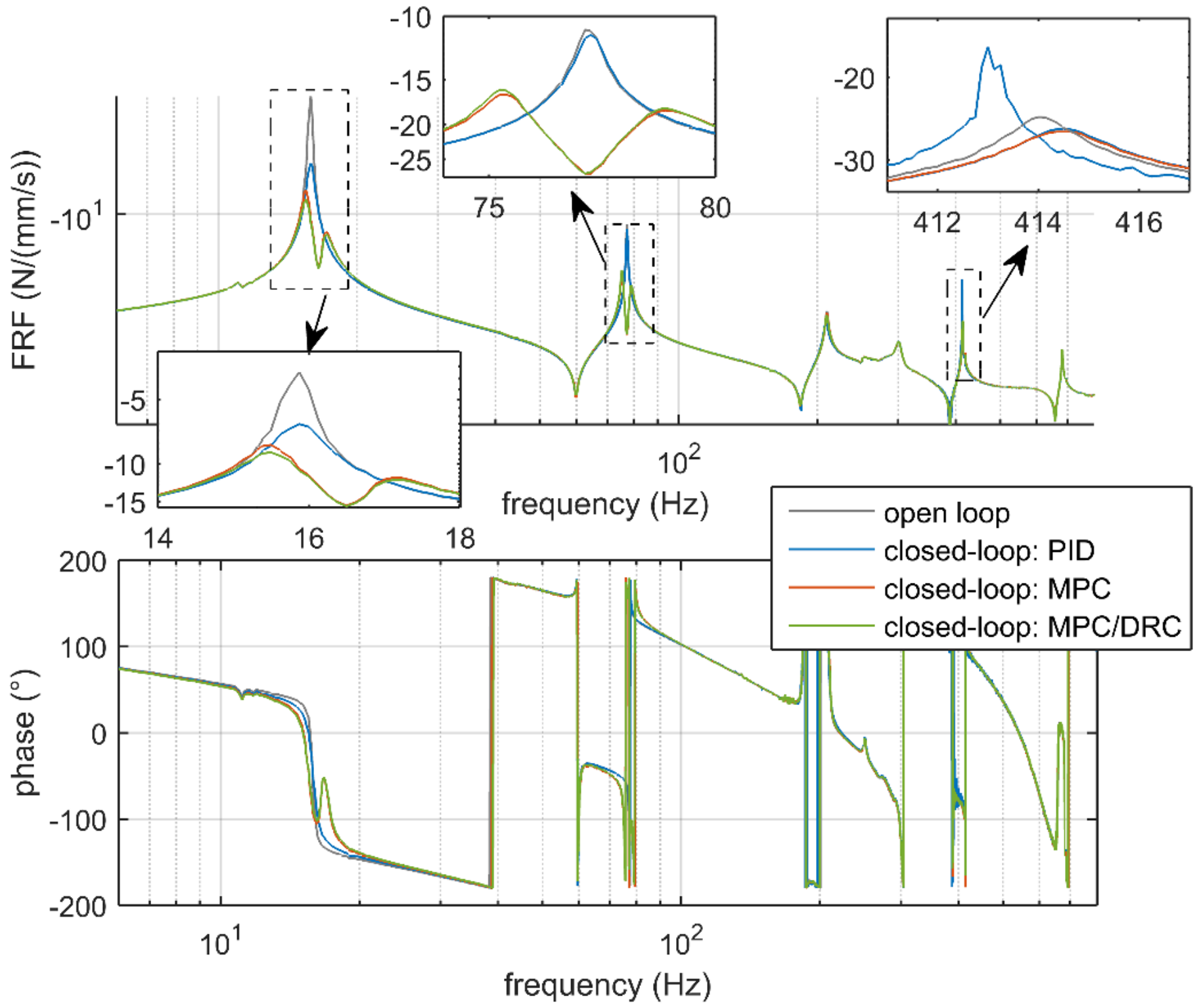

Figure 6.12 Frequency-domain analysis of the closed-loop systems for vibration attenuation performance evaluation [149]

The first observation is that although the PID controller had an acceptable performance in Figure 6.6, it is unable to control any higher mode-shape of the structure without additional tuning of the parameters. For instance, in contrast to both of the methods based on the MPC idea, the second mode shape of the system is undamped employing PID regulator. Moreover, at high frequencies such as in the range of [411 417] Hz, the performance of the closed-loop system based on the PID compensator is even worth than the open-loop uncontrolled case. The second observation is the high vibration attenuation factor for both the repetitive MPCs even at frequencies where the nominal plant model is not valid. This emphasizes the importance of incorporating a broadband disturbance model based on IMP in control synthesis. Then, it can be observed that in the nominal frequency range of reduced-order plant, the MPC based on estimated disturbance a higher suppression level compared to repetitive MPC.

Since the quality of the FFT is the crucial parameter that validates the FRF obtained based on experimental modal analysis, the coherence of the measurements is presented in Figure 6.13.

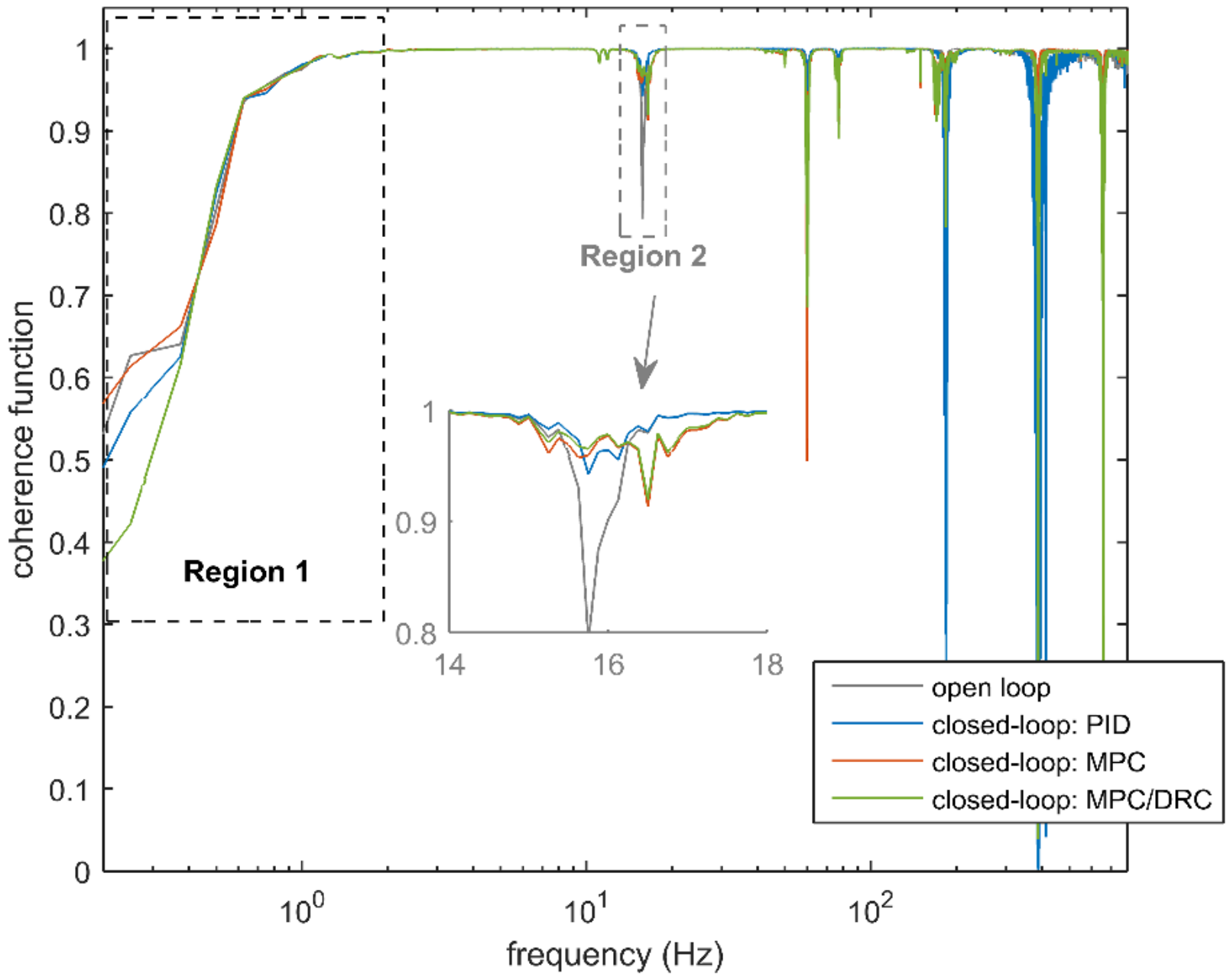

Figure 6.13 Coherence diagram for experimental modal analysis of the closed-loop systems [149]

The coherence function is used as a data quality assessment tool that identifies how much of the output signal is related to the measured input signal. The coherence equal to 1 denotes the perfect measurement. Based on Figure 6.13, three regions are less reliable analyses:

1) Frequencies below 6 Hz: This is the frequency range which the shaker is practically unable to excite. Since the structure under study does not have any resonant frequencies in this range, the deterioration of the result below 6 Hz may be neglected.

2) Around the first natural frequency and other resonance and anti-resonance frequencies: This case is typical behavior of the system in baseband analysis. In order to improve the measurement results, a solution may be a zoom analysis which is mostly used when the modal parameters such as natural frequencies and damping ratios of the system are the primary concern. Here, the focus is lighted on vibration control this step is relinquished. It is observed in the subplot of the coherence diagram between 14 and 18 Hz that due to a significant reduction of vibration amplitude in the resonance state of the structure, the coherence of the closed-loop systems based on MPC is close to one. In contrast to MPC, the PID controller though reducing the vibration amplitude changes the resonance state and thereby needs to be taken care of in new resonant modes. This emphasizes the advantage of MPC from another point of view.

A notable advantage of the repetitive MPC is the spillover rejection when the excitation is in the nominal frequency range of the reduced-order model. In other words, the controller can reject the disturbances of high-order dynamics. With a view to present this quality, FFT analysis ($H_2$ function in [169]) is performed to obtain an estimation of FRF of the control system in an attempt to identify the dominant frequencies of the controller Figure 6.14.

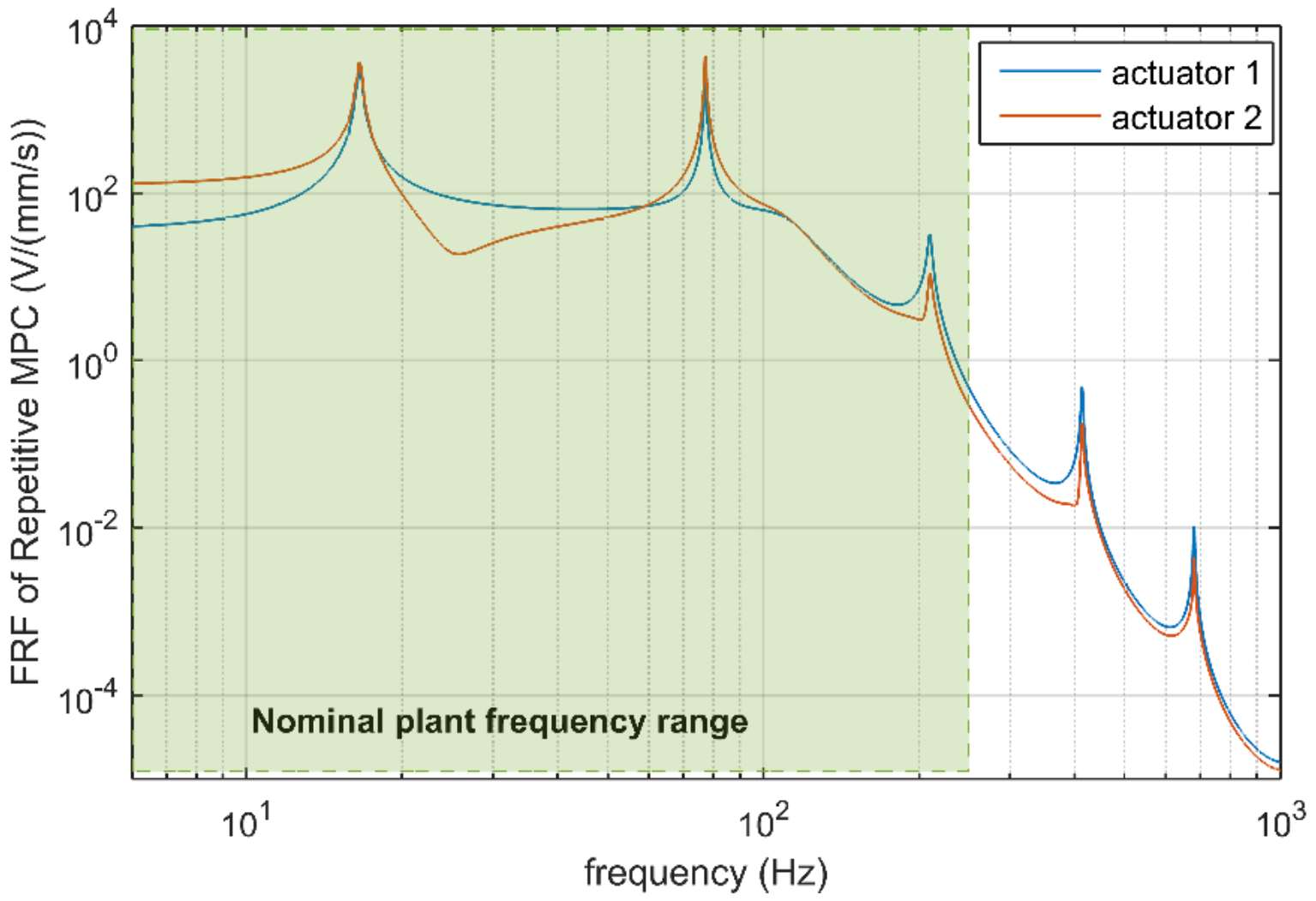

Figure 6.14 FRF of the control system in the nominal frequency range of the internal model [149]

The first observation is that the two primary actuation authorities on the nominal mode shapes of the reduced-order system are on two fundamental natural frequencies. This feature ensures that the control system has captured the key influential mode shapes of the system correctly. Moreover, although the controller can prevent the spillover effect due to the slope of FRF at high frequencies, it can still attenuate the higher-order unmodeled dynamics within the frequency range of the internal model. The latter is a promising feature especially when the order of the controller is limited because of the processing hardware limitations.

By aiming at evaluating the proposed closed-loop configuration in the saturation event, an LQG controller for the output regulation problem is designed. For this purpose, a steady-state Kalman filter is combined with a standard LQRY, and then the design matrices are tuned to achieve a high attenuation level. In the experimental simulations, the disturbance level is increased intentionally such that the control signals stand higher than the saturation level (200 V in magnitude). Final tunings of the LQGY are carried out such that the control systems under comparison experience the same amount of time in saturation which is essential for a fair assessment.

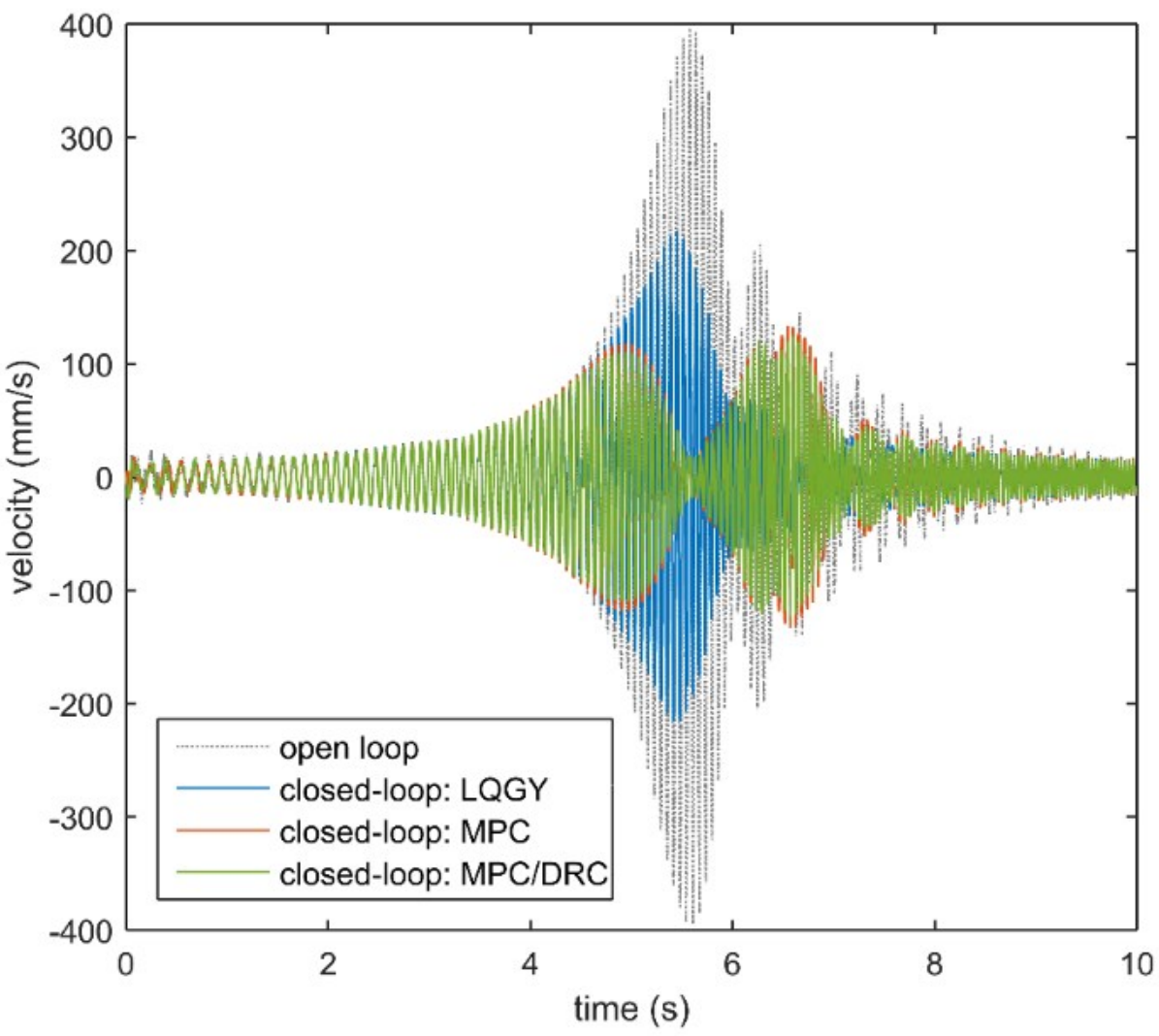

Figure 6.15 Transient behavior of the open-loop system compared to the controlled ones in the saturation incident [149]

Since the LQGY is unable to handle such hard constraints on the control input, a saturation block was added to the SIMULINK model before compiling it into C++ language and uploading the compiled file to DS 1005 PPC board. Then, the three control systems are evaluated in real-time for a mismatch chirp disturbance signal which is active for ten seconds covering the frequency range of [6 25] Hz. The transient responses of the open-loop system against three closed-loop ones are shown in Figure 6.15. It can be seen that MPC-based methods achieved higher attenuation while satisfying the hard constraints. Moreover, the applied control signals on the piezo-actuators are depicted respectively in subplots of Figure 6.16.

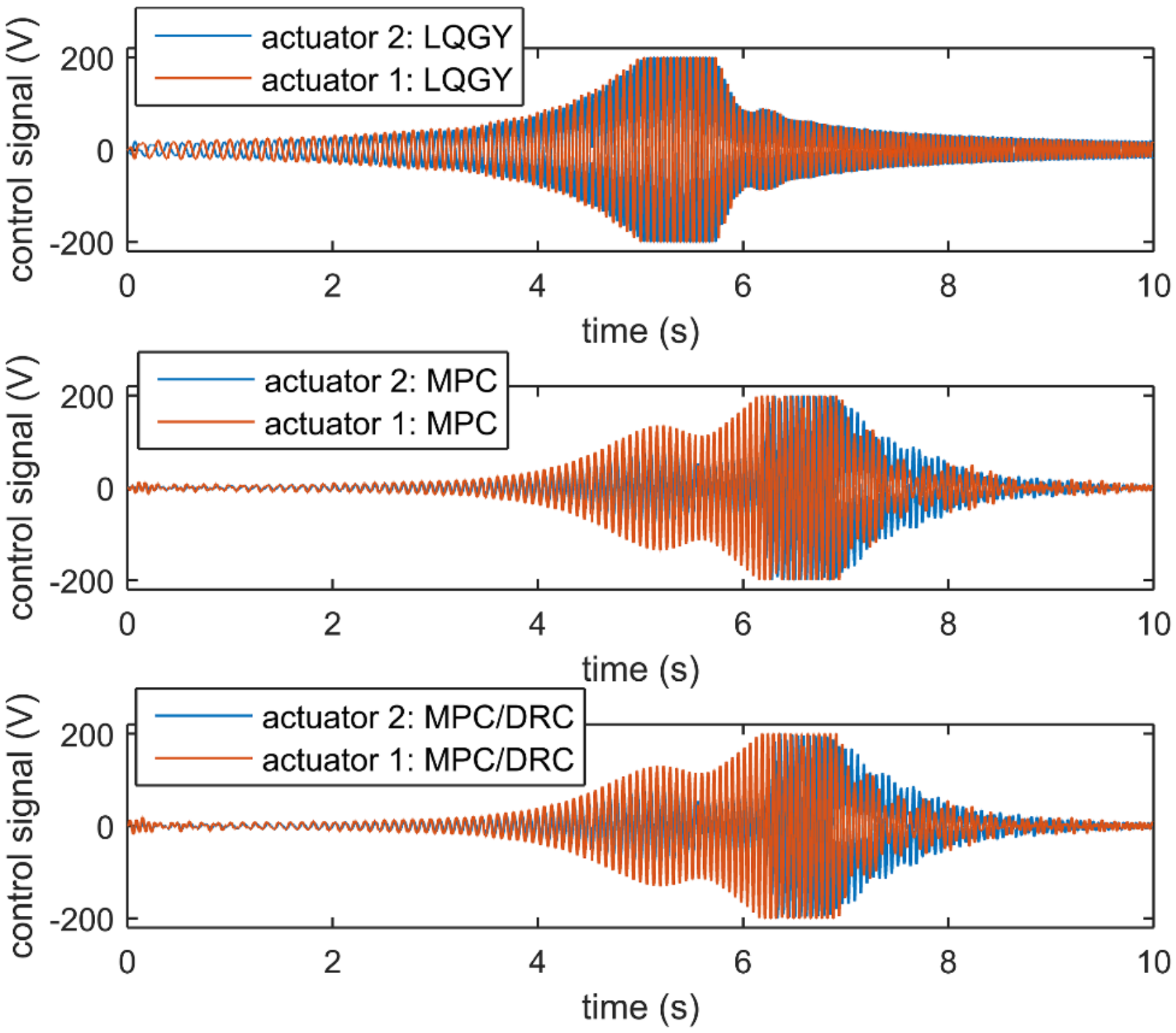

Figure 6.16 Constrained control signals of the MPC-based methods compared to LQGY regulator [149]

## 6.2.2 Final comments

To put in a nutshell, the following points should be highlighted:

- The internal model-based repetitive control paradigm provides an efficient framework for rejecting periodic disturbances in AVC systems. This approach can be conceived as the generalization of the classic integral action used in PID controllers. The integral action, however, rejects only step-type disturbances with zero steady-state error. Moreover, incorporating an internal model of the disturbance signals guarantees the asymptotic rejection of the disturbances. Conventional approaches to robustify MPC systems against disturbances, such as min-max optimization [161], are computationally demanding and then achieve only disturbance attenuation rather than disturbance rejection. In this respect, internal model-based MPC is prominent in terms of both robustness and computational efficiency.
- In comparison with the common MPC algorithms, parameterization of the control inputs by orthonormal basis functions considerably reduces the computational burden of the associated numerical optimization. For example, in the AVC application reported in this chapter, Laguerre parameterization resulted in a >75 % reduction in the number of the optimization variables with respect to the conventional MPC algorithm. Accordingly, fast optimization solvers can be used to benefit from the MPC advantages in AVC systems.
- The over-specification of the controller's output by augmenting a soft saturation element before submitting the control law to the actuator of AVC in morphing structures is undesirable. This eventually excites the unmodeled dynamics of high order nature that are not included in the nominal model of the plant. Similar observations are reported in [170], [171]. This observation is verified by comparing the performance of the proposed control algorithms to the LQG output regulation technique.

- For geometrically linear vibration amplitudes and for the disturbance frequencies limited to the range in which the internal model is valid, the perturbations are guaranteed to be asymptotically rejected. This perception is observed in the time- and frequency-domain results of the experimental implementations.
- By simulating the bounded disturbance signal with abrupt jumps in amplitude, the performance of the control system is proven to be robust against the high-order unmodeled dynamics of the system. Additionally, as a notable behavior, the repetitive MPC shows the capability to cut off the spillover effect of the actuators in the nominal frequency range of the reduced-order model.

In the next chapter, AWC is investigated as another approach that can deal with actuator nonlinearities.

# 7 Dealing with hard constraints: An applied view on analysis and synthesis of AWC with a convex formulation

In this chapter, the windup problem in AVC is studied systematically. Anti-windup methods which are analyzed in this chapter are the so-called two-step paradigm methods in which a *nominal/ideal* control system without considering the windup issues and actuator nonlinearities is developed in the first step; in the second step the nominal controller's input, states, and/or output is manipulated by an anti-windup compensator to deal with the actuator nonlinearities. Instead of evaluating the performance of several anti-windup compensators implemented on independent abstract simulation problems, the same benchmark setup in active-damping control (ADC) is used as in Figure 6.1. The investigated anti-windup schemes (analysis and synthesis) are adapted to the disturbance rejection control. Large attention is given to capture the similarities and differences of the anti-windup methods in dealing with the windup event in a practical context. Therefore, instead of categorizing the (available) methods in literature into static and non-static methods or model recovery and direct linear anti-windup schemes, a logical route is followed to highlight the significance of each method. The mathematical interpretation of the methods is provided for the vibration engineer while delivering forthright implementation algorithms for AVC. The tackled methods are unified on a state-space model obtained from the frequency-domain subspace system identification approach. Practical issues that may arise for each of the analyzed anti-windup methods are mentioned, and detailed guidelines are provided for tuning each of the analyzed anti-windup methods. Finally, in order to compare the compensated system's performance, comprehensive time-domain studies are carried out by separating the transient response of the compensated systems into three modes: linear mode, where the actuation nonlinearity is inactive; the nonlinear mode, where the windup event is in progress, and finally, the output mismatch rejection mode, where the windup incident is over, but performance degradation is still present.

### 7.1.1 Establishing a connection between AWC and AVC

This chapter once again looks into hard constraints in control design this time from a different perspective and deals with the same end-effector regulation problem as before [48]. It should be noted that the simplistic geometry of these structures should not overshadow the intrinsic practical issues of control design such as the non-minimum phase problems of non-collocated configuration and geometrically nonlinear dynamics [172], [173].

In the context of smart structures, it may often occur that the difference between the designed controller's output signal and the actual implemented control signal on the system is nonzero. Other than the fragility of the control system due to the finite-word length in the digital system and round-off errors in binary arithmetics [55], which are mostly permanent, two foremost temporary sources of such imperfections are actuation constraints and control law substitutions. Note that the fragility of the control system with respect to temporal imperfection as the focus of this chapter leaves the analysis of the *fragility* of the controller/observer to the upcoming Chapter 8, i.e., the robustness of the feedback system w.r.t. its gain uncertainties. Gain uncertainties and fragility issues may arise among others due to numerical and quantization errors.

Calamitous influence of substitution appears when in a multi-mode controller each of the modes is responsible for a neighborhood around a distinct linear operating point and repetitive changes between the modes are needed. The offline behavior of switching while the controller is in an online loop introduces overshoot and eventually windup. Although in the case of substitution, a smooth transition between the controller modes may compensate for the mismatch control law, detailed mathematical treatment of this problem is not discussed here. This is justified by considering the application of smart structures in active vibration control where the system model is mostly linearized around a single operating point. However, the actuator nonlinearities such as saturation and hysteresis may play key roles in driving the system states to non-ideal trajectories. Saturation, as a constraint on the amplitude of the control signal, is defined by a particular limit that is eligible to be applied to the active elements. This limit includes the output voltage and electrical current range of a piezo-amplifier, the maximum achievable displacement of a shaker's baffle, and the depolarization voltage of the piezo-actuator patches.

A good deal of control systems in vibrating smart structures use an oversimplified linear model neglecting these nonlinearities. In addition, it is well-established that controllers with sluggish or unstable dynamics eventually suffer from the windup. A bumpless transition in such a case is realizable by compensating the controller input, states, and outputs by means of various anti-windup strategies. In this regard, we intentionally employ a standard approach for the nominal control design so that the main goal, namely unification and comparison of the second step in the decoupled *anti-windup compensator technology* is not overshadowed by the nominal controller. A typical windup scenario in active vibration control appears when the observer-based controller experiences actuator saturation in which the integrated linear state observer is unaware of the actuator nonlinearity chopping off the control signal. A windup in the linear controller's output may happen in this case leading to diverging system output which is catastrophic in the range of space applications.

The subject of decoupled anti-windup analysis and synthesis has reached a mature level. As a result, this chapter by no means is aimed at summarizing the vast well-documented methods on anti-windup compensation. Such an attempt for an applied problem is critical since it may leave out some valuable contributions from the leading scholars on the topic. Instead, the proposed unification scheme is intended to establish a connection between a handful of reliable compensator methods available in the literature and the practical implementation of these methods for the end-effector regulation problem of manipulators with saturated actuators. It should be pointed out that the selected application may be replaced by other plants that are vulnerable w.r.t. actuation nonlinearities such as controlling the flow of gas by positioning of electromechanical actuators on engine throttle which saturates after a certain angle [174]. The especial effort is given for formulating the analysis and synthesis of 'posteriori' algorithms based on convex optimization such as linear matrix inequality (LMI) eigenvalue problem which is a convex minimization subjected to LMI constraints. Additionally, the selected algorithms are mostly based on the stability analysis of Lyapunov-type. The reason behind this perception is that the absolute stability schemes e.g. Circle criterion and Popov stability theorem are mostly limited to single loop systems [175].

It should be noted that the well-known 'priory' strategies based on MPC of Chapter 6.1 that incorporate the hard constraints such as limitations on control effort (amplitude and rate) in synthesis procedure are not of interest in the second part. Such methods in contrast to the two-step paradigm could be regarded as computationally expensive (depending on application) due to the online optimization and are less flexible in terms of linear unconstrained controller synthesis [149]. The latter is because of often conflicting conditions imposed on the control system even before the actuation nonlinearity event. Therefore, knowing that the actuator nonlinearities in real industrial applications are presumed to be seldom to happen in the nominal operating range and fast to be over, the two-step anti-windup paradigm seems to be more suitable in comparison to MPC which is again very dependent on the application of AVC. In other words, the two-step anti-windup paradigm preserves the ideal controller's performance and only compensates the actual control law during (and shortly after) the actuator nonlinearity event.

Accordingly, the methods for analysis and synthesis of anti-windup mechanisms in the modern sense are briefly overviewed in a relatively broad manner to provide constructive tools developed as the typical embodiment of anti-windup augmentation. Instead of an extensive literature review, we have summarized our observations in accordance with implementing different anti-windup scenarios in a practical manner. The author is also well-aware of the extended strategies published in recent years that may serve as attractive frameworks for saturated delayed systems, systems with sampled output, systems with matched and mismatched uncertainties, systems with non-exponentially stable plants, and their discrete counterparts. However, due to the nature of the discussed application, these methods are mostly out of the scope of this dissertation. Additionally, methods that are tailored particularly for PID controllers or for single-input-single-output (SISO) systems are not covered in the more general scope of this dissertation.

The main thrust of this chapter is to unify the available tools in the literature together with highlighting the practical tuning considerations. In the rest of this chapter, the script symbol $\mathcal{I}_n$ stands for identity matrix such that $\mathcal{I}_n \in \mathbb{R}^{n\times n}$ with $\mathbb{R}$ being the set of real-valued numbers. When $n$ is not given, it is intended that the identity matrix has an appropriate dimension in relation to the multiplied and added matrices in the related equations. A similar definition applies for $0_n \in \mathbb{R}^{n\times n}$ and 0, respectively. Additionally, $\mathbb{R}^n_{\geq 0}$, $\mathbb{R}^{n\times n}_{\geq 0}$, and ${}^{+}\mathbb{R}^{n\times n}_{\geq 0}$ represent the space of time-dependent vectors, matrices, and positive definite matrices with dimensions of $n \times 1$, $n \times n$, and $n \times n$ for $t \geq 0$, respectively. Superscript $T$ on vectors and matrices denotes the transpose operator while $A^{-1}$ represents the inverse of a nonsingular matrix $A$. The system in state-space form is represented by $ss(A, B, C, D)$ following the conventional form of definition for state matrix $A$, input matrix $B$, output matrix $C$, and feedthrough term $D$. In the context of linear matrix inequalities, for two square Hermitian matrices with the same dimensions, $A > B$ ($A \geq B$) means that $A - B$ is positive (semi-) definite. The symbols $\prec, \preccurlyeq$ and $\succ, \succcurlyeq$ are element-wise inequality operators. $\mathbb{L}_2$ is Hilbert space of $m$-vector-valued arrays such as $x$ and $y$ on $(-\infty \quad \infty)$ with scalar product $\langle x|y\rangle = \int_{-\infty}^{\infty} x^{cc} y \, \mathrm{d}t$ ($x^{cc}$ is the complex conjugate of $x$). $x \in \mathbb{L}_2$ if $\|x\|_2 \stackrel{\text{def}}{=} \langle x|x\rangle^{1/2}$ is bounded. $\mathbb{L}_{2e}$ is assumed as an extension of $\mathbb{L}_2$ that for any $x \in \mathbb{L}_{2e}$, we also have $x \in \mathbb{L}_2$ and $\forall t >$ T: $x = 0$. $\mathbb{H}_\infty$ is the Hardy space of bounded analytical functions with $\mathbb{H}_\infty$-norm defined as $\|x\|_\infty \stackrel{\text{def}}{=} \sup_{\omega\in\mathbb{R}} \bar{\sigma}\big(G(j\omega)\big)$ with $\bar{\sigma}(.)$ signifying the maximum singular value of the system over frequency $\omega$. A dead zone is defined as $\mathrm{dz}(u) = u - \mathrm{sat}(u)$ for saturation function $\mathrm{sat}(u)$ defined as

$$\mathrm{sat}(u_i) = \begin{cases} u_{i,\min} & \text{if} \quad u_i < u_{i,\min}, \\ u_i \quad \text{if} \quad u_{i,\max} \geq u_i \geq u_{i,\min}, \\ u_{i,\max} & \text{if} \quad u_i > u_{i,\max}, \end{cases}$$

## 7.1.2 Problem formulation and preliminary definitions

### 7.1.2.1 Plant and controller formulation

Consider a finite-dimensional LTI (FDLTI) multi-input multi-output (MIMO) model of a plant ($\mathbf{P}(s)$) in the state space form as

$$\begin{aligned} \dot{x}(t) &= Ax(t) + Bu(t) + Hw(t), \\ y(t) &= Cx(t) + Du(t), \end{aligned} \tag{7.1}$$

where, $x(t) \in \mathbb{R}^n_{\geq 0}$, $u(t) \in \mathbb{R}^{n_u}_{\geq 0}$, $w(t) \in \mathbb{R}^{n_w}_{\geq 0}$, and $y(t) \in \mathbb{R}^{n_y}_{\geq 0}$ are the state, input, square-integrable disturbance (including the lumped parameter variations), and output vectors at time $t \geq 0$, respectively. $\mathbf{P}(s)$ represents the equivalent transfer matrix of (7.1). To avoid notational clutter, the dependency of time-varying vectors is omitted in the rest of the chapter unless it is required to distinguish between time-domain and frequency-/$s$-domains. As a natural assumption, the transfer functions of the system for all input/outputs are assumed to be real rational and analytic in the right-hand side (RHS) of imaginary axis. In addition, $A\colon \mathbb{R}^n_{\geq 0} \to \mathbb{R}^n_{\geq 0}$, $B\colon \mathbb{R}^{n_u}_{\geq 0} \to \mathbb{R}^n_{\geq 0}$, $H\colon \mathbb{R}^{n_w}_{\geq 0} \to \mathbb{R}^n_{\geq 0}$, $C\colon \mathbb{R}^n_{\geq 0} \to \mathbb{R}^{n_y}_{\geq 0}$, and $D\colon \mathbb{R}^{n_u}_{\geq 0} \to \mathbb{R}^{n_y}_{\geq 0}$. For the well-posed system (with an existing and unique solution for $t \geq 0$) in (7.1) with a Hurwitz $A$, a controller is required with linear dynamics that guarantees the stability and performance of the closed-loop system. The interconnection of the controller $\mathbf{C}(s)$ and the plant $\mathbf{P}(s)$ without actuation nonlinearities is written in the form of $u = \hat{u} = y_c$ and $u_c = \hat{u} = y$ such that the linear (assuming $\Psi(.) = \mathcal{I}_{n_u}$)) interconnection of Figure 7.1 is well-posed and internally stable. $\Psi(.): \mathbb{R}^{n_u} \times \mathbb{R} \to \mathbb{R}^{n_u}$ represents the actuation nonlinearity such as saturation. where,

$$\mathbf{P}(s) = \begin{bmatrix} P_{yw} & P_{yu} \\ P_{yw} & P_{yu} \end{bmatrix}. \tag{7.2}$$

Note that $\mathbf{P}(s)$ represents the relationship between input and output by $P_{yu}$ and not $P_{y\hat{u}}$ because in the first step of the two-step paradigm of AWC methods, the controller is developed while neglecting the nonlinearities. The *linear* ideal controller $\mathbf{C}(s)$ that satisfies some user-defined performance indices before the adverse effect of actuation nonlinearity is given as

$$\begin{aligned} \dot{x_c} &= A_c x_c + B_c u_c, \\ y_c &= C_c x_c + D_c u_c. \end{aligned} \tag{7.3}$$

#### 7.1.2.2 Preliminary definitions

First, consider $u^* \in \mathcal{X}$, where $\mathcal{X} \subset \mathbb{R}^{n_u}$ is a set. Then, for an arbitrary vector $u \in \mathbb{R}^{n_u}$, the distance operator is defined as $\text{dist}(u, \mathcal{X}) = \inf_{u^* \in \mathcal{X}} |u - u^*|$. Next, the well-posedness is defined in terms of the existence of the limit: $\lim_{s\to\infty} \left(\mathcal{I}_{n_u} - \mathbf{C}(s)P_{yu}(s)\right)^{-1}$, i.e., the existence of $\left(\mathcal{I}_{n_u} - D_c D\right)^{-1}$ and $\left(\mathcal{I}_{n_y} - D D_c\right)^{-1}$. Moreover, a stable dynamical system $f(f_0, z)$ as a causal operator over $\mathbb{L}_{2e}$ , with $f_0$ and $z$ being the initial condition of the function and the multi-dimensional vector, respectively has the finite increment gain $\kappa_1 > 0$ if $\|f(f_0, z_1) - f(f_0, z_2)\|_2 \le \kappa_1 \|z_1 - z_2\|_2 + \kappa_2$ with $\kappa_2 \ge 0$. In relation to the system (7.1), the finite increment assumption is later used for formulating the actuator nonlinearity mathematically in which $z$ may represent the states of the system as well as the inputs. Additionally, based on the positive real Lemma, the controllable system in (7.1) is referred to as passive if for $P = P^T \in \mathbb{R}^{n\times n} > 0$ and $\delta \in \mathbb{R} > 0$, we have

$$\begin{bmatrix} A^T P + PA & PB - C^T \\ B^T P - C & 2\delta \mathcal{I}_{n_u} - D - D^T \end{bmatrix} \le 0. \tag{7.4}$$

In the standard feedback configuration of Figure 7.1 with included nonlinearity, the system is called finite increment gain $\mathbb{L}_2$ stable if all the closed-loop maps from exogenous signals to all inner variables satisfy the above constraint. In this figure, $\hat{u}$ represents the so-called *achieved actuation* despite the nominal controller output $y_c$.

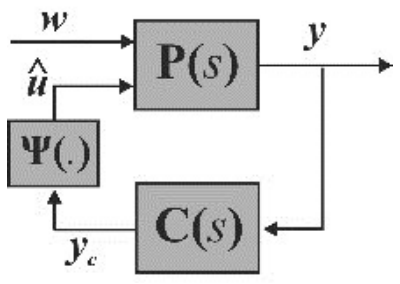

Figure 7.1 Interconnection feedback system with actuator nonlinearity $\Psi(.)$ [176]

Early developments of anti-windup systems were originally bilinear matrix inequality (BMI) that were converted to LMI by fixing a design variable (see [177] and [178]). The LTI anti-windup synthesis is a two-step paradigm that accounts for the linear part of the plant with the predesigned linear controller $\mathbf{C}(s)$ while adding a compensator that minimizes the degradation of local closed-loop performance during and after actuation nonlinearity event. $\mathbf{C}(s)$ is developed under neglecting the actuation nonlinearities and is referred to as the *unconstrained controller*. One of the first controller manipulations in anti-windup direction, independent of the nature of actuation nonlinearities, in the state-space representation was proposed by Hanus based on the conditioning theory [179], [180]. In conditioning technique, depending on the nature of the control problem, tracking (regulation), the reference signals (plant's output signals) are manipulated such that the mismatch between the controller output ($y_c$) and plant input $\hat{u}$ is minimized (see Figure 7.1). In the linear conditioning, an additional compensating transfer function augmented with the closed-loop system is responsible for changing the trajectory of the plant and controller during and after the temporary actuation nonlinearity.

In the application of AVC, the methods that are computationally cheap such as retro-fitting the available designed unconstrained controller are preferred to control redesign strategies. Augmentation of such sort is especially crucial when the unconstrained controller is preliminarily known to be ill-suited for actuation nonlinearity. The regulation problem in terms of DRC is of the main interest, and in this framework, the compensated output signal used for restoring the control law is referred to as *realizable output*. Not surprisingly, the realizable outputs are temporary solutions since the long-time windup events can destroy the control performance as well as the actual output.

**Definition 7.1** The nominal anti-windup problem is defined in terms of finding a system ($\xi^T = [\xi_1^T \quad \xi_2^T\,]^T = \Sigma_{aw}(\Sigma_{aw}(0), x_{aw}, y, y_c)$) where $\xi \in \mathbb{R}^{n_c+n_u}$. In this notation, $\Sigma_{aw}$, $x_{aw}$, $(y, y_c)$, and $\Sigma_{aw}(0)$ represent the LTI anti-windup compensator states, input, and static part (because AWC can also be just a gain matrix known as static AWC), respectively. $\xi$ in ad-hoc framework of LTI anti-windup problem conditions the output of the plant in the form of some realizable output ($u_c = y + \xi_2$) and simultaneously manipulates the controller output before nonlinearity as $u = \Psi(y_c + \xi_1)$. In such a framework, it is assumed that if $\Psi(y_c + \xi_1) = y_c + \xi_1$, for $\Sigma_{aw}(0) = 0$ then $[\xi_1^T \quad \xi_2^T\,]^T = 0$ and otherwise some performance indices on $\text{dist}(u, \mathcal{X})$ and p-norm of the plant's output should be satisfied. For instance, in the case of 2-norm: if $\text{dist}(u, \mathcal{X}) \in \mathbb{L}_2$ then, $(y - y_{nom})(t) \in \mathbb{L}_2$, with $y_{nom}$ being the nominal plant output. Moreover, the set $\mathcal{X}$ can be interpreted as the region inside the control law space that the input nonlinearity such as saturation is still inactive.

**Remark 7.1.** Note that $\text{dist}(u, \mathcal{X}) \in \mathbb{L}_\infty$ is excluded in the scope of the dissertation since the mismatch control law of the anti-windup mechanism may be bounded by means of $(y - y_{nom}) \in \mathbb{L}_\infty$ with a limit cycle which is unacceptable for AVC applications where mostly the windup incidents happen in resonance states of the flexible structure. The interpretation of $\text{dist}(.,.)$ operator for a successful anti-windup scheme (local and global asymptotic stability) can be exquisitely explained in terms of the basin of attraction of the augmented system. Qualitative illustrations in this regard such as extracting the feasible solution space in the presented algorithms for different anti-windup schemes in terms of trajectories of augmented systems may provide useful tools in ranking these methods. This may be especially important for systems with only local exponential stability which is out of the scope of the AVC of lightly damped structures. For the exponentially stable plants, the basin of attraction encapsulates the whole state space.

#### 7.1.2.3 Dealing with input nonlinearities

It is assumed that the constrained control law (e.g., the saturated control signal) is either measurable or estimable. Moreover, it is assumed that the nonlinearity is decentralized and therefore it acts on each control input independently. Violating this assumption may change the *direction* of the implemented control signal trajectory from the actual controller output [181]. Since we are dealing with linear ADC where the system output is a direct superposition of system states (as presenters of the mode shapes), performance degradation may be a result of this change in the direction e.g., injecting skewed control effort on mode shapes that some may also layout of the bandwidth of interest. This mismatch in direction and its effective treatments are also investigated later. The further assumptions on the actuation nonlinearities are separated roughly into two cases to avoid conservatism:

**Case 1.** Let's assume that for a compact set $\mathcal{X}$ there exist positive real numbers $\alpha_1$ and $\alpha_2$ such that the Euclidean norm $(|.|)$ of the nonlinearity satisfies $|\Psi(y_c + v_1) - (y_c + v_1)| \leq \alpha_1 y_c^T(\Psi(y_c + v_1) - v_1)$ and $|\Psi(y_c + v_1) - \Psi(y_c)| \leq \min\{\alpha_1 |v_1|, \alpha_2\}$. Here, $v_1$ is an auxiliary variable with the same dimension as $y_c$ without physical interpretation that helps to separate the type of nonlinearity class of Case 1 from the so-called *sector-nonlinear* function.

**Case 2.** Let's assume that the nonlinearity is memoryless diagonal $\Psi(.) = \text{diag}\{\psi_1, \psi_2, \ldots, \psi_{n_u}\}$. Then, $\Psi \in$ sector $[K_1, K_2]$, with $K_1 = \text{diag}\{K_{11}, K_{12}, \ldots, K_{1n_u}\}$, and $K_2 = \text{diag}\{K_{21}, K_{22}, \ldots, K_{2n_u}\}$, $(K_2 - K_1 > 0)$ if $\forall z_i \in \mathbb{R}, t \geq 0, i = 1,2, \ldots, n_u: K_{1i} z_i^2 \leq z_i \psi_i \leq K_{2i} z_i^2$. $K_{ij}, i = 1,2; j = 1, \ldots, n_u$ are constant values associated with each of the nonlinearity states. It is straightforward to verify that $\Psi(y_c) = \text{sat}(y_c)$ satisfies $K_1 = 0$ and $K_2 = \mathcal{I}_{n_u}$. It should be noted that hysteresis in piezo-actuators can also be treated in the same manner.

**Remark 7.2.** The notation in the algorithms presented in the next section is inconsistent with some of the cited papers in the literature in order to have a unified notation in this dissertation and then comparing the studied anti-windup methods in a unified notation. The current representations are adapted for the output regulation problem of DRC (active damping setup). Unsurprisingly, the articles in the bibliography are not developed in a unified framework, and any comparison eventually requires some notational modifications.

### 7.1.3 Common anti-windup schemes

It is known that the windup event in the internal model control (IMC) may lead to severe destruction of controller performance due to unawareness of the controller from nonlinearity ahead of the control law especially in the case of linear control synthesis. IMC anti-windup scheme is particularly proposed to deal with the inconsistency of actual and implemented control law. In contrast to the conditioning technique, it is intended to keep the real-time plant output close to the ideal controlled plant output which can be formulated as a minimization problem over the control law. It should be pointed out that, in the literature for the linear conditioning technique, the conditioning function augmented with the system can be defined both on the *achieved actuation* $(\hat{u})$ and the difference between the achieved and unconstrained control effort $(\hat{u} - y_c)$. However, the incremental behavior of actuator saturation nonlinearity, as the most common form of treating windup in the literature, suggests employing the mismatch control signal instead of the linear controller's output. In the application of smart structures, the systems are mainly strictly proper, and the instantaneous effect of control law in plant output is infeasible. In consequence, such a minimization should be weighted by a filter to transform the weighted system output into bi-proper form. As a result, this method is mostly limited if the steady-state value of the weighted MIMO system matrix is not diagonal since the convexity of the anti-windup synthesis will be lost [182].

In [182], the off-axis circle criterion is employed to inaugurate the stability of the anti-windup method. In any weighted plant, the filter should be diagonal which otherwise may alter the original plant output's direction. A diagonal filter that can lead to a diagonal weighted steady-state system is in general unavailable and a proper approximation of plant model that guarantees the satisfaction of these two contradictive requirements is essential. Henceforth, a systematical design procedure is not easy to achieve.

The anti-windup stability problem is adequately captured in a mathematical form by Teel and Kapoor in their pioneer paper [183]. In such a framework, the linear unconstrained controller should recover from windup mode if the inconsistency is relieved.

Additionally, as a natural definition towards a systematical algorithm on anti-windup compensator, the $\mathbb{L}_2$ constraint is defined over the transfer function between the mismatch associated with the control law and the mismatch of the plant output. The idea of utilizing the coprime factorization of a stabilizing controller in synthesizing a dynamic anti-windup compensator in linear matrix inequality is pioneered by Saeki and Wada [184].

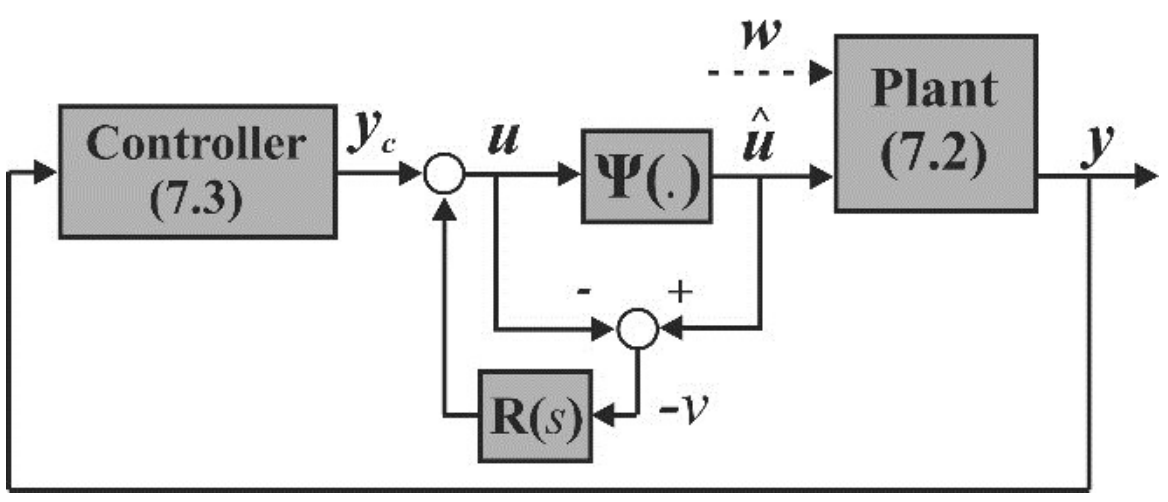

Figure 7.2 Controller output manipulation based on conditioning function $R(s)$ [176]

Their approach was based on the unified coprime factor framework developed in Campo et al. [185] (based on scaled small gain theorem) followed by Kothare et al. [186]. In this agenda, the stability of the transfer function from the output of actuation nonlinearity, as well as the output of the system to the manipulated linear controller output, is guaranteed. In view of Figure 7.2 for this concept, designing an appropriate system that corresponds to this stability requirement is formulated in terms of conditioning the linear controller's output.

**Algorithm 7.1** [183]**.** For a detectable pair $(A, C)$ in (7.1), the nominal and robust (w.r.t. $w$ in (7.1)) anti-windup problem are solvable for actuation nonlinearity of Case 1 (e.g., saturation) with $\dot{x}_{aw} = Ax_{aw} + B(\Psi(y_c + \xi_1) - y_c)$, $\xi_1 = -B^T P x_{aw}$, and $\xi_2 = -Cx_{aw} - D(\Psi(y_c + \xi_1) - y_c)$, where $P = P^T > 0$ such that $A^T P + PA \leq 0$. In the context of **Definition 7.1**, **Algorithm 7.1** refers to the solution for unknown globally Lipschitz function $\xi_1$ that if $\mathrm{dist}(u, \mathcal{X}) \in \mathbb{L}_2$, then the flawed output of the plant under the adverse influence of actuation nonlinearity converges to the ideal trajectory ($y_{nom}$) with the realizable output $y + \xi_2$ in $\mathbb{L}_2$-performance manner.

In **Algorithm 7.1**, the global anti-windup realization without the passivity constraint is tackled. Comparing to the so-called *observer-based* anti-windup technique, instead of introducing additional dynamics to the unconstrained controller, the state equation is revised as $\dot{x_c} = A_c x_c + B_c u_c + L(\Psi(y_c + \xi_1) - y_c)$, which is equivalent to the static anti-windup case in **Algorithm 7.1** ($D = -L$). In such a form, $\mathbb{A}_1$ must be Hurwitz such that a solution for $L$ under the passivity of the closed-loop system is a global solution for static anti-windup compensation [187]. where,

$$\mathbb{A}_1 = \begin{bmatrix} A + BD_c C & BC_c \\ B_c C & A_c \end{bmatrix}. \tag{7.5}$$

The role of $L$ in the stability of the observation error guides the synthesis process. Here, the error refers to the divergence of ideal states of closed-loop systems from their values under actuation nonlinearities. This is the simplified version of the passivity theorem in [188].

On the other hand, in a more general framework of coprime factorization, the control law is constructed based on the compensator $R(s)$ in $u = \left(\mathcal{J}_{n_u} - R(s)\right)^{-1} \mathbf{C}(s)y + \left(\mathcal{J}_{n_u} - R(s)\right)^{-1} R(s)\hat{u}$ (see Figure 7.2). It is trivial that setting $R(s) = \mathbf{P}(s)$ can be treated as a special case of IMC anti-windup scheme [189] and additionally, in connection to conditioning technique in $\mathbb{H}_\infty$ optimization framework, the synthesis of $R(s)$ is closely related to tuning the sensitivity transfer matrix of $\left(\mathcal{J}_{n_u} - R(s)\right)^{-1}$ such that the mismatch control effort is bounded by a value. Based on this concept in the static anti-windup scheme, the control law is manipulated instead as $u = U(s)y + \left(\mathcal{J}_{n_u} - V(s)\right)\hat{u}$, where $\hat{u}$ is the achieved actuation. In this form, the transfer matrices $U(s)$ and $V(s)$ are known as left coprime factorization of $\mathbf{C}(s) = V(s)^{-1}U(s)$, such that $\hat{\mathbf{C}}(s) = [U(s) \quad \mathcal{J}_{n_u} - V(s)]$.

In minimal state-space representation, the entities of $\hat{\mathbf{C}}(s)$ are formulated as $ss(A_c - H_1C_c, B_c - H_1D_c, H_2C_c, H_2D_c)$ and $ss(A_c - H_1C_c, -H_1, H_2C_c, H_2)$, respectively. Then, $H_1 = \Lambda_1(\mathcal{J}_{n_u} + \Lambda_2)^{-1}$ is designed such that it guarantees $A_c - H_1C_c$ to be Hurwitz while $H_2 = (\mathcal{J}_{n_u} + \Lambda_2)^{-1}$. Hence, the design matrices $\Lambda_1 \in \mathbb{R}^{n_c \times n_u}$ and $\Lambda_2 \in \mathbb{R}^{n_u \times n_u}$ should be calculated, accordingly. As pointed out in Doyle et al. [190], since $U(s)$ is stable, $V(s)^{-1}$ (bi-proper and minimum phase) should be conditioned to handle the windup problem.

The dual of the left coprime factorization on the unconstrained controller is the right coprime factorization on the plant with the advantage that the linear conditioning transfer function is almost independent of the controller, and as a result, a nonlinear controller may be applicable in this framework [191]. Defining $\mathbf{P}(s) = N(s)S(s)^{-1}$, the realizable output and conditioned input are defined in Figure 7.3 with $\overline{\mathbf{P}}(s)$ being an estimated model of $\mathbf{P}(s)$ while neglecting the disturbance effects. Accordingly, selecting $S(s) = ss(A + BF, B, F, \mathcal{J}_{n_u})$ for the design variable $F \in \mathbb{R}^{n_u \times n}$ such that $A + BF$ is Hurwitz guarantees the rejection of output mismatch while keeping the conditioning mechanism on control input stable. By setting $F = 0$, this technique guarantees the stability of the conditioning mechanism on the controller's output for the tradeoff of having a long transient anti-windup process. This is unacceptable for elastic manipulators with low structural damping used in practical applications. Two central issues of the framework in [191] are:

1) In terms of global stability there is no concrete tool for analysis and synthesis.

2) The conditioning technique has an inflexible system order that introduces high computational overhead.

In terms of AVC, the latter may be practically infeasible since structural control based on active methods is mostly dealt with in the range of [0 1] kHz. In this frequency range, the elastic active structure may include up to hundreds of mode-shapes and to include such a large order conditioner is computationally intractable. Turner and Postlethwaite [192] proposed a logical route for a static and reduced-order conditioning system that solves the latter deficiency. Their method is one of the first attempts in correctly defining the performance objective of an anti-windup conditioning system to deal with attenuating the energy due to the saturation event using a mismatch output filter (this will be analyzed later). Moreover, by setting $S(s) = \mathcal{J}_{n_u}$, the emitted signal from anti-windup compensator falls into observer-based methods proposed in Kapoor et al. [187]. This feature, although less general, secures that the controller states remain untouched. In AVC applications based on embedded systems where the control unit is programmed and implemented without a flexible configuration, conditioning solely the unconstrained control law may be the only prospect.

**Algorithm 7.2** (LQ-based model recovery AW (MRAW) [193]) Consider the sector nonlinearity (Case 2) for exponentially stable plant (7.1). Independent of the controller's architecture, the global stability of the anti-windup scheme can be achieved while minimizing $J_1 = \int_0^\infty (x_{aw}^T Q_4 x_{aw} + \xi_2^T \mathcal{P}_5 \xi_2) \mathrm{d}t$, where $Q_4 = Q_4^T \in \mathbb{R}^{n \times n} > 0$ and $\mathcal{P}_5 = \mathcal{P}_5^T \in \mathbb{R}^{n_u \times n_u} > 0$ are matrices to be defined by the control engineer, while solving the eigenvalue problem of semi-definite programming (SDP) in Eq.(7.6) subjected to LMI constraints in (7.6).

$$\min \gamma_5,$$

$$\begin{bmatrix} Q_4 A^T + A Q_4 & B U_3 + X_2^T \\ * & X_2 + X_2^T - 2U_3 \end{bmatrix} < 0, \tag{7.6}$$

where $Q_4 = Q_4^T \in \mathbb{R}^{n \times n} > 0$, matrices $X_2$, $X_3$, and diagonal $U_3 > 0$ are the SDP variables. Then, $\xi_2 = K_3 x_{aw} + L_2(\Psi(y_c + \xi_1) - y_c)$ with $K_3 = X_2 Q_4^{-1}$ and $L_2 = X_3 U_3^{-1}$. Additionally, $\mathcal{Q}_{p3}$ and $\mathcal{P}_{p5}$ are free design variables (positive definite).

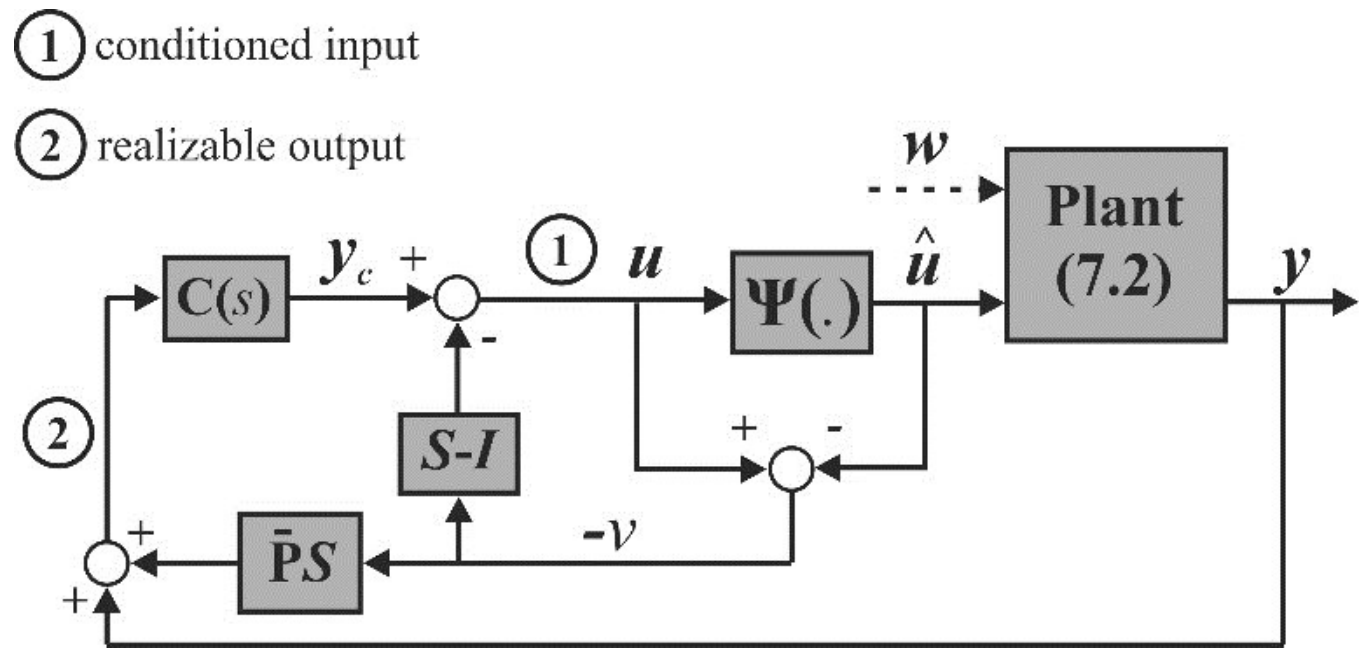

Figure 7.3 Dual of Algorithm 1a: conditioning based on coprime factorization of the plant [176]

Kothare and Morari [194] reduced the sufficient conditions for stability based on the passivity theorem into impressive form in linear matrix inequality. This was one of the first attempts to formulate a reasonably general stability analysis tool compared to [182], [195]. Employing the multipliers theory and classifying the input nonlinearity as Case 2 by satisfying the sector bounded diagonal memoryless elements (in this research also time-invariant), the absolute stability theorem based on [194] can be simplified in the form of some less conservative conditions for stability. As it is discussed later, although the stability may be guaranteed in **Algorithm 7.1**, **Algorithm 7.2**, and upcoming **Algorithm 7.3**, two primary concerns remain to be dealt with: the well-posedness of the interconnection and more accurate performance objectives.

**Algorithm 7.3** (Static Anti-windup) On this basis, the states of the controller are influenced by a static compensator ($\Upsilon_1 \in \mathbb{R}^{n_c \times n_u}$) over the mismatch control effort due to saturation ($\dot{x_c} = A_c x_c + B_c u_c - \Upsilon_1(\hat{u} - y_c)$). Additionally, the output of the linear controller is conditioned with a second static gain matrix ($\Upsilon_2 \in \mathbb{R}^{n_u \times n_u}$) over the same mismatch signal as $y_c = C_c x_c + D_c u_c - \Upsilon_2(\hat{u} - y_c)$ with the aim of a graceful performance degradation during the windup incident. It is noted that, compared to the condition technique in Figure 7.2, the windup is over as soon as $y_c$ falls below $\mathrm{sat}(y_c)$ because there is no feedback loop of a conditioning transfer function directly connected to the saturation block.

The first effective approach tackling this method is presented by Mulder et al. [53] in their attempt to capture the static problem in convex form as shown in Figure 7.4a. It should be noted that compared to [184], which is developed without any constraint on the stability of the open-loop plant, static compensation in **Algorithm 7.3** is limited to exponentially stable systems. Defining $\Delta = \Psi(.) - \mathcal{I}_{n_u}$ in linear fractional transformation (LFT) form of Figure 7.4b, the dynamics of augmented plant, controller, and static anti-windup compensator may be formulated as in (7.7), where $\bar{\mathrm{C}}(s) = ss(A_c, [B_c \quad \mathcal{I}_{n_c} \quad 0_{n_c \times n_u}], C_c, [D_c \quad 0_{n_c \times n_u} \quad \mathcal{I}_{n_u}])$. Reconstruction of Fig. 5a into LFT form of (7.6) facilitates a new interconnection that converts the design problem of static anti-windup compensation into a convex one. The intelligent decision of integrating the design gain matrices ($\Upsilon_1$ and $\Upsilon_2$) in a single augmented system that leaves the nonlinear actuation element aside, similar to treating the uncertainties in standard $\mathbb{H}_\infty$ control synthesis is the main objective of recasting the problem into LFT form.

$$\begin{aligned}
\dot{x}_{au} &= A_{au} x_{au} + (B_{v,au} - B_{\xi,au}\Upsilon)v + H_{au} w, \\
y_c &= C_{c,au} x_{au} + (D_{uv,au} - D_{u\xi,au}\Upsilon)v + D_{uw,au} w, \\
y &= C_{au} x_{au} + (D_{yv,au} - D_{y\xi,au}\Upsilon)v + D_{yw,au} w,
\end{aligned} \tag{7.7}$$

where $\Upsilon^T = [\Upsilon_1^T \quad \Upsilon_2^T]$, and matrices $A_{au}$, $B_{v,au}$, $B_{\xi,au}$, $H_{au}$, $C_{c,au}$, $D_{uv,au}$, $D_{u\xi,au}$, $D_{uw,au}$, $C_{au}$, $D_{yv,au}$, $D_{y\xi,au}$, and $D_{yw,au}$ with appropriate dimensions are the interconnecting variables that can be easily obtained by appropriate use of `sysic` function in MATLAB.

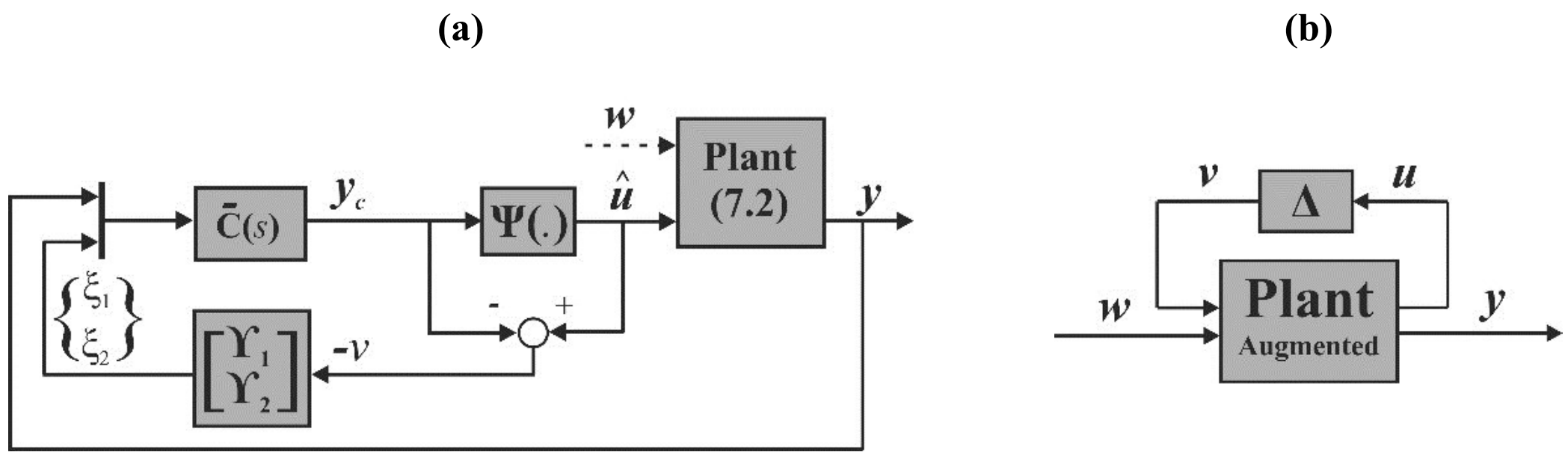

Figure 7.4 (a) Typical embodiment of static anti-windup compensator. (b) LFT form [176]

Based on the Lyapunov theorem, the transfer function from $v$ to $y_c$ is $\mathbb{L}_2$ stable for the sector nonlinearity $\Delta$ in Case 2 and the transfer function form disturbance $w$ to the output $y$ is bounded in terms of $\mathbb{L}_2$ gain by $\gamma_1^*, \gamma_1^* \in \mathbb{R} > 0$, if for positive definite diagonal matrix $\mathcal{M}_1 \in \mathbb{R}^{n_u \times n_u}$, symmetric positive definite matrix $\mathcal{P}_1$ with appropriate dimension, matrix $X_1$ with appropriate dimensions, and $\delta_2 \in \mathbb{R} > 0$, the symmetric LMI (7.8) has a feasible solution.

$$\begin{bmatrix} \mathcal{P}_1 A_{au}^T + A_{au}\mathcal{P}_1 & H_{au} & B_{v,au}\mathcal{M}_1 - B_{\xi,au}X_1 + \mathcal{P}_1 C_{c,au} & \mathcal{P}_1 C_{au}^T & 0 \\ & -\gamma_1 \mathcal{J}_{n_w} & D_{uw,au}^T & D_{yw,au}^T & 0 \\ & & \mathfrak{X} & \mathcal{M}_1 D_{yw,au}^T - X_1^T D_{y\xi,au}^T & \mathcal{M}_1 \\ & & & -\gamma_1^* \mathcal{J}_{n_y} & 0 \\ * & & & & -\delta_2 \mathcal{J}_{n_u} \end{bmatrix} < 0, \quad (7.8)$$

then, for $\mathfrak{X} = -2\mathcal{M}_1 + D_{uv,au}\mathcal{M}_1 + \mathcal{M}_1 D_{uv,au}^T - D_{u\xi,au}X_1 - X_1^T D_{u\xi,au}^T$, we have: $\Upsilon = X_1 \mathcal{M}_1^{-1}$.

In **Algorithm 7.3**, an implicit loop is nontrivially introduced into the anti-windup structure for enhanced performance. As it is discussed later, this algebraic loop can be recast into quadratic programming (QP) problem in the directionality compensation framework of [196]. Nevertheless, the well-posedness of the solution in the real-time analysis is non-trivial. The focal drawback of the static anti-windup scheme in **Algorithm 7.3**, is the infeasibility of (7.8) in several situations due to the quadratic stability Lemma which can be relieved by relaxing the constraint to piecewise quadratic stability as in [197]. As pointed out by Turner and Postlethwaite [192], another drawback of the methods based on convex optimization of SDP such as [53] and [198] is the inability of the method in formulating the output mismatch rejection. In other words, the Lyapunov equation based on quadratic constraints is defined only over minimizing the difference between the saturated system's output and the unconstrained system's output while neglecting the transient behavior. This issue is discussed later in more detail after the introduction of all algorithms.

The augmented controller after anti-windup action converts to the left-coprime factorization in terms of the unconstrained controller as $\overline{\mathbf{C}_1}(s)$: $\dot{x}_c = (A_c + \mathcal{H}_1 C_c)x_c + (B_c + \mathcal{H}_1 D_c)u_c - \mathcal{H}_1\hat{u}$, and $y_c = \mathcal{H}_2 C_c x_c + \mathcal{H}_2 D_c u_c + (\mathcal{J} - \mathcal{H}_2)\hat{u}$ with $\mathcal{H}_1 = -\Upsilon_1(\mathcal{J} + \Upsilon_2)^{-1}$ and $\mathcal{H}_2 = (\mathcal{J} + \Upsilon_2)^{-1}$. Defining $u_o = C_c x_c + D_c u_c$, we have $y_c = \mathcal{H}_2 u_o + (\mathcal{J} - \mathcal{H}_2)\hat{u}$ which is clearly a static algebraic loop (see Figure 7.5a). Note that any static anti-windup ($\mathcal{H}_2 \neq \mathcal{J}$) falls into this category. As pointed out in [199] for $y_c$, $\mathfrak{L}^T = [\mathcal{J} \quad -\mathcal{J}]^T$, $b = [u_{i,max} \quad -u_{i,\min}]$, and internal variable $\lambda \succcurlyeq 0 \in \mathbb{R}^{2n_u}$, the algebraic loop can be presented in the form of the following piecewise linear system

$$\begin{gathered} \mathcal{H}_2\hat{u} - \mathcal{H}_2 u_o + \mathfrak{L}^T\lambda = 0, \\ b - \mathfrak{L}\hat{u} \succcurlyeq 0, \\ \lambda^T(b - \mathfrak{L}\hat{u}) = 0. \end{gathered} \quad (7.9)$$

A multiple linear complementarity problem (MLCP) based on **Definition D.2** in Appendix D. is proposed over (7.9) with matrices given as $\mathcal{M}_{11} = \mathcal{H}_2$, $\mathcal{M}_{12} = \mathfrak{L}^T$, $\mathcal{M}_{21} = -\mathfrak{L}$, and $\mathcal{M}_{22} = 0$ and $a = -\mathcal{H}_2 u_o$ [200].

Note that here only the symmetrical algebraic loop is considered for which an additional constraint should be defined as $\mathcal{H}_2 = \mathcal{H}_2^T > 0$. For asymmetric $\mathcal{H}_2$, although a better performance may be achievable (because of the less conservative solution space), the solution for online MLCP is difficult to calculate. To the modest knowledge of the author, in real-time applications, this is still a challenging problem (see [201]). Therefore, in the context of this dissertation, we are looking at the explicit solution of the symmetric algebraic loop.

In Figure 7.5, $\overline{\mathbf{C}}_2(s)$: $ss(A_c + \mathcal{H}_1 C_c, \mathcal{H}_1, C_c, 0)$, $L_3 = \mathcal{H}_2 - \mathcal{I}$, and $L_4 = \mathcal{H}_2$. Based on [202], the symmetric solution for algebraic equation (7.9) may be written in terms of Karush-Kuhn-Tucker (KKT) conditions as a solution for convex quadratic programming (QP($u_o$)) in (7.10)

$$\begin{gathered} \text{QP}(u_o) = \arg\min_{\hat{u}} \frac{1}{2} \hat{u}^T \mathcal{H}_2 \hat{u} - \hat{u}^T \mathcal{H}_2 u_o, \\ \text{subject to: } \mathfrak{L}\hat{u} \preccurlyeq b. \end{gathered} \tag{7.10}$$

It is proven that the bounded convex problem (7.10) is feasible. In this framework, the implementation of the static anti-windup compensator reduces to Figure 7.5b implying that the algebraic loop in Figure 7.5a is replaced with an online quadratic programming problem. Finally, the gradient projection method can be used for the online implementation of QP [203].

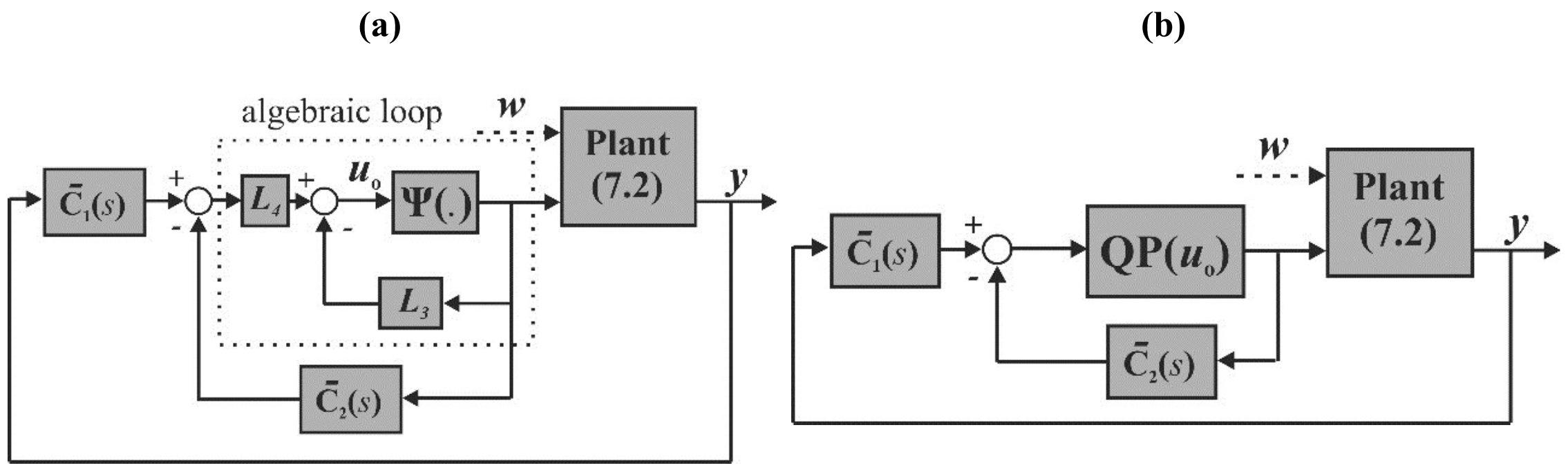

Figure 7.5 (a) Extracting the nontrivial algebraic loop in equivalent system interconnection as in Figure 7.4. (b) Proposed solution based on online quadratic programming in [200].

**Remark 7.3.** The QP problem in addressing algebraic loop in closed-loop systems are commonly known as *directionality compensation*. Recently, Adegbege and Heath [204] proposed a new directionality compensator in the framework of [53]. It should be noted that, the directionality conditioning such as [205] and [206] where explicit constraints are proposed on the structure of static anti-windup system such as $\mathcal{H}_1 = \mathcal{I}$, $\mathcal{H}_2 = D_c^{-T} D_c$ [205] are seemingly conservative and not of the interest of this dissertation.

One notably useful tool in evaluating the performance of a saturated system compensated with a linear conditioning anti-windup technique as well as in providing guidelines to synthesis process in time-domain is proposed by Weston and Postlethwaite [191]. Accordingly, Figure 7.3 is mathematically equivalent to Figure 7.6 with simple algebra. Then, the so-called *linear mode* in relation to the 'intended linear system' in Figure 7.6, refers to the case that $\text{dz}(u) = 0$. As soon as the saturation starts, the *nonlinear mode* is triggered, and since $S(s) - \mathcal{I}_{n_u}$ is in the feedback of 'nonlinear loop,' it may introduce some delay in driving the mismatch control law back to zero even when $y_c$ in magnitude is below the saturation limits. In the absence of the feedback loop, well-posedness is certain by the Lipschitz continuity of the static dead-zone. Stimulation of the nonlinear mode maintains the signal for 'output mismatch filter.' However, *output mismatch rejection mode* does not start till the nonlinear mode is over, i.e., $y_c = \hat{u}$. The latter mode should be interpreted as attenuation of the energy stored in the nonlinear mode after $y_c = \hat{u}$. The restoration of the adequacy of control efforts is carried out as soon as the manipulating signal on plant's output falls below the noise level.

In terms of performance evaluation, the interest of successful anti-windup compensation based on the conditioning technique is defined over the time it takes for output mismatch filter to attenuate the energy stored in the nonlinear loop at the end of the nonlinear mode. Moreover, in the key systematic synthesis of static and dynamic anti-windup schemes such as [53] and [198], respectively, output mismatch rejection mode is mostly neglected. At this point, it should be cleared that although disturbance may be neglected in the stability analysis during the windup event, attentive performance analysis, especially in the post-windup phase, is absolutely necessary.

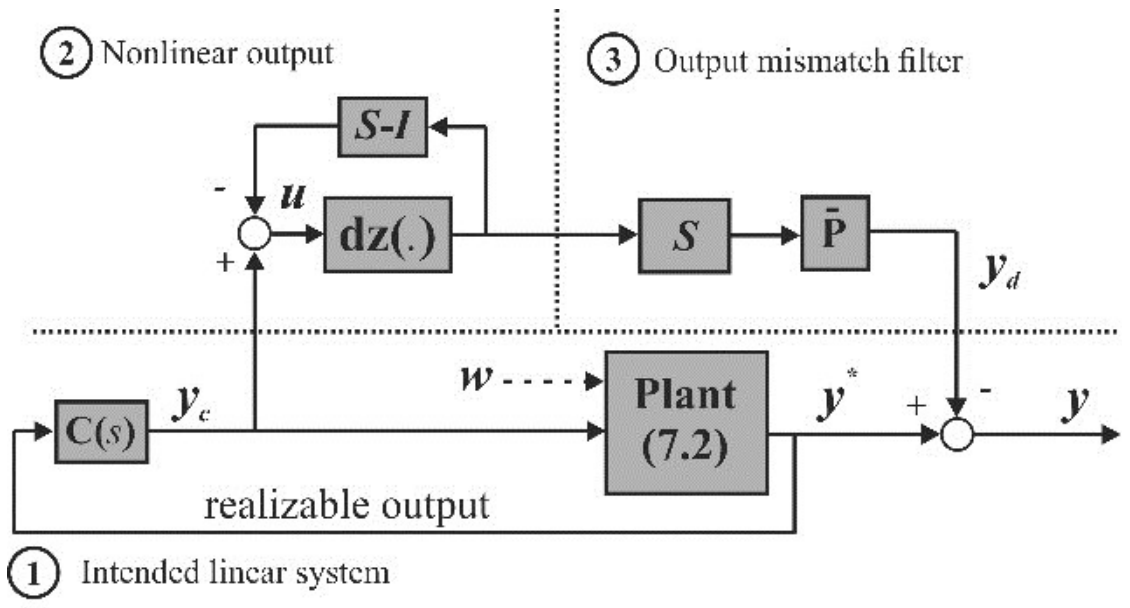

Figure 7.6 Decomposition of the closed-loop system during windup for qualitative performance evaluation [176]

Based on the obtained results in the experimental implementations, it is concluded that the aforementioned three-mode performance evaluation scheme of [54], [191] is a key tool in comparing different algorithms. Additionally, as the focal point of numerical results from the vibration engineer point of view, the scheme in Figure 7.6 provides a crystal clear guidelines in compensator tuning. Accordingly, for the application of vibration control, this tool is followed in having a better view of the stability topographies in the final configuration. For real AVC applications where the system may remain in windup state for a long time, the performance of the anti-windup compensator can be appreciated in accordance with its behavior in output mismatch rejection mode.

**Algorithm 7.4** Synthesizing a reduced-order conditioning transfer matrix $S(s)$ (or static compensator) to achieve minimum post-windup-event performance degradation is addressed in this algorithm. The proof of existence is referred to [192] which is a straightforward procedure based on the Lyapunov stability. In comparison to the case of static compensation which neglects the frequency information of the conditioned system in a tradeoff for zero delays in the nonlinear loop of Figure 7.6, this algorithm benefits from the frequency-dependence feature. Such a tradeoff for vibrating systems with slow dynamics is acceptable since higher frequency ($\gg$ 1 kHz) oscillations are normally dealt with by passive means. Assume that $\theta_1 \in \mathbb{R}_{\geq 0}^{n_u}$ and $\theta_2 \in \mathbb{R}_{\geq 0}^{n_y}$ are generated via $\Theta_1(s)$ and $\Theta_2(s)$ instead of $S(s) - \mathcal{I}_{n_u}$ and $\bar{\mathbf{P}}(s)S(s)$, respectively (see Figure 7.3). Next, based on the trial and error, two low-pass filters $F_1(s)$ and $F_2(s)$ of user-defined order ($n_{F_1}$ and $n_{F_2}$, respectively) with modulus 1 are designed for output mismatch rejection such that $\Theta_1(s) = F_1(s)\widetilde{\Theta}_1$ and $\Theta_2(s) = F_2(s)\widetilde{\Theta}_2$ can be created in terms of real-valued constant matrices $\widetilde{\Theta}_1$ and $\widetilde{\Theta}_2$. In light of Figure 7.3 for $\Theta_1(s)$ and $\Theta_2(s)$, the stability of the nonlinear mode and performance of output mismatch rejection are recast in terms of designing the transfer matrices $S(s) - \mathcal{I}_{n_u} = ss\left(\bar{A}, \left(B_0 + \bar{B}\widetilde{\Theta}\right), \bar{C}_1, \left(D_{01} + \bar{D}_1\widetilde{\Theta}\right)\right)$ and $\bar{\mathbf{P}}(s)S(s) = ss\left(\bar{A}, \left(B_0 + \bar{B}\widetilde{\Theta}\right), \bar{C}_2, \left(D_{02} + \bar{D}_2\widetilde{\Theta}\right)\right)$, respectively for $\widetilde{\Theta} = [\ \widetilde{\Theta}_1^T \quad \widetilde{\Theta}_2^T]^T$. where $\bar{A}$, $B_0$, $\bar{B}$, $\bar{C}_1$, $D_{01}$, $\bar{D}_1$, $\bar{C}_2$, $D_{02}$, and $\bar{D}_2$ are matrices with appropriate dimension depending on order of $F_1(s)$ and $F_2(s)$ that can be obtained by use of `sysic` function in MATLAB. Then, the dynamic anti-windup compensator $\Theta(s) = [\Theta_1^T(s) \quad \Theta_2^T(s)]^T$ is calculated by solving LMI feasibility problem (7.11) for unknown matrix $\mathcal{P}_4 = \mathcal{P}_4^T > 0$, diagonal matrix $U_2 \in \mathbb{R}^{n_u \times n_u}$, matrix $L_1 \in \mathbb{R}^{(n_u+n_y)\times n_u}$, and scalar $\gamma_4 > 0$ [192]

$$\begin{bmatrix} \mathcal{P}_4\bar{A}^T + \bar{A}\mathcal{P}_4 & B_0U_2 + \bar{B}L_1 - \mathcal{P}_4\bar{C}_1^T & 0 & \mathcal{P}_4\bar{C}_2^T \\ & \mathfrak{I}_{22} & \mathcal{I} & U_2D_{02}^T + L_1^T\bar{D}_2^T \\ & & -\gamma_4\mathcal{I} & 0 \\ * & & & -\gamma_4\mathcal{I} \end{bmatrix} < 0, \tag{7.11}$$

$$\mathfrak{I}_{22} = -2U_2 - U_2D_{01}^T - D_{01}U_2 - L_1^T\bar{D}_1^T - \bar{D}_1L_1,$$

which guarantees the $\mathbb{L}_2$ norm of the transfer matrix $\mathcal{T}(s): y_c \to y_d$ to be bounded by $\gamma_4$ with $\widetilde{\Theta} = L_1U_2^{-1}$. It should be noted that one drawback of **Algorithm 7.4** is the nonconvexity of the constraints in a general form which is replaced by a suboptimal solution that may be stuck in local minima. The static case, which solves the strong anti-windup problem optimally, is calculated for constant real-valued matrix $\Theta = [\Theta_1^T \quad \Theta_2^T]^T$ as solution for matrix condition (7.12) assuming $F_1(s) = F_2(s) = \mathcal{I}_{n_u}$. As pointed out in [192], the anti-windup problem is *strongly solvable* if $\mathcal{T}(s)$ is finite gain in terms of $\text{dist}(u, \mathcal{X}) \in \mathbb{L}_2$ in **Definition 7.1**.

$$\begin{bmatrix} \mathcal{P}_4\bar{A}^T + \bar{A}\mathcal{P}_4 & B_0U_2 + \bar{B}L_1 - \mathcal{P}_4\bar{C}_1^T \\ * & \mathfrak{I}_{22}^* \end{bmatrix} < 0,$$
$$\mathfrak{I}_{22}^* = -2U_2 - U_2D_{01}^T - D_{01}U_2 - L_1^T\bar{D}_1^T - \bar{D}_1L_1. \tag{7.12}$$

At this stage, we should clarify that coping with control signal discrepancy in a stable framework and minimizing the norm of the transfer function from unconstrained control law to mismatch output are superior requirements for the true goal of anti-windup systems. This is of course in comparison to putting constraints on disturbance to the output transfer function, e.g., classical $\mathbb{L}_2$ constraint. However, it should be emphasized that based on the discussion on the reformation in Figure 7.6, nontrivially, the effect of disturbance in the nonlinear loop should be investigated at the same importance. Li et al. delicately addressed this problem by adding an extra matrix variable and then employing Projection Lemma [54]. Accordingly, in contrast to **Algorithm 7.1** (see Figure 7.3) and [191], $S(s) - \mathcal{I}_{n_u}$ and $\bar{\mathbf{P}}(s)S(s)$ are replaced with $\mathcal{M}(s)E_1 - \mathcal{I}_{n_u} = ss(A + BF_3, BE_1, F_3, E_1 - \mathcal{I}_{n_u})$ and $\mathcal{N}(s)E_1 = ss(A + BF_3, BE_1, C + DF_3, DE_1)$, respectively for unknown $F_3 \in \mathbb{R}^{n_u \times n}$ and $E_1 \in \mathbb{R}^{n_u \times n_u}$. In light of new structure for general conditioning technique, the anti-windup synthesis for minimizing the $\mathbb{L}_2$ norm from disturbance to mismatch output in Figure 7.6 is summarized in **Algorithm 7.5**.

**Algorithm 7.5** (Disturbance rejection in the nonlinear loop [54]) Let's assume the asymptotically stable transfer function $\mathcal{T}: w \to y_c$ is well-posed and has a minimal realization $ss(A_d, B_d, C_d, D_d)$ where $A_d \in \mathbb{R}^{n_d \times n_d}$. Then, for a symmetric positive definite matrix $\mathcal{P}_9 \in \mathbb{R}^{(n+n_d)\times(n+n_d)}$ solve the SDP (7.13) in terms of positive scalar $\gamma_d$.

$$\min_{\mathfrak{X}_1, \mathfrak{X}_2} \gamma_d$$
$$\mathfrak{X}_1 = \begin{bmatrix} \mathfrak{X}_{111} & W_C + \mathcal{P}_9W_B \\ * & W_D \end{bmatrix} < 0,$$
$$\mathfrak{X}_2 = \begin{bmatrix} A_d^T\mathcal{P}_{922} + \mathcal{P}_{922}A_d & \mathcal{P}_{922}B_d \\ * & -\gamma_d\mathcal{I}_{n_w} \end{bmatrix} > 0, \tag{7.13}$$

where,

$$W_A = \begin{bmatrix} 0 & BC_d \\ 0 & 0 \end{bmatrix} \in \mathbb{R}^{(n+n_d)\times(n+n_d)},$$
$$W_B = \begin{bmatrix} BD_d & 0 \\ B_d & 0 \end{bmatrix} \in \mathbb{R}^{(n+n_d)\times(n_w+n_y)}, \tag{7.14}$$

$$W_C = \begin{bmatrix} 0 & C^T \\ 0 & C_d^T D^T \end{bmatrix} \in \mathbb{R}^{(n+n_d)\times(n_w+n_y)},$$

$$W_D = \begin{bmatrix} -\gamma_d \mathcal{I}_{n_w} & D_d^T D^T \\ D D_d & -\gamma_d \mathcal{I}_{n_y} \end{bmatrix} \in \mathbb{R}^{(n_w+n_y)\times(n_w+n_y)},$$

$$\mathcal{P}_9 = \begin{bmatrix} \mathcal{P}_{911} & \mathcal{P}_{912} \\ \mathcal{P}_{912}^T & \mathcal{P}_{922} \end{bmatrix},$$

$$\mathfrak{X}_{111} = \mathcal{P}_9 \begin{bmatrix} A & 0 \\ 0 & A_d \end{bmatrix} + \begin{bmatrix} A & 0 \\ 0 & A_d \end{bmatrix}^T \mathcal{P}_9 + \mathcal{P}_9 W_A + W_A^T \mathcal{P}_9.$$

Based on the solution for $\gamma_d$ and $\mathcal{P}_9$, solve the LMI condition: $\Pi + \mathcal{H}^T[F_3 \quad E_1]\mathcal{G} + (\mathcal{H}^T[F_3 \quad E_1]\mathcal{G})^T < 0$, for unknown matrices $F_3$, $E_1$, and diagonal positive definite matrix $\mathcal{W}$. where,

$$\mathcal{H} = \begin{bmatrix} B^T & 0_{n_u\times n_d} & -\mathcal{I}_{n_u} & 0_{n_u\times n_w} & D^T \end{bmatrix} \mathrm{diag}\left(\mathcal{P}_9, \begin{bmatrix} \mathcal{W} & 0 \\ 0 & \mathcal{I}_{n_w} \end{bmatrix}, \mathcal{I}_{n_y}\right),$$

$$\mathcal{G} = \begin{bmatrix} \mathcal{I}_n & 0_{n\times n_d} & 0_{n\times n_u} & 0_{n\times n_w} & 0_{n\times n_y} \\ 0_{n_u\times n} & 0_{n_u\times n_d} & \mathcal{I}_{n_u} & 0_{n_u\times n_w} & 0_{n_u\times n_y} \end{bmatrix},$$

$$\Pi = \begin{bmatrix} \mathcal{P}_9 \begin{bmatrix} A & 0 \\ 0 & A_d \end{bmatrix} + \begin{bmatrix} A & 0 \\ 0 & A_d \end{bmatrix}^T \mathcal{P}_9 & \mathcal{P}_9 \begin{bmatrix} 0 & 0 \\ 0 & B_d \end{bmatrix} + \begin{bmatrix} 0 & C_d \\ 0 & 0 \end{bmatrix}^T \begin{bmatrix} \mathcal{W} & 0 \\ 0 & \mathcal{I}_{n_w} \end{bmatrix} & \begin{bmatrix} C^T \\ 0 \end{bmatrix} \\ & \Pi_{22} & 0 \\ * & & -\gamma_d \mathcal{I}_{n_y} \end{bmatrix}, \tag{7.15}$$

$$\Pi_{22} = \begin{bmatrix} \mathcal{W} & 0 \\ 0 & \mathcal{I}_{n_w} \end{bmatrix} \begin{bmatrix} 0_{n_u\times n_u} & D_d \\ 0_{n_w\times n_u} & 0_{n_w\times n_w} \end{bmatrix} + \begin{bmatrix} 0_{n_u\times n_u} & D_d \\ 0_{n_w\times n_u} & 0_{n_w\times n_w} \end{bmatrix}^T \begin{bmatrix} \mathcal{W} & 0 \\ 0 & \mathcal{I}_{n_w} \end{bmatrix}^T - \gamma_d \mathcal{I}_{n_y}.$$

Note that here an additional degree of freedom is exploited into the synthesis process which includes the system's directional characteristics as mentioned before. This added freedom was also investigated by Turner et al. [207], in that only to remove the algebraic loop.

Before progressing to the rest of the results in the dissertation, some additional remarks should be made on algebraic and implicit loops in anti-windup schemes. These remarks by no means are necessary only for **Algorithm 7.5** but for all of the results in this chapter including dynamic cases (see **Algorithm 7.7**). Because we are dealing with continuous LTI plants, it is rather obvious that as long as a measure of the constrained input signal (after a static actuation nonlinearity) is directly fed back to the non-strictly proper controller, an implicit loop will appear. For instance, in **Algorithm 7.5** with $E_1 \neq \mathcal{I}_{n_u}$, despite the superior disturbance rejection performance, the algebraic loop is expected. Some of the key articles that address this issue in a general form are Hu et al., [208] and the works of Syaichu-Rohman and coworkers [196], [209] in which the algebraic loop is treated with QP. In the latter, for the case of saturation nonlinearity, an explicit solution is proposed instead of iterative implicit methods. A summary of theory for resolving the algebraic loop for symmetric saturation nonlinearity ($|u_{i,\min}| = |u_{i,\max}| = u_{i,m}$) is available in (pp. 129 of [54]) and here only the essential steps are reported for strictly diagonally dominant matrix $E_1^{-1}\mathrm{diag}(u_{i,m})$, $i = 1, \dots, n_u$ (refer to Appendix B). In Chapter 6.2, the benchmark problem is defined on a smart beam with two piezo patches as inputs ($n_u = 2$) and as a result $E_1 \in \mathbb{R}^{n_u\times n_u}$ to be calculated

$$E_1 = \begin{bmatrix} e_{1,11} & e_{1,12} \\ e_{1,21} & e_{1,22} \end{bmatrix}, \tag{7.16}$$

For saturated control signal (achieved control) ($\hat{u} = [\hat{u}_1^T \quad \hat{u}_2^T]^T$) with the conventional definition of a sign function (sign(.)):

1) Calculate $y_c - \mathrm{sat}(y_c) = E_1^{-1}\big(\mathrm{sat}(y_c) - \mathrm{sign}\big(E_1^{-1}\mathrm{sat}(y_c)\big)\, u_{i,m}\big)$ for both of the inputs ($i = 1{,}2$). If $y_c - \mathrm{sat}(y_c) = E_1^{-1}$ for $i = 1$ and $i = 2$ have nonzero values and additionally $\mathrm{sign}\big(E_1^{-1}\mathrm{sat}(y_c)\big) = \mathrm{sign}\big(y_c - \mathrm{sat}(y_c)\big)$ then the explicit solution is found, otherwise move to step 2.

2) Take $y_c - \mathrm{sat}(y_c) = 0$, for $i = 1$. Check to see if $\mathrm{dz}(\hat{u}_1 - e_{1,12}\tilde{u}_2) = 0$ for $\tilde{u}_2 = \mathrm{dz}(\hat{u}_2)/e_{1,22}$. If not satisfied go to step 3.

3) Take $y_c - \mathrm{sat}(y_c) = 0$, for $i = 2$. Check to see if $\mathrm{dz}(\hat{u}_2 - e_{1,21}\tilde{u}_1) = 0$ for $\tilde{u}_1 = \mathrm{dz}(\hat{u}_1)/e_{1,11}$.

The above three-step algorithm for extracting an explicit unique solution for piecewise linear algebraic loop based on saturation nonlinearity is first presented in [209] (see pp. 1055). Additionally, we point out the following two remarks on algebraic loop:

a) Three relatively easy solutions to come around this problem in real-time implementations are: 1) to explicitly set feedthrough terms of anti-windup compensator equal to zero. 2) Injecting the signal that generates the implicit loop through a low-pass filter [57]. 3) Defining a unit-delay on the responsible signal if the sampling time of the real-time implementation is significantly smaller than the period of maximum frequency in the nominal range of AVC.

b) The third method is proposed in light of robust stability analysis carried out on the well-posedness of algebraic loops delayed by a *small* multiplicative delay in the feedback channel ($e^{-\varepsilon s}$). Based on the incremental small gain theorem [210], if $\|\Upsilon_2(\mathcal{J} + \Upsilon_2)^{-1}\|_2 < 1$, the closed-loop with piece-wise linear actuator saturation function subjected to mentioned delay is well-posed and robust $\mathbb{L}_2$ incremental finite gain stable or equivalently: $\Upsilon_2 + \Upsilon_2^T + \mathcal{J} > 0$.

**Algorithm 7.6** (Simultaneous controller/anti-windup synthesis) In static anti-windup schemes of **Algorithm 7.3** and **Algorithm 7.4**, the compensator is sequentially designed by incorporating the linear controller's dynamics. The methods based on the detuning of prior unconstrained control are not discussed here and neither simultaneous synthesizing the conditioned controller except for **Algorithm 7.6**. Here we only refer to a combination of multi-objective output feedback controller within LMI framework based on Scherer et al. [211] combined with Kothare and Morari [212]. Similar to **Algorithm 7.3** the *true goal* of the anti-windup compensator in the restoration of the adequate linear behavior of the system after the windup event is neglected here. First, select diagonal positive definite $W_1$ serving as a matrix in Lyapunov equation $V_1(t) > 0$ and then derivate with respect to time: $\dot{V}_1(t) + \big(y_c^T W_1(y_c - \hat{u}) + (y_c - \hat{u})^T W_1 y_c - 2(y_c - \hat{u})^T W_1(y_c - \hat{u})\big) + (y^T y - \gamma_6 w^T w) < 0$, for $\gamma_6 \in \mathbb{R} > 0$ and define $M_2 = W_1^{-1}$. Second, solve the LMI eigenvalue problem $\min_{\mathfrak{I}_1,\mathfrak{I}_2} \gamma_6$ for LMIs [213]

$$\mathfrak{I}_1 = \begin{bmatrix} \mathfrak{I}_{111} & \mathfrak{I}_{112} & H\mathcal{R}_j & \mathfrak{I}_{114} & -B\hat{\Lambda}_2 + \hat{C}^T \\ & \mathfrak{I}_{122} & \mathcal{P}_7 H\mathcal{R}_j & \mathfrak{I}_{124} & -\hat{\Lambda}_1 + C^T\hat{D}^T \\ & & -\gamma_6\mathcal{J} & 0 & 0 \\ & & & -\mathcal{J} & -\mathcal{L}_i D\hat{\Lambda}_2 \\ * & & & & -\hat{\Lambda}_2 - \hat{\Lambda}_2^T \end{bmatrix} < 0,$$

$$\mathfrak{I}_2 = \begin{bmatrix} Q_5 & \mathcal{J} \\ * & \mathcal{P}_7 \end{bmatrix} > 0, \tag{7.17}$$

where,

$$\mathfrak{I}_{111} = AQ_5 + Q_5 A^T + B\hat{C} + \hat{C}^T B^T,$$
$$\mathfrak{I}_{112} = A + B\hat{D}C + \hat{A}^T, \tag{7.18}$$
$$\mathfrak{I}_{114} = \big(CQ_5 + D\hat{C}\big)^T \mathcal{L}_i^T,$$

$$\Im_{122} = \mathcal{P}_7 A + A^T \mathcal{P}_7 + \hat{B}C + C^T \hat{B}^T,$$

$$\Im_{124} = \left(C + D\hat{D}C\right)^T \mathcal{L}_i^T,$$

for unknown matrices $\mathcal{Q}_5$, $\mathcal{P}_7$, $\hat{A}$, $\hat{B}$, $\hat{C}$, $\hat{D}$, $\hat{\Lambda}_1$, and $\hat{\Lambda}_2$ with $\mathcal{L}_i$ and $\mathcal{R}_j$ representing block-diagonal matrices for isolating specific input to output transfer function ($\mathcal{T}_{ij}$). Finally, $\Upsilon_2 = \hat{\Lambda}_2 W_1 - \mathcal{I}$ and $\Upsilon_1 = \mathcal{P}_8^{-1}\left(\hat{\Lambda}_1 - \mathcal{P}_7 B(\mathcal{I} + \Upsilon_2)M_2\right)W_1$ in which for $n_c = n$ [211]

$$\begin{aligned} \mathcal{Q}_6 \mathcal{P}_8^T &= \mathcal{I} - \mathcal{Q}_5 \mathcal{P}_7, \\ D_c &= \hat{D}, \\ C_c &= (\hat{C} - D_c C \mathcal{Q}_5)\mathcal{Q}_6^{-T}, \\ B_c &= \mathcal{P}_8^{-1}\left(\hat{B} - \mathcal{P}_7 B D_c\right), \\ A_c &= \mathcal{P}_8^{-1}\left(\hat{A} - \mathcal{P}_8 B_c C \mathcal{Q}_5 - \mathcal{P}_7 B C_c \mathcal{Q}_6^T - \mathcal{P}_7 (A + B D_c C)\mathcal{Q}_5\right)\mathcal{Q}_6^{-T}. \end{aligned} \tag{7.19}$$

Building on the knowledge of synthesizing linear parameter-varying (LPV) controllers, the feasibility of a fixed-order dynamic anti-windup compensator, in general, is a nonconvex problem. For dynamic anti-windup compensation of an asymptotically stable plant's order, the synthesis process can be converted to a convex problem [198].

**Algorithm 7.7** Unlike **Algorithm 7.1**, in the dynamic anti-windup scheme (see Figure 7.7), the order of the ad-hoc compensator ($\Sigma_{aw}(s)$) is $n_{aw} > 0$ such that

$$\begin{aligned} \dot{x}_{aw} &= A_{aw} x_{aw} + B_{aw}(y_c - \hat{u}), \\ \xi = \begin{bmatrix} \xi_1 \\ \xi_2 \end{bmatrix} &= C_{aw} x_{aw} + D_{aw}(y_c - \hat{u}), \end{aligned} \tag{7.20}$$

where $\xi$ is similarly defined to a static case in **Algorithm 7.3** (**Definition 7.1**) and provides the realizable output for linear controller and conditions the unconstrained control law (in analogy to the static case: $n_{aw} = 0, \Upsilon = D_{aw}$). $A_{aw}$, $B_{aw}$, $C_{aw}$, and $D_{aw}$ are the unknown matrices with appropriate dimensions to be designed.

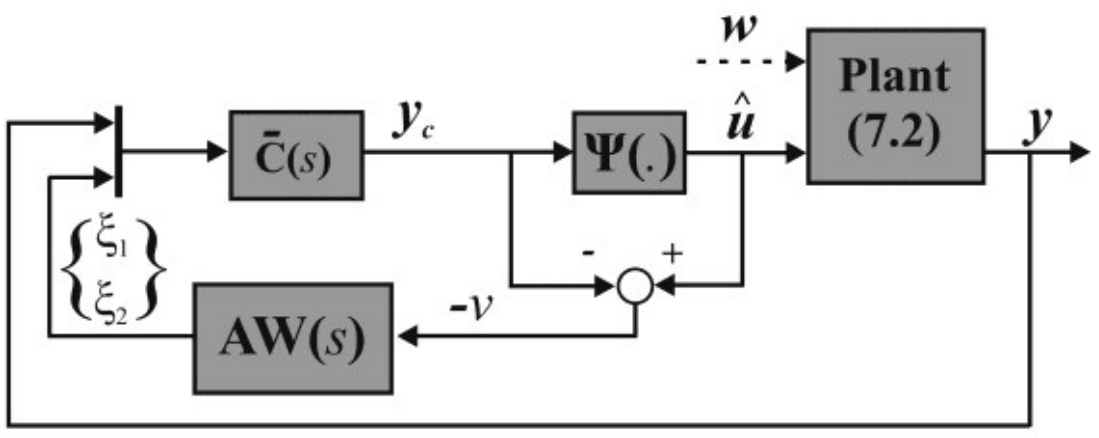

Figure 7.7 Dynamic anti-windup scheme based on the approach in [198]

Given the linear stable plant (7.1), and unconstrained controller (7.3) conditioned by $\xi$, an anti-windup compensator with quadratic performance $\gamma_3 \in \mathbb{R} > 0$ based on the bounded real Lemma (BRL) is achievable. Then, the compensator is intended to guarantee the well-posedness of the interconnection in Figure 7.7 if the matrix conditions have a solution for the LMI eigenvalue problem in Eqs. (7.21) for $\mathcal{Q}_1 = \mathcal{Q}_1^T \in \mathbb{R}^{n+n_c} > 0$ , $\mathcal{P}_3 = \mathcal{P}_3^T \in \mathbb{R}^{n+n_c} > 0$ , and $\gamma_3$ with closed-loop matrices given in (7.22)

$$\begin{bmatrix} \mathcal{P}_{311} A^T + A \mathcal{P}_{311} & H & \mathcal{P}_{311} C^T \\ & -\gamma_3 \mathcal{I}_{n_w} & 0 \\ * & & -\gamma_3 \mathcal{I}_{n_y} \end{bmatrix} < 0, (a) \tag{7.21}$$

$$\begin{bmatrix} \mathcal{Q}_1 A_{cl}^T + A_{cl}\mathcal{Q}_1 & H_{cl,w} & \mathcal{Q}_1 C_{cl}^T \\ & -\gamma_3 \mathcal{I}_{n_w} & 0 \\ * & & -\gamma_3 \mathcal{I}_{n_w} \end{bmatrix} < 0, \text{(b)}$$

$$\mathcal{P}_3 = \mathcal{P}_3^T = \begin{bmatrix} \mathcal{P}_{311} & \mathcal{P}_{312} \\ \mathcal{P}_{312}^T & \mathcal{P}_{322} \end{bmatrix} > 0, (c)$$

$$\mathcal{P}_3 - \mathcal{Q}_1 \geq 0. (d)$$

An interesting observation on Eqs. (7.21), using BRL is that the achievable quadratic performance level $\gamma_3$ has a lower bound of $\max_{s=j\omega,\omega\in\mathbb{R}>0} \left\{ \mathbb{H}_\infty^{P_{yw}(s)}, \mathbb{H}_\infty^{\mathbf{CL}(s):w\to y} \right\}$ in which $P_{yw}$ (given in (7.2)) and $\mathbf{CL}(s): w \to y$ are the transfer function of linear plant and unconstrained closed-loop system, respectively [198]. In addition $\mathbf{CL}(s)$ can be represented in state-space form for state vector $x_{cl} = [x^T \quad x_c^T]^T$ for real-valued matrices $A_{cl}$, $B_{cl,v}$, $B_{cl,\xi}$, $H_{cl,w}$, $C_{c,cl}$, $D_{uv,cl}$, $D_{u\xi,cl}$, $D_{uw,cl}$, $C_{cl}$, $D_{yv,cl}$, $D_{y\xi,cl}$ and $D_{cl,yw}$ with appropriate dimensions as (7.22).

$$\begin{aligned} \dot{x}_{cl} &= A_{cl}x_{cl} + B_{cl,v}(y_c - \hat{u}) + B_{cl,\xi}\xi + H_{cl,w}w, \\ y_c &= C_{c,cl}x_{cl} + D_{uv,cl}(y_c - \hat{u}) + D_{u\xi,cl}\xi + D_{uw,cl}w, \\ y &= C_{cl}x_{cl} + D_{yv,cl}(y_c - \hat{u}) + D_{y\xi,cl}\xi + D_{cl,yw}w. \end{aligned} \tag{7.22}$$

These matrices can be easily obtained by simple linear algebra and block manipulation. As a result, the asymptotic stability constraint of the plant is self-evident. Moreover, by looking at Eqs. (7.21)(a,b) the often infeasibility problem of LMI in Eq.(7.8) is now explainable. Accordingly, even for large $\gamma_1$, the requirement for existence of a quasi-common Lyapunov function for stability measure of linear open loop plant $\mathbf{P}(s)$ and unconstrained closed-loop system $\mathbf{CL}(s)$ is often infeasible. Then, $N_1 \in \mathbb{R}^{(n+n_c)\times n_{aw}}$ and $M_1 \in \mathbb{R}^{n_{aw}\times n_{aw}}$ can be obtained by solving $N_1 N_1^T = \mathcal{P}_3\mathcal{Q}_1^{-1}\mathcal{P}_3 - \mathcal{P}_3$, which has a solution for invertible pair $(\mathcal{P}_3, \mathcal{Q}_1)$ due to the last LMI constraint (7.21), and finally $M_1 = \mathcal{I}_{n_{aw}} + N_1^T\mathcal{P}_3 N_1$. Finally, the anti-windup compensator is constructed by calculating $A_{aw}$, $B_{aw}$, $C_{aw}$, and $D_{aw}$ as the solution for LMI (7.23) and by assuming $U_1 = \delta_3\mathcal{I}$, [198]

$$\Omega_3 + G_1^T \begin{bmatrix} A_{aw} & B_{aw} \\ C_{aw} & D_{aw} \end{bmatrix}^T G_2 + G_2^T \begin{bmatrix} A_{aw} & B_{aw} \\ C_{aw} & D_{aw} \end{bmatrix} G_1 < 0, \tag{7.23}$$

where for $n_{tot} = n + n_c + n_{aw}$ and $n_\xi = n_u + n_c$, $\Omega_3 \in \mathbb{R}^{(n_{tot}+n_u+n_w+n_y)\times(n_{tot}+n_u+n_w+n_y)}$, $G_1 \in \mathbb{R}^{(n_{aw}+n_u)\times(n_{tot}+n_u+n_w+n_y)}$, $G_2 \in \mathbb{R}^{(n_{aw}+n_\xi)\times(n_{tot}+n_u+n_w+n_y)}$, $A_o \in \mathbb{R}^{n_{tot}\times n_{tot}}$, $B_{vo} \in \mathbb{R}^{n_{tot}\times n_u}$, $C_{yo} \in \mathbb{R}^{n_u\times n_{tot}}$, $D_{yvo} \in \mathbb{R}^{n_u\times n_u}$, $G_{11} \in \mathbb{R}^{(n_{aw}+n_u)\times n_{tot}}$, $G_{12} \in \mathbb{R}^{(n_{aw}+n_u)\times n_u}$, $G_{22} \in \mathbb{R}^{(n_{aw}+n_\xi)\times n_u}$, $G_{23} \in \mathbb{R}^{(n_{aw}+n_\xi)\times n_y}$, and $H_w \in \mathbb{R}^{n_{tot}\times n_w}$ are given as

$$\begin{aligned} \Omega_3 &= \begin{bmatrix} \mathcal{Q}_2 A_o^T + A_o\mathcal{Q}_2 & B_{vo}U_1 + \mathcal{Q}_2 C_{yo}^T & H_w & \mathcal{Q}_2 C_{yo}^T \\ & D_{yvo}U_1 & D_{yw} & U_1 D_{yvo}^T \\ & & -\gamma_3\mathcal{I}_{n_w} & D_{yw}^T \\ * & & & -\gamma_3\mathcal{I}_{n_w} \end{bmatrix}, \\ G_1 &= [G_{11}\mathcal{Q}_2 \quad G_{12}U_1 \quad 0 \quad 0], \\ G_2 &= [G_{21} \quad G_{22} \quad 0 \quad G_{23}], \\ \mathcal{Q}_2 &= \begin{bmatrix} \mathcal{P}_3 & N_1 \\ N_1^T & M_1 \end{bmatrix}, G_{21}^T = \begin{bmatrix} 0 & B_{cl,\xi} \\ \mathcal{I}_{n_{aw}} & 0 \end{bmatrix}, \end{aligned} \tag{7.24}$$

$$G_{11} = \begin{bmatrix} 0 & \mathcal{I}_{n_{aw}} \\ 0 & 0 \end{bmatrix},$$

$$G_{22}^T = [0 \quad D_{y\xi,cl}],\, G_{23}^T = [0 \quad D_{y\xi,cl}],$$

$$A_o = \mathrm{diag}(A_{cl}, 0_{n_{aw}}),\, C_{yo} = [C_{cl} \quad 0],$$

$$B_{vo} = \begin{bmatrix} B_{cl,v} \\ 0 \end{bmatrix},\, H_w = \begin{bmatrix} H_{cl,w} \\ 0 \end{bmatrix},\, G_{12} = \begin{bmatrix} 0 \\ \mathcal{I}_{n_u} \end{bmatrix},$$

$$D_{yvo} = D_{y\xi,cl}, D_{yw} = D_{cl,yw}.$$

An important implication is that it is guaranteed to have a solution if matrix conditions are feasible, which is a convenient feature compared to static compensation. **Algorithm 7.7** mostly neglects the notion of the swift transition of the saturated system to the linear unconstrained scenario and as a result similar to **Algorithm 7.1** has a relatively poor post-saturating performance.

**Remark 7.4.** Although many efforts are dedicated to capturing the anti-windup synthesis procedure in the convex optimization problem of SDP such as [53] and [198], the performance of the anti-windup scheme is mostly compared with those of open-loop and unconstrained closed-loop systems. This is mostly because of formulating the problem in $\mathbb{L}_2$ stability of transfer function from disturbance to output subjected to internal stability and well-posedness of the nominal closed-loop system. However, in light of dealing with transient behavior of the saturated system after the windup event, an additional performance index over $\mathcal{T}(s)\colon y_c \to y_d$ in terms of Figure 7.6 in analyzing tool based on separation of windup modes is necessary. This purely deals with saturated system independent of open loop/unconstrained closed-loop performance.

**Algorithm 7.8** Adegbege and Heath very similar to [54] added a nonsingular directionality compensator ($E_2 \in \mathbb{R}^{n_u \times n_u}$) as an extra degree of freedom into the framework of Figure 7.1 [204]. In such a view, $\hat{u}$ is incorporated into a constrained optimization problem to minimize the directionality index: $E_2(u - \hat{u})$. This is formulated as (7.25) for $\mathcal{H}_3 = E_2^T E_2 > 0$.

$$\text{QP}(\hat{u}) = \arg\min_{\hat{u}} \frac{1}{2}(\hat{u} - u)\mathcal{H}_3\hat{u} - \hat{u}^T\mathcal{H}_3 u,$$
$$\text{subject to: } \mathfrak{L}\hat{u} \preccurlyeq b, \qquad (7.25)$$

QP($\hat{u}$) has a unique solution for $\mathcal{H}_3 > 0$ and obviously the case $E_2 = \mathcal{I}$ is already addressed. Based on the sector nonlinearity assumption of Case 2, define $\phi(\bar{u}) = E_2\Psi(u)$ with $\bar{u} = E_2 u$. Then, the conditioning technique based on directionality compensation can be converted from Figure 7.1 into Figure 7.6 with $\widetilde{M} = ss(A + BF_4, BE_2^{-1}, F_4, E_2^{-1})$ and $\widetilde{N} = ss(A + BF_4, BE_2^{-1}, C + DF_4, DE_2^{-1})$. $F_4$, and $E_2$ are not free design variables and should be assigned by the control engineer [214].

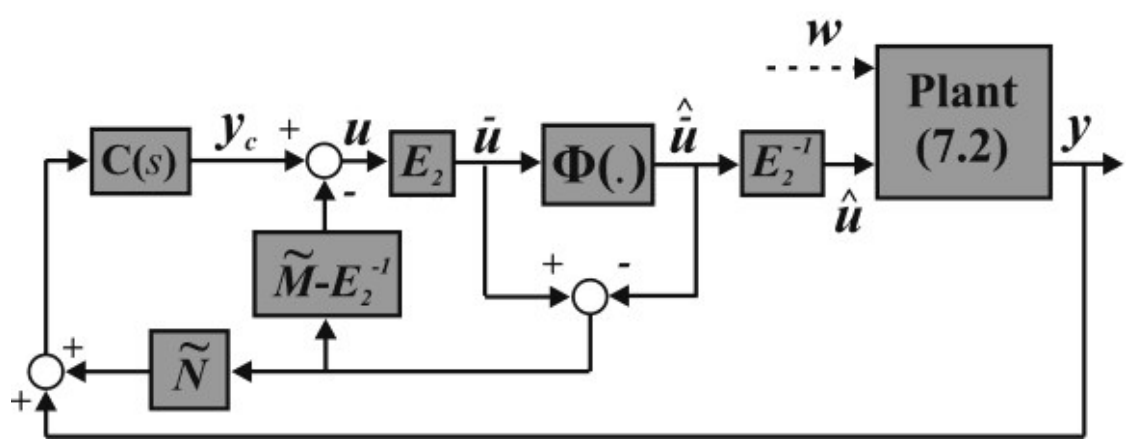

Figure 7.8 Equivalent representation of Figure 7.3 for conditioning based on directionality compensation [176]

In this Regards, two most common choices are $E_2 = \lim_{s\to\infty} \mathbf{C}(s)^{-1}$ also known as "optimization-based conditioning" technique [205] and "Steady-state directionality" [215] both dedicated to decoupling plant's out-put from saturated control law in transient or steady-state. Then, for fixed $E_2$, $T = \mathcal{H}_3^{-1}$, if there exist solutions for $Q_7 = Q_7^T > 0$, $\gamma_7 > 0$, diagonal $U_4$, and matrix $X_4$ such that LMI (7.26) is satisfied, then the plant-order anti-windup in Figure 7.8 is formulated by $F_4 = X_4 Q_7^{-1}$.

$$\mathfrak{I}_3 = \begin{bmatrix} \mathfrak{I}_{311} & BU_4T - X_4^T & 0 & Q_7C^T + X_4^T D^T \\ & -U_4T - TU_4 & \mathcal{I} & TU_4D^T \\ & & -\gamma_7\mathcal{I} & 0 \\ * & & & -\gamma_7\mathcal{I} \end{bmatrix} < 0, \tag{7.26}$$

$$\mathfrak{I}_{311} = AQ_7 + Q_7A^T + BX_4 + X_4^T B^T.$$

In [57], the static anti-windup gain matrix in **Algorithm 7.3** is calculated as a special case of **Algorithm 7.7**.

Gelani et al. resourcefully categorized the anti-windup compensation as direct linear anti-windup (DLAW) and MRAW schemes. In fact, DLAW in the most general form of output regulation problem refers to the linear filter in (7.20) of **Algorithm 7.7** (and correspondingly of course **Algorithm 7.3** and **Algorithm 7.4**) and main contributions in this area are developed in SDP framework only for global stability. The geometrical interpretation of DLAW is explained in [57]. MRAW, on the other hand, follows a dissimilar paradigm in which the anti-windup compensator's dynamics mimics the unconstraint plant similar to **Algorithm 7.1**. In a more general formulation, (7.27) represents the dynamics of MRAW filter.

$$\begin{aligned} \dot{x}_c &= A_c x_c + B_c u_c + \xi_1, \\ y_c &= C_c x_c + D_c u_c + \xi_2, \\ \dot{x}_{aw} &= Ax_{aw} + B(y_c - \hat{u} - \xi_2), \\ \xi_1 &= B_c\big(Cx_{aw} + D(y_c - \hat{u} - \xi_2)\big). \end{aligned} \tag{7.27}$$

Here, $\xi_2$ refers to the degree of freedom to be exploited in the synthesis process as in **Algorithm 7.4**. Note that $\xi_2 = 0$ refers to IMC-based model recovery scheme proposed in [216] for exponentially stable systems. Linear treatment of $\xi_2 = -\rho_1 B^T \mathcal{P}_5 x_{aw}$, where $\rho_1 \in \mathbb{R}^{n_u \times n_u} > 0$ is a diagonal matrix and $\mathcal{P}_5 = \mathcal{P}_5^T \in \mathbb{R}^{n \times n} > 0$ is the solution of LMI: $A^T\mathcal{P}_5 + \mathcal{P}_5 A < 0$ despite its simplicity categorized as an effective method [183]. The disadvantage of MRAW in AVC of lightly damped structures is obvious. However, as an advantage of this line of anti-windup schemes is the fact that the stability of the scheme guarantees driving $x_{aw}$ back to origin, original plant's states are forced to recover automatically. This feature is of course achievable only in case that the plant's model is close enough to real system especially in the nominal frequency range of system where the disturbance signal is active. Moreover, due to flexibility of generating $\xi_2$, unlike DLAW, nonlinear compensation technique is also an option [183], [193]. Since the architecture of MRAW is independent of the controller dynamics, stability and performance of the compensated saturated system are guaranteed even with alternations on the linear control system.

### 7.1.4 Analyzing tools for static and dynamic anti-windup compensators

#### 7.1.4.1 Static case [194]:

The anti-windup compensator together with sector nonlinearity of Case 2, $\widehat{\mathbf{P}}(s)$, and $\widehat{\mathbf{C}}(s)$ is $\mathbb{L}_2$ stable if for $P_1 = P_1^T > 0$ with appropriate dimensions, for diagonal positive definite matrix $X \in \mathbb{R}^{n_u \times n_u}$ and diagonal positive semi-definite matrix $W \in \mathbb{R}^{n_u \times n_u}$, and for scalar $\delta_1 \in \mathbb{R} > 0$, the following symmetric LMI is satisfied

$$\begin{bmatrix} \tilde{A}^T P_1 + P_1\tilde{A} & P_1\tilde{B} - \tilde{C}^T X - \tilde{A}^T\tilde{C}^T W \\ * & \delta_1\mathcal{I}_{n_u} - W\tilde{C}\tilde{B} - \tilde{B}^T\tilde{C}^T W - X\tilde{D} - \tilde{D}^T X - 2X \end{bmatrix} \le 0, \tag{7.28}$$

where $M_{11} = ss(\tilde{A}, \tilde{B}, \tilde{C}, \tilde{D})$ is the state-space realization of $(V(s) - \mathcal{J}_{n_u})P_{u_m\hat{u}}(s) - U(s)P_{y\hat{u}}(s)$. In terms of the multipliers passivity theorem for the stability of the transfer function from $\xi$ to $u$ in absence of disturbance signal, (7.20) is closely related to (7.8). In view of (7.28), obtained from the multiple circle criterion, quadratic disturbance rejection is a BMI for static anti-windup synthesis. More sophisticated interpretations of (7.28) in frequency-domain are expedient for understanding the backbone of formulizing static case. However, this is out of the scope of this dissertation.

#### 7.1.4.2 Dynamic case [198]:

Grimm et al. [198] captured the quadratic stability that entails the $\mathbb{L}_2$ constraint over the transfer function from disturbance to output of the conditioned system for internal stability of the interconnection in Figure 7.7 under the well-posedness of (7.1), (7.3), and (7.20) for Case 2. Similar to **Algorithm 7.3**, this can be recast in terms of Figure 7.4. The internal stability of the system in Figure 7.7 should be defined over the augmented system similar to (7.7) but for the dynamic case ($n_{aw} \neq 0$). A unique representation of such a system in state-space form is shown for $x_{au} = [x^T \quad x_c^T \quad x_{aw}^T]^T \in \mathbb{R}^{n+n_c+n_{aw}}$

$$\begin{aligned} \dot{x}_{au} &= A_{au}x_{au} + B_{au}(y_c - \hat{u}) + H_{au}w, \\ y_c &= C_{c,au}x_{au} + D_{c,au}(y_c - \hat{u}) + D_{uw,au}w, \\ y &= C_{au}x_{au} + D_{au}(y_c - \hat{u}) + D_{yw,au}w, \end{aligned} \tag{7.29}$$

where matrices $A_{au}$, $B_{au}$, $H_{au}$, $C_{c,au}$, $D_{c,au}$, $D_{uw,au}$, $C_{au}$, $D_{au}$, and $D_{yw,au}$ with appropriate dimensions are the interconnecting variables that may be constructed after synthesizing (7.28). Then, the achievable quadratic performance level $\gamma_2 > 0$ is calculated by solving LMI problem (7.30) for $\mathcal{P}_2 = \mathcal{P}_2^T > 0$ and $\delta_3 > 0$ with $U_1 = \delta_3\mathcal{J}$ (for saturation in sector nonlinearity) [198].

$$\begin{gathered} \inf_{\forall \mathcal{P}_2, \gamma_2, \delta_3 : \Omega_1 < 0,} \gamma_2, \\ \Omega_1 = \begin{bmatrix} \mathcal{P}_2 A_{au}^T + A_{au}\mathcal{P}_2 & B_{au}U_1 + \mathcal{P}_2 C_{c,au}^T & H_{au} & \mathcal{P}_2 C_{au}^T \\ & D_{c,au}U_1 + U_1 D_{c,au}^T - 2U_1 & D_{uw,au} & U_1 D_{au}^T \\ & & -\gamma_2\mathcal{J} & D_{yw,au} \\ * & & & -\gamma_2\mathcal{J} \end{bmatrix}, \end{gathered} \tag{7.30}$$

### 7.1.5 Implementation outcomes: Numerical and experimental results

In this section, the experimental rigs (schematically shown in Figure 7.9) that are involved in monitoring the performance of the closed-loop system in real-time based on the benchmark problem are briefly introduced. The controller in form of (7.3) is obtained using a classical linear quadratic Gaussian (LQG) output regulator. We intentionally selected the standard LQG for the nominal control so that the main goal of the study, namely unification and comparison of the second step in the two-step paradigm is not overshadowed by nominal control design's complications. For the sake of brevity technical details of the LQG synthesis are referred to in [217]. In order to design the static/dynamic anti-windup compensator, the obtained control system and the nominal plant in section 7.1.2 are iteratively introduced to each of the algorithms in section 7.1.3, separately. Then, the compensated control system is implemented on the experimental setup in Figure 7.9. In this regard, after constructing the linear controller and anti-windup compensator based on each algorithm in SIMULINK, the model is compiled into dSPACE RTI framework. In order to show the performance of the unconstrained control system, two sets of analyses are carried out in time-domain and frequency-domain. For this purpose, first, the system is excited through the disturbance channel with a sweep sine signal.

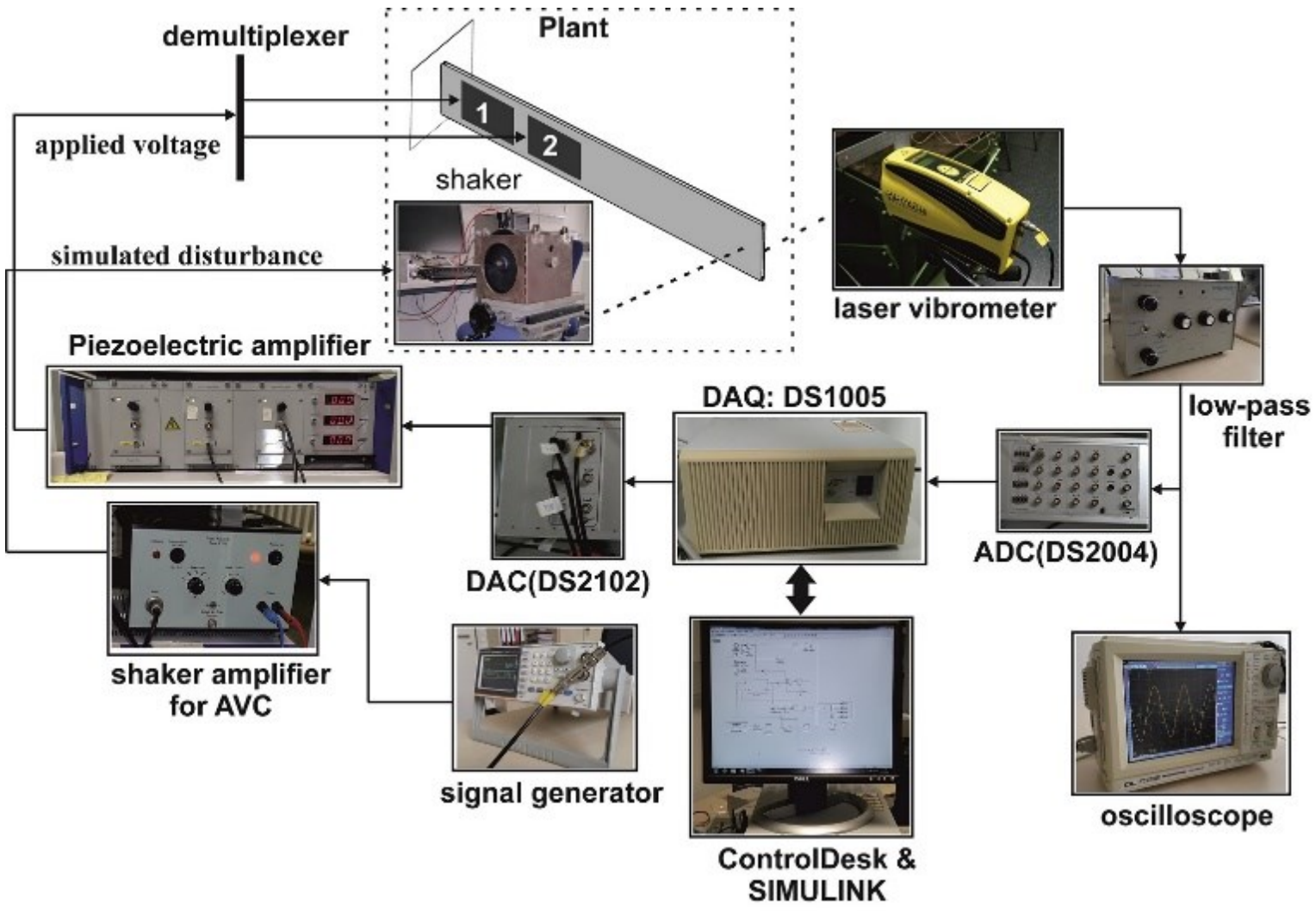

Figure 7.9 Experimental rig for ADC performance evaluation (modified version of Figure 6.1)

Meanwhile, Table 7.1 contains the summary of all algorithms.

Table 7.1 Reviewed algorithm's overview [176]

| Algorithm | Performance | | Stability guarantees | Feasibility guarantees | Mode II optimization | Mode III optimization | Tuning |
|---|---|---|---|---|---|---|---|
| | User-defined | Linear recovery | | | | | |
| Algorithm 7.1 | | √ | Global | Yes | - | - | Eigen values of $A^T P + PA$ |
| Algorithm 7.2 | | √ | Global | Yes | Linear recovery optimized in LQ sense | - | LQ weights |
| Algorithm 7.3 | √ | | Global if feasible | No | User-defined performance optimized | Static | Mapping |
| Algorithm 7.4 | | √ | Global if feasible | No | Linear recovery $\mathbb{L}_2$-optimized | Given by poles | Poles of filter+Mapping |
| Algorithm 7.5 | | √ | Global | Yes | Disturbance rejection $\mathbb{L}_2$-optimized | - | Two-step |
| Algorithm 7.6 | √ | | Global | Yes | User-defined performance optimized | - | Mapping+SP |
| Algorithm 7.7 | | √ | Global | Yes | Linear recovery $\mathbb{L}_2$-optimized+directionality | - | Steady-state directionality+Mapping |

- **Architecture**: This field characterizes whether the anti-windup architecture is user-defined or linear model recovery.
- **Stability guarantees**: This field assumes the plant is asymptotically stable. If the plant has right half-plane poles, no method can provide global stability.
- **Feasibility**: The existence of solution for the formulated SDP of the algorithm.
- **Mode III optimization**: This field records any noteworthy manner in which an anti-windup compensator may have in optimizing the mode III behavior of the system. Static anti-windup compensators immediately are absent from the loop after saturation has cease. Otherwise, Mode III behavior is heavily influenced by the dynamics of the anti-windup compensator.
- **Tuning**: This field captures the parameters used to adjust the anti-windup compensator performance. "Mapping" indicates that the performance is governed by the choice of the closed-loop mapping. Moreover, "two-step" refers to satisfying two sequential LMI constraints in which the result of the first SDP should be used in optimization of the second one. Finally, "SP" indicates that the strictly proper compensator is achievable by presetting the feedthrough term.

The frequency of the excitation signal is transformed from zero to 230 Hz in two cases of the open-loop and closed-loop system with a sampling frequency of 10 kHz. The response of the system for the controlled and uncontrolled cases is shown in Figure 7.10a in time-domain based on the measurement signal generated from LDV, which is further fed to the dSPACE ADC board (DS2004). Accordingly, the applied control signals on each of the piezo-actuators as the output of the PI-E 500 amplifier are depicted in Figure 7.10b.

**(a)** **(b)**

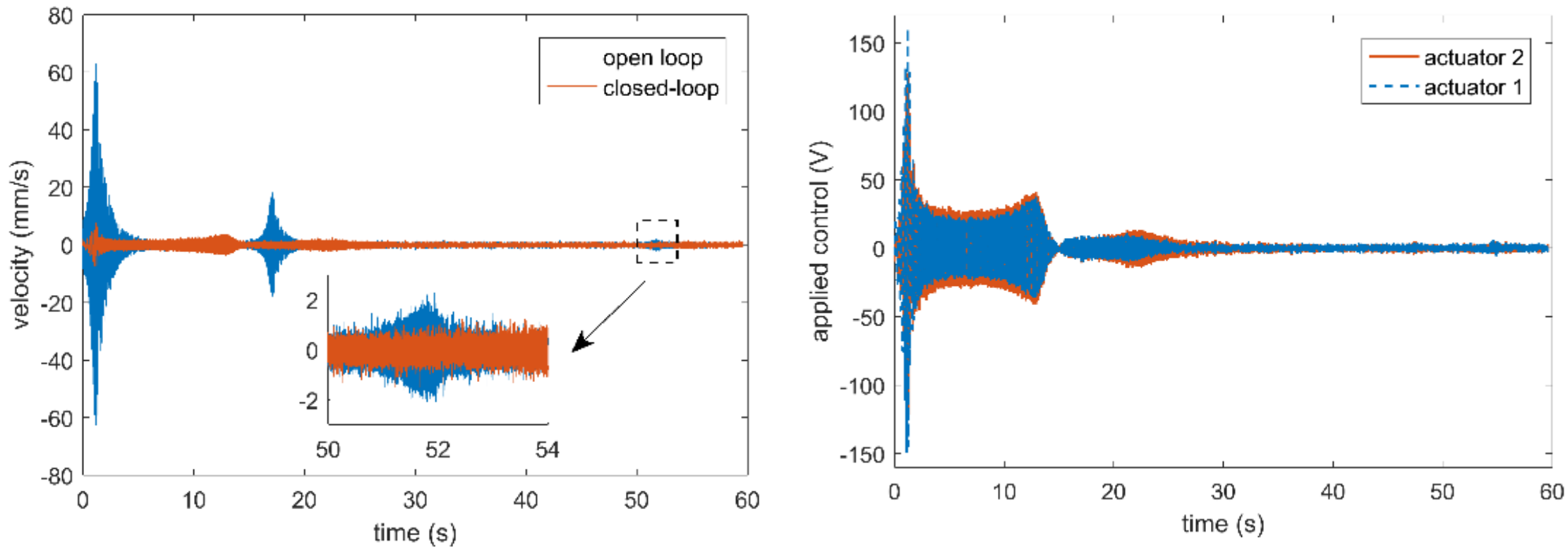

Figure 7.10 (a) Comparison of open-loop and unconstrained closed-loop system. (2) Applied control effort [176]

In order to appropriately capture the performance of the unconstrained control system in suppressing the vibration amplitude, the frequency response function (FRF) of the open-loop system is compared with the closed-loop one in the frequency range of [0 800] Hz in Figure 7.11. The disturbance channel is controlled by Pulse Labshop toolbox while the observer-based controller channel (two patches) is activated per dSPACE system. The DAQ system for controller implementation is triggered prior to the Pulse modal experimental toolbox. Similarly to the time-domain case, the level of the signal generator is intentionally kept within the linear range of actuators. Moreover, as signified by a gray dotted line in Figure 7.11a and Figure 7.11b, it is observed that for saturated actuator which is obtained by increasing the disturbance level, the controller performance degradation can significantly change. Although such an observation is trivial, based on the coherence of the measurement shown in Figure 7.11b, the quality of data in the modal analyses is low because of the nonlinearity involved. Therefore, the comparison of the FRFs of the saturated systems for compensated closed-loop configurations is suppressed in the rest of the section.

**(a)** **(b)**

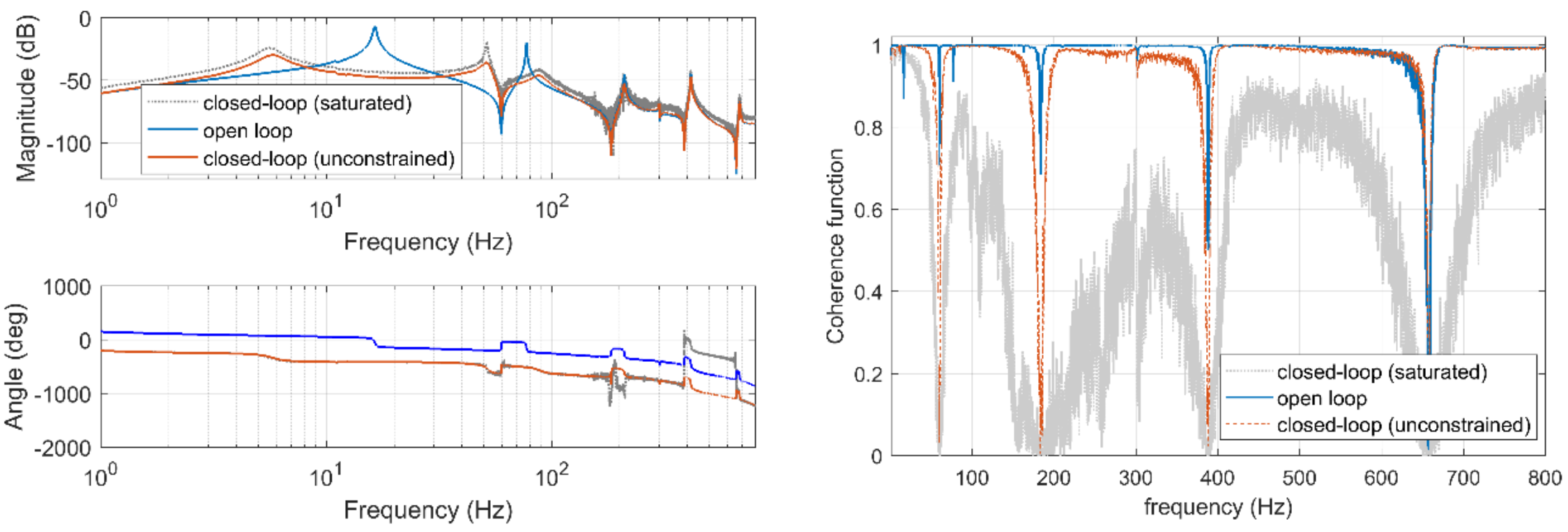

Figure 7.11 (a) Comparison of FRFs of the linear open-loop system, unconstrained closed-loop system, and saturated closed-loop system. (b) Coherence diagram for quality assessment of data in modal analysis [176]

Next, the implementation of the anti-windup scheme using **Algorithm 7.1** is carried out. Accordingly, the appropriate solution for $P$ matrix in Lyapunov equation $A^T P + PA < 0$ is nontrivial. Based on the trial and error, proper choice is obtained by employing `lyap` function in MATLAB such that $A^T P + PA < 5 \times 10^2 C^T C$. For structural control, a systematic optimization of Lyapunov equation for best performance of anti-windup compensator in real-time implementation of saturated system can be carried out by using Hardware-in-Loop (HiL) technique similar to [218]. In order to evaluate the performance of anti-windup system during and after saturation event, the system is excited with a pulse-like signal with the amplitude of 1 and period of 5 sec and signal width of 1 %. The results of the experimental implementation of the designed compensator on the saturated closed-loop is presented in Figure 7.12a while the applied control signals are presented in Figure 7.12b.

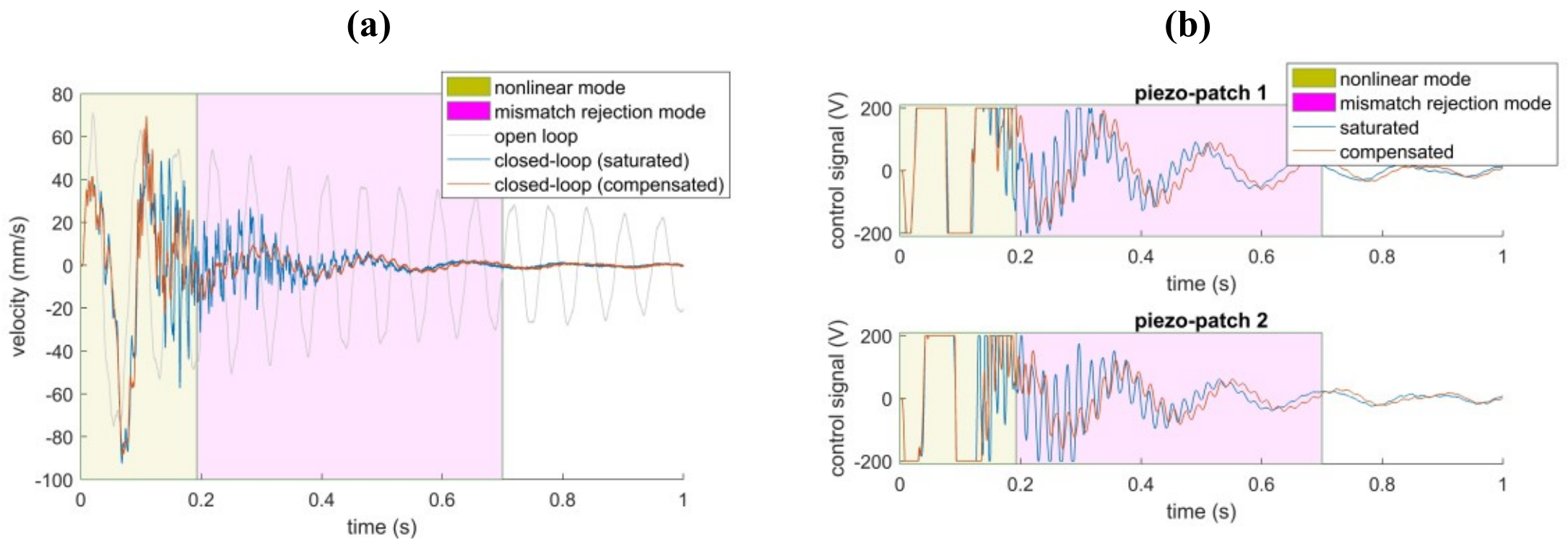

Figure 7.12 (a) Comparison of the system responses based on **Algorithm 7.1**. (b) Control effort on piezo-actuators during each phase in **Algorithm 7.1** [176]

Based on the applied control effort, the performance of the system in both saturated and compensated cases can be divided in two areas: (1) during saturation. (2) after saturation. In order to bring out clear illustrations, only a short time-window is presented for all of the results in the rest of the chapter. Additionally all of the methods during the nonlinear mode, and output mismatch mode are compared later in Figure 7.19 and Figure 7.20.

Following **Algorithm 7.2**, the conditioning anti-windup compensator is synthesized on systems (7.1) and (7.3). In terms of tuning the compensator, a clear connection between the free variables, $\mathcal{Q}_{p3}$ and $\mathcal{P}_{p5}$, and the mathematical model is nontrivial. In practical problems where the nominal model is obtained from the identification method, it is not straightforward to tune the design parameters. Trial and error method in tuning $\mathcal{Q}_{p3}$ and $\mathcal{P}_{p5}$ cannot be carried out because of the large number of parameters involved. Then again, the free parameters are defined as variables for SDP problem after the suppression of the inversion operator. It is observed that the optimal point in SDP problem based on "MOSEK" solver in YALMIP needs to be further retuned. Accordingly, additional constraints on positive definite matrices involved in the optimization process are defined as $Q_4 \preccurlyeq 50$ and $U_3 \succcurlyeq 2000$. In the default setting of the method due to the large values obtained for $Q_4$, $K_3 = X_2 Q_4^{-1}$ is obtained to be negligible. On the other hand, small values on $U_3$ result in chattering because $L_2 = X_3 U_3^{-1}$ is defined over $(\Psi(y_c + \xi_1) - y_c)$ which should be avoided. Similar to Figure 7.12, the results of implementing LQ MRAW based on [193] are presented in Figure 7.13a and Figure 7.13b. It should be reported that in simulation results for **Algorithm 7.2**, (suppressed for the sake of brevity) better performance in nonlinear mode phase can be obtained for the tradeoff of chattering. However, due to the excitation of higher-order mode shapes (spillover effect), such results are not desired in real-time tests.

Static anti-windup based on the multiple-loop circle criterion in [53] is investigated next. Following the **Algorithm 7.3**, some fine tunings are applied in addition to LMI in (7.8) subjected to constraints on the design variables. The solution for the feasibility problem in the default formulation has shown poor performance in recovering from the windup state. Therefore, based on the equivalent Lyapunov equation representing LMI (7.8), (7.10) in [53], variable $\gamma_1 = 1$ is fixed to adequately define the quadratic disturbance rejection leverage by minimizing $\gamma_1^*$. However, defining a convex minimization over $\gamma_1^*$ results in large $\Upsilon$ and chattering problem. Instead, two additional constraints are added to the feasibility problem: $\gamma_1^* \preccurlyeq 2$ and $\delta_2 \preccurlyeq 100$. The upper limits are obtained by trial and error while noting that based on multiplier theory for asymptotic stability of the augmented system in Figure 7.4b, similar to [194], minimization of $\delta_2$ results in a larger stability margin with a tradeoff being the larger values for $\Upsilon$ and requirement of high sampling frequency in the real-time implementation. Accordingly, it is observed that with the processing power available using DAQ system in Figure 7.9, and for the sampling rate of 10 kHz, $\delta_2 = 55.71$ and $\gamma_1^* = 1.32$ are realizable.

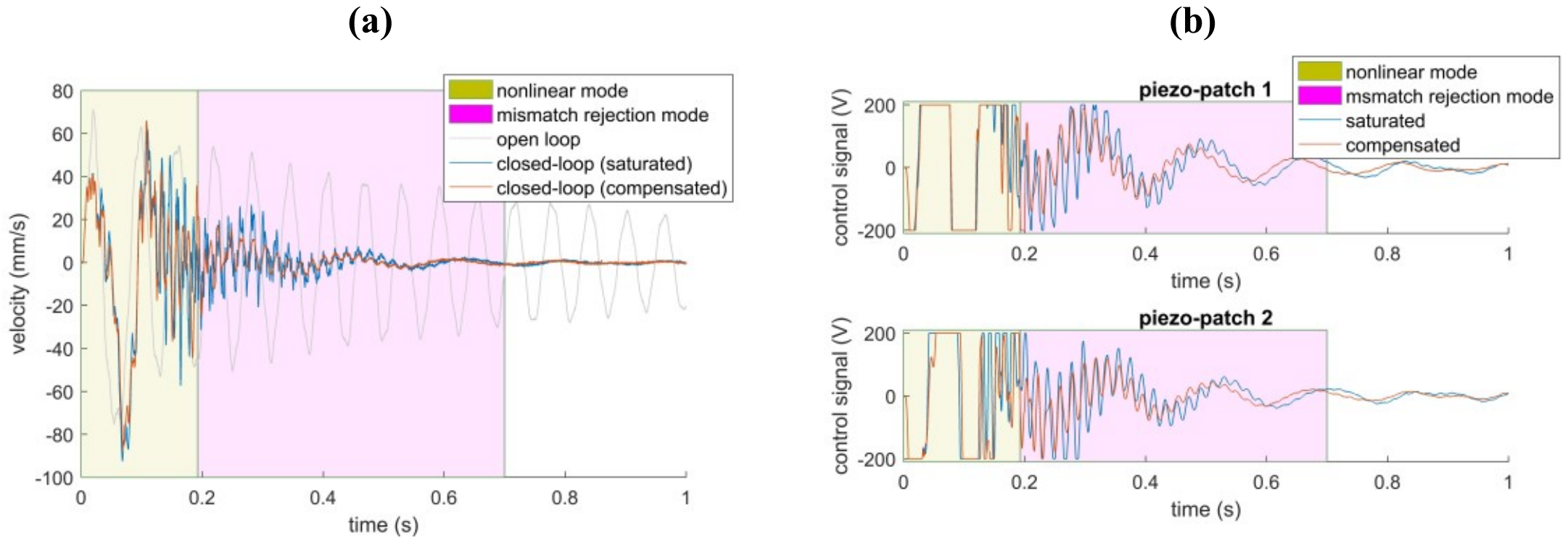

Figure 7.13 (a) Comparison of the system responses based on **Algorithm 7.2**. (b) Control effort during each phase on piezo-actuators in **Algorithm 7.2** [176]

Then, the obtained compensator is implemented on the setup in Figure 7.9 and the comparison of the system responses in time-domain as well as the control signals are presented in Figure 7.14a,b, respectively. In order to benefit from true goal of the anti-windup compensator while keeping a convex synthesis process, the mode decomposition based on [198] can be employed. Accordingly, following the discussion on the general idea shown in Figure 7.6, **Algorithm 7.4** is utilized. Three obvious advantages of such framework are: (1) the dynamic anti-windup compensator has a low order and is computationally efficient. (2) The algebraic loop associated with static anti-windup methods is completely taken care of with the applied filters on the conditioning signals. (3) The gain of transfer function in output-mismatch disturbance rejection is realizable by the control engineer.

Some additional remarks in real-time implementations on setup: (1) since the implementation of SDP problems such as eigenvalue problem in (7.11) is mostly performed in YALMIP or CVX in which strict inequalities are unachievable, it is crucial to add a negative diagonal matrix with appropriate dimensions to RHS of inequality (7.11) such as $\leq -10 \times \mathcal{J}$. (2) Although the performance of the system for the involved unknown variables may be optimized by **Remark 7.4** on $\gamma_4$, fine tuning of the solution is performed by controlling the diagonal matrix elements of $U_2 \succcurlyeq 10^4$. (3) No clear guidelines for the low-pass filters $F_1(s)$ and $F_2(s)$ are available, and therefore, first-order transfer functions are selected and tuned as shown in (7.31) for the best mismatch mode performance.

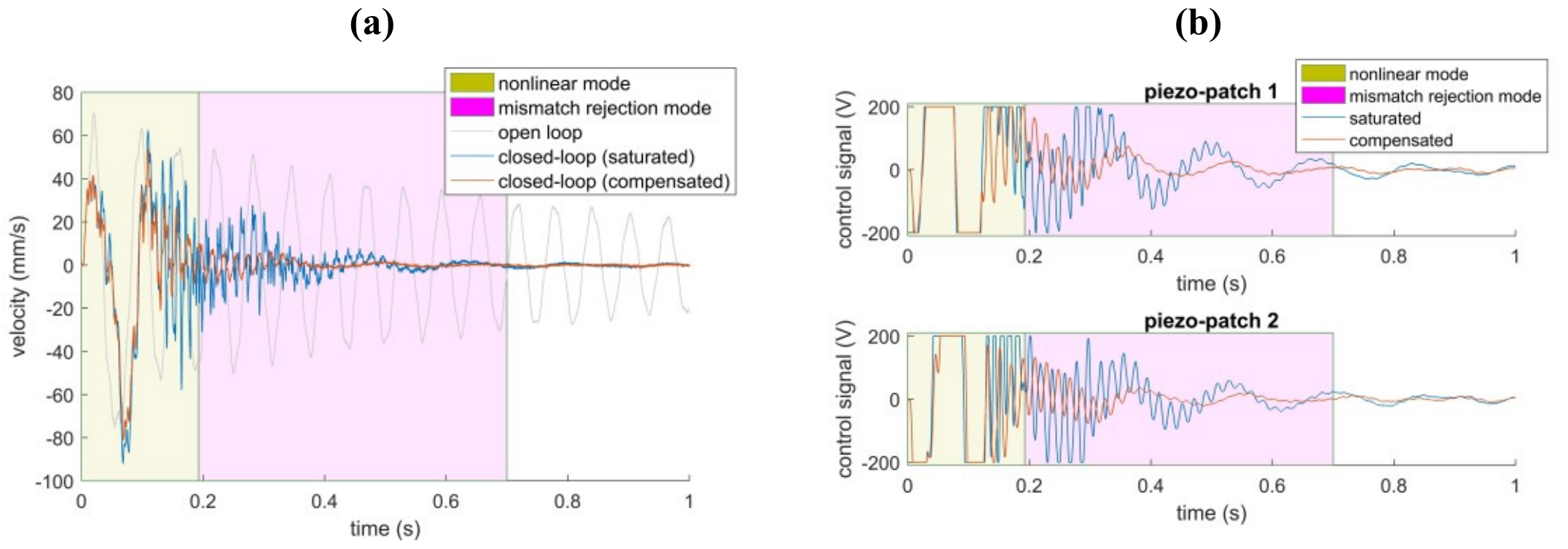

Figure 7.14 (a) Comparison of the system responses based on **Algorithm 7.3**. (b) Control effort during each phase on piezo-actuators in **Algorithm 7.3** [176]

Note that reducing the edge frequency of $F_1(s)$ to values below the natural frequencies of the two fundamental mode shapes and below the edge frequency of $F_2(s)$ deteriorates the energy dissipation in output mismatch rejection phase. It is also observed that selecting the edge frequency of higher values than the nominal frequency range does not contribute to the behavior of the anti-windup system.

$$F_1(s) = \frac{400\times 2\times\pi}{s+(400\times 2\times\pi)},\; F_2(s) = \frac{100\times 2\times\pi}{s+(100\times 2\times\pi)}. \tag{7.31}$$

The transient response of the system during and after windup incident are depicted in Figure 7.15a. Moreover, the conditioned output of the anti-windup compensated controller is compared to the saturated system on each of the control channels in Figure 7.15b.

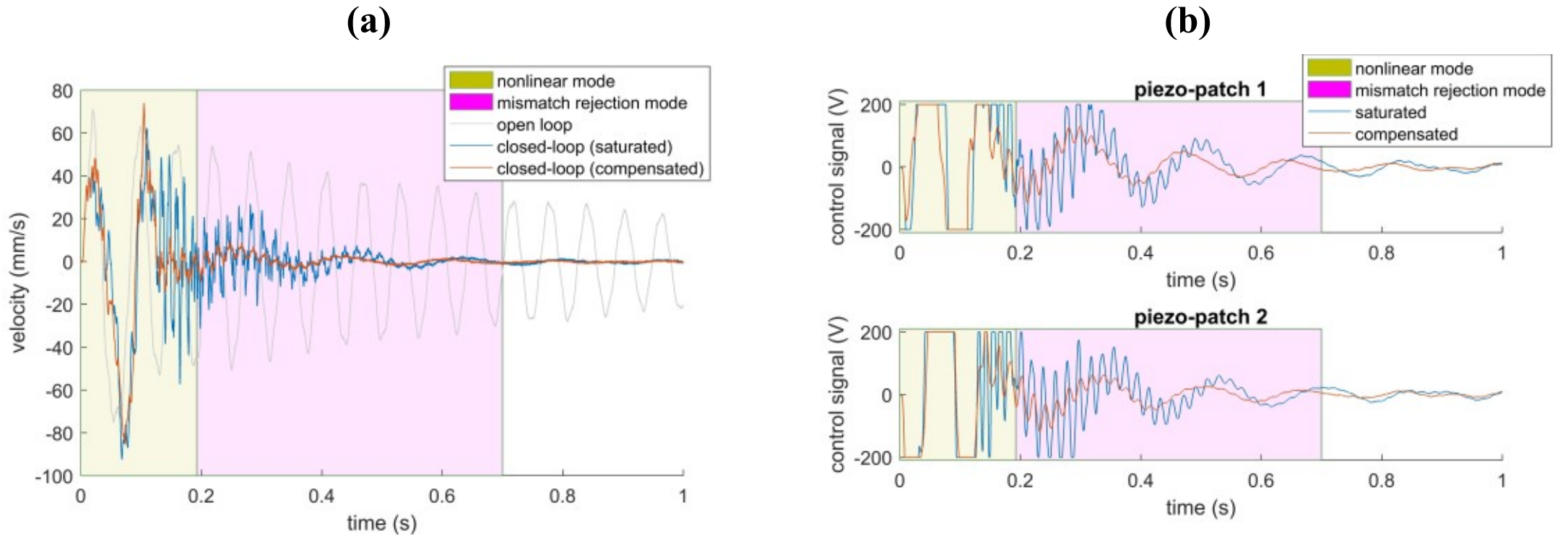

Figure 7.15 (a) Comparison of the system responses based on **Algorithm 7.4**. (b) Control effort during each phase on piezo-actuators in **Algorithm 7.4** [176]

Infeasibility of the second step is typical behavior of anti-windup algorithms that involve two steps in the synthesis process. In order to find a feasible solution for the second LMI, instead of solving the minimization problem over $\gamma_d$ for LMI constraints $\mathfrak{X}_1 < 0$ and $\mathfrak{X}_2 < 0$, $\gamma_d$ is limited also with an upper bound, and then is selected to be $\gamma_d = 3.84 \times 10^4$. Note that the first LMI, (7.13), is feasible for minimum $\gamma_d = 0.46$ using MOSEK solver in YALMIP. However, based on the obtained solutions for $\gamma_d$ and $\mathcal{P}_9$, $\mathfrak{B} \equiv \Pi + \mathcal{H}^T[F_3 \quad E_1]\mathcal{G} + (\mathcal{H}^T[F_3 \quad E_1]\mathcal{G})^T < 0$, becomes infeasible. Hence, first by ensuring the satisfaction of strict positive definiteness by $\mathcal{P}_9 \geq 0.001\mathcal{I}$, and then, by limiting the elements of $\mathcal{P}_9$ as $-1 \preccurlyeq \mathcal{P}_9 \preccurlyeq 1$, the SDP in (7.13) is solved. Next, $\mathfrak{B} < 0$ is satisfied for unknown matrices $F_3$, $E_1$, by assuming $\mathcal{W} = \mathcal{I}$ while reseting the *feasibility radius* such that the decision vector including all unknown parameters (here: $F_3$ and $E_1$) lies within the ball with radius of 2. This radius can be interpreted as the Euclidean norm of the decision variables which otherwise will result in large manipulation of the control signal and realizable output in conditioning systems. Moreover, better performance is achieved in simulation results that have chattering problems and such solutions are unacceptable in real-time implementations. The transient response of the open-loop system is compared with the saturated closed-loop and anti-windup compensated closed-loop system in Figure 7.16a. The control signal applied on each of the active elements is illustrated in sub-figures of Figure 7.16b.

Unfortunately, **Algorithm 7.6** based on [213] and summarized in (7.16) possesses a very poor closed-loop performance even in simulations without saturation limits. It should be noted that although the decomposition $\mathcal{Q}_6\mathcal{P}_8^T = \mathcal{I} - \mathcal{Q}_5\mathcal{P}_7$, is guaranteed to have a solution for nonsingular $\mathcal{I} - \mathcal{Q}_5\mathcal{P}_7$ [211], there is no unique approach (e.g. QR-factorization, LU-factorization, and rank-decomposition) that essentially guarantees the same performance even in an unconstrained controller. It is possible that even for a feasible solution with an inappropriate decomposition $A_c$ may become non-Hurwitz. Moreover, the *simultaneous synthesis* is a vague statement used in [213] for which, first, the unconstrained controller is constructed based on satisfying LMI conditions (7.17), where $\widehat{\Lambda}_i, i = 1,2$ are obtained. Then, a second matrix constraint is needed to be satisfied such that $\Upsilon_i, i = 1,2$ can be calculated. An alternative, as pointed out in the algorithm is to select $M_2$, manually without strict guidelines. The performance evaluation of this algorithm in simulation results is suppressed here.

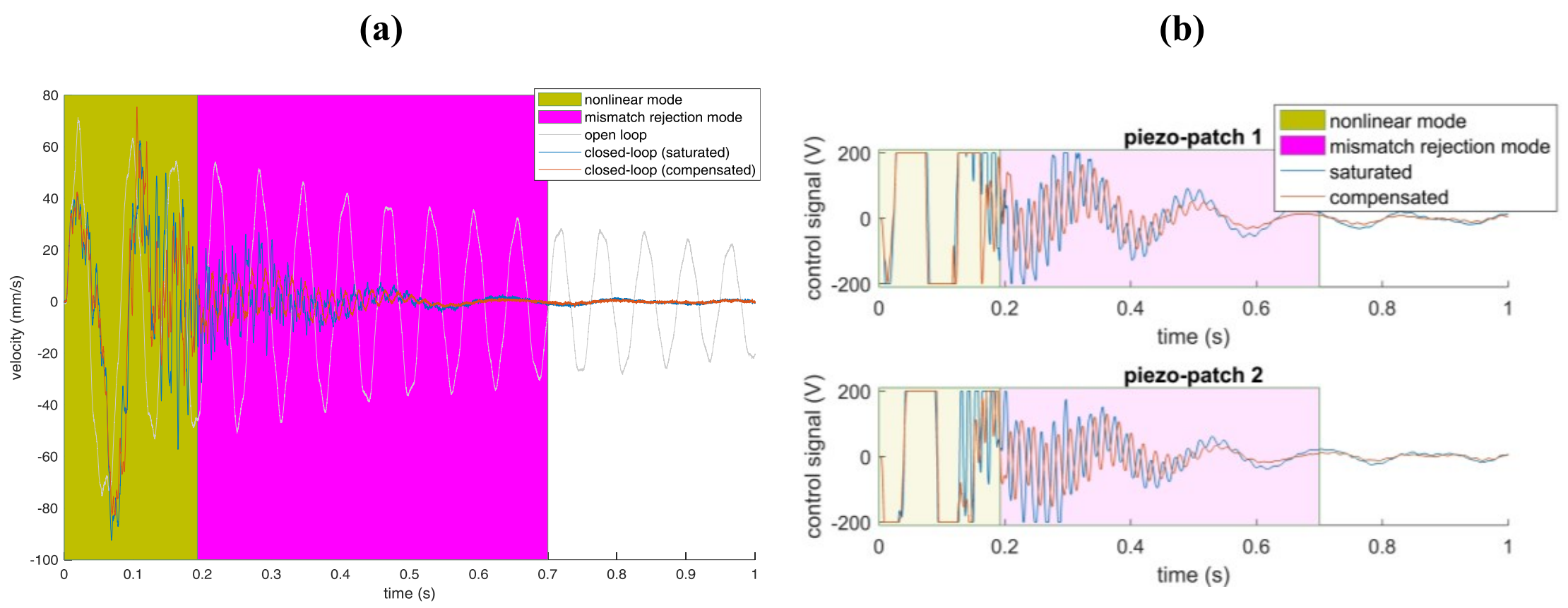

Figure 7.16 (a) Comparison of the system responses based on **Algorithm 7.5**. (b) Control effort during each phase on piezo-actuators in **Algorithm 7.5** [176]

Next, **Algorithm 7.7** is evaluated. Based on the numerical solution of LMI constraints in Eq.(7.21), $\mathcal{P}_3$ and $\mathcal{Q}_1$ are calculated and then based on the Cholesky decomposition on $\mathcal{P}_3\mathcal{Q}_1^{-1}\mathcal{P}_3 - \mathcal{P}_3$, $N_1$ is obtained. In the light of the full-order dynamic anti-windup compensator given in (7.20), LMI (7.23) is satisfied. For this purpose, a key tuning parameter is $\delta_3$ which is selected to be $10^4$ in the experimental implementations. Moreover, since the second LMI is solved independently, as long as the problem is feasible, a strictly proper anti-windup scheme is achievable by presetting the $D_{aw} = 0$. The algorithm is straightforward to implement and the response of the system in two cases of strictly proper and non-strictly proper is presented in Figure 7.17a. Consequently, the applied control signals are shown in Figure 7.17b.

The directionality compensation based on the work of Adegbege and Heath is studied next [204]. Accordingly, using the steady-state directionality based on [215], $E_2$ is calculated and then $T$ is obtained for the LMI (7.26). Note that for $n_u \neq n_y$ (as of the benchmark problem), the optimization-based conditioning technique following Peng et al. is not realizable for obvious reasons [205]. After calculating the conditioning matrix $F_4$ as the result of the feasibility solution of LMI (7.26), instead of implementing the representation of Figure 7.8, it is assumed that $E_2 = \mathcal{J}$.

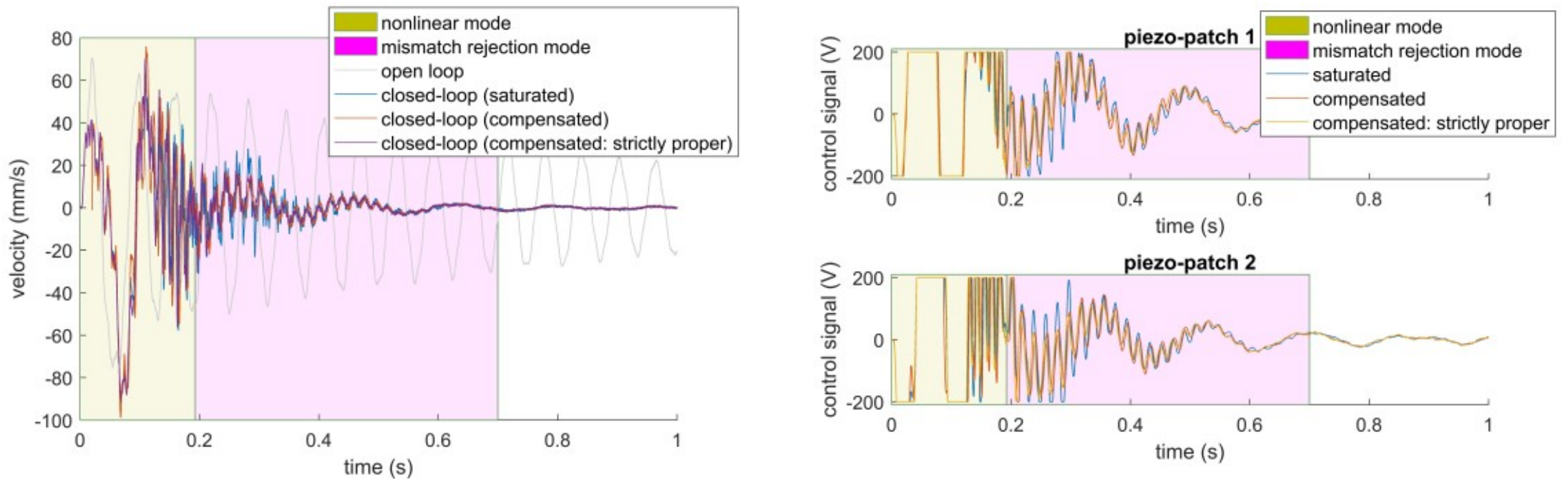

Figure 7.17 (a) Comparison of the system responses based on **Algorithm 7.7**. (b) Control effort during each phase on piezo-actuators in **Algorithm 7.7** [176]

Two points should be made: (1) for non-square systems, based on the decentralized control theory and decentralized nonlinearity involved, it is possible to design the anti-windup compensator on each channel and use $E_2 = \text{diag}(E_{2i}), i = 1, \dots, n_u$ where $E_{2i}$ is calculated on each of the control channels, separately. (2) If the $E_2$ is small enough to reduce the conditioned control signal much below the saturation level, it is observed that the performance of anti-windup system during the event is degraded. LMI (7.26) is a strict inequality which was realized in numerical implementation by enforcing $\mathfrak{I}_3 \leq -0.1\mathcal{J}$ and it is observed that increasing the stability margin leads to more oscillations during the nonlinear mode while having a better performance in the output mismatch rejection phase. Additional constraints on the elements of $U_4$ such as $U_4 \succcurlyeq 2 \times 10^{-4}$ improves the output mismatch rejection quality. Consequently, the implementation results are summarized in Figure 7.18.

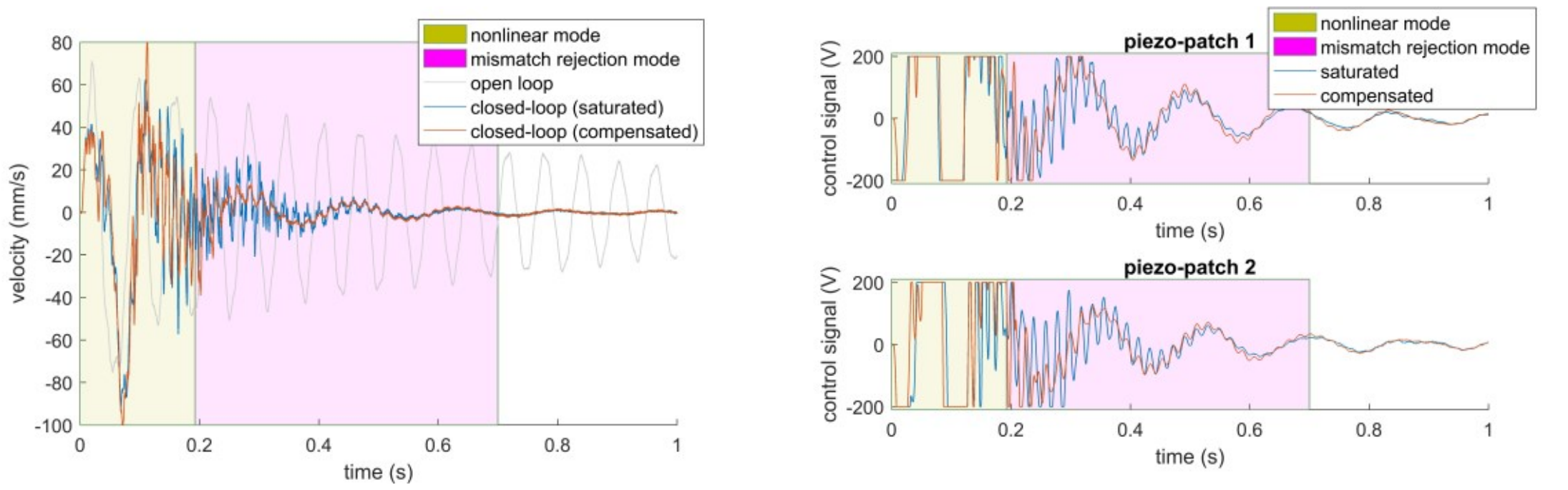

Figure 7.18 (a) Comparison of the system responses based on **Algorithm 7.8**. (b) Control effort during each phase on piezo-actuators in **Algorithm 7.8** [176]

The behavior of the system during the nonlinear mode for all of the algorithms should be compared to the response of the system with unconstrained controller implementation. Since the realization of the ideal controller in real-time is not achievable due to the saturation, the ideal response of the closed-loop system is obtained while suppressing the saturation block in SIMULINK. Then, the divergence of the captured response from this ideal trajectory in the real-time analyses based on the presented algorithms provides a view on anti-windup performance during the saturation event. However, because the response of the system includes a lot of oscillations, direct comparison of algorithms in a single figure is less informative. Therefore, convolution data smoothing with moving average using normalized Gaussian function is carried out on each set of data representing the difference between the ideal system response and the real-time one of the anti-windup-augmented systems. Accordingly, a normalized vector with ≈10 % of the data is used to smooth out the divergence during the nonlinear mode as shown in Figure 7.19.

It can be seen that **Algorithm 7.4** based on the reduced-order anti-windup scheme in [192] has the minimum divergence from the ideal response of the system followed by the static anti-windup method in [53]. Moreover, it is noted that **Algorithm 7.8** has the maximum divergence in comparison to all the implemented algorithms and the saturated case. One should note that a fair comparison between the algorithms is a critical demand that is a complex task for the methods that have lots of free user-defined variables or multiple steps in anti-windup synthesis. Moreover, it is observed that addition of LPFs on anti-windup mechanisms with chattering behavior can significantly improve their performance, however they were not reported in the default formulation of the referred papers and therefore neglected here. As it can be seen in the control signal for the saturated system in Figure 7.12b, 0.08 sec is the time at which the control signal swings from the upper bound to the lower bound where **Algorithm 7.8** based on [204], has the minimum performance degradation after **Algorithm 7.3** and **Algorithm 7.7**. Between **Algorithm 7.1** and **Algorithm 7.2**, the latter based on LQ-based MRAW in [193] has a better performance. Here, by aiming at proving clear illustration, the plots are limited only to one part of the nonlinear mode in an interpretation of windup event based on the work of Weston and Postlethwaite [191].

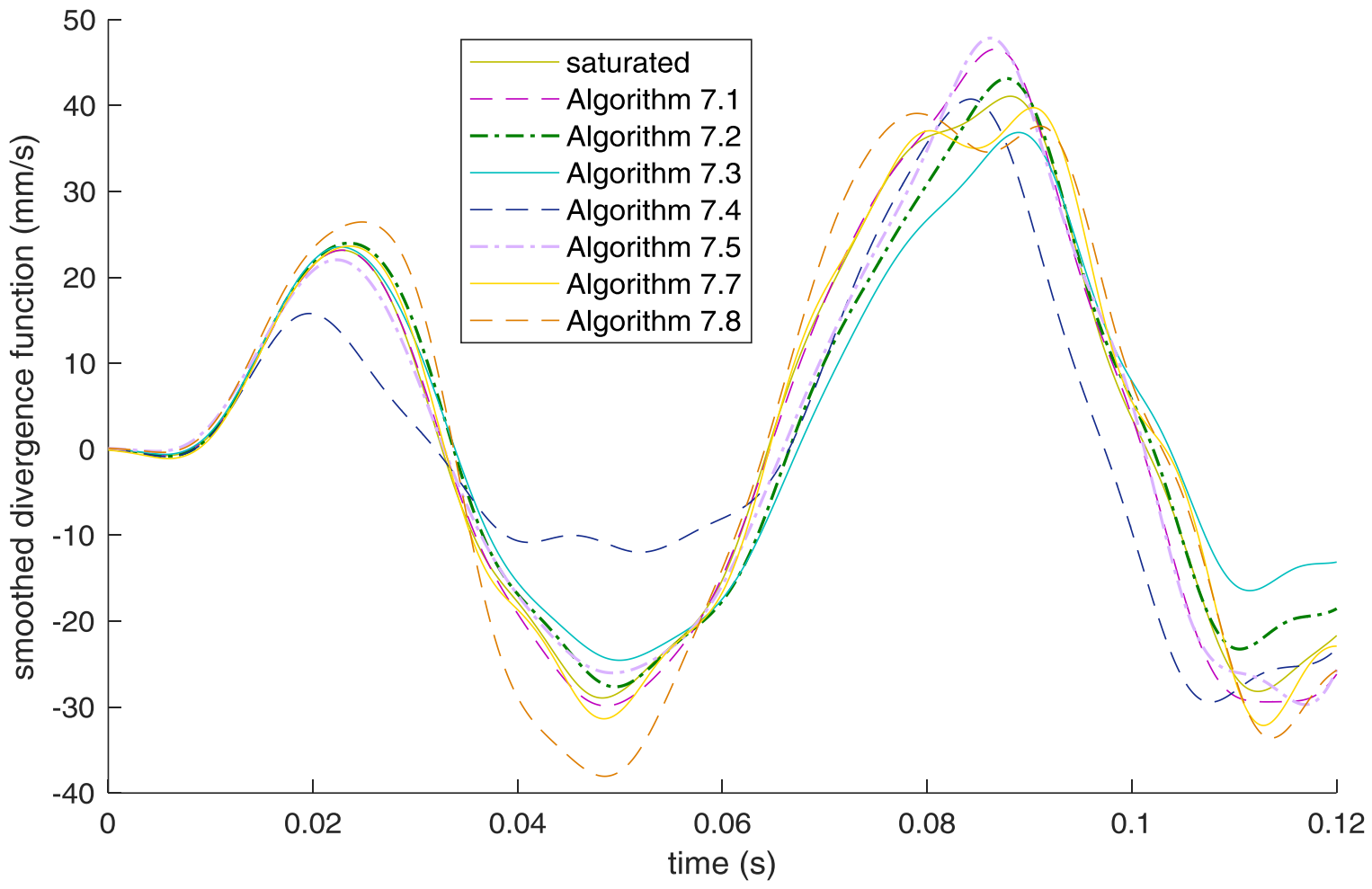

Figure 7.19 Comparison of the divergence of an anti-windup-augmented system from the ideal response during the nonlinear mode [176]

Next, the performance of algorithms in the post-nonlinear phase is investigated. In other words, the energy that is accumulated during the windup event needs to be damped before transmitting into the realizable output. Here in order to separate the nonlinear mode from disturbance rejection mode, the amount of damped energy is compared for all of the algorithms after 0.1926 sec till 0.7 sec during which the nonlinear mode is over, but system trajectory is still different from the linear case. Note that not all of the algorithms return to their linear behavior by 0.7 sec, however, for the sake of quality illustration, the output mismatch rejection is assumed to be ended at 0.7 sec. Since the benchmark problem is a continuous mechanical system, the exact energy of the system cannot be measured in real-time. Instead, assuming that the beam can be modeled as a single degree of freedom system and then integrating over the squared value of the measured velocity at the free end, the kinetic energy can be estimated. Based on the arguments in Figure 7.6, the rate of change of this energy index reflects the performance of the AW compensator in damping the accumulated energy during the nonlinear mode (in the filter $S\bar{P}$). Moreover, in order to solely consider the attenuated energy, all of the algorithms are assumed to have zero initial condition at 0.1926 sec. In this regards, as shown in Figure 7.20, the following can be observed:

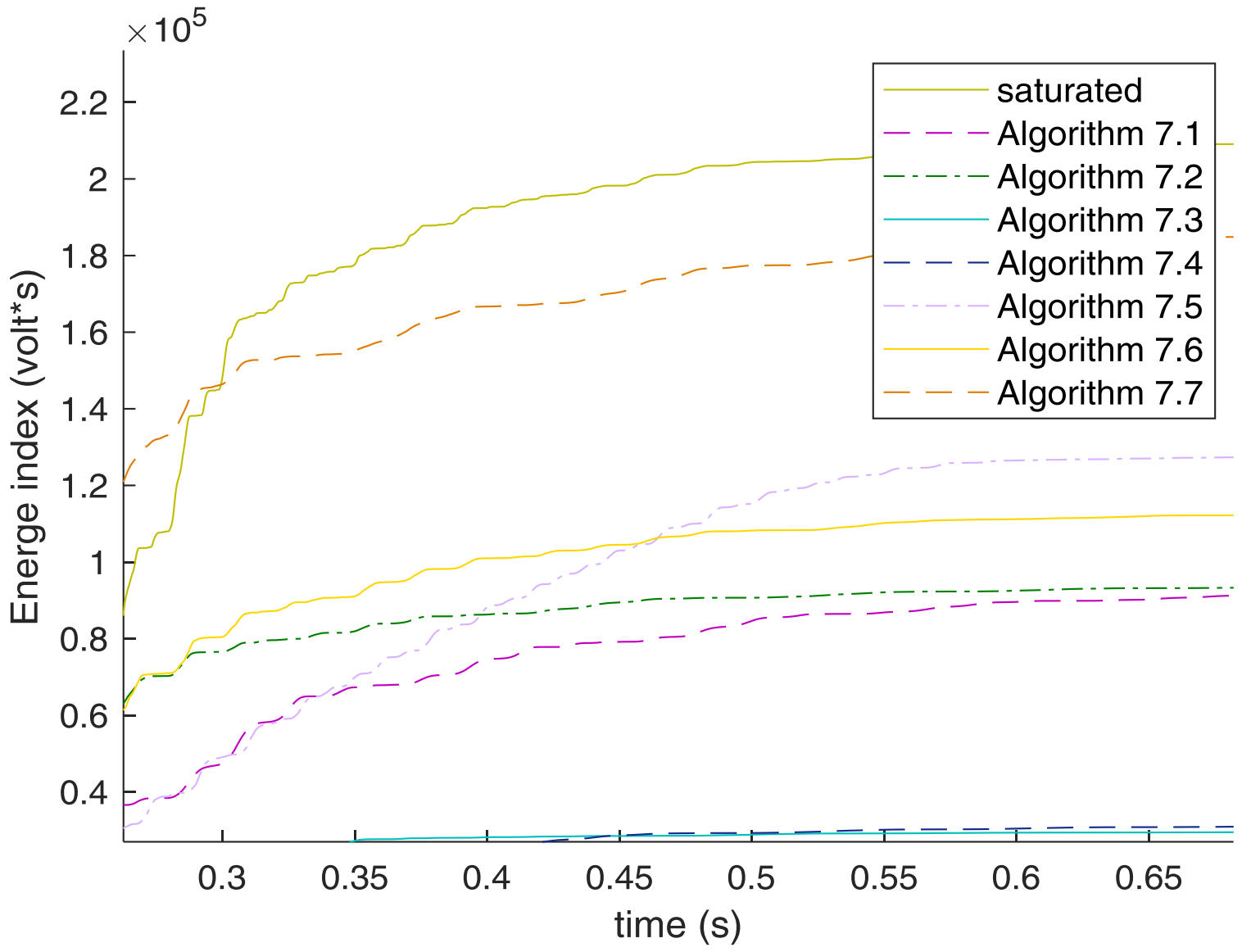

Figure 7.20 Energy index in different algorithms during the mismatch rejection mode [176]

(1) **Algorithm 7.3** and **Algorithm 7.4** keep the minimum energy content during the output mismatch rejection mode. Note that small changes in energy index of some Algorithms such as **Algorithm 7.3** and **Algorithm 7.4** are due to the fact that they fall back into the linear mode faster than other i.e. have better rejection performance.

(2) Model-recovery anti-windup schemes in **Algorithm 7.1** and **Algorithm 7.2** have acceptable energy dissipation resulting in a fast transition of the saturated system trajectory to linear mode while having a simple synthesis and implementation process.

(3) **Algorithm 7.7** has a satisfactory energy level while having a fast transition from the nonlinear to the linear mode.

(4) Perhaps the most critical method in this analysis is **Algorithm 7.5** which is explicitly derived for output mismatch rejection. It can be observed that although in the first half of Figure 7.20, the energy index is kept small but the method requires a long-time transition. At this point, we should clarify that it may be possible to retune the method to achieve a better suppression level, however, for the sake of fair comparison between all of the methods, any major modification of the method is suppressed.

(5) Unsurprisingly, the saturated case without any compensation has the worst output mismatch rejection quality.

# 8 Dealing with the fragility of the system

In the previous two chapters, the transient source of closed-loop system fragility is formulated in terms of actuator saturation using MPC and AWC schemes, however, the permanent source of controller imperfection is left alone. Fragility can be defined as the sensitivity of the control system w.r.t. small perturbations. These perturbations can range from the numerical issues in implementing the controller on an embedded device to perturbations due to neglected constraints in the control system. In this chapter, this issue is analyzed, namely, the fragility of the closed-loop system w.r.t. implementation imperfections is considered. Then, a novel robust non-fragile observer-based controller for the linear time-invariant system with structured uncertainty is introduced. The $H_\infty$ robust stability of the closed-loop system is guaranteed by the use of the Lyapunov theorem in the presence of undesirable disturbance.

For now, the disturbance rejection problem is investigated only by considering the disturbance channel to be known. In other words, no disturbance observer is considered in the process of control design. The result, although conservative (and sub-optimal) will be compared to a more advanced scheme based on disturbance estimation in the upcoming chapters.

For the sake of addressing the fragility problem, independent sets of time-dependent gain-uncertainties are assumed to be existing for the controller and the observer elements. In order to satisfy the arbitrary $H_2$-normed constraints for the control system and to enable automatic determination of the optimal $H_2/H_\infty$ bound of the performance functions in disturbance rejection control, additional necessary and sufficient conditions are presented in a linear matrix equality/inequality framework. The $H_2/H_\infty$ the observer-based controller is then transformed into an optimization problem of a coupled set of LMIs/LME that can be solved iteratively by the use of numerical software such as Scilab. Finally, concerning the evaluation of the performance of the controller, the control system is implemented in real-time on a mechanical system aiming at vibration suppression.

To put in a nutshell, the following contributions are also considered in this chapter: a new robust observer-based control system is proposed that can handle the fragility problem of both of the controller and observer elements in the presence of structured uncertainty in the state, control input, disturbance input, and system output matrices. This uncertainty representation is the most general one compared to the available models in the literature for the continuous system. The solution to the controller/observer design consists of the feasibility and optimization problems that are presented in an LMI framework which can be solved in Scilab. Here, the feasibility problem refers to the problem of finding any solution which satisfies the LMI constraints while the optimization problem refers to minimizing an objective function while guaranteeing that the feasibility problem has a solution.

## 8.1 Problem formulation for the fragility of the control system

Consider the continuous LTI system (8.1)

$$\begin{aligned} \dot{x} &= (A + \Delta A)x + (B + \Delta B)u + (H + \Delta H)w + f, t \geq 0 \\ y &= (C + \Delta C)x, \\ z_2 &= C_1 x + D_1 u, \end{aligned} \tag{8.1}$$

where $x \in \mathbb{R}^n, u \in \mathbb{R}^m, w \in \mathbb{R}^r$, $f \in \mathbb{R}^n$, and $y \in \mathbb{R}^q$ are the state, input, square-integrable external disturbance, norm bounded nonlinearity/un-modeled dynamics, and output vector, respectively. In addition, $A \in \mathbb{R}^{n\times n}, B \in \mathbb{R}^{n\times m}, H \in \mathbb{R}^{n\times r}$, and $C \in \mathbb{R}^{q\times n}$, are state, input, disturbance, and output constant matrices. Moreover, $\Delta A$, $\Delta B$, $\Delta H$, and $\Delta C$ are the associated unknown structured perturbation matrices that are time-dependent. In this equation, $z_2$ represents the $H_2$-index function that stands for arbitrary energy bounded restriction on the states and control input, also $C_1 \in \mathbb{R}^{q_1 \times n}$ and $D_1 \in \mathbb{R}^{q_1 \times m}$, are designer-defined matrices with appropriate dimensions. The dynamics of the proposed non-fragile observer is defined as

$$\dot{\hat{x}} = (A + \Delta A)\hat{x} + (B + \Delta B)u - (L + \Delta L)[y - \hat{y}], t \geq 0$$
$$\hat{y} = (C + \Delta C)\hat{x}, \tag{8.2}$$

where $\hat{x}$ and $\hat{y}$ represent the estimation of the states and output of the system, respectively and $L$ and $\Delta L \in \mathbb{R}^{n \times q}$ are the observer gain and perturbation that is related to the fragility of the observer. In addition, the non-fragile observer-based control effort is assumed to be $u = (K + \Delta K)\hat{x}$, with $K$ and $\Delta K \in \mathbb{R}^{m \times n}$ being the feedback gain and feedback perturbation matrices, respectively. It is presumed that the perturbation matrices, $\Delta L$, and $\Delta K$, are independent of $L$ and $K$ and have the structure of $\Delta K = M_K F_K N_K$, and $\Delta L = M_L F_L N_L$, in which $M_K, N_K, M_L$, and $N_L$ are known matrices with appropriate dimensions. However, the time-varying uncertain matrices $F_K(t)$ and $F_L(t)$ should satisfy $F_K^T F_K \leq I$, and $F_L^T F_L \leq I$, correspondingly. Similarly, $\Delta A = M_A F_A N_A$, $\Delta B = M_B F_B N_B$, $\Delta H = M_H F_H N_H$, and $\Delta C = M_C F_C N_C$, present the structure of the perturbation terms of the system (7.1), in which $F_A^T F_A \leq I$, $F_B^T F_B \leq I$, $F_H^T F_H \leq I$, and $F_C^T F_C \leq I$. Moreover, it is assumed that there exist positive scalars, $a$, $h$, $d$, and $g$, such that $\|\Delta A\| \leq a$, $\|\Delta H\| \leq h$, $\|\Delta C\| \leq d$, and $\|f\| \leq g\|x\|$. By defining the estimation error as $e = x - \hat{x}$, using (8.1) and (8.2), the dynamics of the states and the estimation error are presented as

$$\dot{x} = \big(A + \Delta A + (B + \Delta B)(K + \Delta K)\big)x - (B + \Delta B)(K + \Delta K)e + (H + \Delta H)w + f,$$
$$\dot{e} = \big(A + \Delta A + (L + \Delta L)(C + \Delta C)\big)e + (H + \Delta H)w + f. \tag{8.3}$$

Before presenting the main results, some preliminary Lemmas are introduced that will be used later.

**Lemma 8.1** ($H_\infty$-performance ([219])) *The LTI system $\dot{x} = Ax + Hw, y = Cx$, with $T_{yw} = C(sI - A)^{-1}H$ where s stands for Laplace variable, satisfies the $H_\infty$-norm constraint $\|T_{yw}\|_\infty < \beta$ with Lyapunov function $V(x) = x^T M x, M > 0$ if for $t > 0$,*

$$\dot{V} + y^T y - \beta^2 w^T w < 0. \tag{8.4}$$

**Lemma 8.2** [219] *For real matrices $\Sigma, \Psi, \mathrm{X}$, and symmetric matrix $M$, the following first statement can be guaranteed if and only if the second one holds for a positive scalar $\zeta$ and $\Psi^T\Psi \leq \mathrm{I}$,*

$$M + \Sigma\Psi\mathrm{X} + \mathrm{X}^T\Psi^T\Sigma^T < 0,$$
$$M + \zeta^{-1}\,\Sigma\Sigma^T + \zeta^{-1}\,(\zeta\mathrm{X}^T)(\zeta\mathrm{X}) < 0. \tag{8.5}$$

**Lemma 8.3** [85] *For two arbitrary vectors with appropriate dimension such as p and q, the following inequality is valid for $\epsilon > 0$*

$$p^T q + q^T p \leq \epsilon p^T p + \epsilon^{-1} q^T q. \tag{8.6}$$

**Lemma 8.4** *For a given matrix $S = \begin{bmatrix} S_{11} & S_{12} \\ S_{12}^T & S_{22} \end{bmatrix}$, with symmetric $S_{11}$ and symmetric negative definite $S_{22}$ the following two statements are equivalent*

$$S < 0,$$
$$S_{11} - S_{12} S_{22}^{-1} S_{12}^T < 0. \tag{8.7}$$

**Lemma 8.5** [72] *Consider the LTI system* $\dot{x} = Ax + Hw, y = Cx$. *There exists an observer-based controller such that the closed loop system is admissible if and only if the systems is stabilizable and detectable.*

**Lemma 8.6** ($H_2$-*performance*) *Consider the following stable system*

$$\begin{cases} x = \tilde{A}x + \tilde{B}w \\ z_2 = \tilde{C}x \end{cases} \tag{8.8}$$

*For system* (8.8) *the following two statements are equivalent*

*System* (8.8) *is admissible and the transfer function of the system satisfies* $\left\|T_{z_2w}\right\|_2 < \gamma$.

*There exist* $Q > 0$ *and* $Z > 0$ *such that the following LMIs hold*

$$\begin{bmatrix} \tilde{A}^T Q + Q\tilde{A} & \tilde{C}^T \\ \tilde{C} & -I \end{bmatrix} < 0, \tag{8.9a}$$

$$\begin{bmatrix} Q & Q\tilde{B} \\ \tilde{B}^T Q & Z \end{bmatrix} > 0, \tag{8.9b}$$

$$trace(Z) < \gamma^2 \tag{8.9c}$$

**Theorem 8.1** *The uncertain stabilizable and detectable LTI system of (8.1) together with the observer system of (8.2) is stable and satisfies the* $H_\infty$ *norm constraint* $\left\|T_{yw}\right\|_\infty < \gamma_\infty^*$ *and* $H_2$ *norm constraint* $\left\|T_{z_2w}\right\|_2 < \gamma_2^*$, *if there exist three positive definite symmetric matrices* $P, R \in \mathbb{R}^{n\times n}$ *and* $Z \in \mathbb{R}^{q_1\times q_1}$ , *positive scalars* $\varepsilon_i, i = 1,2,\dots,12$, $\zeta_j, j = 1,2,\dots,9$, *negative scalars* $\zeta_{10}$ *and* $\zeta_{11}$, *matrices* $\hat{P} \in \mathbb{R}^{m\times m}, \hat{L} \in \mathbb{R}^{n\times q}$, *and* $\hat{K} \in \mathbb{R}^{m\times n}$ *such that, while the following LMIs/LME system is satisfied, the objective function is minimized,*

$$\min(\alpha_1\gamma_\infty^* + \alpha_2\gamma_2^*)$$

$$Subject\ to$$

$$\begin{bmatrix} \Phi_{11} & \Phi_{12} & \Phi_{13} \\ & \Phi_{22} & 0 \\ * & & \Phi_{33} \end{bmatrix} < 0, \tag{8.10a}$$

$$\begin{bmatrix} \Psi_{11} & \Psi_{12} & \Psi_{13} & \Psi_{14} \\ & \Psi_{22} & 0 & 0 \\ & & \Psi_{33} & 0 \\ * & & & -\zeta_9 I \end{bmatrix} < 0, \tag{8.10b}$$

$$\begin{bmatrix} P & 0 & PH & PM_H & 0 \\ & R & RH & 0 & RM_H \\ & & N_{33} & 0 & 0 \\ & & & -\zeta_{10} I & 0 \\ * & & & & -\zeta_{11} I \end{bmatrix} > 0, \tag{8.10c}$$

$$\mathrm{trace}(Z) < \gamma_2^*, \tag{8.10d}$$

$$PB = B\hat{P}, \tag{8.10e}$$

*where,*

$$\Phi_{11} = \begin{bmatrix} X_{11} & -B\hat{K} & PH & PM_B & \varepsilon_{12} N_K^T & \varepsilon_1 N_K^T \\ & X_{22} & RH & -PM_B & -\varepsilon_{12} N_K^T & -\varepsilon_1 N_K^T \\ & & X_{33} & 0 & 0 & 0 \\ & & & -\varepsilon_{11} I & 0 & 0 \\ & & & & -\varepsilon_{12} I & 0 \\ * & & & & & -\varepsilon_1 I \end{bmatrix}, \tag{8.11a}$$

$$\Phi_{12} = \begin{bmatrix} P & PBM_K & gP & hP & C^T & PM_B \\ 0 & 0 & 0 & 0 & 0 & 0 \end{bmatrix},$$

$$\Phi_{13} = \begin{bmatrix} 0 & 0 & 0 & 0 & 0 & 0 & 0 \\ R & RM_L & \varepsilon_1 C^T N_L^T & gR & hR & \hat{L} & RM_L \\ 0 & 0 & 0 & 0 & 0 & 0 & 0 \end{bmatrix},$$

$$\Phi_{22} = diag(-\varepsilon_2 I, -\varepsilon_1 I, -\varepsilon_4 I, -\varepsilon_6 I, -\varepsilon_8 I, -\varepsilon_{12} I),$$

$$\Phi_{33} = diag(-\varepsilon_3 I, -\varepsilon_1 I, -\varepsilon_1 I, -\varepsilon_5 I, -\varepsilon_7 I, -\varepsilon_9 I, -\varepsilon_{10} I),$$

$$\Psi_{11} = \begin{bmatrix} N_{11} & -B\hat{K} & C_1^T + K^T D_1^T & \zeta_2 N_K^T & \zeta_4 N_K^T & \zeta_9 N_K^T \\ & N_{22} & -K^T D_1^T & -\zeta_2 N_K^T & -\zeta_4 N_K^T & -\zeta_9 N_K^T \\ & & -I & 0 & 0 & 0 \\ & & & -\zeta_2 I & 0 & 0 \\ & & & & -\zeta_4 I & 0 \\ * & & & & & -\zeta_9 I \end{bmatrix},$$

$$\Psi_{12} = \begin{bmatrix} PM_A & PBM_K & PM_B & PM_B \\ 0 & 0 & 0 & 0 \end{bmatrix},$$

$$\Psi_{13} = \begin{bmatrix} 0 & 0 & 0 & 0 \\ RM_A & \hat{L}M_C & RM_L & RM_L \\ 0 & 0 & 0 & 0 \end{bmatrix},$$

$$\Psi_{14} = \begin{bmatrix} 0 \\ 0 \\ D_1 M_K \\ 0 \end{bmatrix},$$

$$\Psi_{22} = diag(-\zeta_1 I, -\zeta_2 I, -\zeta_3 I, -\zeta_4 I),$$

$$\Psi_{33} = diag(-\zeta_5 I, -\zeta_6 I, -\zeta_7 I, -\zeta_8 I, -\zeta_9 I),$$

$$X_{11} = A^T P + PA + \hat{K}^T B^T + B\hat{K} + C^T C + (\varepsilon_2 + \varepsilon_4 + \varepsilon_5 + \varepsilon_8 + d^2)I + (\varepsilon_{11} K^T N_B^T N_B K),$$

$$X_{22} = A^T R + RA + C^T \hat{L}^T + \hat{L}C + \varepsilon_{10} N_C^T N_C + (\varepsilon_3 + \varepsilon_9 d^2)I,$$

$$X_{33} = (\varepsilon_6 + \varepsilon_7)I - \gamma_\infty^* I,$$

$$N_{11} = A^T P + PA + \hat{K}^T B^T + B\hat{K} + \zeta_1 N_A^T N_A,$$

$$N_{22} = A^T R + RA + C^T \hat{L}^T + \hat{L}C + \zeta_5 N_A^T N_A + \zeta_6 N_C^T N_C + \zeta_7 C^T N_L^T N_L C + \zeta_8 N_C^T N_C, \tag{8.11b}$$

$$N_{33} = Z + \zeta_{11} N_H^T N_H + \zeta_{10} N_H^T N_H,$$

$$\gamma_\infty^* = \gamma_\infty^2,$$

$$\gamma_2^* = \gamma_2^2,$$

*in which,* $0$ *presents zero matrix with appropriate dimension. Also,* $\alpha_k, k = 1,2$ *are the scalar weights that impress the significance of* $H_\infty$*- and* $H_2$*-performance in the designing procedure, respectively. The robust non-fragile observer and controller gain are given by* $L = R^{-1}\hat{L}$*, and* $K = \hat{P}^{-1}\hat{K}$*, respectively.*

**Proof**. Consider the following Lyapunov function

$$V = x^T P x + e^T R e. \tag{8.12}$$

By derivation of Lyapunov with respect to time along the trajectories (8.3) and using **Lemma 8.1** one can obtain

$$
\begin{aligned}
&\dot{V} - \gamma_\infty^2 w^T w + y^T y \\
&\quad = x^T A^T P x + x^T \Delta A^T P x + x^T K^T B^T P x + x^T \Delta K^T B^T P x + x^T K^T \Delta B^T P x \\
&\quad + x^T \Delta K^T \Delta B^T P x - e^T K^T B^T P x - e^T \Delta K^T B^T P x - e^T K^T \Delta B^T P x \\
&\quad - e^T \Delta K^T \Delta B^T P x + w^T H^T P x + w^T \Delta H^T P x + f^T P x + x^T P A x + x^T P \Delta A x \\
&\quad + x^T P B K x + x^T P \Delta B K x + x^T P B \Delta K x + x^T P \Delta B \Delta K x - x^T P B K e \\
&\quad - x^T P \Delta B K e - x^T P B \Delta K e - x^T P \Delta B \Delta K e + x^T P H w + x^T P \Delta H w + x^T P f \\
&\quad + e^T A^T R e + e^T \Delta A^T R e + e^T C^T L^T R e + e^T \Delta C^T L^T R e + + e^T C^T \Delta L^T R e \\
&\quad + e^T \Delta C^T \Delta L^T R e + w^T H^T R e + w^T \Delta H^T R e + f^T R e + e^T R A e + e^T R \Delta A e \\
&\quad + e^T R L C e + e^T R \Delta L C e + e^T R L \Delta C e + e^T R \Delta L \Delta C e + e^T R H w + e^T R \Delta H w \\
&\quad + e^T R f - \gamma_\infty^2 w^T w + x^T C^T C x + x^T \Delta C^T C x + x^T C^T \Delta C x + x^T \Delta C^T \Delta C x, \dot{V} \\
&\quad - \gamma_\infty^2 w^T w + y^T y \\
&\quad = x^T A^T P x + x^T \Delta A^T P x + x^T K^T B^T P x + x^T \Delta K^T B^T P x + x^T K^T \Delta B^T P x \\
&\quad + x^T \Delta K^T \Delta B^T P x - e^T K^T B^T P x - e^T \Delta K^T B^T P x - e^T K^T \Delta B^T P x \\
&\quad - e^T \Delta K^T \Delta B^T P x + w^T H^T P x + w^T \Delta H^T P x + f^T P x + x^T P A x + x^T P \Delta A x \\
&\quad + x^T P B K x + x^T P \Delta B K x + x^T P B \Delta K x + x^T P \Delta B \Delta K x - x^T P B K e \\
&\quad - x^T P \Delta B K e - x^T P B \Delta K e - x^T P \Delta B \Delta K e + x^T P H w + x^T P \Delta H w + x^T P f \\
&\quad + e^T A^T R e + e^T \Delta A^T R e + e^T C^T L^T R e + e^T \Delta C^T L^T R e + + e^T C^T \Delta L^T R e \\
&\quad + e^T \Delta C^T \Delta L^T R e + w^T H^T R e + w^T \Delta H^T R e + f^T R e + e^T R A e + e^T R \Delta A e \\
&\quad + e^T R L C e + e^T R \Delta L C e + e^T R L \Delta C e + e^T R \Delta L \Delta C e + e^T R H w + e^T R \Delta H w \\
&\quad + e^T R f - \gamma_\infty^2 w^T w + x^T C^T C x + x^T \Delta C^T C x + x^T C^T \Delta C x + x^T \Delta C^T \Delta C x,
\end{aligned} \tag{8.13}
$$

assuming the following transformation

$$
\mathcal{L}^T = [x^T \quad e^T \quad w^T], \tag{8.14}
$$

using (8.13) and (8.14), the terms without uncertainty can be presented as $\mathcal{L}^T \Omega_1 \mathcal{L}$, in which $\Omega_1$ is defined as

$$
\Omega_1 = \begin{bmatrix} Y_{111} & -PBK & PH \\ & Y_{122} & RH \\ * & & -\gamma_\infty^2 I \end{bmatrix}, \tag{8.15}
$$

where,

$$
\begin{aligned}
Y_{111} &= A^T P + PA + K^T B^T P + PBK + C^T C, \\
Y_{122} &= A^T R + RA + C^T L^T R + RLC.
\end{aligned} \tag{8.16}
$$

Using **Lemma 8.2** for the uncertain terms of fragility due to controller and observer, one can obtain

$$
\begin{aligned}
&\mathcal{L}^T \begin{bmatrix} \Delta K^T B^T P + PB\Delta K & -PB\Delta K & 0 \\ & -C^T \Delta L^T R - R\Delta LC & 0 \\ * & & 0 \end{bmatrix} \mathcal{L} \\
&\leq \mathcal{L}^T \begin{bmatrix} \varepsilon_1^{-1} PBM_K M_K^T B^T P + \varepsilon_1^{-1}(\varepsilon_1^2 N_K^T N_K) & -\varepsilon_1^{-1}(\varepsilon_1^2 N_K^T N_K) & 0 \\ -\varepsilon_1^{-1}(\varepsilon_1^2 N_K^T N_K) & RM_L M_L^T R + \varepsilon_1^{-1}(\varepsilon_1^2 N_K^T N_K) + \varepsilon_1^{-1}(\varepsilon_1^2 C^T N_L^T N_L C) & 0 \\ 0 & 0 & 0 \end{bmatrix} \mathcal{L} \\
&= \mathcal{L}^T \Omega_2 \mathcal{L},
\end{aligned} \tag{8.17}
$$

then, by utilizing **Lemma 8.4** on the terms with uncertainty in state, disturbance, and output matrices and un-modeled dynamics, one can obtain

$$
\begin{aligned}
x^T \Delta A^T P x + x^T P \Delta A x &\leq x^T (\varepsilon_2 I + \varepsilon_2^{-1} P^2) x, \\
e^T \Delta A^T R e + e^T R \Delta A e &\leq e^T (\varepsilon_3 I + \varepsilon_3^{-1} R^2) e,
\end{aligned} \tag{8.18}
$$

$$f^T P x + x^T P f \leq x^T(\varepsilon_4 I + \varepsilon_4^{-1} g^2 P^2)x,$$

$$f^T R e + e^T R f \leq x^T(\varepsilon_5 I)x + e^T(\varepsilon_5^{-1} g^2 R^2)e,$$

$$w^T \Delta H^T P x + x^T P \Delta H w \leq w^T(\varepsilon_6 I)w + x^T(\varepsilon_6^{-1} h^2 P^2)x,$$

$$w^T \Delta H^T R e + e^T R \Delta H w \leq w^T(\varepsilon_7 I)w + e^T(\varepsilon_7^{-1} h^2 R^2)e,$$

$$x^T \Delta C^T C x + x^T C^T \Delta C x + x^T \Delta C^T \Delta C x \leq x^T(d^2 I + \varepsilon_8 I + \varepsilon_8^{-1} C^T C)x,$$

$$e^T \Delta C^T L^T R + e^T L \Delta C e \leq e^T(\varepsilon_9 d^2 I + \varepsilon_9^{-1} R L L^T R)e.$$

Summation of all of the terms in the right-hand side of all inequalities in (8.18) with the use of (8.14) can be presented as $\mathcal{L}^T \Omega_3 \mathcal{L}$; in which $\Omega_3$ is defined as

$$\Omega_3 = diag(Y_{311}, Y_{322}, Y_{333}), \tag{8.19}$$

where

$$\begin{gathered} Y_{311} = (\varepsilon_2 + \varepsilon_4 + \varepsilon_5 + \varepsilon_8 + d^2)I + \varepsilon_2^{-1} P^2 + \varepsilon_4^{-1} g^2 P^2 + \varepsilon_6^{-1} h^2 P^2 + \varepsilon_8^{-1} C^T C, \\ Y_{322} = (\varepsilon_3 + \varepsilon_9 d^2)I + \varepsilon_3^{-1} R^2 + \varepsilon_5^{-1} g^2 R^2 + \varepsilon_7^{-1} h^2 R^2 + \varepsilon_9^{-1} R L L^T R, \\ Y_{333} = (\varepsilon_6 + \varepsilon_7)I. \end{gathered} \tag{8.20}$$

By use of **Lemma 8.2** and assuming $\|N_L M_C\| < 1$, the uncertain terms with $\Delta L \Delta C$ can be written as

$$e^T \Delta C^T \Delta L^T R e + e^T R \Delta L \Delta C e \leq \mathcal{L}^T \Omega_4 \mathcal{L}, \tag{8.21}$$

where,

$$\Omega_4 = diag(0, \varepsilon_{10}^{-1} R M_L M_L^T R + \varepsilon_{10} N_C^T N_C, 0). \tag{8.22}$$

Substituting the structure of the uncertain terms for $\Delta B$ and $\Delta K$, assuming $\|N_B M_K\| < 1$, and then by subsequent use of **Lemma 8.2** the following two inequalities hold

$$\begin{gathered} x^T K^T \Delta B^T P x + x^T P \Delta B K x - e^T K^T \Delta B^T P x - x^T P \Delta B K e \leq \mathcal{L}^T \Omega_5 \mathcal{L}, \\ x^T \Delta K^T \Delta B^T P x + x^T P \Delta B \Delta K x - e^T K^T \Delta B^T P x - x^T P \Delta B \Delta K e \leq \mathcal{L}^T \Omega_6 \mathcal{L}, \end{gathered} \tag{8.23}$$

in which,

$$\begin{gathered} \Omega_5 = \begin{bmatrix} \varepsilon_{11}^{-1} P M_B M_B^T P + \varepsilon_{11} K^T N_B^T N_B K & -\varepsilon_{11}^{-1} P M_B M_B^T P & 0 \\ -\varepsilon_{11}^{-1} P M_B M_B^T P & \varepsilon_{11}^{-1} P M_B M_B^T P & 0 \\ 0 & 0 & 0 \end{bmatrix}, \\ \Omega_6 = \begin{bmatrix} \varepsilon_{12}^{-1} P M_B M_B^T P + \varepsilon_{12}^{-1}(\varepsilon_{12}^2 N_K^T N_K) & -\varepsilon_{12}^{-1}(\varepsilon_{12}^2 N_K^T N_K) & 0 \\ -\varepsilon_{12}^{-1}(\varepsilon_{12}^2 N_K^T N_K) & \varepsilon_{12}^{-1}(\varepsilon_{12}^2 N_K^T N_K) & 0 \\ 0 & 0 & 0 \end{bmatrix}. \end{gathered} \tag{8.24}$$

Summation of (8.24), (8.22), (8.19), (8.17), and (8.15), using the transformation of (8.14) gives $\dot{V} - \gamma_\infty^2 w^T w + y^T y \leq \mathcal{L}^T \Omega \mathcal{L}$, where $\Omega = \sum_{i=1}^{6} \Omega_i$, and then straight use of Schur complement (8.7) and finally introducing the LME $PB = B\hat{P}$, the LMI (8.10a) can be obtained. It should be mentioned that the same LME will be employed in the rest of the proof. Next, in order to prove the $H_2$-performance, a hybrid state-space form is introduced

$$\dot{x}_{cl} = A_{cl}(\Delta A, \Delta B, \Delta C, \Delta K, \Delta L)x_{cl} + H_{cl}(\Delta H)w + F_{cl}f,$$
$$z_2 = C_1 x + D_1 u = X_{cl}(\Delta K)x_{cl}, \tag{8.25}$$

where $x_{cl}^T = [x^T \quad e^T]$, is the hybrid state-vector and $A_{cl}(\Delta A, \Delta B, \Delta C, \Delta K, \Delta L)$, $H_{cl}(\Delta H)$, and $X_{cl}(\Delta K)$ are the hybrid uncertain state, disturbance, and performance index matrices that are defined as

$$A_{cl}(\Delta A, \Delta B, \Delta C, \Delta K, \Delta L) = A_{cl} + A_{cl}(\Delta),$$
$$H_{cl}(\Delta H) = H_{cl} + H_{cl}(\Delta),$$
$$X_{cl}(\Delta K) = X_{cl} + X_{cl}(\Delta),$$
$$A_{cl} = \begin{bmatrix} A + BK & -BK \\ 0 & A + LC \end{bmatrix},$$
$$A_{cl}(\Delta) = \begin{bmatrix} \Delta A + \Delta BK + B\Delta K + \Delta B\Delta K & -\Delta BK - B\Delta K - \Delta B\Delta K \\ 0 & \Delta A + \Delta LC + L\Delta C + \Delta L\Delta C \end{bmatrix}, \tag{8.26}$$
$$H_{cl} = \begin{bmatrix} H \\ H \end{bmatrix},$$
$$H_{cl}(\Delta) = \begin{bmatrix} \Delta H \\ \Delta H \end{bmatrix},$$
$$X_{cl} = [C_1 + D_1 K \quad -D_1 K],$$
$$X_{cl}(\Delta) = [D_1 \Delta K \quad -D_1 \Delta K].$$

By use of **Lemma 8.5** and (8.9a) for the system of (8.26), the following LMI should be satisfied

$$\begin{bmatrix} A_{cl}^T Q + Q A_{cl}^T & X_{cl}^T \\ X_{cl} & -I \end{bmatrix} + \begin{bmatrix} A_{cl}^T(\Delta) Q + Q A_{cl}(\Delta) & X_{cl}^T(\Delta) \\ X_{cl}(\Delta) & 0 \end{bmatrix} < 0, \tag{8.27}$$

without loss of generality, the positive definite symmetric matrix $Q$ can be formatted as

$$Q = \begin{bmatrix} P & E \\ * & R \end{bmatrix}. \tag{8.28}$$

In this equation, the positive definite symmetric matrices $P$ and $R$ are representing the coupling between $H_\infty$- and $H_2$-performances. It is assumed that $E \in \mathbb{R}^{n \times n}$ is equal to zero in order to prevent the problem to be non-convex. Assuming $E = 0$, and replacing (8.28) in the first matrix of (8.27) and then employing the definitions in (8.26), the deterministic terms of (8.27) can be written as

$$\begin{bmatrix} A_{cl}^T Q + Q A_{cl}^T & X_{cl}^T \\ X_{cl} & -I \end{bmatrix} = \begin{bmatrix} A^T P + PA + \widehat{K}^T B^T + B\widehat{K} & -PBK & C_1^T + K^T D_1^T \\ & A^T R + RA + C^T \widehat{L}^T + \widehat{L} C & -K^T D_1^T \\ * & & -I \end{bmatrix}, \tag{8.29}$$

the uncertain terms in the second matrix of inequality in (8.27) can be reformatted as

$$A_{cl}^T(\Delta) Q + Q A_{cl}(\Delta) = \begin{bmatrix} M_{11} & M_{12} \\ * & M_{22} \end{bmatrix}, \tag{8.30}$$

where

$$M_{11} = P\Delta A + \Delta A^T P + P\Delta BK + K^T \Delta B^T P + PB\Delta K + \Delta K^T B^T P + P\Delta B\Delta K + \Delta K^T \Delta B^T P,$$
$$M_{12} = -P\Delta BK - K^T \Delta B^T P - PB\Delta K - \Delta K^T B^T P - P\Delta B\Delta K - \Delta K^T \Delta B^T P, \tag{8.31}$$
$$M_{22} = R\Delta A + \Delta A^T R + R\Delta LC + C^T \Delta L^T R + RL\Delta C + \Delta C^T L^T R + R\Delta L\Delta C + \Delta C^T \Delta L^T R,$$

by using the consecutive technique of **Lemma 8.2**, one can have

$$\begin{bmatrix} P\Delta A+\Delta A^TP & 0\\ 0 & 0\end{bmatrix} < \begin{bmatrix} \zeta_1^{-1}PM_AM_A^TP+\zeta_1N_A^TN_A & 0\\ 0 & 0\end{bmatrix},$$

$$\begin{bmatrix} PB\Delta K+\Delta K^TB^TP & -PB\Delta K+\Delta K^TB^TP\\ 0 & 0\end{bmatrix} <$$
$$\begin{bmatrix} \zeta_2^{-1}PBM_KM_K^TB^TP+\zeta_2^{-1}(\zeta_2^2N_K^TN_K) & -\zeta_2^{-1}(\zeta_2^2N_K^TN_K)\\ -\zeta_2^{-1}(\zeta_2^2N_K^TN_K) & \zeta_2^{-1}(\zeta_2^2N_K^TN_K)\end{bmatrix},$$

$$\begin{aligned}&\begin{bmatrix} P\Delta BK+K^T\Delta B^TP & -P\Delta BK-K^T\Delta B^TP\\ 0 & 0\end{bmatrix}\\ &\quad< \begin{bmatrix} \zeta_3^{-1}PM_BM_B^TP+\zeta_3K^TN_B^TN_BK & -\zeta_3K^TN_B^TN_BK\\ -\zeta_3K^TN_B^TN_BK & \zeta_3K^TN_B^TN_BK\end{bmatrix},\end{aligned} \tag{8.32}$$

$$\begin{aligned}&\begin{bmatrix} P\Delta B\Delta K+\Delta K^T\Delta B^TP & -P\Delta B\Delta K-\Delta K^T\Delta B^TP\\ 0 & 0\end{bmatrix}\\ &\quad< \begin{bmatrix} \zeta_4^{-1}PM_BM_B^TP+\zeta_4^{-1}(\zeta_4^2N_K^TN_K) & -\zeta_4^{-1}(\zeta_4^2N_K^TN_K)\\ -\zeta_4^{-1}(\zeta_4^2N_K^TN_K) & \zeta_4^{-1}(\zeta_4^2N_K^TN_K)\end{bmatrix},\end{aligned}$$

in addition, the uncertain terms due to the fragility of the observer and output matrices should satisfy the following inequality,

$$\begin{aligned}&R\Delta A+\Delta A^TR+R\Delta LC+C^T\Delta L^TR+RL\Delta C+\Delta C^TL^TR+R\Delta L\Delta C+\Delta C^T\Delta L^TR\\ &\quad<\zeta_5^{-1}RM_AM_A^TR+\zeta_5N_A^TN_A+\zeta_6^{-1}RLM_CM_C^TL^TR+\zeta_6N_C^TN_C+\zeta_7^{-1}RM_LM_L^TR\\ &\quad+\zeta_7C^TN_L^TN_LC+\zeta_8^{-1}RM_LM_L^TR+\zeta_8N_C^TN_C.\end{aligned} \tag{8.33}$$

Applying **Lemma 8.2** on uncertain terms of $H_2$ index function ($z_2$) hands

$$\begin{bmatrix} 0 & 0 & \Delta K^TD_1^T\\ & 0 & -\Delta K^TD_1\\ * & & 0\end{bmatrix} < \begin{bmatrix} \zeta_9^{-1}(\zeta_9^2N_K^TN_K) & -\zeta_9^{-1}(\zeta_9^2N_K^TN_K) & 0\\ -\zeta_9^{-1}(\zeta_9^2N_K^TN_K) & \zeta_9^{-1}(\zeta_9^2N_K^TN_K) & 0\\ 0 & 0 & \zeta_9^{-1}D_1M_KM_K^TD_1^T\end{bmatrix}, \tag{8.34}$$

substituting (8.34), (8.33), (8.32), and (8.29) in (8.27) and then consecutive employment of Schur complement results in LMI (8.10a). The second LMI of

**Lemma 8.6** (9b) for uncertain LTI system (8.25) can be decomposed into deterministic and perturbed terms as

$$\begin{bmatrix} Q & QH_{cl}\\ H_{cl}^TQ & Z\end{bmatrix}+\begin{bmatrix} 0 & QH_{cl}(\Delta)\\ H_{cl}^T(\Delta)Q & 0\end{bmatrix}>0, \tag{8.35}$$

by replacing (8.26) in (8.35), one can acquire the following inequality

$$\begin{bmatrix} P & 0 & PH\\ & R & RH\\ * & & Z\end{bmatrix}+\begin{bmatrix} 0 & 0 & P\Delta H\\ & 0 & R\Delta H\\ * & & 0\end{bmatrix}>0. \tag{8.36}$$

By utilization of the definition of the structured uncertainty for disturbance matrix together with (8.5) for the second matrix in the left-hand side of (8.36), it is easy to get

$$\begin{bmatrix} 0 & 0 & P\Delta H\\ & 0 & R\Delta H\\ * & & 0\end{bmatrix} > \begin{bmatrix} \zeta_{10}^{-1}PM_HM_H^TP & 0 & 0\\ & \zeta_{11}^{-1}RM_HM_H^TR & 0\\ * & & \zeta_{10}N_H^TN_H+\zeta_{11}N_H^TN_H\end{bmatrix}, \tag{8.37}$$

consequently, by substituting (8.37) in (8.36) and then successive use of Schur complement on the resulted LMI, the LMI in (8.10c) is obtained. In addition, the LMI (8.10d) is the direct implementation of

**Lemma 8.6** on the system based on the definition (8.11b). Finally, by assuming $\hat{P}K=\hat{K}$, $RL=\hat{L}$, $\gamma_\infty^*=\gamma_\infty^2$, and $\gamma_2^*=\gamma_2^2$, the proof is complete.

■

**Remark 8.1.** The LMI conditions (8.10a) are affined on the subject to the defined matrices by the introduction of the LME conversion ($PB = B\hat{P}$), which transforms a non-convex problem to a convex one.

**Remark 8.2.** The non-singularity of $\hat{P}$ is guaranteed by the assumption that the input matrix $B$ is a full column rank. In the application of the benchmark AVC shown in Figure 6.1, this assumption is checked and readily satisfied.

**Remark 8.3.**

**Lemma 8.6** is valid for stable systems. Although the stability of the closed-loop system is definite by $H_\infty$-performance, in the case that the nominal system is unstable, the initial guesses for $P$, $\hat{P}$, and $\hat{K}$ should be such that all of the eigen-values of $A + BK$ have a negative real part. Therefore, in the case of an unstable nominal system, the following algorithm should be used.

**Algorithm 8.1** For some arbitrary initial guesses for two positive definite symmetric matrices $P, R \in \mathbb{R}^{n \times n}$, positive scalars $\varepsilon_i, i = 1,2,\dots,12$, matrices $\hat{P} \in \mathbb{R}^{m \times m}, \hat{L} \in \mathbb{R}^{n \times q}$, and $\hat{K} \in \mathbb{R}^{m \times n}$, the LMI/LME of (8.10a) and (8.10e) can be satisfied (feasibility problem). As expected, the resulting controller gain $K_0 = \hat{P}^{-1}\hat{K}$, contents the $H_\infty$-performance and **Remark 8.3**.

Solve the LMI/LME optimization problem by use of the resulting controller gain ($K_0$) of the feasibility problem in the previous step as an initial guess for the $H_2/H_\infty$-performance.

**Remark 8.4.** In the LMI system (8.10a), the controller gain $K$ appears explicitly, even after the transformation of LME which is defined in (8.10a). This is due to the definition of $D_1$ as a design matrix. In order to solve this problem in the general case of unstable nominal system, the following algorithm should be used.

**Algorithm 8.2** Assume $D_1 = 0$, and $\gamma_\infty^* = \gamma_1$, where $\gamma_1$ is the initial condition for the $H_\infty$ norm of the transfer function of the closed-loop system. Then solve the feasibility LMI/LME (8.10a)- (8.10a) with arbitrary initial guesses for $P, R \in \mathbb{R}^{n \times n}$, $\varepsilon_i, i = 1,2,\dots,12$, $\hat{P} \in \mathbb{R}^{m \times m}, \hat{L} \in \mathbb{R}^{n \times q}$, and $\hat{K} \in \mathbb{R}^{m \times n}$. Obtain $K_r = \hat{P}^{-1}\hat{K}$, and $L_r = R^{-1}\hat{L}$, where $r = 0$.

Define $r = r + 1$, and use $K_{r-1}$ and $L_{r-1}$ as the initial guess for solving the general feasibility problem of (8.10a) and (8.11a) for constants $\gamma_\infty^* = \gamma_1$, and $\gamma_2^* = \gamma_2$ , in which $K$ in $\Psi_{11}$ should be replaced with the initial guess from previous step ($K_{r-1}$). Continue this step until $K_r \approx K_{r-1}$ and then go to next step.

Replace $\gamma_\infty^* = \gamma_\infty^* - d\gamma_\infty^*$, and $\gamma_2^* = \gamma_2^* - d\gamma_2^*$, in which $d\gamma_\infty^*$ and $d\gamma_2^*$ are the step size for $H_\infty$and $H_2$-performance functions, respectively. Go to the step 1.

If for desired $\gamma_\infty^*$ and $\gamma_2^*$, step 2 converges to a controller gain $K$ then exit: $K = K_r$, and $L = L_r$.

## 8.2 Experimental and mathematical system

In order to verify the controller/observer performance, they are implemented on the smart beam presented in Figure 6.1. The single measured output signal of the system is obtained by a scanning digital laser Doppler vibrometer (VH-1000-D). The control law of the active vibration system is then realized on SIMULINK platform and then compiled control law is downloaded to the dSPACE DAQ in real-time as shown in Figure 8.1. In this figure, $z_\infty$ represents the designer-defined $H_\infty$-constraint on the transfer function. It should be noted that the proposed controller is developed for arbitrary purposes, i.e. regulation and/or reference tracking, and therefore, the $z_\infty$ is introduced in the formulation as the $H_\infty$-constraint of general transfer function representing the relation of regulation- and/or tracking-error with respect to disturbances and/or reference values, correspondingly. In the context of AVC benchmark problem of this dissertation, since we are dealing with the regulation problem, this transfer function is considered to be from the external disturbance to the measured output ($T_{yw}$). In order to extract the mathematical model of the system, the subspace identification method is used. The interested reader can refer to [20]. With similar logic, $z_2$ is defined as the $H_2$-constraint on an arbitrary transfer function. Additionally, $K$ and $L$ represent the controller and observer gains proposed in **Algorithm 8.2**. For the implementation of the controller, a structure consisting of an aluminum clamped beam with two piezoelectric patches is used. The three mode-shapes of the smart beam are considered as nominal identified reduced order model and the remaining higher-order modes are considered as norm-bounded un-modeled dynamics of the system. For investigation of the robust performance of the uncertain closed-loop system with the designed controller, **Algorithm 8.1** and **Algorithm 8.2** can be used with $C_1 = 0$, and $D_1 = I$. An ultimate goal of the robust non-fragile control system is addressing the implementation practical imperfections by means of evaluating the effectiveness of the control system on the full order system.

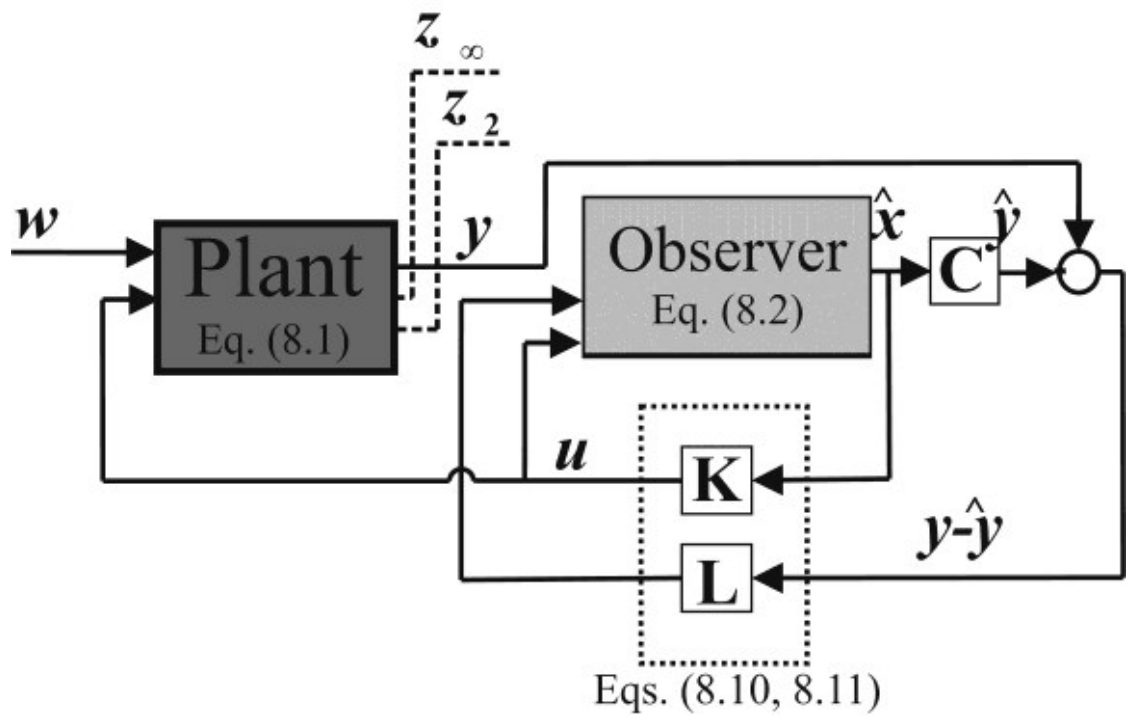

Figure 8.1 Schematic configuration of the control system [55]

In order to design a full (observed) state feedback controller based on the proposed algorithms in the previous section and implement the obtained control law on the experimental setup, an observer-based control system is designed by solving a convex optimization problem (**Algorithm 8.2**). $M_K, N_K, M_L$, and $N_L$ are assumed to be $0.1I$ where $I$ represents identity matrix of appropriate dimension. The optimal solution for LMI/LME is obtained by using Scilab. For this purpose after constructing the control system in SIMULINK and compiling it to dSPACE RTI platform, the system is excited through the disturbance channel with a sweep sine signal. The frequency of the excitation signal is changed from zero to 190 Hz (1193.8 rad/s) in two cases of the open-loop and closed-loop system. The closed-loop system is implemented on the real-time DAQ of the dSPACE with a sampling frequency of 10 kHz.

The responses of the system for the controlled and uncontrolled cases are shown in Figure 8.2 in time-domain based on the measurement signal generated from the Doppler vibro-meter. In addition, the frequency response function (FRF) of the open-loop system is compared with the closed-loop one in the frequency range of $[0 \quad 180]$ Hz. It can be seen that the controller design, based on the reduced-order identified model, suppressed the vibration magnitude within the considered frequency range. In addition, the corresponding control efforts that are generated for piezo-actuator patches by the dSPACE DAC board are shown in Figure 8.3.

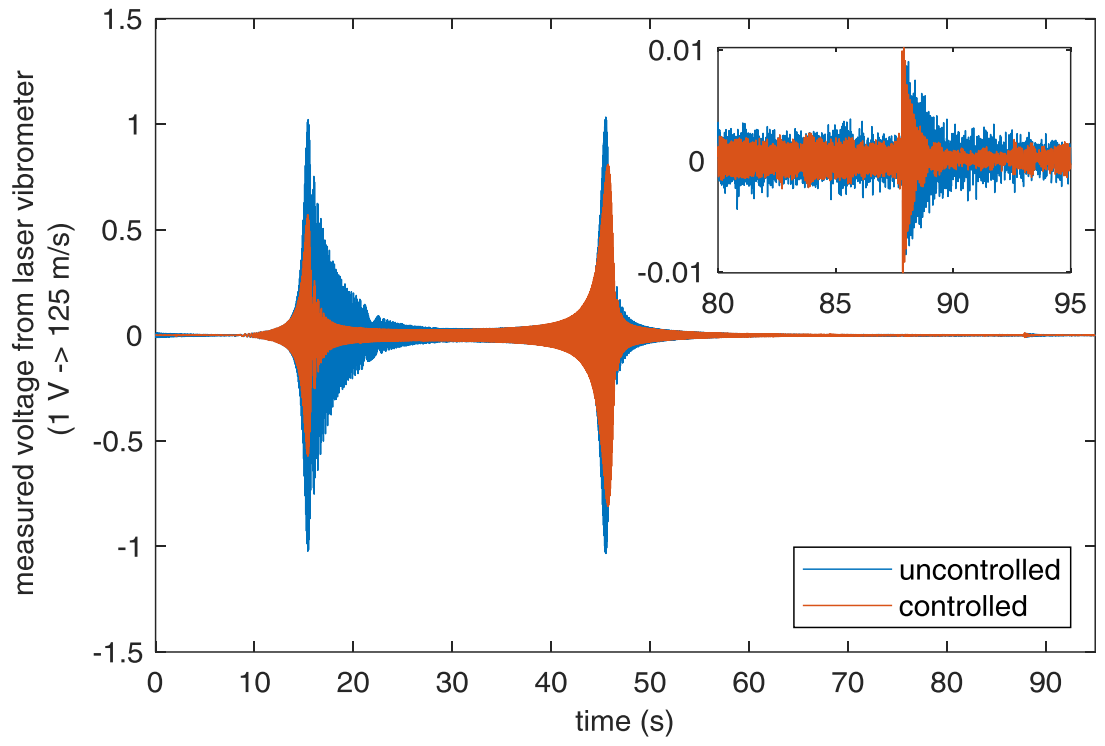

Figure 8.2 (a) Experimental comparison of measured outputs in time-domain [55]

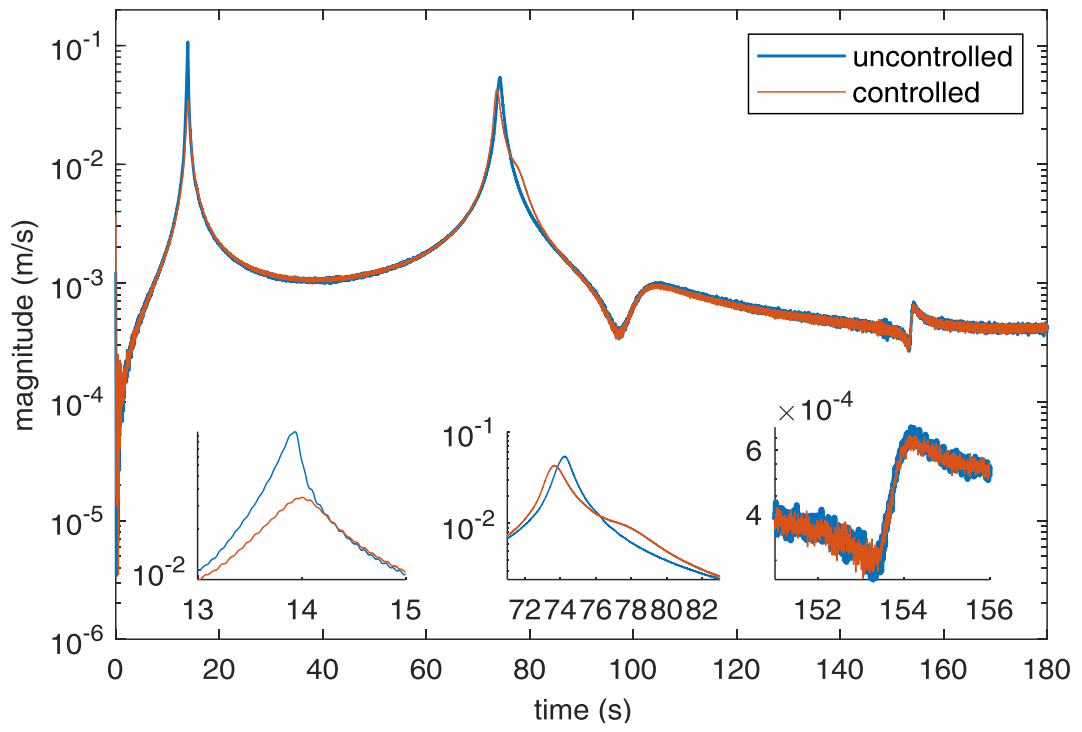

Figure 8.2 (b) Experimental comparison of FRF of the systems [55]

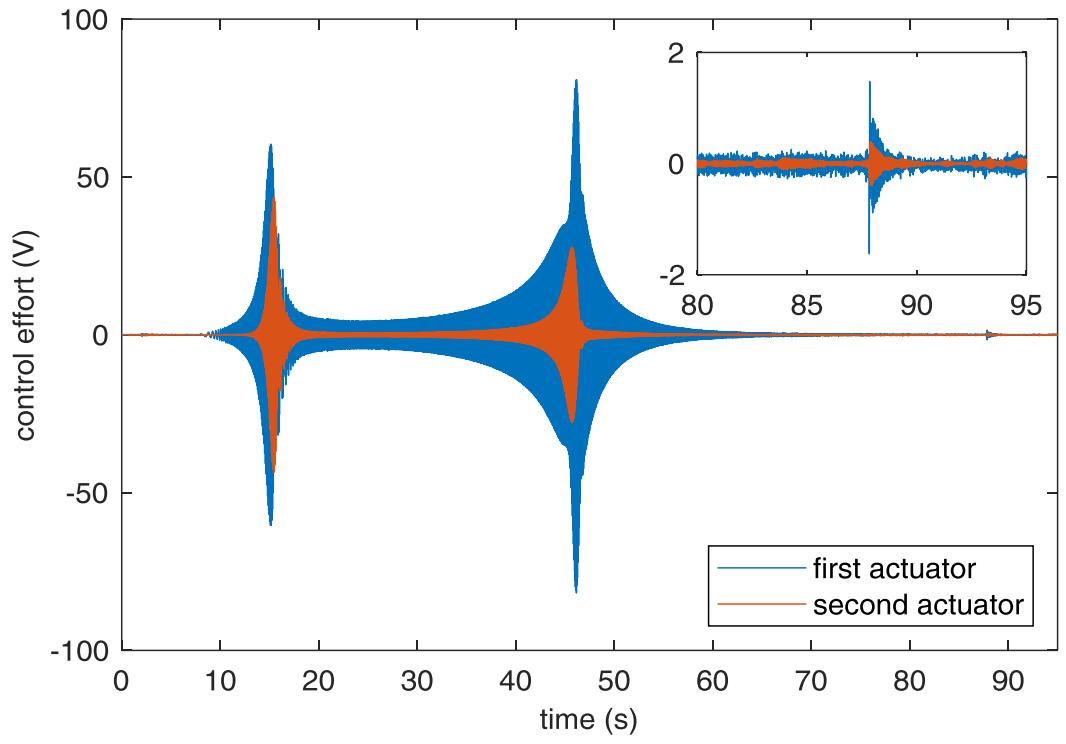

Figure 8.3 Control effort of the piezo-patch actuators [55]

By introduction of $H_2$-norm bounded index function, the control efforts applied to each of the piezo-actuators are limited to a maximum of 80 V without any sudden jump. The experimental results show that the robust control system performs well in attenuating the vibration amplitude in the presence of structured uncertainties in all system matrices. In addition, the observation error of the output is depicted in Figure 8.4.

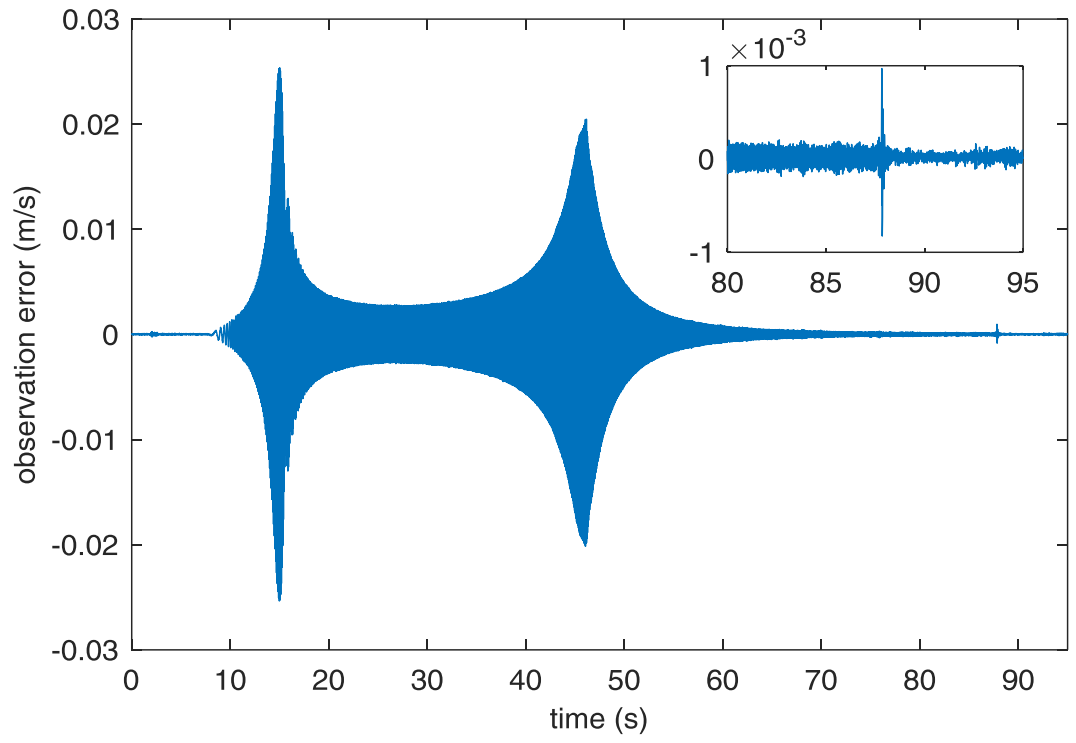

Figure 8.4 Observation error of the observer system [55]

In order to investigate the robust performance of the designed system in high frequency, where the unmodeled dynamics can affect the response of the system, the real structure is excited with a chirp signal with a frequency between 190 and 220 Hz. The measurement output and the applied control effort are presented in Figure 8.5 and Figure 8.6, respectively. It should be mentioned that in order to reject the measurement noise a low-pass filter is designed on the measurement channel.

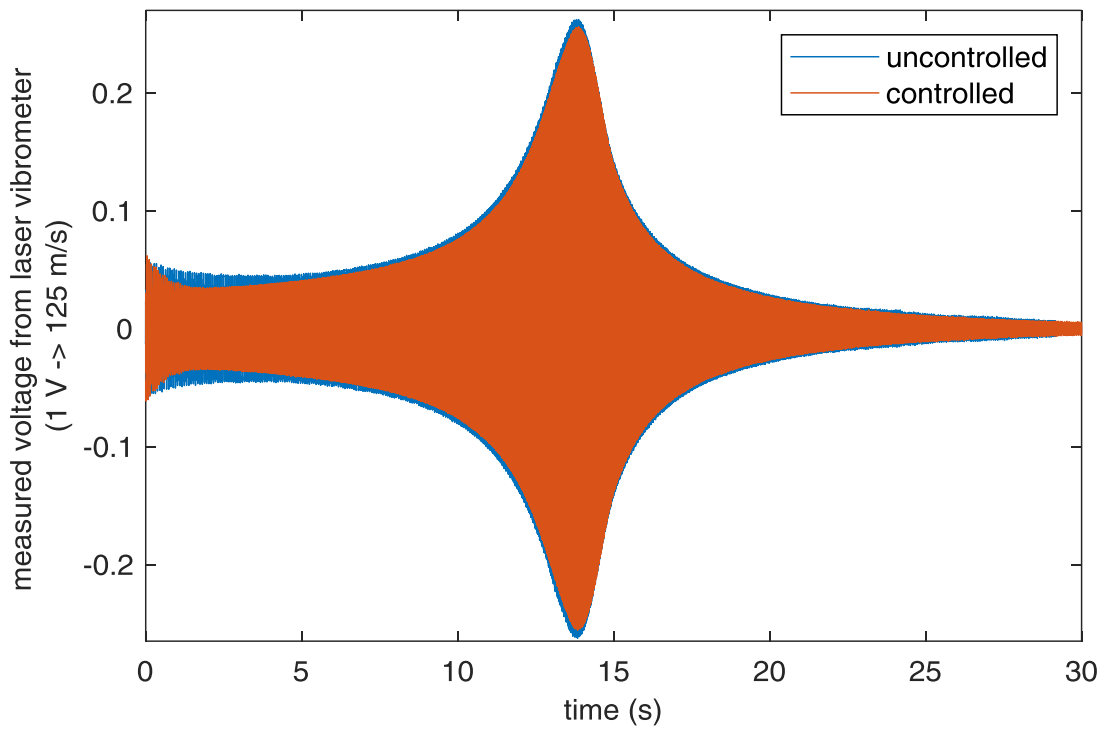

Figure 8.5 Experimental comparison of measured outputs for unmodeled dynamics [55]

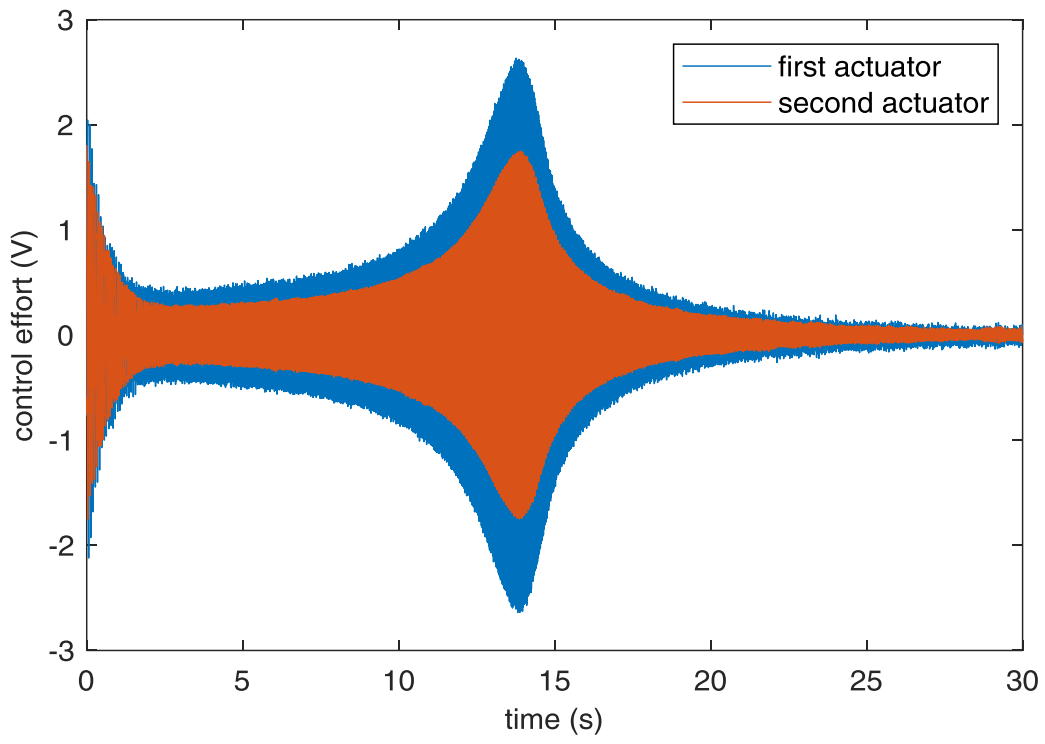

Figure 8.6 Control effort of the piezo-patch actuators for unmodeled dynamics [55]

It can be seen that the observer-based controller can handle the unmodeled dynamics with a limited control effort, however, the performance of the attenuation is reduced significantly in comparison to the previous simulations. This observation can be described based on the conservative effect of the $H_2$-performance function on unmodeled dynamics, for which as can be seen in Figure 8.6, the applied control effort is limited to 3 $V$. In addition, the peak value in Figure 8.6 is representing the higher-order natural frequency of the system which is equal to 204.6 Hz. Finally, in order to study the spillover effect, the structure is excited by mechanical initial displacement. The comparison of the open-loop and closed-loop systems is depicted below.

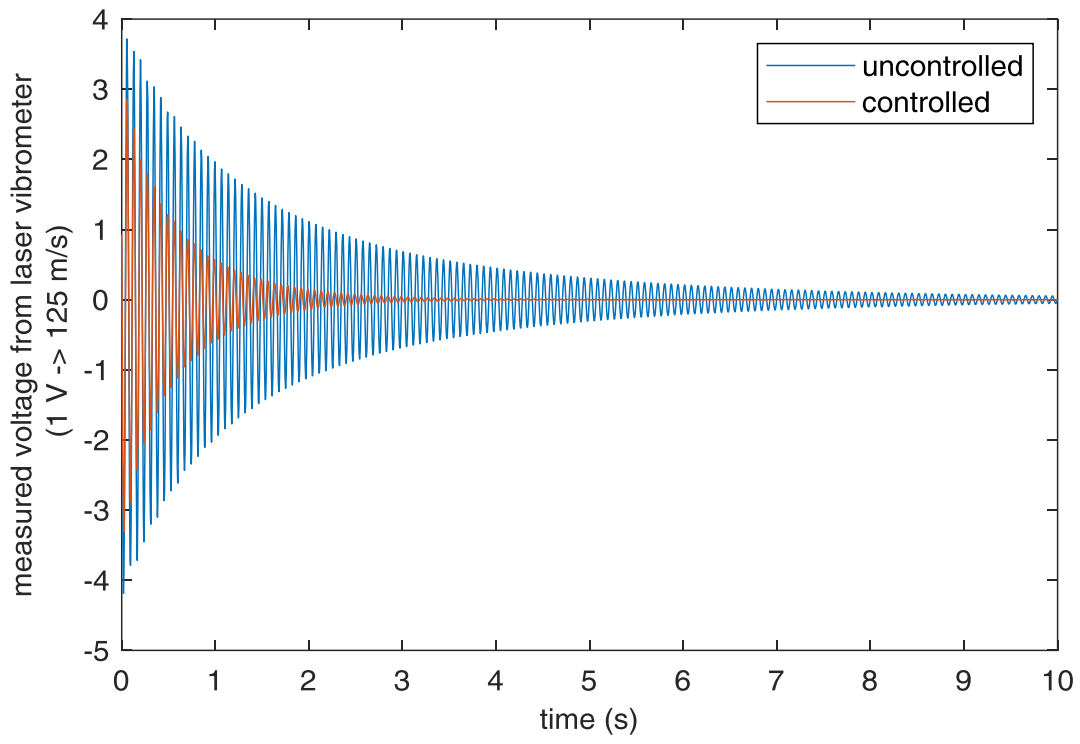

Figure 8.7 Experimental comparison of measured outputs for initial displacement [55]

The fast Fourier transformation (FFT) analysis is performed on the applied control efforts to identify the dominant frequencies of the controller. Figure 8.8 shows that the applied control efforts in frequency-domain between zero and 270 Hz have their significant effect around the first and second eigen-frequencies, however, as one can see, the third mode shape of the structure with natural frequency around 150 Hz (see

Fig. 4b), has no significant effect in the controller input. Moreover, the higher-order dynamic that is studied in Figure 8.6 and Figure 8.7 is excited by the control input which is due to the spillover effect that is not considered in the control design procedure. However, as in natural systems, the lower eigen-frequencies have a dominant impact, which is also confirmed here by the influence of the first and the second eigen-frequencies. Therefore, even if the high-frequency excitation would occur, it would not deteriorate the controller performance due to the negligible influence of the higher modes with respect to the dominant lower frequency vibration suppression.

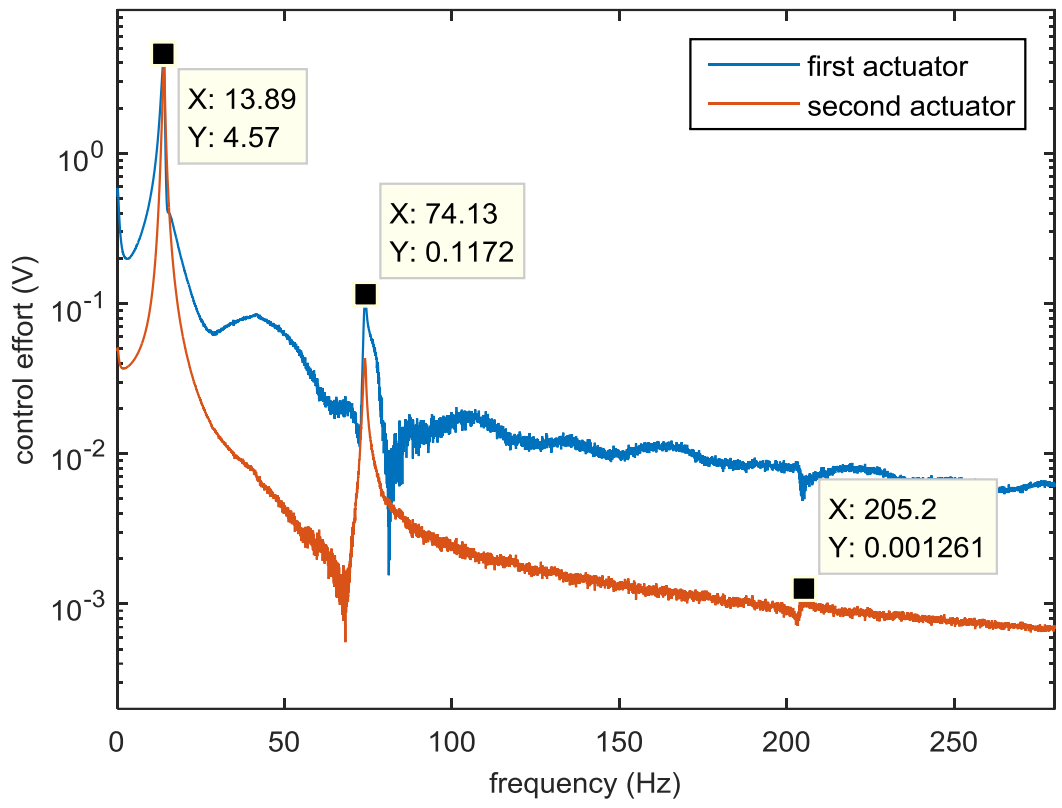

Figure 8.8 FFT of control effort for initial displacement excitation [55]

Finally, for a fair comparison, the proposed method is compared with the results from [71] under the assumption that the input matrix $B$ is full column rank (**Remark 8.2**). In the method proposed by [71], [220], and additional variable ($\rho$) exists which represents the exponential convergence rate of the Lyapunov function. In their work, this variable is multiplied with the design parameter $P$ in Lyapunov equation. They have proposed to calculate the controller gain to reach the maximum $\rho$ in order to have highest rate of exponential stability. However, this cannot be considered in the proposed solution of this chapter as a separate individual design parameter because it makes the problem non-convex. Consequently, the maximum value of $\rho$ is calculated by sarting from small values and increasing it as long as the LMI convex problem stays feasible. Numerically this value is calculated to be $\rho = 0.49$ with $K$ and $L$ shown as:

$$K = \begin{bmatrix} 7.713 & 3.851 & 4.756 & 1.039 & 4.312 & -8.950 \\ -11.54 & 11.61 & -4.176 & -16.42 & -6.539 & 1.464 \end{bmatrix},$$

$$L = [-17.78 \quad 12.26 \quad -0.24 \quad 0.355 \quad 34E-4 \quad -0.1296]^T.$$

With the similar parameters using the proposed control algorithm in this chapter, one obtains:

$$K = \begin{bmatrix} 2.2797 & 1.622 & -10.628 & -1.411 & -12.48 & 26.79 \\ 0.1467 & -9.208 & 8.272 & 30.80 & 17.67 & -4.605 \end{bmatrix},$$

$$L = [25.94 \quad 3.219 \quad 0.1899 \quad -0.4252 \quad 0.2048 \quad 0.3084]^T.$$

The closed-loop eigen values of the system using two control methods, as presented in Table 8.1, represent the better performance of the proposed method compared to [71].

In addition, the comparison of the open-loop response of the system with the closed-loop one based on the proposed method and the method in [71] is presented based on the simulation results. As it can be seen, the proposed controller has a better performance in rejecting the pulse disturbance compared to Figure 8.9. Moreover, the applied control efforts for these two methods together with the realization of the simulated pulse disturbance are depicted in Figure 8.10 (a). It is notable that the proposed approach in **Algorithm 8.2** is doing a better disturbance rejection while simultaneously having a lower control law amplitude applied on the actuators (indicating lower energy consumption). Accordingly, the observation error in Figure 8.10 (b) which also indicates the better performance of the proposed observer.

Table 8.1 Eigen-values of the closed-loop system.

| # | $A$ | $A \pm LC$ | | $A \pm BK$ | |
|---|---|---|---|---|---|
| | | Lien *et al*. (2007) | present | Lien *et al*. (2007) | present |
| 1 | -209.44+450.73i | -184.66+451.01i | -185.44+430.94i | -109.48+455.28i | -200.89+438i |
| 2 | -209.44-450.73i | -184.66-451.01i | -185.44-430.94i | -109.48-455.28i | -200.89-438i |
| 3 | -2.26+467.13i | -11.033+474.93i | -12.536+474.57i | -3.28+468.51i | -3.538+470.01i |
| 4 | -2.26-467.13i | -11.033-474.93i | -12.536-474.57i | -3.28-468.51i | -3.538-470.01i |
| 5 | -0.46+87.6i | -3.691+87.79i | -9.30+82.46i | -0.726+87.96i | -1.487+88.55i |
| 6 | -0.46-87.6i | -3.691-87.79i | -9.30-82.46i | -0.726+87.96i | -1.487-88.55i |

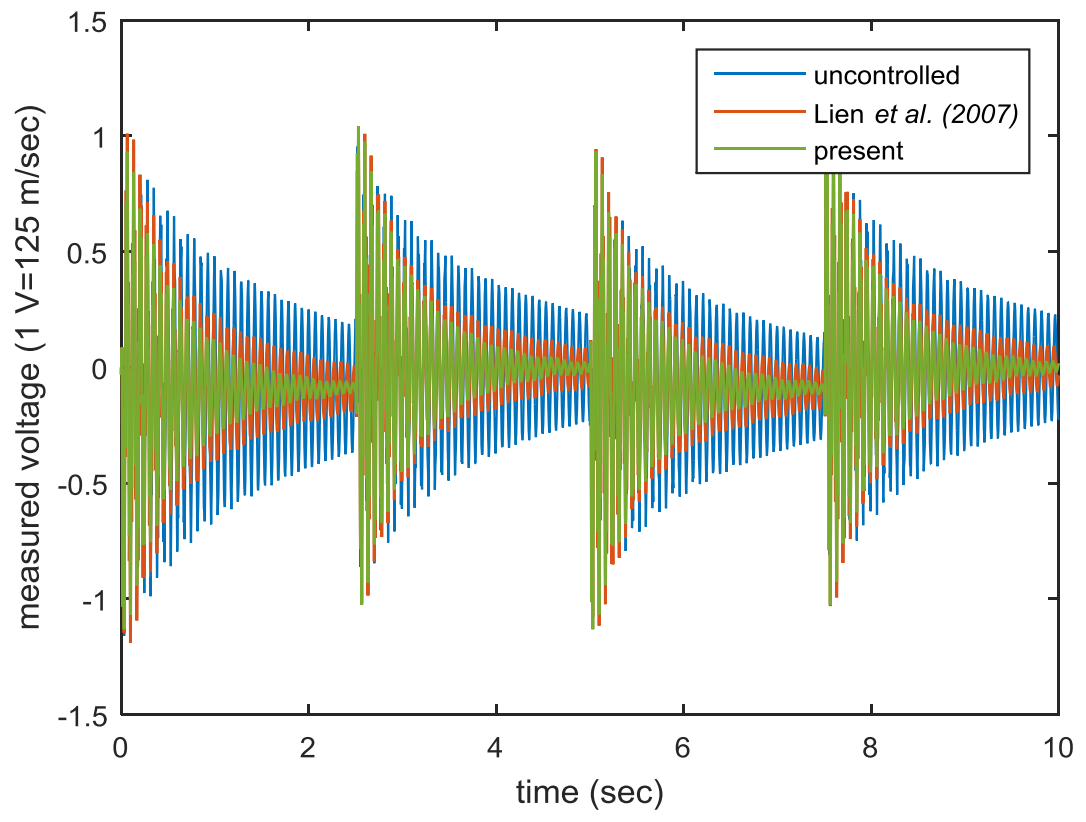

Figure 8.9 Comparison of controlled and uncontrolled system [55]

It should be noted that there is no difficulty regarding the quantification of the bounds of the norms of the constraints in the control problem for the proposed method which makes it easier to tune. In addition, in order to make a fair comparison, and as mentioned before since the method in [71] is developed based on the system without modeling uncertainties, the control implementations are limited to a reduced order system.

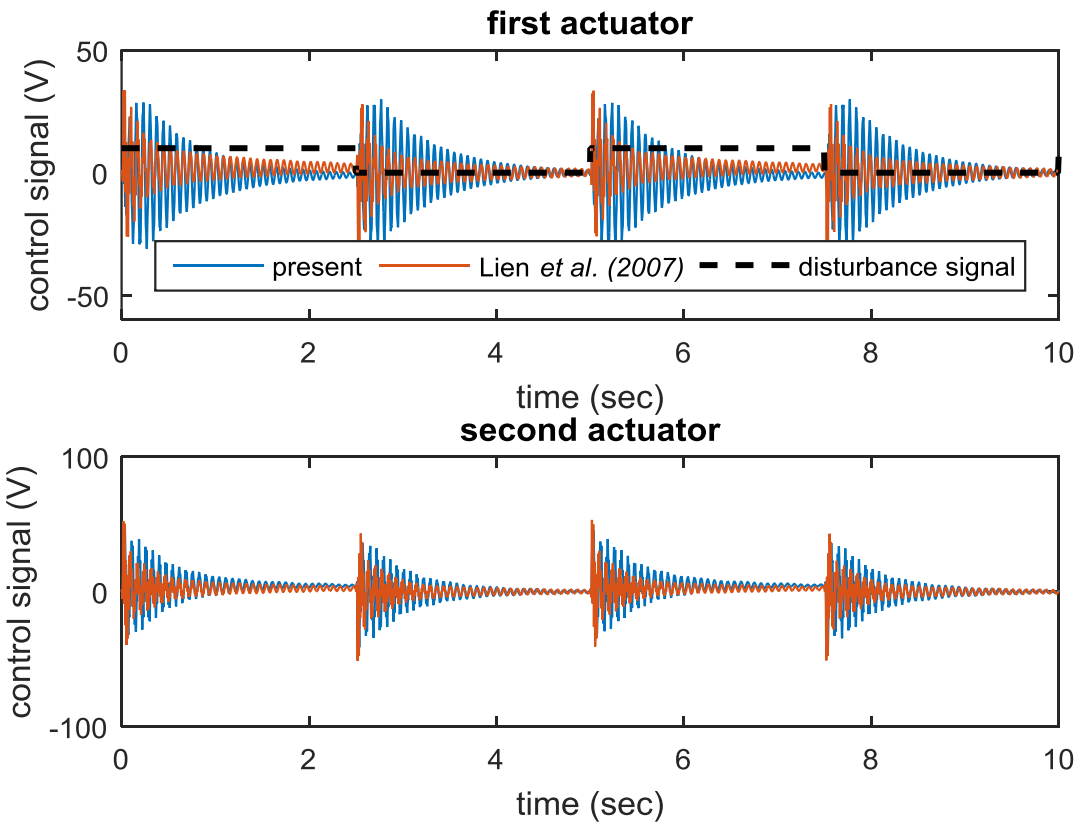

Figure 8.10(a) Control effort [55]

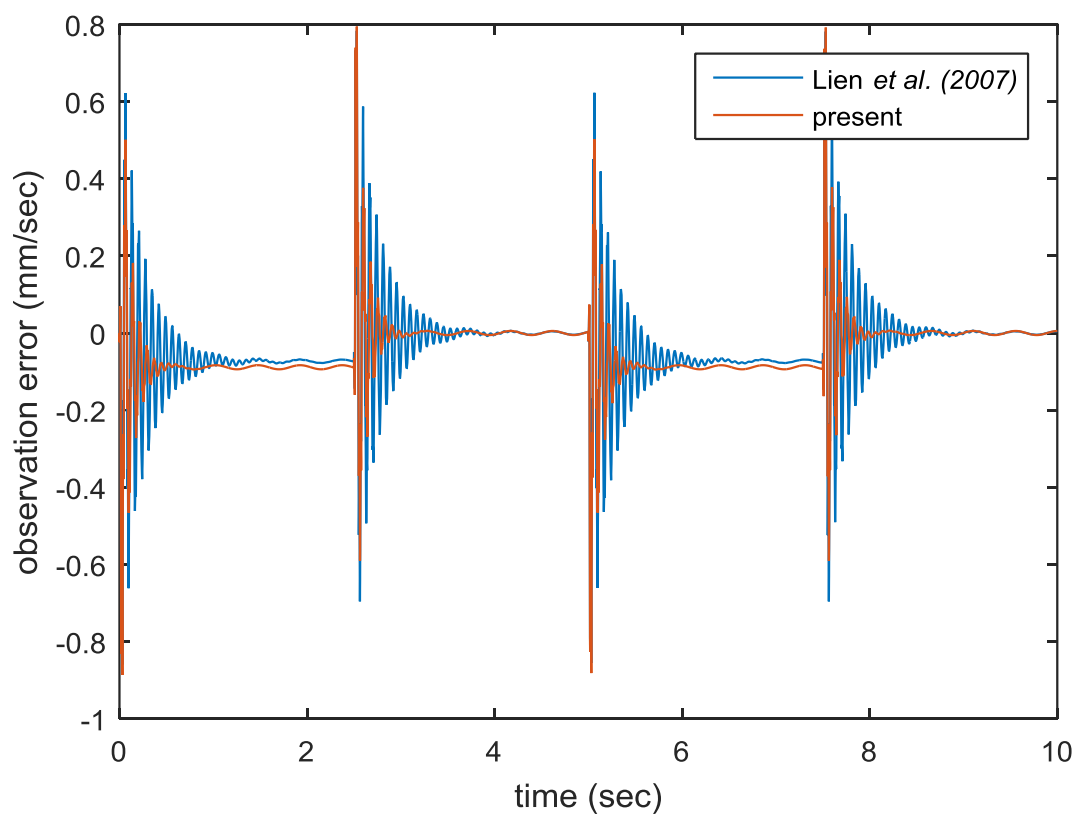

Figure 8.10 (b) Observation error [55]

Obviously one should expect even worse performance when the higher-order mode-shapes (unmodeled dynamics) are excited due to the unknown nature of disturbance signal or even the actuator spillover effect in vibration control.

In this chapter, the fragility of the controller and observer was robustly compensated in the presence of structured nonlinearity. However, the investigations on the unmodeled dynamics indicated the need for improvements. The upcoming chapter is mainly focused on dealing with the unmodeled nonlinear dynamics using an adaptive fuzzy sliding mode controller.

# 9 Dealing with unmodeled dynamics: AFSMC approach

In this chapter, a new observer-based adaptive fuzzy integral sliding mode controller is proposed based on the Lyapunov stability theorem. The plant under study is subjected to a square-integrable disturbance and is assumed to have mismatch uncertainties both in the state- and input matrices. This assumption is borrowed mostly from the previous chapter for the sake of consistency. In addition, a norm-bounded time-varying term ($f(x,t)$) is introduced to address the possible existence of un-modeled/nonlinear dynamics. Based on the classical sliding mode controller, the equivalent control effort is obtained to satisfy the sufficiency requirement of the sliding mode controller and then the control law is modified to guarantee the reachability of the system trajectory to the sliding manifold. The sliding surface is compensated based on the observed states in the form of linear matrix inequality. In order to relax the norm-bounded constraints on the control law and solve the chattering problem of sliding surface with $sign(.)$ operator, a fuzzy logic inference mechanism is combined with the controller. An adaptive law is then introduced to tune the parameters of the fuzzy system online. Finally, by aiming at evaluating the validity of the controller and the robust performance of the closed-loop system, the proposed regulator is implemented on a real-time mechanical vibrating system.

## 9.1 Problem formulation

Consider the open-loop system in the state-space form of

$$\begin{aligned} \dot{x} &= (A + \Delta A)x + (B + \Delta B)u + Hw + f, \\ y &= Cx, \end{aligned} \tag{9.1}$$

where $x \in \mathcal{R}^n$, $u \in \mathcal{R}^m$, and $y \in \mathcal{R}^q$ are the state, input, and output vectors, respectively. It should be noted that the dependencies of these vectors on time are suppressed for the sake of notation simplification. In addition, $f(x,t) \in \mathcal{R}^n$ and $w \in \mathcal{R}^p$ represent the vector of un-modeled/nonlinear dynamics and square-integrable disturbance, respectively. Moreover, $A \in \mathcal{R}^{n\times n}$, $B \in \mathcal{R}^{n\times m}$, $H \in \mathcal{R}^{n\times p}$, and $C \in \mathcal{R}^{q\times n}$ are the state, control input, disturbance input, and output matrices, correspondingly. $\Delta A \in \mathcal{R}^{n\times n}$ and $\Delta B \in \mathcal{R}^{n\times m}$ are the perturbation terms that are considered as modeling sources of uncertainty. The term $Hw$ is the contribution of unknown (but bounded) disturbance ($w(t)$) in system states. The matrix $H$ on the other hand is assumed to be known from the system modeling (white-, grey-, or black-box system identification). In contrast to the previous Chapter 8, it is assumed that the uncertainty matrices and un-modeled dynamics are norm-bounded as $\|\Delta A\| \leq a$, $\|\Delta B\| \leq b$, and $\|f(x,t)\| \leq g_m\|x\|$ with $a$, $b$, and $g_m$ being known positive constants. The uncertainty of disturbance input matrix can be considered as a new source of disturbance and therefore it does not appear in the model of the plant explicitly. By assuming the system (9.1) to be stabilizable with a nominal input matrix of full rank ($B$), the following system is introduced as the dynamics of the observer

$$\begin{aligned} \dot{\hat{x}} &= (A + \Delta A)\hat{x} + (B + \Delta B)u + L(y - \hat{y}), \\ \hat{y} &= C\hat{x}, \end{aligned} \tag{9.2}$$

in which, $\hat{x} \in \mathcal{R}^n$ and $\hat{y} \in \mathcal{R}^q$ are the observed state- and output-vectors, respectively. Also, $L$ symbolizes the observer gain. An alternative representation for observer dynamics is also used later based on the full-order Luenberger observer.

## 9.2 Sliding mode control

### 9.2.1 Sliding surface

The sliding surface is defined as

$$S(t) = B^+\hat{x} + z, \tag{9.3}$$

where, $B^+ \equiv (B^TB)^{-1}B^T$. It is assumed that the initial condition is $z(0) = -B^+\hat{x}(0)$ in order to satisfy $S(0) = 0$. In addition, $z$ can be calculated by solving the following dynamic equation

$$\dot{z} = -(B^+A - B^+LC + K)\hat{x} - B^+Ly, \tag{9.4}$$

with $K \in \mathcal{R}^{m\times n}$ being the designed observed-state feedback gain which moves the trajectory toward the sliding surface and keeps it on the manifold of (9.3). Derivation of the switching function with respect to time and using (9.4) and (9.2) gives

$$\dot{S} = B^+\Delta A\hat{x} + u + B^+\Delta Bu - K\hat{x}. \tag{9.5}$$

Without losing the generality of the problem, let us introduce the uncertain terms as

$$\Delta A = B\Delta A_m + \Delta A_u,$$
$$\Delta B = B\Delta B_m + \Delta B_u, \tag{9.6}$$

where $\Delta A_m = B^+\Delta A$, $\Delta A_u = \Upsilon\Delta A$, $\Delta B_m = B^+\Delta B$, $\Delta B_u = \Upsilon\Delta B$, and $\Upsilon = I - BB^+$. Based on the definition of the uncertain terms, there exist positive constants such that $\|\Delta A_m\| \le a_m$, $\|\Delta A_u\| \le a_u$, and $\|\Delta B_u\| \le b_u$. It should be noted that $(I + \Delta B_m)$ is required to be nonsingular for calculating the equivalent control law, which is guaranteed by assuming $\|\Delta B_m\| \le b_m < 1$. By substituting (9.6) in (9.5), the dynamics of the sliding function can be written as

$$\dot{S} = \Delta A_m\hat{x} + (I + \Delta B_m)u - K\hat{x}. \tag{9.7}$$

The essential constraint of sliding mode is satisfied if $S(t) = 0$ and $\dot{S}(t) = 0$ and therefore, the equivalent control effort of sliding mode can be expressed as

$$u_{eq} = -(I + \Delta B_m)^{-1}[\Delta A_m\hat{x} - K\hat{x}]. \tag{9.8}$$

As it can be seen in (9.8), the sliding mode control effort is only the function of the observed states. By substituting (9.8) in (9.1), the dynamics of the system can be described as

$$\dot{x} = Ax + B\Delta A_m x + \Delta A_u x - B\Delta A_m\hat{x} + BK\hat{x} - \tilde{B}\Delta A_m\hat{x} + \tilde{B}K\hat{x} + Hw + f, \tag{9.9}$$

where $\tilde{B} = \Delta B_u(I + \Delta B_m)^{-1}$. By defining $e = x - \hat{x}$ and using (9.9) and (9.2), the dynamics of the observation error can be calculated as

$$\dot{e} = (A + B\Delta A_m + \Delta A_u - LC)e + Hw + f, \tag{9.10}$$

In the upcoming section, the stability of the error dynamic in (9.10) is proven such that in the presence of the term $Hw + f$, the observer state vector $\hat{x}$ converges to $x$.

## 9.2.2 Stability analysis of the observer-based SMC

**Theorem 9.1** (Sufficient condition) *The uncertain system* (9.1) *with observer system* (9.2) *is quadratically stable and satisfies the* $H_\infty$*-norm bounded function of* $\|T_{yw}\| < \gamma$ *if there exist symmetric positive definite matrices* $X$ *and* $R \in \mathcal{R}^{n\times n}$, *positive constants* $\varepsilon_i, i = 1 \dots 10$, *and* $K, \widehat{K} \in \mathcal{R}^{m\times n}$ *such that the following LMI is satisfied*

$$\begin{bmatrix} \Sigma & \Psi & \Phi \\ & \Pi_1 & 0 \\ * & & \Pi_2 \end{bmatrix} < 0, \tag{9.11}$$

*where*

$$\Sigma = \begin{bmatrix} Y_{11} & -BK & H \\ & Y_{22} & RH \\ * & & -\gamma^2 I \end{bmatrix},$$

$$\Psi = \begin{bmatrix} a_u X & g_m X & \frac{a_m b_u}{(1-b_m)} X & \frac{b_u}{(1-b_m)} \widehat{K}^T & g_m X \\ 0 & 0 & 0 & 0 & 0 \\ 0 & 0 & 0 & 0 & 0 \end{bmatrix},$$

$$\Phi = \begin{bmatrix} 0 & 0 & 0 & 0 & 0 & 0 \\ aB^{+T}B^T & \frac{a_m b_u}{(1-b_m)} I & \frac{a_m b_u}{(1-b_m)} \widehat{K}^T & RB & R & R \\ 0 & 0 & 0 & 0 & 0 & 0 \end{bmatrix}, \tag{9.12}$$

$$\Pi_1 = \mathrm{diag}(-\varepsilon_2 I, -\varepsilon_3 I, -\varepsilon_4 I, -\varepsilon_6 I, -\varepsilon_{10} I),$$

$$\Pi_2 = \mathrm{diag}(-\varepsilon_1 I, -\varepsilon_5 I, -\varepsilon_7 I, -\varepsilon_8 I, -\varepsilon_9 I, -\varepsilon_{10} I),$$

$$Y_{11} = X^T A^T + AX + \widehat{K}^T B^T + B\widehat{K} + \sum_{i=1}^{7} \varepsilon_i I,$$

$$Y_{22} = A^T R + RA + C^T \hat{L}^T + \hat{L} C + \varepsilon_8 a_m^2 I + \varepsilon_9 a_u^2 I,$$

*The controller and the observer gain can be calculated as* $K = \widehat{K} X^{-1}$ *and* $L = R^{-1}\hat{L}$, *respectively. Consequently, the stability of the error dynamic in (9.10) is proven in the presence of the term* $Hw + f$, *i.e. the observer state vector* $\hat{x}$ *converges to* $x$.

**Proof.** Consider the following Lyapunov function

$$V(x, e) = x^T P x + e^T R e. \tag{9.13}$$

Following the quadratic stability Lemma in the previous chapter, the following inequality should be satisfied

$$\dot{V} - \gamma^2 w^T w + y^T y < 0, \tag{9.14}$$

consequently, we can derive

$$
\begin{aligned}
&\dot{V} - \gamma^2 w^T w + y^T y \\
&\qquad = x^T PAx + x^T A^T Px + x^T PB\Delta A_m e + e^T \Delta A_m^T B^T Px + x^T P\Delta A_u x \\
&\qquad + x^T \Delta A_u^T Px + x^T PBKx + x^T K^T B^T Px - x^T PBKe - e^T K^T B^T Px \\
&\qquad - x^T P\tilde{B}\Delta A_m x - x^T \Delta A_m^T \tilde{B}^T Px + x^T P\tilde{B}\Delta A_m e + e^T \Delta A_m^T \tilde{B}^T Px + x^T P\tilde{B}Kx \\
&\qquad + x^T K^T \tilde{B}^T Px - x^T P\tilde{B}Ke - e^T K^T \tilde{B}^T Px + x^T PHw + w^T H^T Px + x^T Pf \\
&\qquad + f^T Px + e^T RAe + e^T A^T Re + e^T RB\Delta A_m e + e^T \Delta A_m^T B^T Re + e^T R\Delta A_u e \\
&\qquad + e^T \Delta A_u^T Re + e^T RHw + w^T H^T Re + e^T Rf + f^T Re - e^T RLCe \\
&\qquad - e^T C^T L^T Re - \gamma^2 w^T w + x^T C^T Cx.
\end{aligned} \tag{9.15}
$$

**Lemma 8.2** in previous chapter can be then employed for the following terms

$$
\begin{gathered}
x^T PB\Delta A_m e + e^T \Delta A_m^T B^T Px \le x^T(\varepsilon_1 P^2)x + e^T\left(\varepsilon_1^{-1} a^2 B^{+T} B^T B B^+\right)e, \\
x^T P\Delta A_u x + x^T \Delta A_u^T Px \le x^T(\varepsilon_2 P^2 + \varepsilon_2^{-1} a_u^2 I)x, \\
x^T Pf + f^T Px \le x^T(\varepsilon_3 P^2 + \varepsilon_3^{-1} g_m^2 I)x, \\
-x^T P\tilde{B}\Delta A_m x - x^T \Delta A_m^T \tilde{B}^T Px \le x^T(\varepsilon_4 P^2 + \varepsilon_4^{-1} a_m^2 b_u^2/(1-bm)^2\,)x, \\
x^T P\tilde{B}\Delta A_m e + e^T \Delta A_m^T \tilde{B}^T Px \le x^T(\varepsilon_5 P^2)x + e^T(\varepsilon_5^{-1} a_m^2 b_u^2/(1-bm)^2)e, \\
x^T P\tilde{B}Kx + x^T K^T \tilde{B}^T Px \le x^T(\varepsilon_6 P^2 + \varepsilon_6^{-1} b_u^2/(1-bm)^2 K^T K)x, \\
-x^T P\tilde{B}Ke - e^T K^T \tilde{B}^T Px \le x^T(\varepsilon_7 P^2)x + e^T(\varepsilon_7^{-1} b_u^2/(1-bm)^2 K^T K)e, \\
e^T RB\Delta A_m e + e^T \Delta A_m^T B^T Re \le e^T(\varepsilon_8 a_m^2 I + \varepsilon_8^{-1} RBB^T R)e, \\
e^T R\Delta A_u e + e^T \Delta A_u^T Re \le e^T(\varepsilon_9 a_u^2 I + \varepsilon_9^{-1} R^2)e, \\
e^T Rf + f^T Re \le x^T(\varepsilon_{10} g_m^2 I)x + e^T(\varepsilon_{10}^{-1} R^2)e,
\end{gathered} \tag{9.16}
$$

using the inequalities in (9.16) and introducing $\mathcal{L}^T = [x^T \quad e^T \quad w^T]$, (9.15) can be reformatted in matrix form as

$$
\dot{V} - \gamma^2 w^T w + y^T y \le \mathcal{L}^T \Omega \mathcal{L}, \tag{9.17}
$$

where

$$
\Omega = \begin{bmatrix} \Omega_{11} & -PBK & PH \\ & \Omega_{22} & RH \\ * & & -\gamma^2 I \end{bmatrix}, \tag{9.18a}
$$

$$
\begin{aligned}
\Omega_{11} &= A^T P + PA + K^T B^T P + PBK + C^T C + \sum_{i=1}^{7} \varepsilon_i P^2 + \varepsilon_2^{-1} a_u^2 I + \varepsilon_3^{-1} g_m^2 I + \frac{\varepsilon_4^{-1} a_m^2 b_u^2}{(1-bm)^2} I \\
&+ \frac{\varepsilon_6^{-1} b_u^2}{(1-bm)^2} K^T K + \varepsilon_{10} g_m^2 I,
\end{aligned} \tag{9.18b}
$$

$$
\begin{aligned}
\Omega_{22} &= A^T R + RA + C^T L^T R + RLC + \varepsilon_1^{-1} a^2 B^{+T} B^T B B^+ + \frac{\varepsilon_5^{-1} a_m^2 b_u^2}{(1-bm)^2} I + \frac{\varepsilon_7^{-1} b_u^2}{(1-bm)^2} K^T K \\
&+ \varepsilon_8^{-1} RBB^T R + \varepsilon_9^{-1} R^2 + \varepsilon_{10}^{-1} R^2 + \varepsilon_8 a_m^2 I + \varepsilon_9 a_u^2 I,
\end{aligned}
$$

after pre- and post-multiplying (9.18a) by diag$(P^{-1}, I, I)$ and defining $P^{-1} = X$, $RL = \hat{L}$, and $KX = \hat{K}$, and then successive use of Schur Complement, LMI (9.11) can be achieved, and therefore, the proof is complete.

■

**Remark 9.1.** In (9.12), $K$ appears explicitly even after the introduction of $X$, therefore the following iterative algorithm should be used in order to find the sub-optimal feasibility solution of LMI (9.11).

**Algorithm 9.1** Consider an initial guess for $K$ and $\gamma$, namely $K_i, \gamma_i, i = 0$ and solve the LMI (9.11), in which all the terms in which $K$ explicitly is present, should be replaced by $K_0$.

Define $i = i + 1$ and use $K_i$ from the previous step in (9.12) to satisfy the feasibility problem (9.11) together with satisfying the linear matrix equality (LME) $K_{i-1}X = \hat{K}$. $X$ is obtained similarily to $\hat{K}$ by solving the LMI.

Replace $\gamma_i = \gamma_i - d\gamma$ with $d\gamma$ being the step size of $H_\infty$-norm bounded constraint of stability and go to step 1.

If for desired $\gamma^*$ the solution for $X$ in step 2 is found, then exit: $K = K_i$. The desired $\gamma^*$ is the minimum value which it can be achieved while keeping the LMI problem still feasible. Based on the values of $\hat{L}$ and $R$ in the last iteration associated with $K = K_i$, the observer gain can be calculated as $L = R^{-1}\hat{L}$.

**Remark 9.2.** The results in **Theorem 9.1** and **Algorithm 9.1** are obtained based on the observer dynamics (9.2) which uses the maximum information that is available about the plant i.e., neglecting unknown $Hw + f$. In the case when the perturbation terms ($\Delta A$ and $\Delta B$) in (9.2) are ignored, the following corollary can be used instead of **Theorem 9.1** and **Algorithm 9.1** to obtain the controller and observer gains.

**Corollary 1.** *The system* (9.1) *with state and input perturbations together with the state observer* $\dot{\hat{x}} = A\hat{x} + Bu + L(y - \hat{y})$ *is quadratically stable and satisfies the* $H_\infty$*-norm bounded function of* $\|T_{yw}\| < \gamma$ *(transfer function from disturbance to output) if there exist* $P, R \in \mathcal{R}^{n\times n}$*, positive constants* $\varepsilon_i, i = 1 \dots 4$*,* $\hat{P} \in \mathcal{R}^{m\times m}$*,* $\hat{L} \in \mathcal{R}^{n\times q}$*, and* $K, \hat{K} \in \mathcal{R}^{m\times n}$ *such that the LMI/LME* (9.19) *is satisfied for* $M_2 = N_2 = 0$*, and the obtained* $\hat{K}$ *satisfies the LMI/LME* (9.20)

$$\begin{bmatrix} \Sigma_{11} & \Sigma_{12} & PM_1 & PH & P & 0 \\ & \Sigma_{22} & RM_1 & RH & 0 & R \\ & & -\varepsilon_1 I & 0 & 0 & 0 \\ & & & -\gamma^2 I & 0 & 0 \\ & & & & -\varepsilon_3 I & 0 \\ * & & & & & -\varepsilon_4 I \end{bmatrix} < 0, \tag{9.19}$$

$$\Sigma_{11} = A^T P + PA + \hat{K}^T B^T + B\hat{K} + C^T C + \varepsilon_1 N_1^T N_1 + \varepsilon_3 g_m^2 I + \varepsilon_4 g_m^2 I,$$

$$\Sigma_{22} = A^T R + RA - C^T \hat{L}^T - \hat{L}C.$$

$$\Sigma_{12} = -PBK,$$

$$PB = B\hat{P}.$$

*Then* $K = \hat{P}^{-1}\hat{K}$ *which should be substituted in* (9.19)

$$\begin{bmatrix} \Lambda_{11} & \Lambda_{12} & PM_1 & PM_2 & PH & P & 0 \\ & \Lambda_{22} & RM_1 & RM_2 & RH & 0 & R \\ & & -\varepsilon_1 I & 0 & 0 & 0 & 0 \\ & & & -\varepsilon_2 I & 0 & 0 & 0 \\ & & & & -\gamma^2 I & 0 & 0 \\ & & & & & -\varepsilon_3 I & 0 \\ * & & & & & & -\varepsilon_4 I \end{bmatrix} < 0, \tag{9.20}$$

$$\Lambda_{11} = A^T P + PA + K^T B^T P + PBK + C^T C + \varepsilon_1 N_1^T N_1 + \varepsilon_2 K^T N_2^T N_2 K + \varepsilon_3 g_m^2 I + \varepsilon_4 g_m^2 I,$$

$$\Lambda_{22} = A^T R + RA - C^T L^T R - RLC + \varepsilon_2 K^T N_2^T N_2 K.$$

$$\Lambda_{12} = -PBK - \varepsilon_2 K^T N_2^T N_2 K,$$

$$PB = B\hat{P}.$$

*The controller and the observer gain can be calculated as $K$ using* (9.19) *and $L = R^{-1}\hat{L}$, respectively.*

**Proof.** Instead of decomposing the perturbation matrices in form of (9.6), $\Delta A = M_1\Gamma_1 N_1, \Delta B = M_2\Gamma_2 N_2$ are used in which $M_1, N_1, M_2$, and $N_2$ are known matrices with appropriate dimensions and $\Gamma_1^T\Gamma_1 \leq I,\ \Gamma_2^T\Gamma_2 \leq I$ are the time varying unknown terms. The equivalent control law can be calculated by following the same treatments as in (9.7) and (9.8) to be equal to $u_{eq} = K\hat{x}$. Then, replacing the equivalent control law in the state equation as well as the error dynamics and using the same Lyapunov function as in (9.13) and following the same steps as in **Theorem 9.1**, an LMI condition can be obtained. In this LMI equation, the terms with input and state perturbation matrices as in equivalent equation (9.16) in **Theorem 9.1**. Next applying the Schur Complement successively and finally introducing the LME condition $PB = B\hat{P}$, the LMI/LME (9.19) and (9.20) are obtained and therefore, the proof is complete.

■

**Remark 9.3.** The LMI conditions (9.19) and (9.20) are affined on the subject to the defined matrices by the introduction of the LME conversion ($PB = B\hat{P}$), which transforms a non-convex problem to a convex one. The non-singularity of $\hat{P}$ is guaranteed by the assumption that the input matrix $B$ is a full column rank.

After obtaining the sufficient condition for quadratic stability of the system on a sliding manifold of (9.3), the reachability condition should be guaranteed by means of ISMC law.

### 9.2.3 Reachability of ISMC

**Theorem 9.2** *(Reachability) Consider the uncertain system (9.1) with the switching function (9.3), for which the controller gain K is obtained by satisfying LMI (9.11) (sufficient condition is guaranteed). Then a stable sliding mode exists for the following SMC control input*

$$u = K\hat{x} - \alpha\frac{S}{\|S\|}, \tag{9.21}$$

*in which, for a positive scalar $\beta$, $\alpha$ is introduced as*

$$\alpha = \frac{1}{1 + b_m}\,[\beta + (a_m + b_m\|K\|)]\|\hat{x}\|, \tag{9.22}$$

**Proof.** Consider the new Lyapunov function as $V(S) = \frac{1}{2}S^T S$. Using (9.7), one can obtain

$$\dot{V} = S^T[\Delta A_m\hat{x} + (I + \Delta B_m)u - K\hat{x}], \tag{9.23}$$

then replacing $u$ from (9.21) into (9.23), $\dot{V}$ can be expressed as

$$\begin{aligned}\dot{V} &= S^T[\Delta A_m\hat{x} + \Delta B_m K\hat{x}] - \alpha\frac{S^T S}{\|S\|} - \alpha S^T\Delta B_m\frac{S}{\|S\|} \\ &\leq \|S\|[\|\Delta A_m\hat{x}\| + \|\Delta B_m K\hat{x}\|] - \alpha\|S\| - \alpha b_m\|S\| \leq -\beta\|S\|,\end{aligned} \tag{9.24}$$

Thus, the proof is complete.

■

**Remark 9.4.** Control law (9.21) has the major issue of difficulty in confirming the upper bound of the controller gain and the observed states. Therefore, a fuzzy system is introduced by replacing the switching term of (9.21) with a fuzzy logic interface.

### 9.2.4 Fuzzy SMC design

The switching function (9.3) can be decomposed as $S = [s_1 \dots s_i \dots s_m]^T$. Allowing for $s_i$ to be the input linguistic variable and $\epsilon_{F,i}$ the output linguistic variable, the fuzzy sets are defined as P (positive), Z (zero), and N (negative) for $s_i$ and PE (positive effort), ZE (zero effort), and NE (negative effort) for $\epsilon_{F,i}$. The fuzzy linguistic rules are introduced as

Rule 1: If $s_i$ is P, the $\epsilon_{F,i}$ is PE.

Rule 2: If $s_i$ is Z, the $\epsilon_{F,i}$ is ZE.

Rule 3: If $s_i$ is N, the $\epsilon_{F,i}$ is NE.

The input membership functions and output membership functions are assumed to be triangle-type and singleton-type, respectively [83]. The output of the defuzzification module can be expressed as

$$\epsilon_{F,j} = \frac{\sum_{k=1}^{3} \mu_{jk} \delta_{jk}}{\sum_{k=1}^{3} \mu_{jk}}, \tag{9.25}$$

where $0 \le \mu_{jk} \le 1$ is the firing strength of rule $1 \le k \le 3$. The center of the output membership functions is presumed to be $\delta_{j1} = \delta_j$, $\delta_{j2} = 0$, and $\delta_{j3} = -\delta_j$ [103]. Utilizing special input membership functions ($\sum_{k=1}^{3} \mu_{jk} = 1$) together with (9.25), the fuzzy controller can be reduced to

$$u_F = (\mu_{j1} - \mu_{j3})\delta_j. \tag{9.26}$$

Now by replacing the second term of integral sliding mode control effort in (9.21) with the new fuzzy sliding mode control law as $u = K\hat{x} + u_F$ and substituting this in (9.23), one can obtain

$$\dot{V} = S^T[\Delta A_m \hat{x} + \Delta B_m K \hat{x} + (I + \Delta B_m) u_F], \tag{9.27}$$

where $u_F = -\zeta[\epsilon_{F,1}, \dots, \epsilon_{F,m}]^T$ and $\zeta = -1/(1 + b_m)$. Then defining $\varphi(\hat{x}, t) = (\Delta A_m + \Delta B_m K)\hat{x} = [\varphi_1 \dots \varphi_m]^T$, the Lyapunov function (9.27) can be reformatted as

$$\dot{V} = \sum_{j=1}^{m} s_j \varphi_j - \zeta \sum_{j=1}^{m} s_j u_{F,j} - S^T \Delta B_m u_F. \tag{9.28}$$

Using (9.26), the following inequality can be stated as for (9.28)

$$\begin{aligned} \dot{V} &\le \sum_{j=1}^{m} |s_j \varphi_j| - \zeta \sum_{j=1}^{m} s_j (\mu_{j1} - \mu_{j3}) \delta_j + \zeta b_m \sum_{j=1}^{m} s_j (\mu_{j1} - \mu_{j3}) \delta_j \\ &\le \sum_{j=1}^{m} |s_j| [|\varphi_j| - |\mu_{j1} - \mu_{j3}| \delta_j], \end{aligned} \tag{9.29}$$

Based on (9.29), for $\dot{V} \le 0$, the following inequality should be satisfied

$$\delta_j > \frac{|\varphi_j|}{|\mu_{j1} - \mu_{j3}|}, j = 1, \dots, m. \tag{9.30}$$

It is proven by [221] that there exists an optimal solution for $\delta_j$ as $\bar{\delta}_j$, which cannot be determined explicitly because of the unknown uncertainty bounds. Hence, the following adaptive law is introduced to address this drawback.

**Theorem 9.3** *For the uncertain system (9.1) with control input $u = K\hat{x} + u_F$, where $K$ is obtained from solving the feasibility problem of (9.11) and $u_F$ from (9.24), $\delta_j$ can be replaced with an adaptive parameter $\hat{\delta}_j$ described as follows*

$$\dot{\hat{\delta}}_j = \beta_j s_j(\mu_{j1} - \mu_{j3}), \tag{9.31}$$

*where $\beta_j$ are positive scalars, then the stable adaptive fuzzy sliding mode controller exists that guarantees the $H_\infty$-performance.*

**Proof.** By defining the estimation error of $\delta_j$ as $\tilde{\delta}_j = \hat{\delta}_j - \bar{\delta}_j$, the following Lyapunov function is used to perform the stability analysis

$$V_1 = V + {}^1\!/_2 \sum_{j=1}^{m} \beta_j^{-1} \tilde{\delta}_j^2. \tag{9.32}$$

Thus using (9.28) and (9.31), we obtain

$$\dot{V}_1 = \sum_{j=1}^{m} s_j \varphi_j - \zeta \sum_{j=1}^{m} s_j u_{F,j} - S^T \Delta B_m u_F + \sum_{j=1}^{m} s_j(\hat{\delta}_j - \bar{\delta}_j)(\mu_{j1} - \mu_{j3}). \tag{9.33}$$

By using the introduced fuzzy control law of (9.26), fuzzy inequality law (9.30), and the definition for $\zeta$, (9.33) can be rewritten as

$$\begin{aligned}\dot{V}_1 &\le \sum_{j=1}^{m} |s_j| \big[|\varphi_j| - \bar{\delta}_j |\mu_{j1} - \mu_{j3}|\big] - \zeta(1 + b_m) \sum_{j=1}^{m} s_j \hat{\delta}_j(\mu_{j1} - \mu_{j3}) + \sum_{j=1}^{m} s_j \hat{\delta}_j(\mu_{j1} - \mu_{j3}) \\ &\le \sum_{j=1}^{m} |s_j| \big[|\varphi_j| - \bar{\delta}_j |\mu_{j1} - \mu_{j3}|\big] < 0,\end{aligned} \tag{9.34}$$

This completes the proof.

■

## 9.3 Experimental implementation of the closed-loop system

In this section, the performance of the observer-based controller is evaluated by the experimental implementation of the control system on a vibrating clamped-free beam shown in the previous chapter. Both of the controller and observer systems are modeled on the SIMULINK platform and then the created model is compiled and uploaded to the DAQ system in real-time (see Figure 6.1).

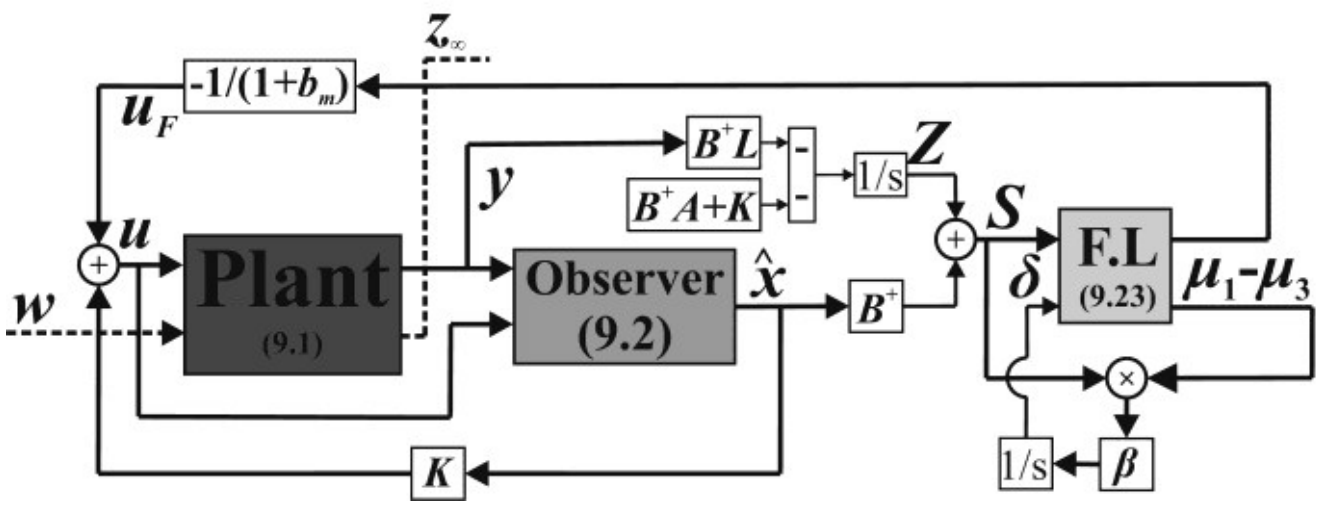

Figure 9.1 Schematic configuration of the control system [103]

In Figure 9.1, $z_\infty$ represents the designer-defined $H_\infty$-constraint on the transfer function, which is considered to be the transfer function from the disturbance ($w$) to the measured output ($T_{yw}$). In order to extract the mathematical model of the system, the subspace identification method is used. The identification process for calculating the system matrices in (9.1) such as $H$ matrix is explained in Chapter 4 [120], [121].

For controller design purposes, a nominal reduced-order system is identified based on the consideration of three mode-shapes of the piezolaminated beam. The higher-order dynamics are considered as the source of uncertainty in form of a norm bounded time-dependent terms. For investigation of the robust performance of the uncertain closed-loop system with the designed controller (9.11), **Algorithm 9.1** can be used. The frequency of the excitation signal is changed from zero to 170 Hz (1068.1 rad/s) in two cases of the open-loop and closed-loop system. The open-loop and the closed-loop systems are implemented on the real-time DAQ of the dSPACE with a sampling frequency of 10 kHz. Investigations are carried out in the time domain by means of the experimental setup shown in Figure 6.1.

**Remark 9.5.** In (9.1) and (9.2), mathematically, $\Delta A$ can be converted to or included in the form of unmodeled dynamics ($f(x,t) \in \mathcal{R}^n$) and the main purpose of this decomposition is the available uncertainty quantification methods in practice. In other words, uncertainty in structural vibrations can essentially be obtained based on the application of DRC. If the disturbance is active in high frequencies, then the effect of unmodeled/nonlinear dynamics in the nominal plant model is more important. The modeling procedure for these nonlinearities is discussed in Chapter 2. In this frequency range, after effective modeling of dynamics of low-frequency nature, the bound of the unknown additive uncertainty function, $f(x,t)$, in plant model can be obtained following the experimental method in [69], [222], [223]. Experimental results in this area are based on generating several nominally identical samples of structural systems, preparing identical and repetitive experimental procedure in exciting the structure, recording the response, and extracting the FRF of the system which hands the uncertainty bounds based on the structure of uncertainty. For instance, this method is employed by Kompella and Bernhard to measure FRF at driver microphones for 57 pickup trucks [224]. Similar results are reported by [69]. In contrast, if the frequency of the disturbance is not higher than the highest natural frequency of the vibrating system, the stochastic finite element method (SFEM) is an appropriate method for calculating the known matrices in mismatch uncertainty of $\Delta A$ and $\Delta B$ [225].

The quantification of uncertainty is not carried out in the dissertation. Instead, the controller is developed by increasing the $a$, $b$, and $g_m$ variables to account for the maximum amount of uncertainty. This results in a conservative (but robust) control system. The response of the system for the controlled and uncontrolled case is shown in Figure 9.2 in time-domain based on the measurement signal generated from the Doppler vibrometer, which is further fed to the dSPACE ADC board (DS2004). In addition, FRF of the open-loop system is compared with the closed-loop one in the frequency range of $[0 \quad 170]$ Hz in Figure 9.2.

Figure 9.2 shows the measured voltage that is collected by the dSPACE ADC board based on the measurement of the laser vibrometer with the sensitivity factor of 1 Volt corresponding to $125\ \mathrm{mm/sec}$. It can be seen that the controller design, based on the reduced-order identified model, suppressed the vibration magnitude within the considered frequency range. In addition, the corresponding control efforts that are generated for piezo-actuator patches by the dSPACE DAC board are shown in Figure 9.3. Figure 9.3 shows that the control efforts applied to each of the piezo-actuators are limited to a maximum of 50 V without any sudden jump. It should be noted that the first actuator refers to the actuator close to the clamped end of the cantilever beam and the second actuator refers to the neighboring actuator in Figure 6.1. The experimental results show that the robust control system performs well in attenuating the vibration amplitude in the presence of structured uncertainties in system matrices. In addition, the observation error of the output ($y - \hat{y}$) in the presence of disturbance (realized as sweep sine signal) is depicted in Figure 9.4.

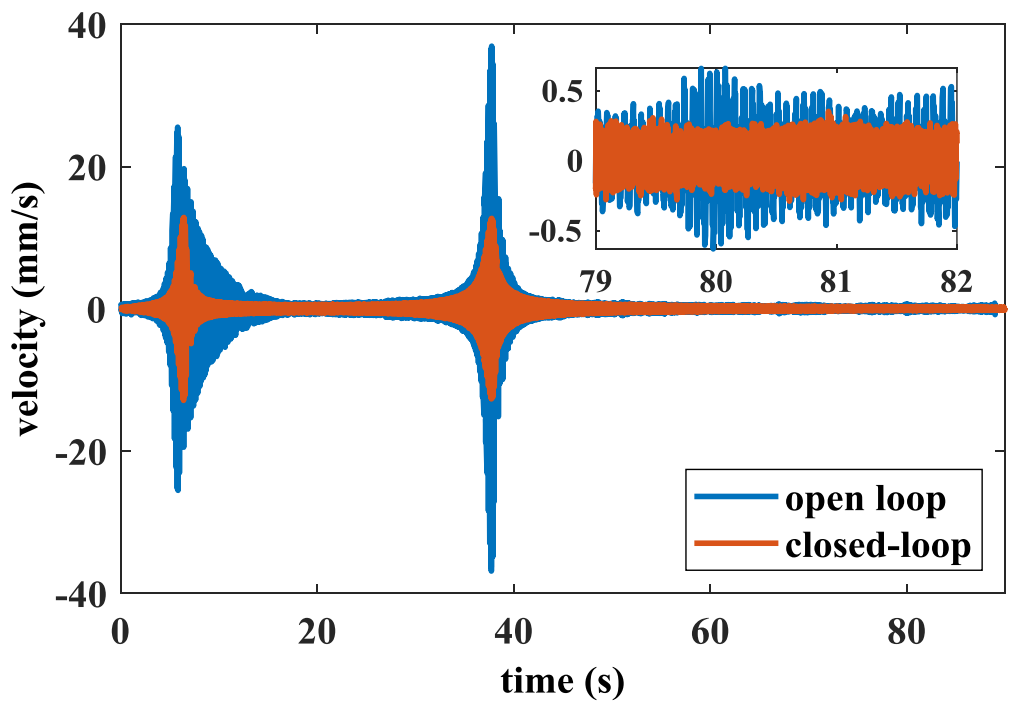

Figure 9.2 (a) Experimental comparison of measured outputs in time-domain [103]

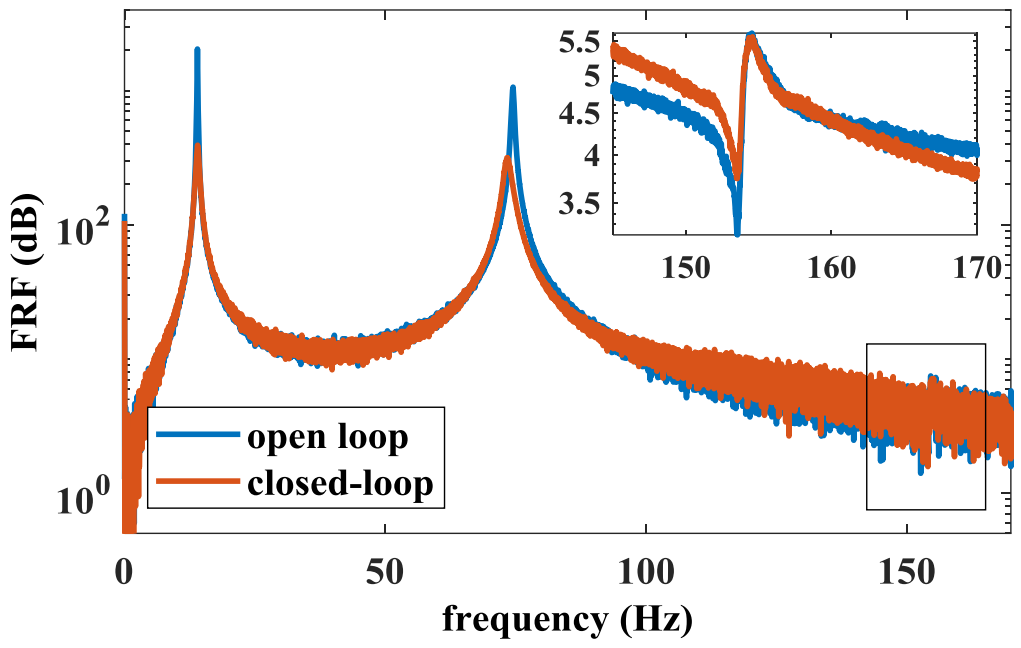

Figure 9.2 (b) Experimental comparison of FRF of the systems [103]

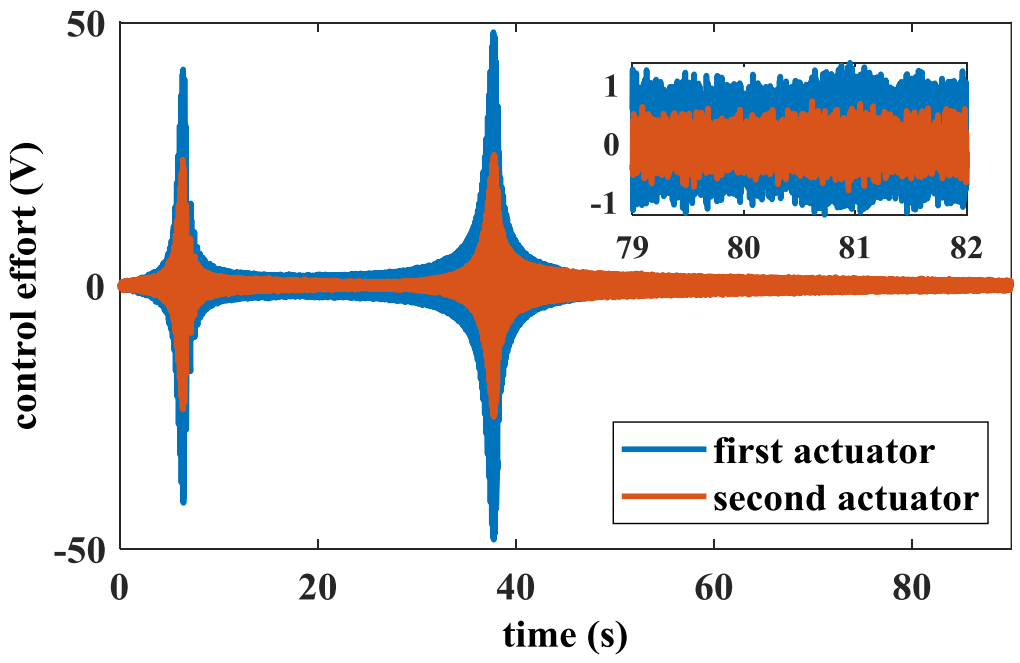

Figure 9.3 Control effort of the piezo-patch actuators [103]

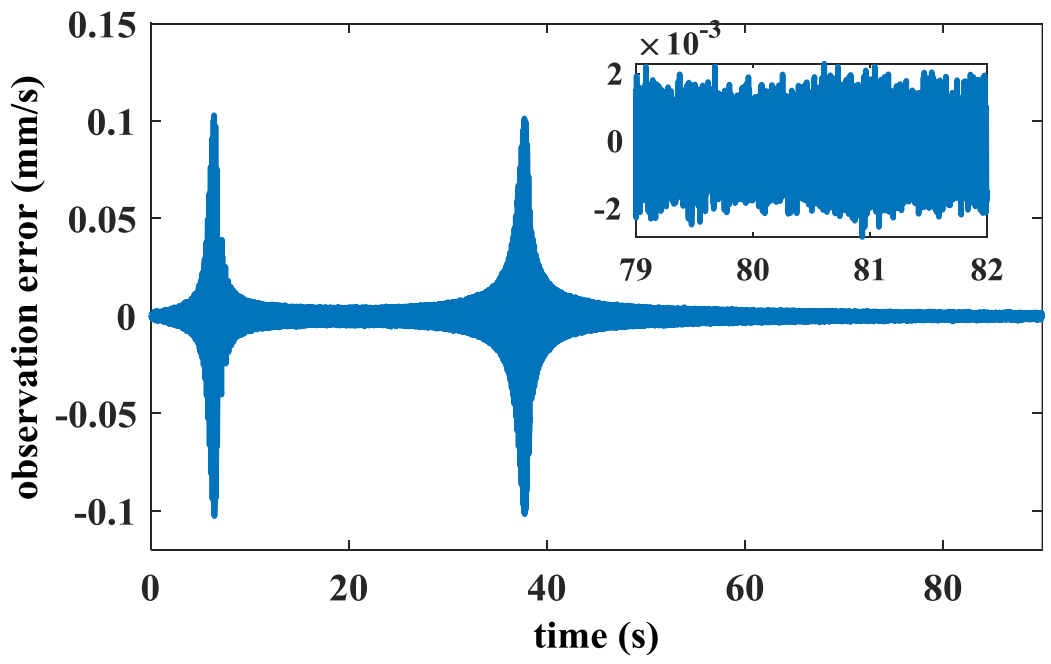

Figure 9.4 Observation error of the observer system [103]

In order to investigate the robust performance of the designed system in high frequency, where the unmodeled dynamics can affect the response of the system, the real structure is excited with a chirp signal with a frequency between 180 and 220 Hz. The measurement output and the applied control effort are presented in Figure 9.5 and Figure 9.6, respectively. It should be mentioned that in order to reject the measurement noise a low-pass fourth-order Butterworth filter with passband frequency equal to 230 Hz is implemented on the measurement channel. It can be seen, in Figure 9.5 and Figure 9.6, that the observer-based controller can handle the unmodeled dynamics with a limited control effort. In addition, the peak value in Figure 9.5 and Figure 9.6 is representing the higher-order natural frequency of the system which is equal to 204.6 Hz. In order to study the spillover effect, the structure is excited by mechanical initial displacement.

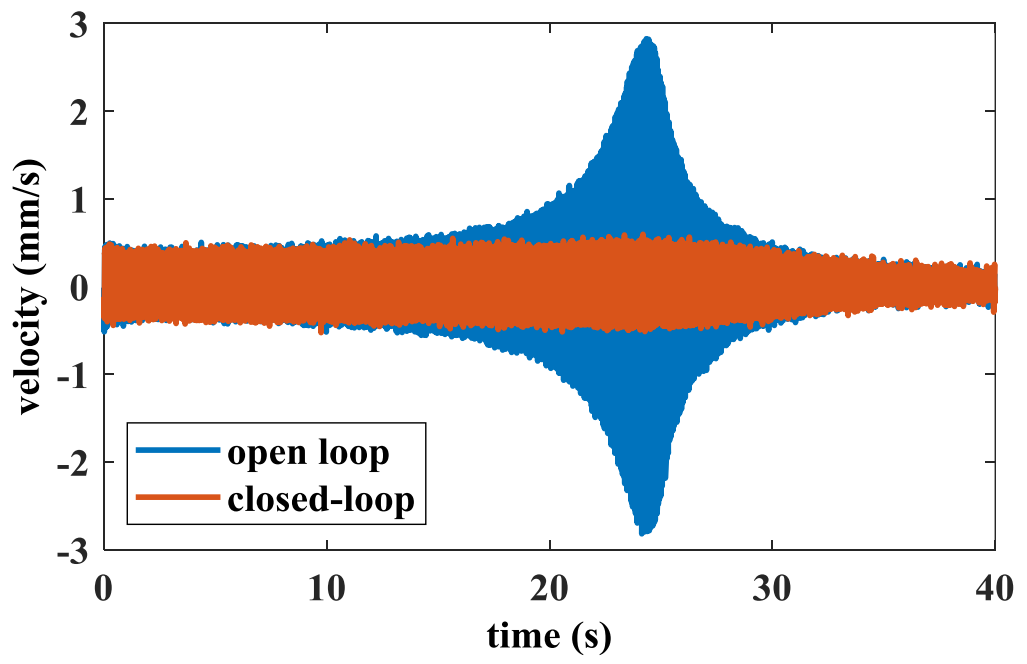

Figure 9.5 Experimental comparison of measured outputs for unmodeled dynamics [103]

The comparison of the open-loop and closed-loop system outputs, i.e. $y$ against $\hat{y}$ is depicted in Figure 9.7 (a). This shows the performance of the controller is suppressing the disturbance in system output i.e., $\|T_{yw}\|_{\infty}$. In addition, the corresponding applied control signals are shown in Figure 9.7 (b) for the two piezo-actuators.

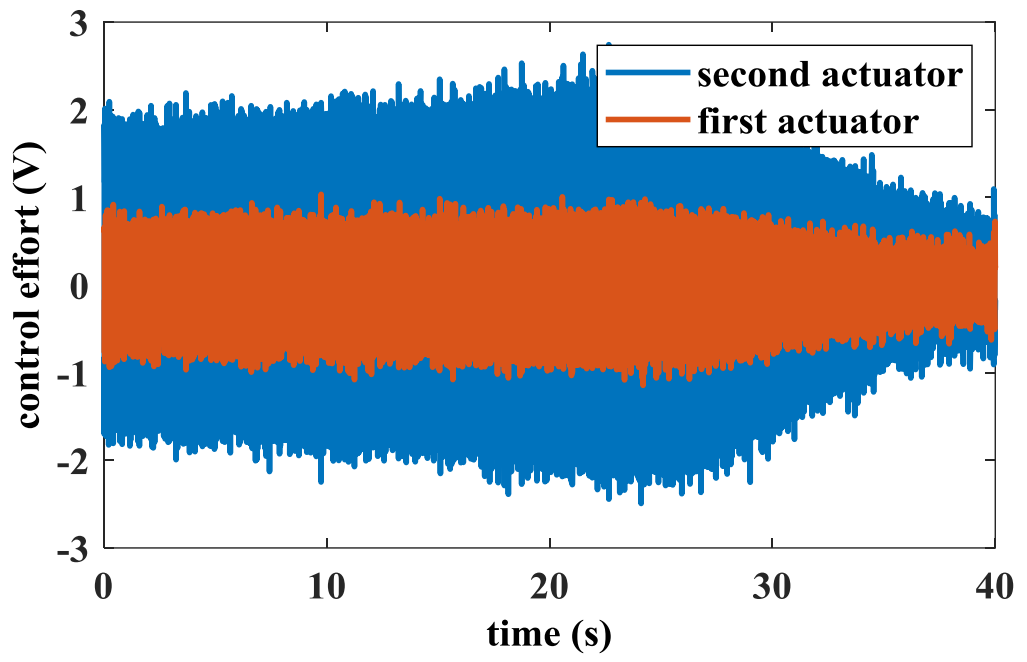

Figure 9.6 Control effort of the piezo-patch actuators for unmodeled dynamics [103]

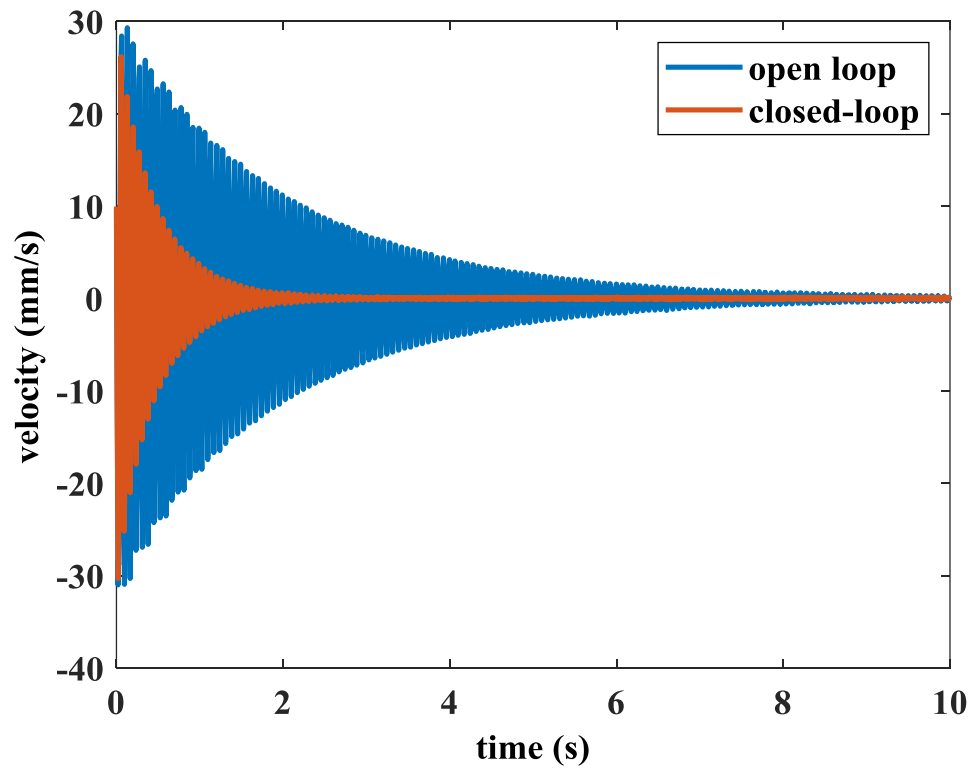

Figure 9.7 (a) Comparison of measured outputs [103]

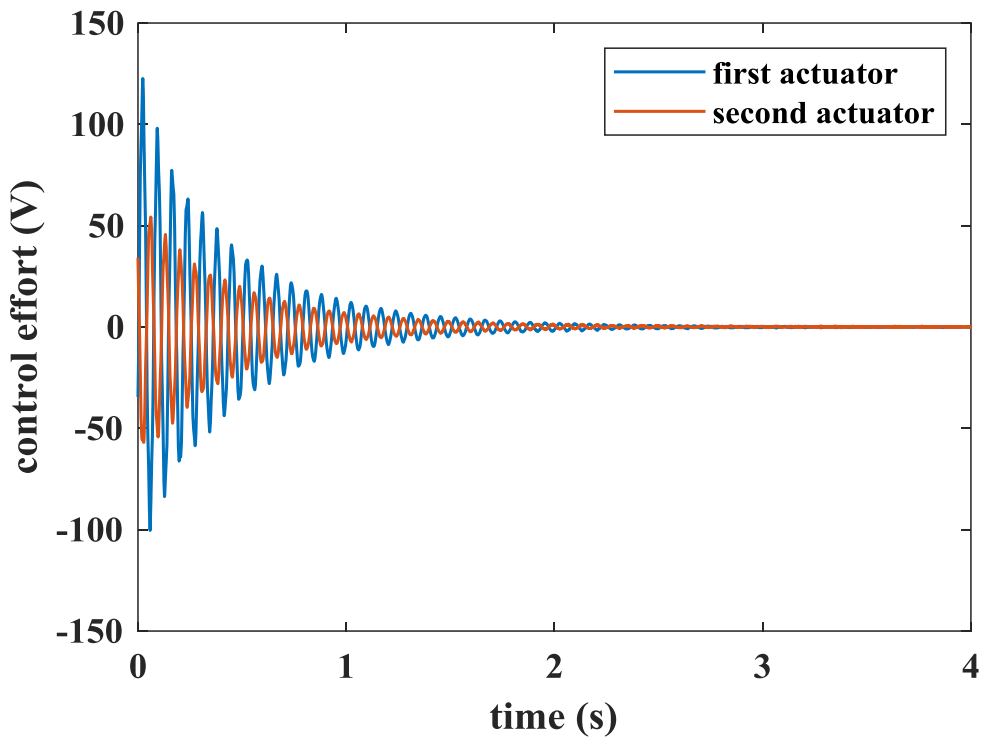

Figure 9.7 (b). Control effort of the piezo-patch actuators [103]

The fast Fourier transformation (FFT) analysis is performed on the applied control efforts to identify the dominant frequencies of the controller.

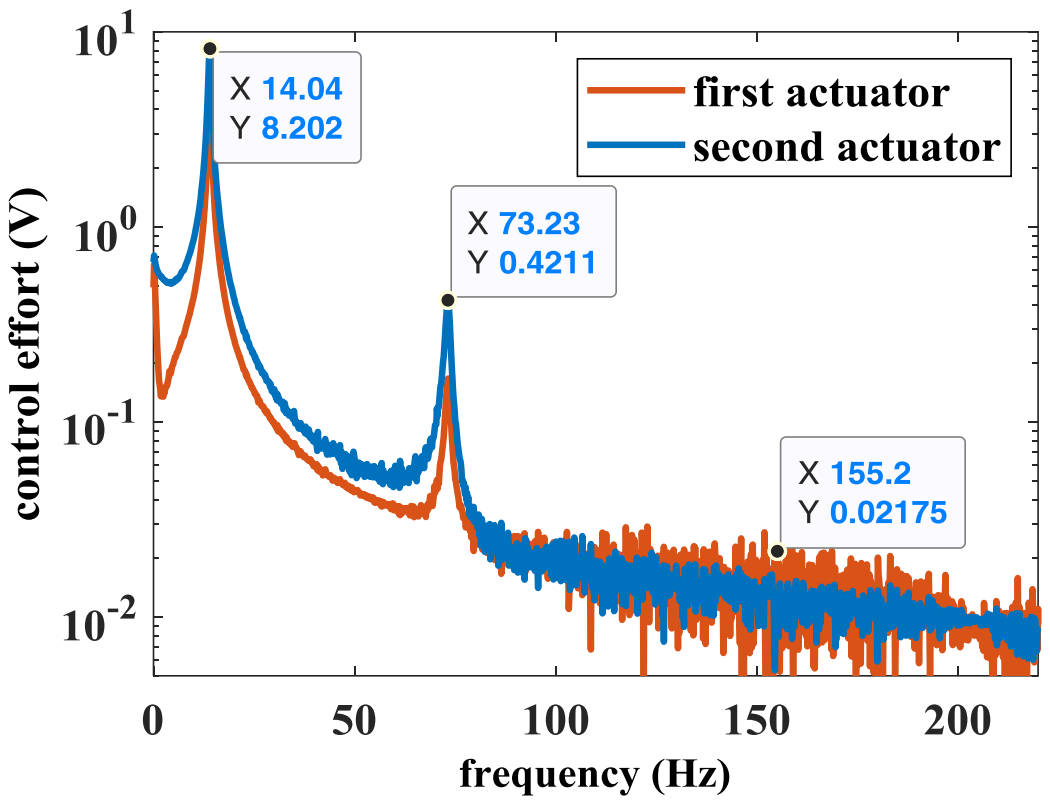

Figure 9.8 FFT of control effort for initial displacement excitation [103]

Figure 9.8 shows that the applied control efforts in frequency-domain between zero and 220 Hz have their significant effect around the first and second eigen-frequencies, however, as one can see, the third mode shape of the structure with natural frequency around 150 Hz, has no significant effect in the controller input. Moreover, the higher-order dynamic that is studied in Figure 9.5 and Figure 9.6 is not excited by the control input which shows the rejection of the spillover effect by the proposed control system. In Figure 9.8, the noisy behavior of the control effort is mainly due to transient and nonlinear distortion and low SNR at high frequencies. It should be noted that no proper modal analysis is performed here to obtain the FRF of the system. Finally, the evolution of the sliding surface, ($S(t)$) in (9.3), for two cases of forced vibration under chirp excitation and free vibration with initial displacement is depicted in Figure 9.9 (a) and Figure 9.10 (b). These two figures indicate a stable switching function from the initial time for the closed-loop system based on the proposed observer-based AFISMC.

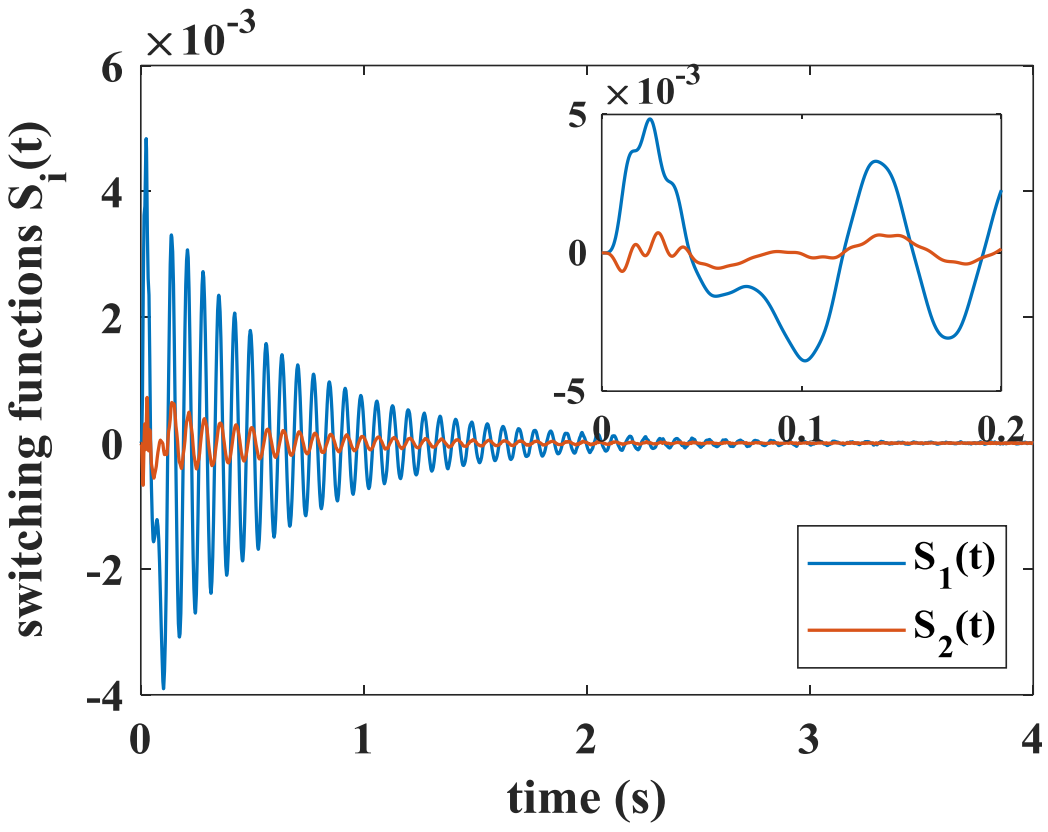

Figure 9.9 (a) Switching functions: free vibration [103]

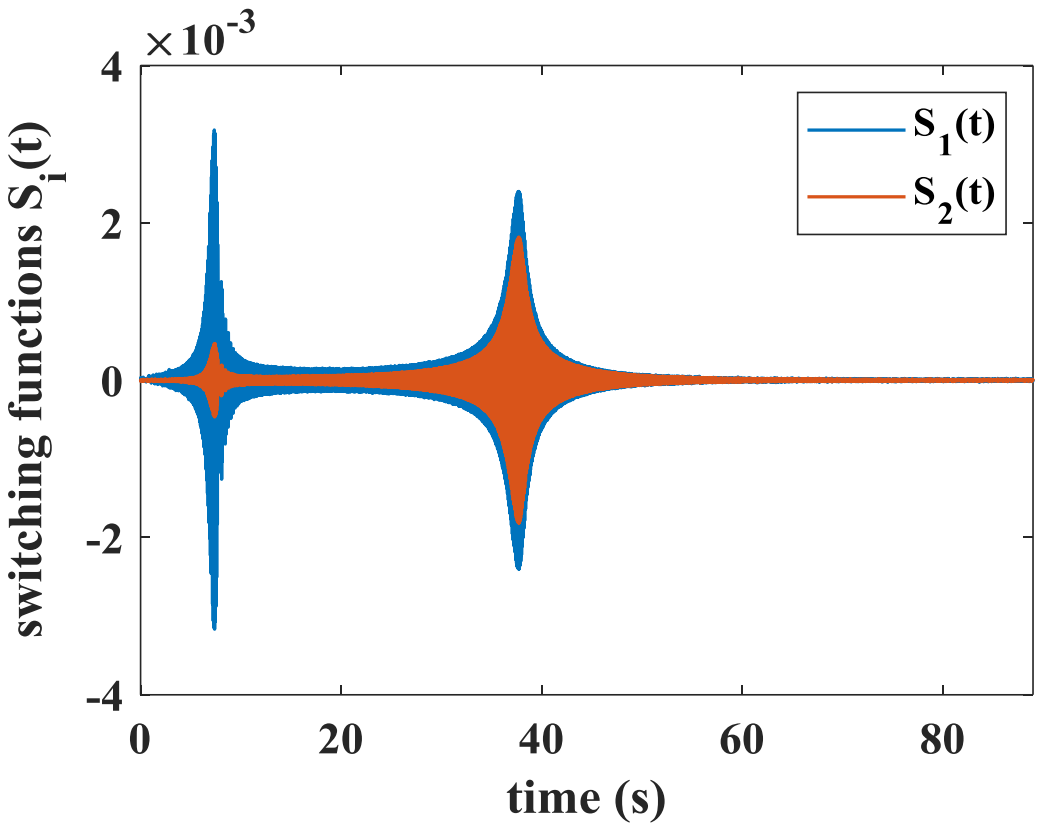

Figure 9.10 (b) Switching functions: forced vibration [103]

The following remarks should be made:

1) The chattering behavior of the SMC is not shown in the numerical/experimental results of Chapter 9.3 because this is known as the main drawback of SMC with $sign(.)$ as the switching function [226].

2) In the formulation of the plant model no additional nonlinearity is considered. However, the proposed sliding-mode controller is capable of dealing with system nonlinearities as a high gain approach. The AVC benchmark setup of the dissertation does not show a significant amount of nonlinearities as proven in Chapter 4.

In this chapter, the extent to which a fuzzy SMC can be used to reject the disturbance in the presence of unknown dynamics is shown. It is assumed in both Chapters 8 and 9 that the nominal part of the plant model is time-invariant. In some applications, due to the nonlinearity of the system or the wide bandwidth of the disturbance, this assumption could be subjected to imperfection. Consequently, the plant model and its order could be time-varying. In the next chapter, this line of uncertainty rejection is tackled with a more suitable approach namely, the *neural-network*-based approach. This will allow us to tackle a more general class of nonlinearity by constantly learning the system model and tuning the corresponding model-based controller accordingly.

# 10 Dealing with unmodeled dynamics: Neural network-based approach

In this chapter, a model-based output feedback recurrent wavelet neural network (RWNN) controller is proposed for a class of nonlinear MIMO systems with time-varying matched/mismatched uncertainties. In contrast to the previous chapter where the controller design heavily relies on the structure of nonlinearity/uncertainty and model accuracy, the proposed solution in this chapter mostly benefits from the learning capability of artificial neural networks. The proposed RWNN emulator adaptively trains to follow (mimic) an ideal state-feedback controller which is designed on the underlying linear model (ULM) of the plant. Simultaneously, the control system employs an adaptive neural network (NN) mechanism to estimate the mismatch between the RWNN controller and this ideal control law and consequently improve the plant's model. As a result, the conservatism associated with the classical robust control methods where the controller is synthesized based on worst-case bounds is dealt with based on the adaptive nature of the RWNN. Moreover, in order to generalize the subjected class of the investigated plants, the echo-state feature of adaptive RWNN is used to contribute to the performance of nonminimum phase systems. Accordingly, in the context of flexible smart structures with non-collocated sensor/actuator configuration, delayed feedback is added in the network which brings about a better match between the model output and the measured output. As a result, even for systems with an unknown Lipschitz constant of lumped uncertainty, the controller trains to compensate with an additional revision of the control law following some Lyapunov-based adaptive stabilizing rules. Additionally, the current approach is proposed as an alternative to the hot topic of nonlinear system identification-based control synthesis where the exact structure of the nonlinearity is required.

## 10.1 A short review of the state-of-the-art in AI-based DRC

The conventional controllers are often synthesized to deliver only a specific control performance. In contrast, the neuro-controllers adapt online to satisfy the time-varying performance objectives in a supervised/unsupervised fashion depending on the available computational power. This distinguishing factor enables the NN to detect and learn extremely involved and nonlinear mappings [92]. Moreover, various sources of nonlinearities in elastic light-weight mechanical structures make it impossible to use simplistic feedforward neural network excluding the tapped delay [93]. In other words, especially in any ANC scheme and AVC of non-collocated input/output (IO) configuration, where the measurement delay is the natural feature of the plant, an adaptive mechanism should be employed in synthesizing the control input signal associated with nonlinear output measurements. Two practical realizations of these nonlinearities in smart structures are the geometrical nonlinearities due to high vibration amplitudes where the linear models are not valid anymore [5] and the non-collocated sensor/actuators configurations where an inherent delay is an inevitable feature of the IO relation [94]. Providing technical details of the nonlinearities mentioned earlier is out of the scope of this dissertation. However, it is worth mentioning that the author is interested in dealing with model-based output feedback nonlinear control design where the effects of these nonlinearities are treated directly subjected to the fact that the quantifications of these nonlinear terms are available in the nominal model of the plant. Recently, methods such as the Reverse Path method [227], [228], and Nonlinear Normal Model method [229], [230] gained much attention for analyzing and quantifying these geometrical nonlinearities. The interested reader is highly recommended to refer to [122] for in-depth technical details. An alternative to such modeling approaches is the combination of semi-analytical techniques, e.g. [5], in structural modeling of nonlinear systems and parameter optimization techniques in the Grey-box System Identification framework [231]. Accordingly, an adaption of the black-box polynomial nonlinear state-space (PNLSS) identification approach i.e., [232] to the geometrically nonlinear system is ongoing research. For AVC purposes, the control design process for such an elaborate nonlinear model is a complicated task. In contrast, the proposed controller in this chapter (see [233]) relies on the underlying linear model (ULM) and treats the nonlinearities in an online framework. Consequently, the comparison of the proposed neuro-controller in this chapter and the nonlinear ones developed based on Grey-box system identification approaches can be used to determine if the latter approach is justifiable. This reasonable comparison not only assesses the justifiability of going through complex modeling procedure in AVC but also can be used as an alternative solution in the case-studies where the uncertainty/nonlinearity detection, characterization, and quantification of the three-step paradigm in [234] is not possible for technical reasons e.g. experimental costs and accessibility issues.

On the other hand, the nonlinear model-free robust control schemes based on the upper-bounds of the norm of the disturbance signals and matched-/mismatched-uncertainties such as high-gain, variable-structure, and fuzzy methods can be used in combination with NN to address nonlinearities in system dynamics [6]. For instance, Jnifene and Andrews proposed a combination of a fuzzy logic controller and neural networks to regulate the end-effector vibration in a flexible smart beam positioned on a two DoF platform [95]. He *et al*. employed a neural network for modeling the dynamics of a flexible robot manipulator subjected to input deadzone. They have used radial basis function NN to capture the deadzone and designed a high-gain observer-based NN controller [96]. Moreover, it is reported that one of the major deficiencies of standard control systems based on linear time-invariant nominal models is the evolution of plant dynamics w.r.t. time and the actuator windup problem both of which add to the nonlinearity of the system [97]. Accordingly, Li et al. introduced a genetic algorithm (GA) based back-propagation neural network suboptimal controller to address the vibration attenuation of a nine DoF modular robot [91].

The non-minimum phase vibrating systems with non-collocated actuator/sensor placements with centralized control configuration may have right-half plane (RHP) zeroes which can significantly restrict the closed-loop performance as reported in [235]. Detailed analysis of the tradeoff imposed on the performance of these nonminimum phase systems at different frequencies (in linear control theory) is previously reported by Freudenberg and Looze [236]. Alternatively, the echo-state feature of adaptive RWNN is investigated in this chapter to contribute to the performance of nonminimum phase systems.

To put it in a nutshell, the following contributions are reported. Instead of worst-case analysis based on the classical robust control methods (as in the previous two Chapters 8 & 9) and on the grounds of following adaptive nonlinear (non-conservative) control synthesis in AVC framework, a network is assigned to identify any dynamics that cannot be fit into the LTI framework, or the identified system fails to capture. This feature together with the *generalization* and *information storing* capabilities of NN opens the possibility of further investigations based on the nonlinear disturbance observer-based control (DOBC) as a hot topic in modern DRC [106]. An ideal controller is derived in terms of the tracking error of the estimated system state, and an adaptive recurrent wavelet neural network (RWNN) is configured to imitate the perfect controller. An advantage of such configuration compared to the asymptotic stabilizing techniques is that the mismatch between the ideal and the realized control law is not left alone. In other words, although the network parameters of RWNN are adaptively tuned following the Lyapunov stabilizing scheme, an additional observer is assigned to identify the error bounds and reject them in the tracking error dynamics. This feature may significantly contribute to the transient performance of the neural controller especially for the application of smart structures where the frequency range of interest may encompass up to hundreds of states which cannot be all considered in the nominal model of the system for obvious reasons. Note that the WNN-control systems benefit simultaneously from the learning capabilities of the artificial NN as well as the identification strength of the wavelet decomposition [237], [238]. Two contributions in terms of applicability of the proposed NN-based control system in real implementations are reported as a) Unlike the state-feedback schemes suggested in the literature (*i.e.,* [239], [240]), and similar to the output feedback neural control in [241], [242], the proposed technique is practical for smart structures where the continuous real system has infinite dynamics (represented with states) which cannot be measured individually. However, following the actuator/sensor placement criteria proposed in the literature (e.g. [20], [129], [163]) that reserve the observability conditions in smart structures, the neural network-based state-observer may provide an accurate measure of system states in real-time. The nonlinear-in-parameters neural network (NLPNN) used in observer design is capable of handling the nonlinearities without *a priori* known dynamics. The modified backpropagation (BP) algorithm is therefore implemented to realize the learning process. For this purpose, the idea in [243] is followed. b) The stabilization of the nominal model of the plant as pointed in [244] encompasses a delay which is separated from the transfer function of IO results in the minimum phase model. This indicates that similar to echo-state NN e.g. [245], the network can capture the delay and the problem of stabilizing a non-minimum phase system (as an alternative to [246]) is a much easier task. It should be noted that the NN-based control methods that include the input nonlinearities such as actuation saturation, deadzone, and output constraints such as [247], [248] are out of the scope of this Dissertation.

In the experimental implementation of the proposed technique of this chapter, a comparison with a standard approach (LQG) is performed. Meanwhile, for some AVC applications (even with complex geometries), systematical methods are available in the literature for providing the measure of IO delay e.g. [49].

## 10.2 Recursive Wavelet Neural Network Controller

The choice recursive network as pointed out in 10.1 is due to its capability to have a memory. Consequently, in the application of AVC where the sensor/actuators are not collocated (see Figure 6.1), the inherent IO delay can be captured in the model and compensated for by the NN-based controller. The choice of wavelet network on the other hand is because of its superior learning capabilities based on wavelet decomposition.

### 10.2.1 Problem statement

Before presenting the problem formulation, it should be noted that the notation that is commonly used in the literature of neural networks is not readily compatible with the problem formulation of control systems in this dissertation. Therefore without loss of generality, this common notation will be adapted for the control problem in this dissertation in this section. Let's consider an $n$-th order multi-input multi-output (MIMO) nonlinear system which has the following state dynamics (10.1). In the state-space representation of the control systems, typically $n = 1$ is assumed.

$$x^{(n)} = f(x) + G(x)u + E(x)d + h(x,u), \tag{10.1}$$

where $x = [x_1, x_2, \dots, x_m]^T \in \Re^m$ is the state vector, $u = [u_1, u_2, \dots, u_p]^T \in \Re^p$ is the control input vector, and $d = [d_1, d_2, \dots, d_q]^T \in \Re^q$ is the disturbance vector. In addition, $f(x) \in \Re^{m\times m}$, $G(x) \in \Re^{m\times p}$, and $E(x) \in \Re^{m\times q}$ are the state, input, and disturbance matrices, respectively. $h(x,u) \in \Re^m$ is an unknown function that represents the unmodeled dynamics of the system, and it is assumed to be represented by a Lipschitz stable function. Inherent delay due to actuator/sensor communications in the closed-loop system is categorized as one example of the various sources of nonlinearity in addition to uncertainty, and measurement noise, which should be compensated by the control algorithm. A poor mathematical model of the system can be dealt with by incorporating $h(x,u)$ in the nominal plant model. The linearization of (10.1) with uncertainties in system matrices around the system equilibrium point in the state space form for $n = 1$ can be represented by

$$\dot{x} = (A + \Delta A)x + (B + \Delta B)u + (H + \Delta H)d + h(x,u),$$
$$y = (C + \Delta C)x, \tag{10.2}$$

where $A \in \Re^{m\times m}$, $B \in \Re^{m\times p}$, $H \in \Re^{m\times q}$, and $C \in \Re^{r\times m}$ are nominal state, input, disturbance, and output matrices, respectively. Additionally, $\Delta A \in \Re^{m\times m}$, $\Delta B \in \Re^{m\times p}$, $\Delta H \in \Re^{m\times q}$, and $\Delta C \in \Re^{r\times m}$ are the unknown and time-dependent uncertainties from the state, input, disturbance, and output matrices, respectively. $h(x,u)$ is the addition of an unknown term that could be nonlinear in general. (10.2) is compatible with the problem formulation notation of previous chapters. When neglecting the model uncertainties, (10.2) becomes

$$\dot{x} = Ax + Bu + Hd + h(x,u),$$
$$y = Cx, \tag{10.3}$$

Next, an ideal control law for tracking problem is constructed in terms of the reference states $x_d$ and system states $x$**.** Accordingly, the tracking controller minimizes the state tracking error $e = x_d - x \in \Re^m$, where $x_d = [x_{d1}, x_{d2}, \ldots, x_{dm}]^T$ is the time-varying reference trajectory that is dependent on time $(x_d(t))$ is suppressed in the notation here and after for the sake of brevity. Note that the system states are not available for measurements and as a result, the reference control law is not realizable in practice. RWNN is partially responsible for tracking this control law using the estimated state vector $\hat{x} \in \Re^m$ which is handed from dynamical state observation mechanism given as

$$\dot{\hat{x}} = A\hat{x} + Bu + Hd + h(\hat{x},u) + G(y - C\hat{x}),$$
$$\hat{y} = C\hat{x}, \tag{10.4}$$

where $G \in \Re^{m\times r}$ is the observer gain and is selected such that $A - GC$ is a Hurwitz matrix. $G$ is guaranteed to exist since $A$ can be selected such that the pair $(A, C)$ is observable. Using the well-known property that a three-layer neural network can approximate nonlinear systems with any degree of nonlinearity [249], the unknown dynamics $h(x,u)$ can be expressed as

$$h(x,u) = Ws(Va) + \Delta, \tag{10.5}$$

where $a = [x\ u] \in \Re^{m+p}$, $V \in \Re^{i\times(m+p)}$, and $W \in \Re^{m\times i}$ are the input vector, the weighting matrices of the input layer and output layer of the neural network, respectively, and $i$ being a design parameter and equals to the number of neurons in the hidden layer of the NN. $\Delta$ is the neural network approximation error while $s(.)$ is the activation function of the network which is defined here and after as a sigmoid function $s(Va)$. Substituting (10.5) in (10.4) yields the neuro-observer dynamics as

$$\dot{\hat{x}} = A\hat{x} + Bu + Hd + Ws(Va) + \Delta + G(y - C\hat{x}),$$
$$\hat{y} = C\hat{x}. \tag{10.6}$$

Next, by assuming that there exists an ideal controller $u^*$ in the form of

$$u^* = Ke \in \Re^p, \tag{10.7}$$

where $K \in \Re^{p\times m}$ is the feedback gain matrix to be calculated. Applying the ideal controller (10.7) into the first equation (10.2), the tracking error dynamics $\dot{e} \in \Re^m$ is derived as $\dot{e} = (A + BK)e - Hd - h(x,u) - \omega(x) + \dot{x}_d - A_n x_d$, with $\omega(x) \in \Re^m$ being the lump uncertainty defined as: $\Delta Ax + \Delta Bu + \Delta Hd$. It should be noted that in the definition of $\omega(x)$, the convergence of observer is presumed which is discussed later. If the feedback gain matrix $K$ is chosen such that $A + BK$ becomes Hurwitz stable, it can be shown that $\lim_{t\to\infty} \| e \| = 0$ for any initial conditions. However, since the uncertainties from (10.2) are always unknown in practical applications, the ideal controller in (10.7) cannot be precisely obtained. Accordingly, an RWNN controller is employed to approximate the ideal controller which has a strong learning ability and high uncertainty tolerance [250]. In the next section, first, the structure of RWNN is briefly introduced, $h(x, u)$ is replaced with $h(\hat{x}, u)$, and then RWNN is adapted for tracking the ideal observed-state feedback control law in (10.7).

## 10.3 RWNN-based Adaptive Controller System

### 10.3.1 Description of the RWNN

Figure 10.1 represents a MIMO recursive wavelet neural network (RWNN), in which $T$ is the time delay that will feedback the important information from the last iteration. RWNN as a subclass of the topologies in the recurrent neural networks (RNN) (refer to [251], [252]) includes self-loops and backward connections which enables the network to commemorate earlier response of the system. The architecture of this RWNN comprises an input layer, a wavelet layer, and an output layer. The wavelet basis function for the wavelet layer is selected as

$$\phi_{ij} = h(z)e^{-\sigma_{ij}^2(z_{rij}-m_{ij})^2}, \tag{10.8}$$

For $i = 1,2,\ldots,L$, where $L$ is the number of inputs in the input layer, and $j = 1,2,\ldots,M$, where $M$ is the number of wavelons in the wavelet layer, and $h(z) = 1 - \sigma_{ij}^2 z_{rij}^2$ is the well-known "Mexican hat" mother wavelet function [253]. WNN can enrich the mapping of IO data using the activation function and achieve faster learning as pointed out in [254], [255]. Additionally, $z = [z_1, z_2, \ldots, z_L]^T \in \Re^L$ is the input vector to the wavelet function, $m_j = [m_{1j}, m_{2j}, \ldots, m_{Lj}]^T \in \Re^L$ is the wavelet dilation parameter, and $\sigma_j = [\sigma_{1j}, \sigma_{2j}, \ldots, \sigma_{Lj}]^T \in \Re^L$ is the wavelet translation parameter. The input to the function with recurrent feedback can be selected as

$$z_{rij}(t) = z_i(t) + r_{ij}\phi_{ij}(t - T), \tag{10.9}$$

where $r_j = [r_{1j}, r_{2j}, \ldots, r_{Lj}]^T \in \Re^L$ is the recurrent weight and $\phi_{ij}(t - T)$ is the part that will assure that the information of the previous time-step is stored, and further weighted in order to help to achieve faster convergence. Therefore, a better track of the nonlinear dynamic behavior of a system can be achieved. The $i$-th multidimensional wavelet basis function is defined as $\Phi_j = \prod_{i=1}^{L} \phi_{ij}$. The multidimensional wavelet basis function in vector form is

$$\Phi(z, m, \sigma, r) = [\Phi_1, \Phi_2, \ldots, \Phi_M]^T \in \Re^M, \tag{10.10}$$

where $m = [m_1^T, m_2^T, \ldots, m_M^T]^T \in \Re^{LM}$, $\sigma = [\sigma_1^T, \sigma_2^T, \ldots, \sigma_M^T]^T \in \Re^{LM}$, and $r = [r_1^T, r_2^T, \ldots, r_M^T]^T \in \Re^{LM}$. The $k$-th output of the RWNN is defined as

$$y_k = \sum_{j=1}^{M} w_{jk}\Phi_j = w_k^T\Phi, \tag{10.11}$$

for $k = 1,2,\ldots,N$ with $N$ being the number of outputs in the RWNN. $w_{jk}$ is the weight connecting the $k$-th output with the $j$-th wavelet layer, and $w_k = [w_{1k}, w_{2k}, \ldots, w_{Mk}]^T \in \Re^M$. The output can also be represented as the product between the weight matrix $W$ and the multidimensional wavelet basis function (10.10) shown as

$$y = W^T \Phi \in \Re^N. \tag{10.12}$$

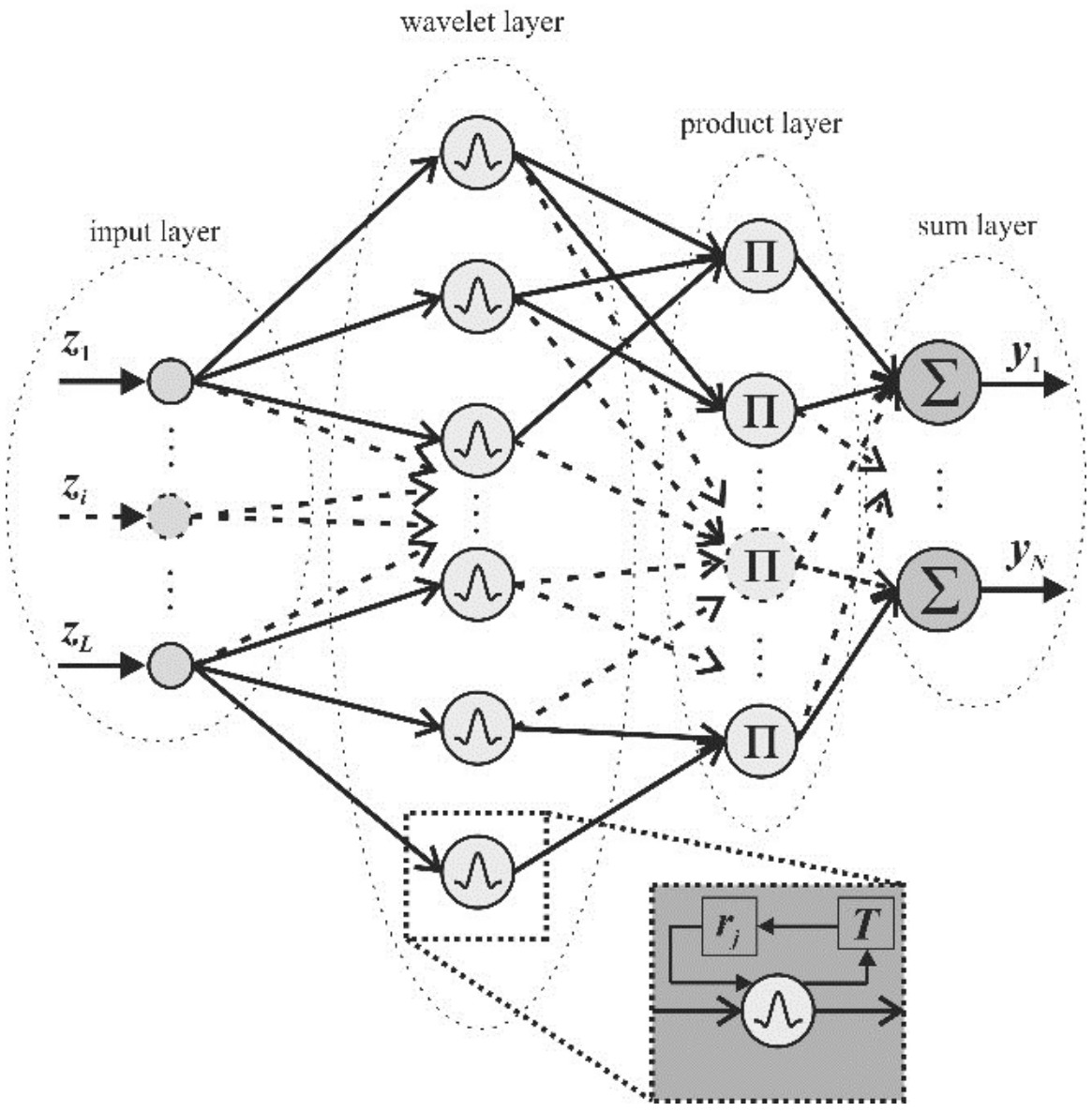

Figure 10.1 Architecture of an RWNN [233]

In control of flexible structures, the neuro-controller follows the ideal controller to address the disturbance rejection problem while compensating for the inherent actuator-induced delay as well as other nonlinearities in the system. This problem is addressed theoretically in the next section.

## 10.3.2 Design of RBAC system

The RWNN-based adaptive control (RBAC) system is shown in Figure 10.2 with the control law

$$u_{ac} = u_{tc} + u_{nc}, \tag{10.13}$$

where $u_{nc} \in \Re^N$ is the RWNN controller output in (10.12), and $u_{tc} \in \Re^N$ is the bounding estimator responsible for compensating the difference between the output of the RWNN controller and the ideal controller as well as guaranteeing smooth and chattering-free compensation [239]. Substituting $u_{ac}$ from (10.13) into $u$ from (10.3) and using (10.7), the error dynamics can be updated. By the universal approximation property (see [256]), there exists an optimal neural controller $u_{nc}^*$ that theoretically can approximate the ideal controller (10.7) such that

$$u^* = u_{nc}(z, m^*, \sigma^*, r^*) + \Delta = W^{*T} \Phi^{*T} + \Delta, \tag{10.14}$$

where $m^* \in \Re^N, \sigma^* \in \Re^{LM}, r^* \in \Re^{LM}, W^* \in \Re^{M \times N}$, and $\Phi^* \in \Re^M$ are the optimal parameters for $m$, $\sigma$, $r$, $W$, and $\Phi$ respectively, such that $\Delta$ is the approximation error vector. Since the optimal parameters for the approximation cannot *a priori* be determined, an estimation neural controller is defined as

$$\hat{u}_{nc} = \widehat{W}^T \Phi(z, \widehat{m}, \hat{\sigma}, \hat{r}) = \widehat{W}^T \widehat{\Phi}, \tag{10.15}$$

where $\widehat{W} \in \Re^{M \times N}$, and $\widehat{\Phi} \in \Re^M$ are the estimated parameters of $W$, and $\Phi$ respectively. Simultaneously, $\widehat{m}, \hat{\sigma}$, and $\hat{r}$ are the estimated parameters of the corresponding variables without the hat operator. The esti-mation error is $\tilde{u}_{nc} = u_{nc}^{*} - \hat{u}_{nc} = \tilde{W}^T\tilde{\Phi} + \tilde{W}^T\hat{\Phi} + \hat{W}^T\tilde{\Phi} + \Delta$, with $\tilde{W} = W^{*} - \hat{W}, \tilde{\Phi} = \Phi^{*} - \hat{\Phi}$. In order to achieve a favorable estimation of a nonlinear function, the adaptive tuning laws will be derived to tune the parameters of the wavelet neural network in an online framework.

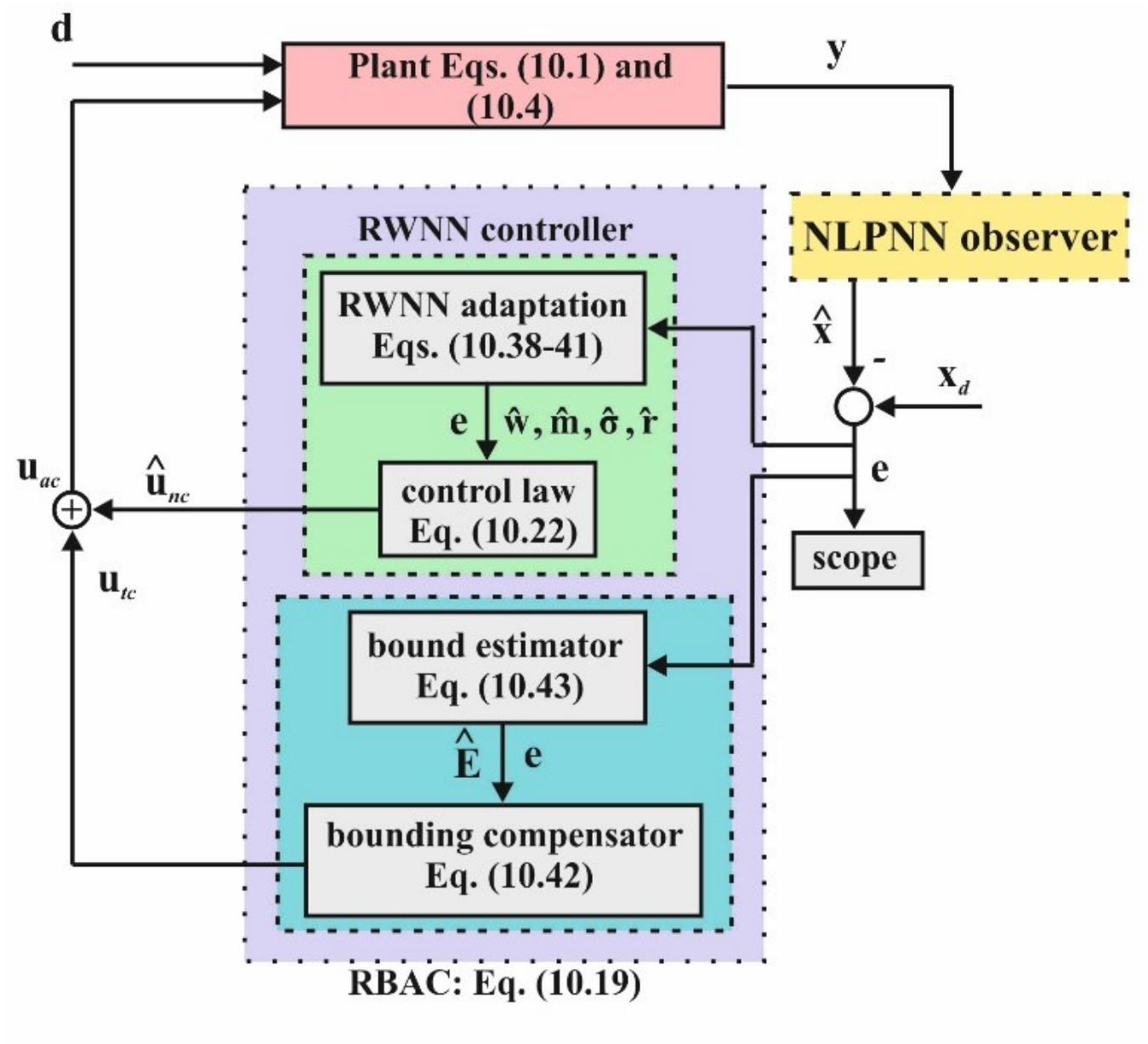

Figure 10.2 Block diagram of the RBAC system [233]

Thus, the Taylor expansion technique is employed over the estimation error of the multidimensional wavelet basis following [239], i.e.

$$\tilde{\Phi} = \Phi_m^T\tilde{m} + \Phi_\sigma^T\tilde{\sigma} + \Phi_r^T\tilde{r} + o, \tag{10.16}$$

where $\tilde{m} = m^{*} - \hat{m}$, $\tilde{\sigma} = \sigma^{*} - \hat{\sigma}$, $\tilde{r} = r^{*} - \hat{r}$ and $o \in \Re^M$ is the vector of higher-order terms,

$$\Phi_m = [\frac{\partial\Phi_1}{\partial m}\frac{\partial\Phi_2}{\partial m}\cdots\frac{\partial\Phi_M}{\partial m}]^T|_{m=\hat{m}} \in \Re^{LM\times M}, \tag{10.17}$$

$$\Phi_\sigma = [\frac{\partial\Phi_1}{\partial\sigma}\frac{\partial\Phi_2}{\partial\sigma}\cdots\frac{\partial\Phi_M}{\partial\sigma}]^T|_{\sigma=\hat{\sigma}} \in \Re^{LM\times M}, \tag{10.18}$$

$$\Phi_r = [\frac{\partial\Phi_1}{\partial r}\frac{\partial\Phi_2}{\partial r}\cdots\frac{\partial\Phi_M}{\partial r}]^T|_{r=\hat{r}} \in \Re^{LM\times M}, \tag{10.19}$$

and $\frac{\partial\Phi_j}{\partial m}$, $\frac{\partial\Phi_j}{\partial\sigma}$, and $\frac{\partial\Phi_j}{\partial r}$ are defined as

$$[\frac{\partial\Phi_j}{\partial m}] = [\underbrace{0\cdots0}_{(j-1)\times L}\frac{\partial\Phi_j}{\partial m_{1j}}\cdots\frac{\partial\Phi_j}{\partial m_{Lj}}\underbrace{0\cdots0}_{(M-j)\times L}]^T, \tag{10.20}$$

$$[\frac{\partial\Phi_j}{\partial\sigma}] = [\underbrace{0\cdots0}_{(j-1)\times L}\frac{\partial\Phi_j}{\partial\sigma_{1j}}\cdots\frac{\partial\Phi_j}{\partial\sigma_{Lj}}\underbrace{0\cdots0}_{(M-j)\times L}]^T, \tag{10.21}$$

$$[\frac{\partial\Phi_j}{\partial r}] = [\underbrace{0\cdots0}_{(j-1)\times L}\frac{\partial\Phi_j}{\partial r_{1j}}\cdots\frac{\partial\Phi_j}{\partial r_{Lj}}\underbrace{0\cdots0}_{(M-j)\times L}]^T. \tag{10.22}$$

Substituting the estimation error ($\tilde{u}_{nc}$) and (10.16) into the updated error dynamics equation, it is straightforward to see

$$\dot{e} = (A - BK)e - Hd - h(x,u) - w(x) + \dot{x}_d - Ax_d + B(\widetilde{W}^T\hat{\Phi} + \hat{W}^T\hat{\Phi}_m^T\tilde{m} + \hat{W}^T\hat{\Phi}_\sigma^T\tilde{\sigma} + \hat{W}^T\hat{\Phi}_r^T\tilde{r} + \varepsilon - u_{tc}), \tag{10.23}$$

where $\varepsilon = \hat{W}^T \mathrm{o} + \widetilde{W}^T\widetilde{\Phi} + \Delta$, denotes the lumped approximation error vector and it is assumed to be bounded by $0 \leq |\varepsilon| \leq E$, with $E$ being a positive constant vector. Since $E$ cannot be obtained in practice, a bound-estimation mechanism must be developed. The estimation error vector of the bound is defined as

$$\tilde{E} = E - \hat{E} \in \Re^N, \tag{10.24}$$

where $\hat{E}$ is the estimated error bound vector. To guarantee that all signals in the closed-loop system are uniformly bounded, a Lyapunov function candidate is defined as

$$V = \frac{1}{2}e^T P_1 e + \frac{1}{2\eta_w} tr(\tilde{W}^T\tilde{W}) + \frac{1}{2\eta_m}\tilde{m}^T\tilde{m} + \frac{1}{2\eta_\sigma}\tilde{\sigma}^T\tilde{\sigma} + \frac{1}{2\eta_r}\tilde{r}^T\tilde{r} + \frac{1}{2\eta_E}\tilde{E}^T\tilde{E} + \int_0^\infty (e^T P_2 e - \gamma_1\omega^T\omega)dt + \int_0^\infty (e^T P_2 e - \gamma_2 d^T d)dt, \tag{10.25}$$

where $\eta_w, \eta_m, \eta_\sigma, \eta_r$, and $\eta_E$ are positive learning rate constants, $P_1$ and $P_2$ are symmetric positive definite matrices with appropriate dimensions, and $\gamma_1$ and $\gamma_2$ are positive real constants. Deriving the Lyapunov function w.r.t. time, and using (10.23), yields

$$\dot{V} = \frac{1}{2}\Omega^T(P_1 e + e^T P_1)\Omega + \Phi^T\tilde{W}B^T P_1 e - \frac{1}{\eta_w}tr(\tilde{W}^T\dot{\hat{W}}) + +\tilde{m}^T(\Phi_m B^T P_1 e - \frac{1}{\eta_m}\dot{\hat{m}}) + \tilde{\sigma}^T(\Phi_\sigma B^T P_1 e - \frac{1}{\eta_\sigma}\dot{\hat{\sigma}}) + \tilde{r}^T(\Phi_r B^T P_1 e - \frac{1}{\eta_r}\dot{\hat{r}}) + (\varepsilon - u_{tc})^T B^T P_1 e - \frac{1}{\eta_E}\tilde{E}^T\dot{\hat{E}} + 2e^T P_2 e - \gamma_1\omega^T\omega - \gamma_2 d^T d, \tag{10.26}$$

where,

$$\Omega = (A - BK)e - Hd - h(\hat{x},u) - w(\hat{x}) + \dot{x}_d - Ax_d, \tag{10.27}$$

with,

$$\hat{\Phi}^T\tilde{W}B^T P_1 e = \sum_{k=1}^N \tilde{w}_k^T\hat{\Phi}b_k^T P_1 e,$$

$$tr(\tilde{W}^T\dot{\hat{W}}) = \sum_{k=1}^N \tilde{w}_k^T\dot{\hat{w}}_k.$$

Here, $b_k$ is the $k$-th column of matrix $B$. The information acquired by NN is saved in its combination weights, which are adaptive and can be adjusted in response to the changes in input signals to the network. The learning laws quantify the rate of these changes. Accordingly, the choice for the learning laws is made as

$$\dot{\hat{w}}_k = \eta_w[\hat{\Phi}b_k^T P_1 e - \mu_w(\hat{w}_k - w_0)], \tag{10.28}$$

$$\dot{\hat{m}} = \eta_m[\hat{\Phi}_m B^T\hat{W}P_1 e - \mu_m(\hat{m} - m_0)], \tag{10.29}$$

$$\dot{\hat{\sigma}} = \eta_\sigma[\hat{\Phi}_\sigma B^T\hat{W}P_1 e - \mu_\sigma(\hat{\sigma} - \sigma_0)], \tag{10.30}$$

$$\dot{\hat{r}} = \eta_r[\hat{\Phi}_r B^T \hat{W} P_1 e - \mu_r(\hat{r} - r_0)], \tag{10.31}$$

where $\mu_w, \mu_m, \mu_\sigma,$ and $\mu_r,$ are small positive constants and $w_0, m_0, \sigma_0,$ and $r_0$, are the initial vectors of $w, m, \sigma,$ and $r,$ respectively. The bounding compensator is chosen as

$$u_{tc} = \hat{E}\tanh(\frac{B^T P_1 e}{\delta}), \tag{10.32}$$

and the bounding estimation law is expressed as below

$$\dot{\hat{E}} = \eta_E[B^T P_1 e\tanh(\frac{B^T P_1 e}{\delta}) - \mu_E(\hat{E} - E_0)], \tag{10.33}$$

where $\tanh(\cdot)$ is the hyperbolic tangent function, $\delta$ and $\mu_E$ are small positive constants, and $E_0$ is the initial estimation vector of $E$. Using (10.28)-(10.33) and after some mathematical manipulations, (10.26) is reformulated to be

$$\begin{aligned} \dot{V} \leq & \frac{1}{2}\Omega^T(P_1 e + e^T P_1)\Omega + \mu_w \sum_{k=1}^{N} \tilde{w}_k^T(\hat{w}_k + w_0) + \mu_m \tilde{m}^T(m - \hat{m}_0) + \mu_\sigma \tilde{\sigma}^T(\hat{\sigma} - \sigma_0) \\ & + \eta_r \tilde{r}^T(r - \hat{r}_0) + 2e^T P_2 e - \gamma_1 \omega^T \omega - \gamma_2 d^T d + \mu_E \tilde{E}^T(\hat{E} - E_0) \\ & + E^T[|B^T P_1 e| - B^T P_1 e\tanh(\frac{B^T P_1 e}{\delta})]. \end{aligned} \tag{10.34}$$

According to [257], the following inequality holds for any $\delta > 0$,

$$0 \leq | B^T P_1 e | - B^T P_1 e\tanh(\frac{B^T P_1 e}{\delta}) \leq \kappa\delta, \tag{10.35}$$

where $\kappa$ is a constant satisfying $\kappa = e^{-(\kappa+1)}$. Using (10.35) and (10.27) and definition of Euclidean norm and after some mathematical manipulations, (10.34) can be written as

$$\begin{aligned} \dot{V} \leq & \frac{1}{2}\mathcal{L}^T \Gamma \mathcal{L} - \frac{\mu_w}{2}\sum_{k=1}^{N}(\|w_k^* - \hat{w}_k\|^2 - \|w_k^* - w_0\|^2 + \|\hat{w}_k - w_0\|^2) - \frac{\mu_m}{2}(\|m^* - \hat{m}\|^2 \\ & - \|m^* - m_0\|^2 + \|\hat{m} - m_0\|^2) - \frac{\mu_\sigma}{2}(\|\sigma^* - \hat{\sigma}\|^2 - \|\sigma^* - \sigma_0\|^2 \\ & + \|\hat{\sigma} - \sigma_0\|^2) - \frac{\mu_r}{2}(\|r^* - \hat{r}\|^2 - \|r^* - r_0\|^2 + \|\hat{r} - r_0\|^2) \\ & - \frac{\mu_E}{2}(\|E^* - \hat{E}\|^2 - \|E^* - E_0\|^2 + \|\hat{E} - E_0\|^2) + \kappa\delta E, \end{aligned} \tag{10.36}$$

with $\mathcal{L}^T = [e^T, d^T, h(\hat{x}, u), \omega^T(\hat{x}), \dot{x}_d, x_d^T]$. Guaranteeing the semi-negative definiteness of the right-hand side of the inequality (10.36), we obtain the bilinear matrix inequality (BMI) on symmetric $\Gamma$. This BMI can be converted to a linear matrix inequality by using the well-known congruence transformation [83]. This is performed by pre- and post-multiplying $\Gamma$ by $diag({P_1}^{-1}, I, I, I, I, I)$. Then, by defining $\hat{P}^{-1} = P_2 = P_1$, and $\hat{K} = K\hat{P}$ the ideal control gain can be obtained as a solution to

$$\hat{\Gamma} = [\Gamma_{ij}] \leq 0. \tag{10.37}$$

with,

$$\begin{aligned} \Gamma_{11} &= \hat{P}A^T + A\hat{P} - \hat{K}^T B^T - B\hat{K} + 4\hat{P}, \\ \Gamma_{12} &= -H_n, \\ \Gamma_{13} &= \Gamma_{14} = -I, \\ \Gamma_{15} &= I, \\ \Gamma_{16} &= -A^T, \\ \Gamma_{22} &= -2\gamma_1 I, \\ \Gamma_{44} &= -2\gamma_2 I, \end{aligned} \tag{10.38}$$

with the rest of the matrix elements being zero matrices with appropriate dimensions. $\Gamma_{ij},\ i,j = 1,\ldots,6$ is linear in parameters, and LMI (10.37) can be solved using standard semi-definite programming software such as MATLAB and Scilab. If $\hat{\Gamma}$ is chosen in a way to force its semi-negative-definiteness, and considering (10.25), the Lyapunov function (10.36) can be obtained as

$$\dot{V} \le -\alpha V + \beta. \tag{10.39}$$

where $\alpha$ and $\beta$ are positive constants given by

$$\alpha = \min\{\frac{1}{\lambda_{max}(\hat{P})}, \mu_l \eta_l\}, \tag{10.40}$$

with $l = w, m, \sigma, r, E$ and,

$$\begin{aligned} \beta \quad &= -\frac{1}{2}[\mu_w \sum_{k=1}^{N} \|w_k^* - w_0\|^2 + \mu_m \|m^* - m_0\|^2 + \mu_\sigma \|\sigma^* - \sigma_0\|^2 \\ &\quad + \mu_r \|r^* - r_0\|^2 + \mu_E \left\|E - {E_0}^2\right\|] + \kappa \delta E. \end{aligned} \tag{10.41}$$

Since $\beta/\alpha > 0$, and the solution of (10.39) satisfies

$$0 \le V(t) \le \frac{\beta}{\alpha} + [V(0) - \frac{\beta}{\alpha}]e^{-at}, \tag{10.42}$$

the RBAC system guarantees that all the signals and parameters $e, W, m, \sigma, E$ are uniformly bounded.

**Remark 10.1.** Within the scope of the AVC, the focus is given on the problem of output regulation, i.e. $x_d = \dot{x}_d = 0$.

**Theorem 10.1** *The asymptotic stability of the tracking error dynamics for the closed-loop system consisting of the plant in (10.2), with asymptotic stable observer dynamics in (10.4), and the recurrent neural network controller with feedback/feedforward ideal control law* $u^* = Ke + B^+\dot{x}_d - B^+Ax_d$, *in which* $B^+ = B^T(BB^T)^{-1}$ *is guaranteed (assuming* $rank(B) = p \ge m$*) while rejecting the effect of matched uncertainties, mismatch disturbance signal, and the unknown dynamics using the controller gain* $K = \hat{P}_1^{-1}\hat{K}$, *which is obtained from satisfying the following LMI/LME conditions. Additionally, the network learning parameters are based on (10.28)-(10.31) with the bound compensator in (10.32) and (10.33). The design parameters in LMI/LME (10.43) are the symmetric positive definite matrices* $P_i, i = 1, \ldots, 4$, *matrices* $\hat{P}_1$ *and* $\hat{K}$ *with appropriate dimensions, and the positive scalars* $\gamma_1, \gamma_2$, *and* $\gamma_3$.

$$\begin{bmatrix} A^T P_1 + P_1 A - K^T B^T - BK + 2(P_2 + P_3 + P_4) & -P_1 H & -P_1 & -P_1 \\ -H^T P_1 & -2\gamma_2 I & 0 & 0 \\ -P_1 & 0 & -2\gamma_3 I & 0 \\ -P_1 & 0 & 0 & -2\gamma_1 I \end{bmatrix} < 0, \tag{10.43}$$

$$P_1 B = B\hat{P}_1. \tag{10.44}$$

**Proof.** The proof is given in Appendix E.

## 10.4 Neuro-adaptive state observer

Owing to the fact that the optimum weights in $V$ and $W$ from (10.5) cannot *a priori* be determined, an estimation of the unknown dynamics is defined as

$$\hat{h}(\hat{x}, u) = \hat{W}s(\hat{V}\hat{a}), \tag{10.45}$$

and the neuro-observer dynamics (10.6) can be revised in the form of the following equation

$$\begin{aligned} \dot{\hat{x}} &= A\hat{x} + Bu + Hd + \hat{W}s(\hat{V}\hat{a}) + G(y - C\hat{x}), \\ \hat{y} &= C\hat{x}. \end{aligned} \tag{10.46}$$

Using (10.3), (10.45), and (10.46) and by adding and subtracting $Ws(Va)$ the state estimation error dynamics $\dot{\tilde{x}}$ becomes

$$\begin{aligned} \dot{\tilde{x}} &= A_c\tilde{x} + \tilde{W}s(\hat{V}\hat{a}) + \epsilon(t), \\ \tilde{y} &= C\tilde{x}. \end{aligned} \tag{10.47}$$

where $A_c = (A - GC)$, $\tilde{W} = W - \hat{W}$, and $\epsilon(t) = W[s(Va) - s(\hat{V}\hat{a})] + \Delta$ is the disturbance term, which for some positive constant $\epsilon$ is assumed to be bounded by $||\epsilon(t)|| \leq \epsilon$. This can be explained by the boundedness of the sigmoid function, and the assumption that the ideal neural network weights $V$ and $W$ are also bounded as $||s(Va)|| \leq s_{max}$, $||W|| \leq W_{max}$, and $||V|| \leq V_{max}$. Following [243], the learning rules in the adaptive neuro-observer construction are proposed as

$$\dot{\hat{V}} = -\eta_1(\tilde{y}^T C A_c^{-1})^T [s(\hat{V}\hat{a})]^T - \rho_1||\tilde{y}||\hat{V}, \tag{10.48}$$

$$\dot{\hat{W}} = -\eta_2(\tilde{y}^T C A_c^{-1}\hat{W}(I - \Lambda(\hat{V}\hat{a})))^T sgn(\hat{a})^T - \rho_2||\tilde{y}||\hat{W}, \tag{10.49}$$

where $\eta_1, \eta_2, \rho_1,$ and $\rho_2$ are small positive learning rates of the backpropagation term and e-modification term that are responsible for incorporating damping in the equations, respectively. Also, $\Lambda(\hat{V}\hat{a}) = diag\{s_i^2(\hat{V}_i\hat{a}_i)\}$, with $s_i$ being the $i$-th element of $s$ and $\hat{V}_i$ the $i$-th row of $V$ and finally, sgn$(\hat{a})$ is the well-known sign function.

## 10.5 Implementation of the observer-based controller on AVC

In the case of flexible manipulators in various applications, the coupled modeling of the multibody system is a complex task [258]. An adaptive output feedback method (e.g. RWNN) is proposed in the previous section in order to take into consideration the time-variability in the plant dynamics.

The identification process in the frequency-domain is carried out whilst neglecting the torsional/in-plane modes (similar to the previous chapters) because the piezo-actuators are unable to contribute to damping of the torsional vibrations of the cantilever beam as well as the in-plane ones. The bandwidth of interest in the parametric model is limited to a maximum of 250 Hz (including three fundamental mode shapes), and vibration due to the higher-order mode shapes of the system up to 1 kHz is assumed to be the source of lumped nonlinear time-varying uncertainty in plant model while the dynamics of higher frequency nature are mostly treated in a passive sense [49]. The interested reader is referred to visit the impressive contributions in [169], [259], and Chapter 9 in [124] for technical details in input excitation signal design, multi-reference modal analysis, and system identification used in this chapter. Nonetheless, it is worth citing that implementations of the subspace method and Maximum Likelihood (ML) technique in the frequency domain are accessible in the Signal Processing Microelectronics (SPM) package [260], [261].

### 10.5.1 Implementation results

The parameter tuning of the RWNN is a nontrivial process that is carried out sequentially while alternatively, a systematic tuning of the network can be performed in real-time following the Hardware-in-Loop (HiL) technique presented in [218]. Accordingly, first, the primary parameters of the RWNN are selected by setting the number of inputs $L = 6$ (six states in the nominal model) in the layer of entry and $M = 6$ as the number of wavelons in the wavelet layer. The latter is a compromise between the computation effort that the DAQ system (see Figure 6.1) should burden and the trade-off between the underfitting and overfitting of the network. In this regard, the symbolic toolbox of MATLAB is used to derive the coefficients of the Taylor series in (10.16) for the selected parameters. One may notice that the wavelet layer is selected to have minimal wevelons while avoiding the underfitting. The reason for this choice is that in practical implementations *task turn-around* time is the key parameter to be controlled. As proof, the experimental implementations are first performed on a dSPACE DS1005 PPC DAQ system featuring a PowerPC 750GX processor running at 1 GHz. However, the implementations failed due to *task overrun* error at sampling time associated with the Nyquist frequency (1.5 msec). Therefore a more powerful processor (DS1006) is used for the following results. In order to give an overview of the processing power required, the final results are obtained with a task turn-around time of 0.3 msec on a single core implementation of DS1006. Then, using the acquired Taylor expansion, the multidimensional wavelet basis is constructed, and the adaptation laws (10.28)-(10.31) are written in the form of some user-defined functions of the SIMULINK platform. The estimation of the neural controller (10.15) is created in a function block of SIMULINK while the delay in the recurrent network is realized by a *memory* block which adds a single integration step delay. The bound compensator in (10.32) is obtained by selecting $\delta = 4$ in hyperbolic tangent function while implementing the bound estimation mechanism (10.33) similar to the adaptation laws as mentioned earlier. It is worth mentioning that higher values of $\delta$ make the bound compensator insignificant and smaller values result in chattering-like behavior which signifies the importance of preserving the inequality in (10.35). At this point, the observed states $\hat{x}$ (following the implementation of (10.48) and (10.49) in SIMULINK model), the network weighting matrices for input and output layers, $W$ and $V$, are calculated and the unknown dynamics is estimated by $h(\hat{x}, u)$. Next, the ideal control is generated as a solution to the LMI (10.37) (see also **Theorem 10.1**) and the state observer gain in the neural observer is assigned based on the neuro-observer dynamics in (10.46). These are given as

$$K = \begin{bmatrix} -61.64 & 21.38 & 21.98 & -7.28 & -14.73 & 3.51 \\ -14.12 & 10.06 & 44.36 & -26.52 & 24.03 & -12.02 \end{bmatrix}, \tag{10.50}$$

$$G^T = [-0.04 \quad 0.06 \quad 0.23 \quad 0.47 \quad -2.55 \quad -0.84], \tag{10.51}$$

for,

$$P_1 = \begin{bmatrix} 5.08 & 0.08 & 0.19 & -0.56 & 0.22 & -0.12 \\ 0.08 & 5.31 & 1.23 & -0.37 & 0.95 & -0.19 \\ 0.19 & 1.23 & 13.12 & -0.18 & -0.25 & 0.25 \\ -0.56 & -0.37 & -0.18 & 12.97 & -0.61 & 0.14 \\ 0.22 & 0.95 & -0.25 & -0.61 & 15.43 & 0.20 \\ -0.12 & -0.19 & 0.25 & 0.15 & 0.20 & 16.94 \end{bmatrix}. \tag{10.52}$$

Finally, the initial values of the adaptation laws are selected to be nonzero values of small quantities (<< 1) as it is observed that the large values of the initial guess on these parameters results in a longer transient convergence period over dynamic equations in the evolution of (10.28)-(10.31). Another set of prominent parameters in the closed-loop behavior is however assigned with $[\eta_w, \eta_m, \eta_\sigma, \eta_r]$ which was after some trial and error selected as 0.3. Although, the nominal simulations (pre-real-time test phase) showed higher values on learning rates might result in faster convergence, in real-time tests, high values on learning rates make the system sensitive to the noise level and unmodeled dynamics. It is worth mentioning that although the experimental results in this section present promising behavior of the network in handling the DRC and unknown dynamic estimations, it is believed that the proposed approach is capable of achieving even better results by fine-tuning the network. This may be a nontrivial task but is achievable by parallel computing time-domain sensitivity analyses such as [262] and HiL optimization e.g. [218].

Here, it is assumed that the initial values of the training parameters are completely unknown and as a result set to very small-valued ($10^{-2}$) matrices with appropriate dimensions. However, the gains on the correction term of the training laws, $\mu_w, \mu_m, \mu_\sigma$, and $\mu_r$, are set all equally to 10 in this experimental investigation. It should be noted that a thorough systematic sensitivity analysis based on the analytical local sensitivity criterion, one-at-a-time (OAT) technique, and the global sensitivity based on the Monte Carlo technique for the control system in closed-loop configuration requires a separate dedicated investigation. Finally, in (10.48) and (10.49), the positive learning rates of the backpropagation term and e-modification term i.e., $\eta_1, \eta_2, \rho_1$, and $\rho_2$ are simply set to 0.1, $10^7$, 0.5, and 0.5, respectively. Additionally, it is observed that for large values of $\eta_2$, the convergence of $\hat{h}(\hat{x}, u)$ in (10.45) requires less transient period. This can also be understood by looking into (10.48) and (10.49). At this stage, the model from SIMULINK is built and a C++ code for the dSPACE module is generated and all the requested model variables are written in a file to be illustrated in ControlDesk NG 5.3 environment within a limited time history.

After constructing the control system as a SIMULINK model and compiling it to the dSPACE RTI platform, the system is excited through the disturbance channel with a chirp signal. The frequency of the excitation is swept from 1 to 250 Hz.

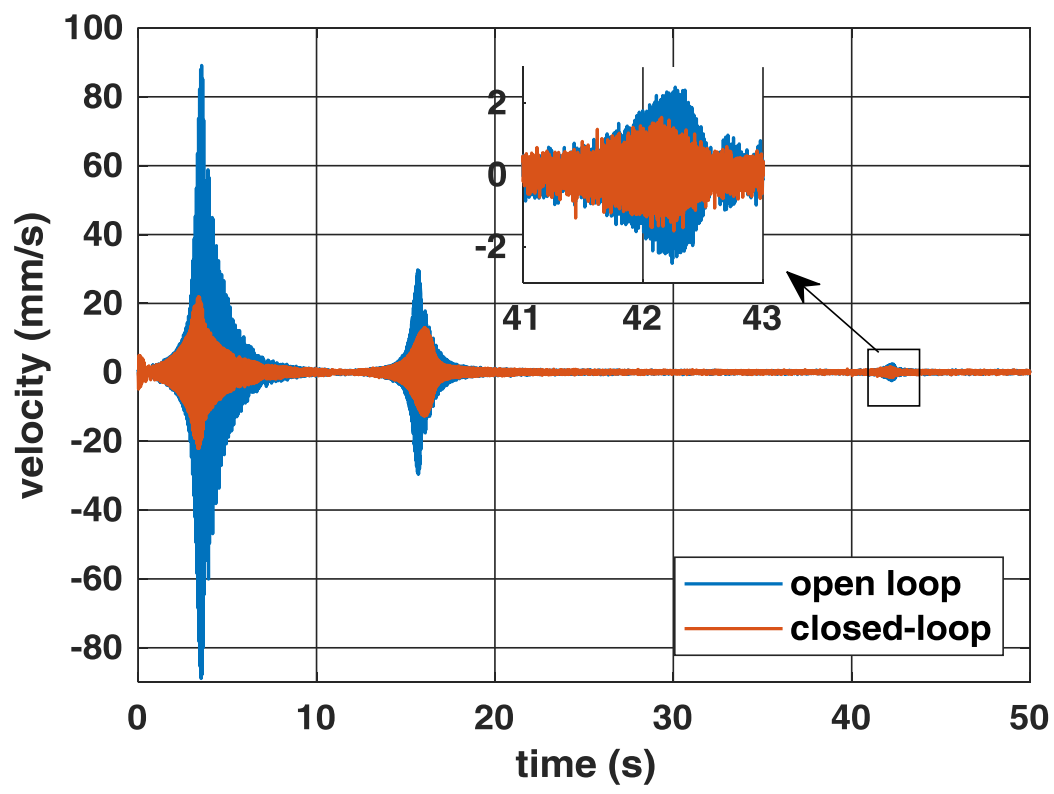

Figure 10.3 Disturbance rejection performance of the closed-loop system [233]

Figure 10.3 shows a thorough evaluation of the vibration attenuation quality in the nominal frequency range. From the captured velocity via laser vibrometer, one can see that the proposed observer-based neural control system suppressed the vibration magnitude within the considered frequency range. The control efforts of the closed-loop system regarding two piezo-actuators are depicted in Figure 10.4.

In order to have a better view of the contribution of RWNN and bound estimating control laws, they are separated into two subplots. The applied control signals on piezo-actuators are in the range of [-55 55] V with smooth variations. In comparison to high-gain approaches (*i.e.*, [55]) and robust control based on the worst-case analysis (*i.e.*, [13]), the bounding compensator observes and addresses the unmodeled dynamics in the system output. The bound estimation mechanism, as shown in Figure 10.5, detects the two regions in which the bounding compensator should intervene with the neural controller to guarantee DRC. By investigating the control law $\hat{u}_{nc}$ based on the RWNN controller in (10.15) and Figure 10.4, the offset of 5 V can be observed (unlike $u_{tc}$). This is due to the static deflection that is generated on the free end of the beam because of the shaker placement configuration. Note that the bounding estimation (10.33) correctly neglects this due to the static nature of the signal that it may generate in the output feedback channel. Additionally, the control effort at the beginning of the excitation takes one second to converge in terms of the adaptation laws. This may be avoided in practical applications by a better definition of the initial conditions of the network parameters. Moreover, the evolutions of the network weights are suppressed here due to the limitations on the dissertation format.

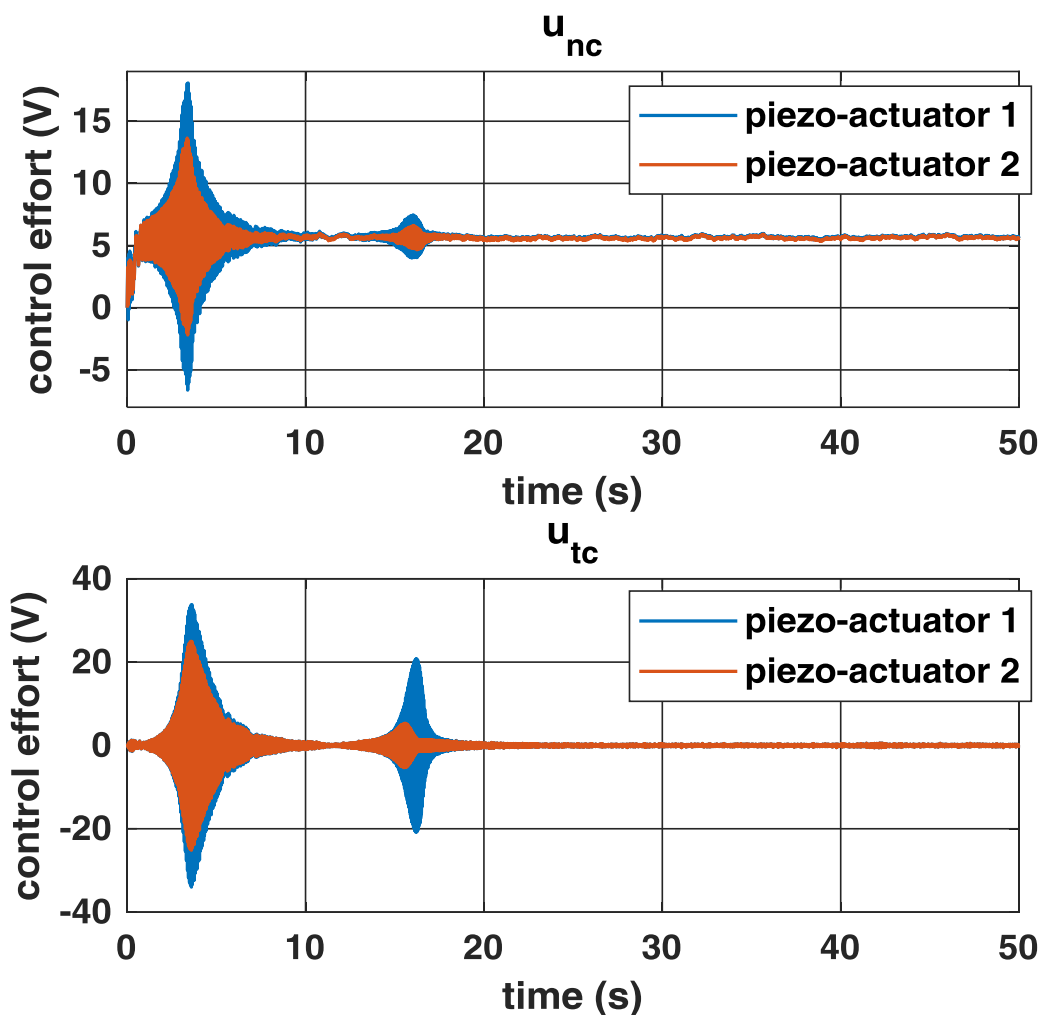

Figure 10.4 Applied control effort of RWNN on the actuators [233]

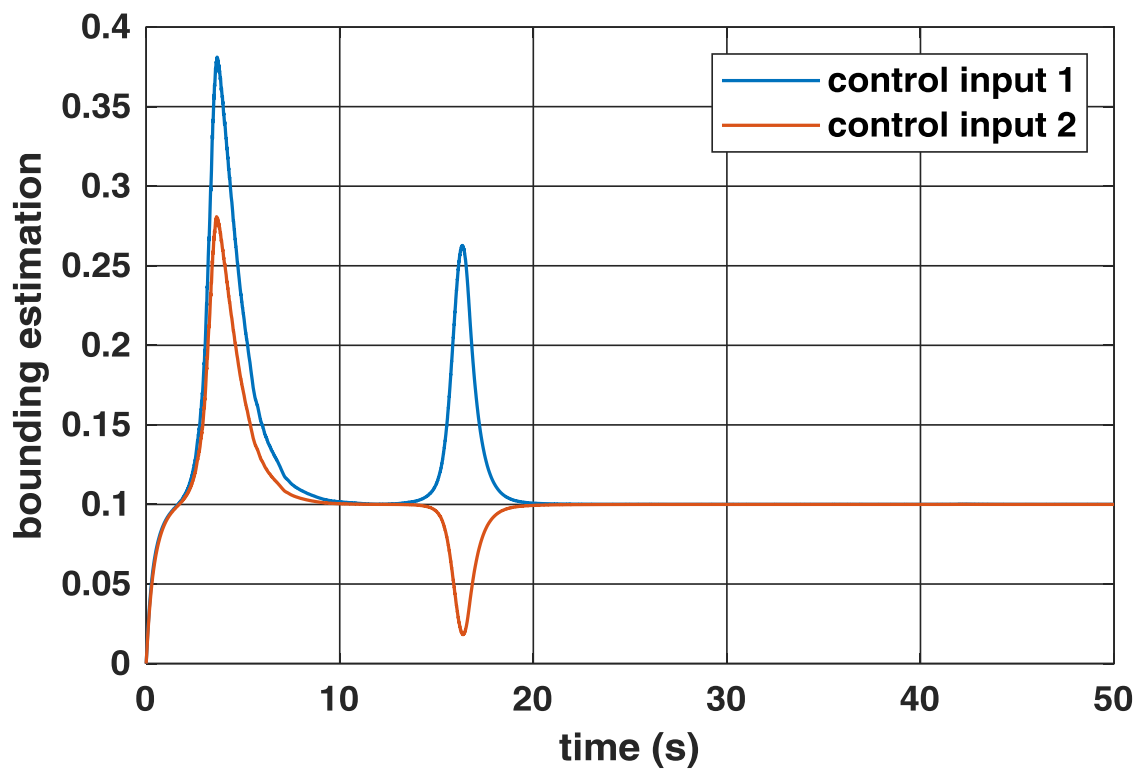

Figure 10.5 Bound estimation result for chirp mismatch disturbance excitation [233]

The output estimation error is also presented in Figure 10.6 which indicates the effectiveness of the state estimation system.

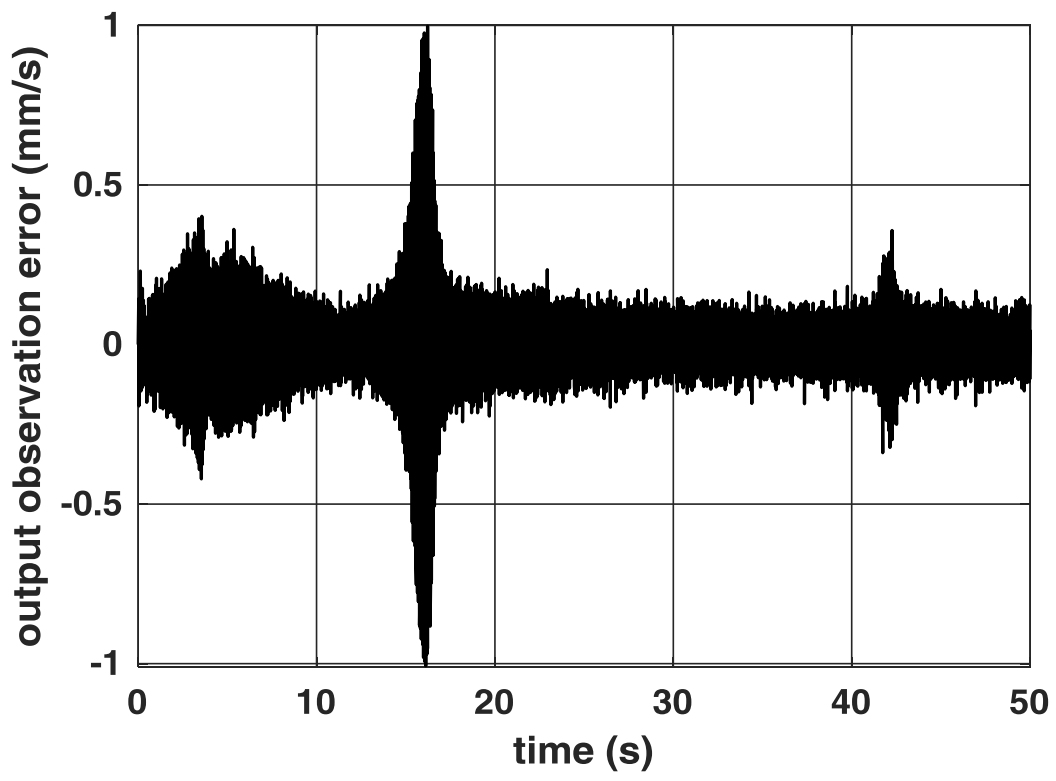

Figure 10.6 Observation error for the plant output [233]

Following, the control system performance is investigated in the frequency domain. Accordingly, the control loop is triggered while the vibration exciter is attached to the LAN-XI DAQ and PULSE LabShop software. The generator in the modal analysis is triggered by a pseudo-random signal in the range of [0 800] Hz. By aiming at assessing the spillover effect, the passband edge of the LPF in the generator measurement channel is fixed at 800 Hz. Consequently, based on FFT, the sampling interval is assigned as 488.3 micro-sec, and the number of Lines is chosen to be 6400 resulting in a sampling frequency of 125 mHz. The leakage error is compensated by using the pseudo-random signal while averaging with a uniform window. Baseband analysis is engaged with 60 Averages. The sampling times are employed on both DAQ systems independently for both open-loop and closed-loop schemes. The frequency responses of the three systems are compared in Figure 10.7.

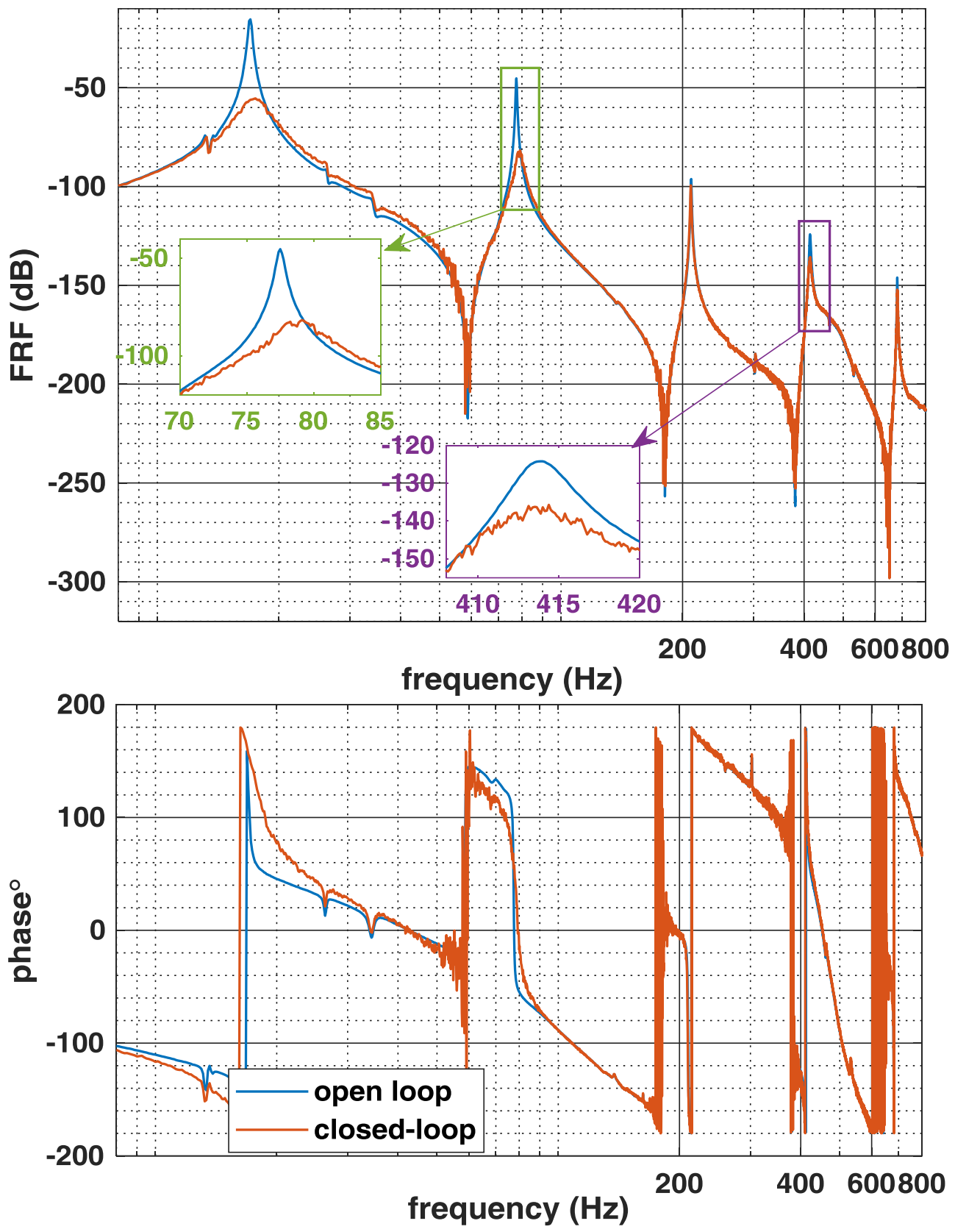

Figure 10.7 Frequency-domain analysis of the closed-loop systems for vibration attenuation performance evaluation [233]

Accordingly, we have compared our proposed method to the open-loop system as well as the closed-loop system based on the Linear Quadratic Gaussian (LQG) regulator. The performance of the two control systems is compared to one another in the frequency range of [0 800] Hz, and we have emphasized the three boxed regions in the vicinity of the natural frequencies. As it may be seen, in the nominal frequency range of the plant model ([0 250] Hz), both the LQG and the proposed observer-based RWNN methods have similar vibration suppression performance. It should be noted that the tuning of LQG is done in such a way that for the free vibration suppression i.e. vibration suppression for an initial step–like disturbance excitation, both of the methods generate the equal amplitude of the control effort (same amount of injected energy). While LQG even in the nominal frequency range of operation is very sensitive to the plant model, the robustness of the proposed RWNN scheme will be shown later. Looking at the higher frequency range where unmodeled dynamics of the system are invoked due to the random disturbance excitations, we can see that while the proposed RWNN method suppresses the vibration, the LQG regulator is completely blind and unable to react to the disturbance signal. In the frequency window of [677,683] Hz (purple box), the closed-loop system based on LQG has even worse performance than the open-loop system without a controller. This emphasizes the importance of the current method for large flexible structures, wherein the bandwidth of AVC (ca. 1 kHz), the system encompasses up to hundreds of mode shapes which cannot be all considered in the nominal plant model.

It is obvious that there exists a frequency shift between the open-loop and closed-loop systems due to a change introduced in the global stiffness of the structure. This frequency shift is more detectable between the first and the second fundamental natural frequencies in Figure 10.7. Here, the FRF of all systems in the frequencies below 8 Hz is suppressed where the shaker is practically unable to excite. Unsurprisingly, since the structure under study does not have any resonant frequencies in this range, the deterioration of the FRF result below 8 Hz may be neglected. Finally, the frequency range of the FRF includes three fundamental natural frequencies that are present in the nominal model of the plant in (10.3) followed by two unmodeled mode shapes. The induced attenuation levels due to the control system associated with these five frequencies are respectively reported as 40.59 dB, 41.77 dB, 3.18 dB, 13.8 dB, and 6.1 dB.

In order to reveal the behavior of the two control schemes, namely the RWNN and the LQG, we have also performed a spectral analysis based on the fast Fourier transformation (FFT) of the control efforts with the same reference disturbance signal as in Figure 10.7. The results as shown in Figure 10.8 discloses the fact that the two methods have different behavior in the two regions of (1) nominal frequency range, (2) unknown dynamics. Due to the structure of our proposed method, the network can detect and reject the unknown dynamics of higher-order nature while LQG is unable to react. The three arrows in the nominal range indicate the fairness of the tuning process of LQG i.e., injection of comparable energy to the system. However, in the range where the reduced-order model is not valid (the two red arrows), the RWNN detects and rejects the excitation of these mode shapes.

In order to see the closed-loop performance in the presence of variations from the system dynamics, we excite the reduced-order plant through the disturbance channel by a sawtooth signal with a frequency of 0.5 Hz. In this regards, the simulation state space model is polluted with 0, 15, 30, and 45 percent matched uncertainty in all system matrices. Figure 10.9 illustrates the output of the open loop system in comparison to the closed-loop systems where there are some matched synthetic uncertainties in the plant model. The control laws associated with each closed-loop simulations in Figure 10.9 are shown in Figure 10.10.

The closed-loop system can handle significant amount of introduced synthetic uncertainties. Next, the impact of $\delta$ in the bounding compensator $u_{tc}$ in (10.32) and the bounding estimator $\hat{E}$ is investigated on the reduced-order model. It is evident that small values of $\delta$ leads in saturation of the hyperbolic tangent function which also depends on the choice of the design parameter $P_1$. In contrary, as we increase the normalization parameter $\delta$, the nonlinear mapping of $tanh(.)$ is more dominant and significant changes can be observed in the nature of the generated control law. Figure 10.11 respectively represents the evolution of $u_{tc}, \hat{E}$, and the absolute error for increasing $\delta$. Accordingly, the simulated disturbance excitation realized with a single period of the sawtooth is set to zero immediately after two seconds.

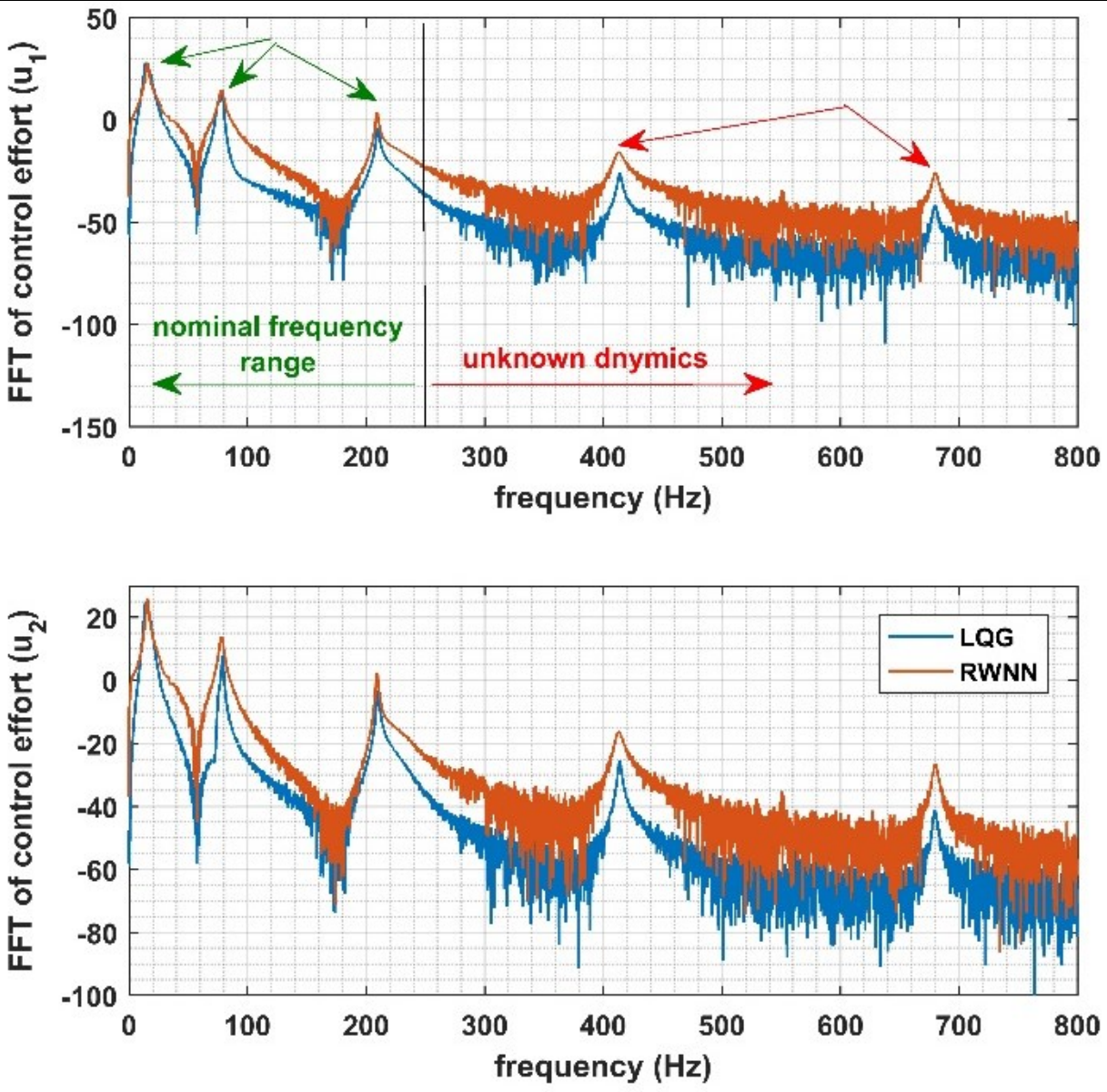

Figure 10.8 Comparison of the frequency content of the control signals for random disturbance excitation in [0 800] Hz [233]

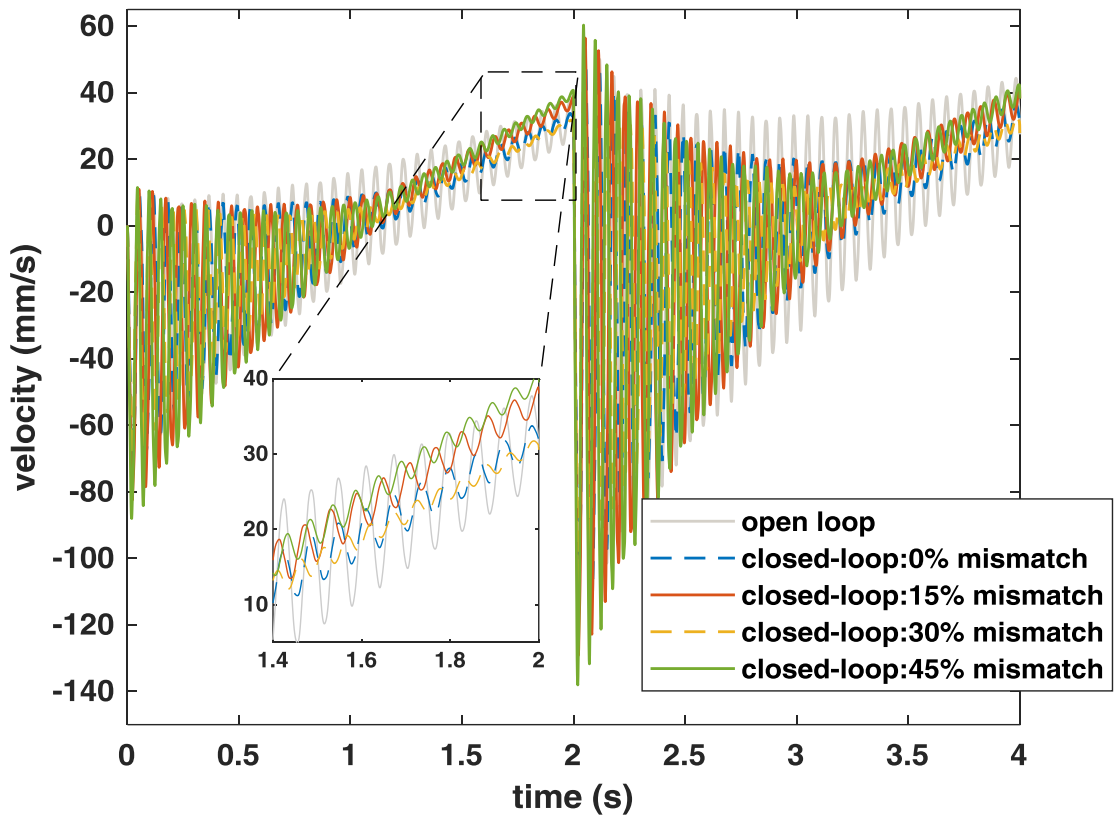

Figure 10.9 Comparison of the system output in simulation for various synthetic matched uncertainty levels in system matrices [233]

In Figure 10.11, it can be observed that for the small normalization parameter $\delta = 0.4$, the chattering problem occurs due to the sudden jumps between the extremums of the hyperbolic tangent. However, for $\delta = 4$ which is used in the rest of the experimental implementations, the chattering disappears and the absolute error converges faster in the benchmark problem. Here the absolute tracking error in the context of the regulation problem is defined by integrating over the absolute value of the output as the deviation from the desired value. For $\delta = 40$ and $\delta = 400$, the error increases and even reaches the open-loop level indicating that the control system is unable to reject the disturbance.

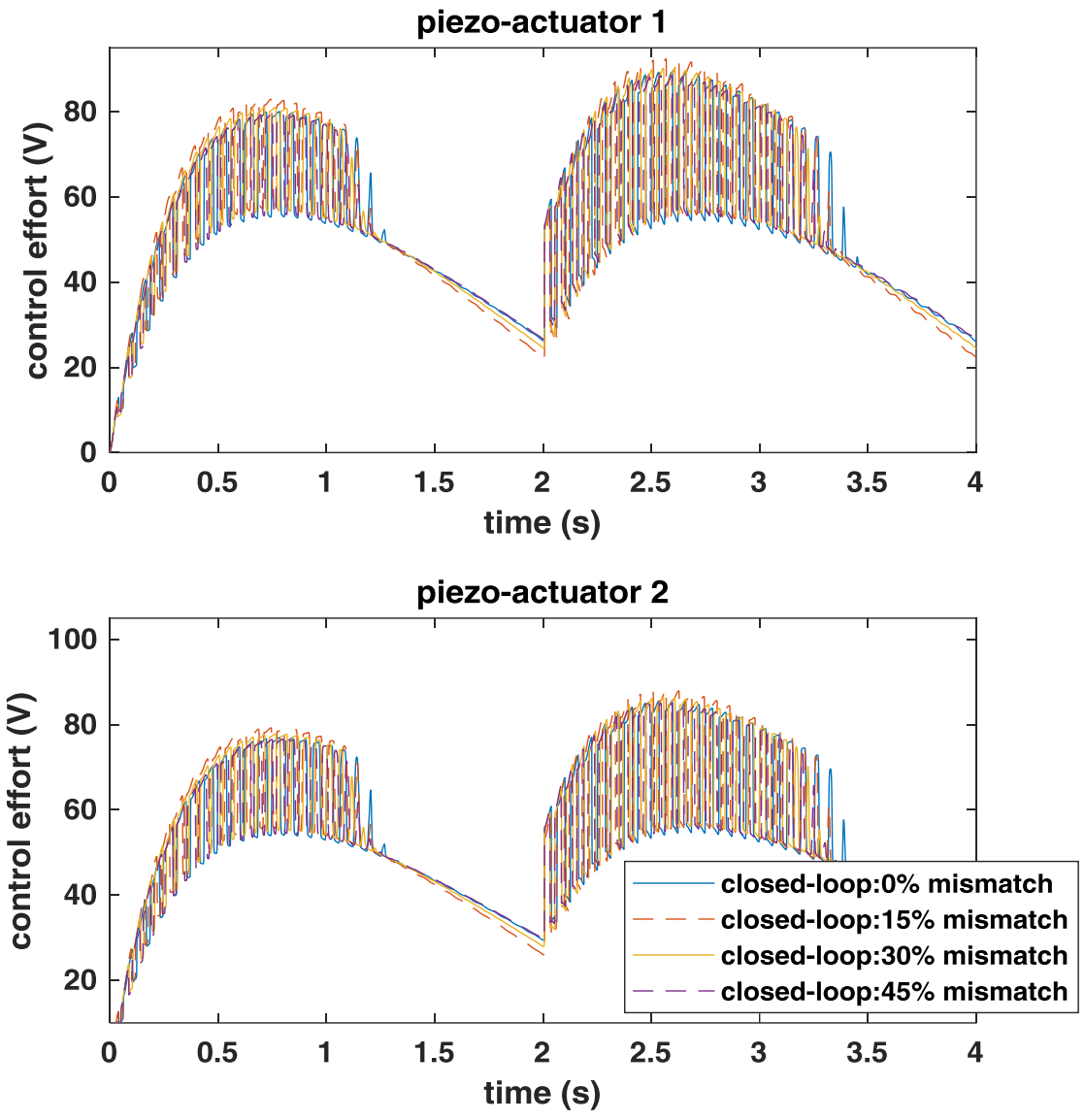

Figure 10.10 Control effort of the closed-loop scheme in the scenarios of synthetic matched uncertainty [233]

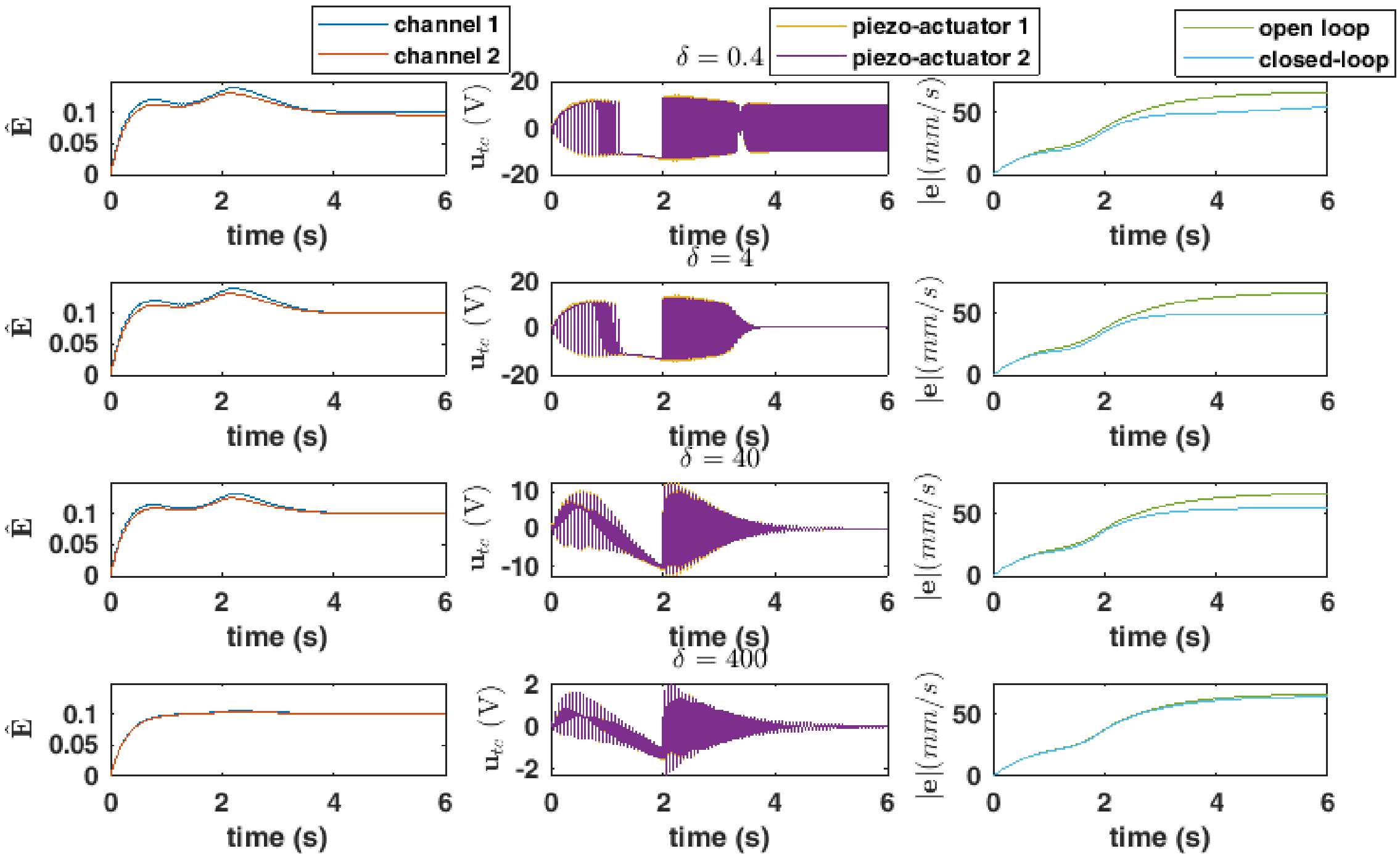

Figure 10.11 The evolution of $u_{tc}$, $\hat{E}$, and the absolute error for increasing $\delta$ [233]

The control law in the latter cases decreases and the bound estimator law is over-trained to an inaccurate convergence level. A possible solution for partially recovering $u_{tc}$ can be obtained by increasing $\eta_E$. For that purpose, Figure 10.12 represents the effect of selecting $\delta = 200$ and $\eta_E = 0.9$ instead of 0.3 in the experimental investigations.

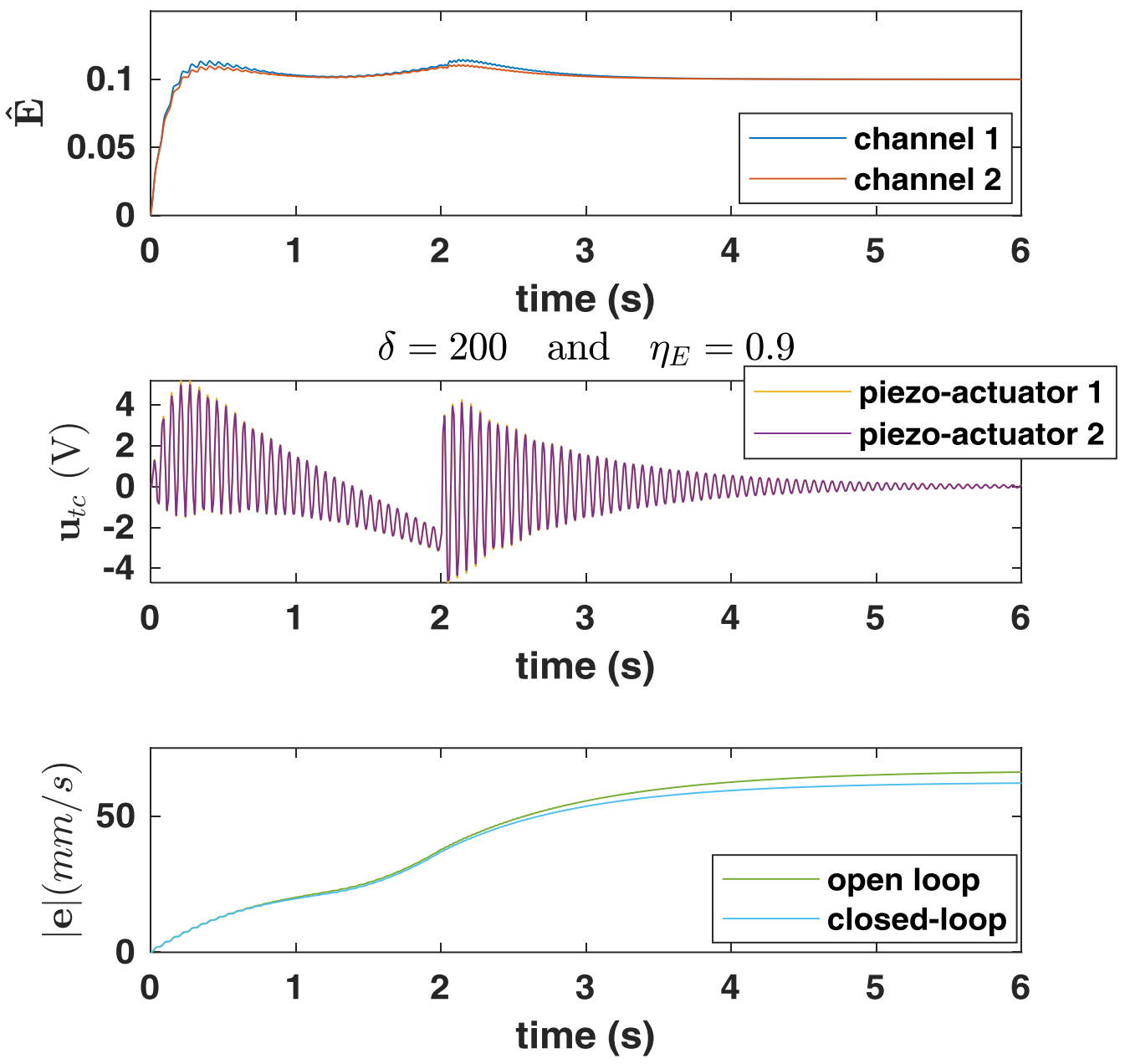

Figure 10.12 $u_{tc}, \hat{E}$ in the case of $\delta = 200$ and $\eta_E = 0.9$ [233]

The dependence of the control law and the disturbance rejection performance on the $\mu_{m,r,\sigma,w}$, $(m_0, r_0, \sigma_0, w_0)$, and $\eta_{m,r,\sigma,w}$ are presented in Figure 10.13, Figure 10.14, and Figure 10.15, respectively.

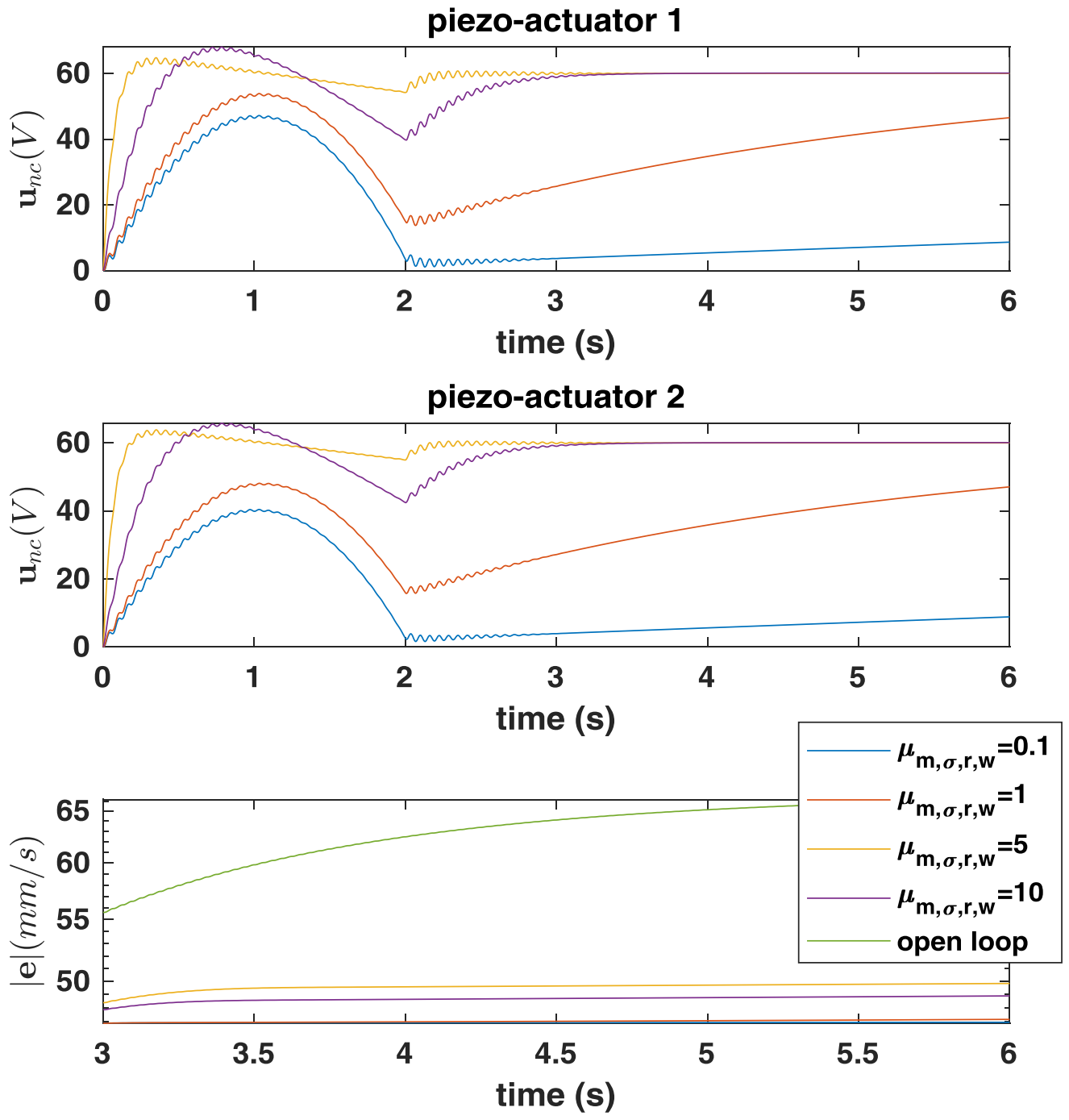

Figure 10.13 Dependence of the control law $u_{nc}$ and absolute cumulative error on $\mu_{m,r,\sigma,w}$ [233]

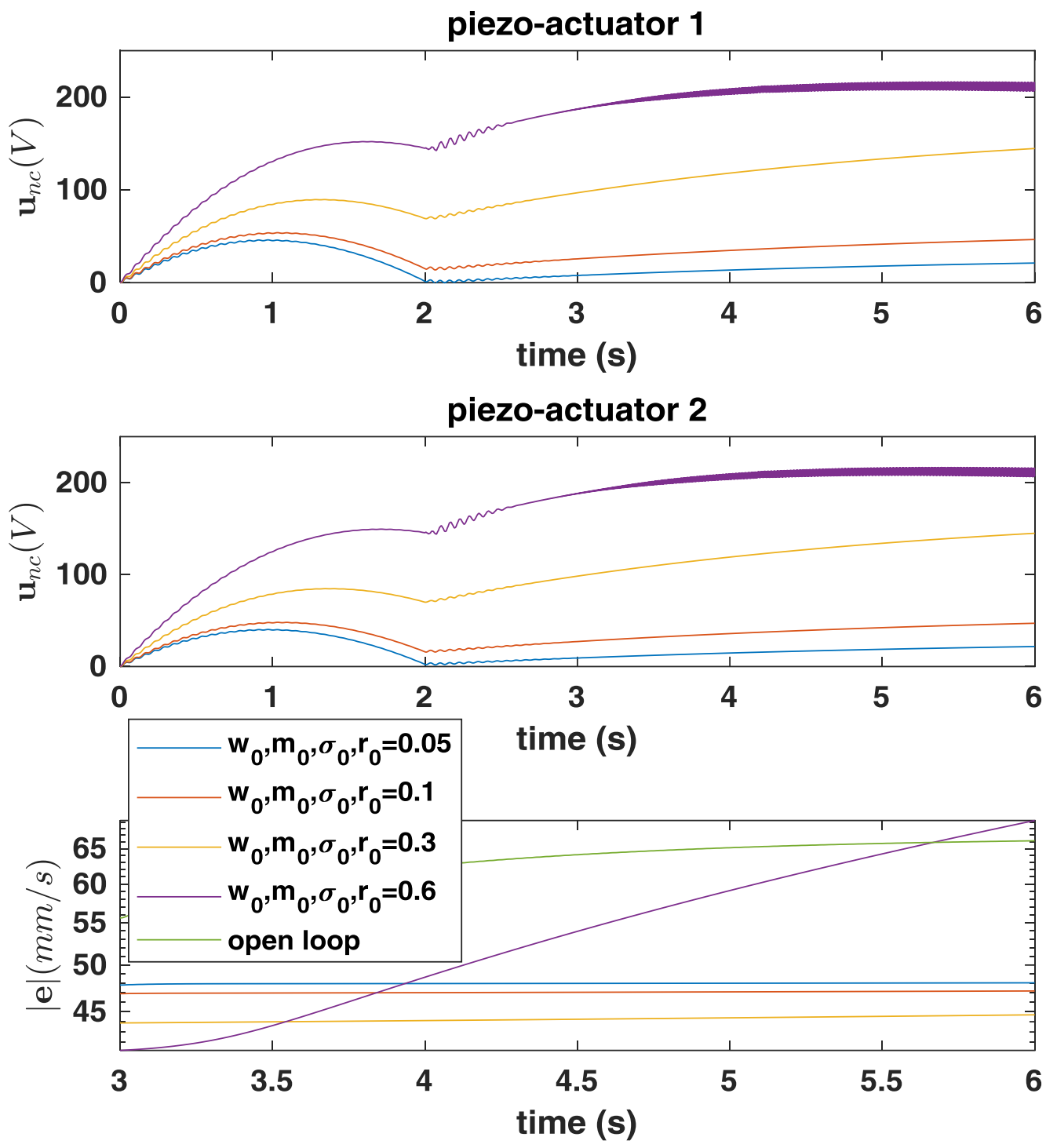

Figure 10.14 Dependence of the control law $u_{nc}$ and absolute cumulative error on $m_0, r_0, \sigma_0$, and $w_0$ [233]

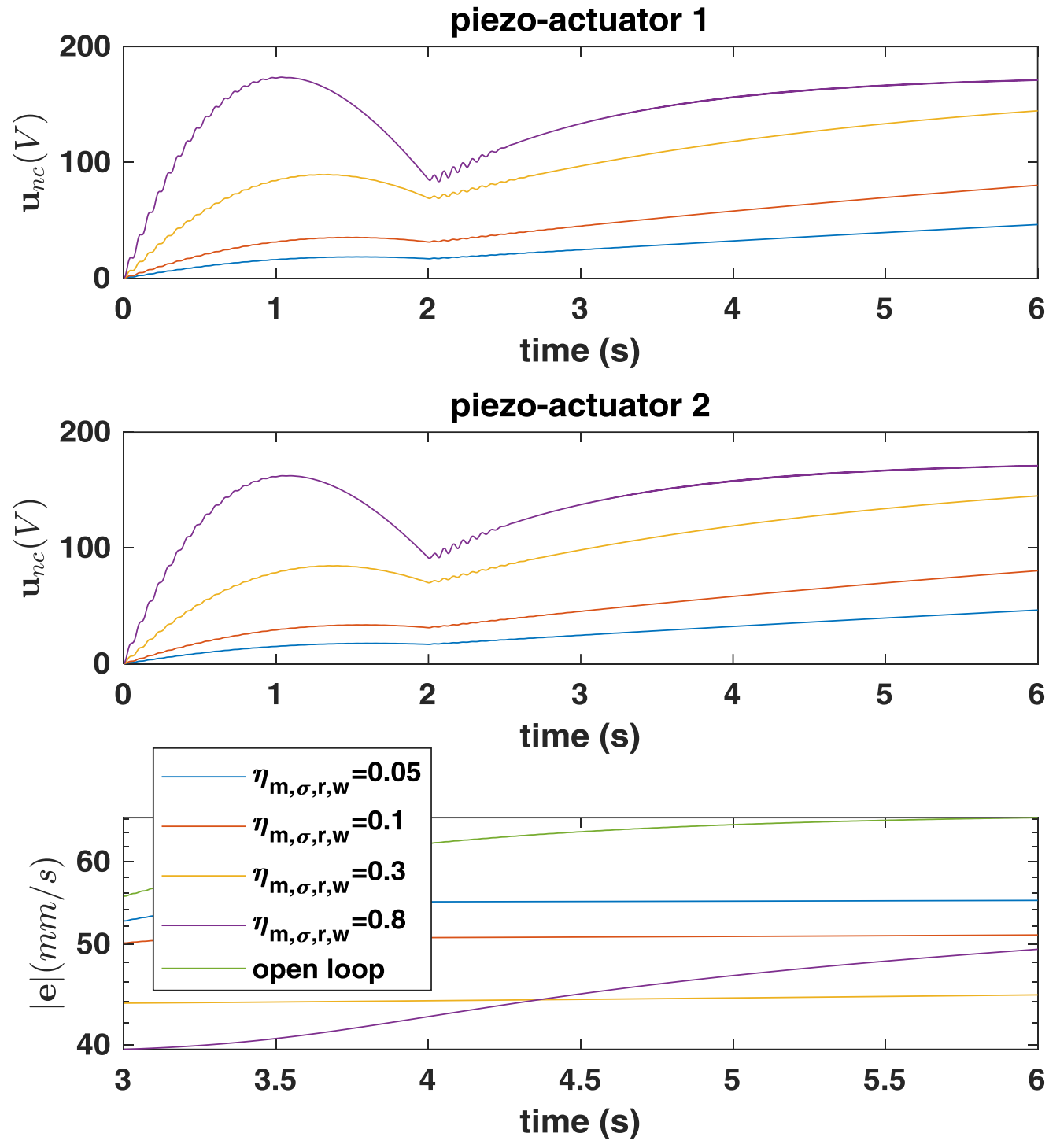

Figure 10.15 Dependence of the control law $u_{nc}$ and absolute cumulative error on $\eta_{m,r,\sigma,w}$ [233]

The following points are observed:

For small values of $\mu_{m,r,\sigma,w}$, the neuro-controller signal becomes small in amplitude, however as it can be seen, $\mu_{m,r,\sigma,w} = 0.1$ and $\mu_{m,r,\sigma,w} = 1$ still has the lowest absolute error level making them the primary choice for disturbance rejection.

By increasing the initial values for the training parameters, the neuro-control law increases and in contrast to the $\mu_{m,r,\sigma,w}$, it results in better disturbance rejection (better tracking) for the price of high energy consumption. This behavior marginally continues up to a certain level after which, as it can be seen in the first two subplots of Figure 10.14 for $m_0, r_0, \sigma_0,$ and $w_0$ equal to 0.6, the closed-loop performance is even worse than the open-loop one. The chattering control law as well as the high amplitude of the regulation signal may lead to a windup problem.

The leaning rates $\eta_{m,r,\sigma,w}$ demonstrates similar behavior as the initialization parameters.

### 10.5.2 Additional remarks about RWNN

Unlike the robust techniques e.g. classical $H_\infty -, \mu -$synthesis, and modern robust methods developed based on Lipschitz conditions of [263], the neural controller is less conservative. It should be noted this concept is intended for morphing mechanical systems since the underlying linear system (ULM) is assumed to be a stable and dominant part of the system response. This feature comes with the price of high computational demand, and in this chapter, the realization of a moderate scale model in active vibration control (AVC) applications is carried out successfully. The proposed adaptive system can accommodate robustness to deal with the mismatch disturbance signals, measurement noise, and tracking problem.

In the context of AVC of large structures with complex geometries, the neuro-controllers are an alternative to future industrial needs especially with the prospect of advances in modern DAQ systems with high computational power and memory. Nonetheless, the systematical network parameter assignment, sensitivity of closed-loop performance w.r.t. network parameters and discontinuous nonlinearities such as actuator saturation, and adaptation of the proposed RBAC system for nonlinear systems with known dynamics can be offered as candidates for future studies.

Two concrete proposals for future works are: 1) to perform a sensitivity analysis in terms of the network parameters. Such a study requires investigation of RWNN performance with respect to network dimension, learning rates, bound compensator parameter, and the choice of the correction gains in learning laws in a Hardware-in-Loop configuration. 2) In contrast to the nonlinear model-based control synthesis based on the polynomial nonlinear state-space system identification technique, the present controlling method mostly relies on the underlying linear model which is faster in terms of nonlinearity/modeling-uncertainty characterization and, quantification. As a result, an interesting comparison study is proposed to compare the adaptive RWNN of this chapter with the aforementioned nonlinear model-based technique.

The focus of the next chapter is on the estimating of the disturbance. More specifically, it is intended to first estimate and then reject the disturbance instead of using the robust techniques in DRC. By this means, the performance of the closed-loop system can be undoubtedly increased. It should be noted that the results of such estimation can be paired with RWNN to achieve an ultimate controller which is robust against unmodeled dynamics, uncertainties as well as external disturbances.

# 11 Dealing with disturbances: robust DOBC techniques

In this chapter, DRC based on unknown input observation (UIO), and disturbance-observer-based control (DOBC) methods are revisited for a class of MIMO systems with mismatch disturbance conditions. In both of these methods, the estimated disturbance is considered to be in the feedback channel. The disturbance term could represent either unknown mismatched signals penetrating the states, or unknown dynamics not captured in the modeling process, or physical parameter variations not accounted for in the mathematical model of the plant. In this chapter, unlike the high-gain approaches and variable structure methods e.g. Chapter 9, a systematic synthesis of the state/disturbance observer-based controller is carried out. For this purpose, first, using a series of singular value decompositions, the linearized plant's model is transformed into disturbance-free and disturbance-dependent subsystems. Then, functional state reconstruction based on the generalized detectability concept is proposed for the disturbance-free part. Then, a DRC based on quadratic stability theorem is employed to guarantee the performance of the closed-loop system. An important contribution offered in this chapter is the independence of the estimated disturbance from the control input which seems to be missing in the literature for disturbance decoupling problems. In the second method, DOBC is reconsidered with the aim of achieving a high level of robustness against modeling uncertainties and matched/mismatched disturbances, while at the same time retaining performance (not losing performance to have conservative robustness). Accordingly, unlike the first method, DRC, a so-called *full information* state observer is developed independently of the disturbance estimation. An advantage of such a combination is that disturbance estimation does not involve output derivatives. Finally, the case of systems with matched disturbances is presented as a corollary of the main results.

## 11.1 A comparison to the state-of-the-art

Research in the area of state observation and disturbance estimation in linear multivariable systems has reached a mature level, and various existence and sufficient conditions have been proposed in the last few decades [98], [99]. In the case of measurable disturbances, feedforward and feedback/feedforward strategies have been used to prevent performance degradation through optimization of some user-defined performance indices. This concept is also referred to as state reconstruction, as in [100], where unknown exogenous input signals are the source of erroneous system data. For such systems, a simple and well-established approach is the sliding mode observer (SMO), which allows for robust reconstruction of the states under unknown inputs [101]–[103]. Considering the disturbance as a bounded unknown mismatched exogenous input instead of opting for the classical stochastic approach based on a statistical distribution of the signal affecting the plant seems to be more practical in control applications and system theory. The mismatched unknown inputs are those signals that penetrate the system states through channels that are independent of the control input signals. Additionally, due to the lack of information about the external disturbances, intuitively high gain controllers are designed in the early development of DRC. It is well-known however that high gain controllers may lead to actuator windup problem and peaking phenomenon [104].

To a large extent, although categorized under a separate control synthesis approach, disturbances (and modeling uncertainties) may be handled by classical (robust) techniques. However, often conflicting performance and stability constraints on the closed-loop system require more advanced alternative approaches in estimating and then rejecting the disturbance. Recently, linear and nonlinear disturbance and uncertainty estimation and attenuation (LDUEA/NDUEA) methods have attracted considerable interest. Disturbance-observer-based control (DOBC) method as one candidate of DUEA, in general, delivers high robustness with respect to modeling uncertainties and matched/mismatched disturbances without performance loss [105]. Additionally, more rigorous stability analysis and synthesis can be obtained through the use of DOBC [106] compared to the methods of extended state observer (ESO), disturbance accommodation control (DAC) [107], and uncertainty and disturbance estimator (UDE) [108]. The unknown input excitations acting on the system also known as the lumped disturbance represent a generalized form of external stimuli that also takes into account the unmodeled dynamics of high-order nature and parameter variations. The case of static matched/mismatched disturbances can be considered as special cases of the control scheme proposed in this chapter [109], [110].

In summary, the following contributions are offered in the upcoming sections:

Two Lyapunov-based methods are proposed to reconstruct the system states in the presence of unknown and unmeasurable input signals. In this regard, the disturbance decoupling problem is revisited, and, based on the concept of unknown input observation (UIO), a reduced-order dynamic system is developed to simultaneously estimate the disturbance and observe the states. In order to transform DRC into a convex feasibility (optimization) problem, the controller synthesis is separated from that of UIO. Compared to most recent researches, one distinct novelty in the proposed combination is the independence of the dynamics of disturbance estimation mechanism from the control input. Consequently, the non-convexity of a systematical DRC synthesis is addressed in this chapter.

The second contribution, compared to the Luenburger state observer, is that the stability of the designed reduced-order UIO for disturbance decoupled subsystem is guaranteed which is missing in the literature. The stability guarantee of the decoupled UIO for systems with modeling uncertainties is a highly advantageous property that may relax the observability condition constraints required for the transformed system. The observability condition, as will be discussed, is a binding constraint in the previous methods due to the uniqueness of the singular value decomposition.

In the last part of the chapter, by aiming at suppressing the time-derivative of the measurement output (inherently polluted by noise) in the existing disturbance estimation mechanism that is based on the concept of DOBC, the UIO synthesis process is broken into two separate sequential problems. In the first, a revised form of the strong functional observer is proposed. This is followed by the second phase where a disturbance estimation and rejection is carried out. Two salient contributions are therefore made: (i) The state observer-based DOBC method, which is missing in the literature, and (ii) The problem formulation accounts for the most general case, namely mismatch disturbance. Additionally, for the first time, the cascaded controller gain is calculated based on convex optimization. Moreover, the limitations on transforming the design procedure into an eigenvalue problem (EVP) in a semi-definite programming (SDP) framework are carefully investigated. Finally, the effect of the measurement noise in output is examined in the disturbance/noise decoupling framework, based on available techniques of fault detection.

## 11.2 System representation and linearization

Consider a multivariable nonlinear dynamic system described in state-space form as

$$\dot{x}(t) = f\big(x(t), u(t)\big) + h\big(x(t)\big)w(t),$$
$$y(t) = l\big(x(t), u(t)\big), \tag{11.1}$$

where, $x(t) \in \Re^{n}_{\geq 0}$, $u(t) \in \Re^{m}_{\geq 0}$, $w(t) \in \Re^{q_w}_{\geq 0}$, and $y(t) \in \Re^{p}_{\geq 0}$ are the state, input, square-integrable disturbance (including un-modeled dynamics of high-order nature and parameter variations in the system elements), and output vectors at time $t \geq 0$, respectively. In addition, in order to express the plant as a nonlinear Lipschitz system, it is assumed that $f(.,.): \Re^{n}_{\geq 0} \times \Re^{m}_{\geq 0} \to \Re^{n}_{\geq 0}$, $h(.): \Re^{n}_{\geq 0} \to \Re^{q_w}_{\geq 0}$, and $l(.,.): \Re^{n}_{\geq 0} \times \Re^{m}_{\geq 0} \to \Re^{p}_{\geq 0}$ are sufficiently smooth and differentiable around the equilibrium points of (11.1). Then, the necessary and sufficient conditions on the asymptotic stability of the observer-based control system can be ascertained from the eigenvalues of the linearized system [263]. For causal plant, $f(0,0) = 0$, with globally Lipschitz continuous dynamics, assuming $h(0) = H \neq 0$, and $l(0,0) = 0$, (11.1) can be linearized in the MIMO form of (11.2). Moreover, the class of systems studied in this chapter satisfies the following assumptions: (*i*) the linearized plant is asymptotically stable, (*ii*) $m > q_w$ and $p > q_w$, and , (*iii*) matrices $H$ and $B$ have full column rank.

$$\dot{x} = Ax + Bu + Hw,$$
$$y = Cx + D_y u, \tag{11.2}$$

where $A = \partial f(x,u)/\partial x$, $B = \partial f(x,u)/\partial u$, $C = \partial l(x,u)/\partial x$, and $D_y = \partial l(x,u)/\partial u$. In (11.2) and throughout the rest of the chapter, to simplify the notation, the dependence of the state space variables on time is omitted. It should be pointed out that the feedthrough term in the linearized output equation is assumed to be zero ($D_y = 0$), since without this assumption the feasibility problem of disturbance estimation may become non-convex (consequently, not as attractive due to the limitations in solving them). Additionally, the disturbance term may appear in the output equation ($y = Cx$) in which case, as long as the condition (*ii*) is satisfied, the system can be transformed into (11.2) [264]. The form of (11.2) as a linearization of (11.1) around the equilibrium point of the plant, where the nonlinear system usually operates, opens the possibility of using global approximation methods such as the Takagi-Sugeno-Kang method of fuzzy inference which may generalize the applicability of the proposed techniques. Alternatively, large departures from the equilibrium could be modeled as uncertainties, and appropriate robust controllers may be synthesized to address these mismatching uncertainties.

## 11.3 Problem formulation and main Theorems

This section starts with a summary of its content, followed by a detailed analysis of existing estimation and disturbance rejection techniques.

In the next subsection, the design problem of UIO-based DRC for the disturbance decoupled system is investigated. Following immediately after, DOBC is investigated. Although DOBC with matching conditions has been addressed in the existing literature, (for example [106], [265]–[268]), the mismatch case for nonlinear systems which satisfy the Lipschitz condition has not yet received much attention, with some exceptions (for example [269], [270]). Most recently, two notable results are reported by [271], [272] to address the problems of variable structure compensator and $H_\infty$ controller for mismatch DOBC. Since, $H_\infty$ controller design is based on the worst case scenario, the performance of the closed-loop system is often compromised to preserve robustness [273]–[275].

In the following, a systematic Lyapunov-based approach is proposed to reconstruct the states using a full-order filter on system outputs and then, the input disturbance is simultaneously estimated and rejected in the plant output.

### 11.3.1 DRC based on disturbance decoupled UIO

Following the UIO method in [246], [276], the state equation can be decoupled by applying singular value decomposition (SVD) on the disturbance matrix ($H$). This operation decouples the system into two subsystems, one of which is independent of the unknown inputs. Accordingly, the SVD of $H$ can be written as $H = Q\Sigma_w R^T = [Q_1 \quad Q_2][\Sigma_1^T \quad 0]^T R^T = Q_1\Sigma_1 R^T$, where $Q \in \Re^{n\times n}$, $\Sigma_w \in \Re^{n\times q_w}$, $R \in \Re^{q_w\times q_w}$, and $\Sigma_1 \in \Re^{q_w\times q_w}$. Then, mapping the state equation by $Q^T$ and defining $Q^T x = \bar{x}$ and $R^T w = \bar{w}$, the decoupled state-space model can be written as in (11.3). This representation, in the terminology of fault observation/detection, is referred to as fault-dependent/fault-free decoupling:

$$\dot{\bar{x}}_1 = \bar{A}_{11}\bar{x}_1 + \bar{A}_{12}\bar{x}_2 + \bar{B}_1 u + \Sigma_1\bar{w}, \tag{11.3}$$

$$\dot{\bar{x}}_2 = \bar{A}_{21}\bar{x}_1 + \bar{A}_{22}\bar{x}_2 + \bar{B}_2 u,$$

where $\bar{A}_{11} \in \Re^{q_w \times q_w}$, $\bar{A}_{12} \in \Re^{q_w \times (n-q_w)}$, $\bar{A}_{21} \in \Re^{(n-q_w)\times q_w}$, $\bar{A}_{22} \in \Re^{(n-q_w)\times(n-q_w)}$, $\bar{B}_1 \in \Re^{q_w \times m}$, and $\bar{B}_2 \in \Re^{(n-q_w)\times m}$. Similar decomposition is applied on output equation ($y = CQ\bar{x} = [\bar{C}_1 \quad \bar{C}_2]\bar{x}$). Then, assuming the non-singular mapping $\mathcal{N}^T = \left[\bar{C}_1^{+T} \quad M^T\right]^T$ with $M \in \Re^{(p-q_w)\times p}$ and $\bar{C}_1^+$ being respectively the arbitrary matrix that guarantees the non-singularity of $\mathcal{N}$ (as a result: the non-singularity of the transformation) and pseudo-inverse of $\bar{C}_1$, the following two conditions can be stated: (*a*) $\bar{x}_1 = \bar{C}_1^+(y - \bar{C}_2\bar{x}_2)$ , (*b*) $My = M\bar{C}_1\bar{x}_1 + M\bar{C}_2\bar{x}_2$. Using condition ($a$), the fault-free equation (11.3) can be decoupled from the fault-dependent states ($\bar{x}_1$) as $\dot{\bar{x}}_2 = \tilde{A}_2\bar{x}_2 + \bar{B}_2 u + \tilde{G}_2 y$, where $\tilde{A}_2 = \bar{A}_{22} - \bar{A}_{21}\bar{C}_1^+\bar{C}_2$ and $\tilde{G}_2 = \bar{A}_{21}\bar{C}_1^+$. In the same fashion, one can obtain $\tilde{y} = \widetilde{H}y = \tilde{C}\bar{x}_2$ with $\widetilde{H} = M\left(\mathcal{I}_p - \bar{C}_1\bar{C}_1^+\right)$ and $\tilde{C} = M\left(\mathcal{I}_p - \bar{C}_1\bar{C}_1^+\right)\bar{C}_2$. It should be noted that based on the assumptions (*ii*) $m > q_w$ and $p > q_w$, it is trivial to show the existence of $M$ for the non-singular mapping $\mathcal{N} \in \Re^{p\times p}$ where $\bar{C}_1^+ \in \Re^{q_w \times p}$. The aim is then to formulate some conditions for the existence of a stable UIO that has a more flexible structure compared to the classical Luenberger's observer or Kalman filter in determining the state estimation error dynamics. Since one of the ultimate goals of this chapter is to include an estimation of the unknown mismatch disturbance input in feedback control law, the flexibility of the UIO is a key requirement. Consequently, the structure of UIO, and therefore the estimation error dynamics, defined in terms of some constraints under which designing the feedback gain for the estimated disturbance signal in DRC is transformable into a convex feasibility problem (see (11.8) compared to (11.7)). Therefore, standard solutions such as the interior point method can be employed. Accordingly, by selecting the dynamics of fault-free functional state-observer as $\dot{\hat{\bar{x}}}_2 = N\hat{\bar{x}}_2 + L\left(\tilde{y} - \tilde{C}\hat{\bar{x}}_2\right) + Ju$ and defining the decoupled state observation residual as $\bar{e}_2 = \bar{x}_2 - \hat{\bar{x}}_2$, one can derive (11.4) similar to [277].

$$\begin{aligned}\dot{\bar{e}}_2 = \left(N - L\tilde{C}\right)\bar{e}_2 + \left(\tilde{A}_2 + \tilde{G}_2 CQ_2 - L\widetilde{H}CQ_2 + L\tilde{C} - N\right)\bar{x}_2 + (\bar{B}_2 - J)u \\ + \left(\tilde{G}_2 C - L\widetilde{H}C\right)Q_1\bar{x}_1. \end{aligned} \tag{11.4}$$

Using (11.4), the estimation of the disturbance can be calculated simultaneously as $\hat{w} = R\Sigma_1^{-1}\left\{\dot{\hat{\bar{x}}}_1 - \bar{A}_{11}\hat{\bar{x}}_1 - \bar{A}_{12}\hat{\bar{x}}_2 - \bar{B}_1 u\right\}$ with $\hat{\bar{x}}_1 = \bar{C}_1^+(y - \bar{C}_2\hat{\bar{x}}_2)$. Then, in contrast to conventional disturbance decoupling methods such as [99], [102], [264], the LME constraints are defined as

$$\begin{aligned}\bar{C}_1^+CB &= \bar{C}_1^+\bar{C}_2 J + \bar{B}_1 \\ \bar{C}_1^+CA &= \bar{C}_1^+\bar{C}_2 L\widetilde{H}C\end{aligned} \tag{11.5}$$

**Definition 11.1.** The extended state reconstruction system is asymptotically stable if: *a*) The pair $(\tilde{A}_2, \tilde{C})$ is observable. *b*) The Lyapunov stability of the estimator is satisfied and *c*) if the matrices $N \in \Re^{(n-q_w)\times(n-q_w)}$, $L \in \Re^{(n-q_w)\times(p-q_w)}$, and $J \in \Re^{(n-q_w)\times m}$ can be calculated such that $\lim_{t\to\infty}\|x - \hat{x}\| = 0$.

**Theorem 11.1** *Decoupled state reconstruction system based on the concept of UIO defined as* $\dot{\hat{\bar{x}}}_2 = N\hat{\bar{x}}_2 + L\left(\tilde{y} - \tilde{C}\hat{\bar{x}}_2\right) + Ju$ *is asymptotically stable in terms of* **Definition 11.1** *and rejects the unknown time-dependent input w in state observation residuals as well as the decoupled fault-free/dependent states if (1) assumptions* (*i*), (*ii*), and (*iii*) *are satisfied, (2) the pair* $(\tilde{A}_2, \tilde{C})$ *is observable, and (3) there exist positive definite matrices* $P_i, i = 1, \ldots, 4$, *matrices N, L, and J with appropriate dimensions, and scalar* $\gamma_1$ *such that the linear matrix inequality (LMI)/LME conditions stated in* (11.6a) *are satisfied*

$$\begin{aligned}\mathbf{\Pi_1} = \left[\Pi_{1ij}\right] \leq 0; i,j = 1,\ldots,6, \\ N < 0, \\ \gamma_1 > 0,\end{aligned} \tag{11.6a}$$

$$\bar{C}_1^{+}CB - \bar{C}_1^{+}\bar{C}_2 J - \bar{B}_1 = 0,$$

$$\bar{C}_1^{+}CA - \bar{C}_1^{+}\bar{C}_2 L\tilde{H}C = 0,$$

*where* $\mathbf{\Pi_1}$ *is a symmetric matrix and* $\Pi_{1ij}$ *are zero matrices with appropriate dimension except for the elements in* (11.5b)

$$\begin{gathered}
\Pi_{111} = N^T - \tilde{C}^T L^T + N - L\tilde{C} + P_1, \\
\Pi_{112} = \tilde{C}^T L^T \bar{C}_2^T \bar{C}_1^{+T} - N^T \bar{C}_2^T \bar{C}_1^{+T}, \\
\Pi_{113} = \tilde{A}_2 + \tilde{G}_2 C Q_2 - L\tilde{H}CQ_2 + L\tilde{C} - N, \\
\Pi_{114} = \tilde{G}_2 C Q_1 - L\tilde{H}CQ_1, \\
\Pi_{115} = \bar{B}_2 - J, \\
\Pi_{123} = \bar{A}_{12} - \bar{C}_1^{+}CAQ_2 + \bar{C}_1^{+}\bar{C}_2 N + \bar{C}_1^{+}\bar{C}_2 L\tilde{H}CQ_2 - \bar{C}_1^{+}\bar{C}_2 L\tilde{C}, \\
\Pi_{124} = \bar{A}_{11}, \\
\Pi_{126} = \Sigma_1 R^T - \bar{C}_1^{+}CH, \\
\Pi_{133} = P_2 + \tilde{A}_2^T + \tilde{A}_2 + \bar{C}_2^T \tilde{G}_2^T + \tilde{G}_2 \bar{C}_2, \\
\Pi_{134} = \tilde{G}_2 \bar{C}_2, \\
\Pi_{135} = \bar{B}_2, \\
\Pi_{166} = -\gamma_1^2 \mathfrak{I}_{q_w}.
\end{gathered} \tag{11.5b}$$

**Proof.** Consider a Lyapunov equation in the form of $V_1(\bar{e}_1, \bar{e}_2, \bar{x}, u, w) = V_{11} + V_{12}$, ($V_{11} = \bar{e}_1^T \bar{e}_1 + \bar{e}_2^T \bar{e}_2 + \bar{x}_2^T \bar{x}_2$, and $V_{12} = \int_0^\infty (\bar{e}_2^T P_1 \bar{e}_2 + \bar{x}_2^T P_2 \bar{x}_2 - \gamma_1^2 w^T w)\mathrm{d}t$,). Then $\dot{V}_1 = \mathbb{L}_1^T \mathbf{\Pi_1} \mathbb{L}_1 \leq 0$, with $\mathbb{L}_1^T = [\bar{e}_2^T \quad \bar{e}_1^T \quad \bar{x}_2^T \quad \bar{x}_1^T \quad u^T \quad w^T]$, will lead to LMI 1 in (11.6a).

Note that the Lyapunov function above is proposed on the grounds of the non-expansive nature of the system (also known as a contractive system) expressed in (11.2). In other words, the $H_\infty$-norm of the linear system in (11.2) can be formulated in an LMI framework using Bounded Real Lemma (BRL). The rest is direct parameter specifications defined in the UIO system construction. It should be noted that there is no guarantee that with the decomposition in (11.2) $\tilde{A}_2$ is Hurwitz and therefore the observability condition of the theorem cannot be relaxed in conventional state observation techniques. Additionally, the proof for the existence of a solution is along similar lines as in [246], and thus is omitted here. This completes the proof.

■

The effect of output measurement noise is treated as a generalization of **Theorem 11.1**. However, in the case that $p = q_n$ ($q_n$ representing the dimension of noise on output) multiple UIO (MUIO) should be designed. The case of matched disturbances is given as **Corollary 11.1** to **Theorem 11.2**.

It should be pointed out that the Lyapunov equation in the proof of **Theorem 11.1** is chosen in the least conservative form. This is because of the fact that the design parameters of UIO are independent of those in DRC (**Theorem 11.2**). In other words, in the most conservative configuration, the quadratic function $V_{11} = \tau_1^T \mathbf{P_1} \tau_1$ with $\tau_1^T = [\bar{x}_1^T \quad \bar{x}_2^T \quad \bar{e}_1^T \quad \bar{e}_2^T]$ should be selected as a Lyapunov candidate in which $\mathbf{P_1}$ is a symmetric positive definite matrix with appropriate dimension. However, due to the coupling introduced by off-diagonal block matrices in $\mathbf{P_1}$, the problem of designing observer and controller parameters simultaneously becomes non-convex (Projection Lemma is inapplicable). In literature, either the off-diagonal block matrices are simply set to zero (decoupling the controller and observer design) (see Remark 3 and 4 of [54]) or additional linear matrix equalities are introduced [71], [103]. However, for UIO in **Theorem 11.1** such LMEs do not exist, and additionally, Congruence transformation is not applicable. Therefore, the decoupled version of the Lyapunov stability theorem is implemented separately for UIO and DRC (similar to the idea in [266]). In such a setting, since the UIO design algorithm is prior to DRC, the term of the control law $u$ introduces zero diagonal elements in $\mathbf{\Pi_1}$ ($\Pi_{155} = 0$) after employing Schur complement Lemma. This is the scenario for which the negative definiteness of $\mathbf{\Pi_1}$ is replaced with $\mathbf{\Pi_1} = [\Pi_{1ij}] \leq 0$. For a Hurwitz nominal system with the assumptions in **Theorem 11.1**, the Lyapunov stability theorem guarantees that: For $w(t) = 0, \forall t \geq 0$ the UIO remains stable for any arbitrary bounded initial condition ($x(0)$). In addition, for $w(t) \neq 0$, the trajectories of the state estimation error dynamics remain bounded as long as $w(t)$ is an $L_2$ bounded function. LMI (11.6a) is a non-strict, which can be converted to a strict feasible LMI through eradicating the existing LME constraints and then reducing the resulted LMI by eliminating any constant null space (page 19 of [219]). Moreover, non-strict LMI constraints do not reduce the generality of the method, since satisfying strict inequality constraints in applied optimization interfaces e.g. YALMIP toolbox package is infeasible.

The functional observer designed based on **Theorem 11.1** *simultaneously* (unlike **Theorem 11.3**) observes the system states and detects the disturbance. This concept is intimately related to UIO in which the observation problem of $\rho + 1$ number of systems is reduced to the design problem of $\rho$ estimators. In view of LME conditions (11.6a), the formulation for the estimated disturbance can be simplified as $\hat{w} = R\Sigma_1^{-1}\{(\bar{C}_1^{+}\bar{C}_2 L\tilde{C} - \bar{A}_{21} - \bar{C}_1^{+}\bar{C}_2 N)\hat{\bar{x}}_2 - \bar{A}_{11}\hat{\bar{x}}_1 + \bar{C}_1^{+} CHw\}$. It should be pointed out that an advantage of the proposed method in comparison to [278], [279] is that the dynamics of the disturbance estimation mechanism (see (11.8)) are independent of the control input. As a result, in order to bridge this gap in the literature, the non-convexity of a systematical DRC synthesis by minimizing the induced norm of the transfer function from disturbance to output can be overcome. One should note that the nominal plant model in [279] includes a stable Lipschitz nonlinear term in the state equation that accounts for un-modeled dynamics and therefore is distinguished from the current results proposed in this chapter. In previous publications, such as in [280], due to this issue, the feedback law in terms of the estimated disturbance is reported as $-(1 - \tanh((\bar{x}_1 - \hat{\bar{x}}_1)/\mathcal{E}_1))\,\hat{w}$ while the state-observer-based feedback gain is calculated by solving an Algebraic Riccati Equation (ARE). In another approach, as an alternative view, [264] used parametric solution of generalized Sylvester equation of [281] to design a reduced order disturbance decoupled observer. Once they reconstructed the state vector, by reformulating the system in (11.2), the fault detection filter in their approach is obtained as expressed in (11.7).

$$
\begin{aligned}
\hat{w} &= R\Sigma_1^{-1}Q_1^{T}\{\Psi_1\hat{\bar{x}}_2 + \Psi_2 y + \Psi_4 u\},\\
\Psi_1 &= \bar{Q}_x(\tilde{A}_2 - \mathcal{L}\tilde{C}) - A\bar{Q}_x,\\
\Psi_2 &= \bar{Q}_x(\tilde{G}_2 - \mathcal{L}\tilde{H}),\\
\Psi_3 &= \bar{Q}_x\bar{B}_2 - B,\\
\bar{Q}_x &= Q\begin{bmatrix} -\bar{C}_1^{+}\bar{C}_2 \\ \mathcal{I} \end{bmatrix},
\end{aligned}
\tag{11.7}
$$

where $\mathcal{L}$ is the observer gain to be designed. Considering the control law as $= K_x\hat{x} + K_w\hat{w}$ , where $\hat{x} = Q[\hat{\bar{x}}_1^T \quad \hat{\bar{x}}_2^T]^T$, will bring $\hat{w}$ to the right-hand side (RHS) of (11.7), and lead to a nonlinear feasibility problem in terms of designing $K_w$. This problem is addressed in this chapter as well. Moreover, in the method proposed, compared to the standard Luenberger's observer, which is designed on the fault-free counterpart of the mapped states, it should be pointed out that although in $\tilde{A}_2 - \mathcal{L}\tilde{C}$, under the assumption of observability of the pair $(\tilde{A}_2, \tilde{C})$, $\mathcal{L}$ can set the eigenvalues of the decoupled state estimation error dynamics in LHS of complex plain, the lack of information on $\tilde{A}_2$ and uniqueness of the SVD may destabilize the observer for relatively small mismatch uncertainty in modeling the system. However, in the proposed method, this problem is addressed by using full information observer $(N - L\tilde{C})$. This trend of observation and rejection of unknown disturbance, which acts on system states and outputs, is referred to as fault detection and isolation (FDI) [106].

**Remark 11.1.** In condition (I) of disturbance decoupling, $\bar{x}_1 = \bar{C}_1^{+}(y - \bar{C}_2\bar{x}_2)$, and in defining the mapping $\mathcal{N}$, an arbitrary matrix $M$ is required to guarantee the non-singularity of the transformation. It follows from **Theorem 11.1** that in order to have an LMI constraint (for possibly a convex optimization problem $\inf_{\Pi_1 \leq 0} \gamma_1^*$ with $\gamma_1^* = \gamma_1^2$), matrix $M$ should be determined prior to the feasibility/optimization phase. The assumption that $M$ is fixed is a conservative way to design the observer. It should be pointed out that defining a new variable $\hat{L} = LM$ postpones the calculation of $M$ while keeping (11.6a) as a convex constraint. This would reduce the conservatism of the obtained solution, however, the lack of an appropriate algorithm in decomposing $\hat{L}$ back into individual variables $L$ and $M$ is the main drawback of this alternative. As a more practical approach in reducing the conservativeness of the UIO design problem, is to construct a bilinear matrix inequality (BMI) that can be defined over the design parameters $M$ and $L$. Convex optimization problem such as (11.6a) is classified as P-hard problem in which P refers to a family of optimization problems (independent of the solution algorithm) that requires a bounded *polynomial time* in reaching a feasible/optimal solution [282], [283]. In contrast, defining $L$ and $M$ in BMI framework is classified as NP-hard which suffers from having a computing *polynomial time*. Therefore, a global solver is required for convergence of non-convex feasibility/optimization subjected to BMI constraints [284]. A practical tool to solve (11.6a) with the revised block matrices (replacing $\tilde{C} = M\tau\bar{C}_2$, $\widetilde{H} = M\tau$, and $\tau = \mathcal{I}_p - \bar{C}_1\bar{C}_1^+$) is to use the well-known branch and bound algorithm using linear programming relaxations (LPR) and convex envelope approximations (BMIBNB). An implementation of this algorithm is available in YALMIP (see sdpsettings) [285].

In addition, in an analogy to the disturbance observation technique based on the geometrical approach for the systems without uncertainty, the state reconstruction, and disturbance rejection has a solution if $\mathrm{Im}(H) \subseteq \mathcal{V}^* \oplus \mathrm{Im}(B)$ where $\mathrm{Im}(.)$ is the span of the transformation, $\mathcal{W} \oplus \mathcal{V}$ represents the subspace summation of $\mathcal{W}$ and $\mathcal{V}$, and finally $\mathcal{V}^*$ is the supremum subspace that is $(A, \mathrm{Im}(B))$-invariant (iff $A\mathcal{V} \subseteq \mathcal{V} \oplus \mathrm{Im}(B)$ for matrices in (11.2)). However, the DRC in this chapter, unlike the geometrical approach reduces to quadratic stability problem instead of setting the transfer function from disturbance to output equal to null [286]–[288]. Koenig and Mammar generalized the conditions in linear UIO to nonlinear systems and developed a reduced-order observer for this purpose (FDI) using the disturbance decoupling technique [276]. Sharma and Aldeen followed the same technique to reconstruct the states as well as estimating the disturbance for nonlinear systems [279], [289]. Accordingly, selecting the DRC law as $u = K_x\hat{x} + K_w\widehat{w}$ for the estimation mechanism in **Theorem 11.1**, (11.8) can be obtained

$$
\begin{gathered}
\widehat{w} = \Omega_{11}\hat{\bar{x}}_2 + \Omega_{12}\hat{\bar{x}}_1 + \Omega_{13}w, \\
\Omega_{11} = R\Sigma_1^{-1}\left(\bar{C}_1^+\bar{C}_2L\tilde{C} - \bar{A}_{21} - \bar{C}_1^+\bar{C}_2N\right), \\
\Omega_{12} = -R\Sigma_1^{-1}\bar{A}_{11}, \\
\Omega_{13} = R\Sigma_1^{-1}\bar{C}_1^+CH, \\
\dot{\hat{\bar{x}}}_1 = \Omega_{21}y + \Omega_{22}\bar{x}_2 + \Omega_{23}w + \Omega_{24}\bar{e}_2, \\
\Omega_{21} = \bar{A}_{11}\bar{C}_1^+ + \bar{B}_2K_xQ_1\bar{C}_1^+ + \bar{B}_1K_w\Omega_{12}\bar{C}_1^+, \\
\Omega_{22} = \bar{A}_{12} - \bar{A}_{11}\bar{C}_1^+\bar{C}_2 + \bar{B}_1K_xQ_2 - \bar{B}_1K_xQ_1\bar{C}_1^+\bar{C}_2 - \bar{B}_1K_w\Omega_{12}\bar{C}_1^+\bar{C}_2 + \bar{B}_1K_w\Omega_{11}, \\
\Omega_{23} = \bar{B}_1K_w\Omega_{13} + \Sigma_1R^T, \\
\Omega_{24} = \bar{B}_1K_xQ_1\bar{C}_1^+\bar{C}_2 - \bar{B}_1K_xQ_2 - \bar{B}_1K_w\Omega_{11} + \bar{B}_1K_w\Omega_{12}\bar{C}_1^+\bar{C}_2, \\
\dot{\hat{\bar{x}}}_2 = \Omega_{31}y + \Omega_{32}\bar{x}_2 + \Omega_{33}w + \Omega_{34}\bar{e}_2, \\
\Omega_{31} = \bar{B}_2K_xQ_1\bar{C}_1^+ + \tilde{G}_2 + \bar{B}_2K_w\Omega_{12}\bar{C}_1^+, \\
\Omega_{32} = \tilde{A}_2 - \bar{B}_2K_xQ_1\bar{C}_1^+\bar{C}_2 + \bar{B}_2K_xQ_2 + \bar{B}_2K_w\Omega_{11} - \bar{B}_2K_w\Omega_{12}\bar{C}_1^+\bar{C}_2, \\
\Omega_{33} = \bar{B}_2K_w\Omega_{13}, \\
\Omega_{34} = -\Omega_{32} + \tilde{A}_2.
\end{gathered} \tag{11.8}
$$

Next, the synthesis of the persistent DRC is reduced to multi-dimensional convex minimization in terms of the system of LMIs/LMEs.

**Theorem 11.2** *The system of* (11.2) *with the feedback control law as* $u = K_x\hat{x} + K_w\widehat{w}$*, where* $\hat{x}$ *and* $\widehat{w}$ *are the observed state vector and the estimated disturbance vector of* **Theorem 11.1** *is asymptotically stable and rejects the disturbance in system states quadratically if 1)* $\bar{B}_1$ *is of full column rank and 2) there exist symmetric matrices* $R_1 \in \mathfrak{R}^{q_w \times q_w}$*,* $R_3 \in \mathfrak{R}^{(n-q_w)\times(n-q_w)}$*,* $R_4 \in \mathfrak{R}^{q_w \times q_w}$*, and* $R_6 \in \mathfrak{R}^{(n-q_w)\times(n-q_w)}$*, matrices* $\hat{R}_1 \in \mathfrak{R}^{m\times m}$*,* $\widehat{K}_x \in \mathfrak{R}^{m\times n}$*, and* $\widehat{K}_w \in \mathfrak{R}^{m\times q_w}$*, and positive scalar* $\gamma_4$ *such that the system of LMI/LME in conditions* (11.9a) *are satisfied*

$$\begin{gathered}\mathbf{\Pi_2} = [\Pi_{2ij}] \le 0; i,j = 1,\dots,4,\\ \gamma_4 > 0,\\ R_1\bar{B}_1 - \bar{B}_1\hat{R}_1 = 0,\end{gathered} \tag{11.9a}$$

*where* $\mathbf{\Pi_2}$ *is a symmetric matrix and* $\Pi_{2ij}$ *are zero matrices with appropriate dimensions except for the elements in* (11.8b)

$$\begin{aligned}
\Pi_{211} &= -\bar{A}_{12}^T R_1 \bar{C}_1^+ \bar{C}_2 + \bar{C}_2^T {\bar{C}_1^+}^T \bar{A}_{11}^T R_1 \bar{C}_1^+ \bar{C}_2 - Q_2^T \widehat{K}_x^T \bar{B}_1^T \bar{C}_1^+ \bar{C}_2 + \bar{C}_2^T {\bar{C}_1^+}^T Q_1^T \widehat{K}_x^T \bar{B}_1^T \bar{C}_1^+ \bar{C}_2\\
&\quad + \bar{C}_2^T {\bar{C}_1^+}^T \Omega_{12}^T \widehat{K}_w^T \bar{B}_1^T \bar{C}_1^+ \bar{C}_2 - \Omega_{11}^T \widehat{K}_w^T \bar{B}_1^T \bar{C}_1^+ \bar{C}_2 - \bar{C}_2^T {\bar{C}_1^+}^T R_1 \bar{A}_{12}\\
&\quad + \bar{C}_2^T {\bar{C}_1^+}^T R_1 \bar{A}_{11} \bar{C}_1^+ \bar{C}_2 - \bar{C}_2^T {\bar{C}_1^+}^T \bar{B}_1 \widehat{K}_x Q_2 + \bar{C}_2^T {\bar{C}_1^+}^T \bar{B}_1 \widehat{K}_x Q_1 \bar{C}_1^+ \bar{C}_2\\
&\quad + \bar{C}_2^T {\bar{C}_1^+}^T \bar{B}_1 \widehat{K}_w \Omega_{12} \bar{C}_1^+ \bar{C}_2 - \bar{C}_2^T {\bar{C}_1^+}^T \bar{B}_1 \widehat{K}_w \Omega_{11} + R_3 + \bar{C}_2^T {\bar{C}_1^+}^T R_4 \bar{C}_1^+ \bar{C}_2,\\
\Pi_{212} &= -\bar{C}_2^T {\bar{C}_1^+}^T \bar{B}_1 \widehat{K}_x Q_1 \bar{C}_1^+ \bar{C}_2 + \bar{C}_2^T {\bar{C}_1^+}^T \bar{B}_1 \widehat{K}_x Q_2 + \bar{C}_2^T {\bar{C}_1^+}^T \bar{B}_1 \widehat{K}_w \Omega_{11}\\
&\quad - \bar{C}_2^T {\bar{C}_1^+}^T \bar{B}_1 \widehat{K}_w \Omega_{12} \bar{C}_1^+ \bar{C}_2,\\
\Pi_{213} &= \bar{A}_{12}^T R_1 \bar{C}_1^+ - \bar{C}_2^T {\bar{C}_1^+}^T \bar{A}_{11}^T R_1 \bar{C}_1^+ + Q_2^T \widehat{K}_x^T \bar{B}_1^T \bar{C}_1^+ - \bar{C}_2^T {\bar{C}_1^+}^T Q_1^T \widehat{K}_x^T \bar{B}_1^T \bar{C}_1^+\\
&\quad - \bar{C}_2^T {\bar{C}_1^+}^T \Omega_{12}^T \widehat{K}_w^T \bar{B}_1^T \bar{C}_1^+ + \Omega_{11}^T \widehat{K}_w^T \bar{B}_1^T \bar{C}_1^+ - \bar{C}_2^T {\bar{C}_1^+}^T R_1 \bar{A}_{11}^T \bar{C}_1^+\\
&\quad - \bar{C}_2^T {\bar{C}_1^+}^T \bar{B}_1 \widehat{K}_x Q_1 \bar{C}_1^+ - \bar{C}_2^T {\bar{C}_1^+}^T \bar{B}_1 \widehat{K}_w \Omega_{12} \bar{C}_1^+ - \bar{C}_2^T {\bar{C}_1^+}^T R_4 \bar{C}_1^+,\\
\Pi_{214} &= -\bar{C}_2^T {\bar{C}_1^+}^T \bar{B}_1 \widehat{K}_w \Omega_{13} - \bar{C}_2^T {\bar{C}_1^+}^T R_1 \Sigma_1 R^T,\\
\Pi_{223} &= \bar{C}_2^T {\bar{C}_1^+}^T Q_1^T \widehat{K}_x^T \bar{B}_1^T \bar{C}_1^+ - Q_2^T \widehat{K}_x^T \bar{B}_1^T \bar{C}_1^+ - \Omega_{11}^T \widehat{K}_w^T \bar{B}_1^T \bar{C}_1^+ + \bar{C}_2^T {\bar{C}_1^+}^T \Omega_{12}^T \widehat{K}_w^T \bar{B}_1^T \bar{C}_1^+,\\
\Pi_{233} &= {\bar{C}_1^+}^T \bar{A}_{11}^T R_1 \bar{C}_1^+ + {\bar{C}_1^+}^T Q_1^T \widehat{K}_x^T \bar{B}_1^T \bar{C}_1^+ + {\bar{C}_1^+}^T \Omega_{12}^T \widehat{K}_w^T \bar{B}_1^T \bar{C}_1^+ + {\bar{C}_1^+}^T R_1 \bar{A}_{11} \bar{C}_1^+\\
&\quad + {\bar{C}_1^+}^T \bar{B}_1 \widehat{K}_x Q_1 \bar{C}_1^+ + {\bar{C}_1^+}^T \bar{B}_1 \widehat{K}_w \Omega_{12} \bar{C}_1^+ + {\bar{C}_1^+}^T R_4 \bar{C}_1^+,\\
\Pi_{234} &= {\bar{C}_1^+}^T \bar{B}_1 \widehat{K}_w \Omega_{13} + {\bar{C}_1^+}^T R_1 \Sigma_1 R^T,\\
\Pi_{244} &= -\gamma_4^2 \mathfrak{I}.
\end{aligned} \tag{11.8b}$$

*Then, the control gains can be obtained as* $K_x = \hat{R}_1^{-1}\widehat{K}_x$ *and* $K_w = \hat{R}_1^{-1}\widehat{K}_w$*.*

**Proof.** Let's propose the Lyapunov candidate as $V_2(\bar{x},u,w) = \bar{x}_1^T R_1 \bar{x}_1 + \int_0^\infty (\bar{x}_2^T R_3 \bar{x}_2 + \bar{x}_1^T R_4 \bar{x}_1 - \gamma_4^2 w^T w)\mathrm{d}t$ and differentiate the Lyapunov function along the system trajectory in (11.2). Due to the independent design variables of $K_x$ and $R_1$, the resulting equation will be nonlinear, and the well-known Congruence transformation cannot be used due to the term $\bar{C}_1^+\bar{C}_2$ in the obtained equation. The LME constraint in (11.9a) is therefore introduced [55]. Following comparable procedure as in the proof of Theorem 1a, $\dot{V}_2 = \mathbb{L}_2^T \mathbf{\Pi_2} \mathbb{L}_2 \le 0$ with $\mathbb{L}_2^T = [\bar{x}_2^T \quad \bar{e}_2^T \quad y^T \quad w^T]$ will lead to the LMI 1 expressed in (11.9a). Existence of solution for LME constraint is guaranteed by satisfying the condition (1) of **Theorem 11.2**. This completes the proof.

■

The two-step method proposed in **Theorem 11.1** and **Theorem 11.2** solves the issues related to the high-gain controllers such as the windup problem in actuator and peaking phenomenon in states and control law. Despite the SMO-based controller mentioned before, the transient response of the control law in DRC is not subjected to peaking problem. However, the proposed disturbance estimator originally requires the calculation of $\dot{\bar{x}}_1$, which includes the derivative of the output. As a result for the real-time implementation, an appropriate low-pass analog filter on the output is required to reject or attenuate noise. Alternatively the results of **Theorem 11.3** or **Theorem 11.5** can be used. Another advantage of the results stated in **Theorem 11.1** and **Theorem 11.2** is the order of observer, where for a plant of order $n$ whole states can be estimated by a functional UIO of order $(n - q_w)$ in comparison to the strong functional observer in **Theorem 11.3** where the total order is $(n + q_w)$ for the auxiliary observer states.

Before proceeding to the strong functional observer, for the sake of completeness, the case in which the output is polluted by noise $n \in \Re^{q_n}_{\geq 0}$ and the state equation is under the effect of input disturbance $(w)$ is considered. Assuming $p \geq q_n + q_w$ then the output equation is formulated as $y = Cx + Fn$ where $F \in \Re^{p\times q_n}$ is a known matrix which can be decomposed as $F = U_n\Sigma_n V_n^T = [U_1 \quad U_2][\Sigma_2^T \quad 0]^T V_n^T = U_1\Sigma_2 V_n^T$ using SVD in which $U_n \in \Re^{p\times p}, \Sigma_n \in \Re^{p\times q_n}$, $V_n \in \Re^{q_n\times q_n}$, and $\Sigma_2 \in \Re^{q_n\times q_n}$. Defining $V_n^T n = \bar{n}$, the noise-dependent and noise-free decoupled output counterparts can be respectively stated as $\bar{y}_1 = U_1^T Cx + \Sigma_2\bar{n}$ and $\bar{y}_2 = U_2^T Cx = \left[\bar{C}_1^* \quad \bar{C}_2^{\,*}\right]\bar{x}$, where $\bar{C}_1^* \in \Re^{(p-q_n)\times q_w}$ and $\bar{C}_2^* \in \Re^{(p-q_n)\times(n-q_w)}$. Accordingly, after decoupling the output equation, $\bar{y}_2$, becomes noise-independent. Unlike the system in (11.2) and the fault-free functional state-observer in (11.4) with penalizing term $(\widetilde{H}y - \tilde{C}\hat{\bar{x}}_2)$, in the case of polluted output measurement, only the noise-free counterpart $(\bar{y}_2)$ is included in UIO dynamics. Henceforth, following a similar approach as in **Theorem 11.1**, first, the state equation is decoupled using the SVD on the disturbance matrix $(H)$. Then, the noise-free output equation is formulated in terms of disturbance-free decoupled states only. Now using $*$ superscript as the associated variable of **Theorem 11.1** in the case that the output is polluted by noise $(n)$, the arbitrary matrix $M$ and matrix $\mathcal{N}$ in previous case change to $M^* \in \Re^{(p-q_w-q_n)\times(p-q_n)}$ and $\mathcal{N}^{*T} = [(\bar{C}_1^*)^T \quad M^{*T}]$, respectively. Then, $\bar{x}_1 = \bar{C}_1^{*+}(\bar{y}_2 - \bar{C}_2^*\bar{x}_2)$ and $\dot{\bar{x}}_2 = \tilde{A}_2\bar{x}_2 + \bar{B}_2 u + \tilde{G}_2^* y$, where $\tilde{G}_2^* = \bar{A}_{21}\bar{C}_1^{*+}U_2^T$. Likewise, $\tilde{y}^* = \widetilde{H}^* y$ with $\widetilde{H}^* = M^*\left(\mathcal{I}_{p-q_n} - \bar{C}_1^*\bar{C}_1^{*+}\right)U_2^T$. The design procedure of the observer $\dot{\hat{\bar{x}}}_2 = N^*\hat{\bar{x}}_2 + L^*\left(\tilde{y}^* - \tilde{C}\hat{\bar{x}}_2\right) + J^* u$ and the controller synthesis according to the control law $u = K_x^*\hat{x} + K_w^*\hat{w}$ are almost identical to **Theorem 11.1** and **Theorem 11.2** in which the design parameters should be resized. For the sake of brevity regenerating, **Theorem 11.1** and **Theorem 11.2** are omitted without any loss of generality.

Next, the matched disturbance case is presented in **Corollary 11.1** (without proof). One should note that in case that the system is under the effect of matched and mismatched disturbances simultaneously ($\dot{x} = Ax + B(u + w_0) + Hw_1$) combination of **Corollary 11.1** and **Theorem 11.1**/**Theorem 11.2** can be used to create a composite control law $(u_c = u^* + u)$ for separate estimation and rejection of $w_0$ and $w_1$, respectively.

**Corollary 11.1** (matched DRC/UIO) *For matched disturbance in* (11.2)*,* $\dot{x} = Ax + B(u + w)$*, a decoupled state estimator* $\dot{\hat{\bar{x}}}_2 = N_u\hat{\bar{x}}_2 + L_u\left(\tilde{y} - \tilde{C}_u\hat{\bar{x}}_2\right) + J_u u$ *is asymptotically stable in terms of* **Definition 11.1** *and rejects the unknown time-dependent input* $w$ *quadratically in state observation residual, output, and the decoupled fault-free/-dependent states with control law* $u = -\hat{w}$ *if assumptions* (*i*), (*ii*) *(without* $m > q_w$*), and* (*iii*) *are satisfied, if the pair* $(\tilde{A}_{u2}, \tilde{C}_u)$ *is observable, and if there exist* $P_{ui}, i = 1,2,3,$ *and scalar* $\gamma_{u1}$ *such that LMI/LME condition* (11.10a) *are satisfied*

$$
\begin{gathered}
\mathbf{\Pi}_{u1} = \left[\Pi_{u1ij}\right] \leq 0; i,j = 1,\dots,6, \\
N_u < 0, \\
\gamma_{u1} > 0, \\
\bar{C}_{u1}^{+}CB - \bar{C}_{u1}^{+}\bar{C}_{u2}J_u - \Sigma_{1u}R_u^T = 0, \\
CA - \bar{C}_{u2}L_u\widetilde{H}_u C = 0,
\end{gathered}
\tag{11.10a}
$$

*where* $\mathbf{\Pi}_{u1}$ *is symmetric matrix and* $\Pi_{u1ij}$ *are zero matrices with appropriate dimensions except for the elements in* (11.9b)

$$
\begin{aligned}
\Pi_{u111} &= N_u^T - \tilde{C}_u^{\,T}L_u^T + N_u - L_u\tilde{C}_u + P_{u1}, \\
\Pi_{u112} &= \tilde{C}_u^T L_u^T \bar{C}_{u2}\bar{C}_{u1}^{+\,T} - N_u^T\bar{C}_{u2}\bar{C}_{u1}^{+\,T},
\end{aligned}
\tag{11.9b}
$$

$$\Pi_{u113} = \tilde{A}_{u2} + \tilde{G}_{u2}CQ_{u2} - L_u\tilde{H}_uCQ_{u2} + L_u\tilde{C}_u - N_u,$$
$$\Pi_{u114} = \tilde{G}_{u2}CQ_{u1} - L_u\tilde{H}_uCQ_{u1},$$
$$\Pi_{u115} = -J_u,$$
$$\Pi_{u123} = \bar{A}_{u12} - \bar{C}_{u1}^{+}\bar{C}_{u2}L_u\tilde{C}_u,$$
$$\Pi_{u124} = \bar{A}_{u11},$$
$$\Pi_{u126} = \Sigma_{1u}R_u^T - \bar{C}_{u1}^{+}CB,$$
$$\Pi_{u133} = \bar{A}_{u22} + \bar{A}_{u22}^{\;T} + P_{u2}\,,$$
$$\Pi_{u134} = \bar{A}_{u21}\bar{C}_{u1}^{+}\bar{C}_{u1},$$
$$\Pi_{u166} = -\gamma_{u1}^2\mathcal{I}_{q_w}.$$

*Using SVD, we write input matrix $B = Q_u\Sigma_uR_u^T = [Q_{u1} \quad Q_{u2}][\Sigma_{1u}^T \quad 0]^TR_u^T$. The control law of the DRC can then be obtained from B together with the transformed decoupled state matrix $\bar{A}_u = Q_u^TAQ_u$, and an arbitrary matrix $M_u \in \Re^{(p-m)\times p}$ as*

$$\begin{aligned} u &= \frac{1}{2}R_u\Sigma_{1u}^{-1}\left(\dot{\hat{\bar{x}}}_1 - \bar{A}_{u11}\hat{\bar{x}}_1 - \bar{A}_{u12}\hat{\bar{x}}_2\right),\\ \tilde{C}_u &= M_u\big(\mathcal{I}_p - \bar{C}_{u1}\bar{C}_{u1}^{+}\big)\bar{C}_{u2},\\ \tilde{H}_u &= M_u\big(\mathcal{I}_p - \bar{C}_{u1}\bar{C}_{u1}^{+}\big)\bar{C}_{u2},\\ \tilde{G}_{u2} &= \bar{A}_{u21}\bar{C}_{u1}^{+},\\ \tilde{A}_{u2} &= \bar{A}_{u22} - \bar{A}_{u21}\bar{C}_{u1}^{+}\bar{C}_{u2}. \end{aligned} \tag{11.9c}$$

It is obvious that as long as the quadratic stability based on (11.10a) is fulfilled, the two-step DRC (**Theorem 11.1**/**Theorem 11.2**) reduces to a single-step procedure as stated in **Corollary 11.1**. However, asymptotic stability cannot be fulfilled because $\Pi_{u122} = \Pi_{u144} = \Pi_{u155} = 0$. In the matched case, and since $u = -\hat{w}$, one expects to solve this issue for $\Pi_{u155}$. But, because $J_u$ in $\Pi_{u115}$ is a design variable, the obtained matrix inequality obtained after substituting (11.9c) and using Schur complement Lemma is nonlinear. As an alternative approach to full information UIO design based on disturbance/noise decoupling technique for DRC, DOBC schemes for perturbed systems are investigated in **Theorem 11.3**/**Theorem 11.5**. This is presented in the next section.

### 11.3.2 DOBC based on full information state reconstruction

Instead of UIO in **Theorem 11.1**, and assuming that the transfer matrix between the unknown input and system output has a relative degree of one, i.e. $rank(C) = rank(CH) = p$, let us propose the dynamics of the states observer to be $\dot{s} = \Gamma s + \Phi(y - \hat{y}) + \Delta u$, $\hat{x} = s - \theta y$ in which $s \in \Re_{\geq 0}^r$ is the vector of the observer states. In order to simplify the design process, it is assumed that $r = n$. Additionally, $\Gamma \in \Re^{n\times n}$, $\Phi \in \Re^{n\times p}$, $\Delta \in \Re^{n\times m}$, and $\theta \in \Re^{n\times p}$ are unknown matrices to be determined such that the conditions in **Definition 11.1** are satisfied for the related full-order system matrices (instead of a decoupled system in the previous section). Defining the observation residual as $\varepsilon = x - \hat{x}$, the dynamics of the state estimation error can be determined as $\dot{\varepsilon} = (\Gamma - \Phi C)\varepsilon$. Additionally the following conditions should be satisfied $a_1)\ \eta H = 0$, $a_2)\ \eta B - \Delta = 0$, and $a_3)\ \eta A - \Gamma\eta = 0$, in which $\eta = \mathcal{I}_n + \theta C$. Then, based on $(a_1)$, $\theta = \theta_1 + Y\theta_2$ with $\theta_1 = -H(CH)^+$, $\theta_2 = \mathcal{I}_p - (CH)(CH)^+$, and $Y \in \Re^{n\times p}$ is an unknown matrix to guarantee the satisfaction of $(a_1)$ (assumption (*ii*)). Note that this assumption is only used here and not in the results of the previous chapter. Then, using the calculated $Y$ and following $(a_{2,3})$, it is easy to obtain: $\eta = \eta_1 + Y\eta_2$ where $\eta_1 = \mathcal{I}_n - H(CH)^+C$, $\eta_2 = \big(\mathcal{I}_p - (CH)(CH)^+\big)C$ and $\Delta = \big\{\mathcal{I}_n + \big[-H(CH)^+ + Y\big(\mathcal{I}_p - (CH)(CH)^+\big)\big]C\big\}B$. Assuming $\Gamma = \eta A - XC$, one can obtain $X = \Gamma\theta \in \Re^{n\times p}$ which changes the state estimation error dynamics to $\dot{\varepsilon} = (\eta A - (X + \Phi)C)\varepsilon$. Next, using **Theorem 11.3**, the stability of the state estimation error dynamics will be guaranteed by calculating $\eta$ and $X$, properly. The UIO technique in the generalized framework of (strong) $x$-detectable observer is addressed to avoid the appearance of $\dot{y}$ in disturbance estimation ($\hat{w}$). In view of observability/detectability, an observer is called $x$-detectable if for an arbitrary bounded initial condition and $u(t) = 0 \Rightarrow x(t) \longrightarrow 0$ and it is strong $x$-detectable if $\forall(x_0, u(t)) \Rightarrow x(t) \longrightarrow 0$. It is proven that there exist an UIO for system (11.2) *iff* it is strongly $x$-detectable ([290]: Theorem 3.2; [291]).

**Theorem 11.3** (Strong functional observer (SFO)) *The UIO of the second scheme with dynamics of* $\dot{s} = \Gamma s + \Phi(y - \hat{y}) + \Delta u$, $\hat{x} = s - \theta y$ *asymptotically observes the system states of* (11.2) *while keeping the transfer function from disturbance and control input in state estimation error dynamics equal to null if* $C$ *is full rank and if for a symmetric positive definite matrix* $\delta_1 \in \Re^{n\times n}$ *and* $\delta_1' \in \Re^{n\times n}$, *the constraint of* (11.11a) *has a solution for* $\hat{X} \in \Re^{n\times p}$ *and* $\hat{\Phi} \in \Re^{n\times p}$

$$A^T\eta^T\delta_1 - C^T\left(\hat{X} + \hat{\Phi}\right)^T + \delta_1\eta A - \left(\hat{X} + \hat{\Phi}\right)C + \delta_1' = 0. \tag{11.11a}$$

*Then,* $X = \delta_1^{-1}\hat{X}$ *and* $\Phi = \delta_1^{-1}\hat{\Phi}$ *can be calculated.*

**Proof.** The proof is trivial and is suppressed. The necessary condition can be proven by following a similar approach to [100].

■

The term strong functional observer is introduced by [292] for state observation of any system with unknown disturbances without any knowledge of the input. Consequently, the observer does not necessarily have to mimic the plant's state-space model, e.g. the form of the observer in **Theorem 11.3**.

**Remark 11.2.** The process of determining $Y$ should be carried out prior to **Theorem 11.3**. In other words, $Y$ is the variable that guarantees the satisfaction of some additional LMEs which are omitted in this Theorem and otherwise leads to an infeasible convex problem. If $p = q_w$, there exists a unique solution for the intermediate matrix $Y \in \Re^{n\times p}$ that guarantees the satisfaction of $(a_1)$. If $q_w > p$, then a solution for $Y$ may be calculated by means of Least Square (LS) algorithm. Not surprisingly, in contrast to the results in the previous section, condition (*ii*), $p > q_w$, can be relieved. This is because of the fact that in DOBC based on SFO (and revised SFO), no disturbance decoupling is included. In other words, $M \in \Re^{(p-q_w)\times p}$ requires $p > q_w$. It is obvious that for $p > q_w$ in **Theorem 11.3**, there exist more than one feasible solution for matrix $Y$ that satisfies condition $a_1$ and subsequently $a_{2,3}$. If calculating $Y$ is not possible, then, even for a bounded $w$, the state estimation error dynamics change to $\dot{\varepsilon} = (\Gamma - \Phi C)\varepsilon + (\eta A - \Gamma\eta)x + (\eta B - \Delta)u + \eta H w$ and due to $\Gamma\eta$ the optimization problem becomes non-convex. In order to address this issue for systems with $p \neq q_w$, the structure of SFO is revised to $\dot{s} = As + \Phi(y - \hat{y}) + \Delta u$, $\hat{x} = s - \theta y$. Assuming that $\Gamma = A$, in order to make the SFO synthesis algorithm less conservative, LMEs $a_i, i = 1,2,3$ are released and instead **Theorem 11.4** (without proof) returns $\Phi$, $\Delta$, and $\theta$. In this case, the transfer function from control input and disturbance to state estimation error dynamics is minimized in the sense of induced $H_2$ norm, in contrast to **Theorem 11.3** in which, the transfer functions are set to null.

**Theorem 11.4** (Revised SFO) *SFO with revised dynamics as* $\dot{s} = As + \Phi(y - \hat{y}) + \Delta u$, $\hat{x} = s - \theta y$ *observes the system states of* (11.2) *in the sense of* **Definition 11.1** *while minimizing the induced* $H_2$*-norm of the transfer functions from control input and unknown* $L_2$ *bounded disturbance signal (*$\|w, u\|_2 \neq 0$*) to state estimation error dynamics if* $C$ *is full rank, if the plant is asymptotically stable, and if for symmetric positive definite matrix* $\delta_1^* \in \Re^{n\times n}$, *matrices* $\Phi$, $\Delta$, and $\theta$ *with appropriate dimensions, and positive scalars* $\gamma_{5u}$ *and* $\gamma_{5w}$, *the constraint of* (11.10b)*has a feasible solution (for user defined positive weighting scalars* $\alpha_w$ *and* $\alpha_u$*)*

$$\inf_{\boldsymbol{\Pi}_{\mathbf{3a}}} \alpha_w\gamma_{5w} + \alpha_u\gamma_{5u},$$

$$\boldsymbol{\Pi}_{\mathbf{3a}} = \left[\Pi_{3aij}\right] < 0; i,j = 1,\ldots,4, \tag{11.10b}$$

$$\Pi_{3a11} = A^T + A - C^T\Phi^{\mathrm{T}} - \Phi C + \mathcal{I}_n,$$

$$\Pi_{3a12} = \eta A - A\eta,$$
$$\Pi_{3a13} = \eta A - \Gamma\eta,$$
$$\Pi_{3a14} = \eta H,$$
$$\Pi_{3a22} = A\delta_1^* + \delta_1^* A,$$
$$\Pi_{3a23} = \delta_1^* B,$$
$$\Pi_{3a24} = \delta_1^* H,$$
$$\Pi_{3a33} = -\gamma_{5u}^2 \mathfrak{I}_m,$$
$$\Pi_{3a44} = -\gamma_{5w}^2 \mathfrak{I}_{q_w},$$

The result in **Theorem 11.3** in contrast to full-order Luenberger's observer is a stronger form of the functional state reconstruction mechanism and in the case of unknown input excitation is referred to as SFO. Unlike the reduced-order observer that simultaneously calculates the gain matrices for unknown input estimation and states reconstruction, in this method, an individual mechanism is assigned for disturbance estimation. Accordingly, in view of the proposed method in [268], [274], [293], the auxiliary state $z \in \Re_{\geq 0}^{q_w}$ is defined to recast the UIO problem in the framework of a second observation mechanism in terms of the disturbance and states. This mechanism is defined as $\dot{z} = -\beta H(z + \beta\hat{x}) - \beta(A\hat{x} + Bu)$ while $\widehat{w} = z + \beta\hat{x}$ where $\beta \in \Re^{q_w \times n}$. Contrary to the result of **Theorem 11.1**, the estimation order is equal to $q_w$. By defining the dynamics of auxiliary disturbance estimation residuals as $\varepsilon_w = w - \widehat{w}$ and derivation with respect to time and defining $\dot{w} = \nu$, (11) can be obtained

$$\varepsilon_w = (\beta\Gamma - \beta A - \beta\Phi C)\varepsilon_w - 2\beta\theta CAx + \nu. \tag{11.12}$$

Next, the disturbance observer-based compensation gain is redesigned by including an estimated measure of the bounded unknown input in the feedback channel to create a composite controller that counteracts the mismatched lumped disturbances in the input channel. Accordingly, persistence disturbance rejection is carried out by means of static observer-based state/disturbance feedback.

**Theorem 11.5.** (Static observer-based control) *For the state-space representation in* (11.2)*, the control law defined as feedback of the reconstructed states and estimated disturbance (*$u = T_x\hat{x} + T_w\widehat{w}$ *with* $T_x \in \Re^{m\times n}$ *and* $T_w \in \Re^{m\times q_w}$ *being the unknown controller gains to be calculated) can reject the disturbance in output quadratically and simultaneously design the estimation gain* $\beta$ *if both the disturbance (*$w$*) and its derivative (*$\dot{w}$*) are* $L_2$ *bounded, if plant is controllable, and also if there exist symmetric positive definite matrices* $\delta_2 \in \Re^{q_w\times q_w}$ *and* $\delta_3 \in \Re^{n\times n}$ *, matrices* $\hat{T}_x \in \Re^{m\times n}$*,* $\hat{T}_w \in \Re^{m\times q_w}$*,* $\hat{\beta} \in \Re^{q_w\times n}$*, and* $\hat{\delta}_3 \in \Re^{m\times m}$ *and scalars* $\gamma_6$ *and* $\gamma_7$ *such that the LMI/LME conditions* (11.13a) *are satisfied.*

$$\begin{gathered}\mathbf{\Pi_{3b}} = [\Pi_{3bij}] \leq 0; i,j = 1,\dots,5,\\ \gamma_{6,7} > 0,\\ \delta_3 B - B\hat{\delta}_3 = 0,\end{gathered} \tag{11.13a}$$

*where* $\mathbf{\Pi_{3b}}$ *is a symmetric matrix and* $\Pi_{3bij}$ *are zero matrices with appropriate dimensions except for the elements in* (11.12b)

$$\begin{gathered}\Pi_{3b11} = -H^T\hat{\beta}^T - \hat{\beta}H,\\ \Pi_{3b12} = \hat{\beta}(\Gamma - A - \Phi C),\\ \Pi_{3b13} = -2\hat{\beta}\theta CA - \hat{T}_w^T B^T,\\ \Pi_{3b15} = \delta_2,\\ \Pi_{3b23} = -\hat{T}_x^T B^T\\ \Pi_{3b33} = C^T C + \delta_3 A + A^T\delta_3 + B\hat{T}_x + \hat{T}_x^T B^T,\\ \Pi_{3b34} = \delta_3 H + B\hat{T}_w,\\ \Pi_{3b44} = -\gamma_6^2\mathfrak{I}_{q_w}, \Pi_{3b55} = -\gamma_7^2\mathfrak{I}_{q_w}.\end{gathered} \tag{11.12b}$$

*Then,* $\beta = \delta_2^{-1}\hat{\beta}$, $T_x = \hat{\delta}_3^{-1}\hat{T}_x$, *and,* $T_w = \hat{\delta}_3^{-1}\hat{T}_w$.

**Proof.** Define the quadratic Lyapunov equation in terms of disturbance estimation residuals and real system states as $W_1(\varepsilon_w, x, w, \dot{w}) = \varepsilon_w^T\,\delta_2\varepsilon_w + x^T\delta_3 x + \int_0^\infty (y^T y - \gamma_6^2 w^T w - \gamma_7^2 v^T v)\mathrm{d}t$ and then transforming the scalar $\dot{W}_1$ in the form of $\dot{W}_1 = \mathbb{L}_3^T \boldsymbol{\Pi}_{3b}\mathbb{L}_3 \leq 0$ with $\mathbb{L}_3^T = [\varepsilon_w^T \quad \varepsilon^T \quad x^T \quad w^T \quad v^T]$ will lead to LMI 1 in (11.12a). The existence of a solution for LME constraint is guaranteed by satisfying condition (1) of **Theorem 11.2**. The bounded-input bounded-output (BIBO) stability with effective design is achieved for the system of (11.2) under the assumptions of (*i*), (*ii*), and (*iii*). Therefore, the proof is complete.

■

**Remark 11.3.** The LMI constraints defined in all of the proposed Theorems and the **Corollary 11.1** include BRL as the basis of robust process control in an LMI framework. BRL as the worst-case performance measure is utilized to calculate the $H_\infty$ norm of the transfer function $G(s) = C(s\mathcal{I}_n - A)H$, $(\sup_{\mathrm{Re}(s)} \bar{\sigma}(G(s)) = \sup_{\omega\in\Re} \bar{\sigma}(G(j\omega)))$ in which $\bar{\sigma}(.)$ is the maximum singular value of $G$, under the assumption that $\|w\|_2 \neq 0$. Therefore, based on the concept of SDP and EVP, a minimizing problem on the maximum singular value of $G$ can be defined as $\inf_{\mathrm{LMI\,(i)}} \sup_{\omega\in\Re} \bar{\sigma}(G(j\omega))$ (Similar to results in (11.11a)) [pages 1-25 of [294], 48]. LMI (i) represent Eqs. (11.6a), (11.9a), (11.10a), and (11.13a) .

It should be mentioned that in [271] compared to [270] a more general DOBC framework is proposed that includes both matched and mismatched disturbance in the nominal state-space model (in analogy to (11.2): $\dot{x} = Ax + B(u + w_0) + Hw_1$). In such a representation, the matched disturbance is assumed to be generated by an external system of known dimension and dynamics. Then, the composite DOBC and $H_\infty$-controller are used to form a control law as $u(t) = -\hat{w}_0 + Kx(t)$. Accordingly, the idea of internal model control (IMC) is employed in creating a counteracting control effort with the matched disturbance. However, the mismatch case still remains to be handled by $H_\infty$-controller. In contrast to this idea, in this chapter both of the proposed methods are based on including the mismatch disturbance estimation in feedback channel ($u = K_x\hat{x}(t) + K_w\hat{w}(t)$) which is more general. Additionally, two important contributions here are: 1) [271] assumes that the states of the system are available in constructing the control law ($u(t) = -\hat{w}_0 + Kx(t)$). However, here, careful attention is paid to reconstruct and observe the states based on the concept of UIO and disturbance decoupling problem. Accordingly, it is assumed that only measured output signal is available and more generally, the effect of noise is incorporated in derivation of generalized solution in previous section) In the state reconstruction based on UIO and disturbance/noise decoupling, the matched disturbance case is investigated as **Corollary 11.1** without additional assumptions on dynamics of the exogenous system generating $w_0$. Such a view covers the modeling parametric uncertainties as well as the common harmonic disturbances. Moreover, this chapter similar to [266], is proposed in a two-stage design procedure. However, in [266], in contrast to the results presented here, first the nonlinear controller is designed based on measurable disturbance assumption and satisfying performance requirements and then, based on DOBC concept the disturbance is estimated and replaced in control law. A limiting assumption, that is relieved in this chapter, is then the nature of unknown signal which is assumed to be harmonic in [266]: ($\dot{\zeta} = A_\zeta\zeta, w = C_\zeta\zeta$).

**Remark 11.4.** It should be noted that the asymptotic stability of the closed-loop system based on **Theorem 11.5** is only guaranteed if the steady-state value of $v(t)$ satisfies: $\lim_{t\to\infty} v(t) = 0$. However, as it is reported in [293], the proposed approach has a robust performance in tracking disturbances with fast dynamics. Moreover, even though the high-frequency inputs may be estimated in this method, in most the applications due to bandwidth limitation of the actuators, it is challenging to reject the unknown input using the estimation of disturbance in the feedback channel.

**Remark 11.5.** It should be pointed out that, since the input matrix ($B$) and disturbance matrix ($H$) are distinct in mismatch case, $m \neq q_w$ in general, the control law in (11.13a) or (11.9a) cannot guarantee the complete elimination of the unknown input in system output. In other words, the null-ness of the transfer function from disturbance to output is reduced to a quadratic rejection problem. Additional constraints based on subspace theory will lead to conservative results.

**Remark 11.6.** Both of the methods preserve the nominal performance of the controller ($u = K_x\hat{x}$ in (11.9a) and $u = T_x\hat{x}$ in (11.13a)) if the unknown input stops to drive the system since the inner loop for estimation of disturbance ($w$) will automatically become zero.

## 11.4 Numerical Example

The system we consider here as the numerical example is an abstract sixth-order plant with three control inputs, one disturbance channel, and three output measurement signals. The state-space matrices of the plant in the form of (11.2) are presented in (11.14)

$$A = \begin{bmatrix} -586.8 & -25.9 & -70.8 & -99.6 & 59.2 & -94.8 \\ -25.9 & -847.7 & 43.3 & -118.5 & -82.1 & -269.2 \\ -70.8 & 43.3 & -669.9 & -12.1 & 45 & 45.3 \\ -99.6 & -118.5 & -12.1 & -781 & 22.7 & -146.1 \\ 59.2 & -82.1 & 45 & 22.7 & -766.5 & -63.9 \\ -94.8 & -269.2 & 45.3 & -146.1 & -63.9 & -933.4 \end{bmatrix} \times 10^{-3},$$

$$[B \quad H] = \begin{bmatrix} -1255.7 & 591.8 & 133.4 & 949 & 717.8 & 1652.9 \\ 489 & -366.7 & -676.1 & -1943.3 & 0 & 4 \\ 767.7 & 273.3 & 0 & 0 & -476.7 & 0 \\ 1039.1 & 166.5 & 0 & 382.9 & -1701.5 & 0 \end{bmatrix}^T \times 10^{-3}, \tag{11.14}$$

$$C = \begin{bmatrix} -336.6 & -234.3 & -890.5 & 0 & 0 & -295.1 \\ -136.6 & 0 & 244.3 & -505 & 946.8 & 556.2 \\ 0 & 36.2 & 0 & 0 & 0 & 0 \end{bmatrix} \times 10^{-3}.$$

Then, using **Theorem 11.1**, the unknown input observer matrices are obtained as (11.15). It should be pointed out that the auxiliary intermediate variable $M$ is calculated based on metaheuristic method such that conditions in LME constraints (11.5) of **Theorem 11.1** are satisfied.

$$N = \begin{bmatrix} -9055.5 & -76.1 & -82.5 & 135.3 & -64.2 \\ -76.1 & -9644.1 & 8.8 & 828 & 695.9 \\ -82.5 & 8.8 & -12592.3 & -241.7 & -429.2 \\ 135.3 & 828 & -241.7 & -12548.9 & 487.5 \\ -64.2 & 695.9 & -429.2 & 487.5 & -12226.9 \end{bmatrix} \times 10^{-3},$$

$$L = \begin{bmatrix} 732.77 & -145.68 & -21.32 & -7.21 & -36.41 \\ -277.12 & 40.23 & 5.29 & 1.11 & 14.54 \end{bmatrix}^T,$$

$$J = \begin{bmatrix} 714.6 & 133.4 & 1231.4 & -537.3 & 1652.9 \\ -385.3 & -676.1 & -1986 & 189.9 & 4 \\ 187.8 & 0 & -196.6 & 397.2 & 0 \end{bmatrix}^T \times 10^{-3}, \tag{11.15}$$

$$M = \begin{bmatrix} 299.9 & 498.6 & 386.9 \\ 655.7 & 638 & 113.1 \end{bmatrix} \times 10^{-3}.$$

After observing the decoupled disturbance-free states ($\hat{\bar{x}}_2$) and using $\hat{\bar{x}}_1 = \bar{C}_1^{+}(y - \bar{C}_2\hat{\bar{x}}_2)$ to observe the faulted (disturbance-dependent) states, unknown input disturbance is estimated and the system states are reconstructed based on the SVD of disturbance matrix ($H = Q\Sigma_w R^T$) as: $\hat{x} = Q\hat{\bar{x}}$. Then, using **Theorem 11.2** appropriate controller matrices are calculated ($K_x$ and $K_w$) and the results are shown in (11.16)

$$K_x = \begin{bmatrix} 978.8 & 236.3 & 555.6 & 954.1 & 457.8 & 878.2 \\ 115.3 & 2685.2 & 104.5 & 434.4 & 983.4 & 493 \\ -37.2 & 686.1 & 744.6 & 2920.9 & -16.5 & -1091.1 \end{bmatrix} \times 10^{-3},$$
$$K_w = [158.2 \quad 638.5 \quad -507.4]^T \times 10^{-3}, \tag{11.16}$$

Now based on the concept of SFO, to obtain a full-order unknown input state observer, LME condition (11.11a) in **Theorem 11.3** is solved using MATLAB LMI toolbox and the resulted matrices are shown in (11.17)

$$\Gamma = \begin{bmatrix} -0.665 & 1.713 & 10.934 & -9.4 & 17.254 & 12.017 \\ -1.713 & -1.023 & 2.889 & -6.248 & 11.604 & 6.679 \\ -11.076 & -2.889 & -1.028 & -24.109 & 45.883 & 23.404 \\ 9.216 & 6.248 & 24.121 & -0.439 & -0.144 & 8.112 \\ -16.984 & -11.604 & -46.008 & 0.634 & -0.877 & -14.651 \\ -12.404 & -6.679 & -23.212 & -7.937 & 14.721 & -1.092 \end{bmatrix},$$
$$\Phi = \begin{bmatrix} 2.038 & -0.145 & -3.648 & 7.54 & -14.137 & -8.304 \\ -5.026 & -3.568 & -13.296 & 0 & 0 & -4.406 \\ -1.611 & -0.938 & -3.093 & -0.975 & 1.828 & -0.108 \end{bmatrix}^T, \tag{11.17}$$
$$\Delta = \begin{bmatrix} -67.4 & 158.2 & -1631.1 & -121.1 & -2929.4 & 1182.3 \\ 581.7 & 44.7 & 441.4 & -958.8 & 914.2 & 296.9 \\ 644.8 & 28.8 & -635.1 & -587.2 & -885.6 & -168.4 \end{bmatrix}^T \times 10^{-3},$$
$$\theta = \begin{bmatrix} -848.7 & 936.2 & 2998.3 & 2211.5 & 4184.7 & 757.2 \\ 701.4 & -101.1 & -599.5 & -249 & -1714.2 & -154.4 \\ -654 & 107.8 & -178 & -1269 & -1313.8 & -1021.4 \end{bmatrix}^T \times 10^{-3},$$

Then, using the obtained UIO of **Theorem 11.3** with matrices in and employing **Theorem 11.5**, the disturbance estimation mechanism can be calculated simultaneously with the static feedback gain control law ($u = T_x\hat{x} + T_w\hat{w}$). The associated matrices $\beta$, $T_x$, and $T_w$ are calculated and presented in (11.18)

$$\beta = [5657.8 \quad -1913.1 \quad -932.3 \quad 384.2 \quad 2507.1 \quad 2141],$$
$$T_x = \begin{bmatrix} 778.5 & 660.3 & 70.3 & 572.7 & -444 & 417.9 \\ -51 & 317.9 & -29.4 & 144 & 52.4 & 549 \\ 1134.6 & -195.8 & 5.7 & -291.8 & -681.2 & -1083.1 \end{bmatrix} \times 10^{-3}, \tag{11.18}$$
$$T_w = [110.6 \quad 237.4 \quad -1787.4]^T \times 10^{-3},$$

In order to evaluate the performance of UIO of a decoupled system (**Theorem 11.1**) in disturbance estimation and compare the estimation quality with the two-step feasibility problem of **Theorem 11.5**, the plant is excited through the disturbance channel with a non-stationary signal. Figure 11.1 represents the disturbance estimation comparison of the two methods. As it can be seen, both of the methods are estimating the unknown input signal which consists of four parts: 1) Chirp signal with damped amplitude (swept-frequency $[0.1 \quad 10]$ Hz in 6 sec). 2) Sudden jump disturbance between 5 and 6 sec. 3) saw-tooth harmonic signal with frequency 0.5 Hz. 4) Realization of white Gaussian noise signal between 12 to 15 sec. It is observed that disturbance estimation based on independent dynamics in (11.12) and **Theorem 11.5** has a slightly better performance than simultaneous disturbance estimation mechanism based on disturbance/noise decoupling problem. This performance difference can be quantitatively measured by integrating over the absolute value of estimation error. As can be seen in Figure 11.2, DOBC stays in lower values for non-stationary unknown signal compared to UIO/DRC of **Theorem 11.1**/**Theorem 11.2**. Moreover, in the range of $[12 \quad 15]$ sec, in which the realization of white Gaussian signal with a sampling time of 1 msec is active, the estimation error based on UIO/DRC increases significantly. However, it should be noted that for large systems with the various source of erroneous signals, the disturbance estimation based on the decoupling method is more computationally efficient.

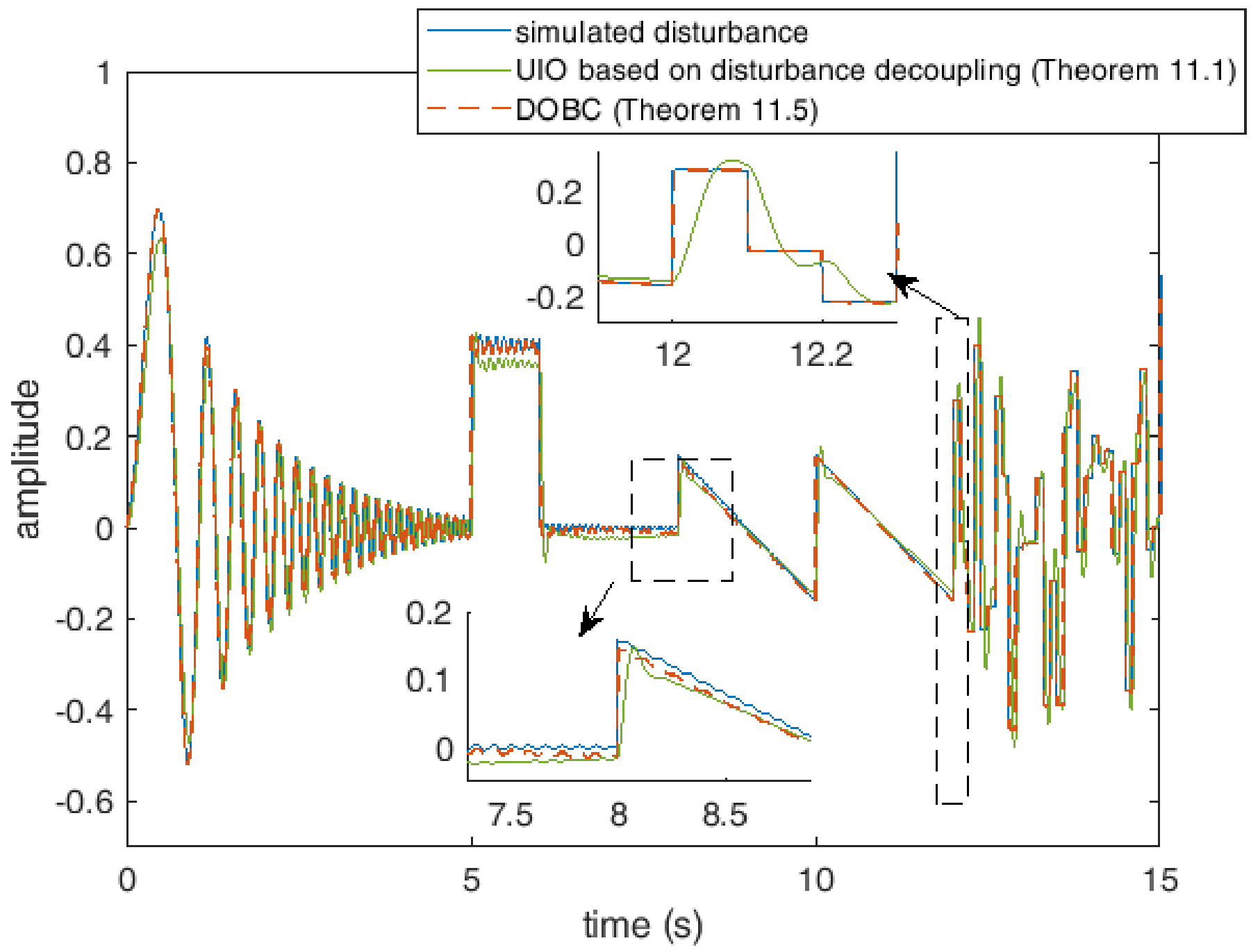

Figure 11.1 Comparison of the performance of disturbance estimation mechanism [295]

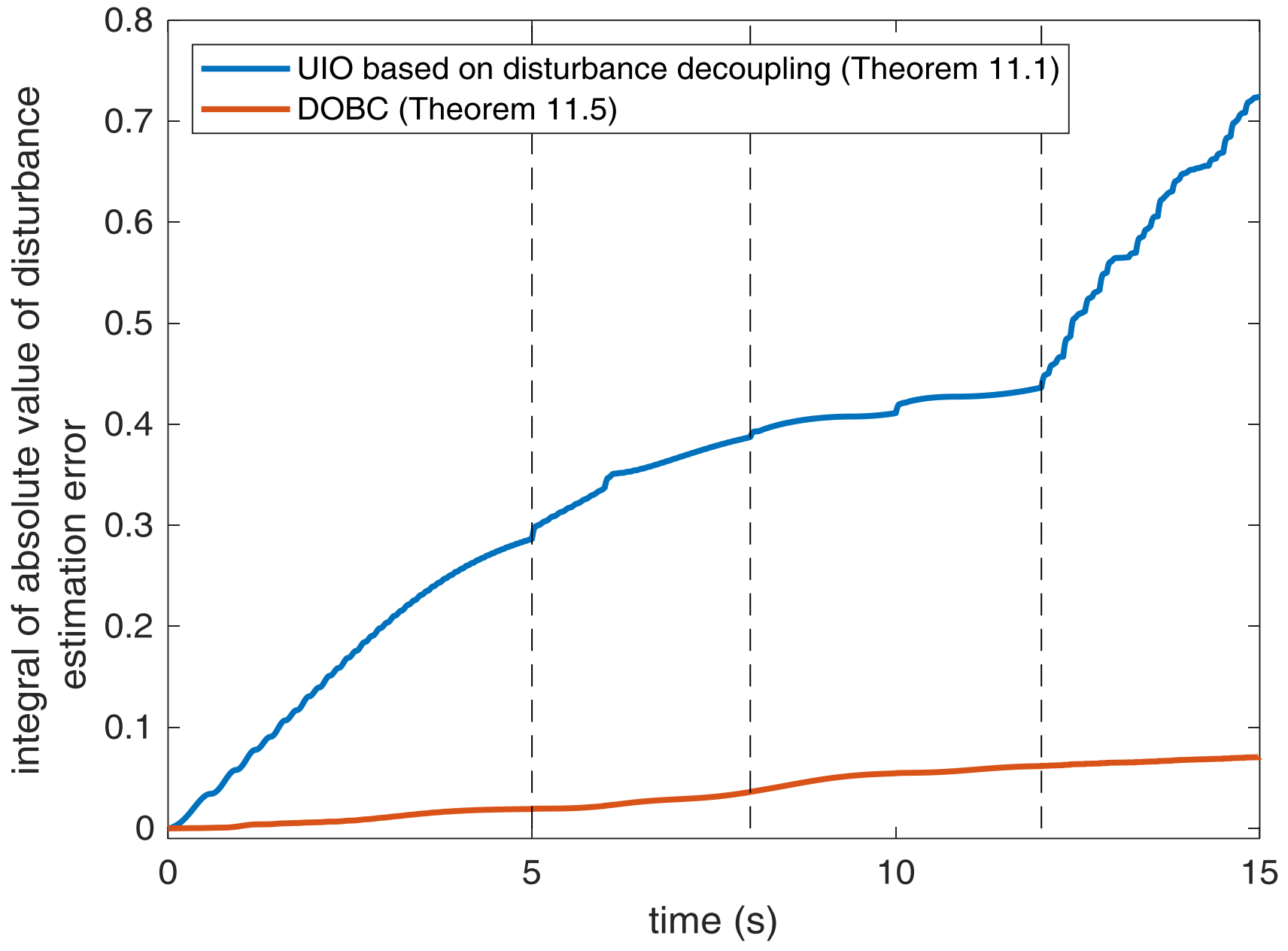

Figure 11.2 Comparison of disturbance estimation error by means of integral of absolute value [295]

Next, in order to assess the performance of disturbance rejection in both methods, the closed-loop systems after the implementation of the DRC (**Theorem 11.2**) and DOBC (**Theorem 11.5**), with given matrices in (11.16) and (11.18), are excited with a sinusoidal signal of frequency 1.59 Hz in a time window of $[0 \quad 5]$ sec. The three output signals of the three systems are compared in Figure 11.3: 1) open-loop system; 2) closed-loop system based on UIO/DRC of decoupling problem (**Theorem 11.1**/**Theorem 11.2**) 3) SFO/DOBC based on **Theorem 11.3**/**Theorem 11.5**.

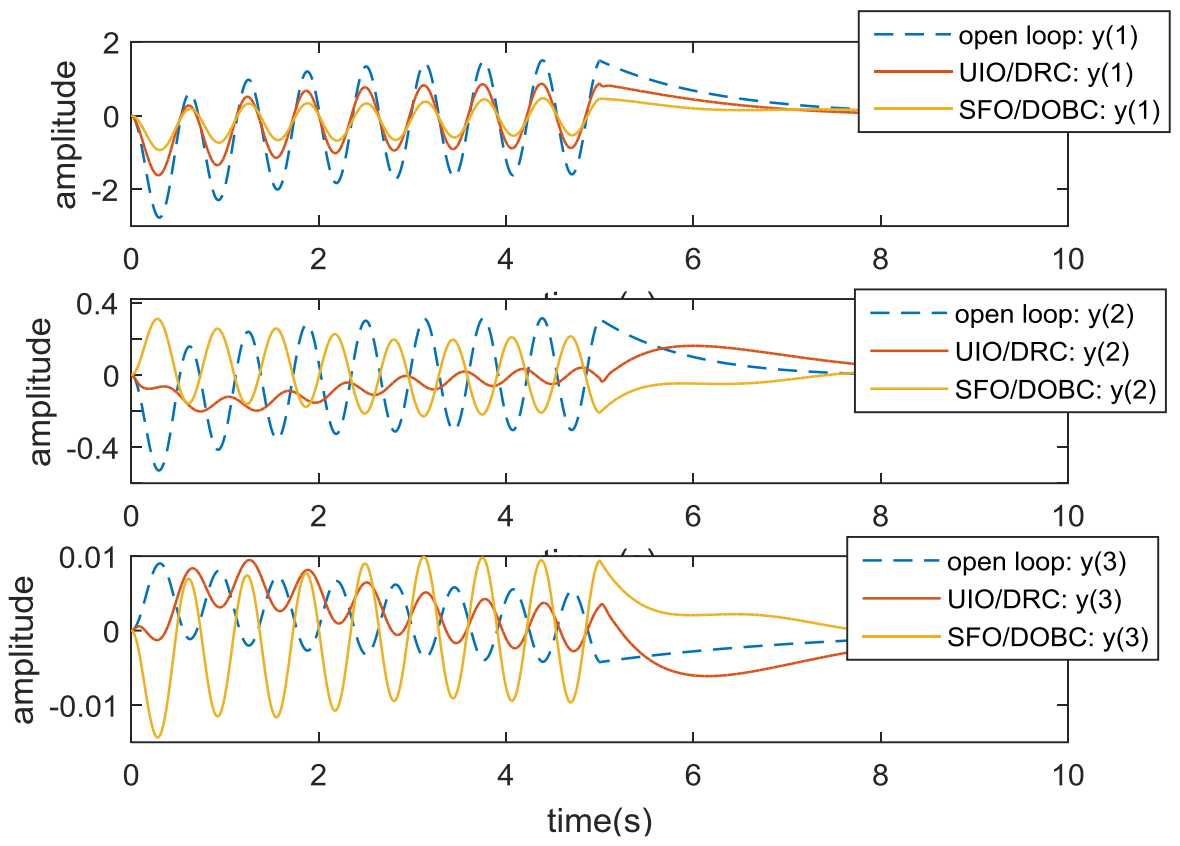

Figure 11.3 Comparison of the performance of the disturbance rejection

# 12 Finite Element-based SiL closed-loop testing

In the previous chapters, the performances of all novel control systems are investigated in the real-time setup. The main advantage of such tests among others is that the controller performance is tested in real-time where modeling uncertainties, feedback delays, inherent system couplings, nonlinearities of various sort persist. However, for more complex systems that a control engineer may have limited access to the actual system due to cost issues, the benchmark problem is not an option. An example of such applications could be aerospace and large structures. Consequently, a platform where a closed-loop system with the listed benefits above is available which however is easily accessible and is of low cost is missing. The final chapter of this dissertation aims at providing a solution for control engineers to explore the extent of control theory in semi-real applications, namely using SiL.

A new framework for running the FE packages inside an online loop together with MATLAB is introduced to provide a platform where closed-loop systems such as control algorithms can be simulated with high fidelity numerical models. Contrary to the HiL such as the ones shown previously e.g., Figure 6.1, in the proposed SiL, the FE package represents a simulation platform replicating the real system which can be out of access due to several strategic reasons, e.g., costs and accessibility. Practically, SiL for sophisticated structural design and multi-physical simulations provides a platform for preliminary tests before prototyping and mass production. This feature may also reduce the new product's costs significantly and may add several flexibilities in implementing different instruments with the goal of shortlisting the most cost-effective ones before moving to real-time experiments for the civil and mechanical systems. The proposed SiL interconnection is proposed but not limited to ABAQUS as long as the host FE package is capable of executing user-defined commands in the FORTRAN language. The focal point of this chapter is on using the compiled FORTRAN subroutine as a messenger between ABAQUS/CAE kernel and MATLAB Engine. In order to show the generality of the proposed scheme, the limitations of the available SiL schemes in the literature are addressed here. Additionally, all technical details for establishing the connection between FEM and MATLAB are provided.

## 12.1 State-of-the-art in SiL for controller implementation purposes

Leaving out the system identification techniques as it was analyzed in Chapter 4, the superiority of FEM as a modeling approach is due to the fact that the analytical/semi-analytical solutions for coupled systems are mostly limited to simple geometries as discussed in both of Chapters 2/3. In order to put the FE model in a computationally affordable form with a limited number of dynamical states, a post-processing step is defined including a model reduction in terms of modal coordinates. This in turn makes the design/analyses tractable especially if the optimization or control of the structure is desirable.

In this chapter, a new Software-in-the-Loop scheme (SiL) is constructed to address the effects of the feedback loop in the time-domain analysis with application in control theory and online parameter optimization. The SiL presented in this chapter is **not** limited to any particular physical domain and can be used for dynamics and vibration, structural health monitoring, fluid mechanics, thermo-elastic analysis, and electro-/magneto-domains. As long as the problem under study can be defined in a step module (see [164], [296], [297]) of the commercial FE package of ABAQUS, it can be categorized as one of the applications of this chapter. The italic terms here and after referring to the standard commands, e.g. *step*, in ABAQUS GUI. As another advantage of this scheme in comparison to those available in the literature ([84], [298]–[303]), MATLAB toolboxes for robust control, global optimizations, neural networks, and fuzzy systems are accessible in the SiL which significantly increases the applicability of the method to general engineering problems. In this regards, the proposed technique is a candidate for non-parametric modeling of continuous nonlinear multi-domain structures with complex geometries where the real system is not accessible for measurements [122], [234]. Moreover, the proposed SiL can be used for the modeling of common benchmark problems in structural dynamics as a test platform for new control and optimization methods before moving to real-time measurements [304]. Concerning the computation time, time-variability of the analysis, and the validation and verification of the proposed approach, detailed investigations are carried out.

Because the SiL approach is computationally demanding, it is not recommended for problems with simple geometries. Accordingly, analytical solutions (see Chapters 2 and 3) or the methods that include system/parameter identification, model reduction in combination with offline design are recommended to be employed instead (as shown in Chapter 4). However, for multi-physics problems without analytical models, systems with complicated geometries, benchmark problems where the real-time setup is out of access, and industrial problems where FE solutions are trusted, the presented method is an alternative to the field tests which are costly. ABAQUS is nominated over other packages as the dynamic simulator of the SiL configuration due to its capabilities in compiling FORTRAN subroutines. However, the proposed scheme can be applied to other commercial packages such as NASTRAN as long as the software provides a pool for including external commands.

The smart structural design based on the general FE approach is previously studied in the literature. For instance, Lim et al. [305] used 3D finite elements for modeling a multi-input-multi-output (MIMO) smart plate with discrete piezo-patches and designed an optimal controller on the reduced-order model by solving the algebraic Riccati equation (ARE). The performance of the designed controller for the vibration suppression of the clamped plate is presented for both the steady-state and the transient cases. Ray et al. in a similar problem to this chapter, used a FORTRAN subroutine to implement sensitivity enhancing control (SEC) for damaged smart beam [298]. Karagülle et al. used ANSYS to integrate a PID control action into the solution of FE [299]. Similar results are reported by [84]. Following the same trend, Rahman and Alam compared their experimental active vibration control (AVC) of a cantilever beam based on a PID controller with ABAQUS using 1D Finite element formulation [300]. Recently, Gao et al. used the ABAQUS UAMP subroutine to include an active controller in the model of an aircraft's vertical fin under dramatic buffet loads. They included a finite element model of macro fiber composite (MFC) actuators in ABAQUS implicit. The results are matched with those obtained from the experimental implementation of the controller on the prototype of the fin [301]. The main limitation of implementing the control algorithm in SiL using standard ABAQUS scripting is the challenges that are introduced for matrix computations such as solving ARE for the optimal controller synthesis. The platform for ABAQUS coding is an application programming interface (API) that is realized per Python object-oriented language, and Python is not a well-established language for control algorithms [302]. This problem is therefore tackled in the course of this chapter. More recently, Orszulik and Gabbert presented an attractive interface for establishing a connection between the Simulink and ABAQUS for active vibration control of a cantilever beam [303], [306].

It should be indicated that the proposed method in this chapter is by no means limited to structural optimization and structural vibration control, however, the applications of different SiL schemes in the literature are mostly concentrated on AVC [5], [103], [307]. Henceforth, the author is mostly attentive in developing a new mechanism for uncertainty quantification in mechanical structures with complex geometries that can be used as an alternative to conventional methods [68]. Uncertainty quantification regarding the unmodeled dynamics of high order nature is classically dealt with as a lumped stable bounded time-varying function. In terms of controller synthesis (classical robust methods), such a view leads to conservative results. However, analytical modeling of simpler geometries under large vibration amplitudes hands the structure of uncertainty e.g. quadratic or cubic terms [21], [111], [308]. Next, by use of the parametric identification methods on the data obtained from the approaches that provide time-dependent responses of the nonlinear system (such as the SiL proposed here) makes an efficient tool for nonlinear system identification. This grey-box parameter identification approach as the center of intensive research in the identification community (see for instance [232], [309]) can be viewed as an alternative to some of the nonlinear system identification techniques based on discrete modeling of continuous mechanical structures reported in [122]. Such a nonlinear nominal model can be later used for model-based controller synthesis in contrast to robust control methods which operate based on a worst-case analysis. One should note that at the current stage in the literature, the lumped structural uncertainty quantification is obtained based on statistical analyses of an enormous number of experimental setups which is mostly limited because of the costs and difficulties regarding providing multiple identical systems [69]. Before moving to the methodology section in order to make the chapter more readable, a brief description of the numerical problem in association with the coupling scheme is presented as follows:

The ABAQUS/MATLAB is coupled to solve a time-dependent loop, namely closed-loop simulation in control theory. To this end, a regulation problem is investigated for AVC. In order to realize the feedback control as shown in Figure 12.1, methodology in the next section is proposed. To keep the numerical example tractable an output feedback linear quadratic Gaussian (LQG) controller is synthesized on the nominal model of the system. The details of obtaining the reduced-order model are additionally explained

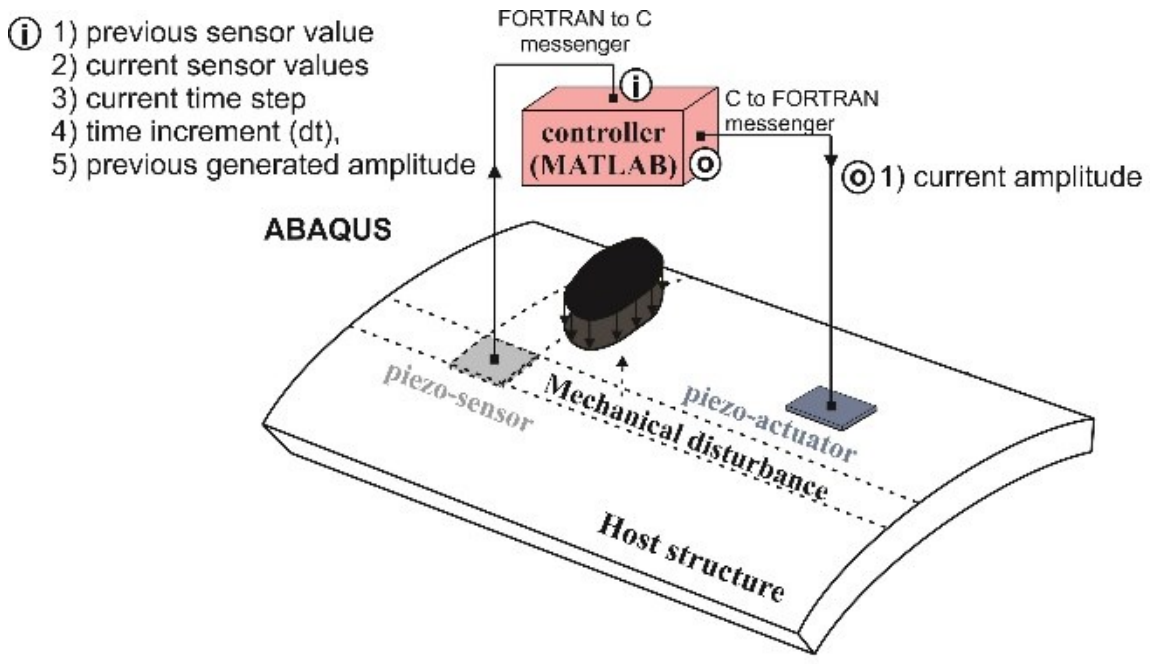

Figure 12.1 The schematic definition of the SiL benchmark [164]

## 12.2 Methodology of SiL based on the Fortran messengers

In the proposed coupling scheme, the SiL is outlined that uses Python and FORTRAN as two messengers between the ABAQUS Kernel and MATLAB engine. For this purpose, the Python script is used to define the model geometry/material and model interactions e.g. boundary conditions, while the time-dependent external loads are realized by the FORTRAN UAMP subroutine. Sensor elements (later referred to as `SENUi`) are defined on measurable physical variables e.g. electric potential in piezo-sensor of Figure 12.1 with the predefined frequency of data extraction from a particular nodal set in *history* output (`seti`) by using the command: `*Output, history, sensor, name=SENUi, frequency *Node Output, nset=seti`. All the terminologies here and after in `Courier New` font represent the commands used in Python and their syntax. Moreover, the time-varying amplitude of the user-defined load is generated by the command: `*Amplitude, name, definition=USER, variables`. At this set of monitored points (`seti`) the solution-dependent data are extracted from the active elements. For instance, in sub-problem (b), `seti` represents the nodal group on the top surface of the piezo-actuator having the same applied electric potential.

UAMP subroutine determines the value of amplitude in a time-dependent manner. This dependence is defined in terms of a function that takes the sensor values and current time increment for controlling the amplitude of a stimulus in *load* module of ABAQUS/CAE (see Figure 12.1). Since all information passed to the subroutine is updated at each time increment, ABAQUS is set to stall and waits for an update on the previous value. In this stage, user-defined solution-dependent state variables (sensor values) are needed to be sent to the MATLAB engine. This action requires a C compiler that takes the value, calls the MATLAB engine to open, executes a function (in an m-file), brings back the updated amplitude, closes the MATLAB engine, and finally waits for a recall. "Engine applications" are standalone C programs that permit the user to call MATLAB from another environment and then use it as the computation machine. These standalone programs in the proposed SiL framework are called within UAMP subroutine. The C compiler is therefore responsible for the Engine applications while the FORTRAN compiler handles UAMP.

Moreover, the input array in the MATLAB block of Figure 12.1 is passed from engine application to UAMP subroutine and vice versa. For this purpose, `ENGOPEN` routine calls MATLAB computational engine, `MXCREATEDOUBLEMATRIX` creates an array, and `MXCREATEDOUBLESCALAR` creates a scalar double. All the routines in `Consolas` font here and after representing the commands in the FORTRAN subroutine. `MXDESTROYARRAY` is used to deallocate the memory occupied by the specified array, `MXGETPR` sets the pointer to the first component of the real data, `MXCOPYREAL8TOPTR` copies real values from the FORTRAN array into the MATLAB matrix. Furthermore, `ENGPUTVARIABLE` is employed to write an array to the MATLAB engine and assign a variable name for that array, while `ENGEVALSTRING` evaluates the expression contained in a string for the engine session started by `ENGOPEN`, `ENGGETVARIABLE` reads the array from the engine session. Finally, `MXCOPYPTRTOREAL8` is used to copy real values from the MATLAB array into the FORTRAN array, and `ENGCLOSE` routine terminates the current MATLAB session. This interconnection is presented in the flowchart of Figure 12.2. In order to provide the technical details for the interested reader while maintaining the readability of the chapter, the coupling script is separated from the problem definition/simulations and moved to the Appendix F.

## 12.3 Problem definition

### 12.3.1 Preliminary action items for FE modeling

Before proceeding to the main problem formulation, the cylindrical panel shown in Figure 12.1 is used as the geometry of the structure. The main reason for selecting such geometry in the scope of this chapter is that: 1) Defining an additional circular cylindrical coordinate system for the *material orientation* in orthotropic or anisotropic cases is required which makes the problem more general. 2) Non-conventional *assembly* and *interaction* constraints are required in sequential structure optimization.

#### 12.3.1.1 Mesh convergence and state-space modeling

It is assumed that the host structure has a mid-plain radius of 0.5 m, a thickness of 0.03 m, a length of 1 m, and an arc angle of 60°. Additionally, the piezo patches are assumed to have a thickness of 5 mm. The host structure is made of single-layer orthotropic steel with a density of $\rho = 7800\ \mathrm{kg/m^3}$, while the actuator and sensor are fabricated from PZT4 and $Ba_2NaNb_5O_{15}$, respectively. The elasticity matrix of the host, actuator, and sensor as well as the piezoelectric and dielectric matrices of the patches are presented in Appendix 2 of [164].

To minimize the computational burden without losing accuracy mesh convergence analyses (MCAs) are carried out and the results are suppressed for the sake of brevity. A detailed analysis maybe found in [164].

Next, following [166] and by appropriate selection of the modal coordinates [129], [310], the reduced-order dynamical equation of motion for each actuator/sensor pair can be written as (12.1a).

$$\dot{x} = Ax + Bu, \qquad y = Cx + Du, \tag{12.1a}$$

where,

$$A = \begin{bmatrix} 0 & \Omega \\ -\Omega & -2Z\Omega \end{bmatrix}, B = [B_{m_1} \quad B_{m_2}]^T, C = [C_{m_1} \quad C_{m_2}], D = 0. \tag{12.1b}$$

with $\Omega = (M_m^{-1} K_m)^{1/2}$, $Z = \text{diag}(\xi_i), i = 1, \ldots, n$ in which $M_m$, $K_m$, $\xi_i$, and $n$ are the modal mass, stiffness matrices, damping ratio associated with mode number $i$, and the total number of mode-shapes considered in the modeling process, respectively. Additionally, $x \in \mathfrak{R}^{2n}$, $u \in \mathfrak{R}^{n_u}$, and $y \in \mathfrak{R}^{n_y}$ are the state, input, and output vectors, respectively while $A \in \mathfrak{R}^{2n \times 2n}$, $B \in \mathfrak{R}^{2n \times n_u}$, and $C \in \mathfrak{R}^{n_y \times 2n}$ are the state, control input, and output matrices, correspondingly. $B_{m_{1,2}}$ and $C_{m_{1,2}}$ are the modal coupling matrices of the piezo-actuator/-sensor, respectively. Finally, $\mathfrak{R}$ symbolizes the set of all real numbers.

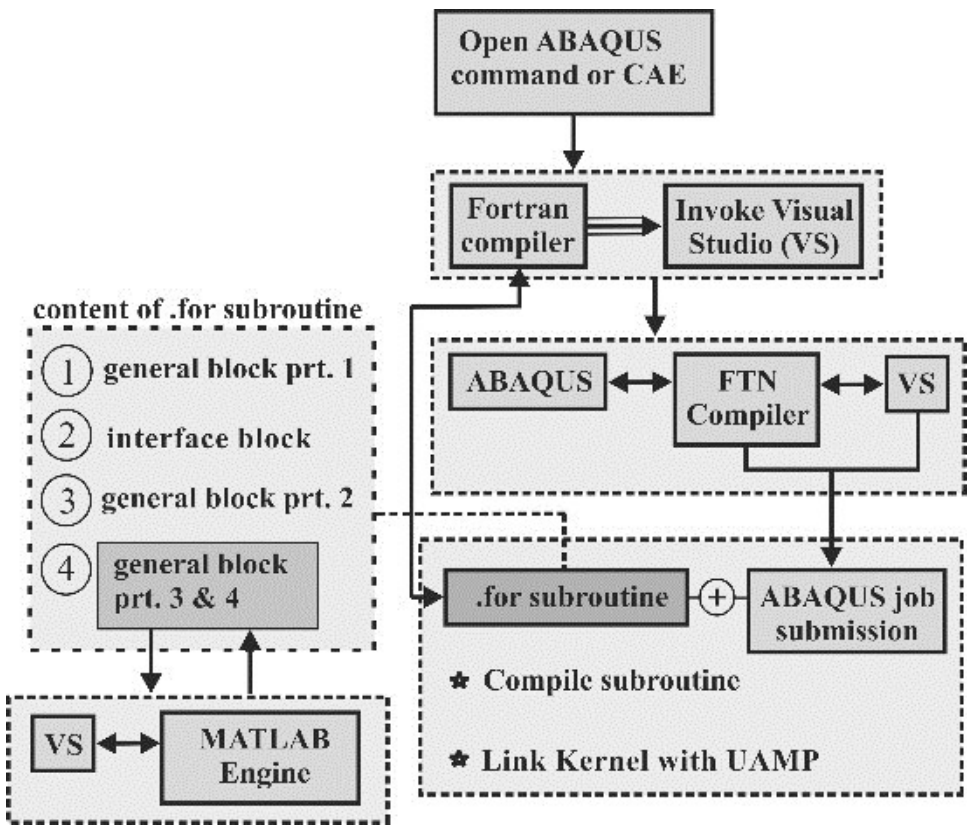

Figure 12.2 ABAQUS kernel and MATLAB engine simulation workflow [164]

Technically, the state space model in (12.1a) is created by using the modal electrical potential coupling matrices which are calculated in ABAQUS/CAE and extracted by Python script using `session.writeFieldReport` command [297].

An optimization problem can be defined as briefly explained in the next chapter to find the optimal sensor actuator position.

#### 12.3.1.2 Actuator/senor placement objective

Optimal placement for plate and panel structures was investigated previously in [20], [310]–[312] based on a genetic algorithm (GA). Kumar and Narayanan [313] have applied their LQR controller-based criteria to find the optimal location of piezoelectric actuators/sensors in vibration control of plates using GA. In the paper by [314], the maximization of the controllability Gramian in combination with GA was used as the criterion for optimal placement of a clamped plate. A similar approach with modal controllability and observability Gramians and GA was also employed by Sadri et al. [315], [316]. In this chapter, the objective function is defined in terms of $H_2/H_\infty$ norm of input to output gains [163], [164]. Without loss of generality the optimal sensor actuator placement for the collocated symmetric case study is given as Figure 12.3. The two configurations are selected for 1) The non-optimal configuration with collocated actuators/sensors. 2) The optimal location is based on the multi-modal sensor/actuator criteria.

#### 12.3.1.3 Problem formulation in closed-loop systems

In this section, the application of the real-time SiL is shown for AVC of two piezo-laminated panels in Figure 12.3. Sensor values as predefined time-varying physical measures in the *output history* of the system response are collected in nodal level are sent to be processed inside "Engine application" (Refer to Appendix F.). The control law (feedback signal) is generated by using an arbitrary linear/nonlinear controller synthesized in MATLAB. As long as the structure of the controller can be formulated in an m-file in the form of a system of ordinary differential equations (ODE), the proposed coupling scheme can handle the implementation of the controller inside a loop with ABAQUS/CAE. For the simulation example, the designed controller is parameterized in a function that takes a vector of inputs (see Figure 12.1) including the previous output of the plant, sensor values, current time step, time increment, and amplitude generated from MATLAB in the previous step. This vector is imported as the input data for the fixed-step ODE solver. As a result, it is essential to create a set that records the amplitude of the plant inside FORTRAN and provides the data at each increment for MATLAB. To keep the problem tractable, an optimal controller is designed based on the reduced-order nominal model. The model is obtained using system identification which is detailed in [164].

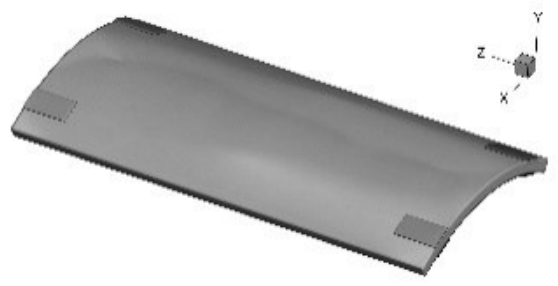

Configuration 1

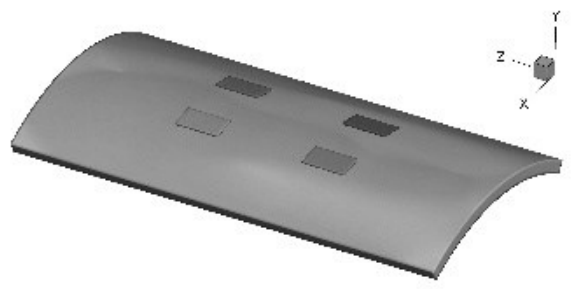

Configuration 2

Figure 12.3 Symmetrical configurations for optimal (2) and non-optimal (1) actuator/sensor placement [164]

Before moving to the closed-loop performance evaluation, it should be pointed out that since the transient simulations are performed in *dynamic implicit* scheme, the time integration damping generated by ABAQUS is automatically introduced due to Hilber-Hughes-Taylor as an extension of Newmark's $\beta$-method time integration. The parameters corresponding to the transient fidelity are selected to be $\alpha = -0.05$, $\beta = 0.275625$, and $\gamma = 0.55$ such that the numerical energy dissipation is kept minimal. This operator has the advantage that it is unconditionally stable for a linear system [297]. Next, the vibration suppression performance is investigated in the frequency range of the reduced-order system. Accordingly, the panel is excited by a uniformly distributed time-varying pressure that acts over all partitions simultaneously with a chirp profile: magnitude of $10^6\ N$ and frequency swept between [1350 2500] Hz within 1 sec. The open-loop and closed-loop system responses are compared at an observation nodal point in the center of the host layer's top surface in Figure 12.4 for configurations 1 and 2, respectively. The numerical values of the system matrix for the reduced-order model as well as the output feedback controller are given in Appendix G.

**(a) configuration 1** **(b) configuration 2**

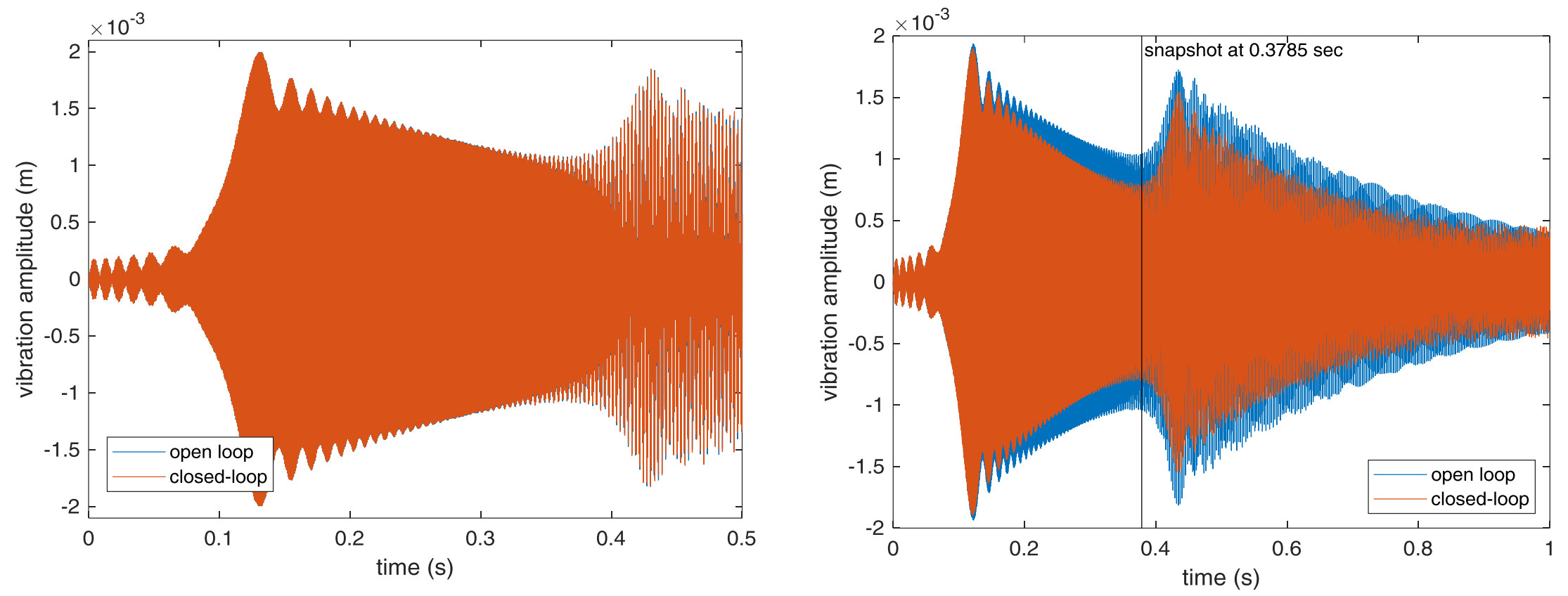

Figure 12.4 Comparison of the open-loop and closed-loop systems in time-domain: (a) non-optimal configuration. (b) optimal configuration [164]

It is obvious that due to the placement of active elements, applying the same weighting for two configurations leads to widely different results; one of which is unstable. The optimal configuration 2 suppresses the vibration in the nominal frequency range while the non-optimal configuration has no significant effect in suppressing the vibrations. Additionally, the spatial vibration suppression of the system can be observed for any specific time-step e.g. the snapshot of the open-loop and closed-loop systems in Figure 12.5 for $t = 0.3785$ sec which is an advantage of the SiL method.

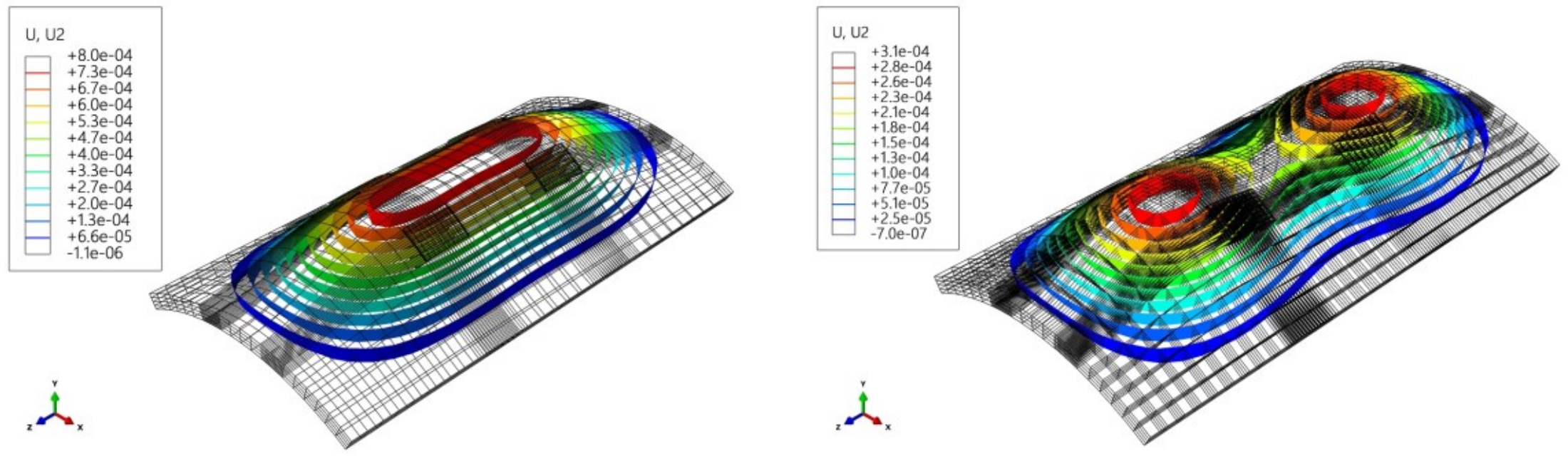

Figure 12.5 Comparison of spatial displacement of open-loop and closed-loop systems in contour plot of isosurfaces in the deformed panel at 0.3785 sec snapshot [164]

# 13 Conclusions and outlook

## 13.1 Conclusions

In order to perfect the disturbance rejection performance in applied problems, the controller development is probably the least time-consuming phase. This is a counter-intuitive statement but is justified due to the nature of the problem. Accordingly, the scope of the external stimuli makes a great difference in formulating mathematics. While in Chapter 6, a repetitive scheme is appropriate due to the nature of disturbance being formulated by the internal model principle, in Chapters 5, 8, and 9 only a robust scheme is followed based on the worst-case analysis since no assumption other that BIBO is considered. Consequently, if neither the nature nor the bandwidth of disturbance is known *a priori*, the problem formulation can be different for the control development than when the linear response of the system can be deduced from data. This is not limited to bandwidth since the amplitude of the disturbance also plays a crucial role in the plant model. Accordingly, all systems are nonlinear unless the external excitations are below a certain level, otherwise, we may not realize a linearization. This was shown in Chapters 2, 4 correspondingly, and the formulated problem evolves accordingly as shown in chapters 8-10.

Although model-based techniques may have the potential to replace the standard techniques of the industry e.g., PID, they need to be perfected for the case study. Namely, the literature is saturated with white-box-based models which outperform the model-free approaches but are not applicable because the white-box models do not readily represent real-time data. The author believes only data-driven models which do not postpone the validation phase to the late phases of controller development/implementation can make a positive impact. Chapter 4, beautifully proposes a walkthrough of data-driven nonlinear modeling which could be more time-consuming and delicate even from the controller development. It is important for the author to mention that a model is not superior to the other but depending on what bandwidth and amplitude range the control inputs as well as noise and disturbances are active, one model may outperform the other. Therefore the control engineer should be attentive regarding what is the problem being really solved.

The author believes that standardization of data-driven approaches towards being an industry-trusted solution is on its way. Accordingly, AI-based models, controllers, and estimators such as the one proposed in **10** are going to replace the classical solution for high-end technological systems such as those used in aerospace applications or in autonomous driving. Here, the extent of techniques such as polynomial nonlinear state-space models, reverse-path methods in modeling, and deep-learning networks in feedback system design is undeniably the future of the system theory.

## 13.2 Outlook

Recently, methods such as the Reverse Path method [227], [228], and Nonlinear Normal Model method [229], [230] gained much attention for analyzing and quantifying these geometrical nonlinearities. The interested reader is highly recommended to refer to [122] for in-depth technical details. An alternative to such modeling approaches is the combination of semi-analytical techniques, e.g. [5], in structural modeling of nonlinear systems and parameter optimization techniques in the Grey-box System Identification framework [231]. Accordingly, the adaption of the black-box polynomial nonlinear state-space (PNLSS) identification approach i.e., [232] to the geometrically nonlinear system is ongoing research.

The nonlinear model-free robust control schemes based on the upper-bounds of the norm of the disturbance signals and matched-/mismatched-uncertainties such as high-gain, variable-structure, and fuzzy methods can be used in combination with NN to address nonlinearities in system dynamics [6]. For instance, He et al. employed a neural network for modeling the dynamics of a flexible robot manipulator subjected to input deadzone. They have used radial basis function NN to capture the deadzone behavior and designed a high-gain observer-based NN controller [96]. The non-minimum phase systems with centralized control configuration may have right-half plane (RHP) zeroes, which can significantly restrict the closed-loop performance as reported in [235]. It is proposed as an outlook to employ/adapt the echo-state feature of adaptive RWNN in this dissertation to contribute to improving the performance of nonminimum phase systems.

The interpretation of dist(.,.) operator for a successful anti-windup scheme (local and global asymptotic stability) can be exquisitely explained in terms of the basin of attraction of the augmented system. Qualitative illustrations in this regard such as extracting the feasible solution space in the presented algorithms for different anti-windup schemes in terms of trajectories of augmented systems may provide useful tools in ranking these methods. This may be especially important for systems with only local exponential stability which is out of the scope of the AVC of lightly damped structures. For the exponentially stable plants, the basin of attraction encapsulates the whole state space.

Recently, new anti-windup schemes such as directionality compensation [204], nonlinear scheduling anti-windup scheme [317], the anticipation of actuator saturation [318], and deferred action approach [319] are proposed that require to be investigated separately on the problem and tuned carefully based on the real-time response of the system.

Also, evaluation of the methods for other finite increment gain nonlinearities such as hysteresis could be useful for other applications. Some preliminary research has been reported here [172], [320].

More sophisticated post-processing analyses can be performed by using the phase portrait of the trajectory based on the system states in modal coordinates e.g., the contribution of the fundamental natural mode of the vibrating system to the windup event. This requires excitation of a certain frequency by a harmonic sinusoidal function close to a natural frequency and assessing the contribution of each mode-shape to a windup event and then observing the trajectory of the system returning to the rest point.

Two concrete proposals for future works regarding Chapter 10 are: 1) to perform a sensitivity analysis in terms of the network parameters. Such a study requires investigation of RWNN performance with respect to network dimension, learning rates, bound compensator parameter, and the choice of the correction gains in learning laws in a Hardware-in-Loop configuration. 2) In contrast to the nonlinear model-based control synthesis based on the polynomial nonlinear state-space system identification technique, the present controller method mostly relies on the underlying linear model which is faster in terms of nonlinearity/modeling-uncertainty characterization and, quantification. As a result, an interesting comparison study is proposed to compare the adaptive RWNN of this chapter with the aforementioned nonlinear model-based technique.

Two open problems regarding Chapter 11 are: (1) the design of the intermediate variable which guarantees the satisfaction of the additional linear matrix equality constraints defined to reject the effect of disturbance and suppress the direct effect of control input in state estimation error dynamics (instead of bilinear matrix inequality constraint), and (2) The appearance of the first derivative of the output measurement in disturbance estimation. For the class of systems with an equal number of outputs and unknown input signals (1) can be addressed (as of this chapter). To address (2), a possible solution, as followed in the second part of this chapter, is to separate the disturbance estimation dynamics from UIO. This approach separates the observation problem into two sub-problems and an obvious disadvantage is that after designing the strong functional state observer (SFO), the feasibility of the second sub-problem (disturbance estimation) is not guaranteed. Then again, if the plant is under the effect of multiple erroneous signals, the order of the observer/estimator will increase. Moreover, if the general governing dynamics of the system have multiple equilibrium points in the framework of Rajamani satisfaction of linear matrix inequality conditions of all linearized systems is a challenging task. It should be noted that since the general view of DRC in this chapter is for mismatch cases, feedforward methods are neglected as an alternative to the inclusion of estimated disturbance in the feedback channel which is most undesirable in practice. The implementation of the main results in the methodology section on two vibrating mechanical smart structures in real-time is ongoing research [13]. Moreover, systematic sensitivity analyses on the closed-loop performance in terms of the tuning parameters based on the one-at-a-time (OAT) technique are proposed as a potential future research topic. Additionally, considering the current investigations in nonlinear system identification [122], the semi-analytical modeling of geometrically nonlinear systems [111], and the PNLSS [309], the adaptation/generalization of the main results of this chapter is an interesting topic. In contrast to the robust techniques based on worst-case scenarios, such a view is less conservative because the nonlinearity of the plant is only affecting the control design when they appear in system output.

# Appendix A.

The constants in (2.9) are calculated as

$$
\begin{aligned}
I_0 &= h_a\rho_a + h_c\rho_c + h_s\rho_s, \\
I_1 &= h_a z_a\rho_a + h_s z_s\rho_s, \\
I_2 &= \frac{h_a^3\rho_a}{12} + \frac{h_c^3\rho_c}{12} + \frac{h_s^3\rho_s}{12} + \rho_a h_a z_a^2 + \rho_s h_s z_s^2.
\end{aligned} \tag{A.1}
$$

By employing the definition of $N_{xx}$ for the coupled three-layered media, we can write

$$
N_{xx} = \int_{-\frac{h}{2}}^{\frac{h}{2}} \sigma_{xx} dz = \int_c \sigma_{xx} dz + \int_s \sigma_{xx} dz + \int_a \sigma_{xx} dz, \tag{A.2}
$$

then, using Eqs. (2.2) and (2.4) and carrying out the integration in $z$-direction the following equation can be obtained

$$
\begin{aligned}
N_{xx} = {} & J_{xx1}\left(\frac{\partial w_0}{\partial x}\right)^2 + J_{xx2}\frac{\partial^2 w_0}{\partial x^2} + J_{xx3}\left(\frac{\partial w_0}{\partial y}\right)^2 + J_{xx4}\frac{\partial^2 w_0}{\partial y^2} + J_{xx5}\left(\frac{\partial w_0}{\partial x}\right)\left(\frac{\partial \dot{w}_0}{\partial x}\right) + J_{xx6}\frac{\partial^2 \dot{w}_0}{\partial x^2} \\
& + J_{xx7}\left(\frac{\partial w_0}{\partial y}\right)\left(\frac{\partial \dot{w}_0}{\partial y}\right) + J_{xx8}\frac{\partial^2 \dot{w}_0}{\partial y^2},
\end{aligned} \tag{A.4}
$$

and similarly

$$
\begin{aligned}
N_{yy} = {} & J_{yy1}\left(\frac{\partial w_0}{\partial x}\right)^2 + J_{yy2}\frac{\partial^2 w_0}{\partial x^2} + J_{yy3}\left(\frac{\partial w_0}{\partial y}\right)^2 + J_{yy4}\frac{\partial^2 w_0}{\partial y^2} + J_{yy5}\left(\frac{\partial w_0}{\partial x}\right)\left(\frac{\partial \dot{w}_0}{\partial x}\right) \\
& + J_{yy6}\frac{\partial^2 \dot{w}_0}{\partial x^2} + J_{yy7}\left(\frac{\partial w_0}{\partial y}\right)\left(\frac{\partial \dot{w}_0}{\partial y}\right) + J_{yy8}\frac{\partial^2 \dot{w}_0}{\partial y^2},
\end{aligned} \tag{A.5}
$$

$$
N_{xy} = J_{xy1}\left(\frac{\partial w_0}{\partial x}\right)\left(\frac{\partial w_0}{\partial y}\right) + J_{xy2}\frac{\partial^2 w_0}{\partial x \partial y}, \tag{A.6}
$$

$$
\begin{aligned}
M_{xx} = {} & K_{xx1}\left(\frac{\partial w_0}{\partial x}\right)^2 + K_{xx2}\frac{\partial^2 w_0}{\partial x^2} + K_{xx3}\left(\frac{\partial w_0}{\partial y}\right)^2 + K_{xx4}\frac{\partial^2 w_0}{\partial y^2} + K_{xx5}\left(\frac{\partial w_0}{\partial x}\right)\left(\frac{\partial \dot{w}_0}{\partial x}\right) \\
& + K_{xx6}\frac{\partial^2 \dot{w}_0}{\partial x^2} + K_{xx7}\left(\frac{\partial w_0}{\partial y}\right)\left(\frac{\partial \dot{w}_0}{\partial y}\right) + K_{xx8}\frac{\partial^2 \dot{w}_0}{\partial y^2},
\end{aligned} \tag{A.7}
$$

$$
\begin{aligned}
M_{yy} = {} & K_{yy1}\left(\frac{\partial w_0}{\partial x}\right)^2 + K_{yy2}\frac{\partial^2 w_0}{\partial x^2} + K_{yy3}\left(\frac{\partial w_0}{\partial y}\right)^2 + K_{yy4}\frac{\partial^2 w_0}{\partial y^2} + K_{yy5}\left(\frac{\partial w_0}{\partial x}\right)\left(\frac{\partial \dot{w}_0}{\partial x}\right) \\
& + K_{yy6}\frac{\partial^2 \dot{w}_0}{\partial x^2} + K_{yy7}\left(\frac{\partial w_0}{\partial y}\right)\left(\frac{\partial \dot{w}_0}{\partial y}\right) + K_{yy8}\frac{\partial^2 \dot{w}_0}{\partial y^2},
\end{aligned} \tag{A.8}
$$

$$
M_{xy} = K_{xy1}\left(\frac{\partial w_0}{\partial x}\right)\left(\frac{\partial w_0}{\partial y}\right) + K_{xy2}\frac{\partial^2 w_0}{\partial x \partial y}, \tag{A.9}
$$

where,

$$
J_{xx1} = \frac{1}{2}\left[c_{11}^c h_c + c_{11}^{*s} h_s + c_{11}^{*a} h_a + \left(e_{31}^{*s} - G_p e_{31}^{*a}\right)\frac{e_{31}^{*s}}{\epsilon_{33}^{*s}} h_s\right],
$$

$$
J_{xx2} = -\left[c_{11}^{*s} h_s z_s + c_{11}^{*a} h_a z_a + \left(e_{31}^{*s} - G_p e_{31}^{*a}\right)\frac{e_{31}^{*s}}{\epsilon_{33}^{*s}} h_s z_s\right],
$$

$$
J_{xx3} = \frac{1}{2}\left[c_{12}^c h_c + c_{12}^{*s} h_s + c_{12}^{*a} h_a + \left(e_{31}^{*s} - G_p e_{31}^{*a}\right)\frac{e_{32}^{*s}}{\epsilon_{33}^{*s}} h_s\right],
$$

$$J_{xx4} = -\left[c_{12}^{*s}h_s z_s + c_{12}^{*a}h_a z_a + \left(e_{31}^{*s} - G_p e_{31}^{*a}\right)\frac{e_{32}^{*s}}{\epsilon_{33}^{*s}}h_s z_s\right],$$

$$J_{xx5} = -\frac{e_{31}^{*a}e_{31}^{*s}}{\epsilon_{33}^{*s}}G_d h_s,$$

$$J_{xx6} = -\frac{e_{31}^{*a}e_{31}^{*s}}{\epsilon_{33}^{*s}}G_d h_s z_s,$$

$$J_{xx7} = -\frac{e_{31}^{*a}e_{32}^{*s}}{\epsilon_{33}^{*s}}G_d h_s,$$

$$J_{xx8} = -\frac{e_{31}^{*a}e_{32}^{*s}}{\epsilon_{33}^{*s}}G_d h_s z_s,$$

$$J_{yy1} = \frac{1}{2}\left[c_{12}^{c}h_c + c_{12}^{*s}h_s + c_{12}^{*a}h_a + \left(e_{32}^{*s} - G_p e_{32}^{*a}\right)\frac{e_{31}^{*s}}{\epsilon_{33}^{*s}}h_s\right],$$

$$J_{yy2} = -\left[c_{12}^{*s}h_s z_s + c_{12}^{*a}h_a z_a + \left(e_{32}^{*s} - G_p e_{32}^{*a}\right)\frac{e_{31}^{*s}}{\epsilon_{33}^{*s}}h_s z_s\right],$$

$$J_{yy3} = \frac{1}{2}\left[c_{22}^{c}h_c + c_{22}^{*s}h_s + c_{22}^{*a}h_a + \left(e_{32}^{*s} - G_p e_{32}^{*a}\right)\frac{e_{32}^{*s}}{\epsilon_{33}^{*s}}h_s\right],$$

$$J_{yy4} = -\left[c_{22}^{*s}h_s z_s + c_{22}^{*a}h_a z_a + \left(e_{32}^{*s} - G_p e_{32}^{*a}\right)\frac{e_{32}^{*s}}{\epsilon_{33}^{*s}}h_s z_s\right],$$

$$J_{yy5} = -\frac{e_{32}^{*a}e_{31}^{*s}}{\epsilon_{33}^{*s}}G_d h_s,$$

$$J_{yy6} = -\frac{e_{32}^{*a}e_{31}^{*s}}{\epsilon_{33}^{*s}}G_d h_s z_s,$$

$$J_{yy7} = -\frac{e_{32}^{*a}e_{32}^{*s}}{\epsilon_{33}^{*s}}G_d h_s,$$

$$J_{yy8} = -\frac{e_{32}^{*a}e_{32}^{*s}}{\epsilon_{33}^{*s}}G_d h_s z_s,$$

$$J_{xy1} = \frac{1}{2}[2c_{66}^{c}h_c + 2c_{66}^{*s}h_s + 2c_{66}^{*a}h_a],$$

$$J_{xy2} = -[2c_{66}^{*s}h_s z_s + 2c_{66}^{*a}h_a z_a],$$

$$K_{xx1} = \frac{1}{2}\left[c_{11}^{*s}h_s z_s + c_{11}^{*a}h_a z_a + \left(e_{31}^{*s}z_s - G_p e_{31}^{*a}z_a\right)\frac{e_{31}^{*s}}{\epsilon_{33}^{*s}}h_s\right],$$

$$K_{xx2} = -\left[\frac{1}{12}c_{11}^{c}h_c^3 + c_{11}^{*s}\left(z_s^2 h_s + \frac{1}{12}h_s^3\right) + c_{11}^{*a}\left(z_a^2 h_a + \frac{1}{12}h_a^3\right) + \left(e_{31}^{*s}z_s - G_p e_{31}^{*a}z_a\right)\frac{e_{31}^{*s}}{\epsilon_{33}^{*s}}h_s z_s \right.$$
$$\left. + \left(\frac{h_s^3}{12}\frac{(e_{31}^{*s})^2}{\epsilon_{33}^{*s}} + \frac{h_a^3}{12}\frac{(e_{31}^{*a})^2}{\epsilon_{33}^{*a}}\right)\right],$$

$$K_{xx3} = \frac{1}{2}\left[c_{12}^{*s}h_s z_s + c_{12}^{*a}h_a z_a + \left(e_{31}^{*s}z_s - G_p e_{31}^{*a}z_a\right)\frac{e_{32}^{*s}}{\epsilon_{33}^{*s}}h_s\right],$$

$$K_{xx4} = -\left[\frac{1}{12}c_{12}^{c}h_c^3 + c_{12}^{*s}\left(z_s^2 h_s + \frac{1}{12}h_s^3\right) + c_{12}^{*a}\left(z_a^2 h_a + \frac{1}{12}h_a^3\right) + \left(e_{31}^{*s}z_s - G_p e_{31}^{*a}z_a\right)\frac{e_{32}^{*s}}{\epsilon_{33}^{*s}}h_s z_s \right.$$
$$\left. + \left(\frac{h_s^3}{12}\frac{e_{31}^{*s}e_{32}^{*s}}{\epsilon_{33}^{*s}} + \frac{h_a^3}{12}\frac{e_{31}^{*a}e_{32}^{*a}}{\epsilon_{33}^{*a}}\right)\right],$$

$$K_{xx5} = -\frac{e_{31}^{*a}e_{31}^{*s}}{\epsilon_{33}^{*s}}G_d h_s z_a,$$

$$K_{xx6} = -\frac{e_{31}^{*a}e_{31}^{*s}}{\epsilon_{33}^{*s}}G_d h_s z_s z_a,$$

$$K_{xx7} = -\frac{e_{31}^{*a}e_{32}^{*s}}{\epsilon_{33}^{*s}}G_d h_s z_a,$$

$$K_{xx8} = -\frac{e_{31}^{*a}e_{32}^{*s}}{\epsilon_{33}^{*s}}G_d h_s z_s z_a,$$

$$K_{yy1} = \frac{1}{2}\left[c_{12}^{*s}h_s z_s + c_{12}^{*a}h_a z_a + \left(e_{32}^{*s}z_s - G_p e_{32}^{*a}z_a\right)\frac{e_{31}^{*s}}{\epsilon_{33}^{*s}}h_s\right],$$

$$K_{yy2} = -\left[\frac{1}{12}c_{12}^{c}h_c^3 + c_{12}^{*s}\left(z_s^2 h_s + \frac{1}{12}h_s^3\right) + c_{12}^{*a}\left(z_a^2 h_a + \frac{1}{12}h_a^3\right) + \left(e_{32}^{*s}z_s - G_p e_{32}^{*a}z_a\right)\frac{e_{31}^{*s}}{\epsilon_{33}^{*s}}h_s z_s + \left(\frac{h_s^3}{12}\frac{e_{31}^{*s}e_{32}^{*s}}{\epsilon_{33}^{*s}} + \frac{h_a^3}{12}\frac{e_{31}^{*a}e_{32}^{*a}}{\epsilon_{33}^{*a}}\right)\right],$$

$$K_{yy3} = \frac{1}{2}\left[c_{22}^{*s}h_s z_s + c_{22}^{*a}h_a z_a + \left(e_{32}^{*s}z_s - G_p e_{32}^{*a}z_a\right)\frac{e_{32}^{*s}}{\epsilon_{33}^{*s}}h_s\right],$$

$$K_{yy4} = -\left[\frac{1}{12}c_{22}^{c}h_c^3 + c_{22}^{*s}\left(z_s^2 h_s + \frac{1}{12}h_s^3\right) + c_{22}^{*a}\left(z_a^2 h_a + \frac{1}{12}h_a^3\right) + \left(e_{32}^{*s}z_s - G_p e_{32}^{*a}z_a\right)\frac{e_{32}^{*s}}{\epsilon_{33}^{*s}}h_s z_s + \left(\frac{h_s^3}{12}\frac{(e_{32}^{*s})^2}{\epsilon_{33}^{*s}} + \frac{h_a^3}{12}\frac{(e_{32}^{*a})^2}{\epsilon_{33}^{*a}}\right)\right],$$

$$K_{yy5} = -\frac{e_{32}^{*a}e_{31}^{*s}}{\epsilon_{33}^{*s}}G_d h_s z_a,$$

$$K_{yy6} = -\frac{e_{32}^{*a}e_{31}^{*s}}{\epsilon_{33}^{*s}}G_d h_s z_s z_a,$$

$$K_{yy7} = -\frac{e_{32}^{*a}e_{32}^{*s}}{\epsilon_{33}^{*s}}G_d h_s z_a,$$

$$K_{yy8} = -\frac{e_{32}^{*a}e_{32}^{*s}}{\epsilon_{33}^{*s}}G_d h_s z_s z_a,$$

$$K_{xy1} = \frac{1}{2}\left[2c_{66}^{*s}z_s h_s + 2c_{66}^{*a}z_a h_a\right],$$

$$K_{xy2} = -\left[\frac{1}{6}c_{66}^{c}h_c^3 + 2c_{66}^{*s}\left(z_s^2 h_s + \frac{1}{12}h_s^3\right) + 2c_{66}^{*a}\left(z_a^2 h_a + \frac{1}{12}h_a^3\right)\right]$$

$$c_1 = \frac{\left(j^4 K_{yy4}L_x^4 + i^2 j^2\left(K_{xx4} + 2K_{xy2} + K_{yy2}\right)L_x^2 L_y^2 + i^4 K_{xx2}L_y^4\right)\pi^4}{4L_x^3 L_y^3},$$

$$\begin{aligned} c_2 = \frac{1}{L_x^3 L_y^3 m(-4j^2 + m^2)n(-4i^2 + n^2)}16i^2 j^2 \Big(L_x^2 m^2 \Big(i^2 L_y^2\left(J_{xx4} + J_{yy2} - K_{xx3} + K_{xy1}\right) \\ + L_x^2\left(j^2\left(2J_{yy4} + 3K_{yy3}\right) - K_{yy3}m^2\right)\Big) \\ + L_y^2 \Big(i^2\left(2J_{xx2} + 3K_{xx1}\right)L_y^2 \\ + L_x^2\left(j^2\left(J_{xx4} + J_{yy2} + K_{xy1} - K_{yy1}\right) + \left(J_{xy2} - K_{xy1}\right)m^2\right)\Big)n^2 - K_{xx1}L_y^4 n^4\Big)\pi^2, \end{aligned} \quad \text{(A.10)}$$

$$c_3 =$$

$$
\begin{aligned}
1/c_3^*\Bigg(&4\pi^2 ijklmn\big((-1)^{i+k+n}-1\big)\big((-1)^{j+l+m}\\
&-1\big)\Big(j^2L_x^2\Big(L_y^2\Big(i^2\big(J_{xy2}+K_{xx3}+K_{xy1}+K_{yy1}\big)-\big(J_{xy2}+K_{yy1}\big)(k^2+n^2)\Big)\\
&+K_{yy3}L_x^2(l^2+m^2)\Big)\\
&+L_y^2n^2\Big(L_y^2\big(i^2K_{xx1}+2k^2(J_{xx2}+K_{xx1})\big)\\
&+L_x^2\Big(l^2\big(J_{xx4}+J_{xy2}+J_{yy2}+K_{xy1}\big)+m^2\big(J_{xy2}-K_{xy1}\big)\Big)\Big)-i^2J_{xy2}l^2L_x^2L_y^2\\
&-i^2J_{xy2}L_x^2L_y^2m^2+i^2k^2K_{xx1}L_y^4-i^2K_{xx3}l^2L_x^2L_y^2-i^2K_{xx3}L_x^2L_y^2m^2\\
&+J_{xx4}k^2L_x^2L_y^2m^2+J_{xy2}k^2l^2L_x^2L_y^2+J_{xy2}k^2L_x^2L_y^2m^2+J_{yy2}k^2L_x^2L_y^2m^2\\
&+2J_{yy4}l^2L_x^4m^2-k^4K_{xx1}L_y^4-k^2K_{xy1}l^2L_x^2L_y^2+k^2K_{xy1}L_x^2L_y^2m^2-K_{xx1}L_y^4n^4\\
&-K_{yy3}L_x^4(l^2-m^2)^2\Big)\Bigg),
\end{aligned}
$$

$$
\begin{aligned}
c_3^* = \Big(&L_x^3L_y^3(i-k-n)(i+k-n)(i-k+n)(i+k+n)(j-l-m)(j+l-m)(j-l+m)(j+l\\
&+m)\Big),
\end{aligned}
$$

$$
\begin{aligned}
c_4 = \left(\frac{\pi^2}{36ijL_x^3L_y^3}\right)\Bigg(&24\big(j^4J_{yy4}L_x^4+i^2j^2\big(J_{xx4}+J_{xy2}+J_{yy2}-K_{xx3}\big)L_x^2L_y^2+i^4(J_{xx2}+K_{xx1})L_y^4\big)\\
&-8\big(j^4\big(J_{yy4}+4K_{yy3}\big)L_x^4\\
&+i^2j^2\big(J_{xx4}-J_{xy2}+J_{yy2}+K_{xx3}+2K_{xy1}-2K_{yy1}\big)L_x^2L_y^2\\
&+i^4(J_{xx2}+K_{xx1})L_y^4\big)(2\cos(j\pi)+1)\\
&+\cos(3i\pi)\Big(3\big(j^4J_{yy4}L_x^4+i^2j^2\big(J_{xx4}-2J_{xy2}+J_{yy2}-4K_{xx3}\big)L_x^2L_y^2\\
&+i^4(J_{xx2}+4K_{xx1})L_y^4\big)\\
&-\Big(j^4\big(J_{yy4}+4K_{yy3}\big)L_x^4\\
&+i^2j^2\Big(J_{xx4}+2J_{xy2}+J_{yy2}+4\big(K_{xx3}+2K_{xy1}+K_{yy1}\big)\Big)L_x^2L_y^2\\
&+i^4(J_{xx2}+4K_{xx1})L_y^4\Big)(2\cos(j\pi)+1)\Big)\\
&+3\cos(i\pi)\Big(-9j^4J_{yy4}L_x^4-3i^2j^2\big(3J_{xx4}+2J_{xy2}+3J_{yy2}-4K_{xx3}\big)L_x^2L_y^2\\
&-3i^4(3J_{xx2}+4K_{xx1})L_y^4\\
&+\big(3j^4\big(J_{yy4}+4K_{yy3}\big)L_x^4\\
&+i^2j^2\big(3J_{xx4}-2J_{xy2}+3J_{yy2}+4K_{xx3}+8K_{xy1}-4K_{yy1}\big)L_x^2L_y^2\\
&+i^4(3J_{xx2}+4K_{xx1})L_y^4\big)(2\cos(j\pi)+1)\Big)\Bigg),
\end{aligned}
$$

$$
\begin{aligned}
c_5 = \Big(1/\big(ijL_x^3L_y^3(j^2-4m^2)(i^2-4n^2)\big)\Big)\,8m^2n^2\Big(&i^2L_y^2\big(L_x^2\big(j^2\big(J_{xy2}+K_{xx3}+K_{xy1}+K_{yy1}\big)\\
&-2\big(J_{xy2}+K_{xx3}\big)m^2\big)+2K_{xx1}L_y^2n^2\big)\\
&+2\Big(L_x^4m^2\big(j^2K_{yy3}+J_{yy4}m^2\big)\\
&+L_x^2L_y^2\Big((-j^2)\big(J_{xy2}+K_{yy1}\big)+\big(J_{xx4}+2J_{xy2}+J_{yy2}\big)m^2\Big)n^2+J_{xx2}L_y^4n^4\Big)\Big)\pi^2,
\end{aligned}
$$

$$
c_6 = -\left(\frac{\big(3j^4J_{yy3}L_x^4+i^2j^2\big(3J_{xx3}-2J_{xy1}+3J_{yy1}\big)L_x^2L_y^2+3i^4J_{xx1}L_y^4\big)\pi^4}{64L_x^3L_y^3}\right),
$$

$$
c_7 = -\left(\frac{3\big(j^4J_{yy7}L_x^4+i^2j^2\big(J_{xx7}+J_{yy5}\big)L_x^2L_y^2+i^4J_{xx5}L_y^4\big)\pi^4}{64L_x^3L_y^3}\right),
$$

$$
c_8 = -\left(\frac{\Big(i^2L_y^2\big(J_{xx3}L_x^2m^2+J_{xx1}L_y^2n^2\big)+j^2\big(J_{yy3}L_x^4m^2+J_{yy1}L_x^2L_y^2n^2\big)\Big)\pi^4}{16L_x^3L_y^3}\right),
$$

$$c_9 = \frac{\left(j^4 K_{yy8} L_x^4 + i^2 j^2 \left(K_{xx8} + K_{yy6}\right) L_x^2 L_y^2 + i^4 K_{xx6} L_y^4\right)\pi^4}{4 L_x^3 L_y^3},$$

$$c_{10} = \left(\frac{\pi^2}{36 ij L_x^3 L_y^3}\right)\Bigg(24\left(j^4 J_{yy8} L_x^4 + i^2 j^2 \left(J_{xx8} + J_{yy6} - K_{xx7}\right) L_x^2 L_y^2 + i^4 \left(J_{xx6} + K_{xx5}\right) L_y^4\right)$$
$$- 8\left(j^4 \left(J_{yy8} + 4K_{yy7}\right) L_x^4 + i^2 j^2 \left(J_{xx8} + J_{yy6} + K_{xx7} - 2K_{yy5}\right) L_x^2 L_y^2\right.$$
$$\left. + i^4 \left(J_{xx6} + K_{xx5}\right) L_y^4\right)(2\cos(j\pi) + 1)$$
$$+ \cos(3i\pi)\Big(3\left(j^4 J_{yy8} L_x^4 + i^2 j^2 \left(J_{xx8} + J_{yy6} - 4K_{xx7}\right) L_x^2 L_y^2 + i^4 \left(J_{xx6} + 4K_{xx5}\right) L_y^4\right)$$
$$- \Big(j^4 \left(J_{yy8} + 4K_{yy7}\right) L_x^4 + i^2 j^2 \Big(J_{xx8} + J_{yy6} + 4\left(K_{xx7} + K_{yy5}\right)\Big) L_x^2 L_y^2$$
$$+ i^4 \left(J_{xx6} + 4K_{xx5}\right) L_y^4\Big)(2\cos(j\pi) + 1)\Big)$$
$$+ 3\cos(i\pi)\Big(-9 j^4 J_{yy8} L_x^4 - 3 i^2 j^2 \left(3\left(J_{xx8} + J_{yy6}\right) - 4K_{xx7}\right) L_x^2 L_y^2$$
$$- 3 i^4 \left(3 J_{xx6} + 4K_{xx5}\right) L_y^4$$
$$+ \left(3 j^4 \left(J_{yy8} + 4K_{yy7}\right) L_x^4 + i^2 j^2 \left(3J_{xx8} + 3J_{yy6} + 4K_{xx7} - 4K_{yy5}\right) L_x^2 L_y^2\right.$$
$$\left. + i^4 \left(3J_{xx6} + 4K_{xx5}\right) L_y^4\right)(2\cos(j\pi) + 1)\Big)\Bigg),$$

$$c_{11} = \frac{1}{L_x^3 L_y^3 m(-4j^2 + m^2) n(-4i^2 + n^2)}\Big(8 i^2 j^2 \Big(L_x^2 m^2 \Big(i^2 \left(2J_{yy6} - K_{xx7}\right) L_y^2$$
$$+ L_x^2 \left(j^2 \left(2J_{yy8} + 3K_{yy7}\right) - K_{yy7} m^2\right)\Big)$$
$$+ L_y^2 \left(j^2 \left(2J_{xx8} - K_{yy5}\right) L_x^2 + i^2 \left(2J_{xx6} + 3K_{xx5}\right) L_y^2\right) n^2 - K_{xx5} L_y^4 n^4\Big)\pi^2\Big),$$

$$c_{12} = \frac{1}{L_x^3 L_y^3 m(-4j^2 + m^2) n(-4i^2 + n^2)}\Big(8 i^2 j^2 \Big(L_x^2 m^2 \Big(i^2 \left(2J_{xx8} - K_{xx7}\right) L_y^2$$
$$+ L_x^2 \left(j^2 \left(2J_{yy8} + 3K_{yy7}\right) - K_{yy7} m^2\right)\Big)$$
$$+ L_y^2 \left(j^2 \left(2J_{yy6} - K_{yy5}\right) L_x^2 + i^2 \left(2J_{xx6} + 3K_{xx5}\right) L_y^2\right) n^2 - K_{xx5} L_y^4 n^4\Big)\pi^2\Big),$$

$$c_{13} = -\left(\frac{\Big(i^2 L_y^2 \left(J_{xx7} L_x^2 m^2 + J_{xx5} L_y^2 n^2\right) + j^2 \left(J_{yy7} L_x^4 m^2 + J_{yy5} L_x^2 L_y^2 n^2\right)\Big)\pi^4}{16 L_x^3 L_y^3}\right),$$

$$c_{14} = \frac{1}{c_3^*}\Bigg(2ijklmn\Bigg((-i^2) K_{xx7} l^2 L_x^2 L_y^2 + i^2 k^2 K_{xx5} L_y^4 - k^4 K_{xx5} L_y^4 + 2J_{yy8} l^2 L_x^4 m^2 + 2J_{xx8} k^2 L_x^2 L_y m^2$$
$$- i^2 K_{xx7} L_x^2 L_y^2 m^2 - K_{yy7} L_x^4 (l^2 - m^2)^2$$
$$+ L_y^2 \left(2J_{yy6} l^2 L_x^2 + \left(i^2 K_{xx5} + 2k^2 \left(J_{xx6} + K_{xx5}\right)\right) L_y^2\right) n^2 - K_{xx5} L_y^4 n^4$$
$$+ j^2 L_x^2 \Big(K_{yy7} L_x^2 (l^2 + m^2) + L_y^2 \Big(i^2 \left(K_{xx7} + K_{yy5}\right) - K_{yy5}(k^2 + n^2)\Big)\Big)\Bigg)\pi^2 \big(-1$$
$$+ (-1)^{j+l+m}\big)\left(-1 + (-1)^{i+k+n}\right)\Bigg),$$

$$c_{15} = \frac{1}{ij L_x^3 L_y^3 (j^2 - 4m^2)(i^2 - 4n^2)}\Big(8 m^2 n^2 \Big(i^2 L_y^2 \left(L_x^2 \left(j^2 \left(K_{xx7} + K_{yy5}\right) - 2K_{xx7} m^2\right) + 2K_{xx5} L_y^2 n^2\right)$$
$$+ 2\Big(L_x^4 m^2 \left(j^2 K_{yy7} + J_{yy8} m^2\right) + L_x^2 L_y^2 \Big((-j^2) K_{yy5} + \left(J_{xx8} + J_{yy6}\right) m^2\Big) n^2$$
$$+ J_{xx6} L_y^4 n^4\Big)\Big)\pi^2\Big),$$

$$c_{16} = -\frac{I_0 L_x L_y}{4} - \frac{I_2 \left(j^2 L_x^2 + i^2 L_y^2\right)\pi^2}{4 L_x L_y},$$

$$c_{17} = \frac{4 I_1 i L_y \left(-1 + \cos(i\pi)^3\right)\left(2 + \cos(j\pi)\right)}{9 j L_x} - \frac{4 I_1 j L_x \left(2 + \cos(i\pi)\right)\left(4 + 2\cos(j\pi)\right)}{9 i L_y},$$

$$c_{18} = \frac{-8I_1 i^2 j^2 \left(L_x^2 m^2 + L_y^2 n^2\right)}{L_x L_y m(-4j^2 + m^2)n(-4i^2 + n^2)},$$

$$c_{19} = \frac{8I_1 m^2 n^2 \left(L_x^2 (j^2 - 2m^2) + L_y^2 (i^2 - 2n^2)\right)}{ijL_x L_y (j^2 - 4m^2)(i^2 - 4n^2)},$$

$$c_{20} = \frac{2I_1 ijklmn \left(j^2 L_x^2 - L_x^2 (l^2 + m^2) + L_y^2 (i^2 - k^2 - n^2)\right)\left(-1 + (-1)^{j+l+m}\right)\left(-1 + (-1)^{i+k+n}\right)}{L_x L_y (j - l - m)(j + l - m)(j - l + m)(j + l + m)(i - k - n)(i + k - n)(i - k + n)(i + k + n)}.$$

# Appendix B.

We refer to the impressive attempt of Kothare and Morari [194] to briefly investigate the multipliers theory for providing a less conservative tool for analyzing the closed-loop system in presence of sector nonlinearity. The interpretation is to consider the stability of a static anti-windup compensator in which the controller's output and input are conditioned with design matrices $\Lambda_1$ and $\Lambda_2$ for $\widehat{\mathbf{P}}(s)$ in (B.1). Accordingly, instead of analyzing the stability of the original system, for the sector nonlinearity of Case 2, a weighted transformed system, the stability of which guarantees the stability of the original configuration is analyzed based on the passivity theorem. Without going into the details of selecting the *stability multipliers*, for a rearranged form of Figure 7.1 as shown in Figure B.1, the stability condition can be obtained.

$$\widehat{\mathbf{P}}(s) = \begin{bmatrix} P_{yw} & P_{y\hat{u}} \\ P_{yw} & P_{y\hat{u}} \\ 0 & P_{u_m\hat{u}} \end{bmatrix},$$

$$M = \begin{bmatrix} \left(V(s) - \mathcal{J}_{n_u}\right)P_{u_m\hat{u}}(s) - U(s)P_{y\hat{u}}(s) & U(s)P_{y\hat{u}}(s) \\ -P_{y\hat{u}}(s) & P_{yw}(s) \end{bmatrix}, \quad \text{(B.1)}$$

where $u_m$ is the measure or estimation of $\hat{u}$.

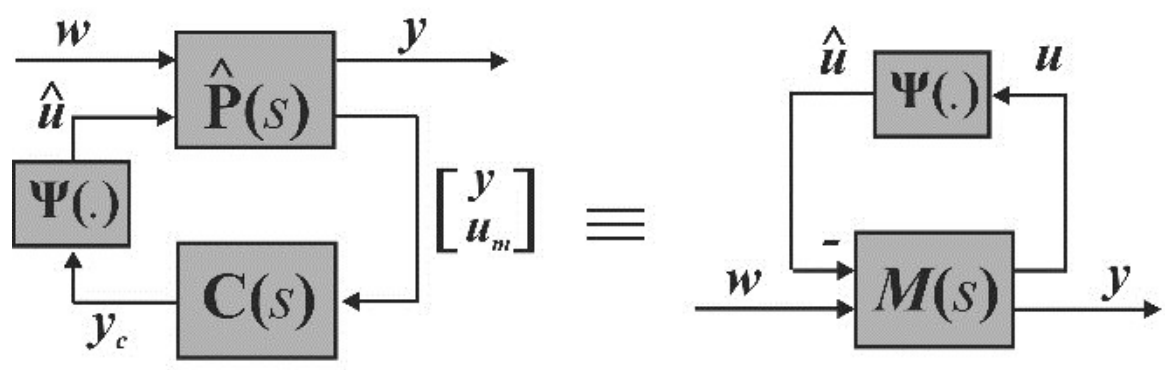

Figure B.1 Rearranged form of Figure 7.1 for stability analysis based on multipliers theory.

# Appendix C.

Assume a square matrix $\mathcal{V} \in \mathbb{R}^{n\times n}$ with its elements denoted as $v_{ij}$, $i, j = 1, \dots, n$. Then, matrix $\mathcal{V}$ is referred to as strictly diagonally dominant if the following holds for all $i = 1, \dots, n$.

$$v_{ii} > \sum_{\substack{j=1 \\ j\neq i}}^{n} |v_{ij}|, \tag{C.1}$$

For more details refer to [321].

# Appendix D.

**Definition D.1** [322] Linear complementarity problem (LCP) is defined over a known vector $q \in \mathbb{R}^l$ and a known matrix $M \in \mathbb{R}^{l \times l}$ in terms of finding a solution for $z \succcurlyeq 0 \in \mathbb{R}^l$ satisfying

$$\begin{aligned} q + Mz &\succcurlyeq 0, \\ z^T(q + Mz) &= 0. \end{aligned} \tag{D.1}$$

LCP$(q, M)$ is called feasible if the first (D.1) is satisfied regardless of the second (D.1) and is called solvable if both equations are simultaneously satisfied.

**Definition D.2** (multiple LCP (MLCP) [323]) For given matrices $\mathcal{M}_{11} \in \mathbb{R}^{m \times m}$, $\mathcal{M}_{12} \in \mathbb{R}^{m \times l}$, $\mathcal{M}_{21} \in \mathbb{R}^{l \times m}$, and $\mathcal{M}_{22} \in \mathbb{R}^{l \times l}$ and two vectors $a \in \mathbb{R}^m$ and $b \in \mathbb{R}^l$, a solution for vectors $v \in \mathbb{R}^m$ and $\lambda \succcurlyeq 0 \in \mathbb{R}^l$ satisfying

$$\begin{aligned} \mathcal{M}_{11} v + \mathcal{M}_{12} \lambda + a &= 0, \\ \mathcal{M}_{21} v + \mathcal{M}_{22} \lambda + b &\succcurlyeq 0. \\ \lambda^T(\mathcal{M}_{21} v + \mathcal{M}_{22} \lambda + b) &= 0. \end{aligned} \tag{D.2}$$

For nonsingular $\mathcal{M}_{11}$ as it is pointed out in Adegbege and Heath [200], MLCP is convertible to LCP.

# Appendix E.

Let's define the revised feedback/feedforward ideal control law as $u^* = Ke + B^+\dot{x}_d - B^+Ax_d$, with $B^+ = B^T(BB^T)^{-1}$. Then, the actual implemented control law $u$ can be parametrized in terms of $u^*$ and the derivation from the ideal control law for the RWNN control law as $u = u^* + (u_{nc}^* - u_{tc} - u_{nc})$. Defining the tracking error as before, its dynamics can be represented as

$$\dot{e} = (A - BK)e - h(x,u) - Hd - \omega(x) + B(u_{nc}^* - u_{tc} - u_{nc}). \tag{E.1}$$

In comparison to the previous results, the terms $\dot{x}_d - Ax_d$ disappear in the tracking error dynamics based on the revised ideal control law. It should be noted that in the output regulation problem of AVC $\dot{x}_d = x_d = 0$. Then, the revised tracking error dynamics is given as

$$\dot{e} = (A - BK)e - h(x,u) - Hd - \omega(x) + B[\tilde{W}^T\hat{\Phi} + \hat{W}^T\Phi_m^T\tilde{m} + \hat{W}^T\Phi_\sigma^T\tilde{\sigma} + \hat{W}^T\Phi_r^T\tilde{r} + \varepsilon - u_{tc}]. \tag{E.2}$$

Next, by defining a new Lyapunov function as

$$\begin{aligned} V = {} & \frac{1}{2}e^TP_1e + \frac{1}{2\eta_w}tr(\tilde{W}^T\tilde{W}) + \frac{1}{2\eta_m}\tilde{m}^T\tilde{m} + \frac{1}{2\eta_\sigma}\tilde{\sigma}^T\tilde{\sigma} + \frac{1}{2\eta_r}\tilde{r}^T\tilde{r} + \frac{1}{2\eta_E}\tilde{E}^T\tilde{E} \\ & + \int_0^\infty (e^TP_2e - \gamma_1\omega^T\omega)dt + \int_0^\infty (e^TP_3e - \gamma_2 d^Td)dt \\ & + \int_0^\infty (e^TP_4e - \gamma_2 h^Th)dt, \end{aligned} \tag{E.3}$$

and derivating w.r.t. time, and substituting the learning rules from (10.28)-(10.31), the bound estimator/compensator from (10.32) and (10.33), we obtain

$$\begin{aligned} \dot{V} \leq {} & \frac{1}{2}e^T[(A^TP_1 + P_1A - K^TB^TP_1 - P_1BK + 2P_2 + 2P_3 + 2P_4)e \\ & -d^T(H^TP_1)e - e^T(P_1H)d - h^T(P_1)e - e^T(P_1)h - \omega^T(P_1)e \\ & -e^T(P_1)\omega] - \gamma_1\omega^T\omega - \gamma_2 d^Td - \gamma_3 h^Th + \beta, \end{aligned} \tag{E.4}$$

where $\beta$ is calculated according to(10.41). Defining $\mathcal{L}^T = [e^Td^Th^T\omega^T], \dot{V} \leq \frac{1}{2}\mathcal{L}^T\Gamma\mathcal{L} + \beta$ , where

$$\Gamma = \begin{bmatrix} A^TP_1 + P_1A - K^TB^TP_1 - P_1BK + 2P_2 + 2P_3 + 2P_4 & -P_1H & -P_1 & -P_1 \\ -H^TP_1 & -2\gamma_2 I & 0 & 0 \\ -P_1 & 0 & -2\gamma_3 I & 0 \\ -P_1 & 0 & 0 & -2\gamma_1 I \end{bmatrix} < 0. \tag{E.5}$$

Since the matrix inequality constraint is nonlinear in parameters, we introduce the linear matrix equality $P_1B = B\hat{P}_1$, while using the transformation $\hat{K} = \hat{P}_1K$ which results in (10.43). The feasibility of LME (10.44) is guaranteed by full column rank condition over $B_n$ [71]. The LMI condition (10.43) permits enforcing the negative definiteness condition in comparison to (10.37) and this completes the proof.

# Appendix F.

In order to arrange a ready-to-use interface for the reader, the revised form of standard UAMP is broken into some script blocks. This increases the readability of the script in comparison to a batch entrance. Let's name the revised UAMP subroutine "general block" which starts with general block **part 1** in Figure F.1.

general block **part 1**

```
      Subroutine uamp(
C        passed in variables
     *   ampName, time, ampValueOld, dt, nProps, props, nSvars, svars,
     *     lFlagsInfo, nSensor, sensorValues, sensorNames,
     *     jSensorLookUpTable,
C        to be defined (if needed)
     *     ampValueNew,
     *     lFlagsDefine,
     *     AmpDerivative, AmpSecDerivative, AmpIncIntegral,
     *     AmpIncDoubleIntegral)
      include 'aba_param.inc'
```

Figure F.1 The general FORTRAN subroutine's first block: subroutine declaration and UAMP's variable definition

For brevity, the standard UAMP is not explained here, and the interested reader may refer to ABAQUS user subroutine reference manual. Next, in contrast to standard UAMP, an interface should be defined including the signatures of some functions and additional subroutines. The interface in FORTRAN is declared with the `interface` keyword after general block **part 1**, and then ends before defining the class of `sensorValues` in the program block of Figure F.2 by `end interface` keyword. This block in the rest of the appendix is referred to as "interface block".

interface block **part 1**

```
interface
    function ENGOPEN (command) bind(C,name="ENGOPEN")
       integer(INT_PTR_KIND()) :: ENGOPEN
       character, dimension(*), intent(in) :: command
    end function ENGOPEN
end interface
```

Figure F.2 The first part of the interface block as the language-binding-spec attribute

At this point, the MATLAB engine should be called inside the main subroutine, and since the FORTRAN compiler is used for ABAQUS/CAE kernel, any function/subroutine inside the interface block should be bound with the language-binding-spec attribute, using the keyword `bind`. Such an entity in the FORTRAN processor is treated as its conforming object in the companion C compiler. Note that the engine applications in the visual studio (VS) environment are straightforward to be compiled. However, the harvested features of VS in ABAQUS require the interface block which serves as the recognition platform between the case-sensitive commands of FORTRAN and C. As an example, `ENGOPEN` routine is presented as the first entity to interface block in Figure F.2.

Such a function needs the declaration of the class of its input and output variables. To increase the readability, the rest of the interface block is presented at the end of the Appendix. After ending the interface block, the variables together with their classes are defined in a similar manner to MATLAB function definition syntax and then followed by general block **part 2** which is partially shown in Figure F.3.

general block **part 2**

```
! Time vector parameters
double precision timestep1, timestep2
C     time indices
        parameter (iStepTime        = 1,
     *           iTotalTime       = 2,
     *           nTime            = 2)
timestep1 = time(iStepTime)
timestep2 = timestep1 + dt
```

Figure F.3 General block for time indices and various information flags

The general block **part 2** may contain the time indices, definition/activation of various information flags, the definition of sensor values at the end of the previous increment (`sensorValues`), and description of the array of the solution-dependent state variables (`svars`). These variables should be written in FORTRAN language intrinsic data types. Then, `iGetSensorID('SENUi', jSensorLookUpTable)` would deliver the user-defined solution-dependent state variables which should be passed through MATLAB engine.

Note that independent of the numerical example Figure 12.1, the current steps are general for other applications mentioned in the introduction section as long as they can be formulated in terms of a time-dependent step module of ABAQUS with a series of external loads or time-varying boundary conditions. At this point, the MATLAB engine is opened, a double array of the desired size is defined (`forvar`) for the variables that are going to be processed in the MATLAB engine (`matlabsession`), a time vector is created in MATLAB, and the vector is copied from FORTRAN array to MATLAB array using `MXGETPR`. This completes the general block **part 3** as illustrated in Figure F.4.

general block **part 3**

```
integer*8 matlabsession
matlabsession = ENGOPEN('matlab')
! Check point 1:
T = MXCREATEDOUBLEMATRIX(Mi, Ni, 0)
Call MXCOPYREAL8TOPTR(forvar, MXGETPR(T), Ni)
status = ENGPUTVARIABLE(ep, 'TT', T)
if (status .ne. 0) then
   write(6,*) 'ENGPUTVARIABLE failed: Check point 1'
   stop
endif
```

Figure F.4 The general program for opening the engine and evaluating the MATLAB function for the vector defined in UAMP subroutine

Next, a MATLAB function (m-file) that controls the execution of the online post-processing algorithm is called. This function (`matlabfunc`) is a gateway to the MATLAB toolboxes such as signal processing, control design, fuzzy systems, etc. The m-file initiates the commands at the beginning of each time increment while the ABAQUS/CAE kernel is in a wait state. User-defined functions can be called only if the current directory includes a copy of the function, however, since the MATLAB engine still needs to be called from ABAQUS, the only possibility is to add the directory containing the m-file to the permanent MATLAB directory using “pathtool”. Since the function called in the global MATLAB directory cannot save the variables, it is recommended to create a dummy vector in UAMP and assign the generated results from MATLAB iteratively to its elements. This action has three advantages: 1) the simulations can be terminated using some if-conditions on `ampValueNew` (see Figure F.1) in the case of violation of a physical constraint or performance index. In the application of AVC, as shown in the numerical example of Figure 12.1, this criterion can be the maximum control effort generated from the controller exceeding the piezo-patch depolarization voltage or maximum displacement of a shaker baffle. 2) During the simulation, the visualization module can be executed over the resulted “.odb” file for observing the generated amplitudes in UAMP. 3) An additional advantage is the possibility to have access to these variables in Tecplot Software.

At this point, the MATLAB function is called from FORTRAN subroutine, the resulted array in MATLAB is read by the messenger of FORTRAN interface, and the obtained array is copied from MATLAB array to FORTRAN array as in general block **part 4** in Figure F.5.

general block **part 4**

```
! Check point 2:
if (ENGEVALSTRING(matlabsession, 'out1 = matlabfunc(TT);') .ne. 0) then
   write(6,*) 'ENGEVALSTRING failed: check point 2'
   stop
endif
out2 = ENGGETVARIABLE(matlabsession, 'out1')
call MXCOPYPTRTOREAL8(MXGETPRS(out2), out3, No)
```

Figure F.5 The general program for bringing back the MATLAB function's output using the FORTRAN messenger

It is evident that variables `out1`, `out2`, and `out3` are classified in general block **part 2** (suppressed for the sake of brevity). Finally, the amplitude is updated in the FORTRAN subroutine based on the values of `out3`, and the array is deallocated as `call MXDESTROYARRAY(T)` for the new increment.

Additional notes:

- Since the routines in the interface block are mixed-case, the declarations in Fig. F.6 are needed before the interface.

  Pre-interface block

  ```
  !DEC$ OBJCOMMENT LIB:"libeng.lib"
  !DEC$ OBJCOMMENT LIB:"libmx.lib"
  !DEC$ OBJCOMMENT LIB:"libmat.lib"
  ```

  Figure F.6 Intel-style pre-processing directives

  where the `OBJCOMMENT LIB` directive postulates a library in an object heading. In this case, the "linker" looks for the character constant `.lib` by the `OBJCOMMENT` directive in the command line of the script. The reason behind this obligatory declaration is that FORTRAN is not case-sensitive. However C is, and the routines in MATLAB libraries are mixed-case. As a result, Intel-style pre-processing directives are required. The three libraries mentioned above are available by installing the MATLAB software.
- If C code correctly links with the external libraries but "unresolved external symbol reference" error appears, assembler output should be checked. For Microsoft Windows operating systems (OS), some changes should be applied: low case, up case, and mixed case. A list of such changes is available for the linker in the header file "fintrf.h" provided by Mathworks which contains the declaration of the pointer type needed by the MATLAB/FORTRAN interface. Since this header file is not readable for the ABAQUS compiler, those changes should be found and applied manually from the assembler output. One should note that only the Windows linker has this feature. For instance, in this appendix, `mxCopyReal8ToPtr` is replaced with `MXCOPYREAL8TOPTR730` along with some other modifications. These changes are different from one OS to another. This is a common approach when one needs to write an assembly program to interface with a C application using decorated function names. Then, in order to figure out the problematic name mangling, an empty shell subroutine in C is recommended to be written. The produced assembly output gives the correct form to be referred to in subroutine that should be applied manually.

As it can be seen in the interface block **part 2**, additional subroutines are `MXDESTROYARRAY`, `MXCOPYREAL8TOPTR`, and `MXCOPYPTRTOREAL8`, which all require `DECORATE` attribute on a mixed-language application. Additionally, unlike the functions, no return values should be assigned to them.

interface block **part. 2**

```
      function MXCREATEDOUBLEMATRIX (a1,b1,c1)
     *         bind(C,name="MXCREATEDOUBLEMATRIX")
     integer*8 :: a1,b1,c1
     integer*8 :: MXCREATEDOUBLEMATRIX
     intent(in) :: a1,b1,c1
     end function MXCREATEDOUBLEMATRIX

     function MXCREATEDOUBLESCALAR (a2)
     *         bind(C,name="MXCREATEDOUBLESCALAR")
     real*8 :: a2
     integer*8 :: MXCREATEDOUBLESCALAR
     intent(in) :: a2
     end function MXCREATEDOUBLESCALAR
```

```
        Subroutine MXDESTROYARRAY (a3)
!DEC$ ATTRIBUTES DECORATE, ALIAS:"MXDESTROYARRAY" :: MXDESTROYARRAY
        integer*8, dimension(*) :: a3
        end Subroutine MXDESTROYARRAY

        function MXGETPR(a4) result(ptr) bind(C, name='MXGETPR')
        import
        implicit none
        integer*8, dimension(*), intent(in) :: a4
        integer*8 :: ptr
        end function MXGETPR

        Subroutine MXCOPYREAL8TOPTR (a5,b5,c5)
!DEC$ ATTRIBUTES DECORATE, ALIAS:"MXCOPYREAL8TOPTR" :: MXCOPYREAL8TOPTR
        real*8 a5(*)
        integer*8 b5
        integer*8 c5
        end Subroutine MXCOPYREAL8TOPTR

        function ENGPUTVARIABLE (a6,b6,c6)
     *          bind(C,name="ENGPUTVARIABLE")
        integer*8, intent(in) :: a6
        character, dimension(*), intent(in) :: b6
        integer*8, dimension(*), intent(in) :: c6
        end function ENGPUTVARIABLE

        function ENGEVALSTRING (a7,b7)
     *          bind(C,name="ENGEVALSTRING")
        integer(INT_PTR_KIND()) :: ENGEVALSTRING
        integer*8, intent(in) :: a7
        character, dimension(*), intent(in) :: b7
        end function ENGEVALSTRING

        function ENGGETVARIABLE (a8,b8)
     *          bind(C,name="ENGGETVARIABLE")
        integer*8 :: ENGGETVARIABLE
        integer*8, intent(in) :: a8
        character, dimension(*), intent(in) :: b8
        end function ENGGETVARIABLE

        Subroutine MXCOPYPTRTOREAL8 (a9,b9,c9)
!DEC$ ATTRIBUTES DECORATE, ALIAS:"MXCOPYPTRTOREAL8" :: MXCOPYPTRTOREAL8
        integer*8 a9
        real*8 b9
        integer*8 c9
        end Subroutine MXCOPYPTRTOREAL8

        function engClose (a10) bind(C,name="engClose")
        integer(INT_PTR_KIND()) :: engClose
        integer*8, intent(in) :: a10
        end function engClose
```

Figure F.7 The second part of the interface block after Figure F.2

# Appendix G

In this appendix, the numerical values of the material properties of the piezo-actuated panel in Chapter 12.3.1.3 are given. For the host structure:

$$C_h = \begin{bmatrix} 28.3 & 12.1 & 12.1 & 0 & 0 & 0 \\ 12.1 & 28.3 & 12.1 & 0 & 0 & 0 \\ 12.1 & 12.1 & 28.3 & 0 & 0 & 0 \\ 0 & 0 & 0 & 8.1 & 0 & 0 \\ 0 & 0 & 0 & 0 & 8.1 & 0 \\ 0 & 0 & 0 & 0 & 0 & 8.1 \end{bmatrix} [\mathrm{GPa}],$$

and the piezo-actuator/sensor (density: $\rho_a = 5300 \frac{\mathrm{kg}}{\mathrm{m}^3}, \rho_s = 7500 \frac{\mathrm{kg}}{\mathrm{m}^3}$)

$$C_a = \begin{bmatrix} 23.9 & 10.4 & 5.2 & 0 & 0 & 0 \\ 10.4 & 24.7 & 5.2 & 0 & 0 & 0 \\ 5.2 & 5.2 & 13.5 & 0 & 0 & 0 \\ 0 & 0 & 0 & 6.5 & 0 & 0 \\ 0 & 0 & 0 & 0 & 6.6 & 0 \\ 0 & 0 & 0 & 0 & 0 & 7.6 \end{bmatrix} \times 10[\mathrm{GPa}]$$

$$C_s = \begin{bmatrix} 13.9 & 7.8 & 7.43 & 0 & 0 & 0 \\ 7.8 & 13.9 & 7.43 & 0 & 0 & 0 \\ 7.43 & 7.43 & 11.5 & 0 & 0 & 0 \\ 0 & 0 & 0 & 2.56 & 0 & 0 \\ 0 & 0 & 0 & 0 & 2.56 & 0 \\ 0 & 0 & 0 & 0 & 0 & 3.06 \end{bmatrix} \times 10[\mathrm{GPa}]$$

$$E_a = \begin{bmatrix} 4.3 & 0 & 0 \\ -0.4 & 0 & 0 \\ -0.3 & 0 & 0 \\ 0 & 0 & 0 \\ 0 & 0 & 2.8 \\ 0 & 3.4 & 0 \end{bmatrix} [\mathrm{C/m^2}] \quad E_s = \begin{bmatrix} 15.1 & 0 & 0 \\ -5.2 & 0 & 0 \\ -5.2 & 0 & 0 \\ 0 & 0 & 0 \\ 0 & 0 & 12.7 \\ 0 & 12.7 & 0 \end{bmatrix} [\mathrm{C/m^2}]$$

$$\epsilon_a = \begin{bmatrix} 1.96 & 0 & 0 \\ 0 & 2.01 & 0 \\ 0 & 0 & 0.28 \end{bmatrix} [\mathrm{nF/m}] \quad \epsilon_s = \begin{bmatrix} 6.5 & 0 & 0 \\ 0 & 6.5 & 0 \\ 0 & 0 & 5.6 \end{bmatrix} [\mathrm{nF/m}]$$

where $E_a, E_s, \epsilon_a$, and $\epsilon_s$ represent the piezoelectricity matrices (stress coefficients) and the dielectric matrices for actuator and sensor, respectively.

# About the author

**Personal information**

Name: Atta Oveisi

Birth: 13/09/1988 in Tabriz, Iran

Nationality: German-Iranian

**Educational history**

04/2014-04/2021: Ph.D. candidate in Mechanical Engineering, Ruhr University Bochum, Germany.

09/2010-03/2013: M.Sc. in Mechanical Enginnering, Iran University of Science and Technology, Tehran, Iran.

09/2006-09/2010: B.Sc. in Mechanical Enginnering, University of Tabriz, Tabriz, Iran.

**Work history**

01/2019-Now: Development engineer, Daimler AG, Stuttgart, Germany.

03/2014-12/2018: Scientific assistant, Ruhr University Bochum, Germany.